\documentclass[11pt,twoside,openright]{book}
\usepackage[english]{babel}
\usepackage[utf8x]{inputenc}
\usepackage{breakurl}
\usepackage{bold-extra}
\usepackage{float}
\usepackage{graphicx}
\usepackage{longtable}
\usepackage{lscape}
\usepackage{amssymb}
\usepackage[nointegrals]{wasysym} % symbols
\usepackage{gensymb}
\usepackage{natbib}
\usepackage{fancyhdr}
\usepackage{framed}
\usepackage{subcaption}
\usepackage{minibox}
\usepackage{textcomp}
\usepackage{makeidx}
\usepackage{longtable}
\usepackage{wrapfig}
\usepackage{booktabs}
\usepackage{geometry}
\usepackage{gensymb}
\usepackage[titletoc,title]{appendix}
\usepackage{color}
\usepackage{url}
\usepackage{fourier-orns}
\usepackage{amsmath}
\usepackage{alltt, xcolor}
\usepackage{scrextend}
\usepackage{blindtext}
\usepackage{enumerate} 
\usepackage{tablefootnote}
\usepackage{txfonts}
\usepackage{ulem}
\usepackage[labelfont=bf]{caption}
\usepackage{lettrine} % Para letras capitulares
\usepackage{lipsum}
\usepackage{multirow}
\usepackage{titlesec}
\usepackage{sectsty} % Para cambiar el font de chapter y numero de chapter
\usepackage[T1]{fontenc}
\usepackage{Alegreya} %% Option 'black' gives heavier bold face 
\usepackage{multicol} % Uso de más de una columna.
\usepackage{booktabs}
\usepackage[referable]{threeparttablex}
\usepackage{tablefootnote}
\usepackage{tabularx}
\usepackage{wasysym} % Símbolos de planetas
\usepackage{hyperref} % Para que las \ref tengan otro nombre
\usepackage{pdfpages}
\usepackage{multicol}
\usepackage{ragged2e}
\usepackage{caption}
\usepackage{xurl}    % Divide url en dos líneas en cualquier caracter
\usepackage{marvosym}
\usepackage{footmisc}
\usepackage{multicol}
\usepackage[printonlyused]{acronym}
\usepackage{cleveref}
\usepackage{color}
\usepackage{courier}
\usepackage{setspace}
\usepackage{subcaption}
\usepackage{afterpage}
\usepackage{changepage}
\usepackage{yfonts}
\usepackage{appendix}
\usepackage{adjustbox}
\usepackage{listings}

\bibpunct{(}{)}{;}{a}{}{,}
\hypersetup{%
  colorlinks = true,
  linkcolor  = blue}
\hypersetup{colorlinks,citecolor=blue}
\counterwithout*{footnote}{chapter} % No resetea numeración footnote cada capítulo
\newcommand{\tR}[1]{\textcolor[rgb]{0.8,0.0,0.0}{#1}}

\newcommand{\tB}[1]{\textcolor[rgb]{0.0,0.0,1.0}{#1}}
\newcommand{\tGray}[1]{\textcolor{gray}{#1}}

\makeindex
\crefname{section}{§}{§§}
\Crefname{section}{§}{§§}
\begin{document}

\bibliographystyle{My_apj} 
 % My_apj es una copia de apj.bst que lleva en comentarios todas las llamadas a la función name.or.dash. Con ello se evita que se ponga un guión en los autores repetidos en la bibliografía.
\begin{titlepage}
\renewcommand{\thefootnote}{\Roman{footnote}} 
{\fontfamily{ptm}\selectfont

%%%%% ANTE-PORTADA %%%%%

\begin{center}
\begin{LARGE}
  \textsc{\textbf{Universidad Complutense de Madrid}}\\\vspace{.4cm}
\end{LARGE}
\begin{Large}
  \textbf{Facultad de Ciencias Físicas}\\\vspace{.4cm}
\end{Large}
\begin{figure} [H]
  \vspace{1cm}
  \centering
  \includegraphics[width=.3 \textwidth]{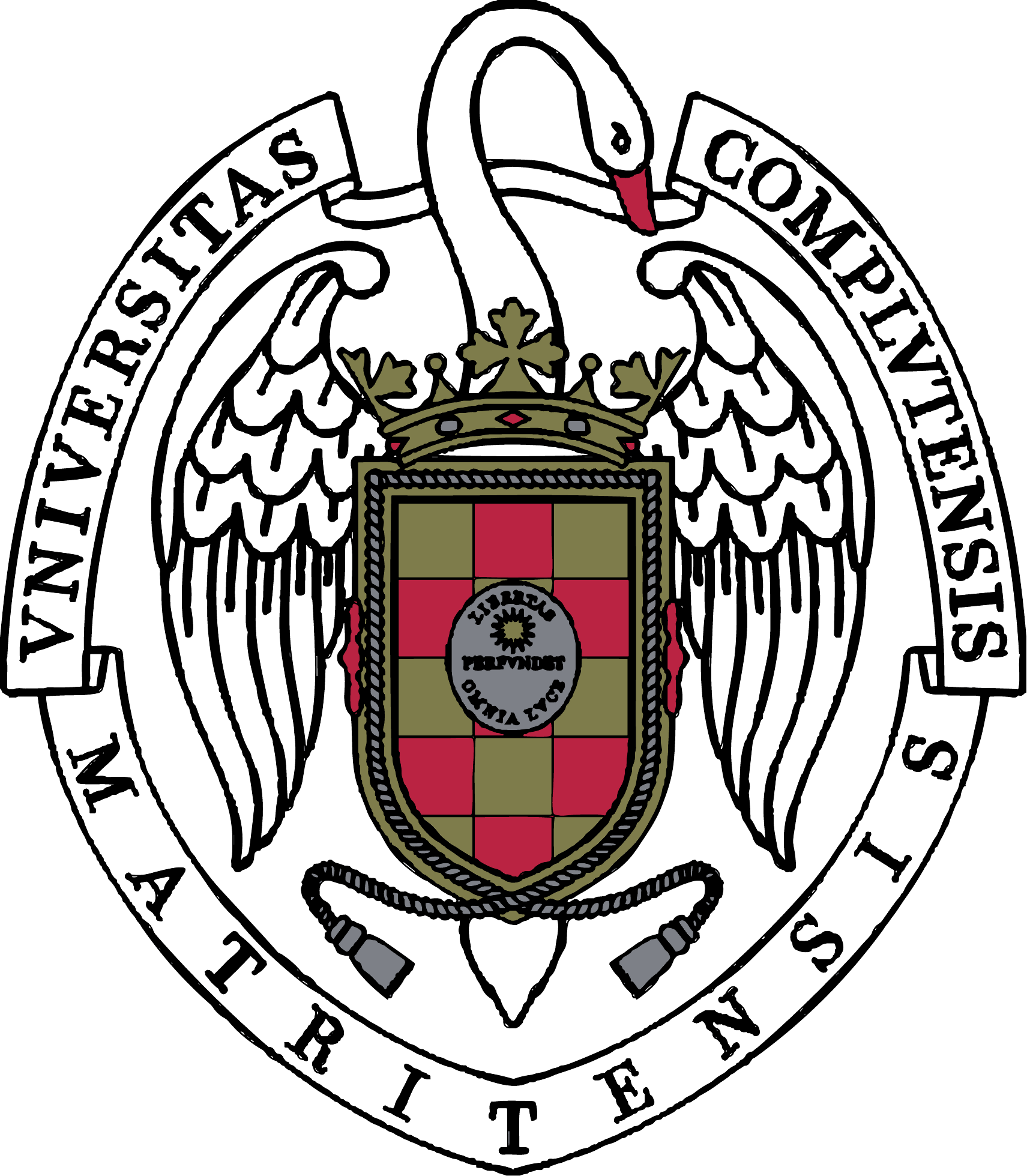}
\end{figure}
\begin{huge}
  \textsc{\textbf{Tesis Doctoral}}\\\vspace{1cm}
\end{huge}
\begin{LARGE}
\textbf{Multiplicidad de sistemas estelares en la vecindad solar,\\binarias de alta separación y estrellas con planetas}\\
\vspace{.8cm}
\textbf{Multiplicity of stellar systems in the solar neighbourhood,\\wide binaries, and planet-hosting stars}\\
\end{LARGE}
\vspace{1.2 cm}
\begin{large}
  Memoria para optar al grado de doctor presentada por:\\\vspace{0.5 cm}
\end{large}
\begin{huge}
  \textbf{Francisco Javier González Payo}\\
\end{huge}
\vspace{0.8 cm}
\begin{Large}
  \textbf{Directores:} \\\vspace{0.4 cm}
  Dr. José Antonio Caballero Hernández\\\vspace{0.2 cm}
  Dra. Miriam Cortés Contreras\\\vspace{0.4 cm}
\end{Large}
\end{center}\vfill 

\clearpage\hbox{}\thispagestyle{empty}\newpage % Deja página en blanco 

}

\end{titlepage}

%%%%% PORTADA %%%%%

\begin{titlepage}
\renewcommand{\thefootnote}{\Roman{footnote}} 
{\fontfamily{ptm}\selectfont

\begin{center}
\begin{LARGE}
  \textsc{\textbf{Universidad Complutense de Madrid}}\\\vspace{.4cm}
\end{LARGE}
\begin{Large}
  \textbf{Facultad de Ciencias Físicas}\\\vspace{0.2cm}
\end{Large}
\begin{figure}[H]
  \vspace{0.4cm}
  \centering
  \includegraphics[width=.25 \textwidth]{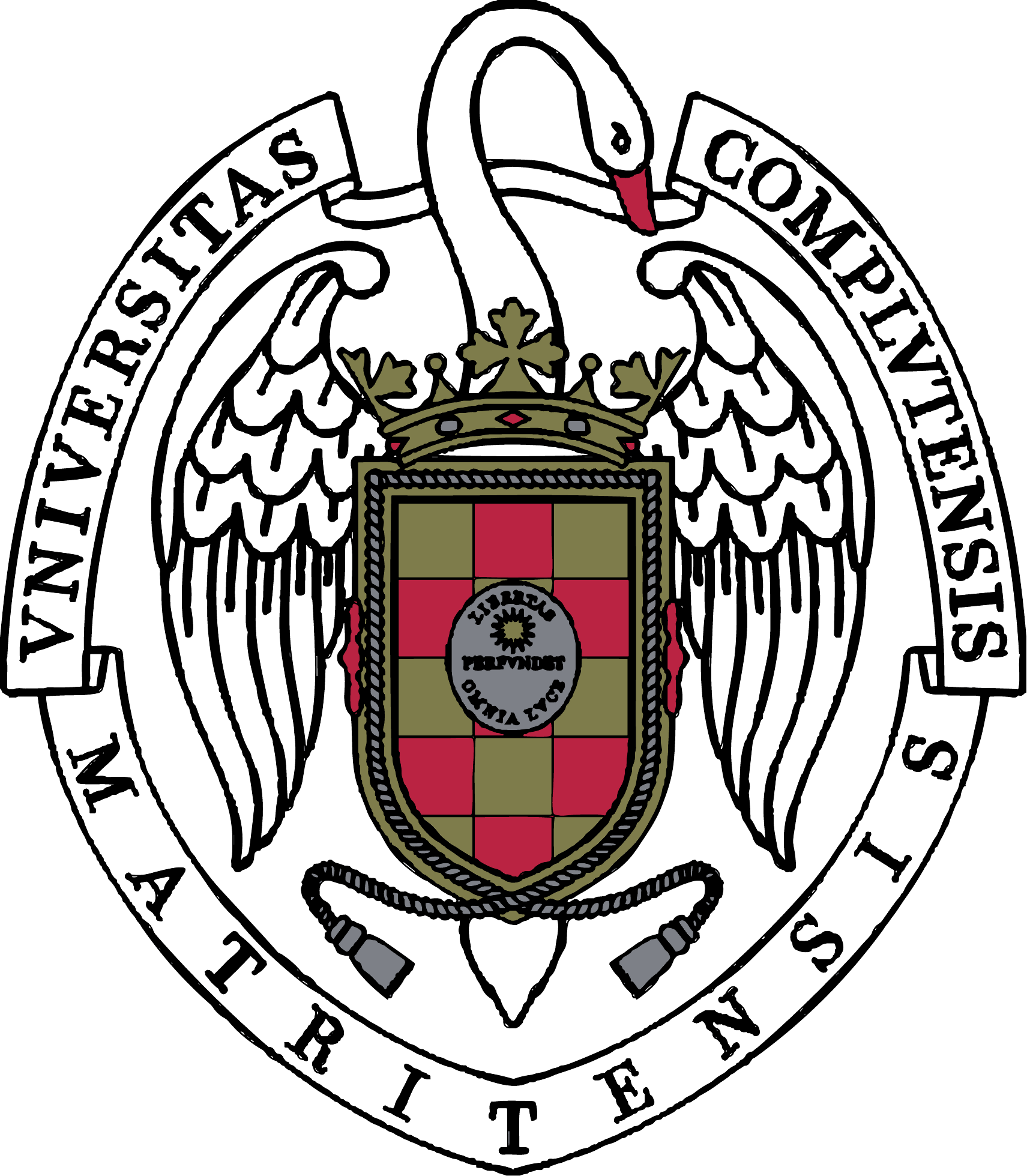}
\end{figure}
\begin{huge}
  \textsc{\textbf{Tesis Doctoral}}\\\vspace{0.3cm}
\end{huge}
\begin{LARGE}
\textbf{Multiplicidad de sistemas estelares en la vecindad solar,\\binarias de alta separación y estrellas con planetas}\\
\vspace{.8cm}
\textbf{Multiplicity of stellar systems in the solar neighbourhood,\\wide binaries, and planet-hosting stars}\\
\end{LARGE}
\vspace{1.2 cm}
\begin{large}
  Programa de Doctorado en Astrofísica\\\vspace{0.4 cm}
  Memoria para optar al grado de doctor presentada por:\\\vspace{0.5 cm}
\end{large}
\begin{huge}
  \textbf{Francisco Javier González Payo}\\
\end{huge}
\vspace{0.7 cm}
\begin{Large}
  \textbf{Directores:} \\\vspace{0.3 cm}
  Dr. José Antonio Caballero\\
  Dra. Miriam Cortés Contreras\\\vspace{0.8 cm}

\textbf{Madrid, Julio 2025}\\[1.5 cm]
\end{Large}
\end{center}\vfill 

}

\end{titlepage}

\clearpage\hbox{}\thispagestyle{empty}\newpage % Deja página en blanco 

\includepdf[pages=-]{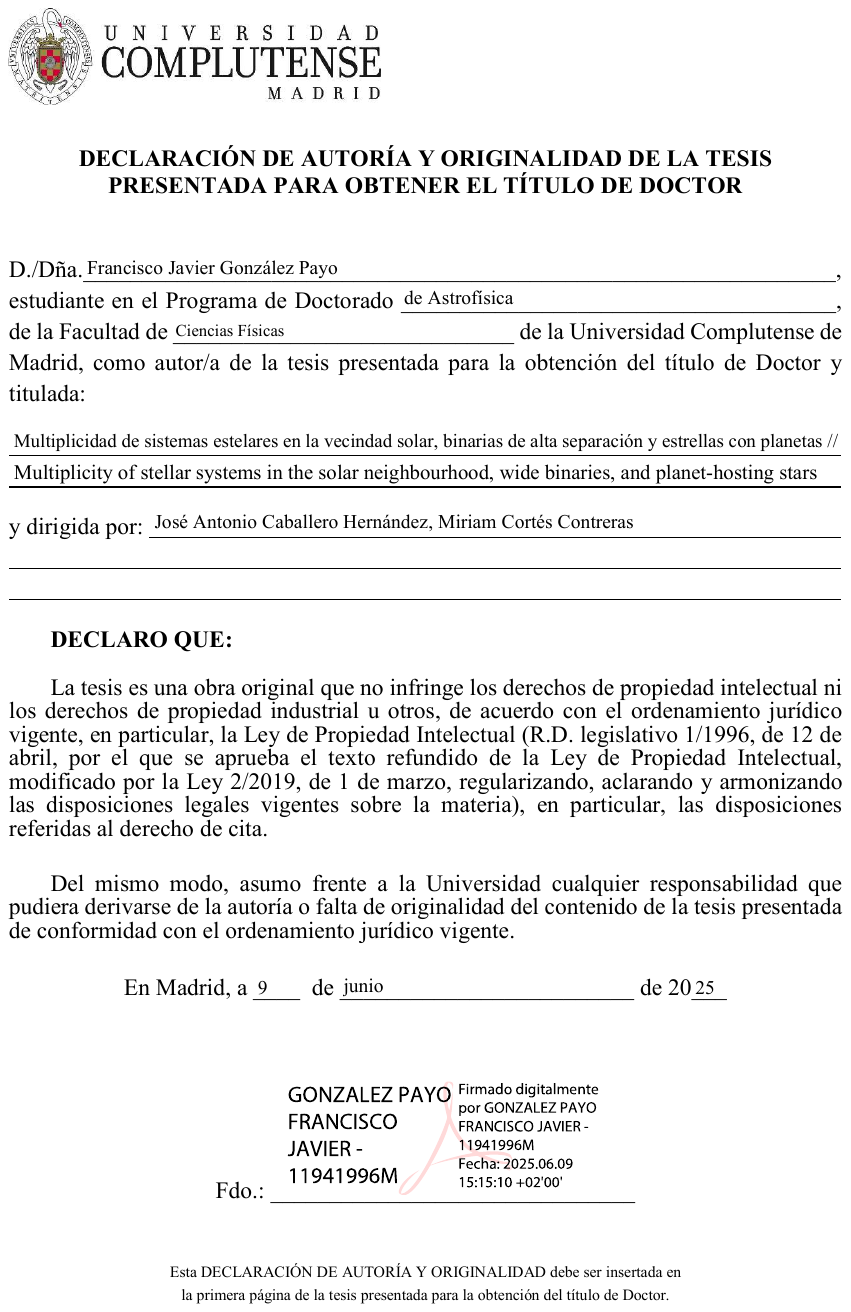}

\clearpage\hbox{}\thispagestyle{empty}\newpage % Deja página en blanco 

\setcounter{page}{1}

\setcounter{footnote}{0}

%%%%% DEDICATORIA %%%%%

\thispagestyle{empty}
\vspace*{7cm}
%\chapter{}
\begin{flushright}
\textit{A mi padre.}
\\
\vspace{0.1cm}
\textit{Desde más allá de las estrellas vela por nosotros.}
\end{flushright}

\clearpage\hbox{}\thispagestyle{empty}\newpage % Deja página en blanco 

\thispagestyle{empty}

\vspace*{4cm}

\begin{figure*}[h]
  \centering
  \includegraphics[width=0.30\linewidth, angle=0]{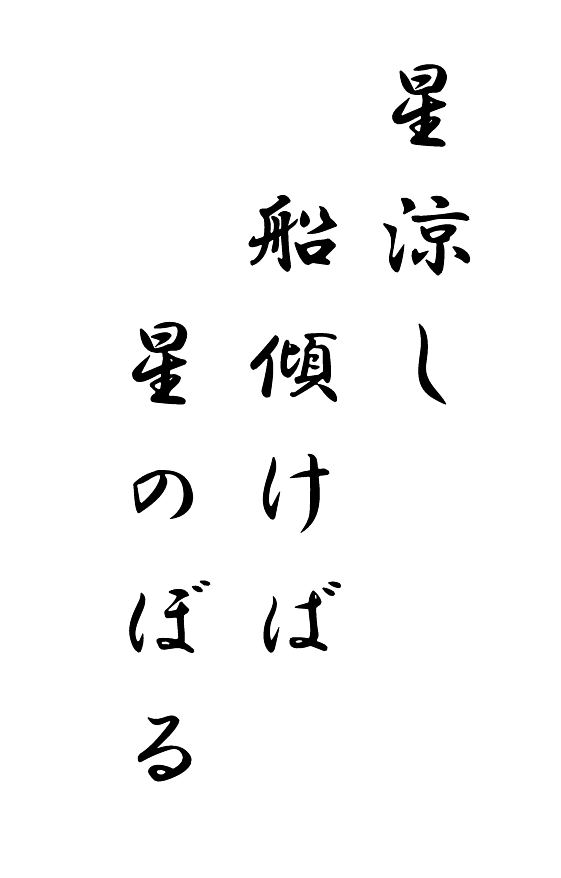}
  \label{haiku}
\end{figure*}
\begin{center}

\textit{If you tilt the boat, you will climb to the stars}\\
\vspace{0.5cm}
Seison Yamaguchi (1892--1988)

\end{center}

%%%%% COMANDOS VARIOS %%%%%

\setlength{\parskip}{2mm} % separación entre párrafos

 \makeatletter
 \newcommand\Desmesurado{\@setfontsize\Desmesurado{25}{42}}
 \makeatother
 \makeatletter
 \newcommand\MasDesmesurado{\@setfontsize\MasDesmesurado{50}{42}}
 \makeatother
  \makeatletter
 \newcommand\Enano{\@setfontsize\Enano{5.4}{5.5}}
 \makeatother

\renewcommand{\chaptername}{}
\titleformat{\chapter}[display]
  {\normalfont\MasDesmesurado\centering\color{black}}{\centering\chaptertitlename\ \thechapter}{80pt}{\bfseries\Desmesurado\color{black}}
\titlespacing*{\chapter} 
  {0pt}{50pt}{40pt}

\hypersetup{linkcolor=blue} % Pone color azul en los links de tablas y figuras.

%%%%% AGRADECIMIENTOS %%%%%

\chapter*{Agradecimientos}    
\label{ch:Agradecimientos}
\vspace{1.2cm}
\pagestyle{fancy}
\fancyhf{}
\lhead[\small{\textbf{\thepage}}]{\textbf{Agradecimientos}}
\rhead[\small{\textbf{Agradecimientos}}]{\small{\textbf{\thepage}}}
\addcontentsline{toc}{chapter}{Agradecimientos}

Son numerosas las personas a las que debo expresar mi enorme agradecimiento por su inmensa contribución a la elaboración de esta tesis. Algunas me han brindado su apoyo técnico y científico, otras su ayuda material o logística, y una tercera categoría, la que más valoro, su respaldo moral y emocional.

En primer lugar, debo reconocer y agradecer la dedicación incansable de mis dos directores de tesis, el Dr. José A. Caballero y la Dra. Miriam Cortés Contreras, quienes han invertido innumerables horas en impulsar este trabajo. Sin su guía y apoyo, aunque pueda parecer un tópico, esta tesis no habría visto la luz. Asimismo, debo mencionar a otras personas que han contribuido significativamente a mejorar los contenidos y publicaciones de este trabajo, como la Dra. María Rosa Zapatero Osorio, la Dra. Maricruz Gálvez Ortiz y el Dr. Carlos Cifuentes. Su orientación y comentarios críticos no solo han mejorado la calidad del contenido, sino que también han aportado nuevas perspectivas que han permitido un análisis más profundo y riguroso. También deseo expresar mi agradecimiento a las revisoras de esta tesis, la Dra. Rachel Matson y la Dra. Céline Reylé, por sus observaciones detalladas y sugerencias constructivas.

Mis compañeros de UNIE, quienes prácticamente consideraban este mi segundo doctorado como un logro compartido, y en parte así es, también merecen mi reconocimiento. En especial, quiero agradecer a la Dra. Ana M. Ruiz Leo por facilitar mi tiempo y ayudarme en lo que he necesitado dentro y fuera de la sala de profesores de la Universidad. 

No puedo dejar de mencionar a mis compañeros estudiantes de doctorado en el Centro de Astrobiología (CAB - CSIC/INTA), por orden alfabético, Abel, Amadeo, Diego, Eva, Jaime, José Luis, Luis, Olga, Pedro y Raquel. Todos ellos, al igual que yo, se han enfrentado a los desafíos que implica realizar un doctorado de calidad y me han acogido como a uno más, a pesar de que nuestra diferencia de edad es de treinta años. 

Mi gratitud también se extiende a todas aquellas personas del CAB que me han brindado su ayuda en diferentes tareas informáticas, como Antonio y Sergio, quienes siempre han estado dispuestos a dedicarme el tiempo que necesitara. Su paciencia, disposición y conocimientos han sido de gran valor para la resolución de problemas técnicos, permitiéndome avanzar con mayor eficacia en el desarrollo de este trabajo. Por supuesto también a Margie, por su valioso apoyo logístico, siempre pendiente de gestionar la autorización de seguridad y de encontrarme un sitio en mis múltiples visitas al CAB.

Por último, quiero dedicar un profundo agradecimiento a la persona más importante en mi vida, Belén, mi mujer, que siempre me ha animado a continuar y me ha facilitado el poder dedicar tantas horas en casa a la elaboración de esta tesis. Su amor y apoyo incondicional han sido fundamentales para superar los retos y dificultades que este trabajo conlleva.
\begin{center}
\aldine
\end{center}

\clearpage\hbox{}\thispagestyle{empty}\newpage % Deja página en blanco 

\hypersetup{linkcolor=black} % Pone color negro en el índice.
\setcounter{tocdepth}{3}

\thispagestyle{plain}
\vspace{9cm}
\lhead[\small{\textbf{\thepage}}]{\textbf{Contents}}
\rhead[\small{\textbf{Contents}}]{\small{\textbf{\thepage}}}
\setlength{\parskip}{0mm}{\tableofcontents}

\newpage

%%%%% ABSTRACT %%%%%

\setlength{\parskip}{2mm} % separación entre párrafos

\large
\thispagestyle{empty}
\begin{center}
\renewcommand{\thefootnote}{\arabic{footnote}}
\textbf{\Large{Abstract}}\\
\end{center}
\normalsize

This doctoral thesis presents a comprehensive study on stellar multiplicity in the solar neighborhood (d\,$<$\,10\,pc) and in multiple systems with planets (d\,$<$\,100\,pc). These systems are characterised using data from the Washington Double Star catalogue, \textit{Gaia} DR3, and a thorough literature review.

One of the main achievements is the creation of the most complete and homogeneous sample of multiple systems within 10\,pc of the Sun. The multiplicity fractions and companion star fractions are determined with reduced uncertainties, achieving greater statistical reliability compared to previous studies, thanks to the completeness of the sample. Additionally, an analysis of orbital periods ranging from one day to millions of years suggests that the log-normal cumulative distribution could be considered a 21st-century revision of Öpik's law, representing a significant contribution within such a complete sample.

Another key advance is the detailed study of wide binary systems ($\rho \ge$ 1000 arcsec) using \textit{Gaia} DR3 astrometry. This work increases the sample size by more than an order of magnitude and improves astrometric precision. Newly identified astrometric companions, not previously catalogued, include ultra-cool dwarfs at the M-L boundary and a hot white dwarf, allowing a better distinction between physical binary systems and unrelated members of young kinematic groups.

The thesis examines the impact of multiplicity on exoplanetary systems within 100\,pc. New companion stars in known planetary systems are identified, and separations for over 200 stellar pairs are measured, collecting key parameters from 276 exoplanets. Compared to a large sample of single stars with exoplanets, significant trends emerge, such as a higher proportion of massive planets with short orbital periods and greater orbital eccentricities in multiple systems. Nearly 22\% of exoplanetary systems have stellar companions, with a significant influence ($>$\,4$\sigma$) from planets with high eccentricities in multiple systems where the projected physical separation is small relative to the planet’s semi-major axis. A slight trend ($>$\,2$\sigma$) is also observed, indicating that high-mass planets ($M>$\,40 M$_\oplus$) in multiple systems tend to orbit closer than those around single stars. These trends contribute to the ongoing debate on the influence of multiplicity on planetary evolution.

Finally, the thesis conducts a historical review of the first observations of multiple stellar systems, ana\-lysing the catalogues of the 17th-century astronomer Giovanni Battista Hodierna. It demonstrates that Hodierna published the first list of multiple systems more than a century earlier than previously believed, redefining the early history of double and multiple star astronomy.

Overall, this research represents a significant advancement in the understanding of multiple stellar systems and their relationship with exoplanets. By combining \textit{Gaia} DR3 data, a meticulous compilation of previous information, and a historical perspective, novel and statistically robust results are obtained, with important implications for models of stellar and planetary formation and evolution. This thesis paves the way for new improved
studies with \textit{Gaia} DR4.

\vfill
\noindent\textbf{Keywords:} techniques: photometric --- astrometry --- surveys --- virtual observatory tools --- subdwarfs --- stars: low-mass --- binaries: general --- binaries: visual --- astronomical databases: miscellaneous ---  planetary systems --- history and philosophy of astronomy \\

%%%%% RESUMEN %%%%%

\large
\clearpage\hbox{}\thispagestyle{empty}\newpage % Deja página en blanco
\setcounter{footnote}{0}
\thispagestyle{empty}
\begin{center}
\textbf{\Large{Resumen}}\\
\end{center}
\normalsize

Esta tesis doctoral presenta un estudio exhaustivo sobre la multiplicidad estelar en la vecindad solar (d\,$<$\,10\,pc) y en sistemas múltiples con planetas (d\,$<$\,100\,pc). Estos sistemas se caracterizan a partir de datos del catálogo Washington Double Star, \textit{Gaia} DR3 y un análisis pormenorizado de la literatura. 

Uno de los principales logros es la elaboración de la muestra más completa y homogénea de sistemas múltiples a menos de 10\,pc del Sol. Se determinan las fracciones de multiplicidad y de estrellas compañeras con incertidumbres reducidas y con una mayor fiabilidad estadística con respecto a estudios previos gracias a la completitud de la muestra. Además, se realiza un análisis de periodos orbitales desde un día a millones de años que sugiere que la distribución acumulativa log-normal del ajuste podría considerarse una revisión de la ley de Öpik del siglo XXI, lo que resulta un aporte ciertamente novedoso en una muestra tan completa.

Otro avance clave es el estudio detallado de sistemas binarios amplios ($\rho \ge$ 1000\,arcsec) utilizando la astrometría de \textit{Gaia} DR3. Este trabajo logra un incremento de más de un orden de magnitud en el tamaño de la muestra y una mejora en la precisión astrométrica. Se identifican nuevos compañeros astrométricos no catalogados previamente, incluyendo enanas ultrafrías en el límite M-L y una enana blanca caliente, lo que permite diferenciar mejor entre sistemas binarios físicos y miembros de grupos cinemáticos jóvenes no relacionados.

La tesis examina el impacto de la multiplicidad en sistemas exoplanetarios dentro de 100 pc. Se identifican nuevas estrellas compañeras en sistemas planetarios conocidos y se miden separaciones para más de 200 pares estelares, recopilando parámetros clave de 276 exoplanetas. Comparando con una muestra numerosa de estrellas simples con exoplanetas, se identifican tendencias significativas, como una mayor proporción de planetas masivos con periodos orbitales cortos en sistemas múltiples y mayor excentricidad orbital. Casi el 22\% de los sistemas exoplanetarios tienen compañeros estelares, con una influencia significativa ($>$\,4$\sigma$) por planetas con altas excentricidades orbitales en sistemas múltiples con pequeñas ratios entre la separación proyectada y el semieje mayor de la órbita planetaria, además de una ligera tendencia ($>$\,2$\sigma$) a que los planetas de alta masa ($M>$\,40 M$_\oplus$) en sistemas múltiples orbiten más cerca que en estrellas individuales. Estas tendencias contribuyen al debate sobre la influencia de la multiplicidad en la evolución planetaria.

Finalmente, este trabajo realiza una revisión histórica de las primeras observaciones de sistemas estelares múltiples, analizando los catálogos del astrónomo del siglo XVII Giovanni Battista Hodierna. Se demuestra que Hodierna publicó la primera lista de sistemas múltiples más de un siglo antes de lo que se creía, redefiniendo la historia temprana de la astronomía de estrellas dobles y múltiples.

En conjunto, esta investigación supone un avance significativo en el conocimiento de los sistemas estelares múltiples y su relación con los exoplanetas. Gracias a la combinación de datos de \textit{Gaia} DR3, una compilación meticulosa de información previa y un enfoque histórico, se obtienen resultados novedosos y estadísticamente sólidos que tienen importantes implicaciones para los modelos de formación y evolución estelar y planetaria. Esta tesis prepara el camino para los nuevos estudios mejorados que se esperan con \textit{Gaia} DR4. 

\vfill
\noindent\textbf{Palabras clave:} técnicas: fotometría --- astrometría --- surveys --- herramientas de observatorio virtual --- estrellas: baja masa --- binarias: general --- binarias: visual --- bases de datos astronómicas: misceláneo --- sistemas planetarios --- historia y filosofía de la ciencia de astronomía \\

\normalsize

\clearpage\hbox{}\thispagestyle{empty}\newpage % Deja página en blanco 

\hypersetup{linkcolor=blue} % Pone color azul en los links de tablas y figuras.

%%%%% CAPITULO 1 %%%%%

\chapter{Introduction} 
\label{ch:introduction}
\vspace{2cm}
\pagestyle{fancy}
\fancyhf{}
\lhead[\small{\textbf{\thepage}}]{\textbf{Section \nouppercase{\rightmark}}}
\rhead[\small{\textbf{Chapter~\nouppercase{\leftmark}}}]{\small{\textbf{\thepage}}}

\begin{flushright}
\small{\textit{``Two stars keep not their motion in one sphere''}}

\small{\textit{Henry IV, Part 1, Act 5, Scene 4}}

\vspace{-2mm}
\small{\textit{-- William Shakespeare}}
\end{flushright}
\bigskip

\lettrine[lines=3, lraise=0, nindent=0.1em, slope=0em]{M}{ultiple stars have been observed} since ancient times, although objects within a multiple system were not initially considered as connected, as the prevailing belief was that it was merely a coincidental occurrence. Many visible stars could be classified as multiples when observed with the naked eye, but probably less than half of them actually are. It is clear that visible stars represent only a very small fraction of all stars in the Milky Way, but the known quantity of visual binaries is sufficient to underscore the significance of multiplicity. The orbital periods of binary systems vary greatly, ranging from mere hours to centuries, with some potentially spanning millions of years. The proximity or distance between the two components can vary significantly, affecting their evolutionary paths. The presence of a binary companion can drastically alter a star's evolution, especially if the orbital period is short. Although many differences between isolated and binary stars are understood, unresolved questions persist, complicating our understanding of stellar evolution \citep{eggleton06}.

Multiplicity is not limited to binaries, as many systems contain three or more stars, forming complex gravitational relationships \citep{evans68,tokovinin97,murdin01}. The frequency of such systems varies with stellar mass, being more common among massive stars \citep[][and references therein]{li24,cortescontreras17b}. In fact, recent data sets \citep{bordier24,li24} indicate that the mean multiplicity increases as a function of stellar mass, with more massive stars more likely to be in triple or higher order configurations. Stellar and dynamical interactions, such as mass transfer and ejections, impact the multiplicity \citep{preece24}. Advances in spectroscopy \citep[e.g.][]{chini12,baroch21} and space telescopes (e.g. \textit{Hipparcos}, \citealt{perryman97}; \textit{Gaia}, \citealt{gaiacollaboration16a}; \textit{James Web} Space Telescope, \citealt{mcelwain23}; \textit{Chandra} X-ray Telescope, \citealt{weisskopf00}; XMM-\textit{Newton}, \citealt{mason01b}), reveal many new companions, highlighting multiplicity's role in star formation, stellar dynamics, and life cycle understanding.

\section{History of the observation of multiple stars}
\label{sec:history}

\begin{figure}
  \centering
  \includegraphics[width=1\linewidth, angle=0]{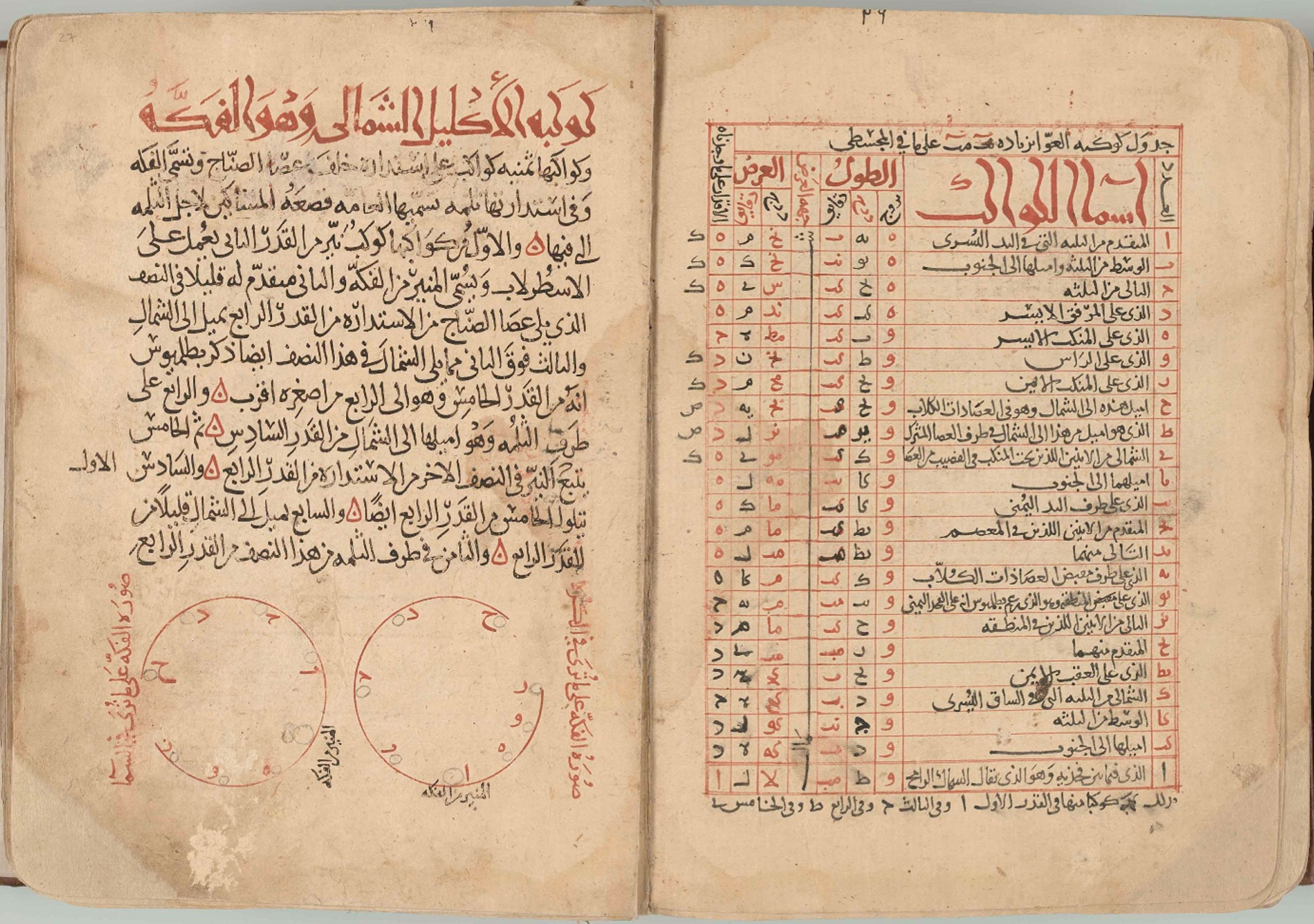}
  \includegraphics[width=1\linewidth, angle=0]{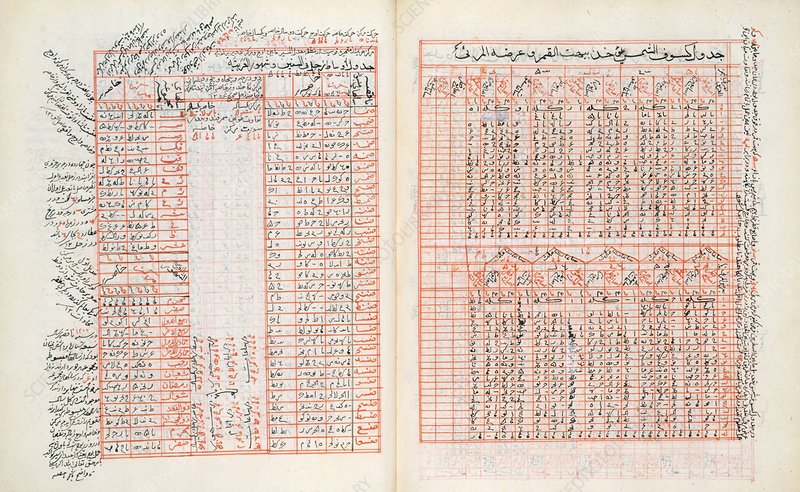}  
     \caption[\textit{Book of Constellations of the Fixed Stars} from Abd Al-Rahman Al Sufi, published in 986, and \textit{Ulugh Beg’s Catalogue of Stars} from Ulugh Beg, published in 1437.]{\textit{Top}: \textit{Book of Constellations of the Fixed Stars}, from Abd Al-Rahman Al Sufi, published in 986. The left hand page describes Corona Borealis constellation, while the right hand page tabulates the stars in the preceding constellation Bo\"otes (Credit: Museum of Islamic Art, Doha, Qatar). \textit{Bottom}: \textit{Ulugh Beg’s Catalogue of Stars} published in 1437, from Ulugh Beg. Page from an edition of ``Zij-i Sultani'' (Credit: Royal Astronomical Society / Science Photo Library). 
}
   \label{fig:old_catalogues}
\end{figure}

Ptolemy (c. 100--c. 170 AD) was the first astronomer to assign the term \textit{diplous} ($\delta\iota\pi\lambda \textit{o}\hat{\upsilon}\varsigma$) to a double star, specifically \textit{Ainalrami}, $\nu^{\text{01}}$ Sgr\footnote{All abbreviations used for constellations can be found in Appendix~\hyperref[ch:Appendix_B]{B}.}, and $\nu^{\text{02}}$ Sgr. These stars were observed with the naked eye, recorded in the Almagest, and are now identified as an optical double. In Arab astronomy, numerous star names collectively denote two or more adjacent stars that are easily distinguishable with the naked eye. In contrast, the widely recognised naked-eye pair \textit{Mizar} ($\zeta$ UMa) and \textit{Alcor} (g~UMa) in the Big Dipper received distinct names, likely due to their differing brightness levels, which did not depict them as a discernible ``pair'' of stars \citep{heintz78}. The Middle Ages did not contribute new knowledge about multiple stars, even though two of the great ancient star catalogues had been published (see Fig.~\ref{fig:old_catalogues}). The first was the \textit{Kitab Suwar al-Kawakib} or \textit{Book of Constellations of the Fixed Stars} by the Persian astronomer Abd Al-Rahman Al Sufi (903--986), also known in the Western world as Azophi. The second was \textit{Ulugh Beg’s Catalogue of Stars}, published in 1437 by Mirza Muhammad Taraghay bin Shahrukh (1394--1449), better known as Ulugh Beg, a Timurid sultan as well as an astronomer and mathematician. However, neither of these catalogues contains any reference to multiple systems.

\begin{figure}[h]
  \centering
  \includegraphics[width=0.95\linewidth, angle=0]{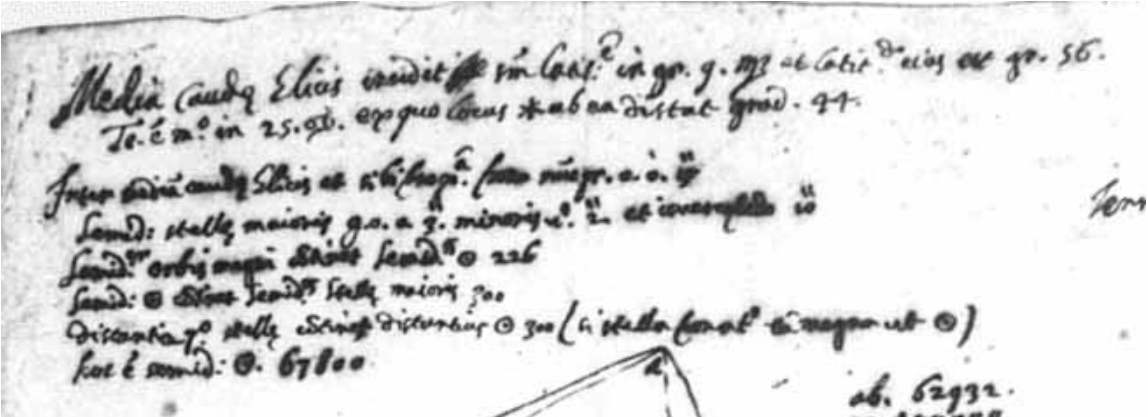}
     \caption[Detailed record of the observation of \textit{Mizar} in Galileo's handwriting.]{Detailed record of the observation of \textit{Mizar} in Galileo's handwriting. No date is mentioned, but there are good reasons to believe it was made on 15 January 1617. Ms. Gal. 70 c. 10r, Biblioteca Nazionale Centrale Firenze. The translation, performed by Thomas Winter (University of Nebraska), is: ``The middle star of the \textit{Tail of Elix} [\textit{Great Bear}] falls, by [ecliptical] longitude, on the 9th degree of Virgo, and its latitude is 56. Earth is now in Cancer 25, out of which position the star is 44 degrees distant. Between the middle star of the \textit{Tail of Elix} and the star closest to it [\textit{Mizar} B], I now put 0.0.15''. The semidiameter [radius] of the large star, 0.0.3''; of the smaller, 2''; the interval, 10''. The semidiameter of the great orb [Earth's orbit] contains 226 solar semidiameters. The solar semidiameter contains 300 semidiameters of the large star. So the distance of the star contains 300 solar distances, if the star is posited to be as big as the Sun, that is, 67\,800 solar semidiameters.'' \citep{ondra04}.
}
   \label{fig:galileo}
\end{figure}

It was not until the seventeenth century, with the invention of the telescope, that interest in double stars was rekindled. This led to the discovery of a new visual binary, \textit{Mizar} A and B, attributed to Benedetto Castelli (1577--1643). Castelli requested Galileo (1564--1642) to observe it in 1616, although the discovery was later inaccurately attributed to Giovanni Battista Riccioli (1598--1671) in 1650 \citep{ondra04}. Castelli, a former student of Galileo, likely requested it due to his master's renowned astronomical observations. Even in the present day, Galileo's detailed studies of multiple-star systems remain notable. Indeed, Galileo meticulously recorded his observations, determining the apparent angular diameters of \textit{Mizar}'s component stars as 6\,arcsec and 4\,arcsec, with a separation of 15\,arcsec. The record of his observation is still preserved today (Fig.~\ref{fig:galileo}). Similarly, he sketched the Trapezium region, mapping stars separated by 15\,arcsec and noting their relative sizes, although absolute measurements were not provided. Remarkably, both his measurements of \textit{Mizar} and the Trapezium's sketch closely align with modern astronomical data. Galileo's measurements of \textit{Mizar} match the expected Airy disk sizes for the telescopes of his era, demonstrating 2-arcsec accuracy \citep{graney09}. Galileo proposed using these star associations to measure stellar parallax, supporting the Copernican model of Earth's orbit around the Sun and aiding in determining stellar distances. Consequently, the search for double stars began for this very purpose.

\begin{figure}[H]
 \begin{subfigure}{.5\textwidth}
  \centering
  \includegraphics[width=0.95\linewidth, angle=0]{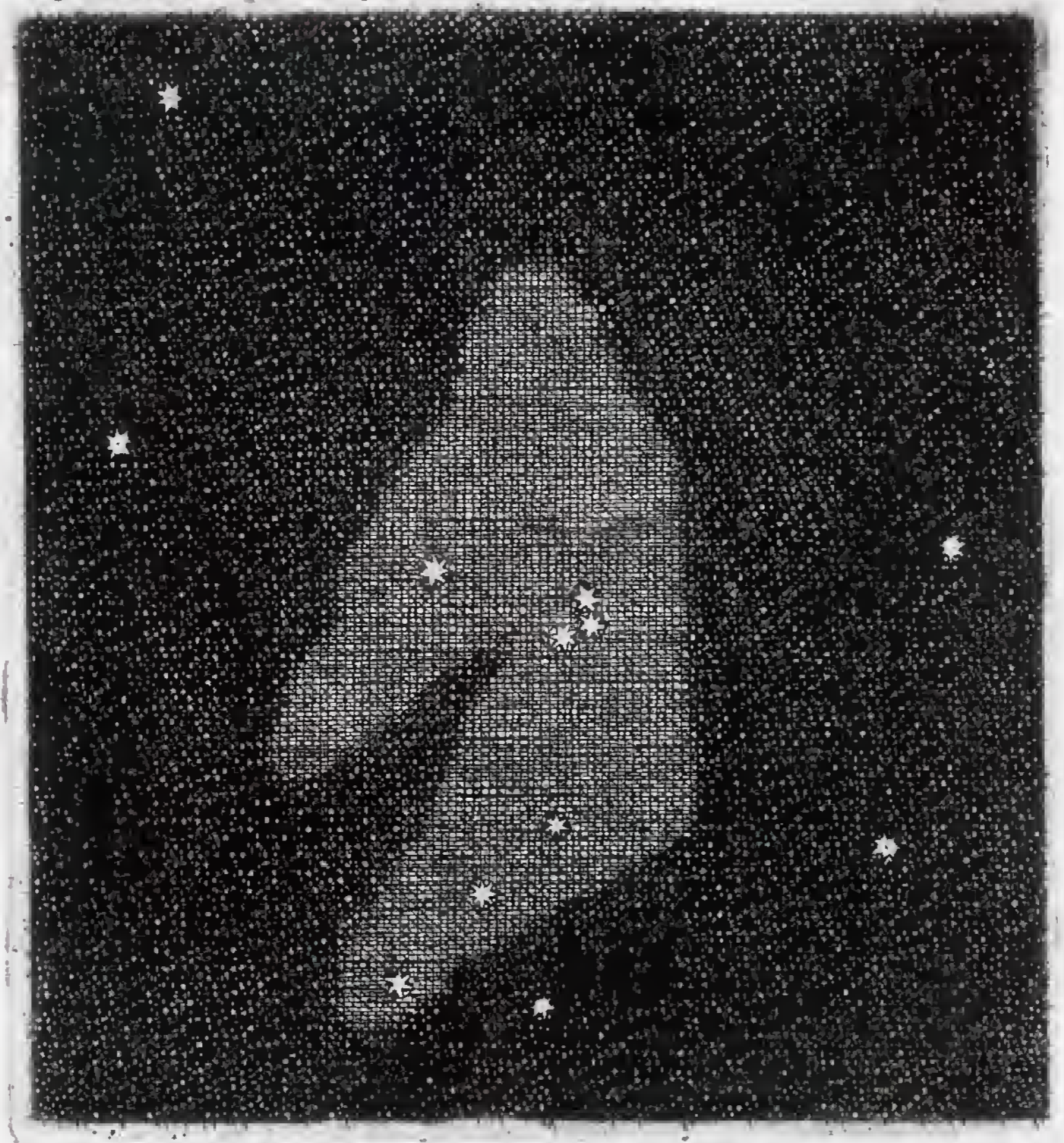}
 \end{subfigure}
  \begin{subfigure}{.5\textwidth}
  \centering
  \includegraphics[width=0.95\linewidth, angle=0]{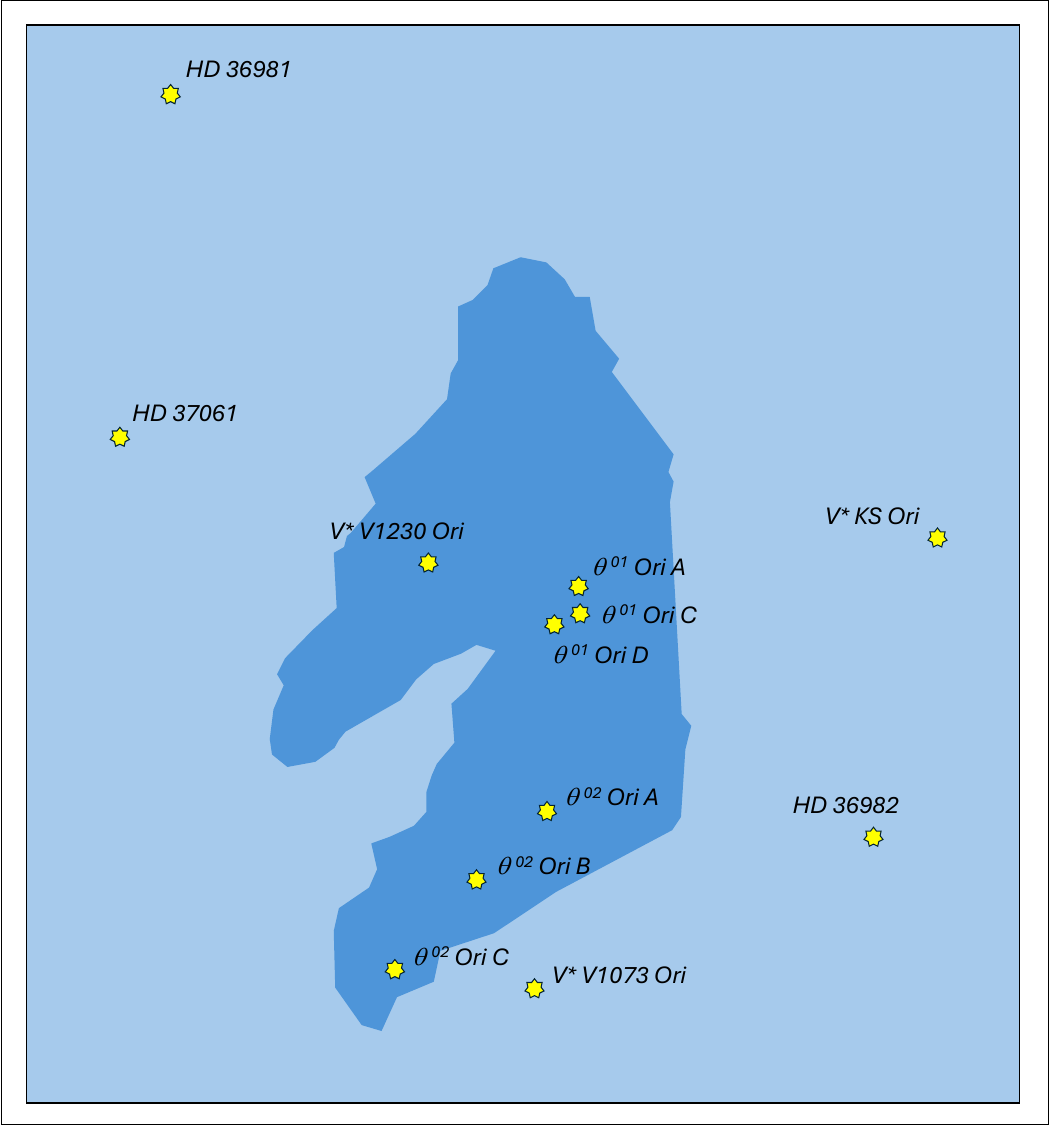}
 \end{subfigure} 
     \caption[Engraving showing the Orion nebula with the triple star $\theta^{\text{01}}$~Orionis, contained in \textit{Systema Saturnium} from Christiaan Huygens (1659) and the interpretation of the current names of the stars.]{\textit{Left panel}: Engraving showing the Orion nebula with the triple (later considered as quadruple) star $\theta^{\text{01}}$~Orionis (\textit{Trapezium}), contained in \textit{Systema Saturnium} from Christiaan Huygens (1659). \textit{Right panel}: Interpretation of the current names of the stars (outlined by the author).
}
   \label{fig:huygens}
\end{figure}

Mid-seventeenth century astronomers were very active in the search for new objects in the sky, discovering new double stars and multiple star groups. Dutch astronomer Christiaan Huygens (1629--1695) published in 1659 the book \textit{Systema Saturnium} \citep{huygens1659} in which he presented an engraving announcing that $\theta$ Ori, the central star of Orion's sword, was actually a group of three very close stars surrounded by an extended nebulous region, as shown in Fig.~\ref{fig:huygens} \citep{biro17}. This number of stars was later extended to four in 1673 by Jean Picard (1620--1682). 
But a lesser-known Sicilian astronomer, Giovanni Battista Hodierna (1597--1660), was the first in history to present a list of twelve stars with companions, making his work the third to mention multiple stars, after Ptolemy and Castelli/Galileo. We will focus on the figure of Hodierna in Chapter~\ref{ch:hodierna}.

In 1664, the English astronomer Robert Hooke (1635--1703), while observing the comet \textit{Hevelius}, detected \textit{Mesarthim} ($\gamma$\,Ari), that was the closest pair (8\,arcsec) ever discovered \citep{argyle19}. Some years later, two French Jesuit Fathers, Jean de Fontaney (1643--1710) and Jean Richaud (1633--1693) discovered to be double the star \textit{Acrux} ($\alpha$ Cru) in the Southern Cross in 1685 from Cape Town (South Africa), and $\alpha$\,Cen in 1689 from Pondicherry (India), respectively \citep{henroteau28,kameswararao84,kochhar91}. During the eighteenth century, more astronomers were responsible to provide new discovered pairs such as Gottfried Kirch, James Bradley, Nathaniel Pigott, Nevil Maskelyne, Jean-Dominique Maraldi, Giuseppe Piazzi, and others that helped to increase the list of known multiple stars.

\begin{figure}[H]
 \begin{subfigure}{.45\textwidth}
  \centering
  \includegraphics[width=0.83\linewidth, angle=0]{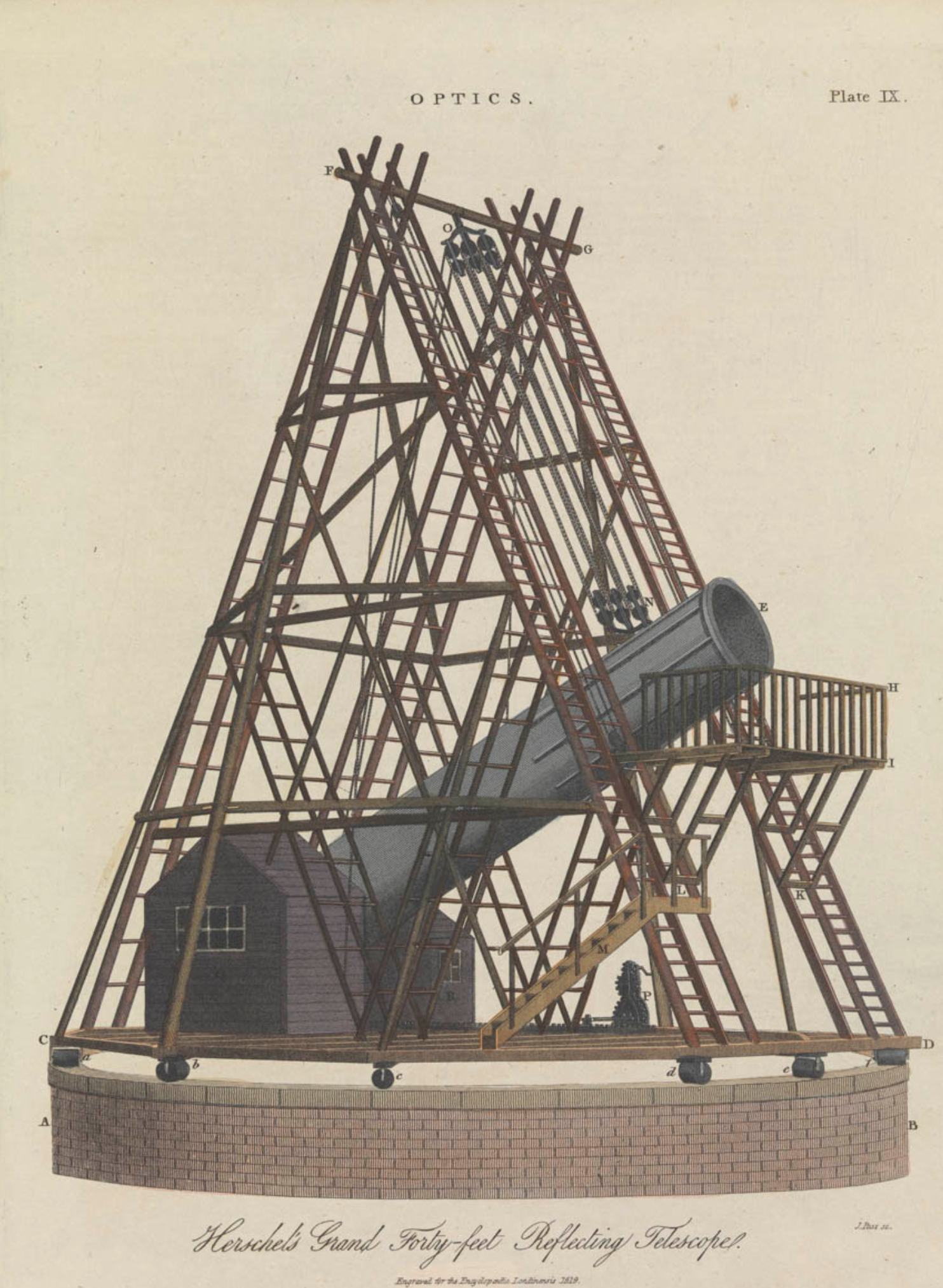}
 \end{subfigure}
  \begin{subfigure}{.55\textwidth}
  \centering
  \includegraphics[width=0.92\linewidth, angle=0]{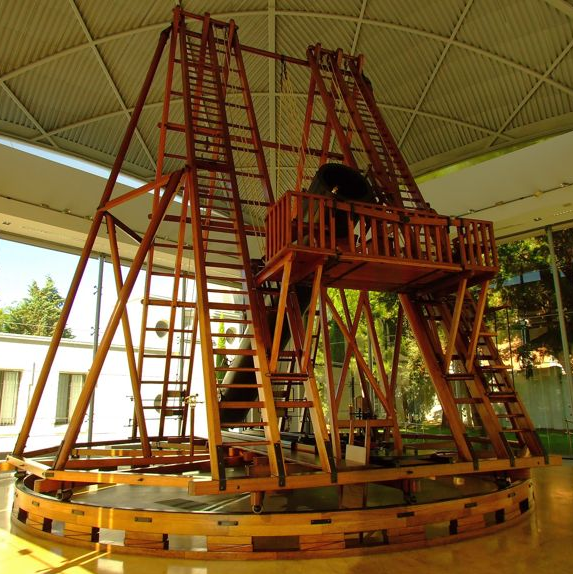}
 \end{subfigure}
 \caption[William Herschel's 40 feet Newtonian reflector telescope (1789).]{William Herschel finished building his 40 feet Newtonian reflector telescope in 1789 with grants totalling £4000 from King George III. It was the largest, most powerful telescope in the world and attracted visitors from far and wide. \textit{Left panel}: Engraving from the National Maritime Museum, Greenwich, London. \textit{Right panel}: Reconstruction of William Herschel's telescope carried out in 2004. Originally built in London between 1796 and 1802, it was a 25-foot instrument and was installed at the Royal Astronomical Observatory of Madrid. However, when Napoleon's troops occupied it in 1811, the telescope was destroyed, with only the mirror preserved (Credit: Observatorio Astronómico Nacional).
}
   \label{fig:herschel}
\end{figure}

The first catalogue of double stars was published in 1779 by another Jesuit priest, Christian Mayer (1719--1783), court astronomer and professor of physics and mathematics at Heidelberg. In a previous book, he listed seventy-two double star systems discovered by himself and other astronomers, though without much detail. However, in 1781 he presented another work, entitled \textit{Astronomisches Jahrbuch} for 1784 \citep{astronomischen1781}, which included a longer list of objects along with their corresponding position angles \citep{tenn13}. He compiled the list during 1777--1778, although his instruments were not well-suited for distinguishing genuine binary systems from coincidental alignments of stars. Mayer speculated about the possibility of them being physical systems (``These stars could be small suns revolving around larger suns'') as predicted by Isaac Newton \citep{niemela01}. Nevertheless, in the following century, William~F. Herschel (1738--1822) questioned that idea, and considered that multiple systems could have a gravitational link only when their orbital motion were proved. Herschel wanted to observe binary stars to calculate their parallaxes to determine the distances to stars, driven by his obsession to discover the structure of the Universe using his telescope \citep{fracastoro88,niemela01}, which was the largest in the world at that time (Fig.~\ref{fig:herschel}).

Herschel, with the invaluable help of his sister Caroline (1750--1848), one of the most neglected figures in the history of astronomy due to being a woman at that time, began compiling his first catalogue of double stars in 1782 composed of 269 doubles, 227 of which he claimed to be the first to see. He added 434 more to the list in 1785, all discovered by them. In the beginning of the nineteenth century, Herschel published the famous article \textit{Observations on the two lately discovered celestial bodies} \citep{herschel1802} where he announced the discovery of binary stars, real star systems governed by the laws of universal gravitation \citep{niemela01}. This marked the first time science confirmed Newton’s laws were valid beyond the solar system, initiating a revolution, though Herschel failed to measure parallax.

Friedrich von Struve (1793--1864) resumed the cataloguing of pairs, making very precise measurements of the relative positions of 2714 double stars, publishing them in \textit{Stellarum duplicium et multiplicium mensurae micrometricae} \citep{struve1837}. He was the first astronomer to measure the parallax of the star \textit{Vega} ($\alpha$ Lyr), although Friedrich Bessel (1784--1846) had been the first to do it, in this case with 61 Cyg \citep{batten77}.

\begin{figure}[H]
  \centering
  \includegraphics[width=0.6\linewidth, angle=0]{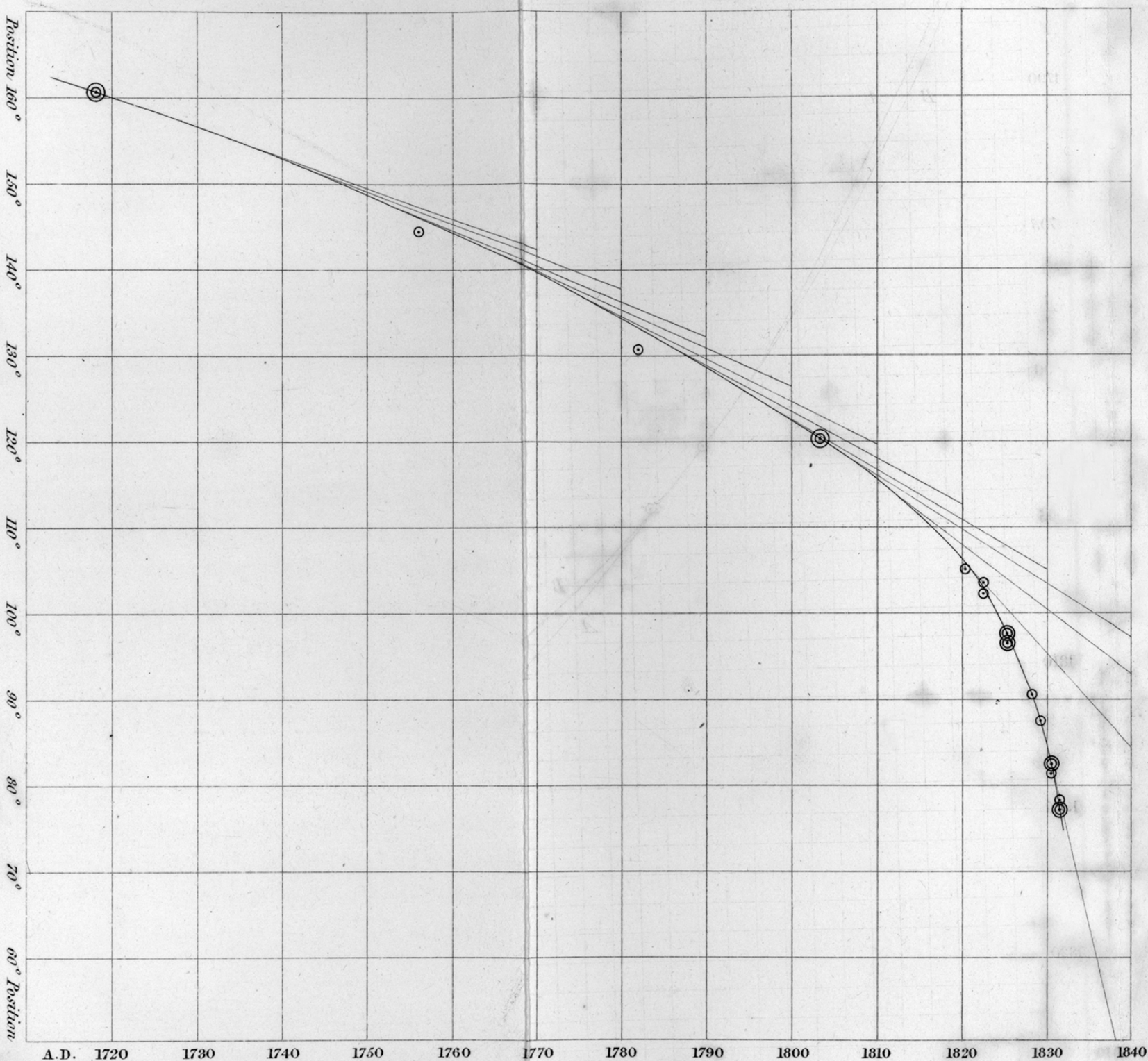}
     \caption[Herschel's graphical method applied to the double star $\gamma$ Vir.]{Herschel's graphical method applied to the double star $\gamma$ Vir. The position angle is plotted against the time. The observations marked by double circles carry greater reliability. By drawing tangents, he obtains angular velocities at uniformly spaced intervals along the curve \citep{hankins06}.
}
   \label{fig:graphmethod}
\end{figure}

Between 1821 and 1823, John W. Herschel (1792--1871), son of William Herschel, collaborated with Sir James South (1785--1867) in re-examining his father's double stars using two refractor telescopes, one with a focal length of 7 feet and the other one of 5 feet. In 1831, he was knighted by William IV, and two years later, he received a medal from the Royal Society for his memoir \textit{On the investigation of the orbits of revolving double stars} \citep{herschel1833}. This recognition marked the culmination of his father's discovery of gravitational stellar systems, achieved through the invention of a graphical method that allowed the eye to visualise the two component stars of the binary system orbiting in accordance with Newton's law \citep{clerke1908}. He argued that his approach, relying on human judgment rather than mathematical analysis, yielded superior outcomes compared to computation, particularly given the inherent uncertainty in the data \citep{hankins06}. An example can be seen in Fig.~\ref{fig:graphmethod}.

In the following decades, many remarkable figures such as Friedrich W. Argelander, Ercole Dembowski, Johann H. von Mädler, William R. Dawes, Angelo Secchi, Admiral William H. Smyth, and many others, contributed discoveries of new binary pairs. But the forerunner of binary stars in the modern era was undoubtedly the American astronomer Sherburne W. Burnham (1838--1921), who started as amateur discovering about 1500 pairs. He first submitted his list to the Smithsonian Institution, but it was rejected. In 1874, it was scheduled to be printed at the United States Naval Observatory (USNO), but the typesetting was interrupted before it was completed, and the archive was lost. Finally, in 1886, the Smithsonian Institution offered to publish it, but Burnham had become discouraged and declined the offer \citep{eggen53}. Burnham worked at Lick Observatory for four years, starting in 1888 \citep{frost21}. After leaving in 1892, he revised the manuscript of his catalog for five years, and the Carnegie Institute published it later \citep{eggen53} as \textit{A general catalogue of double stars}, also known as \textit{Burnham Double Star Catalogue} \citep[BDS,][]{burnham1906} with 13\,665 entries. The official recognition from the Astronomical Societies to the binary stars arrived in 1922, when the International Astronomy Union (IAU) created the Commision 26 for ``Double and Multiple Stars'' \citep{iau1922}.

Another American astronomer, Robert G. Aitken (1864--1951), performed a systematic search of the northern sky for pairs, compiling the famous \textit{Aitken Double Star Catalogue} (ADS), published in 1932 in two volumes, entitled \textit{New general catalogue of double stars within 120$\degree$ of the North Pole} \citep{aitken1932}. This catalogue contains the data of 17\,180 double stars north of declination --30$\degree$. The same task was performed for the Southern Hemisphere by the Scottish astronomer Robert T. A. Innes (1861--1933). His catalogue, called \textit{Southern Double Stars} \citep[SDS,][]{innes1899}, had the same format as the ADS, and together, they covered the entire sky \citep{tenn13}.

Hamilton M. Jeffers (1893--1976) continued Aitken's work, and Willem H. van den Bos (1896--1974) continued the Innes' one. They published together a new catalogue called \textit{Index catalogue of visual double stars} \citep[IDS,][]{jeffers63} containing 64\,247 pairs. Charles E. Worley (1935--1997) transferred the IDS database from the Lick Observatory to the United States Naval Observatory (USNO) in 1965 \citep{douglass98}, where it was started to be considered by the IAU as the international source of double star data \citep{argyle19}.

\begin{figure}[H]
  \centering
  \includegraphics[width=0.85\linewidth, angle=0]{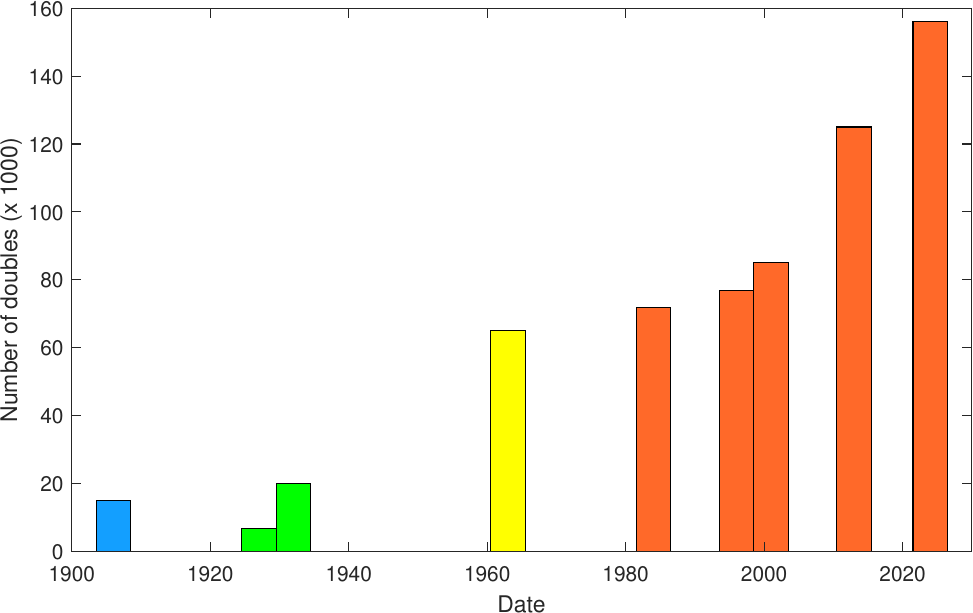}
     \caption[Evolution of the size of major double star catalogues and the WDS between 1906 and 2024.]{Evolution of the size of major double star catalogues and the WDS between 1906 and 2024. Blue for BDS (1906: north of --31\degree); green for SDS (1927: south of --19\degree) and ADS (1932: north of --30\degree); yellow for IDS (1963: all-sky); and red for WDS (1984, 1996, 2001, 2013, 2024: all-sky). Data source: The United States Naval Observatory webpage: \url{https://crf.usno.navy.mil/wdstext/}.
}
   \label{fig:pairs}
\end{figure}

In 1984, the \textit{Washington Double Star} catalogue \citep[WDS,][]{mason01} was created, incorporating the original data from IDS along with subsequent updates, reaching 73\,160 pairs. The catalogue has been expanded over time with numerous measurements, primarily derived from the \textit{Hipparcos} \citep{perryman97} and \textit{Tycho} \citep{hog00} catalogues, along with results from speckle interferometry and various other sources. A distinctive 1--3 letter discovery code is employed to designate the observer responsible for reporting the information.

Today it is maintained by the USNO, and it is the world's principal database for astrometric double and multiple star information. The WDS catalogue contains positions (J2000), discoverer designations, epochs, position angles, separations, magnitudes, spectral types, proper motions, and, when available, \textit{Durchmusterung}\footnote{\textit{Durchmusterung} or \textit{Bonner Durchmusterung} (BD) is an astrometric star catalogue of the whole sky, published by the Bonn Observatory in Germany from 1859 to 1863, with an extension published in Bonn in 1886.} numbers, and notes for the components of the registered multiple stars. The new versions, initially available on CD, and later accessible via the Internet, progressively increased the amount of registered data, reaching the current total of about 157\,000 pairs. Fig.~\ref{fig:pairs} shows the evolution over time of the size of major double star catalogues and the WDS. 

\section{Stellar multiplicity}
\label{sec:stellar_multiplicity}

A double star, or in general a multiple star, is a system composed of two or more stars located in a close proximity to each other in the sky as an observer can see them by naked-eye or with a telescope. 
Except for stars coincidentally aligned along the same line of sight but spatially distant, these entities are gravitationally bound as a system. In certain instances, they maintain enough separation to drift through space collectively, whereas in others, they actively orbit the common center of gravity of the system \citep{mullaney05}.
Therefore, in terms of real bound stars, a binary system is defined as one containing two stars that orbit in closed trajectories around their common center of gravity due to their mutual gravitational interaction \citep{batten73}. Another characterisation, proposed by \citet{lipunov89}, asserts that two stars form a binary system if they travel within a confined spatial region. Many systems, particularly those with substantial separations, often encompass additional objects, thereby forming multiple systems. These multiple systems involve three or more stars organised into varying hierarchical tiers \citep{tokovinin97,tokovinin08,eggleton08,duchene13}.

Wide binaries are entities characterised by orbits that maintain considerable separation, resulting in minimal reciprocal gravitational influence. For a system to be considered a binary, a gravitational bond must exist between both stars, or, in the case of a multiple system, all involved entities must sustain a state of gravitational equilibrium. The components within extensive multiple systems exhibit substantial separations, leading to relatively low gravitational energies. 

\citet{duchene13} defined stellar multiplicity as ``an ubiquitous outcome of the star-formation process''. It is worth noting that a thorough characterisation of the properties of each component in a system can help determine whether specific processes were essential to the system's formation or evolution. These key properties include proper motion, distance, and radial velocity. Binary star systems, representing the simplest form of stellar multiplicity, are composed of two stars bound by gravity and orbiting a shared centre of mass. These systems offer fundamental insights into stellar dynamics and evolution, serving as critical benchmarks for understanding more complex stellar arrangements. The study of binaries dates back to \citet{kepler1609}, who first formulated his laws of planetary motion by observing the motions of binary stars.

In the field of stellar astrophysics, stellar multiplicity holds immense significance. Approximately half of the solar-type stars in the solar neighborhood exist as components of multiple systems \citep{duquennoy91,raghavan10}. When these stars come into close proximity, their interactions can give rise to a multitude of astrophysical phenomena, including Type Ia supernovae, X-ray binaries, and gravitational wave sources \citep{demarco17}. The intricate interplay between stellar evolution and multiplicity is responsible for the emergence of most of these phenomena.

\subsection{Classification of binary stars}
\label{sub:classification_binary_stars}

Binary stars can be classified either by their detection method or by their separation distance. Classification based on separation distance is particularly common due to its significant implications for the evolution and dynamics of binary systems, as well as for any planetary systems that may exist around them. Systems with smaller separations often experience stronger gravitational interactions, influencing their orbital parameters and internal structure over time.

In contrast, systems with larger separations tend to behave more like independent stars, with minimal gravitational influence on each other. However, because the boundaries of separation distances are not well-defined, many studies instead focus on classifying binary stars according to their detection method.

\subsubsection{According to their detection method}
\label{sec:according_detection_method}

There are different types of binaries based on how they are observed or detected, either through naked-eye observation, or using techniques such as interferometry, spectroscopy, or astrometry, among others \citep{heintz78,mullaney05}. Any binary star can be classified at the same time in several of the following classes:

\begin{itemize}

\item \textit{Optical doubles:} An optical binary is a set of two stars that lie along nearly the same line of sight, typically separated by more than 5 arcsec. They have similar coordinates but are located at very different distances, or their proper motions are different, or both. Additionally, stars often exhibit unequal brightness, which can reflect differences in distances, although this is not a defining criterion \citep{argyle12}. 

\begin{figure}[H]
  \centering
  \includegraphics[width=0.82\linewidth, angle=0]{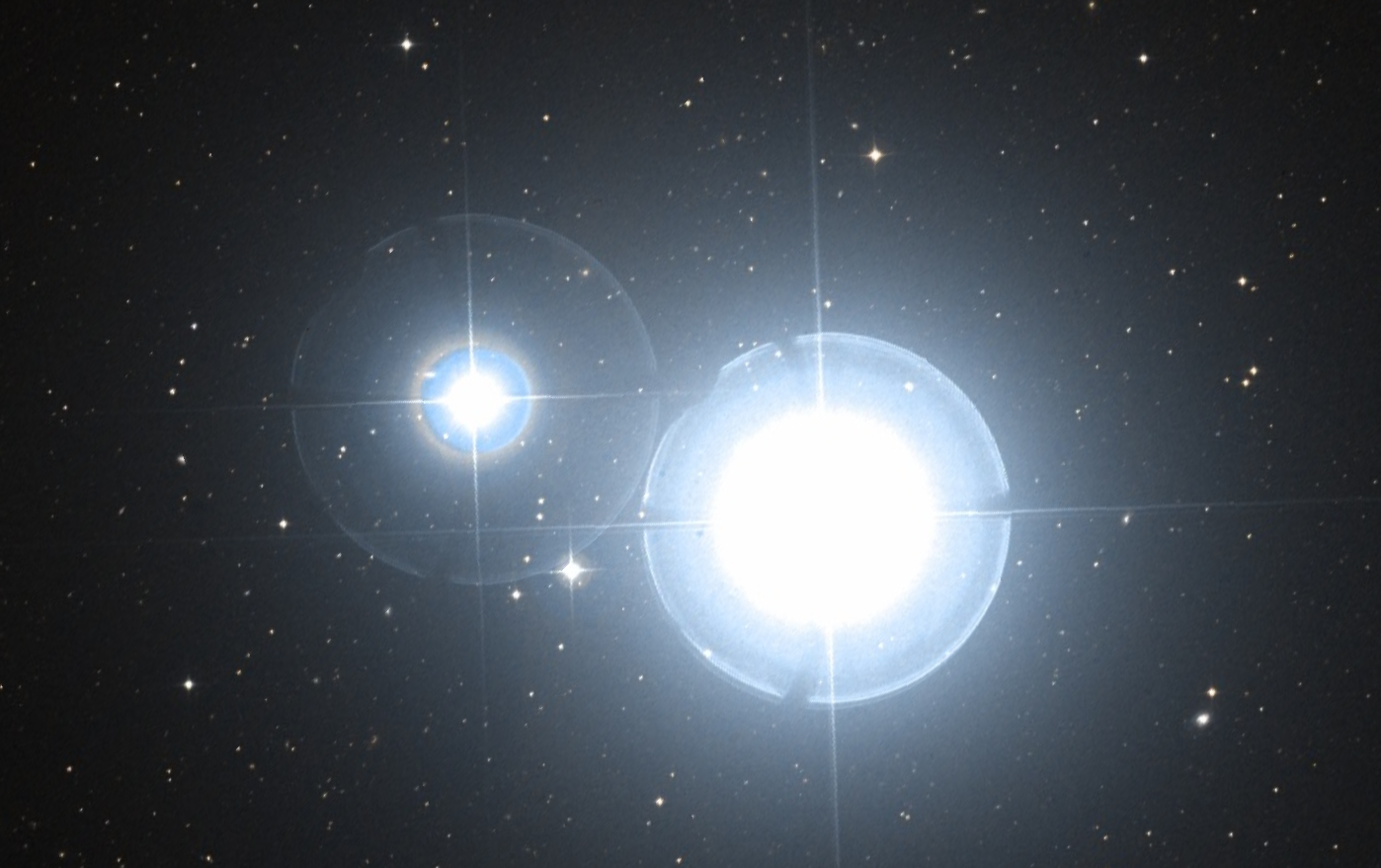}
     \caption[Optical pair composed of \textit{Mizar} ($\zeta$ UMa) and \textit{Alcor} (g UMa).]{Optical pair composed of \textit{Mizar} ($\zeta$ UMa) and \textit{Alcor} (g UMa). Image obtained from \textit{Aladin} \citep{bonnarel00} with data of DSS2 \citep{lasker96}.
}
   \label{fig:mizaralcor}
\end{figure}

Optical binaries are not true binary stars because they lack orbital motion around each other, as they are not gravitationally bound. Instead, they are pairs of stars that appear close together in the sky, despite being situated at different distances. This phenomenon results from alignments of stars, creating an illusion. Optical doubles are considered less common than genuine binary systems. Notably, within the constellation Ursa Major, the stars \textit{Mizar} and \textit{Alcor} exemplify an optical double configuration as shown in Fig.~\ref{fig:mizaralcor}. In fact, \textit{Mizar} is located at 26.2\,pc, while the distance to \textit{Alcor} is at 25\,pc. Another example is \textit{Albireo} A/B ($\beta^{\text{01}}$ Cyg / $\beta^{\text{02}}$ Cyg).

\item \textit{Visual binaries:} A visual binary is a gravitationally bound binary star system \citep{argyle12} composed of two or more visible stars, observable with the naked eye or through a telescope, that orbit around their common centre of gravity. Their orbital periods range from a few years to several centuries. Systems with shorter periods can be detected in different ways, depending on the telescope aperture, resolution function, and atmospheric conditions \citep{mullaney05}. Some examples of naked-eye double stars are: $\kappa^{\text{01}}$ Tau / $\kappa^{\text{02}}$ Tau, $\theta^{\text{01}}$ Tau / $\theta^{\text{02}}$ Tau, $\alpha^{\text{01}}$ Lib / \textit{Zubenelgenubi} ($\alpha^{\text{02}}$ Lib), $\epsilon^{\text{01}}$ Lyr / $\epsilon^{\text{02}}$ Lyr, \textit{Anser} ($\alpha$ Vul) / 8 Vul, \textit{Prima Giedi} ($\alpha^{\text{01}}$ Cap) / \textit{Algedi} or \textit{Giedi} ($\alpha^{\text{02}}$ Cap). 

\item \textit{Common proper motion binaries:} Common proper motion binaries are pairs of stars that move through space with nearly identical proper motions. They are typically widely separated systems, often thousands of astronomical units apart, since their detection requires them to be spatially resolved and to have measurable proper motions. Unlike close binaries, these systems are not bound by tight orbits but still share a common space velocity. This shared motion suggests a common origin, such as formation from the same molecular cloud.

Proper motion is the angular displacement of a star across the sky, measured in milliarcseconds per year (mas/a). If two stars exhibit the same proper motion vector, it may indicate a physical association. Common proper motion binaries can be identified using astrometric data from surveys such as \textit{Gaia}, whose precise measurements have significantly increased the number of known systems of this kind \citep[e.g.][]{hartman20,kervella22}. Their long orbital periods make the detection of orbital motion challenging. However, if the gravitational attraction between the stars is strong and orbital motion becomes measurable, this method of identification may be less suitable, as the proper motions of the components can differ. Nevertheless, these systems are crucial for studying stellar formation and Galactic dynamics. They also serve as testbeds for stellar evolution theories by allowing comparisons between component stars of similar age and origin.

Some of these binaries are thought to be remnants of disrupted star clusters or stellar associations. Others may eventually become unbound due to tidal interactions with the Galactic gravitational field. They can be distinguished from chance alignments by requiring consistent parallax and radial velocity measurements. Spectroscopic follow-up can indicate their radial velocity and it is often necessary to confirm that both stars are truly co-moving.

Examples of such systems include \textit{Sirius} A/B, 61 Cyg A/B, and GJ 15 A/B.

\item \textit{Interferometric binaries:} Observation and measurement of binary systems was one of the earliest contributions of interferometry to astrophysics \citep{morgan78}. The original observations were based on ideas from \citet{fizeau1867}. He established that better results in measuring stellar diameters could be achieved by partially covering the aperture of telescopes, allowing only two small sub-apertures. However, problems related to the instability of the mechanical system delayed successful observations until 1974, when Antoine Labeyrie combined the light from two independent telescopes located at the Observatoire de la Côte d’Azur in France \citep{labeyrie75}.

The principle of stellar interferometry involves combining individual telescopes to create a stellar interferometer. In this setup (see Fig.~\ref{fig:stellar_interferometry}), the resolution is no longer determined by the diameter of each telescope but by the distance between them, known as the baseline (B). Instead of capturing images of stars directly, an interferometer captures the interference pattern, known as interference fringes, generated by combining light from two or more telescopes. Interference fringes emerge when light waves undergo constructive interference, resulting in alternating light and dark bands. These fringes encode valuable information about the size, shape, and brightness distribution of the star. Binary stars are resolved when the interference fringes produced by each component can be distinguished, allowing astronomers to measure their angular separation and relative brightness. Starlight reaches one telescope before another due to their spatial separation. To compensate for these differences in light travel distances, the starlight is channelled from each telescope through vacuum tubes into a laboratory setting. Inside the lab, the path lengths are equalised using optical delay lines, which consist of movable carts travelling along precision rails. These carts continuously adjust their positions to maintain phase synchronization among the telescopes. Starlight reflects off mirrors mounted on the delay line carts and is directed to an instrument that combines light from multiple telescopes, capturing interference fringes using a camera. Therefore, for example, connecting two 8-meter telescopes separated by a baseline of approximately 130 meters, such as those in ESO's Very Large Telescope Interferometer (VLTI), improves the angular resolution by a factor determined by the baseline-to-telescope diameter ratio (130/8 $\approx$ 16), achieving approximately 3 milliarcsecond (mas) resolution in the near-infrared range. As a result, a wide range of stars can be resolved, revealing shapes that are not necessarily circular \citep{glindemann11}.

\begin{figure}[H]
  \centering
  \includegraphics[width=0.75\linewidth, angle=0]{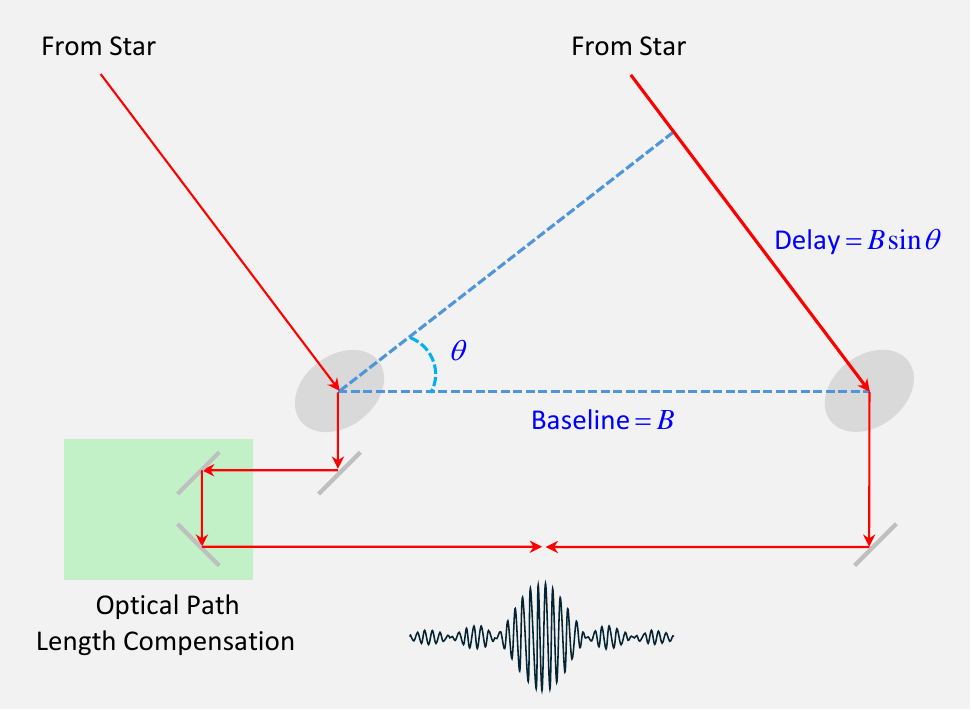}
     \caption[Basic schema showing the basics of interferometry with two telescopes.]{Basic schema showing the basics of interferometry with two telescopes (Based on the figure provided by the Center for High Angular Resolution Astronomy, \url{https://www.chara.gsu.edu/public/basics-of-interferometry/}).
}
   \label{fig:stellar_interferometry}
\end{figure}

In addition to measuring stellar diameters, interferometry is used to resolve close binaries. Therefore, interferometric binaries are close binaries detected using the technique of interferometry \citep{halbwachs15,boffin16}, which exploits the interference and diffraction effects resulting from the wave nature of light. A particular form of this technique, based on the analysis of large numbers of short exposures that freeze the variation of atmospheric turbulence, is speckle interferometry \citep{labeyrie70,hariharan92,al-wardat02}, which can restore the spatial power spectrum of an object up to frequencies limited by the aperture of the astronomical instrument. This technique allows measurements of binaries with separations as small as 0.01 arcsec. 

Examples of interferometric binaries are HD 98800B \citep{boden05}, HD 27935 \citep{halbwachs16}, and 12 Per \citep{mcalister76}.

\item \textit{Astrometric binaries:} Some stars appeared to be orbiting around an empty space or an invisible companion. It became evident that mathematics could be applied to infer the mass of the missing companion. To calculate the position and characteristics of such a companion, the position of the visible star is carefully measured, and the variations in its movement due to the gravitational effect of the secondary star are studied to detect the periodic shifts in position (see Fig.~\ref{fig:sirius}). These astrometric measurements and analyses gave rise to the name for this type of binary system.

\begin{figure}[H]
 \begin{subfigure}{.67\textwidth}
  \centering
  \includegraphics[width=1\linewidth, angle=0]{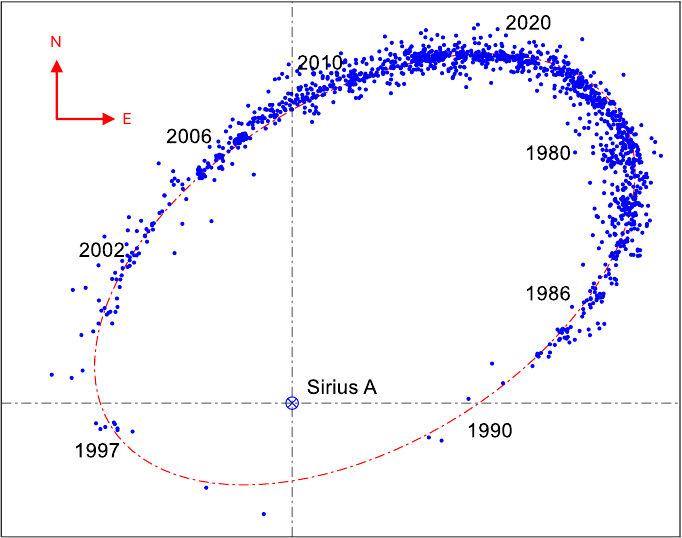}
 \end{subfigure}
  \begin{subfigure}{.37\textwidth}
  \centering
  \includegraphics[width=0.71\linewidth, angle=0]{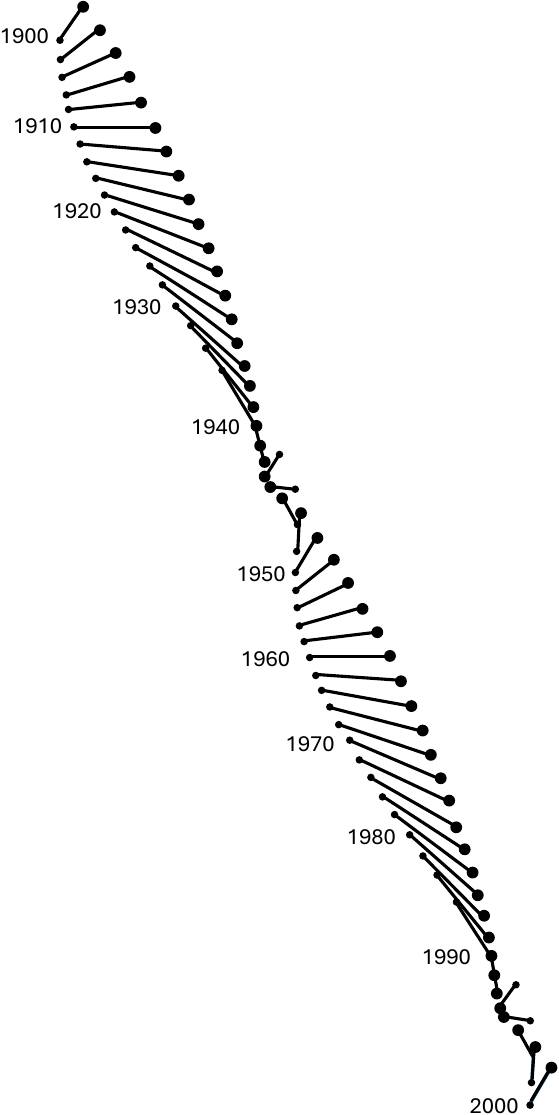}
 \end{subfigure} 
     \caption[Multiple observations of proper motion of \textit{Sirius} B, and the characteristic wobble of astrometric binaries.]{\textit{Left panel:} Multiple observations of proper motion of \textit{Sirius} B from 1980 to 2022. From Earth, the separation between \textit{Sirius} A and B varies between 3 and 11\,arcsec on a period of 50.1 years. \textit{Sirius} was first detected as an astrometric binary though is now considered a visual system (data observations kindly provided by Dr. Rachel Matson from USNO). \textit{Right panel:} \textit{Sirius} A and B characteristic wobble of astrometric binaries.
}
   \label{fig:sirius}
\end{figure}

Astrometric binaries were first detected with long-focus photographic telescopes \citep{vandekamp43}. These close binaries are now detected through precision photographic or CCD astrometry, which involves imaging a target star against a more distant star-field that serves as a fixed reference frame. By repeating this task over time, the star's path across the sky can be measured, revealing a wobbling motion caused by the gravitational influence of the companion. A similar technique is used with a multi-channel astrometric photometer, which can detect the movement of nearby stars in just a few hours or days. Measurements are typically performed on stars within about 10\,pc.

The presence of a companion star is easier to detect if it is very massive, as it produces a noticeable shift in the visible star's position. If the companion is less massive, more precise measurements are required. From this analysis, other data can be obtained, such as the orbital period, ellipticity, inclination angle of the orbit, and the mass of the companion \citep{asada04}. Kepler's laws can also be applied to determine the characteristics of the binary system.

Notorious examples of astrometric binaries are: \textit{Sirius} A--B ($\alpha$ CMa), as can be seen in Fig.~\ref{fig:sirius}, and \textit{Procyon} A--B \citep[$\alpha$ CMi,][]{liebert13}.

\item \textit{Spectroscopic binaries:} Spectroscopic binaries are detected through the periodic shifting of spectral lines due to the Doppler effect. The inventor of this technique was the German astrophysicist Hermann Karl Vogel (1841--1907) obtaining periodic Doppler shifts in the components of \textit{Algol} \citep{vogel1890}. The separation between the companions is usually very small, resulting in high orbital velocities. If the system has components along the line of sight, this radial velocity can be measured with a spectrometer by observing the shift in the stars' spectral lines. This is why binaries detected with this method are known as spectroscopic binaries. Typically, these systems cannot be resolved as visual binaries, even with the most powerful telescopes. The lines shift toward the blue end as the primary star moves toward Earth in its orbit, and toward the red end as it moves away. If both stars have nearly the same luminosity, a double set of lines will appear, shifting back and forth in opposite directions as both stars orbit around their common centre of gravity.

Depending on the difference in apparent brightness and the resolution capabilities of the spectrograph, the Doppler shift can be detected either from the brighter component (single-lined SB1 system) or within a combined spectrum of both components (double-lined SB2 system). The generation of the Doppler shift for SB1 and SB2 systems, caused by varying radial velocities along our line of sight, is illustrated in Fig.~\ref{fig:sb1sb2}. In SB2 systems, the luminosity difference between the components is minimal, resulting in a split in the spectral line. In contrast, in SB1 systems, only the absorption line of the brighter component is observed, shifting around a neutral position. Remarkable examples include orbits around imperceptible black holes or extrasolar planets, which subtly shift the spectrum of the orbited star by just a few tens of meters per second.

\begin{figure}[H]
  \centering
  \includegraphics[width=0.7\linewidth, angle=0]{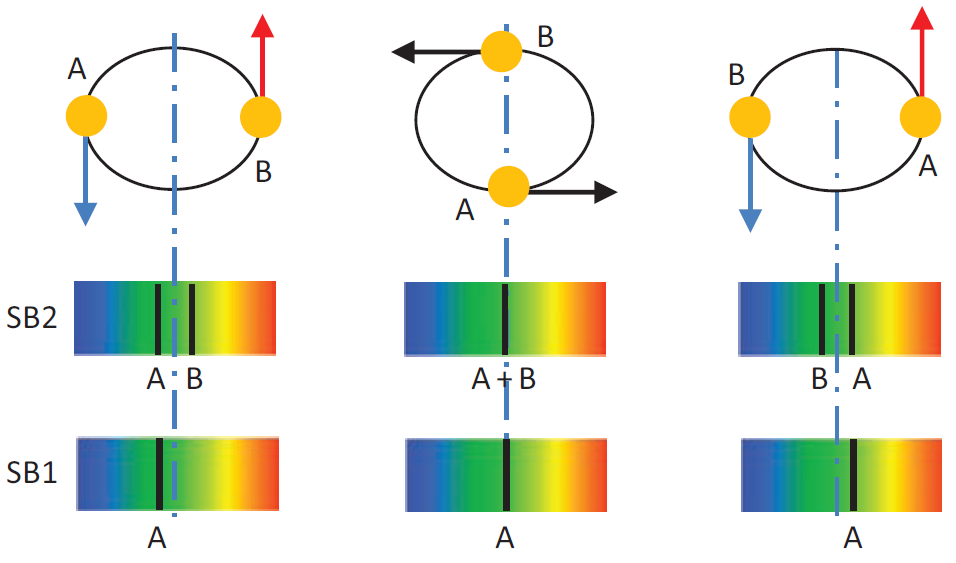}
     \caption[Effect of Dopper-shift in SB1 and SB2 spectra.]{Effect of Dopper-shift in SB1 and SB2 spectra \citep{walker17}.
}
   \label{fig:sb1sb2}
\end{figure}

There can be instances with three stars in a spectroscopic system, known as triple-line spectroscopic triple systems (SB3, sometimes denoted as ST3), such as GJ~3916, GJ~4383 \citep{baroch21}, and LP~655--43 \citep{winters20}. Additionally, there are instances with four stars, forming quadruple-line spectroscopic quadruple systems (SB4), such as BD--225866 \citep{shkolnik08}. These systems exhibit multiple sets of spectral lines corresponding to the different stellar components, allowing for more complex orbital dynamics and providing rich data for studying stellar interactions and the dynamics of multiple star systems.

It is difficult to find spectroscopic binaries that are also visual binaries. Only a few cases are known, and the reason is that the separation between the spectroscopic binary components is typically too small to be resolved as visual binaries, but they exhibit high orbital velocities that can be detected spectroscopically. On the other hand, visual binaries have a larger separation, allowing their companions to be easily resolved, but their orbital velocities are too small to be measured through spectroscopic methods. Therefore, the most common condition for a binary system to be both visual and spectroscopic is that it should be relatively close to Earth.

We know many spectroscopic binaries such as CC Eri, $\eta$ And \citep{strassmeier93}, 6~Dra, $\chi$ Aur, HD 201193, 24 Aqr \citep{pourbaix04}, $\xi$ Cyg \citep{parsons98}, Schulte 15 \citep{kobulnicky14}, $\theta$ Cep, $\zeta$ Sct, \citep{batten78}, HD 18328 \citep{batten89}, TYC 7330-913-1 \citep{collins18}, etc.

\item \textit{Eclipsing binaries:} This type of binaries is detected when the orbital plane of the stars is aligned with our line of sight from Earth, causing them to eclipse each other once per orbit and resulting in a noticeable decrease in the system's brightness. The variation in brightness of an eclipsing binary is described by its light curve, and depending on its shape, there will be different type of eclipsing binaries. A primary eclipse occurs when the primary star (the brightest one) is obscured by the secondary star (the faintest one), while the secondary eclipse happens when the secondary star is eclipsed by the primary. If both stars have the same brightness, the eclipses will be symmetrical. However, if there is a significant difference in brightness, the secondary eclipse may be harder to detect. If the orbits are circular, the primary and secondary eclipses will occur at regular intervals, whereas elliptical orbits would cause the eclipses to be unevenly spaced. Eclipsing binaries are relatively common, with about 29\% of known variable stars being eclipsing binaries \citep{christy22}. 

\begin{figure}[H]
  \centering
  \includegraphics[width=0.9\linewidth, angle=0]{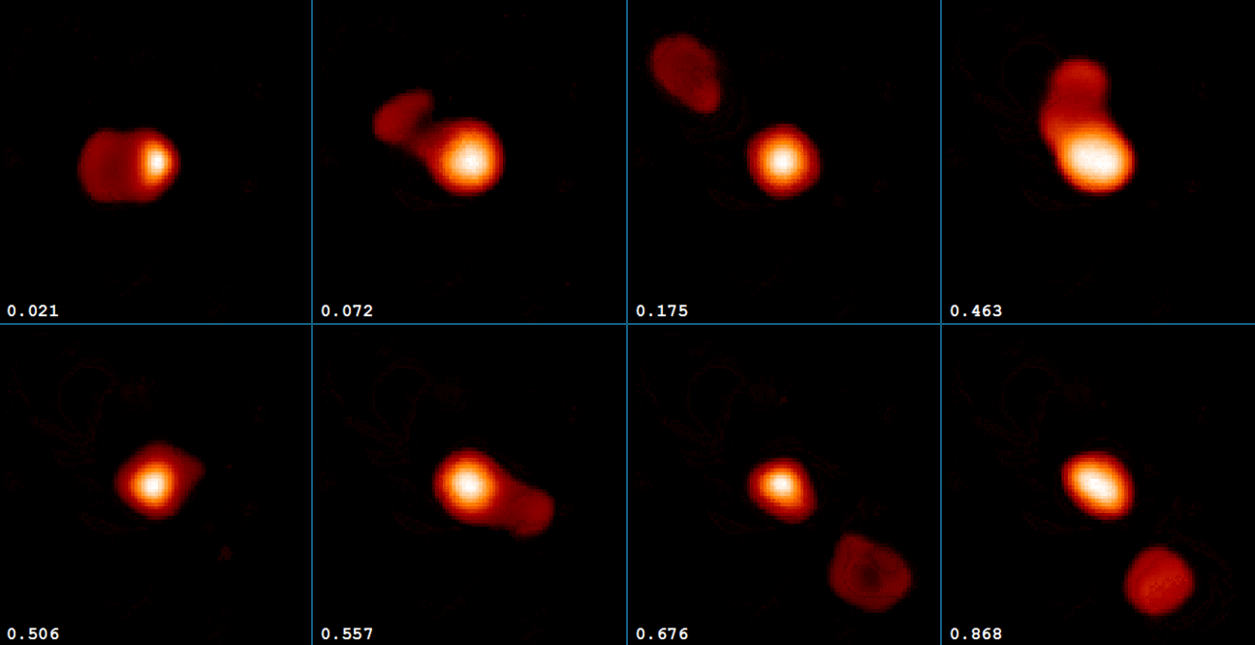}
     \caption[The eclipsing binary \textit{Algol} Aa1 / \textit{Algol} Aa2.]{\textit{Algol} ($\beta$ Per) is a confirmed three-star system (plus two more possible companions, \citealt{jetsu21}). In the image, the eclipsing binary \textit{Algol} Aa1 / \textit{Algol} Aa2. Their orbital plane contains the line of sight to the Earth. This set of images \citep{baron12} were obtained by CHARA \citep{mcalister88} interferometer in the near-infrared \textit{H}-band, sorted according to orbital phase. 
}
   \label{fig:algol}
\end{figure}

Examples of eclipsing binaries are: \textit{Algol} \citep[$\beta$ Per,][see Fig.~\ref{fig:algol}]{klagyivik13}, RV Crt \citep{pojmanski02}, RU Lep \citep{alfonsogarzon12}, BN Peg \citep{hoffman09}, V796 Cyg \citep{ijspeert21}, \textit{Sheliak} \citep[$\beta$ Lyr,][]{kabath09}.

\item \textit{Cataclysmic binaries:} Cataclysmic variable stars are binary stars that consist of two components; a white dwarf primary, and a mass transferring secondary. They are systems in which the sporadic ejection of mass from the extended atmosphere of one star onto the surface of the other leads to various behaviours, the most notable being violent outbursts of radiation. During these outbursts, the system's luminosity can increase by a factor of 100 (equivalent to five magnitudes) within a single day. Unlike novae, cataclysmic binaries experience periodic outbursts, allowing an external observer to witness multiple eruptions over several years \citep{mobberly09}. The most intense outbursts, however, are novae, where a star brightens by an enormous factor, typically around ten million times, within a few days, gradually fading over months or years \citep{smith06}. 

\begin{figure}[H]
  \centering
  \includegraphics[width=0.85\linewidth, angle=0]{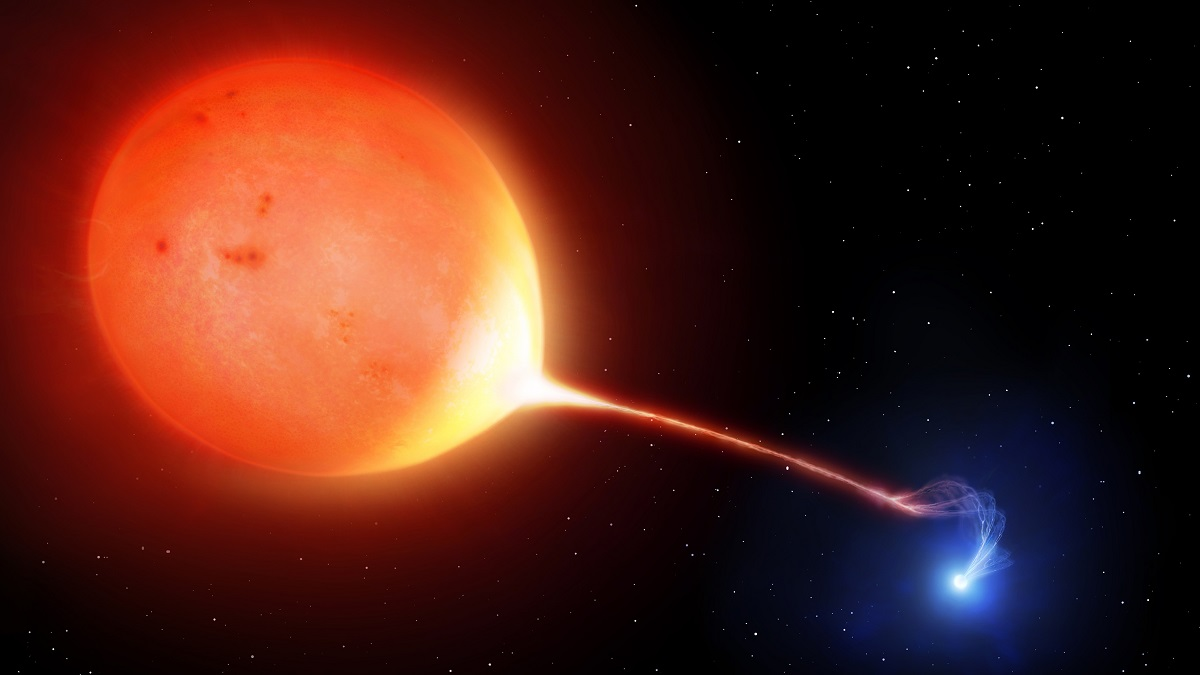}
     \caption[Artistic representation of a cataclysmic binary system including a red dwarf orbiting a white dwarf star.]{Artistic representation of a cataclysmic binary system including a red dwarf orbiting a white dwarf star. Credit: Mark Garlick / Science Photo Library / Getty.
}
   \label{fig:cataclysmic}
\end{figure}

The \textit{Atlas of Cataclysmic Variables} \citep{downes01}, compiled and finalised in February 2006, is a catalogue that contains 1830 cataclysmic variable stars or objects suspected of being cataclysmic variables based on their characteristics. It serves as an essential resource for astronomers and researchers studying this class of variable stars, which exhibit erratic changes in brightness due to various processes like mass transfer between a white dwarf and a companion star. 

Some examples are: 1ES 0851+39.2 \citep{samus03}, HE 0409--3029 \citep{downes01}, CTCV J0333--4451 \citep{tappert04}, PG 0248+054 \citep{green86}, V1033 Cas \citep{bird07}.

\item \textit{Other types of binaries:} In 2003, a double pulsar composing a binary system was discovered from the Parkes radio telescope in Australia. PSR J0737--3039 is the only known double pulsar. It consists of two neutron stars emitting electromagnetic waves in the radio wavelength in a relativistic binary system with an orbital period of 2.4 hours \citep{burgay03}. Additionally, some other exotic types of binaries have been discovered: A pulsar-white dwarf system, PSR B1620--26 \citep{arzoumanian96}; a pulsar-neutron star system, PSR B1913+16 \citep{johnston94}; and a pulsar and a main-sequence star, PSR J0045--7319 \citep{bell94}.
\end{itemize}

\subsubsection{According to their separation}
\label{sub:according_separation}

The most useful classification of binaries is based on the separation between their two components \citep{batten73}: close and wide binaries. This type of classification is not straightforward. In fact, close binaries are the origin of many significant phenomena, such as Type Ia supernovae \citep[e.g.][]{iben84,webbink84,maoz14}, which are of great importance because they have been used to demonstrate and measure the accelerated expansion of the Universe. Additionally, the recent detection of gravitational waves has provided a novel opportunity to observe the mergers of binary systems involving black holes \citep{abbot16} and neutron stars \citep{abbot17}. On the other hand, wide binaries are more susceptible to gravitational perturbations that can easily disrupt them \citep[e.g.][]{correaotto17,tian20}. This characteristic makes wide binaries a unique tool for investigating both the visible and invisible structures within the Galaxy.

The parameter that defines this classification is the separation between the stars in the binary system. Sometimes, the term ``close'' refers to binary systems where the components are close enough to influence or distort one another. Using this definition, all visual binaries and many spectroscopic binaries would be classified as ``wide'', and only those systems with short periods would be considered as ``close'' (Table~\ref{tab:different_periods_binaries} presents examples illustrating how widely can vary the periods of binary systems). The concept of close systems gained prominence in 1966 when two astrophysicists, Miroslav Plavec and Bohdan Paczy{\'n}ski, independently proposed a new perspective based on the influence one star exerts on the evolution of its companion \citep{paczynski66,plavec67}.

\begin{table}[H]
 \centering
 \caption{Binary systems with different orbital periods}
 \scriptsize
 \begin{tabular}{cc@{\hspace{10mm}}cc@{\hspace{10mm}}cc@{\hspace{10mm}}cc}
 \noalign{\hrule height 1pt}
 \noalign{\smallskip}
 
 \multicolumn{2}{c}{\textit{Very short period}} & \multicolumn{2}{c}{\textit{Intermediate period}} & \multicolumn{2}{c}{\textit{Long period}} & \multicolumn{2}{c}{\textit{Very long period}}\\ 
  \noalign{\smallskip}
 \hline
 \noalign{\smallskip}
  System & Period & System & Period & System & Period & System & Period \\
 \noalign{\smallskip}
 \hline
 \noalign{\smallskip} 
    HM Cnc & 5.4 minutes & \textit{Algol} & 2.87 days & \textit{Sirius} A/B &  50 years & $\epsilon$ Ind A/Bab & 89\,000 years \\
    W UMa & 8 hours & $\beta$ Lyr & 12.9 days & $\alpha$ Cen A/B & 79.8 years & Proxima / $\alpha$ Cen & 511\,000 years \\
    V1309 Sco & 1.4 days & X Per & 250 days & p Eri A/B & 468 years & \textit{Fomalhaut} A/C & 51\,800\,000 years \\
 \noalign{\smallskip}

 \noalign{\hrule height 1pt}
 \end{tabular}
 \label{tab:different_periods_binaries}
\end{table}
\normalsize

Alternative approaches to classifying binary stars were proposed by \citet{krat44}, \citet{struve51}, \citet{kopal55}, and \citet{sahade60}. Among them, Kopal’s classification is grounded in the concept of the Roche lobe \citep{roche1873}, which delineates the region around each star where its gravitational influence prevails. Depending on whether neither, one, or both stars fill their respective Roche lobes, binary systems are classified as detached, semi-detached, or contact. This physically motivated scheme is useful for tracing interaction-driven evolutionary processes and relies primarily on three parameters: the component masses, radii, and orbital separation.

Plavec and Paczyński built on this physical framework but introduced a more comprehensive classification that explicitly incorporates stellar evolution. While Kopal’s scheme considers structural and orbital parameters, it does not distinguish systems based on the evolutionary state or luminosity of the stars. Plavec addressed this by combining two key dimensions: the evolutionary status of each star, represented by its position on the Hertzsprung-Russell diagram, and its geometrical configuration within the binary (i.e., detached, semi-detached, or contact, following Kopal). This hybrid system enabled a more nuanced categorization by including luminosities along with masses, radii, and separations, bringing the total to seven parameters. However, Plavec emphasized that certain binary configurations cannot be continuously interpolated from one type to another by merely adjusting these numerical values, highlighting the complexity of evolutionary effects in close binaries.

It can be argued that, to date, there is no universally accepted classification system based on stellar separation \citep{batten67}. The distinction between close and wide is often based solely on their angular separation, but this strongly depends on the authors and the objectives of the studies. Therefore, there is no consensus or universal classification in this regard either. However, \citet{cifuentes25} state that the classification of close and wide pairs should be considered in a dynamic manner rather than as a fixed distinction, taking into account the detection limits and spatial resolution of the observational facilities used. A practical approach to defining wide pairs is to consider those that can be resolved under natural seeing conditions (1–2,arcsec) without relying on advanced techniques such as Adaptive Optics (AO) or Lucky Imaging (LI), both defined in Sect.~\ref{sec:direct_imaging}. Given its spatial resolution of 2-4,arcsec, the 2MASS survey provides a useful reference for identifying such wide pairs. In this study, we adopt this definition while acknowledging that the terms ``wide'' and ``close'' may sometimes refer to specific separations, especially when citing previous works.

Among wide binaries, we focus on a particularly extreme subset known as ``ultrawide'' or ``very wide'' binaries, characterised by exceptionally large separations. The definition of what constitutes an ultrawide binary varies across the literature: some authors adopt purely distance-based criteria such as 0.1\,pc \citep{tolbert64,kraicheva85,abt88,caballero09}, 1\,pc \citep{jiang10,caballero10}, or even between 1 and 8\,pc \citep{shaya11,kirkpatrick16,gonzalezpayo21}, while others define them based on binding energy thresholds or in the context of theoretical frameworks involving Galactic dynamics \citep[10$^{\text{33}}$\,J as in][]{caballero09,caballero10}. As such, the concept of ultrawide binaries remains fluid and context-dependent, but it is crucial for understanding the limits of binary formation and survival in the Galactic environment.

\subsection{Detection methods}
\label{sec:detection_methods}

Detecting binary stars is a fundamental aspect of astrophysics. While some binary stars can be directly observed as separate objects, many remain hidden due to their close separations, extreme brightness differences, or orientation relative to Earth. To overcome these challenges, astronomers use a variety of detection methods, each suited to different types of binaries and observational conditions.

\subsubsection{Direct imaging}
\label{sec:direct_imaging}

Direct imaging methods are among the most fundamental techniques for detecting binary stars, as they allow astronomers to directly observe both stellar components in a system. However, their effectiveness depends on the separation of the stars, their brightness contrast, and the observational conditions. Several advanced techniques have been developed to improve the resolution and clarity of images, making it possible to detect even close or faint binary companions.

Each of these direct imaging techniques plays a crucial role in detecting and studying binary stars. While wide binaries can often be detected with conventional optical telescopes, closer or fainter companions require advanced methods such as speckle interferometry, adaptive optics, and infrared imaging. By combining these approaches, astronomers can gain a more complete understanding of binary star systems, their orbital characteristics, and their evolution over time.

\begin{itemize}
    \item \textit{Optical imaging:} The most straightforward approach is optical imaging, where a telescope captures the light from a binary system and attempts to resolve the two stars as distinct points of light. This method is effective for wide binaries, where the stars are sufficiently far apart for conventional telescopes to distinguish them individually. A classic example is \textit{Sirius}, the brightest star in the night sky, which is actually a binary system consisting of a main-sequence star (\textit{Sirius} A) and a white dwarf (\textit{Sirius} B). Despite its proximity, \textit{Sirius} B was not directly observed until 1862 due to its faintness compared to \textit{Sirius} A. Modern telescopes can now easily resolve such systems, but for closer binaries, additional techniques are required.

    \item \textit{Micrometric method:} The micrometric method for detecting binary stars is one of the oldest techniques used in observational astronomy. It relies on direct visual observations using a filar micrometer, an instrument attached to a telescope to precisely measure the angular separation between two stars. A telescope equipped with a filar micrometer is used to measure the angular separation of the binary system. The micrometer consists of a set of fine adjustable threads (wires) in the eyepiece. By aligning these threads with the two stars, astronomers measure the angular distance between them. The position angle (the orientation of the binary system relative to a reference direction, usually north) is also recorded. By regularly measuring the separation and position angle over time, astronomers can track the orbital motion of the stars. These data allow them to determine the system's orbital parameters, such as eccentricity and period.

    A filar micrometer can be built with two webs, needlepoints, or reticle lines that are positioned at the focal plane. The filar micrometer shown in the left part Fig.~\ref{fig:filarmicrometer} is built with two webs in focus with the image, and magnified. Once the image is aligned at the focal plane of the eyepiece, the telescope drives are adjusted to center the object, Mars, between the movable (M) and centerline (T) webs, as well as the fixed (F) and centerline (T) webs. The separation is then measured using the micrometer thimble and spindle and recorded in either millimeters (mm) or fractional inches.

    \begin{figure}[H]
    \centering
     \includegraphics[width=0.8\linewidth, angle=0]{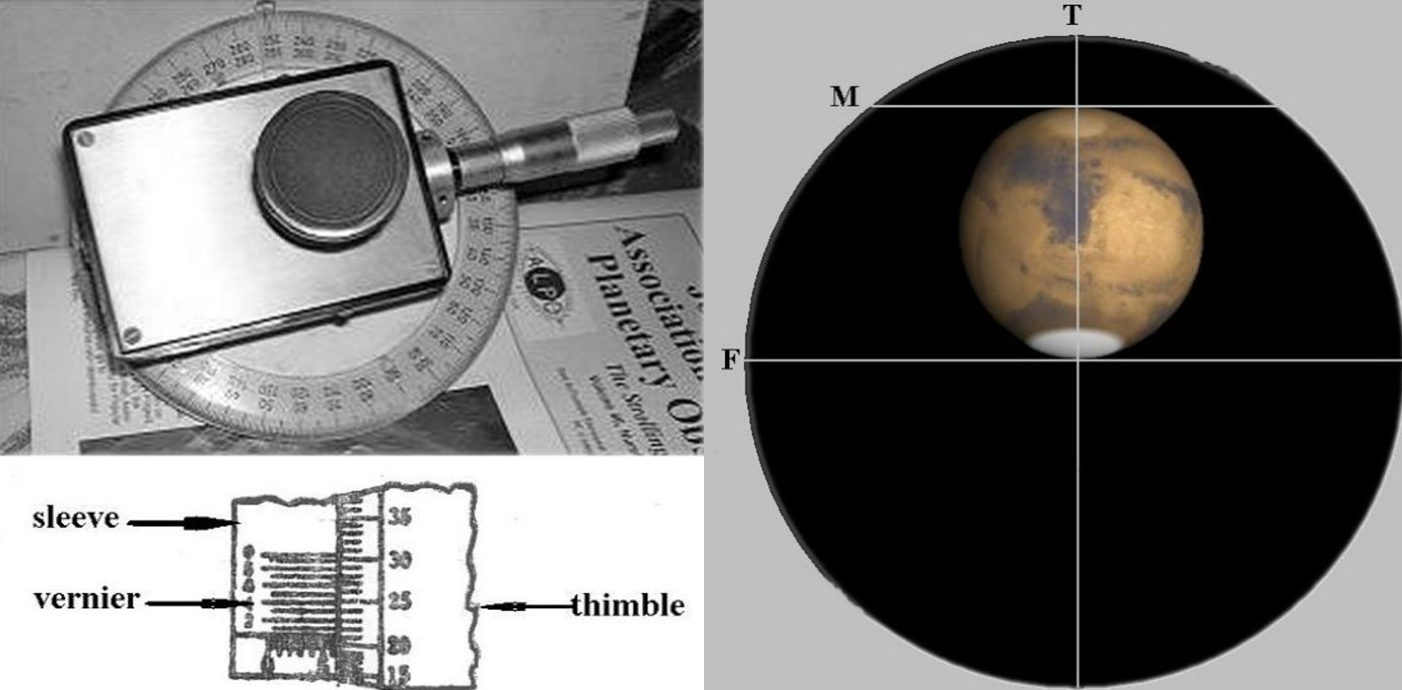}
     \caption[A Bi-Filar micrometer produced by Ron Darbinian.]{\textit{Left panel:} A Bi-Filar micrometer produced by Ron Darbinian. It was available with either a 12mm or 27mm eyepiece mounted on the black box housing the webs and their holding mechanism. On the right, the silver handle serves as the micrometer thimble and spindle, allowing adjustments to the movable web. Behind the black box, a protractor is positioned, capable of rotating 360\,deg to measure the position angle of double stars \citep{byrne84}. \textit{Right panel:} Appearance of Mars in the micrometer eyepiece. (Credit: \url{https://www.alpo-astronomy.org/jbeish/Observing_Mars_10.html}).}
     \label{fig:filarmicrometer}
    \end{figure}    
    
    \item \textit{Speckle interferometry:} This method helps to overcome the limitations imposed by Earth's atmosphere. Normally, atmospheric turbulence causes a blurring effect that limits the resolving power of ground-based telescopes, making it difficult to distinguish very close binary stars. Speckle interferometry, proposed by \citet{labeyrie70}, solves this problem by taking many rapid, short-exposure images, effectively ``freezing'' the momentary distortions caused by the atmosphere. The\-se images are then processed computationally to reconstruct a much sharper view of the system. This technique has been used to confirm the binary nature of \textit{Betelgeuse}, a red supergiant in Orion \citep[e.g.][]{kluckers97}, as well as to study stars in globular clusters where many close binaries exist. It allows astronomers to resolve binary systems with angular separations as small as 0.02 arcseconds, that is the current limit using $\sim$8\,m telescopes, which is equivalent to resolving a coin from a distance of several kilometres.
    
    \item \textit{Adaptive optics (AO):} Another powerful technique for direct imaging is adaptive optics, a technology that dynamically corrects for atmospheric distortion in real time. Using a deformable mirror that changes shape thousands of times per second, AO can compensate for the blurring effects of turbulence \citep{roddier99}, producing images with resolutions comparable to those obtained from space telescopes. This method is particularly valuable for detecting faint companions close to bright stars, as the extreme brightness difference would otherwise make the secondary star invisible. One of the most famous cases where AO played a crucial role is the study of Gliese 229, a binary system where the companion, Gliese 229 B, was identified as one of the first confirmed brown dwarfs \citep{nakajima95}. Since brown dwarfs do not generate enough fusion to shine brightly in visible light, they are extremely difficult to detect using conventional imaging techniques, but adaptive optics allowed astronomers to separate it from its brighter primary star.
    
    \item \textit{Lucky imaging (LI):} Lucky imaging is an advanced astronomical technique used to enhance the resolution of ground-based telescopes by minimising the effects of atmospheric turbulence. It is particularly useful for detecting close binary stars and resolving faint companions near bright stars. Its principle is based on the fact that the Earth's atmosphere distorts light from celestial objects, causing images to appear blurry when captured with long exposures. Lucky imaging takes advantage of rapid, high-speed imaging to capture thousands of short-exposure frames (typically a few milliseconds each). Most of these frames are distorted, but a small fraction ($\sim$1--10\%) will be nearly unaffected by atmospheric turbulence \citep{fried78}. These ``lucky'' frames are then selected and combined to produce a final, sharp image.

    The advantages are evident: LI helps distinguish very close binary stars that would otherwise appear as a single star in traditional observations. Also, by tracking the positions of binary components over time, astronomers can determine orbital parameters and masses. And, finally, the identification of faint companions is easier since the technique enhances the detection of low-mass stars, brown dwarfs, and exoplanets orbiting bright stars. The Nordic Optical Telescope (NOT) and the \textit{William Herschel} Telescope (WHT) have successfully used LI to resolve binary stars in crowded stellar fields. The AstraLux camera on the Calar Alto Observatory's 2.2m telescope has detected faint white dwarf companions around bright main-sequence stars.
    \item \textit{Coronography:} Coronagraphy is a high-contrast imaging technique originally developed for studying the solar corona by blocking direct starlight. This method has been adapted for astronomical observations to detect faint companions, such as binary stars, exoplanets, and circumstellar disks, by suppressing the overwhelming glare of a primary star.   A coronagraph consists of an optical mask or occulter that blocks the light from a bright central star, allowing astronomers to detect fainter objects in its immediate vicinity. Modern coronagraphs use adaptive optics and post-processing techniques to further enhance contrast and remove residual starlight.

    Coronagraphs enable the identification of stellar companions that would otherwise be hidden in the glare of a bright primary star. This is particularly useful for detecting sub-arcsecond separation binaries. By suppressing the primary star’s light, astronomers can obtain more accurate photometry and spectroscopy of low-mass companions, including brown dwarfs and white dwarfs in binary systems. Observations over time allow the tracking of companion movement, helping refine orbital solutions and determine dynamical masses.

    It has some limitations: Coronagraphy is less effective for detecting very close binaries with separations below the instrument's inner working angle. It also requires high-precision optics and post-processing techniques to suppress residual speckle noise. another limitation is that it works best in the infrared, where contrast between a primary star and a low-mass companion is higher.

    \item \textit{Space imaging:} Space-based telescopes have revolutionised the detection and characterisation of binary star systems by eliminating atmospheric distortions and providing access to wavelengths that are challenging to observe from the ground. These advantages make space imaging one of the most effective methods for studying binary and multiple stellar systems. The reason is that ground-based telescopes suffer from atmospheric turbulence, which limits angular resolution and contrast, making it difficult to resolve close binary systems. Space telescopes, free from these distortions, provide sharper images and more precise astrometric measurements. 
    
    The advantages of space imaging are their high angular resolution to resolve binaries with very small separations, their stable, long-term astrometry to detect binary motion. Their broad wavelength coverage, including ultraviolet and infrared, to study different stellar components is also a bonus.

    Limitations of space imaging are cost and accessibility, field of view (many space telescopes focus on specific targets rather than wide-field surveys), and resolution limits because despite high precision, some extremely close binaries may still require interferometric techniques.

    The most used techniques in space imaging for binary detection are: direct imaging, interferometry, coronagraphy, astrometric wobble detection, and spectroscopy. Examples of space missions are \textit{Hubble} Space Telescope\footnote{\url{https://science.nasa.gov/mission/hubble/}}, \textit{Spitzer} Space Telescope \citep{werner04}, and the \textit{James Webb} Space Telescope \citep{mcelwain23}.

    \item \textit{Multiwave imaging:} Multiwavelength imaging is a powerful technique for detecting and characterising binary and multiple stellar systems. By observing a system at different wavelengths-ranging from X-rays to radio-astronomers can gain deeper insights into the physical properties of the stars, their interactions, and the presence of companions that might be otherwise undetectable in a single wavelength band.

    One of the main advantages of multiwavelength imaging in binary detection is its ability of detecting hidden companions because they may be too faint in optical wavelengths but can be more easily detected in infrared (IR) due to their thermal emission. Also, X-ray observations can reveal young or active stars in a binary, as they often exhibit strong coronal activity.

    Different wavelengths provide clues about the age and composition of stars. For instance, ultraviolet (UV) observations can highlight hot, massive stars, while infrared can reveal cool, low-mass stars and dust disks. White dwarf companions, which are difficult to detect optically, can be found in the UV. Infrared and submillimeter imaging help identify disks around binary stars, providing insights into planetary formation and disk evolution in multiple systems.

    Other advantages are that by comparing observations at different wavelengths, astronomers can infer the presence of additional components in hierarchical multiple systems. And young stellar objects and protostellar binaries can be identified in star-forming regions using infrared and millimeter imaging.
    
\end{itemize}

\subsubsection{Astrometric methods}
\label{sec:astrometric_methods}

\begin{figure}[H]
    \centering
     \includegraphics[width=0.7\linewidth, angle=0]{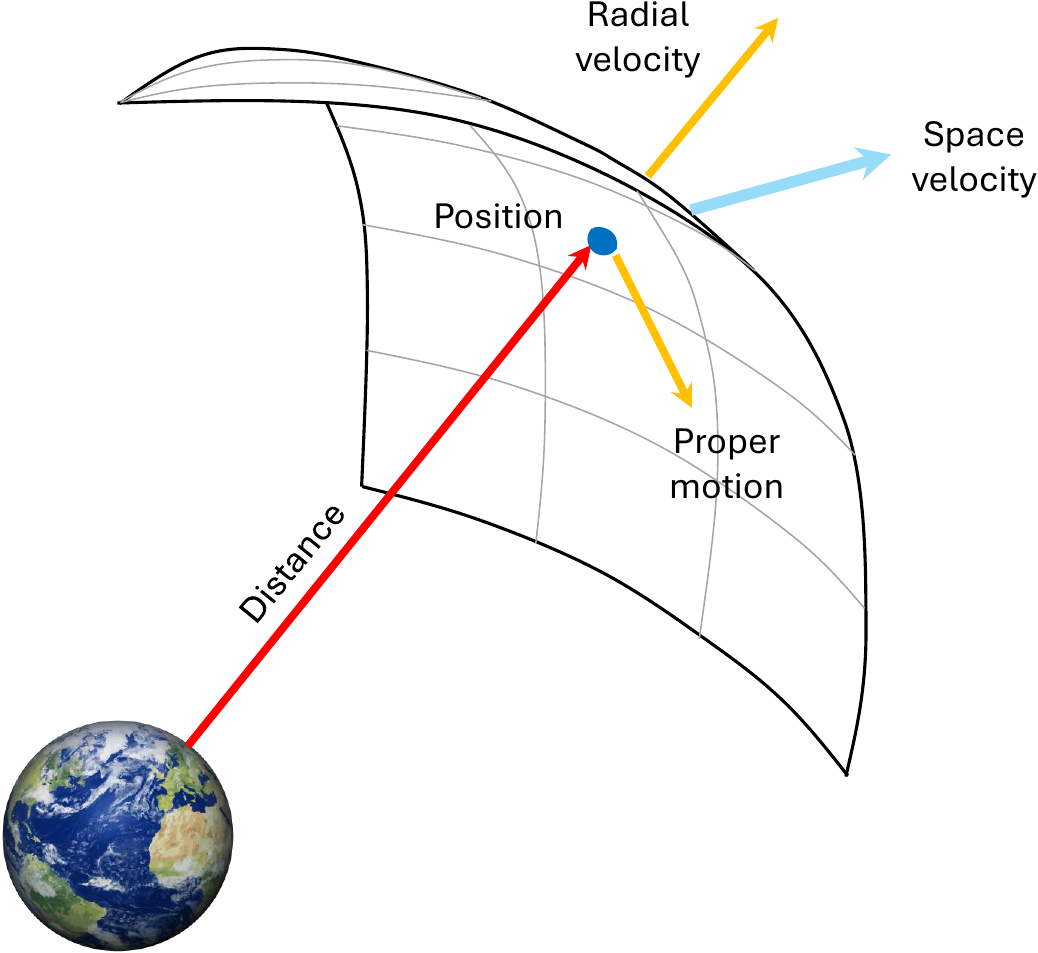}
     \caption[Graphical explanation of proper motion and radial velocity.]{Graphical explanation of proper motion and radial velocity: The curved grid represents a portion of the celestial sphere. Proper motion is how an object appears to move sideways across the sky. Radial velocity is how an object appears to move towards or away from us (plotted by the author based on the figure from  \url{https://cosmosatyourdoorstep.com/2018/02/19/stars-and-proper-motion/}).}
     \label{fig:propermotion}
    \end{figure}

Astrometry, the precise measurement of a star’s position in the sky over time, is a powerful tool for detecting binary stars, especially when one of the components is too faint or too close to be directly observed. The space velocity of a star, its true motion through space relative to the Sun, can be decomposed into radial velocity that is the motion along the line of sight, measured via Doppler shift, and proper motion that is the motion across the sky, measured in angular change over time (see Fig.~\ref{fig:propermotion}). A star’s motion across the sky, described by its proper motion, can reveal subtle deviations from a straight trajectory. When a visible star has an unseen companion, its gravitational influence can cause the primary star to exhibit a small, periodic motion superimposed on its proper motion, known as an astrometric wobble. By precisely tracking these positional changes over time and analysing deviations from uniform proper motion, astronomers can infer the presence of a binary companion. From this, they can determine the orbital parameters and even estimate the mass of the unseen object. One of the earliest successful applications of astrometry in binary detection was the discovery of \textit{Sirius} B, the white dwarf companion to \textit{Sirius} A. Long before it was visually confirmed in 1862, astronomers had noticed that \textit{Sirius}’s motion showed irregularities inconsistent with that of a solitary star. These unexplained deviations in its proper motion led to the prediction of an unseen companion, which was later confirmed with improved telescopes.

Modern astrometric studies have improved with space-based observatories. The \textit{Hipparcos} mission, launched by the European Space Agency (ESA) in 1989, provided precise astrometric measurements for over 100\,000 stars \citep{perryman97}, leading to the detection of numerous binary systems. However, the most significant advancement in astrometric binary detection has come from \textit{Gaia}, the ESA mission launched in 2013 (see Sect.~\ref{sec:gaia}). \textit{Gaia} has revolutionised our understanding of binary stars by tracking the minute movements of over a billion stars with unprecedented precision. Through its data, astronomers have identified thousands of binary systems, including cases where the companion is a dim red dwarf, a brown dwarf, or even an unseen exoplanetary-mass object.

Astrometric methods are particularly useful for detecting long-period binaries, where the stars orbit each other over decades or centuries. A key example is Proxima Cen, the closest known star to the Sun. Astrometric studies have provided strong evidence that Proxima Cen is part of a wide binary or even a triple system with $\alpha$ Cen A and B. Similarly, the motion of Barnard’s Star, one of the nearest stars to Earth, has been extensively studied for signs of unseen companions.
Additionally, astrometry has been instrumental in identifying black hole binaries where the visible star orbits an unseen, massive object. In 2022, \textit{Gaia} data revealed Gaia BH1, a binary system where a Sun-like star exhibits an astrometric wobble indicative of an orbiting black hole approximately ten times the mass of the Sun \citep{elbadry23}. This discovery marked one of the first detections of a black hole through purely astrometric means.
By continuously refining measurement precision, astrometry remains a crucial method for detecting and characterising binary stars, particularly in cases where one of the components is faint, invisible, or non-luminous. As data from \textit{Gaia} continues to be analysed, even more hidden binary systems will likely be uncovered, deepening our understanding of stellar evolution and dynamics.

\subsubsection{Spectroscopic methods}
\label{sec:spectroscopic_methods}

Spectroscopic methods are among the most effective techniques for identifying and characterising binary stars, particularly spectroscopic binaries, which are too close to be resolved as separate objects through direct imaging. Instead of observing two distinct stars, astronomers detect the presence of a binary system by analysing shifts in the spectral lines of the star due to the Doppler effect. As the two stars orbit their common centre of mass, their radial velocities (motion along the observer’s line of sight) change periodically, causing their spectral lines to shift back and forth. The technique has been instrumental in identifying exoplanets, white dwarfs, neutron stars, and black holes.

We already mentioned the type of binaries called spectroscopic binaries (SB1, SB2, ST3) in Sect.~\ref{sec:according_detection_method}. In many cases, only one star in the binary system is bright enough for its spectrum to be detected, making it a single-lined spectroscopic binary (SB1). The unseen companion can be inferred by the periodic shifts in the observed star’s spectral lines. The amplitude and period of these shifts allow astronomers to determine the orbital parameters and place constraints on the mass of the unseen companion. A well-known example of an SB1 system is \textit{Helvetios} (51~Peg), which was initially thought to be a single star but was later found to exhibit periodic Doppler shifts in its spectrum. While this particular case led to the first confirmed discovery of an exoplanet (51~Peg b, \citealt{mayor95}), the same method is widely used for identifying stellar companions, including unseen white dwarfs, neutron stars, and black holes. Another example is HD 114762 \citep{latham89}, a SB1 system where radial velocity measurements suggested the presence of a companion with a minimum mass near the boundary between a brown dwarf and a low-mass star.

In some binary systems, the spectra of both stars are visible, making them double-lined spectroscopic binaries (SB2). In these cases, astronomers observe two sets of spectral lines that shift in opposite directions as the stars orbit each other. By analysing these shifts, astronomers can determine the mass ratio of the two stars and obtain more complete orbital information. A classic example of an SB2 system is \textit{Mizar}~A, a component of the famous \textit{Mizar}-\textit{Alcor} (79-80~UMa) system in the Big Dipper. In 1889, \textit{Mizar} A was the first spectroscopic binary ever discovered when astronomers noticed its spectral lines periodically splitting and merging, indicating the presence of two stars orbiting each other. Another well-known SB2 system is \textit{Algol} ($\beta$ Per), an eclipsing and spectroscopic binary where both stars’ spectra contribute to the observed shifts. First presented in 1881 by astronomer Edward Charles Pickering as an eclipsing binary \citep{astronomicalregister1881}, Hermann Carl Vogel confirmed this hypothesis in 1889 by detecting periodic changes in \textit{Algol}'s spectrum, attributed to the Doppler effect \citep{batten89b}, which allowed the inference of variations in the radial velocity of the binary system.

\subsubsection{Photometric methods}
\label{sec:photometric_methods}

Photometric methods involve the precise measurement of a star's brightness over time. These techniques are especially useful for detecting eclipsing binaries, where one star periodically passes in front of the other, causing a measurable dip in brightness. By analysing these variations in luminosity, astronomers can determine key orbital parameters, estimate stellar radii, and even infer the presence of unseen companions.

\begin{itemize}
    \item \textit{Eclipsing binaries and light curves:} When two stars in a binary system orbit each other along a line of sight aligned with Earth, they can undergo mutual eclipses, where one star blocks the light of the other. This produces characteristic periodic dips in the system’s brightness, which can be recorded as a light curve, a graph showing the star’s brightness as a function of time, see Fig.~\ref{fig:lightcurve}. Light curves of eclipsing binaries typically show two dips per cycle: The primary eclipse occurs when the brighter star is eclipsed by the fainter one, causing a significant drop in brightness. The secondary eclipse happens when the dimmer star is obscured by the brighter one, resulting in a smaller decrease in luminosity. The shape and depth of these dips provide essential information about the stars' relative sizes, brightness, and inclination of the orbit.

    Additionally, the analysis of light curves allows astronomers to determine the orbital period and eccentricity of the system. If the stars are nearly equal in brightness, the depth of both eclipses may be similar, whereas a large difference in luminosity results in a more pronounced primary eclipse. By combining photometric data with spectroscopic measurements, researchers can also derive the masses of both stars, leading to a better understanding of their evolutionary stages and physical properties.

    \begin{figure}[H]
    \centering
    \includegraphics[width=1\linewidth, angle=0]{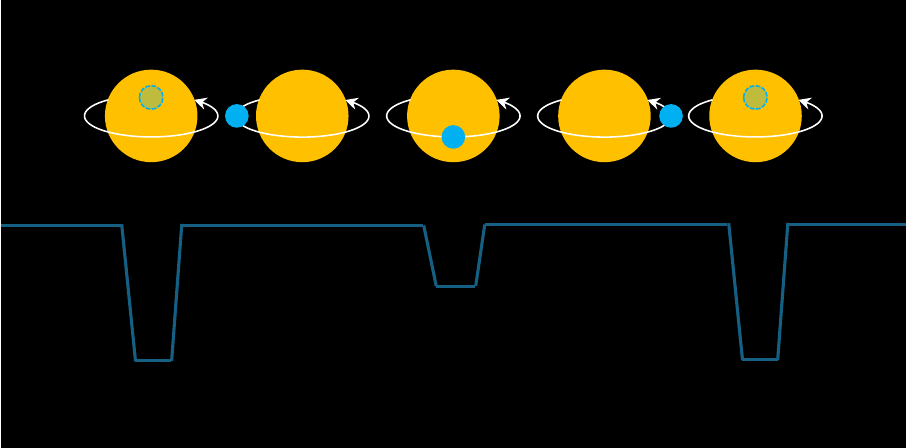}
    \caption[Light curve of eclipsing binary stars.]{Light curve of eclipsing binary stars. Based on fig.~17-24 from \citet{freedman08}.
    }
    \label{fig:lightcurve}
    \end{figure}    

    \item \textit{Ellipsoidal variability and reflection effects:} Not all photometric variations in binary systems are due to eclipses. Some binaries show periodic brightness changes due to tidal distortions and reflection effects: In ellipsoidal variables, the gravitational interaction between the two stars deforms them into slightly elongated shapes. As they orbit, the projected area facing Earth changes, leading to variations in brightness. In some close binaries, one star reflects light from its companion, creating a periodic increase in brightness as the reflecting side becomes more visible. An example of an ellipsoidal variable is HM Cnc (RX J0806.3+1527), a binary system composed of two white dwarfs \citep[e.g.][]{macfarlane15}. Data from some X-ray satellites such as \textit{Chandra} X-Ray Observatory, XMM-\textit{Newton} and the \textit{Swift} Gamma-Ray Burst Mission, suggest that the orbital period of the two stars is decreasing at a rate of 1.2 milliseconds per year, getting closer by approximately 60 centimetres per day, having an ultra-short-period orbit of just 5.4 minutes \citep{hakala03,israel04,esposito14}. At this rate, they can be expected to merge in approximately 340\,000 years.

\end{itemize}

\subsubsection{High-energy observation methods}
\label{sec:High-energy_observation_methods}

High-energy astrophysics provides some important tools for detecting and studying binary star systems, particularly those containing compact objects such as neutron stars and black holes. They are mainly used from modern space observatories like \textit{Chandra}, XMM-\textit{Newton}, \textit{Fermi}, and \textit{Swift} to uncover the nature of binary systems that remain invisible in optical wavelengths. Some examples where these methods helped to discover X-ray binaries are Cygnus X-1, one of the first identified black hole candidates. It is a high-mass X-ray binary \citep{webster72,bolton72} where a massive star transfers material onto an invisible companion, later confirmed to be a black hole; and Sco X-1, the first discovered X-ray source beyond the Solar System, identified as a low-mass X-ray binary with a neutron star accreting from a less massive companion \citep{shklovsky67}. For pulsar binaries we can mention PSR B1913+16 (The Hulse-Taylor Pulsar), a binary system containing two neutron stars; or PSR J0348+0432, a pulsar orbiting a white dwarf, used to test general relativity in strong gravitational fields. The \textit{Fermi} Gamma-ray Space Telescope and the \textit{Neil Gehrels Swift} Observatory are key instruments for detecting pulsars in binary systems.

In this category the relativistic jets and radio observations can be included. Some high-energy binary systems produce powerful relativistic jets-narrow beams of high-speed particles emitting across the electromagnetic spectrum. These jets are often detected in X-rays and radio waves. An example of jet-producing binary systems is
SS 433 \citep{hillwig04,cherepashchuk21}, a microquasar system where a compact object (likely a black hole) launches relativistic jets. This was one of the first sources to show clear Doppler-shifted emission lines from an astrophysical jet. The Very Large Array (VLA) and the Event Horizon Telescope (EHT) contribute to studying relativistic jets in binary systems.

\subsubsection{Overview of detection methods}
\label{sec:overview_detection_methods}

In recent decades, substantial technological and methodological advances have enhanced our ability to detect and characterise binary and multiple star systems. Innovations in instrumentation, data analysis, and theoretical modeling have not only expanded the range of detectable systems but also improved the precision of fundamental stellar parameters. This section presents a comparative overview of the principal detection techniques for binaries, highlighting their respective strengths, limitations, and effective ranges in terms of orbital separation.

Binary detection methods can broadly be categorised by the type of observable they rely on: spectroscopic, astrometric, photometric, and imaging techniques. No single method is universally applicable across all binary separations and system configurations; instead, each technique is optimised for particular regimes. 

\begin{itemize}
    \item \textit{Spectroscopic methods:} Spectroscopic techniques are highly effective for detecting close binaries with separations less than $\sim$10 au \citep{abt90,latham02,melo03,sana12,moe19,kounkel19}. By measuring periodic Doppler shifts in the spectral lines of one or both components, these methods allow for the derivation of orbital parameters and minimum masses. However, they are limited to systems with favorable orbital inclinations and often cannot resolve individual components.
    \item \textit{Astrometric Methods:} Astrometry offers a complementary approach, detecting the motion of a star’s position on the sky due to the gravitational influence of an unseen companion. Thanks to space-based missions like \textit{Gaia}, this method is now effective for separations ranging from $\sim$1 to over 20\,000\,au  \citep{belokurov20,mazzola20,dieterich10}. While highly sensitive to wide systems, its ability to detect close binaries is constrained by angular resolution and observational baselines.
    \item \textit{Photometric methods:} Photometry can identify eclipsing binaries and transiting planets by detecting periodic dimming in a star’s light curve. This method is best suited for systems with orbital planes aligned along our line of sight and can yield precise orbital periods and relative radii. Missions like TESS and Kepler have greatly expanded this technique’s reach \citep{offner23}, although it remains inherently biased toward edge-on systems.

    \item \textit{Interferometric and imaging techniques:} High-resolution imaging, including speckle imaging \citep{tokovinin20}, Lucky Imaging \citep{law08,bergfors10,janson12}, adaptive optics \citep{shatsky02,close03,derosa14,ward15}, and aperture masking interferometry \citep{kraus08}, is effective for resolving binaries at intermediate separations (10 to 1000\,au). For even closer systems (1 to 300\,au), long-baseline interferometry provides angular resolutions down to milliarcseconds \citep{raghavan10,rizzuto13,sana14}.
    \item \textit{Wide binaries and proper motion companions:} At separations greater than 300 au and up to $\sim$20\,000\,au, wide binaries can be identified by common proper motion. The confirmation of gravitational binding often relies on astrometric consistency over long timescales \citep{lepine07a,elbadry18,winters19,hartman20,gonzalezpayo23}. As already stated, their detection requires them to be spatially resolved and to have measurable proper motions.
    \item \textit{Complementary and indirect techniques:} Other methods, such as excess luminosity \citep{bardalezgagliuffi14,niu23} or spectral deblending \citep{gullikson16}, can indicate binarity but often cannot constrain orbital parameters or separations. In many cases, combining multiple methods yields the most comprehensive results \citep{winters19,raghavan10,tokovinin14a}.
\end{itemize} 

The convergence of ground-based and space-based technologies, from optical interferometers and adaptive optics to missions like \textit{Gaia}, TESS, \textit{Hubble} Space Telescope, and \textit{James Webb} Space Telescope, has significantly expanded the detection space. These advances have enabled discoveries of binary systems across the electromagnetic spectrum, including systems with ultra-cool dwarfs or planetary companions \citep{erskine07,mahadevan08,mann11,gallenne23}. Emerging frontiers include gravitational wave detection from interacting binaries and precise dynamical mass determinations through high-angular-resolution imaging. These efforts are driving a renaissance in binary star studies, offering new insights into stellar formation and evolution, and providing critical empirical constraints for theoretical models \citep{guinan07}. Table~\ref{tab:high_spatial_resolution_instruments} shows the properties of the current and future main high spatial resolution instruments. 

In summary, each detection method occupies a particular niche in the multidimensional parameter space of binary stars. Continued synergy among techniques is essential to achieving a holistic understanding of stellar multiplicity.

\begin{table}[H]
 \centering
 \caption[Properties of some representative high spatial resolution instruments chronologically ordered.]{Properties of some representative high spatial resolution instruments chronologically ordered.}
 \scriptsize
 \begin{tabular}{l@{\hspace{2mm}}l@{\hspace{2mm}}c@{\hspace{1mm}}c@{\hspace{3mm}}p{1.1cm}@{\hspace{2mm}}p{5.2cm}}
 \noalign{\hrule height 1pt}
 \noalign{\smallskip}
 Name & Location & Year & Resolution &  Technique & Reference$^{\text{(a)}}$ \\
 &  &  & (arcsec) &   &  \\ 
 \noalign{\smallskip}
 \hline
 \noalign{\smallskip} 
 \multicolumn{6}{c}{\textit{Instruments}}  \\ 
 \noalign{\smallskip}
 \hline
 \noalign{\smallskip} 
    NPOI & Lowell Obs. Arizona, USA & 1994 & 0.0017 & IF, O & \url{https://lowell.edu/research/telescopes-and-facilities/npoi/} \\
    \noalign{\smallskip}
    Keck AO system & Keck I \& II Obs. Hawaii, USA & 1999 & 0.04 & AO, IF, O, NIR & \url{https://www2.keck.hawaii.edu/optics/ao/} \\
    \noalign{\smallskip} 
    NAOMI & WHT. La Palma, Spain & 1999 & 0.6 & NIR, C, AO, O & \url{https://www.ing.iac.es//~docs/wht/ingrid/wht-ingrid-2/index.html} \\
    \noalign{\smallskip}
    GRAVITY (VLTI) & VLT, Cerro Paranal. Chile & 2001 & 0.002 & IF, O, NIR & \url{https://www.eso.org/sci/facilities/paranal/instruments/gravity.html} \\
    \noalign{\smallskip} 
    NIRC2 & Keck II Obs. Hawaii, USA & 2001 & 0.035 & NIR, C, AO & \url{https://www2.keck.hawaii.edu/inst/nirc2/}\\
    \noalign{\smallskip} 
    OSIRIS & Keck I Obs. Hawaii, USA & 2005 & 0.05 & NIR, AO & \url{https://www2.keck.hawaii.edu/inst/osiris/} \\
    \noalign{\smallskip}
    FastCAM & \textit{Carlos Sánchez} Telescope. Spain;  & 2006 & 0.1 & LI, O & \url{https://iac.es/en/projects/fastcam} \\
     & Nordic Optical Telescope. Spain & & & & \\
    \noalign{\smallskip}
    AstraLux & Calar Alto. Spain & 2007 & 0.1 & LI, O & \url{https://www2.mpia-hd.mpg.de/ASTRALUX/} \\
    \noalign{\smallskip}    
    VEGA (CHARA Array) & Mt. Wilson Obs. California, USA & 2007 & 0.0003 & IF, O & \url{https://www.chara.gsu.edu/instrumentation/vega} \\
    \noalign{\smallskip}    
    ACAM & WHT. La Palma, Spain & 2009 & 0.6 & LI, O & \url{https://www.ing.iac.es/Astronomy/instruments/acam/index.html} \\
    \noalign{\smallskip}
    ALMA & Atacama, Chile & 2011 & 0.5 & IF & \url{https://www.almaobservatory.org/en/home/} \\
    \noalign{\smallskip}
    GeMS & Gemini ST. Cerro Pachón, Chile & 2011 & 0.05 & AO, C, NIR & \url{https://www.gemini.edu/instrumentation/gems} \\
    \noalign{\smallskip}    
    SPHERE & VLT. Cerro Paranal, Chile & 2014 & 0.03 & NIR, C, AO, O & \url{https://www.eso.org/sci/facilities/paranal/instruments/sphere.html} \\
    \noalign{\smallskip}
    SCExAO & Subaru Obs. Hawaii, USA & 2015 & 0.02 &  AO, NIR, O & \url{https://www.naoj.org/Projects/SCEXAO/scexaoWEB/000home.web/indexm.html} \\
    \noalign{\smallskip}    
    Alopeke & Gemini NT. Hawaii, USA & 2018 & 0.02 & SPI, O & \url{https://www.gemini.edu/instrumentation/alopeke-zorro} \\
    Zorro & Gemini ST, Cerro Pachón. Chile & 2018 & 0.02 & SPI, O & \url{https://www.gemini.edu/instrumentation/alopeke-zorro} \\
    GPI 2.0 & Gemini ST. Cerro Pachón, Chile & 2025 & 0.02 & C, AO, NIR & \url{https://www.gemini.edu/instrumentation/future-instruments/gpi2} \\
    \noalign{\smallskip}      
    ELT (planned) & Cerro Armazones, Chile & 2028 & 0.005 & C, AO, O, NIR & \url{https://elt.eso.org/} \\
    \noalign{\smallskip}
    TMT (planned) & Mauna Kea, Hawaii, USA & 2030 & 0.005 & C, AO, O & \url{https://www.tmt.org/} \\
    \noalign{\smallskip}
    GMT (planned) & Las Campanas, Chile & 2030 & 0.01 & C, AO, O, NIR & \url{https://giantmagellan.org/} \\
    \noalign{\smallskip}
 \hline
 \noalign{\smallskip}       
 \multicolumn{6}{c}{\textit{Space Telescopes}} \\ 
 \noalign{\smallskip}
 \hline
 \noalign{\smallskip}   
    \textit{Hubble} & Low Earth orbit & 1990 & 0.03 & SO, SPI, UV, IR, O & \url{https://science.nasa.gov/mission/hubble/} \\
    \noalign{\smallskip} 
    \textit{Gaia} & L2 point & 2013 & 0.00001 & SO, A, O & \url{https://sci.esa.int/web/gaia/} \\
    \noalign{\smallskip}
    \textit{James Webb} & L2 point & 2021 & 0.1 & SO, NIR, MIR & \url{https://webbtelescope.org/home} \\
    \noalign{\smallskip}
 \noalign{\hrule height 1pt}
 \end{tabular}
 \label{tab:high_spatial_resolution_instruments}
\begin{justify}
\scriptsize{\textbf{\textit{Notes. }}}
\scriptsize{$^{\text{(a)}}$ A: Astrometry; AO: Adaptive Optics; C: Coronography; IF: Interferometry; IR: Infrared; LI: Lucky Imaging; MIR: Mid-Infrared; NIR: Near Infrared; O: Optical; SO: Space Observatory; SI: Submilimetric Interferometry; SPI: Speckle Imaging.}
\end{justify} 
\end{table}
\normalsize

\subsection{Concepts used for multiple star systems}
\label{sec:concepts_multiple_star_systems}

Throughout this thesis, a series of concepts about multiple star systems will be used, which will be defined here beforehand to avoid different interpretations from other authors.

\subsubsection{Primary star and companion stars}
\label{sec:primary_companion}

The term primary star can refer to either the brighter or the more massive component of a binary system, depending on context. In most visual binary observations, the primary is defined as the brighter star in the observed wavelength band, typically the one that appears more luminous from Earth. However, this brightness-based designation is wavelength-dependent, and the relative brightness of the two stars can change across different spectral bands. When the stars have nearly equal brightness, the primary is often assigned based on historical convention, typically following the discoverer's designation \citep{aitken64}. In contrast, some authors define the primary star as the more massive component, particularly in the context of interacting binaries or when dynamical information (e.g., from spectroscopy or astrometry) is available. The more massive star generally dominates the gravitational dynamics of the system, and the secondary (or companion) is considered to orbit around it, or more precisely, around the system’s barycentre. The term companion is also widely used in modern astronomy, especially in exoplanet studies, to refer to nearby objects, whether gravitationally bound or not.

In some binary or multiple star systems, however, the designation of the primary star is not always straightforward. For example, in eclipsing binaries or systems with significant mass transfer, the initially less massive star can gain mass and potentially outshine the original primary star. In such scenarios, the brightest or most massive star may change over time, complicating the definition of the ``primary'' star. This highlights the dynamic nature of stellar interactions and the evolving characteristics of stars within a system, particularly when mass exchange or tidal effects become significant.

Although the previously mentioned is the generalised agreement, this is not always the case. For example, in the Washington Double Star catalogue \citep{mason01}, we find pairs where the primary is not the brightest star. This happens when the secondary star was identified as the primary in the past or even when, in hierarchical systems, the star considered primary is the one that gravitationally dominates the system. Due to these reasons, the primary may be fainter than the secondary or secondaries, but the nomenclature has been maintained for historical consistency.

\subsubsection{Separation, angular position and epoch}
\label{sec:separation_angular_position_epoch}

The separation ($\rho$) and position  angle ($\theta$) between the stars, and the time of observation (Epoch) are often used in spherical or polar coordinate systems to describe the positions, orbits, and motions of the stars. An example is shown in Fig.~\ref{fig:example_rho_theta}.

\begin{itemize}
  \item \textit{Epoch:} In stellar observations, an epoch refers to a specific moment in time used as a reference for astronomical measurements, particularly positions and motions of celestial objects. Because stars move due to proper motion, precession, nutation, and other effects, their coordinates change over time. 
  
  Star catalogues list positions at a given epoch. It is expressed in Julian years as ``J'' followed by the year in decimal format, where the decimal fraction represents the portion of the year that has elapsed (e.g. J2000.0, which corresponds to 1 January 2000, at 12:00 TT). To convert any date to Julian years the following equation must be used:
  \begin{equation}
      J=A+\frac{M-\text{1}}{\text{12}}+\frac{D}{\text{365.25}}
  \end{equation}
  where $A$ is the year, $M$ is the month and $D$ is the day of the date to convert.

  Alternatively, once can determine the epoch in Julian years from the Julian date ($JD$) following:
  \begin{equation}
      J = \text{2000} + \frac{JD - \text{2451545.0}}{\text{365.25}}
  \end{equation}

  \begin{figure}[H]
  \centering
  \includegraphics[width=0.75\linewidth, angle=0]{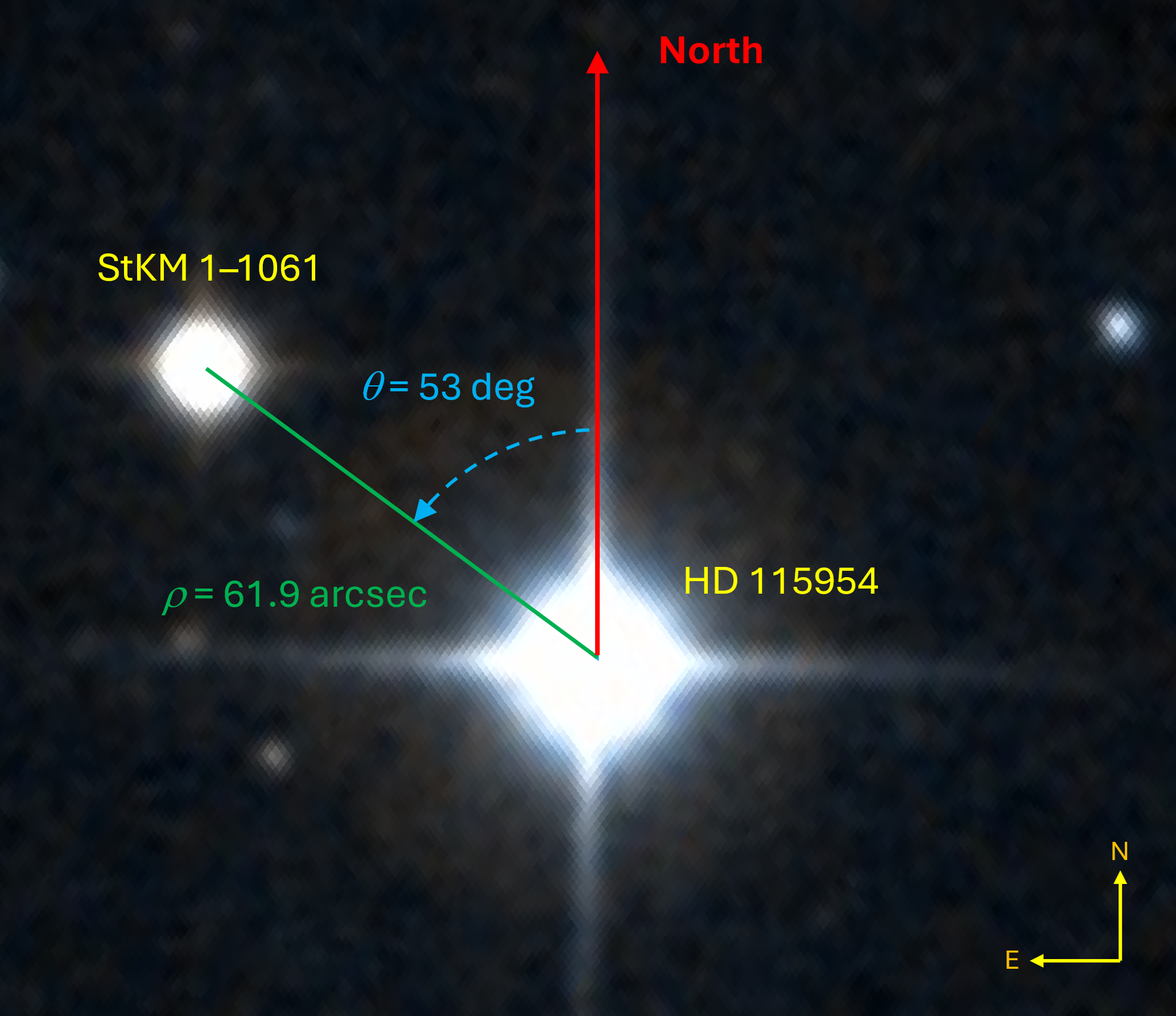}
     \caption[Example to show separation and angular position.]{In the system WDS 13199+3822/UC 184, the primary star is HD~115954, and StKM 1--1061 is the secondary star because the first one is brighter ($G=$\,8.2\,mag vs $G=$\,11.0\,mag) and, therefore more massive. The line separating the primary from the secondary star is the angular distance or separation. Measuring the angle from the north line of the primary star clockwise toward the line separating both stars results in the position angle. All the data were obtained at epoch J2015.5. Image from DSS2 \citep{lasker96}.}
   \label{fig:example_rho_theta}
  \end{figure}
  
  \item \textit{Separation ($\rho$)}: Angle between two stars projected in the sky. Measured in milliarcsec (mas), arcsec, arcmin, or degrees.
  
  \item \textit{Position angle ($\theta$)}: It is measured in counterclockwise degrees (counterclockwise may be incorrect if the orientation of the image is reversed) from the line connecting celestial north to the brightest star, passing through the line separating the primary and secondary stars.
\end{itemize}

\subsubsection{Orbital mechanics}
\label{sec:orbital_mechanics}

\begin{itemize}
    \item \textit{Barycentre}: It is the centre of mass that is the point around which stars in a multiple system orbit. This depends on the relative masses of the stars.
    \item \textit{Kepler's laws of motion}: These govern the orbits of stars in multiple star systems. Johannes Kepler (1571--1630) introduced his first two laws in his \textit{Astronomia Nova} \citep{kepler1609}, and the third in his \textit{Harmonices Mundi} \citep{kepler1619}. \\ 
    \vspace{-5mm}
    \begin{itemize}
        \item \textit{First law}: The planets move in ellipses with the Sun at one focus. \\
        \vspace{-5mm}
        \item \textit{Second law}: The line joining the Sun and a planet sweeps out equal areas in equal intervals of time (law of areas).\\
        \vspace{-5mm}
        \item \textit{Third law}: The square of the period of revolution $P$ of a planet is proportional to the cube of the semi-major axis of the orbit $a$. For all planets the ratio $P^\text{2}/a^\text{3}$ is equal to a constant.
        \begin{equation}    
             \frac{P^\text{2}}{a^\text{3}}=\frac{\text{4}\pi^\text{2}}{G(M_\text{1}+M_\text{2})}
        \end{equation}
    If $M_\text{2}\ll M_\text{1}$ then $M_\text{2}$ may be neglected and we can use Kepler's third law for planets.      
    \end{itemize}
    The point in an orbit where two stars are closest to each other is known as the \textit{periastron}, while the point where they are farthest apart is called the \textit{apastron}. These two points are connected by the major axis or the line of \textit{apsides} \citep{lipunov89}.
    \item \textit{Orbital resonance}: Occurs when two orbiting bodies exert a regular, periodic gravitational influence on each other due to their orbital periods being related by a ratio of small integers (e.g., 2:1 or 3:2). Orbital resonance in multiple star systems can have significant implications for their long-term stability, orbital evolution, and even the formation of planets within them:
    \begin{itemize}
        \item \textit{Dynamical stability or chaos:} Orbital resonances can either stabilise or destabilise a multiple-star system. If the resonance leads to periodic gravitational interactions that reinforce each other, it can drive chaotic evolution and eventually lead to close encounters, ejections, or even collisions between stars. Alternatively, resonances can help maintain long-term stability by locking stars into synchronised orbits, preventing close approaches.
        \item \textit{Energy and angular momentum exchange:} Resonances facilitate the transfer of angular momentum between the stars, potentially altering their orbits significantly over time. This can cause orbital eccentricities to increase or decrease, leading to phenomena like Kozai-Lidov oscillations \citep{kozai62,lidov62}, where one body's orbit periodically becomes highly eccentric due to interactions with a distant companion \citep[e.g.][]{sidorenko18,wang23}.
        \item \textit{Effects on planet formation and survival:} Resonances in multiple-star systems can shape protoplanetary disks, influencing planet formation by creating gaps or instabilities. They may destabilise planetary orbits, causing ejections or migrations, but can also stabilise them in protected configurations.
        
        \item \textit{Tidal evolution and stellar spin:} Strong tidal forces in resonant systems can synchronise stellar spins or cause spin-orbit misalignments, leading to tidal locking and changes in internal structure and activity over time \citep{burkart14}.
        
        \item \textit{Possible formation of exotic systems:} In extreme cases, orbital resonances in a multiple-star system might lead to the formation of exotic configurations, such as hierarchical triples, compact binaries, or even merging events leading to new stellar objects \citep[e.g.][]{toonen16,trani22}.
    \end{itemize}

\end{itemize}

\subsubsection{Orbital elements}
\label{sec:orbital_elements}

The orbital elements are the parameters needed to define an orbit uniquely. In celestial mechanics, these elements are typically considered for two-body systems following Keplerian orbits. While there are various mathematical approaches to describe the same orbit, specific schemes that use a set of six parameters are commonly employed in astronomy and orbital mechanics. Being strict, a real orbit and its elements change over time because of gravitational perturbations by other celestial bodies and the effects of general relativity. In our analysis, we use a Kepler orbit, that is an idealised, mathematical approximation of the orbit at a particular time.

In an inertial frame of reference, two orbiting bodies follow separate paths, each with the common centre of mass as their focal point. However, when observed from a non-inertial frame centred on one of the bodies, only the trajectory of the other body is visible. Keplerian elements are used to describe these non-inertial trajectories. An orbit can have two sets of Keplerian elements depending on which body is chosen as the reference. The reference body, typically the more massive one, is called the primary, while the other is referred to as the secondary. The primary is not always the more massive body, and even when the masses are equal, the orbital elements still vary based on the choice of primary.

\begin{figure}[H]
     \centering
     \includegraphics[width=0.7\linewidth, angle=0]{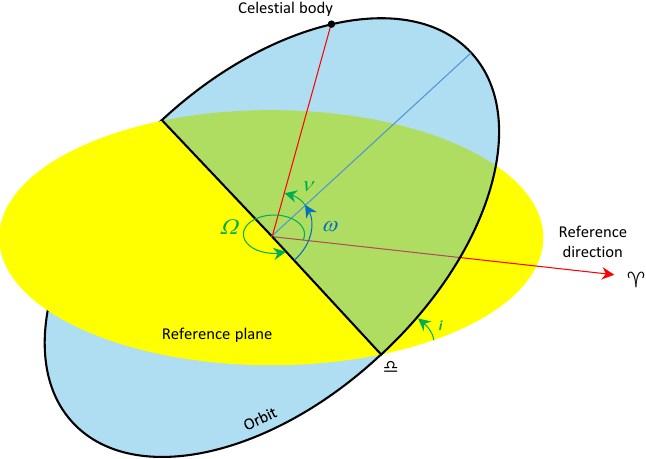}
     \caption[Orbital elements.]{Orbital elements: Here, the orbital plane (blue) intersects a reference plane (yellow). The intersection is called the line of nodes, as it connects the reference body (the primary) with the ascending and descending nodes. The reference body and the vernal point (\Aries) establish a reference direction and, together with the reference plane, they establish a reference frame. Here it is represented too the point where the orbit passes from south to north through the reference plane, symbolised by \Libra.
     }
    \label{fig:orbital_elements}
    \end{figure}

To provide a complete dynamical description of a system, it is necessary to know the period $P$, which has come to be considered an additional orbital element. The below quantities are typically used to define a binary orbit \citep{batten73}. They are also represented in Fig.~\ref{fig:orbital_elements}.

\begin{itemize}
    \item \textit{Orbital period (P)}: The orbital period is the time a celestial body (such as a planet, moon, or satellite) takes to complete one full orbit around another body. It depends on the masses of the bodies involved and the size of the orbit. It is generally expressed in days in the case of spectroscopic and eclipsing binaries, or years in the case of visual binaries. Sometimes it is used instead the parameter \textit{true anomaly ($\nu$, $\theta$, or f)} at time $t_{\text{0}}$ that defines the position of the orbiting body along the ellipse at a specific time, or ``epoch'', expressed as an angle from the periapsis (defined some lines below).
    
    \item \textit{Semi-major axis of the orbit (a)}: It is the longest radius of an elliptical orbit, extending from the center of the ellipse to the farthest edge. It represents half of the longest diameter of the orbital ellipse and is a key parameter in describing the size of the orbit. Usually expressed in astronomical units (au).
    
    \item \textit{Eccentricity of the orbit (e)}: The eccentricity of an orbit is a dimensionless parameter that defines the shape of the orbit by measuring its deviation from a perfect circle. It is given by:
    \begin{equation}
        e=\frac{r_a-r_p}{r_a+r_p}
    \end{equation}
    where $r_a$ is the \textit{apoapsis} distance (farthest point from the central body), and $r_p$ is the \textit{periapsis} distance (closest point to the central body). It is a dimensionless number between 0 (circle) and 1 (parabola). 

    \item \textit{Inclination of the orbital plane (i)}: It is the angle between the orbital plane of a celestial body (such as a planet, moon, or satellite) and a chosen reference plane. For planets in the Solar System, the reference plane is typically the ecliptic plane (Earth's orbital plane around the Sun). For satellites orbiting a planet, the reference plane is usually the planet's equatorial plane.
    
    \item \textit{Longitude of the ascending node ($\Omega$)}: Position angle measured from north through east of the line of nodes joining the intersections of the orbital and tangent planes, and measured in the latter.
    
    \item \textit{Argument of periapsis ($\omega$)}:  
    The angle, measured in the orbital plane, between the ascending node and the periapsis (the point of closest approach to the central body). It defines the orientation of the orbit’s closest approach relative to the reference direction. It is measured counterclockwise from the ascending node to the periapsis, ranging from 0$\degree$ to 360$\degree$.
    \begin{itemize}
        \item If $\omega=$\,0$\degree$, the periapsis is aligned with the ascending node.
        \item If $\omega=$\,90$\degree$, the periapsis is at the highest point above the reference plane.
        \item If $\omega=$\,180$\degree$, the periapsis is aligned with the descending node.
    \end{itemize}
    This angle is conventionally measured differently depending on the observational method. Visual observers (observers who study binary systems by resolving their components individually in images) typically report $\omega$ for the secondary, or fainter, component of the system, while observers of eclipsing and spectroscopic systems (those who determine orbital parameters using light curves or spectra) usually provide the value for the primary component. In any given system, the values for the primary and secondary differ by 180$\degree$.
    
    \item \textit{Time of periastron passage ($T_\text{0}$)}: The periastron passage refers to the moment when a star, planet, or other celestial body in an elliptical orbit around another body is at its closest approach (periastron) to the central object. 

\end{itemize}

\begin{table}[h]
 \centering
 \caption[Summary of information obtainable from binary systems.]{Summary of information obtainable from binary systems \citep{batten73}.}
 \footnotesize
 \scalebox{1}[1]{
 \begin{tabular}{@{\hspace{2mm}}l@{\hspace{3mm}}lc@{\hspace{1mm}}c@{\hspace{5mm}}c@{\hspace{3mm}}c@{\hspace{2mm}}}
 \noalign{\hrule height 1pt}
 \noalign{\smallskip}
Element &  & Visual binary & \multicolumn{2}{c}{Spectroscopic binary} & Eclipsing binary \\
 & & & One spectrum & Two spectra & \\
 \noalign{\smallskip}
 \hline
 \noalign{\smallskip}
Period & $P$ & Yes & Yes & Yes & Yes \\
 \noalign{\smallskip}
Semi-major axis & $a$ & Apparent & $a_\text{1}\sin i$ & $a\sin i$ & No \\
 \noalign{\smallskip}
Eccentricity & $e$ & Yes & Yes & Yes & Yes \\
 \noalign{\smallskip}
Argument of periapsis & $\omega$ & Yes & Yes & Yes & Yes \\
  \noalign{\smallskip}
Time of periastron passage  & $T$ & Yes & Yes & Yes & Yes \\
  \noalign{\smallskip}
Inclination & $i$ & Yes & No & No & Yes \\
 \noalign{\smallskip}
Longitude of ascending node & $\Omega$ & Yes$^{(a)}$ & No & No & No \\
 \noalign{\smallskip}
Mass (primary) & $m_\text{1}$ & \multirow{2}{*}{$m_\text{1}+m_\text{2}^{(b)}$} & \multirow{2}{*}{$f(m)$} & $m_\text{1}\sin^\text{3}i$ & \multirow{2}{*}{No} \\ 
Mass (secondary) & $m_\text{2}$ &  &  & $m_\text{2}\sin^\text{3}i$ &  \\
 \noalign{\smallskip}
\multirow{3}{*}{Radii} & \multirow{2}{*}{$R_\text{1}$} & \multirow{3}{*}{No} & Estimated  & Estimated & \multirow{2}{*}{$r_\text{1}=R_\text{1}/a$} \\
 & \multirow{2}{*}{$R_\text{2}$} &  & from spectrum & from spectrum & \multirow{2}{*}{$r_\text{2}$} \\
 &  &  & and luminosity & and luminosity & \\ 
   \noalign{\smallskip}
\multirow{2}{*}{Fractional luminosity} & $L_\text{1}$ & \multirow{2}{*}{Yes} & Estimated & Estimated & \multirow{2}{*}{Yes} \\
 & $L_\text{2}$ &  & from spectrum & from spectrum &  \\
   \noalign{\smallskip}
\multirow{2}{*}{Spectral types} & \multirow{2}{*}{SpT} & \multirow{2}{*}{Yes} & \multirow{2}{*}{Yes} & \multirow{2}{*}{Yes} & If several \\
 &  &  &  &  & colours available \\
  \noalign{\smallskip}
\multirow{2}{*}{Limb darkening} & $u_\text{1}$ & \multirow{2}{*}{No} & \multirow{2}{*}{No} & \multirow{2}{*}{No} & \multirow{2}{*}{Possible} \\
 & $u_\text{2}$ &  &  &  &  \\
 \noalign{\smallskip}
 \noalign{\hrule height 1pt}
 \end{tabular}
}
 \label{tab:binaries_parameters}
\begin{justify}
\scriptsize{\textbf{\textit{Notes. }}}
\scriptsize{$^{\text{(a)}}$ Ambiguous without radial-velocity observations.}
\scriptsize{$^{\text{(b)}}$ If parallax known.}
\end{justify} 
\end{table}

The parameters that can be determined depend on the type of binary system being observed. For instance, in a visual binary, it is possible to determine the orbital period ($P$), inclination ($i$), argument of periapsis ($\omega$), eccentricity ($e$), and the time of periastron passage ($T_\text{0}$). However, it is impossible to distinguish between the ascending and descending nodes of the orbit without accompanying radial-velocity measurements. Spectroscopic binary observations, on the other hand, provide different information. While the orbital period can be easily determined, neither the longitude of the ascending node ($\Omega$) nor the inclination ($i$) can be derived from spectroscopy alone. In the case of an eclipsing binary, the period ($P$) is readily measurable, and the inclination ($i$) can be determined since it relates to the minimum projected separation between the two stars' centers, which influences the shape and depth of the light curve. However, no information about the absolute size of the orbit can be gathered from the light curve alone, and the semi-major axis ($a$) remains unknown without spectroscopic data \citep{batten73}. Table~\ref{tab:binaries_parameters} shows a summary of information obtainable from binary systems. 

\subsubsection{Thiele-Innes method}
\label{sec:thiele_innes}

When astronomers observe a binary star system, they see only the projection of the true 3D orbit onto the sky plane. The challenge is that the observed orbit differs from the true orbit due to inclination and other orbital parameters. To fully describe the orbit of a binary star, the seven orbital elements described in Sect.~\ref{sec:orbital_elements} are needed. However, from observations, we only obtain sky-projected positions. The problem was how to derive the full orbital parameters from these limited observations in a direct way.

\citet{thiele1883} and later Innes developed a linearisation method to simplify the problem by introducing a set of constants that replace the usual angular orbital elements ($i$, $\omega$, $\Omega$). These constants allow to express the observed positions in a linear form, making it easier to fit the observations directly without solving for the angles $i$, $\omega$, and $\Omega$ first. In the late 19th and early 20th centuries, computations were done manually or with mechanical calculators, and this method reduced computational complexity. The Thiele-Innes method allowed direct solutions without needing iterative approximations. 

The Thiele-Innes method is used to compute the orbital parameters of visual binary stars based on angular observations, such as the projected separation and position angle of the components as seen from Earth. In visual binaries, we do not have direct access to the three-dimensional positions or velocities of the stars; instead, we rely on two-dimensional measurements of their relative motion projected onto the plane of the sky. It transforms these angular measurements into orbital elements by applying Kepler's laws and geometric considerations.

The method describes the relative orbit using parameters that relate observed angular positions to the underlying orbital configuration. Three key parameters often used are $a_\text{1}$ and $a_\text{2}$ (the semi-major axes of the orbits of the primary and secondary components, projected onto the plane of the sky), and $\omega$ (the argument of periastron).

Although Kepler's equations govern the motion of the stars, in the case of visual binaries, the challenge is to relate the observed angular positions, which are inherently two-dimensional, to the full three-dimensional orbit. The Thiele-Innes method provides a framework to perform this transformation, allowing us to infer the full orbital parameters from projected observational data. The relative positions of the two components of the binary star are expressed by the following equations:
\begin{equation}
    \rho = \sqrt{a_\text{1}^\text{2} + a_\text{2}^\text{2} + \text{2} a_\text{1} a_\text{2} \cos(\omega + \theta)}
\end{equation}
\begin{equation}
    \theta = \arctan\left(\frac{a_\text{1} \sin(\omega + \theta)}{a_\text{1} \cos(\omega + \theta) + a_\text{2}}\right)
\end{equation}
where $\rho$ is the angular separation and $\theta$ is the position angle.

The Thiele-Innes method resolves the orbit of a visual binary system using the following fundamental equations for the coordinates of the two stars' orbits, based on angular observations $\rho$ and $\theta$:

\begin{equation}
    \frac{d^\text{2} \rho}{dt^\text{2}} = \frac{\text{1}}{m_\text{1} + m_\text{2}} \left( \frac{G (m_\text{1} + m_\text{2})}{r^\text{3}} \right) \left( \rho \right)
\end{equation}
where $r$ is the distance between the two stars, and $m_\text{1}$ and $m_\text{2}$ are the masses of the two components.

The following system of equations involves the coordinates of the binary system and determines the positions of the two stars in the binary orbit, and its solutions provide the orbital parameters like $a_\text{1}$, $a_\text{2}$, $\omega$, which can then be used to construct the orbit and predict the system's future positions:

\begin{equation}
  \begin{bmatrix}
  A & B & C \\
  D & E & F \\
  G & H & I
  \end{bmatrix}
  \begin{bmatrix}
  x_\text{1} \\
  x_\text{2} \\
  x_\text{3}
  \end{bmatrix}
  =
  \begin{bmatrix}
  y_\text{1} \\
  y_\text{2} \\
  y_\text{3}
  \end{bmatrix}
\end{equation}

Through fitting the astrometric observations (angular separation and position of the stars), the Thiele-Innes method adjusts these orbital parameters to model the orbital motion and the separation of the binary components. Though not as commonly used today due to advances in measurement technology, this method remains fundamental in the historical study of visual binary star systems.

\subsubsection{Types of systems and orbits}
\label{sec:types_systems_orbits}

\begin{itemize}

    \item \textit{Simple binary systems:} These are the most simple stellar systems, composed by two stars. Except for relativistic effects (e.g. two spinning neutron stars), they are completely stable.

    \item \textit{Hierarchical systems:} Hierarchical multiple star systems are configurations where stars are arranged in nested orbits, typically organised in a series of stable, gravitationally bound subsystems. These systems often consist of two or more stars, and the arrangement can vary in complexity depending on the number of components and their orbital relationships.

    \begin{figure}[H]
    \centering
    \includegraphics[width=0.6\linewidth, angle=0]{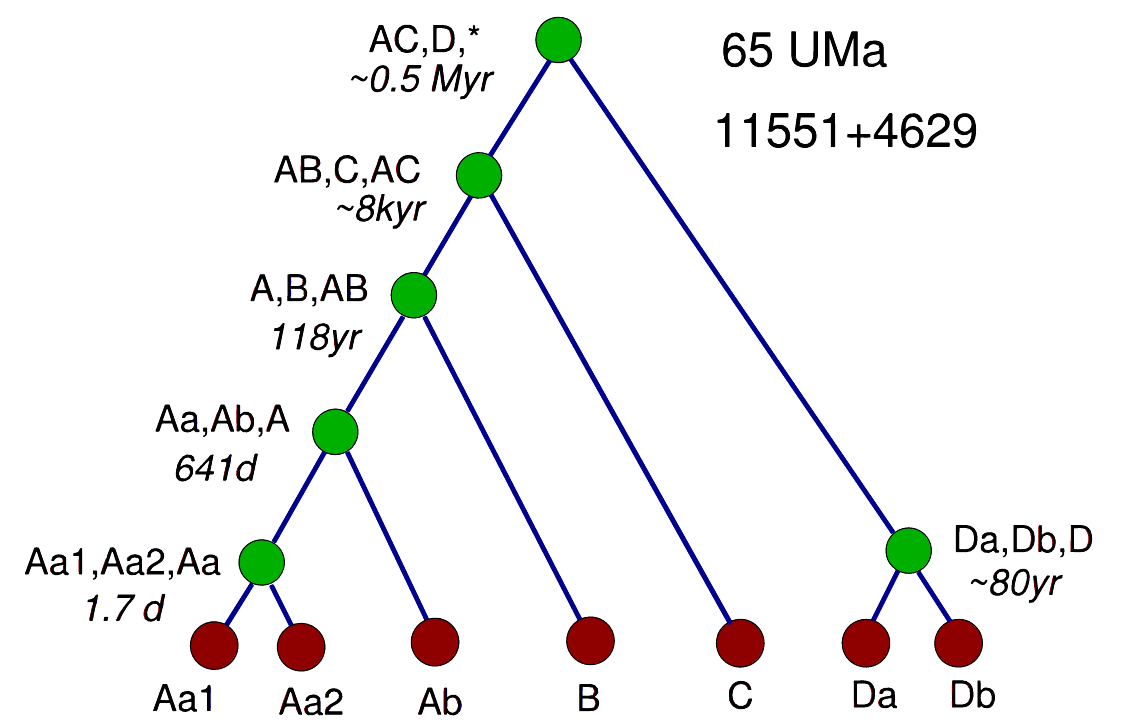}
     \caption[Structure of the 5-tier hierarchical system 65 UMa (WDS 11551+4629, HR 4560, DN UMa).]{Structure of the 5-tier hierarchical system 65 UMa (WDS 11551+4629, fig.~2 of \citealt{tokovinin21b}).}
    \label{fig:hierarchical_system}
    \end{figure}     

    Complex hierarchies containing more than four stars can be considered as combinations of basic binary, triple, and similar systems \citep{tokovinin21b}. Fig.~\ref{fig:hierarchical_system} illustrates the structure of the unique 5-tier system 65 UMa (WDS 11551+4629); no other 5-tier hierarchies have been discovered so far. Its structure is almost like that of a planetary system, except that the outermost component, D, is itself a binary. A hierarchical system with N tiers can have up to 2$^\text{N}$ components, assuming all possible levels are occupied. This condition is met in 2+2 quadruple systems. The 65 UMa system contains only 7 components, while the maximum number of stars in a 5-tier hierarchy is 32 ($\text{2}^{\text{5}}$). This pattern is common: usually, only a portion of the available levels is occupied.

    \item \textit{Non-hierarchical systems:} Non-hierarchical orbits refer to systems where two or more stars interact with each other without a clear, dominant star that governs the motion of the others. These systems are more complex than hierarchical ones, where a central star dominates the motion of others in a well-defined hierarchy.

    In non-hierarchical systems, all stars exert significant gravitational influence on one another, creating a dynamic where no single star dominates the others. This can lead to more complex orbital motions, often resulting in chaotic or unpredictable behavior over time. These systems tend to be less stable than hierarchical ones. Over time, the gravitational interactions between the stars can cause shifts in their orbits, potentially leading to ejections or mergers. However, some non-hierarchical systems can remain stable under certain conditions, particularly if the stars are in particular orbital configurations.    Unlike hierarchical systems, where orbits are generally predictable and stable, the stars in non-hierarchical systems often follow elliptical (with high eccentricity) or chaotic orbits. The mutual gravitational interactions between the stars make their orbits less predictable and more dynamic.

    \item \textit{Stable and unstable orbits:} Depending on the masses and distances, orbits can either remain stable over long periods or be unstable, leading to possible ejections of stars or rearrangement of the system's configuration.

    \item \textit{S-type and p-type orbits:} When the classification is done with stars having planets orbiting around them, we can have two different types of orbits. A s-type (satellite-type) orbit is when a planet orbits one star in a binary or multiple system. Some examples are HD~25171 \citep{moutou11}, HD~21019 \citep{stassun19}, or $\beta$~Cnc \citep{lee14}. A p-type (planetary-type) orbit is when a planet orbits around both stars in a binary system (circumbinary orbit). 

    \begin{figure}[H]
    \centering
    \includegraphics[width=0.8\linewidth, angle=0]{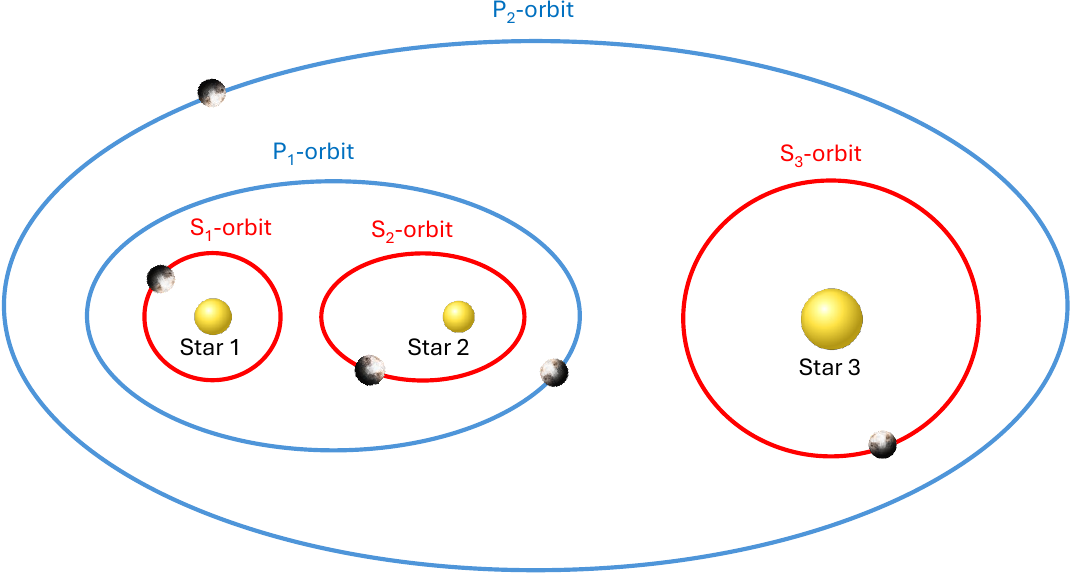}
    \caption[Triple system with s-type (red) and p-type (blue) orbits.]{Triple system with s-type (red) and p-type (blue) orbits (based on fig.~1 from \citealt{busetti18}).}
    \label{fig:s-p_systems}
    \end{figure}      

    The orbits of p-type systems have more probability to have unstable orbits than in s-type systems due to the variable gravitational field of the central binary, the lack of conservation of energy and angular momentum, and the presence of disruptive resonances (e.g. \citealt{quarles18,zoppetti20,bromley21}).
    
    There are less than 20 p-type systems discovered. New detections continue, especially with advancements in space telescopes like TESS (Transiting Exoplanet Survey Satellite, \citealt{ricker15}). The number of p-type systems will likely grow as observational techniques improve and as we get better at detecting planets in multi-star systems. Some examples are RR~Cae \citep{qian12}, Kepler-16 \citep{doyle11}, and HD~202206 \citep{benedict17}. Fig.~\ref{fig:s-p_systems} shows different configurations of s-type and p-type orbits.    
    
\end{itemize}

\subsection{Star formation and evolution}
\label{sec:star_formation_evolution}

The study of stellar multiplicity has significant implications for our understanding of star formation. In fact, stellar multiple systems are ubiquitous products of the star formation processes \citep[e.g.][]{duquennoy91,fischer92,raghavan10,duchene13,offner23}. Current models suggest that all stars form mainly from collapsing molecular clouds with masses between $\text{10}^\text{4}$ and $\text{10}^\text{7}$\,M$_\odot$ \citep{goldreich74,williams97} where the fragmentation of the initial cloud leads to the formation of binary or higher-order systems. Observations of young stellar clusters, like the Trapezium Cluster in the Orion Nebula, have provided valuable insights into the early stages of multiple star formation \citep[e.g.][]{lada03,gravitycollaboration18}.

The current theories regarding the formation of multiple systems can be categorised into three primary groups \citep{offner23}: models proposing the formation through the fragmentation of a core or filament \citep{konyves15,pineda15}, fragmentation of a massive accretion disk \citep{thompson88,gammie01,tokovinin20b}, or via dynamic interactions \citep[e.g.][]{bate97,ostriker99,stahler10,lee11}. The latter model also has the potential to reconfigure the hierarchy and multiplicity of systems formed through the preceding fragmentation channels. Moreover, stellar multiplicity holds significant importance in both stellar evolution and stellar populations. Interactions among stars within multiple systems can trigger mass transfer, tidal interactions, and even stellar mergers, all of which can profoundly influence the evolutionary trajectory of individual stars and the overall characteristics of the system. Studies have shown that the presence of nearby companions can influence the mass accretion rates of young stars, impacting their subsequent evolution \citep[e.g.][]{dario14,zagaria22}.

\begin{figure}[H]
\centering
\includegraphics[width=1\linewidth, angle=0]{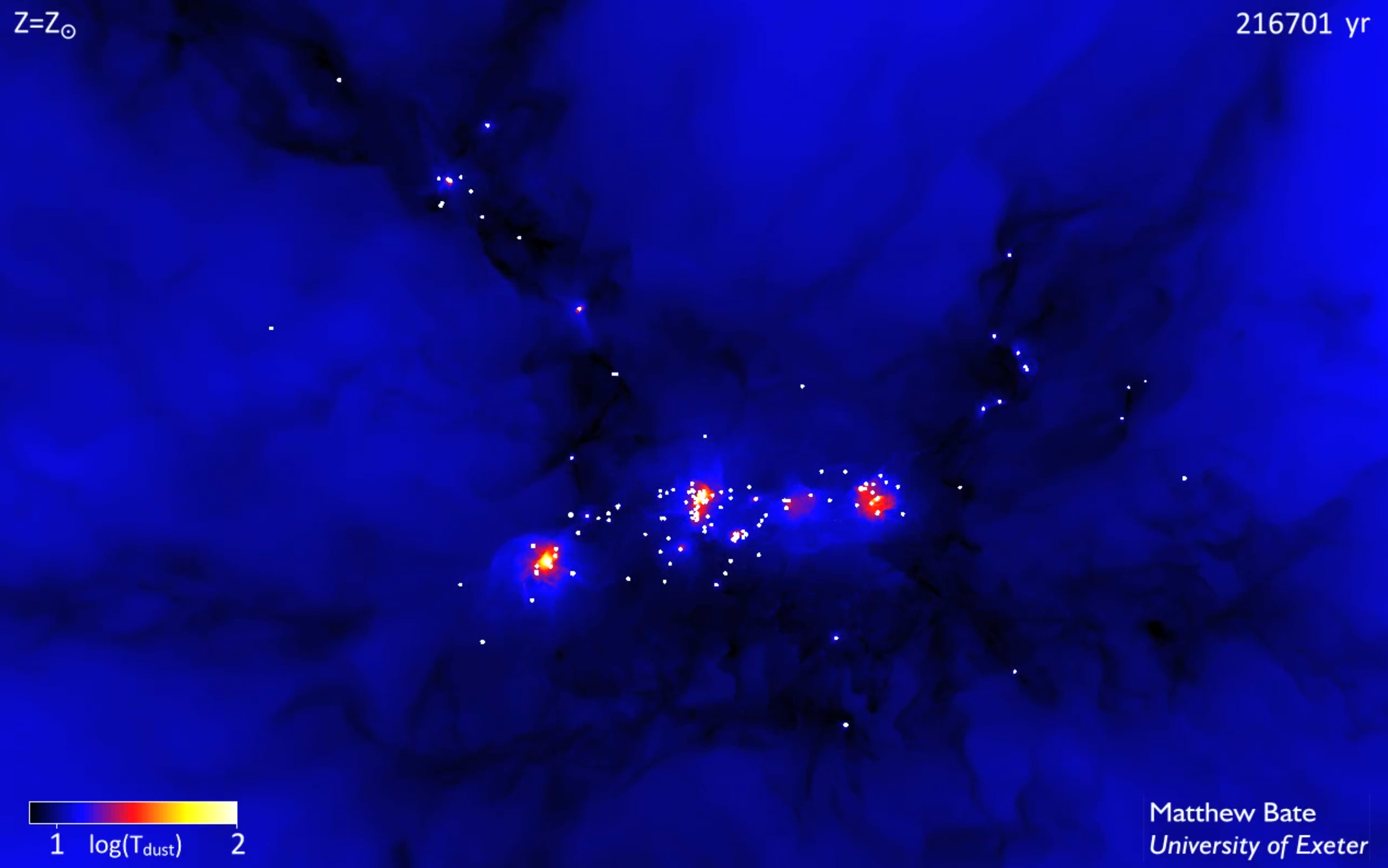}
\caption[Frame of a solar metallicity star formation simulation movie showing dust temperature in different colours.]{Frame of a solar metallicity star formation simulation movie showing dust temperature in different colours (credit: Matthew Bate, \url{http://www.astro.ex.ac.uk/people/mbate/Cluster/clusterMetallicity2.html}).}
\label{fig:star_formation}
\end{figure} 

The examination of post-interaction systems, particularly those found within globular clusters, offers valuable insights that help constrain theories of stellar evolution. Furthermore, comprehending stellar multiplicity enhances the study of exoplanets \citep[e.g.][]{bonavita07,mugrauer19,fontanive21,gonzalezpayo24}, as gravitational interactions within multiple-star systems can impact the formation, stability \citep{wiegert97,busetti18}, and potential habitability of planets \citep{quintana06,quintana07,eggl18}.

\subsection{Multiplicity fraction}
\label{multiplicity_fraction}

The multiplicity fraction is a key parameter in astrophysics that quantifies the fraction of stellar systems that contain multiple stars (binaries, triples, etc.) rather than being single stars. This parameter is useful to help test models of star formation by determining whether stars tend to form alone or in groups. Higher multiplicity fractions in massive stars suggest different formation mechanisms compared to lower-mass stars. In terms of evolution, dynamical interactions over time (e.g., stellar encounters) can break apart or modify multiple star systems. Observing multiplicity across different stellar populations helps understand dynamical processes in clusters and the field.

To quantitatively evaluate the multiplicity of stellar or substellar objects, we use the notation introduced by \citet{batten73}, where $f_n$ represents the fraction of primaries with \textit{n} companions. Thus, the multiplicity frequency or multiplicity fraction (\textit{MF}) is:
\begin{equation}
    MF=\frac{\displaystyle\sum_{n=\text{1}}f_n}{\displaystyle\sum_{n=\text{0}}f_n}=\frac{B+T+Q+...}{S+B+T+Q+...}
    \label{eq:multiplicity_fraction}
\end{equation}
where \textit{S}, \textit{B}, \textit{T}, \textit{Q} denote the number of single, binary, triple, and quadruple systems, respectively \citep{reipurth93}. Another usual indicator is the Companion Star Fraction (\textit{CSF}), that measures the average number of companions per system:
\begin{equation}
    CSF=\frac{\displaystyle\sum_{n=\text{1}}(n-\text{1})f_n}{\displaystyle\sum_{n=\text{0}}f_n}=\frac{B+\text{2}T+\text{3}Q+...}{S+B+T+Q+...}
    \label{eq:company_star_fraction}
\end{equation}
\textit{MF} is a number between 0 and 1, but \textit{CSF} is a number greater than 0 (even greater than 1). Uncertainties in the measurement in the \textit{MF} follow binomial statistics, while Poisson statistics govern the \textit{CSF} \citep{burgasser03c,sana14,moe17}. Measurements of \textit{MF} are less prone to being influenced by the detection of all subsystems compared to \textit{CSF}, which clarifies the preference for utilising \textit{MF} in comparing theoretical predictions with observational data \citep{reipurth93}. 

The fraction of higher-order multiple systems, \textit{HF}, represents the proportion of systems that are not just simple binaries but instead consist of three or more stars, often referred to as higher-order multiple systems. \textit{HF} provides a useful statistical measure of the prevalence of complex star systems. It is:

\begin{equation}
    HF=\text{1}-\displaystyle\sum_{n=\text{1}}f_n
    \label{eq:high_order_system_fraction}
\end{equation}
The calculation of stellar multiplicity involves various uncertainties that can affect the accuracy of the obtained results. The main sources of uncertainty are:

\begin{itemize}
    \item \textit{Observational limitations:} The ability of instruments to resolve close stars is limited, which can lead to an underestimation of stellar multiplicity \citep[e.g.][] {kuntzer16,borodina21}.
    \item \textit{Selection effects:} Observational samples may be biased toward certain types of stars or systems, affecting the representativeness of results. \citet{eggleton09} discussed selection effects and observational uncertainties in determining stellar multiplicity.
    \item \textit{Theoretical models:} Simplifications and assumptions in models can introduce discrepancies between predictions and actual observations. \citet{stanway20} evaluated the impact of uncertainty in binary parameters on the properties of stellar populations.
    \item \textit{Uncertainty propagation:} When multiple measurements are combined to calculate properties of stellar systems, individual uncertainties propagate, affecting the final precision. The uncertainty propagation technique allows estimation of how individual measurement uncertainties influence the combined result \citep{andrae10}.
\end{itemize}

To estimate uncertainties, the 95\% confidence interval is often computed using the Wilson expression \citep{wilson1927}. This method is particularly useful when proportions are close to the extreme values (0 or 1) or when the sample size is relatively small, as it provides more accurate and less biased intervals compared to the standard normal approximation:

\begin{equation}
CI_{\rm Wilson} = \frac{k+\frac{\kappa^\text{2}}{\text{2}}}{n+\kappa^\text{2}} \pm \frac{\kappa \sqrt{n}}{n+\kappa^\text{2}} \sqrt{\hat{\epsilon}(\text{1}-\hat{\epsilon})+\frac{\kappa^\text{2}}{\text{4}n}},
\end{equation}

where \textit{n} is the number of trials, \textit{k} is number of observed successes, $\hat{\epsilon} = k/n$, and $\kappa$ is the number of standard deviations corresponding to the wished confidence interval for a normal distribution.
For a confidence interval of $\text{100}\cdot(\text{1} - \alpha)$\%,
\begin{equation}
\kappa = \Phi^{-\text{1}} (\text{1}-\alpha/s) = \sqrt{\text{2}}{\rm erf}^{-\text{1}} (\text{1}-\alpha),
\end{equation}
where erf$^{-\text{1}}$ is the inverse of the error function.

Also, for cases where the population of trials is very small, and assuming Poisson statistics and a 95\% confidence interval, the Wald interval \citep{agresti98} can be used. It has the advantage of adjusting the standard error by incorporating a correction for small-sample variability:
\begin{equation}
CI_{Wald}= \lambda \pm \text{1.96}\sqrt{\lambda/n}, 
\end{equation}
where $\lambda$ is the number of successes in $n$ trials. In this work, we frequently used the Wald interval. There are some reasons:
\begin{itemize}
    \item \textit{Uncertainty calculation:} The Wald interval provides a way to quantify the range of plausible values for the multiplicity fraction, taking into account the sample size and the number of multiple systems.
    \item \textit{Poisson statistics assumption:} The analysis assumes that the distribution of stellar systems follows Poisson statistics, which is common in astronomical studies where events (in this case, the identification of multiple systems) are independent.
    \item \textit{Simplicity:} The Wald interval is relatively easy to compute, facilitating its application in studies with a large number of stellar systems. 
    \item \textit{Comparison with other studies:} Using the Wald interval allows for comparisons with other studies that also employ this methodology, making it easier to assess the consistency of the findings.
    
\end{itemize}

However, there are more advanced methods, such as Markov Chain Monte Carlo (MCMC) techniques, provide more precise uncertainty estimates by accounting for the full probability distributions of the parameters. Despite this, the Wald interval remains useful for obtaining a quick and simple estimation of uncertainties in the multiplicity fraction.

These multiplicity indicators have been intensively used during the elaboration of this thesis for comparing with previously published works, trying to establish the multiplicity of systems depending mainly on their spectral types and metallicities.

\section{Resources}
\label{sec:resources}

\subsection{\textit{Gaia}}
\label{sec:gaia}

All works published during the elaboration of this thesis were mainly based on the catalogue containing the data obtained during the \textit{Gaia} mission \citep{gaiacollaboration16a}, especially the Data Releases 2 and 3, described in Sect.~\ref{sec:gaia_data_releases}. \textit{Gaia} is a space observatory of the European Space Agency (ESA) designed for astrometry: measuring the positions, distances and motions of stars with unprecedented precision, and the positions of exoplanets by measuring attributes about the stars they orbit such as their apparent magnitude and color. The mission aimed to construct by far the largest and most precise 3D space catalogue ever made, totalling approximately one billion astronomical objects, mainly stars, but also planets, comets, asteroids and quasars, among others in the Milky Way and beyond. Through photometry, the stellar properties of this sample can be deduced. From spectroscopy, \textit{Gaia} allowed the extraction of some 150 million radial velocities for the brightest stars. The nominal operations were scheduled for five years. 

\subsubsection{\textit{Gaia} mission}
\label{sec:gaia_mission}

The \textit{Gaia} mission\footnote{\textit{Gaia}, ESA's billion star surveyor, \url{https://www.esa.int/Science_Exploration/Space_Science/Gaia/}} was a groundbreaking space observatory operated by the European Space Agency (ESA) with the primary goal of creating a highly precise 3D map of astronomical objects throughout the Milky Way Galaxy in unprecedented detail. This map includes data that represent roughly 1\% of the total stars in the Milky Way. The \textit{Gaia} mission emerged from the growing need for precise astrometric data to answer fundamental questions about the structure, formation, and evolution of the Milky Way Galaxy. In the late 1990s, the European Space Agency (ESA) began exploring the possibility of a space observatory dedicated to mapping the positions, distances, and motions of stars with unprecedented accuracy. The concept for the \textit{Gaia} mission evolved through extensive scientific and technical studies. 

The construction of the \textit{Gaia} spacecraft involved collaboration among numerous European aerospace companies, research institutions, and space agencies. The spacecraft's payload included two telescopes with focal plane arrays of charge-coupled devices (CCDs) designed to capture precise measurements of star positions and brightness. \textit{Gaia} embarked on its journey with Arianespace, launching atop a Soyuz ST-B rocket equipped with a Fregat-MT upper stage. The liftoff took place at the Ensemble de Lancement Soyouz in Kourou, French Guiana, precisely on 19 December 2013 at 09:12 UTC (06:12 local time). Following a 43-minute ascent, the satellite separated from the rocket's upper stage at 09:54 UTC\footnote{BBC News. ``\textit{Gaia} `billion-star surveyor' lifts off'' by J. Amos, \url{https://www.bbc.com/news/science-environment-25426424}}. The spacecraft set its course toward the Sun-Earth Lagrange point L2, situated approximately 1.5 million kilometres away from Earth, reaching its destination on 8 January 2014. The L2 point offers \textit{Gaia} a stable gravitational and thermal environment conducive to its mission objectives. \textit{Gaia} follows a Lissajous orbit at L2, avoiding the Earth's obstruction of the Sun, which could otherwise limit solar energy absorption and disrupt thermal equilibrium. Upon launch, a 10-meter-diameter sunshade was unfurled, ensuring that all telescope components remain cool and that \textit{Gaia} operated efficiently using solar panels on its surface. Through careful selection of materials and design, \textit{Gaia} was capable of functioning within temperature ranges spanning from --170\,\degree\,C to 70\,\degree\,C.

After reaching its operational orbit, \textit{Gaia} underwent a series of commissioning and calibration procedures to ensure the accuracy and reliability of its measurements. The spacecraft's complex data processing pipeline was developed to handle the vast amount of data collected during its mission lifetime. \textit{Gaia}'s observing strategy involved scanning the entire sky multiple times over several years to gather data on the positions, distances, motions, and characteristics of stars and other celestial objects.

Since the start of its mission, \textit{Gaia} released multiple data sets containing billions of precise astrometric measurements. These data sets have revolutionised our understanding of the Milky Way Galaxy and beyond. Scientists have used \textit{Gaia} data to study the distribution of stars, the structure of the Milky Way, the dynamics of stellar populations, the properties of exoplanets, and the physics of the Universe on both cosmic and local scales.

The \textit{Gaia} mission represented a landmark achievement in the field of astrometry and has laid the groundwork for future generations of space-based observatories. Its legacy will continue to shape our understanding of the Universe for years to come, providing invaluable data for astronomers and scientists worldwide. The mission stood as a testament to international collaboration, technological innovation, and scientific exploration, offering unprecedented insights into the cosmos and inspiring new discoveries in the quest to unravel the mysteries of the Universe.

\subsubsection{\textit{Gaia} ESA archive}
\label{sec:gaia_esa_archive}

The \textit{Gaia} ESA archive contains deduced positions, parallaxes, proper motions, radial velocities, and brightness measurements. Complementary information on multiplicity, photometric variability, and astrophysical parameters is provided for a large fraction of sources\footnote{Welcome to the \textit{Gaia} ESA Archive, \url{https://gea.esac.esa.int/archive/}}. The \textit{Gaia} data is made available through the \textit{Gaia} ESA archive. The \textit{Gaia} ESA archive is physically located at the ESAC Science Data Centre (ESDC), Madrid (Spain), and is developed and operated by the ESA Science Archives Team. In its last version it contains 941.5 TB of information, and it is widely consulted by more than 32\,000 users per month\footnote{ESAC Science Data Centre, \url{https://www.cosmos.esa.int/web/esdc}}.

\begin{figure}[H]
  \centering
  \includegraphics[width=1\linewidth, angle=0]{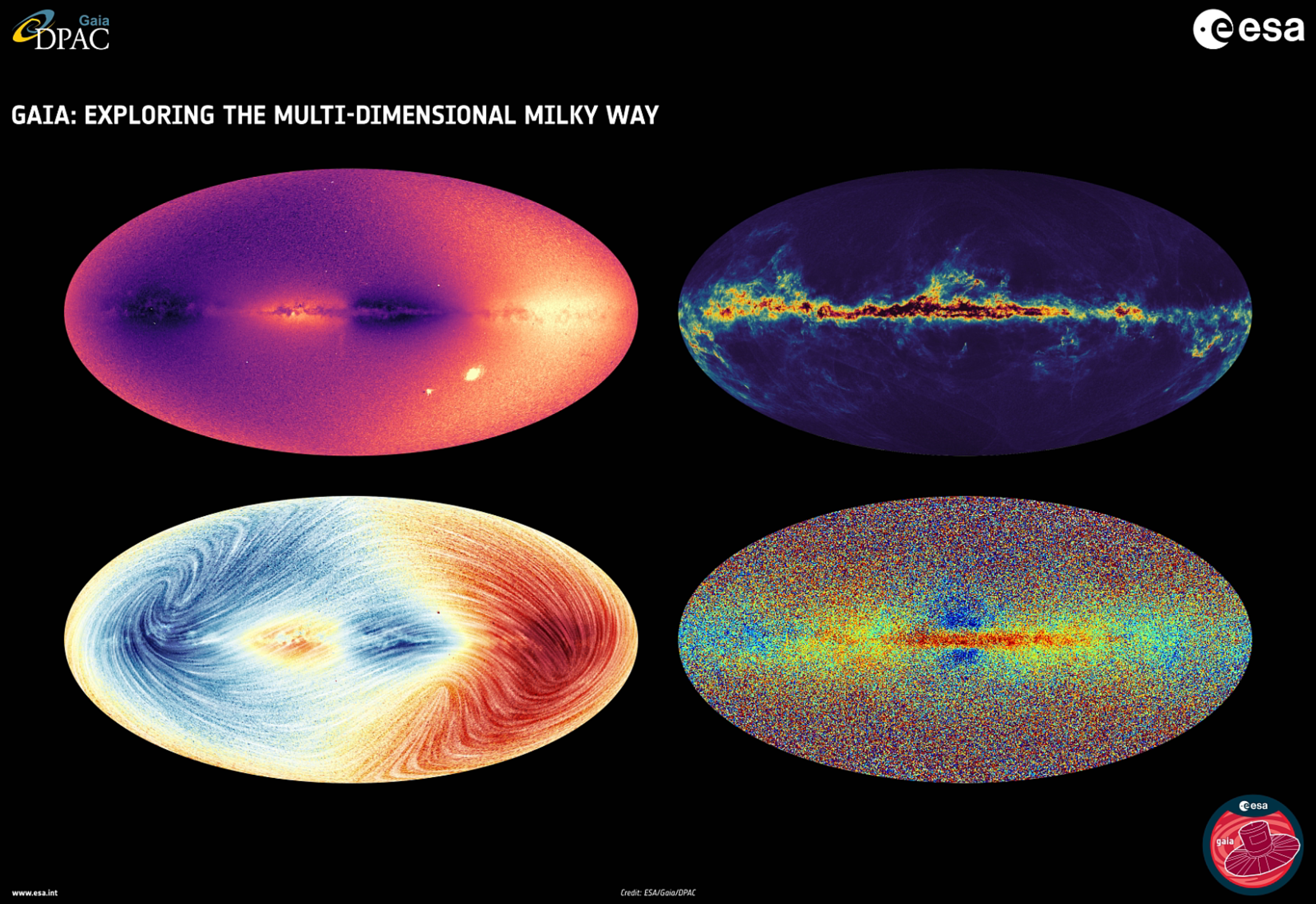}
     \caption[The four sky maps made with the \textit{Gaia} DR3 catalogue.]{The four sky maps made with the \textit{Gaia} DR3 catalogue: Radial velocity map (\textit{top left}) shows us the speed at which more than 30 million objects in the Milky Way (mostly stars) move towards or away from us; Radial velocity and proper motion map (\textit{bottom left}) shows the velocity field of the Milky Way for ~26 million stars. Blue indicates the parts of the sky where the average motion of stars is towards us and red shows the regions where the average motion is away from us; Interstellar dust map (\textit{top right}) tells us what is in between the stars. The space between stars is not empty but instead filled with dust and gas clouds, out of which stars are born; Chemical map (\textit{bottom right}) shows what stars are made of, and it can tell us about their birthplace and their journey afterwards, and therefore about the history of the Milky Way. Credit: ESA/Gaia/DPAC.
}
   \label{fig:gaiamaps}
\end{figure}

Since its launch \textit{Gaia} has generated unprecedented volumes of scientific data. The \textit{MareNostrum} supercomputer\footnote{Barcelona Supercomputer Center (BSC): \textit{MareNostrum}, \url{https://www.bsc.es/marenostrum/marenostrum/}}, located in Barcelona, Spain, has been instrumental in processing this data, enabling scientists to analyse the positions, distances, movements, and compositions of celestial objects with high precision. The supercomputer's capabilities allow for the execution of complex algorithms necessary for astrometric calculations, which are fundamental to achieving \textit{Gaia}'s objectives. \textit{MareNostrum} supports the identification of new celestial objects and helps refine existing data models. For example, it has aided in the analysis of variable stars and the detection of new planetary systems, contributing to the mission's success in uncovering the dynamics of the Milky Way and its surroundings. As the \textit{Gaia} mission transitions from data collection to data processing and analysis phases, \textit{MareNostrum} will continue to play a pivotal role. The upcoming \textit{Gaia} Data Release 4 (DR4), expected in 2026, will involve processing an anticipated 500 TB of data products, which highlights the importance of \textit{MareNostrum} in handling such extensive datasets and ensuring the accuracy and reliability of the findings.

\subsubsection{\textit{Gaia} data releases}
\label{sec:gaia_data_releases}

The \textit{Gaia} data sets are released in different stages increasing the amount of information. Therefore, the first one (DR1) was released on 14 September 2016, based on 14 months of observations made through September 2015, as was announced by ESA in a press release\footnote{ESA: ``\textit{Gaia}'s billion-star map hints at treasures to come'', \url{https://sci.esa.int/web/gaia/-/58272-gaia-s-billion-star-map-hints-at-treasures-to-come}}. DR1 included positions and magnitudes in a single photometric band for 1.1 billion stars using only \textit{Gaia} data, positions, parallaxes and proper motions for more than 2 million stars based on a combination of \textit{Gaia} and Tycho-2 data, light curves and characteristics for about 3000 variable stars, and positions and magnitudes for more than 2000 extragalactic sources used to define the celestial reference frame. \textit{Gaia} DR2\footnote{\textit{Gaia} Data Release 2 (DR2), \url{https://www.cosmos.esa.int/web/gaia/dr2/}} was released on 25 April 2018, using the data of the observations performed between 25 July 2014 and 23 May 2016. It collects positions, parallaxes and proper motions for 1.3 billion stars and positions of an additional 300 million stars in the magnitude range $g=\text{3}-\text{20}$, red and blue photometric data for about 1.1 billion stars, single colour photometry for an additional 400 million stars, and median radial velocities for about 7 million stars between magnitude 4 and 13. It also contains data for more than 14\,000 Solar System objects. 

\begin{table}[h]
 \centering
 \caption[Number of sources in \textit{Gaia} DR1, DR2, and DR3.]{Number of sources in \textit{Gaia} DR1, DR2, and DR3 (\url{https://www.cosmos.esa.int/web/gaia/dr3/}).}
 \footnotesize
 \scalebox{1}[1]{
 \begin{tabular}{l@{\hspace{4mm}}c@{\hspace{4mm}}c@{\hspace{4mm}}c}
 \noalign{\hrule height 1pt}
 \noalign{\smallskip}
 & \textit{Gaia} DR3 & \textit{Gaia} DR2 & \textit{Gaia} DR1 \\
 \noalign{\smallskip}
 \hline
 \noalign{\smallskip}
Total number of sources & 1\,811\,709\,771 & 1\,692\,919\,135 & 1\,142\,679\,769  \\
Number of sources with full astrometry & 1\,467\,744\,818 & 1\,331\,909\,727 & 2\,057\,050  \\
Number of 5-parameter sources & 585\,416\,709 &  &   \\
Number of 6-parameter sources & 882\,328\,109 &  &   \\
Number of 2-parameter sources & 343\,964\,953 & 361\,009\,408 & 1\,140\,622\,719  \\
Sources with mean \textit{G} magnitude & 1\,806\,254\,432 & 1\,692\,919\,135 & 1\,142\,679\,769  \\
Sources with mean \textit{GBP}-band photometry & 1\,542\,033\,472 & 1\,381\,964\,755 & -  \\
Sources with mean \textit{GRP}-band photometry & 1\,554\,997\,939 & 1\,383\,551\,713 & -  \\
Sources with radial velocities & 33\,812\,183 & 7\,224\,631 & -  \\
Sources with mean \textit{GRVS}-band magnitudes & 32\,232\,187 & - & -  \\
Sources with rotational velocities & 3\,524\,677 & - & -  \\
Mean $BP$/$RP$ spectra & 219\,197\,643 & - & -  \\
Mean RVS spectra & 999\,645 & - & -  \\
Sources with object classifications & 1\,590\,760\,469 & - & -  \\
Stars with emission-line classifications & 57\,511 & - & -  \\
Sources with astrophysical parameters from $BP$/$RP$ spectra & 470\,759\,263 & 161\,497\,595 & -  \\
Sources with astrophysical parameters assuming an unresolved binary & 348\,711\,151 & - & -  \\
Sources with spectral types & 217\,982\,837 & - & -  \\
Sources with evolutionary parameters (mass and age) & 128\,611\,111 & - & -  \\
Hot stars with spectroscopic parameters & 2\,382\,015 & - & -  \\
Ultra-cool stars & 94\,158 & - & -  \\
Cool stars with activity index & 1\,349\,499 & - & -  \\
Non-single stars (astrometric, spectroscopic, eclipsing, orbits, trends) & 813\,687 & - & -  \\
Non-single stars - orbital astrometric solutions & 169\,227 & - & -  \\
Non-single stars - orbital spectroscopic solutions (SB1 / SB2) & 186\,905 & - & -  \\
Non-single stars - eclipsing binaries & 87\,073 & - & -  \\
Solar system objects & 158\,152 & 14\,099 & -  \\
 \noalign{\smallskip}
 \noalign{\hrule height 1pt}
 \end{tabular}
}
 \label{tab:gaia_releases}
\end{table}

The third data release was split into two different parts due to some uncertainties in the data pipeline. So, the first part, released on 3 December 2020, was denominated Early Data Release 3 (EDR3) and included improved positions, parallaxes and proper motions. The full DR3\footnote{\textit{Gaia} Data Release 3 (DR3), \url{https://www.cosmos.esa.int/web/gaia/dr3/}} of \textit{Gaia}, published on 13 June 2022, included all the EDR3 data plus other relevant data such as Solar System data, variability information, non-single stars, etc. Table~\ref{tab:gaia_releases} shows the number of different type of sources included in each \textit{Gaia} dataset released up to now. A specific data relase, \textit{Gaia} Focused Product Release from October 2023 was focused on $\omega$~Cen to contain more that 500\,000 stars from that region. Fig.~\ref{fig:gaiamaps} shows the four sky maps made with the \textit{Gaia} DR3 catalogue.

The next version, DR4, is planned for the five-year nominal mission. It will contain full astrometric, photometric and radial-velocity catalogues, variable-star and non-single-star solutions, source classifications plus multiple astrophysical parameters for stars, unresolved binaries, galaxies, quasars, and exoplanets, with details of all of them of epochs and transit data. Most of these measurements are expected to be 1.7 times more precise than DR2, being proper motions up to 4.5 times more precise. The last version will be DR5 and will contain more than ten years of data, and its precision will be 1.4 times more precise than DR4, with proper motions 2.8 times more precise than DR4. %The release will not occur before three years after the end of the mission.

Focusing in the last released version, \textit{Gaia} DR3, it provides detailed astrometric solutions for around 1468 million sources \citep{gaiacollaboration23b,babusiaux23}. The uncertainty in their positions, parallaxes, and proper motions ranges from about 0.015 mas to 2 mas. Another 344 million sources have less accurate solutions, with positional uncertainty ranging from 1 to 4 mas. 

\subsubsection{Relevance of \textit{Gaia} mission}
\label{sec:relevance_gaia}

The \textit{Gaia} mission represents a significant advancement over previous astronomical missions due to its unique instrumentation and comprehensive approach to stellar observations. Unlike earlier missions, \textit{Gaia} employs a dual photometric instrument, which includes the Blue Photometer ($BP$) and Red Photometer ($RP$), enabling the measurement of luminosity across a broad spectral range of 320-1000 nm for stars brighter than magnitude 20. In comparison, \textit{Hipparcos} \citep{perryman97}, its catalogue \textit{Tycho}, and its extended catalogue \textit{Tycho-2} \citep{hog00} collected information of stars with magnitude up to $\sim$12.4. \textit{Hipparcos}, primarily focused on astrometry and lacked the sophisticated spectroscopic capabilities of \textit{Gaia}. 

\textit{Gaia}'s Radial Velocity Spectrometer (RVS) not only provides precise measurements of radial velocities but also complements the proper-motion measurements from its astrometric instruments. Furthermore, the RVS uses a distinct methodology to estimate object magnitudes based on data collected from the RP spectrum, thus improving accuracy in classification and observation prioritization \citep{debruijne17}. This capability allows for detailed determination of stellar properties such as temperature, mass, age, and elemental composition, marking a notable enhancement in multi-colour photometry techniques compared to its predecessors. 

The operational framework of \textit{Gaia} is also more sophisticated than earlier missions. Mission operations are managed from the European Space Operations Centre (ESOC) and encompass a wide range of activities, including telemetry acquisition, spacecraft health monitoring, and anomaly management. This holistic approach to mission management ensures that \textit{Gaia} can adapt to various challenges, such as potential instrument failures, by employing redundancy in critical systems—an improvement over the more limited operational strategies of past missions. \textit{Gaia}'s data processing and analysis also set it apart. The mission benefits from a dedicated software pipeline and extensive cross-checking by independent scientific teams, significantly enhancing data reliability and accuracy. In contrast, earlier missions had fewer resources for data validation, which sometimes led to less reliable results.

\subsubsection{End of \textit{Gaia} mission}
\label{sec:end_mission_gaia}

After more than a decade of continuous sky scanning, \textit{Gaia} concluded its scientific operations on 15 January 2025, as its cold gas propellant neared depletion. In the weeks following the cessation of its primary mission, \textit{Gaia} is scheduled to undergo a series of onboard technology tests. Subsequently, the spacecraft will be transitioned from its operational position near the Sun-Earth L2 point to a heliocentric orbit, effectively removing it from Earth's gravitational influence. The final decommissioning and passivation of \textit{Gaia} are planned for 27 March 2025\footnote{Last starlight for ground-breaking \textit{Gaia}, \url{https://www.esa.int/Science_Exploration/Space_Science/Gaia/Last_starlight_for_ground-breaking_Gaia}}. Despite the end of its data collection phase, the mission's impact will continue. The \textit{Gaia} team is focused on processing the extensive dataset gathered over the mission's lifespan. The fourth data release (DR4), encompassing the first 5.5 years of observations, is anticipated in mid-2026. The final data release (DR5), which will include the complete 10.5 years of data, is expected by the end of 2030\footnote{Sky-scanning complete for ESA's Milky Way mapper \textit{Gaia}, \url{https://www.universetoday.com/170452/the-gaia-missions-science-operations-are-over/}}.

\textit{Gaia}’s extensive catalogue of stars has transformed our knowledge of the Milky Way’s structure and evolution, and its scientific impact will continue to grow as researchers delve deeper into the mission’s data. The \textit{Gaia} science team is not slowing down; they are actively preparing for the fourth data release (DR4), expected before mid-2026. With data gathered over five and a half years, DR4 promises not just more information but also improvements in data precision and scope compared to previous releases.

Following the full transmission of \textit{Gaia}’s data to Earth, efforts will shift to compiling the fifth and final data release (DR5). This release will include observations collected over a span of 10.5 years, representing the most extensive dataset yet. However, DR5 is not anticipated to be ready before the late 2020s. So far, less than a third of \textit{Gaia}’s full dataset has been made public, and the complete, fully processed archive may not be available until sometime in the 2030s due to the intensive computational and human effort required\footnote{Goodnight, \textit{Gaia}! ESA spacecraft shuts down after 12 years of Milky Way mapping, \url{https://www.space.com/goodnight-gaia-star-tracking-spacecraft-shuts-down-jan152025}}.

\subsection{Washington Double Star (WDS) catalogue}
\label{sec:WDS_catalogue}

\subsubsection{Main catalogue}
\label{sec:main_catalogue}

WDS catalogue \citep{mason01} is one of the sources we have based on our investigations on multiplicity. As defined in their webpage\footnote{\url{https://crf.usno.navy.mil/wds/}}, ``The Washington Double Star Catalog (WDS) maintained by the United States Naval Observatory is the world's principal database of astrometric double and multiple star information. The WDS Catalog contains positions (J2000), discoverer designations, epochs, position angles, separations, magnitudes, spectral types, proper motions, and, when available, Durchmusterung numbers and notes for the components of 149\,104 systems based on 1\,899\,327 means as of Mon Apr 28 01:47:34 PM EDT 2025''.

\begin{figure}[H]
  \centering
  \includegraphics[width=0.9\linewidth, angle=0]{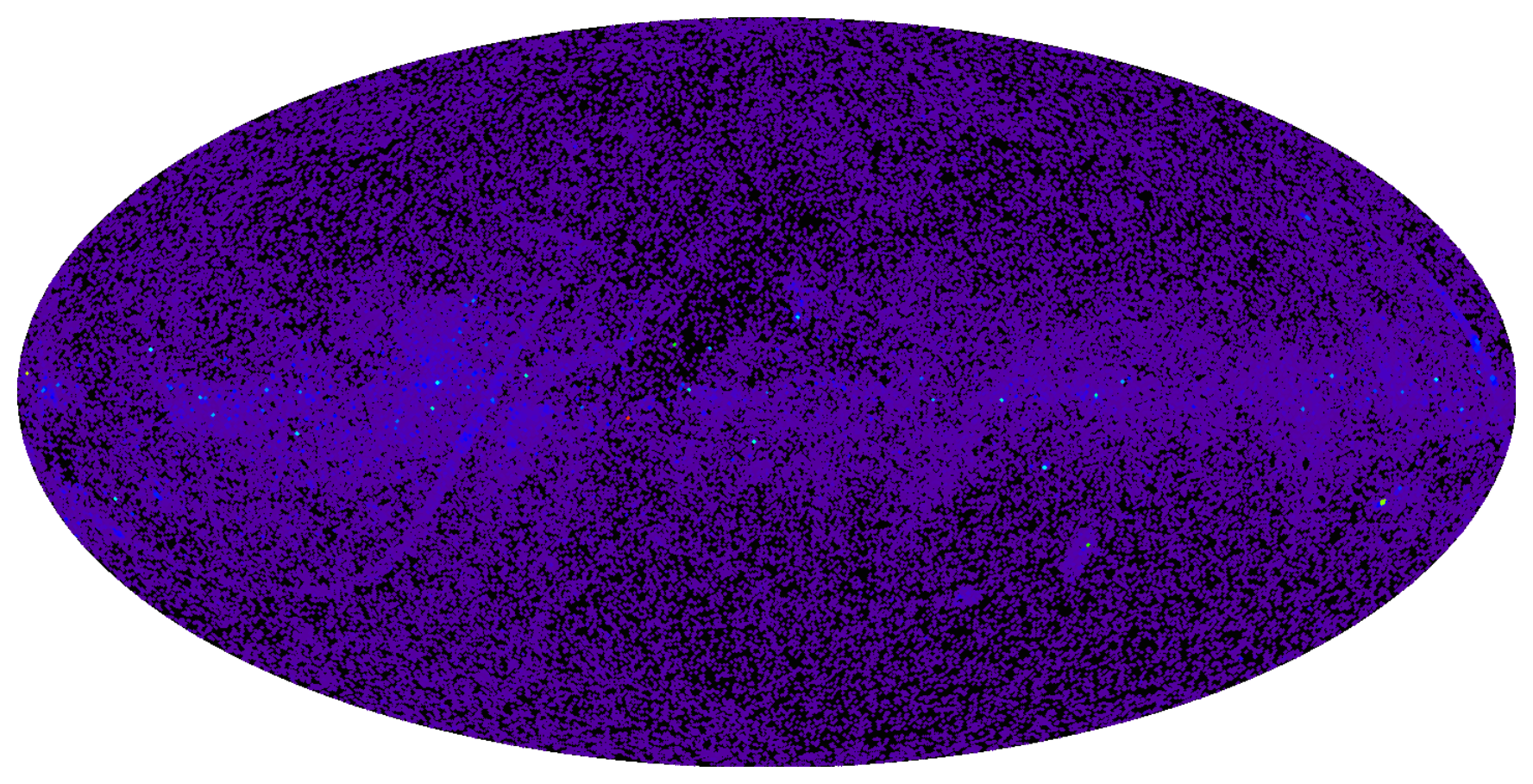}
     \caption[WDS catalogue objects distribution.]{WDS catalogue objects distribution. The colours indicate the density of objects per square degree: Increasing the density, the colours change in the sequence from violet, blue, green, yellow, orange to red. Source: \url{https://cdsarc.u-strasbg.fr/viz-bin/Cat?B/wds}.}
   \label{fig:wds_distribution}
\end{figure}

The catalogue also includes multiple stars. In general, a multiple star with $n$ components will be represented by entries in the catalogue for $n-\text{1}$ pairs of stars. The distribution of objects contained in the WDS catalogue across the sky is quite uniform, as shown in Fig.~\ref{fig:wds_distribution}.

WDS catalogue is sustained through contributions from both professional and amateur astronomers worldwide (the incorporated pairs have discovery codes mentioning the discoverer), including not only modern observations but also early measurements. Observations of double and multiple star systems are submitted regularly by contributors (i.e. \citealt{lepine07a,caballero09,tokovinin08,shaya11}), including data collected from ground-based telescopes, space missions (\textit{Gaia}, i.e. \citealt{gonzalezpayo23,solano23}), and archival research. The catalogue is continuously updated to reflect new measurements, discoveries, and improved data accuracy, making it one of the most comprehensive resources for double star research.

\subsubsection{WDS Supplemental (WDSS) catalogue}
\label{sec:wdss}

In addition to the main catalogue, there is a supplementary database known as ``The Washington Double Star Supplemental Catalog'' (WDSS)\footnote{\url{https://crf.usno.navy.mil/data_products/WDS/Supplement/wdss.html}}, designed to record binary and multiple systems identified in large all-sky surveys. Although not all \textit{Gaia} data have been incorporated yet, many binaries discovered using \textit{Gaia} (e.g., \citealt{tian20,elbadry21}) are already included. As of now, the WDSS contains 2\,428\,660 unique systems (not listed in the WDS), based on 11\,249\,794 individual measurements (according to the website).

Since its initial release in mid-2017, the WDSS has expanded rapidly, largely driven by data from \textit{Gaia} DR2 and DR3. It also serves as a testing ground for potential new formats for the main WDS, so its structure should be considered provisional. The growing number of entries in the WDS has revealed limitations in its current format—for instance, the WDS designation system, based on arcminute-precision coordinates, lacks the accuracy needed to uniquely identify pairs in crowded fields. Additionally, summary lines in the WDS have limited space for cross-referencing primary star names, and offer no room for secondary coordinates or designations. Fields for component labels, magnitudes, spectral types, and proper motions are also constrained.

\subsubsection{Sixth Catalog of Orbits of Visual Binary Stars (ORB6)}
\label{sec:orb6}

The Sixth Catalog of Orbits of Visual Binary Stars (ORB6)\footnote{\url{https://crf.usno.navy.mil/wds-orb6}} is maintained by the USNO and continues the series of visual binary star orbit compilations previously published by \citet{finsen34}, \citet{finsen38}, \citet{worley63}, \citet{finsen70}, \citet{worley83}, and, most recently, by \citet{hartkopf01} in the Fifth Catalog of Orbits of Visual Binary Stars. This version was removed from the web in August 2007, having been fully replaced by the Sixth Catalog.

The Sixth Catalog is continously updated, and currently holds 10\,983 orbits. Each orbit has been assigned a grade on a 1--5 scale, consistent with previous catalogues, though the grading scheme has been revised as described below. Additionally, ephemerides are provided for all orbits with complete elements, along with plots incorporating all relevant data from the current WDS database. ORB6 provides also, for all known visual binary orbits and some astrometric orbits, the seven orbital elements described in Sect.~\ref{sec:orbital_elements}.

\subsection{Virtual Observatory}
\label{sec:virtual_observatory}

Astronomy is undergoing a transformation driven by an unprecedented increase in data production. Advances in telescopes, detectors, and computing technologies now enable astronomical instruments to generate vast amounts of images and catalogues, reaching terabyte scales. These datasets will cover the entire electromagnetic spectrum, from gamma rays and X-rays to visible light, infrared, and radio waves \citep{surace11}. With the decreasing cost of storage and the widespread use of high-speed networks, the concept of seamlessly interconnected multi-terabyte online databases is becoming a reality. More catalogues will be linked, search tools will become increasingly sophisticated, and research conducted using online data will rival that done with traditional telescopes. Additionally, technological progress, driven by Moore’s law \citep{moore65}, is paving the way for new survey telescopes capable of imaging the entire sky every few days and producing petabyte-scale datasets \citep{szalay01}. Today, many huge amounts of data are obtained from different architectures, generally related to a particular instrument: \textit{Hubble} Space Telescope\footnote{\url{https://registry.opendata.aws/hst}}, \textit{Chandra} X-ray Observatory \citep{evans24}, \textit{Gaia} \citep{gaiacollaboration23b}, etc., and it would require the combination of these enormous datasets from multiple instruments. These advancements are fundamentally reshaping how astronomy is practised, with far-reaching implications for the organisation and collaboration within the astronomical community.

\subsubsection{Definition}
\label{sec:VO_definition}

The Virtual Observatory (VO) is an advanced framework that combines various astronomical data ar\-chives and software tools, leveraging computer networks to enhance research in astronomy and related fields. The VO must enable the FAIR principles for data: Findable, Accessible, Interoperable and Re-usable for the astronomers’ needs. By providing a collaborative environment, the VO enables researchers to access, analyse, and synthesise extensive datasets from ground-based and space-borne observatories, promoting the interdisciplinary research, thus transforming the landscape of astronomical research. In the past years, the concept of the VO has gained significant momentum as a solution for managing, analysing, distributing, and ensuring interoperability of astronomical data. The VO has been designed to be used in the future, not as a specific software package but as a complete framework with co-operating data services, allowing the integration of new visualization and analysis tools, and interfaces. So the priority of the VO projects are the full mutual compatibility of the modules to allow a creative diversity\footnote{What is the VO?, \url{https://webtest.ivoa.info/about/what_is_vo/}}. The VO minimises ambiguities in data representation, thereby enhancing interoperability across different datasets and platforms\footnote{NASA Astronomical Virtual Observatories, \url{https://heasarc.gsfc.nasa.gov/navo/summary/vo_glossary.html}}. As technology continues to evolve, the VO is positioned to play a pivotal role in democratising access to astronomical data and fostering innovative research collaborations worldwide. 

\subsubsection{History}
\label{sec:VO_history}

The concept of the VO emerged around the year 2000, concurrently in the United States and Europe, as a response to the increasing need for better access to astronomical data and the growing volume of data generated by various observatories. Initial discussions and organisational meetings were held at prominent institutions, including the California Institute of Technology and the European Southern Observatory (ESO) in Germany, where key events such as ``Virtual Observatories of the Future''\footnote{\textit{Virtual Observatories of the Future}, Caltech campus, Pasadena, California, USA 13--16 June 2000}, and ``Mining the Sky''\footnote{\textit{Mining the Sky}, MPA/ESO/MPE Joint Astronomy Conference, Garching bei München, 31 July 31 -- 4 August 2000} \citep{banday01} laid the groundwork for international collaboration in the astronomical community. 

\begin{figure}[H]
  \centering
  \includegraphics[width=0.7\linewidth, angle=0]{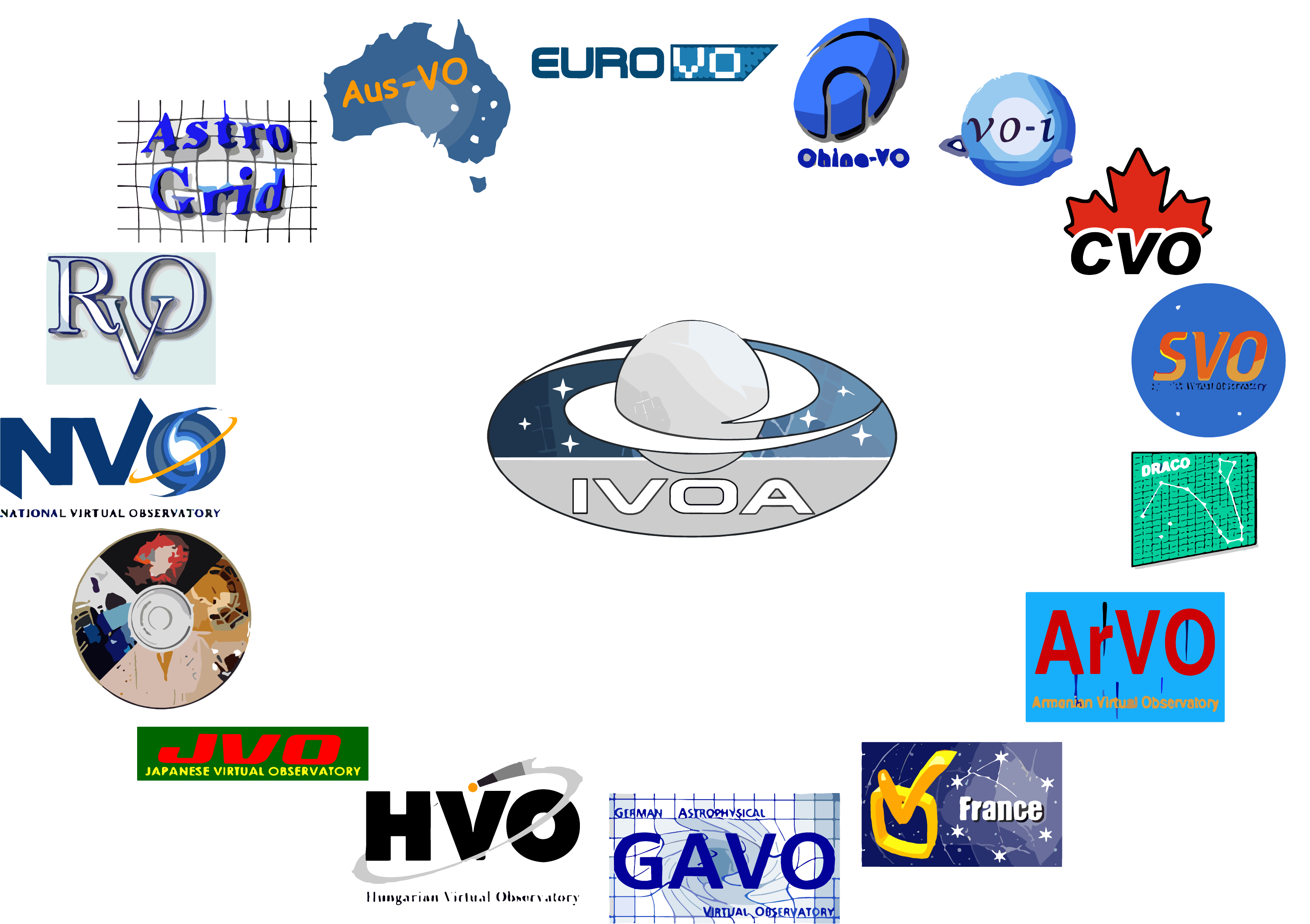}
     \caption[The International Virtual Observatory Alliance (IVOA).]{The International Virtual Observatory Alliance (IVOA).} 
   \label{fig:VO_alliance}
\end{figure}

Following these foundational meetings, the International Virtual Observatory Alliance (IVOA)  was established in June 2002 during a meeting\footnote{IVOA meeting 1, \url{https://wiki.ivoa.net/internal/IVOA/IvoaRepMin/ivoa-fm1-20020613.pdf}} in Garching, Germany. This alliance aimed to unify various national VO projects and facilitate the development of common data access protocols and standards. By 2021, the IVOA had expanded to include 21 national projects and programs across six continents (see Fig.~\ref{fig:VO_alliance}), allowing astronomers worldwide to discover and access data from hundreds of telescopes and facilities through a single query system. The IVOA has also spearheaded the development of key standards and protocols to facilitate the integration and accessibility of diverse astronomical data, despite challenges posed by varying data schemas from multiple sources \citep{mcglynn12}.

As described in the IVOA P. Padovani interop meeting of 2010: ``The final goal of the VO is to facilitate and foster astronomical research and astronomers are its ultimate users and scientific requirements should now drive the IVOA process'' \citep{surace11}. 

\subsubsection{VO standards and tools}
\label{VO_standards}

The exchanges and formats are based on the Extensible Markup Language (XML) format, developed by the World Wide Web Consortium\footnote{World Wide Web Consortium, \url{http://www.w3.org/XML}}. IVOA defined the data models to describe most data that can be provided by the astronomical community. Data models definitions must include their defined VO format. For example, one of most used formats to exchange data is \textit{VOTable}. It is one of the first VO standards defined as an XML standard for the interchange of data represented as a set of tables. It was initially designed to be close to the FITS binary table format and has been set as an exchange format for tabular data. The structure is a set of metadata to describe the fields and a set of separated data \citep{ochsenbein00}.

In order to allow the user to access the data whatever the tool the astronomer is using, data access protocols were defined by IVOA, for example: Simple Image Access (SIA)\footnote{Simple Image Access, Version 2.0. IVOA Recommendation 23 December 2015, \url{https://www.ivoa.net/documents/SIA/20151223/REC-SIA-2.0-20151223.pdf}}, Simple Line Access Protocol (SLAP)\footnote{Simple Line Access Protocol, Version 2.0. IVOA Working Draft 9 April 2019, \url{https://www.ivoa.net/documents/SLAP/20190409/WD-SLAP-2.0-20190409.pdf}}, Single Spectrum Access (SSA)\footnote{Simple Spectral Access Protocol, Version 1.1. IVOA Proposed Recommendation, 6 July 2011, \url{https://www.ivoa.net/documents/SSA/20110706/PR-SSA-1.1-20110706.html}}, Table Access Protocol (TAP)\footnote{Table Access Protocol, Version 1.1. IVOA Recommendation 2019-09-27: \url{https://www.ivoa.net/documents/TAP/20190927/REC-TAP-1.1.pdf}}, ConeSearch\footnote{Simple Cone Search, Version 1.1. IVOA Working Draft 2020-08-28, \url{https://www.ivoa.net/documents/ConeSearch/20200828/WD-ConeSearch-1.1-20200828.html}}, or VO Query Language (VOQL). The most used VO query language is Astronomical Data Query Language \citep[ADQL,][]{yasuda04}.

This is the detailed and comprehensive list of the VO tools that were extensively used throughout the elaboration of this Ph.D. thesis:

\begin{itemize}

    \item \textit{VizieR} \citep{ochsenbein00}: This tool provides the most complete library of published astronomical catalogues --tables and associated data-- with verified and enriched data, accessible via multiple interfaces. Query tools allow the user to select relevant data tables and to extract and format records matching given criteria.

    \item \textit{Simbad} \citep{wenger00}: The \textit{Simbad} astronomical database provides basic data, cross identifications, bibliography and measurements for astronomical objects outside the solar system.    

    \item \textit{Aladin Sky Atlas} \citep{bonnarel00}: \textit{Aladin} is an interactive sky atlas allowing the user to visualise digitised astronomical images or full surveys, superimpose entries from astronomical catalogues or databases, and interactively access related data and information from the \textit{Simbad} database, the \textit{VizieR} service and other archives for all known astronomical objects in the field (Fig~\ref{fig:VO_aladin}).  
    
    \begin{figure}[H]
      \centering
    \includegraphics[width=0.8\linewidth, angle=0]{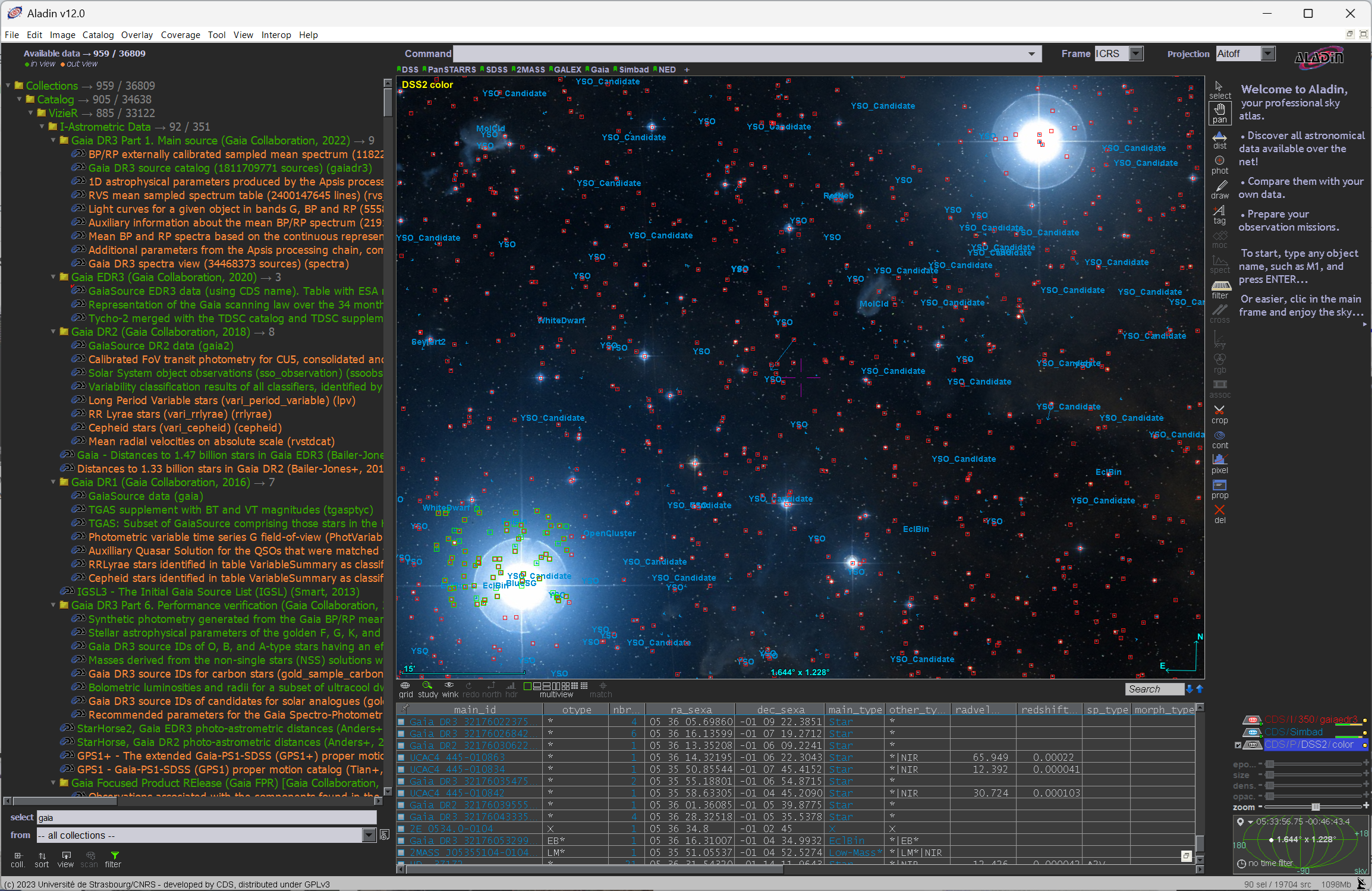}
      \caption{\textit{Aladin} Desktop, main application of the \textit{Aladin} Sky Atlas suite.} 
      \label{fig:VO_aladin}
    \end{figure}

    \item \textit{Topcat} \citep{taylor05}: The tool is an interactive graphical viewer and editor for tabular data. Its aim is to provide most of the facilities that astronomers need for analysis and manipulation of source catalogues and other tables, even non-astronomical data. It understands a number of different astronomically important formats (including FITS, VOTable and CDF) and more formats can be added. It is especially good at interactive exploration of large tables.
    
    \item \textit{VO Sed Analyser \citep[VOSA,][]{bayo08}}: VOSA performs automatically the following tasks:
    \begin{itemize}
        \item Read user photometry-tables.
        \item Query several photometric catalogues accessible through VO services (increases the wavelength coverage of the data to be analysed).
        \item Query VO-compliant theoretical models (spectra) and calculate their synthetic photometry.
        \item Perform a statistical test to determine which model reproduces best the observed data (optionally fitting at the same time the optimal interstellar extinction).
        \item Provide the likelihood of the model parameters (and the interestellar extinction).
        \item Use the best-fit model as the source of a bolometric correction.
        \item Provide the estimated bolometric luminosity for each source.
        \item Generate a Hertzsprung-Russell diagram with the estimated parameters.
        \item Provide an estimation of the mass and age of each source.
    \end{itemize}
    
    \item \textit{SVO Discovery Tool\footnote{\url{http://sdc.cab.inta-csic.es/SVODiscoveryTool/jsp/searchform.jsp}}}: The SVO Discovery Tool is designed as a VO tool capable of automatically searching for scientific content related to astronomical objects. On the one hand, given a list of objects of interest provided by the user, the tool can return the required physical parameters found across \textit{VizieR}, \textit{Gaia}, UCAC5, and HSOY, as well as simultaneously locate spectral and image services. Similarly, it can also search for photometry of the objects in that list through the VOSA tool.
\end{itemize}

Beyond this Ph.D. thesis there are many more tools, not used here. Some remarkable ones are:

\begin{itemize}
    \item \textit{Astroquery} \citep{ginsburg19}: It consists of a set of tools for querying astronomical web forms and databases.
    \item \textit{Cassis} \citep{vastel15}: Web or standalone client for spectrum support of multiple spectrum formats. It uses line list databases for line identification from VAMDC, use SSA, EPN-TAP, SAMP. \textit{Cassis} has also an \textit{Aladin} plug-in to handle data from \textit{Lofar}, \textit{Alma}, \textit{XMM} and, in preparation, \textit{SKA}.
    \item \textit{CDS X-match service}\footnote{\url{http://cdsxmatch.u-strasbg.fr/}}: The CDS X-match service is a tool allowing astronomers to efficiently cross-identify sources between very large catalogues (up to 1 billion rows) or between a user-uploaded list of positions and a large catalogue. 
    \item \textit{Filter profile service}\footnote{\url{http://svo2.cab.inta-csic.es/theory/fps/}}: It is a repository of filter information for the VO.
    \item \textit{STILTS} \citep{taylor06}: The \textit{STIL Tool Set} is a set of command-line tools based on the \textit{Starlink Tables Infrastructure Library} (\textit{STIL}). It deals with the processing of tabular dat. The package has been designed for astronomical tables such as source catalogues. Some of the tools are generic and can work with multiple formats, and others are specific to the VOTable format. \textit{STILTS} is the command-line counterpart of the GUI table analysis tool \texttt{Topcat}. The package is robust, fully documented, and designed for efficiency, especially with very large datasets.
\end{itemize}

\subsubsection{New and independent initiatives: The OAJ case}
\label{sec_vo_initiatives}

Some observatories and scientific entities are developing their own datasets and tools based on standard VO definition. An interesting example is the \textit{Observatorio Astrofísico de Javalambre} (OAJ) in Spain that created new data sets:

\begin{itemize}
    \item \textit{Javalambre-Physics of the Accelerated Universe Astrophysical Survey (J-PAS):}\footnote{\url{https://www.j-pas.org/}} This ambitious project aims to map 8500 square degrees of the northern sky using 59 filters, including broad, intermediate, and narrow bands \citep{benitez14}. The goal is to create a comprehensive 3D map of the universe, detailing the positions and distances of hundreds of millions of galaxies. 
    \item \textit{Javalambre Photometric Local Universe Survey (J-PLUS):}\footnote{\url{https://www.j-plus.es/}} Using 12 unique optical filters, J-PLUS focuses on observing the local universe. The third data release (DR3) encompasses 3192 square degrees and catalogues approximately 47.4 million objects, providing invaluable data for various astrophysical studies \citep{delpino24}. 
\end{itemize}

To ensure that the vast amounts of data generated are accessible and usable by the global scientific community, the Centro de Estudios de Física del Cosmos de Aragón (CEFCA) has developed the CEFCA Catalogues Portal\footnote{\url{https://archive.cefca.es/catalogues/}}. This platform adheres to the FAIR principles by implementing VO services that facilitate seamless data discovery and retrieval. Efforts include the publication of the CEFCA Catalogues Publishing Registry and the enhancement of data provenance information, ensuring that datasets are well-documented and easily accessible for research purposes \citep{civera24}. 
Through these initiatives, the OAJ not only provides cutting-edge astronomical data but also plays a pivotal role in advancing VO standards, promoting collaborative and transparent scientific research.

\subsubsection{VO challenges and limitations}
\label{VO_challenges}

The integration of simulation data and services into the VO environment presents significant challenges that make effective data management and utilization difficult \citep{djorgovski03}:
\begin{itemize}
    \item The increasing volume, quality, and complexity of astronomical datasets. Datasets can contain billions of data vectors in parameter spaces of tens or hundreds of dimensions. Upcoming projects and sky surveys will generate data volumes measured in petabytes.
    \item The heterogeneity of the data and measurement errors. A typical VO dataset may consist of $\sim$10$^9$ data vectors in $\sim$100 dimensions.
    \item Selection effects and censored data. Surveys will be affected by selection effects that operate explicitly on some parameters but also project onto other data dimensions through correlations of these properties. Some selection effects may be unknown.
    \item The intrinsic clustering properties (functional form, topology) of the data distribution in the parameter space of the observed attributes. Object classes form multivariate ``clouds'' in parameter space, but these clouds, in general, need not be Gaussian.
    \item Classification and extraction of desired subpopulations.
    \item Understanding the relationships between observed properties within these subpopulations.
    \item Linking astronomical data to astrophysical models.
    \item Design and implementation of clustering algorithms that should involve close collaboration between astronomers, computer scientists, and statisticians.
    \item Interoperability and reusability of algorithms and models in a wide variety of problems posed by an enriched data environment.
    \item Discovery and interpretation of multivariate correlations in massive datasets.
    \item Development of effective and powerful data visualization.
    \item Scientific verification and evaluation, testing, and monitoring of any newly discovered object clas\-ses, physical clusters discovered by these methods, and other astrophysical analyses of the results.
\end{itemize}
To overcome the challenges faced by VO techniques, different approaches can be considered: Interdisciplinary collaboration, development of faster and smarter algorithms, dimensionality reduction techniques, effective data visualization, consideration of data heterogeneity and complexity, non-parametric data modelling, interoperability and reusability, scientific verification and evaluation, incorporation of scientific knowledge, smart sampling methods, and multiresolution analysis, among others.

\section{Objects studied in this thesis}
\label{sec:objects_studied}

\begin{figure}[H]
  \centering
  \includegraphics[width=0.65\linewidth, angle=0]{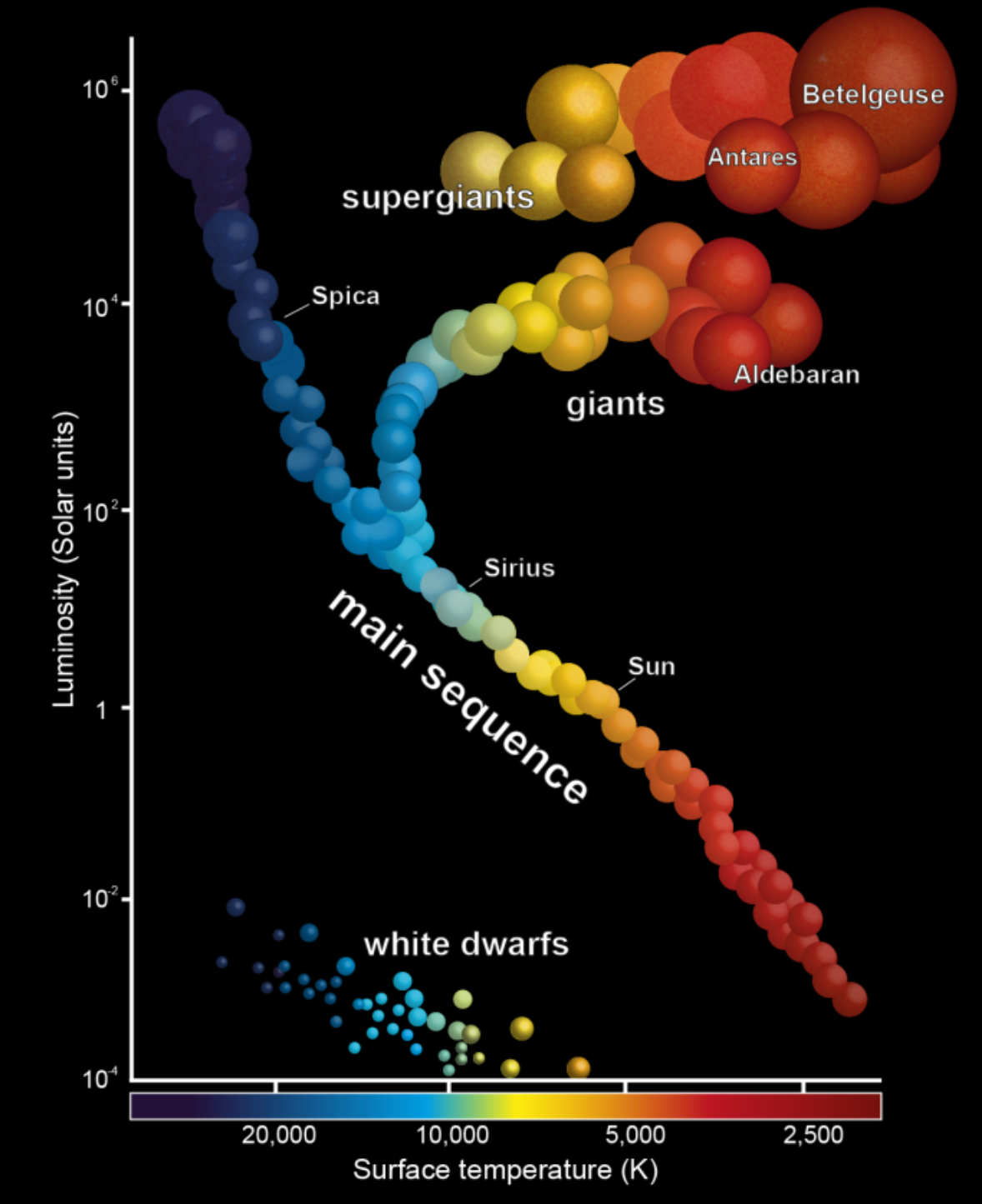}
     \caption[The Hertzprung-Russell diagram.]{The Hertzprung-Russell diagram (credit: \url{https://www.schoolsobservatory.org/learn/space/stars/classification}).} 
   \label{fig:H-R_diagram}
\end{figure}

During this thesis many types of stars have been studied as part of the multiple systems that are analysed. They can be easily located in a Hertzprung-Russell diagram as shown in Fig.~\ref{fig:H-R_diagram}. 

The Hertzsprung-Russell diagram, often shortened to the H-R diagram or HRD, is a graphical representation depicting the correlation between the absolute magnitudes or luminosities of stars and their stellar classifications or effective temperatures. Based on its original mass, each star undergoes distinct evolutionary phases determined by its internal composition and energy generation mechanisms. These phases entail changes in the star's temperature and luminosity, causing it to migrate across various regions of the H-R diagram throughout its evolution. This demonstrates the profound utility of the H-R diagram, enabling astronomers to discern a star's internal makeup and evolutionary phase solely by assessing its placement on the diagram. This scatter plot was developed independently by Ejnar Hertzsprung, first in 1905 \citep{hertzsprung1909}, although the diagrams did not appear until 1911, and by Henry Norris Russell in 1913 \citep{russell1914} who was able to demonstrate the effect more dramatically. This was because he conducted measurements of the parallaxes of numerous stars at Cambridge, allowing him to plot absolute magnitude against spectral type for a multitude of data points \citep{gingerich13}. This diagram marked a significant advancement in our comprehension of stellar evolution.

This section provides an overview of the stellar classifications explored within this study, alongside considerations of substellar objects and extrasolar planets, commonly referred to as exoplanets. It delves into the categorisation of stars, encompassing a range of spectral types, including but not limited to O, B, A, F, G, K, M, and L stars. Additionally, it addresses the properties of substellar objects, such as brown dwarfs, which bridge the gap between stars and planets in terms of mass and luminosity. Furthermore, the discussion extends to the field of exoplanets, highlighting the exploration and characterisation of planets beyond our Solar System, offering valuable insights into planetary diversity and formation mechanisms in direct relation with the multiplicity of star systems.

\subsection{Main sequence stars}
\label{sec:main_sequence_stars}

A main sequence star is a star that is in the main phase of its life cycle. This name provides from the prominent diagonal band from upper left to lower right, called main sequence, in an H-R diagram. Following the Yerkes spectral classification \citep{morgan43}, these stars have the designation of class V to distinguish their luminosity class. They are the most common type of stars in the Universe due to the main sequence is the longest stage of a star's life \citep{arnett96} during which it fuses hydrogen into helium in its core. This process releases huge amounts of energy, which creates the heat and light that we observe from the star. The main sequence stars are primarily composed of hydrogen and helium, with trace amounts of other elements. Their size vary widely, from smaller stars like red dwarfs to massive stars like class O-stars. Size correlates with temperature and luminosity; larger stars are hotter and more luminous, while smaller stars are cooler and less luminous \citep{huang63,eker18,wang18}. The temperature of a main sequence star ranges from around 2000\,K for cool red dwarfs to over 30\,000\,K for hot blue giants. In terms of luminosities, the main sequence stars have a wide range depending on their size and temperature. This type of stars can last for millions to billions of years, depending on its mass. Smaller, cooler stars have longer lifetimes, while larger, hotter stars have shorter lifetimes \citep{romano05,groh19}.

Main sequence stars form from collapsing clouds of gas and dust, called molecular clouds, through a process known as stellar formation. Gravity pulls these clouds together, causing them to contract and heat up. As the temperature and pressure increase, nuclear fusion reactions are initiated in the core, primarily converting hydrogen into helium. Once these reactions stabilise, the star enters the main sequence phase. Throughout their main sequence phase, stars steadily burn hydrogen in their cores, converting it into helium through nuclear fusion. This process releases energy, which creates outward pressure that balances the force of gravity trying to collapse the star. As the star ages, the composition of its core changes as more hydrogen is converted into helium. Eventually, the core runs out of hydrogen fuel, leading to changes in the star's structure and behaviour. When a main sequence star exhausts its core hydrogen fuel, it begins to evolve into a different stage of its life cycle. The exact evolution depends on the mass of the star. For smaller stars like our Sun, they will expand into red giants before eventually shedding their outer layers to form planetary nebulae. The core left behind may become a white dwarf. Larger stars may undergo more dramatic transformations, such as supernova explosions or the formation of neutron stars or black holes. Table~\ref{tab:example_stars} shows some examples of main sequence stars with their spectral types, radius, masses, luminosities and temperatures.

\subsection{Giant and subgiant stars}
\label{sec:giant_stars}

A main sequence star with a mass between 0.5 and 8\,M$_\odot$ evolves into a subgiant when it exhausts the hydrogen in its core, disrupting hydrostatic equilibrium. This occurs because hydrogen fusion in the core ceases, halting the energy production that maintained radiation pressure against gravity. At this point, the star begins to expand. The subgiant phase is transient and lasts relatively little time (a few hundred million years for a star like the Sun). Then, the star continues evolving into a red giant due to the continuous core collapse, further increasing temperature and pressure, and the enhanced fusion in the hydrogen shell, generating more energy and causing the outer layers to expand even further. Therefore, a giant star, also known simply as a giant, refers to a stellar body with a significantly greater radius and luminosity compared to a main-sequence star of the same surface temperature. 

These stars occupy positions above the main sequence on the H-R diagram, typically falling into luminosity classes 0/Ia+ (hypergiants or extremely luminous supergiants), Ia (luminous supergiants), Iab (intermediate-size luminous supergiants), Ib (less luminous supergiants), II (bright giants), III (normal giants), and IV (subgiants). The name of ``giant'' was introduced by Ejnar Hertzsprung around 1905 to distinguish stars with disparate luminosities despite having similar temperatures or spectral types. The evolution of the star in the H-R diagram is: In the main sequence, the star is in the diagonal band ranging from hot blue stars to cool red stars.
As a subgiant, it moves to the right in the H-R diagram.
As a red giant, it increases in luminosity, positioning itself in the upper right region of the diagram.

Giant stars are celestial bodies that have exhausted the hydrogen fuel in their cores having converted it to helium \citep{ridpath04}, and have expanded in size and luminosity compared to their main sequence counterparts. These massive stars are crucial in the life cycle of galaxies, as they play significant roles in enriching interstellar material with heavier elements and in shaping the dynamics of stellar populations. Giant stars come in various spectral types and sizes, with red giants and red supergiants being the most common. These stars exhibit characteristics such as pulsations, strong stellar winds, and the eventual shedding of their outer layers to form planetary nebulae. Observations of giant stars across different wavelengths, from radio to X-rays, provide valuable insights into their physical properties and evolutionary stages.

\begin{table}
 \centering
 \footnotesize
 \caption[Examples of different types of stellar and substellar objects and their main properties. ]{Examples of different types of stellar and substellar objects and their main properties.}
 \scalebox{1}[1]{
 \begin{tabular}{lccccc}
 \noalign{\hrule height 1pt}
 \noalign{\smallskip}
Star & Spectral & Radius & Mass & Luminosity & Temperature \\
 & type & (R$_\odot$) & (M$_\odot$) & (L$_\odot$) & (K) \\
 \noalign{\smallskip}
 \hline
 \noalign{\smallskip}
 \multicolumn{6}{c}{\textit{Super-giant, giant and sub-giant stars}}\\
 \noalign{\smallskip}
 \hline
 \noalign{\smallskip} 
\textit{Alnitak} ($\zeta$ Ori) & O9.7\,Ib & 20 & 33 & 250\,000 & 29\,000 \\
Schulte 19 & B0\,Iab & 246 & 110 & 1\,660\,000 & 13\,700 \\
\textit{Rigel} ($\beta$ Ori) & B8\,Ia & 78 & 21 & 185\,000 & 12\,100 \\
11 Per & B8\,IV & 3.2 & 3.77 & 210 & 14\,550 \\
22 And & F5\,II & 17 & 6.1 & 1436 & 6\,270 \\ 
\textit{Wezen} ($\delta$ CMa) & F8\,Ia & 200 & 17 & 50\,000 & 6\,200 \\
HD 134606 & G6\,IV & 1.16 & 1.05 & 1.16 & 5\,576 \\ 
$\epsilon$ Sge & G3\,III & 6.01 & 3.1 & 138 & 4\,978 \\
\textit{Pollux} ($\beta$ Gem) & K0\,III & 9.1 & 1.91 & 32.7 & 4\,586 \\
\textit{Aldebaran} ($\alpha$ Tau) & K5\,III & 45.1 & 1.16 & 439 & 3\,900 \\ 
\textit{Antares} ($\alpha$ Sco) & M1.5\,Iab-Ib & 680 & 12 & 75\,900 & 3\,600 \\
\textit{Betelgeuse} ($\alpha$ Ori) & M2\,Ia-Iab & 887 & 20 & 100\,000 & 3\,500 \\
UY Sct & M4\,Ia & 909 & 25 & 340\,000 & 3\,365 \\
 \noalign{\smallskip}
 \hline
 \noalign{\smallskip}
 \multicolumn{6}{c}{\textit{Main sequence stars}}\\
 \noalign{\smallskip}
 \hline
 \noalign{\smallskip} 
BI 253 & O2\,V & 12 & 100 & 800\,000 & 50\,000 \\
\textit{Acrab} ($\beta^{\text{01}}$ Sco) & B5\,V & 6.3 & 15 & 31\,600 & 28\,000 \\
\textit{Sirius A} ($\alpha$ CMa) & A1\,V & 1.71 & 2.02 & 25 & 9\,940 \\
\textit{Vega} ($\alpha$ Lyr) & A0\,V & 2.4 & 2.1 & 40 & 9\,600 \\
$\beta$ Pic & A6\,V & 1.8 & 1.75 & 8.7 & 8\,052 \\
\textit{Porrima} ($\gamma$ Vir) & F1\,V & 1.3 & 1.56 & 6 & 6\,757 \\
\textit{Sun} & G2\,V & 1 & 1 & 1 & 5\,780 \\
$\alpha$ Men & G7\,V & 0.96 & 0.96 & 0.81 & 5\,570 \\
\textit{Ran} ($\epsilon$ Eri) & K2\,V & 0.74 & 0.82 & 0.34 & 5\,084 \\
$\epsilon$ Ind A & K5\,V & 0.71 & 0.76 & 0.21 & 4\,649 \\
Gliese 229A & M1\,V & 0.69 & 0.58 & 0.016 & 3\,700 \\
vB 10 & M8\,V & 0.12 & 0.09 & 0.0005 & 2\,508 \\
 \noalign{\smallskip}
 \hline
 \noalign{\smallskip}
 \multicolumn{6}{c}{\textit{White dwarfs}}\\
 \noalign{\smallskip}
 \hline
 \noalign{\smallskip} 
\textit{Sirius B} ($\alpha$ CMa B) & DA1.9\,C & 0.0084 & 1.018 & 0.056 & 25\,000 \\
GALEX J100559.1+224932 A & DA$\sim$ & 0.0210 & 0.28 & ... &21\,100 \\
 $o^{\text{02}}$ Eri B & DA2.9\,C & 0.014 & 0.573 & 0.013 & 16\,500 \\
 GALEX J100559.1+224932 B & DA$\sim$ & 0.0174 & 0.27 & ... &10\,500 \\
\textit{Maru} (L 97--3 A) & DQ\,D & ... & 0.58 & ... & 10\,205 \\
\textit{Procyon B} ($\alpha$ CMi B) & DQ\,Z & 0.012 & 0.60 & 0.00049 & 7\,740 \\
 EGGR 372 & DQ9\,P & 0.0098 & 0.81 & 0.00009 & 5\,590 \\
 EGGR 45 & DZ11 & 0.014 & 0.45 & 0.00006 & 5\,180 \\
  \noalign{\smallskip}
 \hline
 \noalign{\smallskip}
  \multicolumn{6}{c}{\textit{Ultra-cool dwarfs (non brown dwarfs)}}\\
 \noalign{\smallskip}
 \hline
 \noalign{\smallskip}
 LSR J1835+3259 & M8.5\,V & 0.052 & 0.01 & ... & 2\,800 \\
 LP 731--58 & M6.5\,V & 0.089 & 0.12 & 0.0008 & 2\,800 \\
 SPECULOOS-3 & M6.5\,V & 0.134 & 0.1 & 0.08 & 2\,800 \\ 
 Teegarden's Star & M7\,V & $\sim$0.08 & 0.1 & 0.00073 & 2\,700 \\
 Gliese 752 B & M8\,V & 0.089 & 0.1 & 0.00034 & 2\,600 \\
 TRAPPIST-1 & M8\,V & 0.121 & 0.09 & 0.00055 & 2\,516 \\
 Scholz's star & M9.5\,V & 0.095 & 0.1 & 0.0002 & 2\,400 \\
 \noalign{\smallskip}
 \hline
 \noalign{\smallskip}
 \multicolumn{6}{c}{\textit{Brown dwarfs}}\\
 \noalign{\smallskip}
 \hline
 \noalign{\smallskip} 
\textit{Teide 1} & M8\,V & $\sim$0.1 & 0.052 & 0.001 & 2\,600 \\
2MASS J05233822--1403022 & L2.5 & 0.09 & 0.07 & 0.00014 & 2\,074 \\
2MASS J09373487+2931409 & L0 & 0.09 & 0.06 & 0.0002 & 2\,000 \\
Gliese 229B & T7 & 0.045--0.055 & 0.02--0.06 & ... & 1\,200 \\
\textit{Ahra} (L 97--3 B) & Y1 & $<$0.05 & $<$0.012 & ... & 400 \\
WISEA J085510.74--071442.5 & Y1 & 0.1 & 0.003 & 0.0000001 & 250 \\
 \noalign{\smallskip}
 \noalign{\hrule height 1pt}
 \end{tabular}
}
 \label{tab:example_stars}
\end{table}

\subsection{Ultra-cool dwarfs}
\label{sec:ultra-cool_dwarfs_definition}

Objects similar to stars with effective temperatures below 2700\,K (down to about 250\,K, the coldest known up to now, according to \citealt{luhman14}) are denominated ``ultra-cool dwarfs'' (UCD). Ultra-cool dwarfs include objects of spectral types M7--9, L, T, and Y. However, not all of them are main sequence stars. Main sequence stars maintain hydrostatic equilibrium by fusing hydrogen in their cores. In this sense, ultra-cool dwarfs of spectral types M7 to M9 are indeed main sequence stars, although they have extremely low luminosities and surface temperatures below 2500\,K. Some examples are Teegarden's Star \citep[M7\,V,][]{alonsofloriano15} or TRAPPIST-1 \citep[M8\,V,][]{gizis00a} that are ultra-cool main sequence stars. As temperature decreases beyond M9, we encounter L, T, and Y-type objects, which are often brown dwarfs rather than stars:

\begin{itemize}
    \item \textit{L dwarfs (1300--2200 K):} Some may be stars, but many are brown dwarfs that do not sustain stable hydrogen fusion. Examples: DENIS J025503.3-470049 \citep{martin99}, 2MASSW J1507476--162738 \citep{kirkpatrick00}, WISEP J180435.40+311706.1 \citep{kirkpatrick11}.
    \item \textit{T dwarfs (500--1300 K):} Cool brown dwarfs with methane in their spectra. Examples: Gliese 229B \citep{nakajima95}, 2MASS J12545393--0122474 \citep{leggett00}, 2MASSI J0415195--093506 \citep{burrows02}, 2MASS J09393548--2448279 \citep{tinney05}. 
    \item \textit{Y dwarfs (<500 K):} Even colder objects, similar to giant planets. Examples: WISE J154151.65--225024.9, WISE J182831.08+265037.7, WISE J140518.39+553421.3 \citep{kirkpatrick11}, WISE J035934.06--540154.6 \citep{kirkpatrick12}, WISEA J085510.74--071442.5 \citep{tinney14}.
\end{itemize}

The theoretical mass limit between main sequence stars and brown dwarfs is around ~0.075\,M$_\odot$ ($\sim$78\,M$_{\text{Jup}}$), corresponding to spectral types near M8--L2. Objects below this threshold cannot sustain stable hydrogen fusion and are therefore not main sequence stars. 

This diverse category, introduced by \citet{kirkpatrick95}, encompasses stars of exceptionally low mass, and brown dwarfs (that we will see below), including the coldest known stars until spectral type T6.5. It means that many UCDs but not all are still main sequence stars. All these sources are substellar objects incapable of sustaining hydrogen fusion but deuterium. They constitute approximately 15\% of the astronomical objects in the solar neighbourhood \citep{bochanski10,cantrell13,gillon16,kirkpatrick19}. According to core-accretion theory, the diminutive masses of these ultra-cool dwarfs and the compact sizes of their protoplanetary disks suggest the existence of a considerable but, as yet undiscovered population of terrestrial planets orbiting them, ranging from metal-rich Mercury-sized worlds to more hospitable volatile-rich Earth-sized ones (e.g. TRAPPIST-1, \citealt{gillon17}; SPECULOOS-3, \citealt{gillon24}).

Those UCDs that fuse hydrogen do so very slowly and therefore have substantially longer lifespans compared to other types of low-mass stars. Estimates suggest their lifetimes are at least several hundred billion years, with the smallest among them potentially living for about 12 trillion years. Therefore, all UCDs are still in the early stages of their life cycles. Models indicate that, at the conclusion of their lifespans, the least massive (0.25\,M$_\odot$) of these stars will evolve into blue dwarfs rather than expanding into red giants \citep{adams05}.

UCDs play a significant role in the study of stellar formation and evolution, as their unique characteristics (interesting structure, life time, membership in young stellar groups, etc.) offer insights into the mechanisms underlying the development of low-mass stars and the formation of planetary systems. Therefore, they are good star evolution tracers.

\begin{figure}[H]
  \centering
  \includegraphics[width=0.95\linewidth, angle=0]{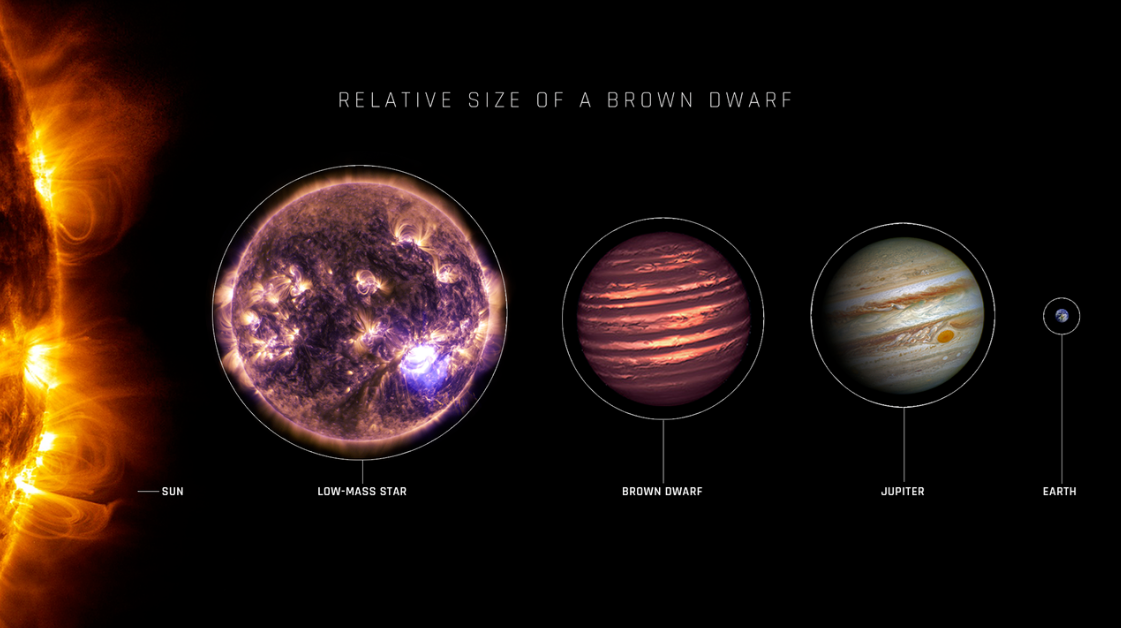}
     \caption[Relative size of a brown dwarf.]{Relative size of a brown dwarf. Credit: NASA, ESA, SDO, NASA-JPL, Caltech, Amy Simon (NASA-GSFC). Designer: Elizabeth Wheatley (STScI).} 
   \label{fig:BD_comparison}
\end{figure}

A brown dwarf, whose existence was first confirmed in 1995 \citep{nakajima95,rebolo95}, is an interesting type of UCD and an object that is very recurrently found in multiple systems as this thesis shows. The heating of a star's core depends on the gravitational energy it releases, which, in turn, is determined by the star's mass and the degree to which it contracts. Consequently, stars with lower masses must contract to achieve higher densities so that their cores can reach the minimum temperature required for hydrogen fusion. For stars with masses below 0.1\,M$_\odot$, the core density becomes high enough for free electrons to occupy the lowest energy Fermi states. The most energetic states of this degenerate electron population provide pressure support against gravitational contraction, preventing the star's radius from decreasing significantly below R$_\text{Jup}$. When objects have a mass less than 0.072 M$_\odot$ (for solar metallicity, since mass is strongly dependant with metallicity, especially for low-mass stars, according to \citealt{burrows89}, and \citealt{kroupa97}), degeneracy pressure stops contraction before the critical hydrogen fusion temperature is achieved. This results in hydrostatic equilibrium being reached, but not thermal equilibrium. These ``failed stars'' \citep{burgasser08b} are brown dwarfs. They are substellar objects that are more massive than the biggest gas giant planets, but less than the least massive main-sequence stars (see size comparison in Fig.~\ref{fig:BD_comparison}). Their mass is between 13 and 80\,M$_\text{Jup}$, even though other authors establish different mass ranges.

\subsection{White dwarfs}
\label{sec:white_dwarfs_definition}

A white dwarf is a type of star that represents the final stage in the evolution of most low and intermediate mass stars, including those with initial masses up to $\sim$8.5--9\,M$_\odot$ with low metallicity, or up to $\sim$9.5--11\,M$_\odot$ with solar metallicity \citep{siess07}. In fact, they are among the most common stellar objects in the Milky Way. These stars are the remnant cores of stars that have exhausted their nuclear fuel and undergone red giant phases with significant mass loss. As a result, they contract to very small sizes, typically similar to that of Earth, while retaining a mass comparable to that of our Sun. While most massive stars become supernovae as they reach the end of their lives, a star with a mass less than the former mentioned limit will eventually evolve into a white dwarf. Approximately 97\% of the stars in the Milky Way will undergo this transformation, ultimately becoming white dwarfs \citep{fontaine01}.

Inside a white dwarf, the matter is under immense pressure, leading to pressure ionization, where atoms are fully ionised and electrons are free to move throughout the star like in a metal \citep{althaus10}. Due to the high densities and the uncertainty principle, these electrons possess very high momenta, and the effects of special relativity must be considered when studying their properties. The internal structure of a white dwarf is largely governed by electron degeneracy pressure, a quantum mechanical effect that arises when electrons are packed into a small volume \citep{hoard11}. This degeneracy pressure is mostly independent of temperature, which is why white dwarfs cannot prevent cooling by contracting like normal stars.

White dwarfs are typically composed primarily of carbon and oxygen, the products of helium burning in their progenitor stars' cores. They also possess thin outer layers of hydrogen and/or helium. Having exhausted their nuclear fuel, white dwarfs no longer generate energy through fusion. They simply radiate their stored thermal energy into space and gradually cool down over billions of years. As they cool, the ions in their core can crystallize, releasing latent heat that affects the cooling rate \citep{vanhorn15}. White dwarfs have extremely high surface gravities, reaching values of the order of $log\,g$$\approx$\,8 \citep{hoard11}, causing heavier elements to sink towards the core and lighter elements to float to the surface. This gravitational settling is a key factor in determining the observed atmospheric composition.

Actual white dwarf models, which consider special relativistic effects in the degenerate electron equation of state, were constructed by \citet{chandrasekhar31a,chandrasekhar31b}. During this analysis, \citet{chandrasekhar31b} made the groundbreaking discovery that white dwarfs had a maximum mass, commonly referred to as the ``Chandrasekhar limit'', of 1.4565\,M$_\odot$, but in reality, the value depends on the metallicity. Therefore, if the white dwarf is primarily composed of helium, it does not form through the typical stellar evolution of single stars. Helium white dwarfs usually have lower masses (less than 0.5\,M$_\odot$), and are produced in binary systems where the evolution of the progenitor star is interrupted by mass transfer \citep{vanhorn15}. For white dwarfs primarily composed of carbon, the zero-temperature white dwarf models by \citet{hamada61} for pure carbon composition reach a maximum mass of 1.396\,M$_\odot$ when electrostatic interactions are included. For a white dwarf primarily composed of oxygen, the mentioned models foresee a limit similar to white dwarfs mainly composed of carbon. Finally, precise calculations for pure iron white dwarfs give a maximum mass of 1.112 M$_\odot$, again using the models from \citet{hamada61}. Although we know white dwarfs with estimated masses as between 0.17\,M$_\odot$ \citep{kilic07} and 1.33\,M$_\odot$ \citep{kepler07}, the mass distribution is strongly concentrated around 0.6\,M$_\odot$, with the majority falling within the range of 0.5--0.7\,M$_\odot$ \citep{kepler07}. The estimated radii of observed white dwarfs typically range from 0.8\% to 2\% of the radius of the Sun \citep{shipman79}. This corresponds roughly to the Earth's radius, which is approximately 0.9\% of the solar radius. It is important to highlight that white dwarfs with masses exceeding the Chandrasekhar limit are unstable and will collapse, potentially leading to events such as Type Ia supernovae in binary systems \citep{nomoto84,hillebrandt00,maoz14}.

White dwarfs are often found in binary systems. In close binaries, a white dwarf can accrete matter from its companion star, leading to phenomena like dwarf novae and classical novae. Accretion can also pollute the white dwarf's atmosphere with elements from the companion or the surrounding environment \citep{vanhorn15}.

\begin{table}[H]
 \centering
 \caption[Primary spectral types in white dwarfs.]{Primary spectral types in white dwarfs (updated by \citealt{gray09} from the original table from \citealt{sion83}).}
 \footnotesize
 \scalebox{1}[1]{
 \begin{tabular}{l@{\hspace{3mm}}l}
 \noalign{\hrule height 1pt}
 \noalign{\smallskip}
Symbol & Characteristics \\
 \noalign{\smallskip}
 \hline
 \noalign{\smallskip}
 DA    &  Only Balmer lines, no He {\tiny I} or metals present \\
 DB    &  He {\tiny I} lines; no H or metals present \\
 DC    &  Continuous spectrum, no lines deeper than 5\% in any part of the electromagnetic spectrum \\C
 DO    &  He {\tiny II} strong; He {\tiny I} or H present\\ 
 DZ    &  Metal lines only; no H or He \\
 DQ    &  Carbon features, either atomic or molecular, in any part of the electromagnetic spectrum  \\
 \noalign{\smallskip}
 \hline
 \noalign{\smallskip}
 Additional symbols \\
 \noalign{\smallskip}
 \hline
 \noalign{\smallskip}
 P     & Magnetic white dwarfs with detectable polarization \\ 
 H     & Magnetic white dwarfs without detectable polarization \\
 X     & Peculiar or unclassifiable spectrum \\
 E     & Emission lines are present \\
 ?     & Uncertain assigned classification; a colon (:) may also be used \\
 V     & Optional symbol to denote variability \\
 d     & Circumstellar dust \\
 C {\tiny I}, C {\tiny II}, O {\tiny I}, O {\tiny II} & Added within parentheses to hot DQ star types to indicate presence of these atomic species  \\
 \noalign{\smallskip}
 \noalign{\hrule height 1pt}
 \end{tabular}
}
\label{tab:white_dwarfs}
\end{table}

The spectral types of white dwarfs are peculiar and very different from those of main sequence stars. White dwarfs are classified based on the dominant elements observed in their atmospheres as shown in Table~\ref{tab:white_dwarfs}. The system is based on the proposal from \citet{luyten52}, later developed by \citet{greenstein60}, where \citet{kuiper41} was the first person to systematically attempt classifying white dwarf spectra \citep{sion83}. The system, known as the primary spectral type, classifies the white dwarfs according to the presence of certain lines. White dwarfs are broadly divided into two main groups: those exhibiting Balmer lines (hydrogen-rich white dwarfs, designated as DAs) and those that do not, collectively referred to as non-DAs. Within the non-DA category, various subtypes are identified based on the features observed in their spectra. White dwarfs showing helium absorption lines (He {\tiny I} or He {\tiny II}) are classified as DB and DO, respectively. Those with carbon features, whether atomic or molecular, are labelled DQ, while those displaying metallic lines such as Ca {\tiny II} or Fe {\tiny II} are called DZ. White dwarfs with very weak spectral features or none at all, resulting in a continuous spectrum, are classified as DC. It is also frequent to observe multiple elements in the spectrum of a white dwarf, resulting in composite classifications. For instance, a DA white dwarf that shows weaker helium lines is classified as DAB, while one exhibiting additional metallic lines is designated as DAZ. Additionally, certain phenomena, such as the presence of a magnetic field or spectral variability, are indicated by appending a secondary label, H or V, respectively, to the primary spectral type \citep{garciazamora23}.

\subsection{Exoplanets}
\label{sec:exoplanets}

\subsubsection{Definition}
\label{sec:definition_exoplanets}

\begin{figure}[H]
  \centering
  \includegraphics[width=0.6\linewidth, angle=0]{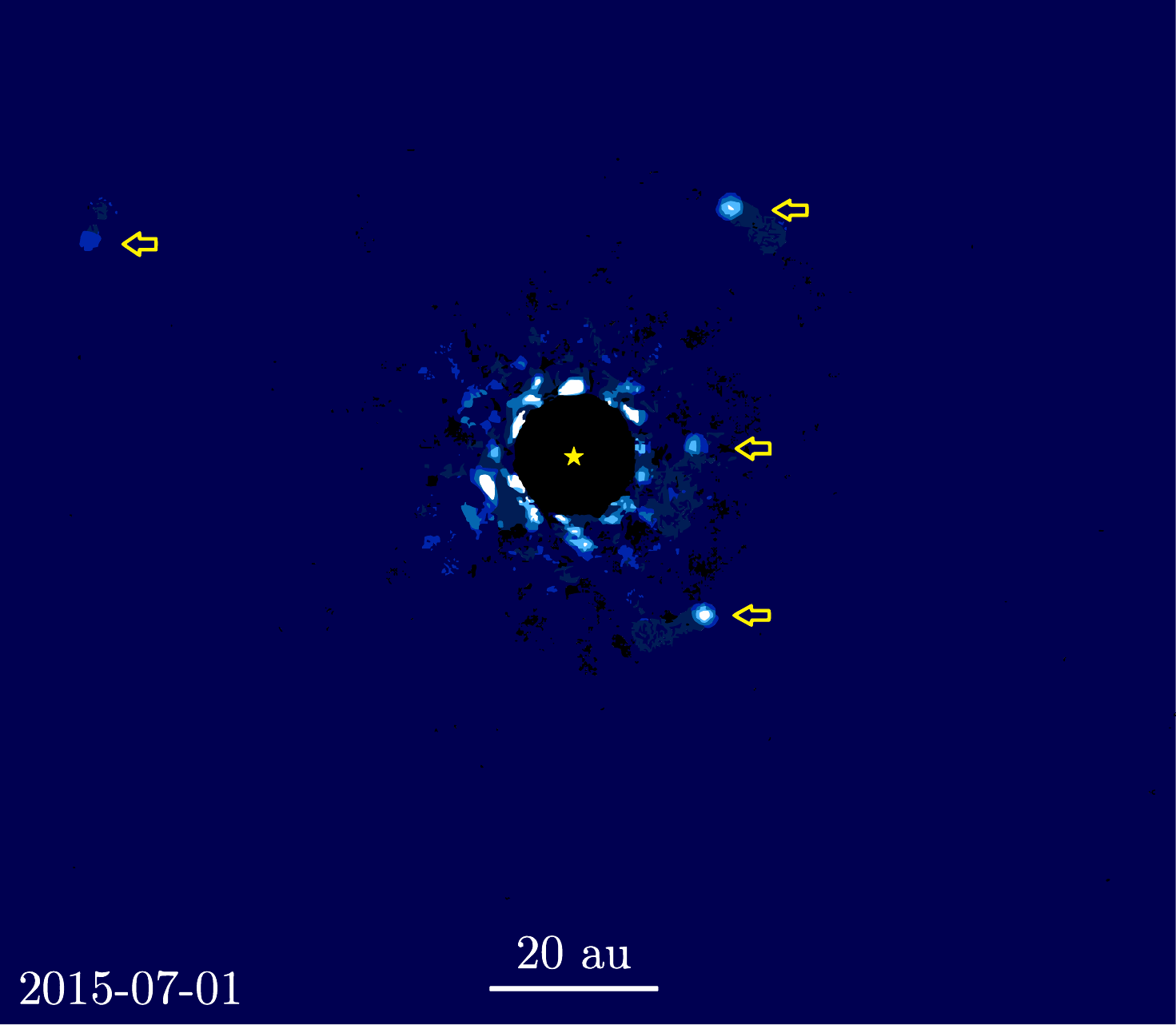}
     \caption[HD 218396 system with four super-Jupiters.]{The HD 218396 system harbors four super-Jupiters (marked with arrows) orbiting with periods that range from decades to centuries. Image taken with the Keck Telescope. Credit: Jason Wang (Caltech) and Christian Marois (NRC Herzberg).
} 
   \label{fig:exoplanets}
\end{figure}

An exoplanet, also known as an extrasolar planet, refers to a planet that orbits around other star, beyond our Solar System. Although the first indications of an exoplanet were recorded as early as 1917 \citep{vanmaanen1917}, they were not initially recognised as such. It was not until 1992 that the first confirmation of an exoplanet detection occurred \citep{wolszczan92}. Subsequently, in 2003, a different exoplanet, initially identified in 1988 \citep{campbell88}, was officially confirmed. As of 4 June 2025, there have been 7487 exoplanets confirmed in 5130 planetary systems, with 1047 systems hosting multiple planets \citep[Encyclopaedia of exoplanetary systems\footnote{\url{https://exoplanet.eu/home/}},][]{schneider11}. The \textit{James Webb} Space Telescope \citep{mcelwain23} is anticipated to uncover additional exoplanets and provide deeper insights into their characteristics, including their composition, environmental conditions, and potential habitability \citep{beichman14,beichman17,kalirai18,gialluca21}. An image of the star HD\,21839 with four exoplanets can be seen in the Fig.~\ref{fig:exoplanets}.

In 2001, the International Astronomical Union (IAU) Working Group on Extrasolar Planets released a position statement outlining a working definition of an ``exoplanet'': An exoplanet is a planetary-mass object that orbits a star (or stellar remnant) outside the Solar System. To be classified as an exoplanet, the object must meet specific criteria, including:
\begin{itemize}
    \item \textit{Orbiting a star or stellar remnant:} It must be bound gravitationally to a star or a stellar remnant, rather than free-floating in space (which would classify it as a rogue planet).
    \item \textit{Mass limit:} It must have a mass below the deuterium fusion boundary ($\sim$13\,M$_\text{Jup}$), distinguishing it from brown dwarfs.
    \item \textit{Planetary characteristics:} The object should have a formation history or composition similar to planets, as opposed to stars or stellar remnants.
\end{itemize}

This definition underwent  modifications\footnote{\url{https://w.astro.berkeley.edu/~basri/defineplanet/IAU-WGExSP.htm}} in 2003: ``Rather than try to construct a detailed definition of a planet which is designed to cover all future possibilities, the Working Group on Extrasolar Planets (WGESP) has agreed to restrict itself to developing a working definition applicable to the cases where there already are claimed detections, e.g., the radial velocity surveys of companions to (mostly) solar-type stars, and the imaging surveys for free-floating objects in young star clusters. As new claims are made in the future, the WGESP will weigh their individual merits and circumstances, and will try to fit the new objects into the WGESP definition of a `planet', revising this definition as necessary. This is a gradualist approach with an evolving definition, guided by the observations that will decide all in the end. Emphasising again that this is only a working definition, subject to change as we learn more about the census of low-mass companions, the WGESP has agreed to the following statements:''

\begin{itemize}
    \item Objects with true masses below the threshold mass for thermonuclear fusion of deuterium, currently calculated to be 13 Jupiter masses, for sources of solar metallicity that orbit stars or stellar remnants are ``planets'' regardless of their origin. The minimum mass or size required for an exoplanetary entity to qualify as a planet ought to align with the standards applied within our Solar System.
    \item Substellar objects with actual masses surpassing the threshold for deuterium thermonuclear fusion are classified as ``brown dwarfs'', independently of their formation process or location.
    \item Objects drifting freely within young star clusters that possess masses below the threshold required for deuterium thermonuclear fusion are not categorised as ``planets'' but rather as ``sub-brown dwarfs'', or any other suitable designation.
\end{itemize}

Although these criteria are based purely on the deuterium-burning mass or on the formation mechanism, they constitute a starting point for a reasonable working definition of a planet. Likely, this definition will evolve as our knowledge of exoplanets continues to expand.

\subsubsection{Methods for exoplanet detection}
\label{sec:methods_exoplanets_detection}

Every planet emits an exceedingly dim illumination compared to its host star. For instance, a star similar to the Sun shines approximately a billion times brighter than the reflected light emanating from any of its orbiting planets. Apart from the inherent challenge of discerning such faint luminosity, the glare from the parent star obscures it. Consequently, only a few exoplanets have been directly observed, with even fewer successfully distinguished from their host stars. Instead, astronomers have typically relied on indirect techniques to detect exoplanets. Currently, various indirect methods have been proven successful. We mention here the most efficient different methods (represented in Fig.~\ref{fig:exop_detection}.):

    \begin{figure}[H]
        \centering
        \includegraphics[width=1\linewidth, angle=0]{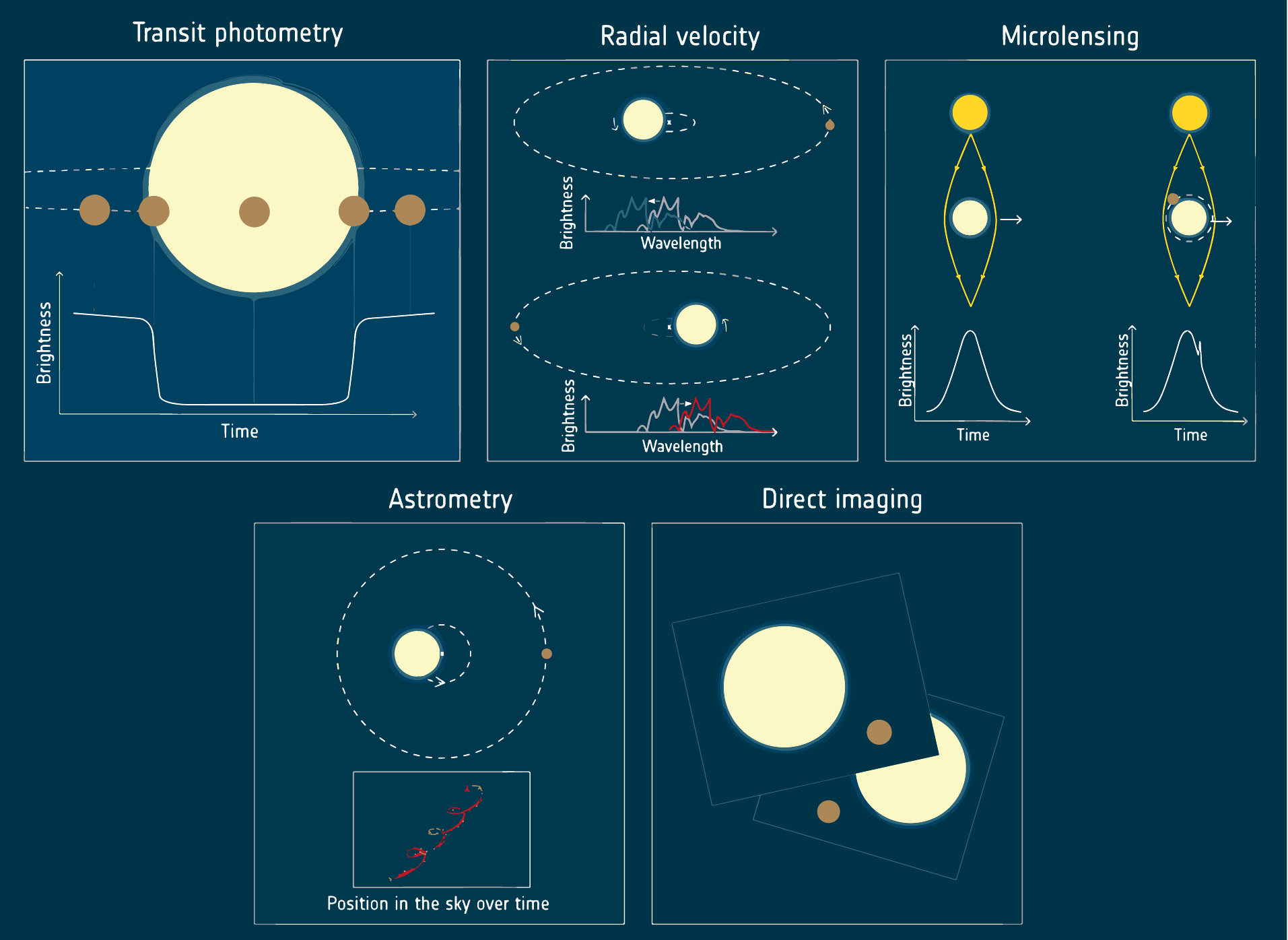}
         \caption[Exoplanet detection methods.]{Exoplanet detection methods. Credit: The European Space Agency  (\url{https://www.esa.int/ESA_Multimedia/Images/2019/12/Exoplanet_detection_methods}).} 
       \label{fig:exop_detection}
    \end{figure}    

\begin{enumerate}
    \item \textit{Radial velocity method:} The movement of a planet as it orbits around a star induces a reciprocal motion in the star around the combined centre of mass of the star and the planet. Consequently, this leads to the cyclic disturbance of three characteristics of the star, all of which have been observed: changes in radial velocity, shifts in angular position on the celestial sphere, and alterations in the timing of periodic reference signals \citep{perryman11}. The velocity of the star around the system's centre of mass is significantly smaller compared to that of the planet, primarily because the radius of its orbit around the centre of mass is extremely small. Nevertheless, modern spectrometers, such as the HARPS \citep[High Accuracy Radial Velocity Planet Searcher,][]{mayor03} instrument at the ESO 3.6-meter telescope in La Silla Observatory, Chile, the HIRES \citep[High Resolution Echelle Spectrometer,][]{vogt94} spectrometer at the Keck telescopes, EXPRES \citep[EXtreme PREcision Spectrometer,][]{jurgenson16} at the Lowell Discovery Telescope, or CARMENES, \citep[Calar Alto high-Resolution search for M dwarfs with Exoearths with Near-infrared and optical Echelle Spectrographs,][]{quirrenbach14} are capable of detecting velocity variations as low as 3 m/s or even less.
    
    \item \textit{Transits method:} While the radial velocity method provides information about the mass of a planet, the transit method can determine the radius of the planet. When a planet transits in front of the disk of its host star, the observed brightness of the star decreases by a small amount. This number depends on the relative sizes of the star and the planet. This method presents two significant inconveniences. Firstly, the visibility of planetary transits depends on the fortuitous alignment of the orbit of the planet with the perspective of astronomers. The probability of such alignment occurring is determined by the ratio of the diameter of the star to the orbital diameter, although for smaller stars the size of the radius of the planet is also important. Approximately 10\% of planets with close orbits experience this alignment, a proportion that diminishes for those with larger orbits. For instance, for a planet orbiting a Sun-sized star at 1,au, the likelihood of a random alignment resulting in a transit is 0.47\%. A further challenge of the transit method is its bias toward detecting close-in planets, as the probability of transit decreases significantly with increasing orbital distance. 
    
    Consequently, this method cannot definitively rule out any given star as a potential planetary host. Nevertheless, through the comprehensive scanning of vast areas of the sky, encompassing thousands or even hundreds of thousands of stars simultaneously, transit surveys can uncover a greater number of extrasolar planets compared to the radial-velocity method \citep{hidas05}. Space missions such as CoRoT \citep{baglin06}, Kepler \citep{borucki10}, and TESS \citep{ricker15} have  revolutionised exoplanet discovery by detecting the tiny dips in starlight caused by planets transiting their host stars. They enabled the identification of thousands of exoplanet candidates, many Earth-sized and in habitable zones.

    \item \textit{Direct imaging method:} The light reflected by planets is extremely faint compared to that of their stars, making their detection challenging as they are eclipsed by the halo of light of the host star. If planets orbit at a sufficiently distant distance, even though the reflected light is weak, they could be detected through the thermal emission they produce. The scenario in which detecting exoplanets is easier is when the system is closer to us and the planets are much larger than Jupiter, they are widely separated from their star, and they are hot enough to emit infrared radiation, in which they are much brighter than in visible wavelengths. Coronagraphs are often used to block the light from the star, making the planet more visible. 
    
    This direct imaging method requires significant optothermal stability \citep{brooks15}. Sometimes, multiple observations at different wavelengths are needed to verify if the planet is indeed a brown dwarf. The direct imaging method is often used to reliably measure the orbits of the studied planets. It is evident that this method is more effective, unlike the transit method, when the orbits of these planets are perpendicular to our line of sight, meaning we can observe the planet throughout its entire orbit without occultations behind the star host. Planets detected with this method can fall into two categories: planets orbiting around much more massive stars than the Sun and that they are young enough to have protoplanetary disks; and brown dwarfs orbiting dim stars or at least 100\,au away from their stars. The direct imaging method is also commonly used to detect rogue planets.
    
    \item \textit{Gravitational microlensing method:} Gravitational microlensing occurs when the gravitational field of a star works as a magnifying lens, amplifying the light from a distant background star. This phenomenon occurs only under nearly perfect alignment between the two stars. These lensing events are very brief, typically lasting for weeks or days, and they are produced due to the relative motion of the two stars and Earth with respect to each other. Over the past decade, more than a thousand of such events have been observed. If the foreground star responsible for lensing hosts a planet, the own gravitational field of the planet can contribute measurably to the lensing effect. However, due to the rarity of such alignments, continuous monitoring of a vast number of distant stars is necessary to detect planetary microlensing contributions at a reasonable rate.
    
    This method proves most effective for planets located between Earth and the galactic center, where a plethora of background stars is available. The main projects for detection through microlensings are: OGLE\footnote{Optical Gravitational Lensing Experiment, \url{https://ogle.astrouw.edu.pl/}} \citep{udalski92}, KMT\footnote{Korea Microlensing Telescope Network, \url{https://kmtnet.kasi.re.kr/kmtnet/}} \citep{henderson14}, and MOA\footnote{Microlensing Observations in Astrophysics, \url{http://www2.phys.canterbury.ac.nz/moa/}} \citep{alcock95,alcock97} surveys, and some pulsars. There are many more initiatives such as PLANET\footnote{Probing Lensing Anomalies NETwork, \url{https://planet.iap.fr/}}, that joined to $\mu$FUN\footnote{Microlensing Follow-Up Network, \url{https://cgi.astronomy.osu.edu/microfun/}} collaboration in 2009. This approach enables almost uninterrupted, round-the-clock surveillance through a global network of telescopes, offering the chance to detect microlensing signals from planets with low masses (similar to Earth). This strategy proved successful in the detection of the first low-mass planet on a wide orbit \citep{beaulieu06}.

    \item \textit{Astrometry method:} The astrometry method consists of a detailed measuring and recording the stars position in the sky, and observing the evolution in time of such position. If there is a planet orbiting the star, the star will move in a small elliptical orbit due to the gravitational influence of the planet. The more massive the star is, the smaller the orbit will be. This method is the oldest one for searching for extrasolar planets, probably used even earlier than it was used by William Herschel in the late 18th century. The first registered astrometric calculation for an exoplanet was made by William Stephen Jacob in 1855 for 70 Oph \citep{jacob1855}, even though there were later measurements by \citet{see1896}. Since then, other astronomers have made calculations for this star and others, leading to the discovery of several planets around the star Lalande 21185 in 1996 \citep{gatewood96}, but eventually, all these discoveries were refuted \citep{see1896,sherrill99}. The reason is that the changes in stellar positions are too small compared to atmospheric variations, making the technique highly debated within the scientific community. As a result, all planets less than 0.1\,M$_\odot$ discovered before 1996 using this method were not considered valid. The shift occurred in 2002 when the \textit{Hubble} Space Telescope (HST) characterised a planet around the star Gliese 876 \citep{gatewood96}, previously discovered using this technique.

    A potential benefit of the astrometric method is its high sensitivity to planets with large orbits, making it a valuable complement to other techniques that are more effective at detecting planets with smaller orbits. However, this method demands very long observation periods (years or even decades) since planets far enough from their star to be detected via astrometry also take a long time to complete an orbit. Planets orbiting one of the stars in binary systems are easier to detect because they cause noticeable perturbations in the stars' orbits. Nonetheless, follow-up observations are required with this method to determine which star the planet is orbiting. By 2022, the use of radial velocity combined with astrometry, particularly with the help of \textit{Gaia}, has led to the detection and characterisation of many Jovian planets \citep{feng21,li21,feng22,winn22}, such as the closest Jupiter analogues, $\epsilon$~Eri b and $\epsilon$~Ind~Ab \citep{feng19,llopsayson21}. Moreover, radio astrometry with the Very Long Baseline Array\footnote{National Radio Astronomy Observatoty: The Very Large Array, \url{https://public.nrao.edu/telescopes/vla/}} (VLBA) has been instrumental in discovering planets around TVLM~513--46546 and EQ~Peg~A \citep{curiel20,curiel22}.

\end{enumerate}

Table~\ref{tab:exoplanet_detection_methods} shows a comparison of the different exoplanet detection methods, remarking the suitable planets to be applied, their advantages, challenges, and the number of planets discovered with each method.

\begin{table}[H]

\centering
\caption[Comparison of exoplanet detection methods.]{Comparison of exoplanet detection methods.}
\footnotesize
\begin{tabular}{p{1.8cm}p{1.8cm}p{3.6cm}p{4cm}p{2.5cm}}
\noalign{\hrule height 1pt}
\noalign{\smallskip}
Method & Type & Advantages & Challenges & Number of \\
Method & of planet &  &  & discoveries \\
\noalign{\smallskip}
\hline
\noalign{\smallskip}
Radial & All & Detects planets over wide & Less sensitive to small planets, & Hundreds (HARPS, \\
velocity &  & distances, determines mass & challenging for distant ones & CARMENES) \\
\noalign{\smallskip}
\hline
\noalign{\smallskip}
Transits & All & High discovery rate, & Requires alignment with  & Thousands \\
 &  & detailed atmospheric study & our line of sight & (Kepler, TESS) \\
\noalign{\smallskip}
\hline
\noalign{\smallskip}
Direct & Large, young & Direct observation, & Only effective for large, & Few (e.g., \\
imaging & planets & atmospheric study & distant planets & HR 8799 system) \\
\noalign{\smallskip}
\hline
\noalign{\smallskip}
Gravitational & Distant & Can detect planets & Limited data, rare events & Few (OGLE,  \\
microlensing & planets &  at great distances &  &  KMT, MOA) \\
\noalign{\smallskip}
\hline
\noalign{\smallskip}
Astrometry & Planets in & Detects distant,  & Requires extreme precision, & Few \\
 & wide orbits & wide-orbit planets & difficult for small planets &  \\
\noalign{\smallskip}
\noalign{\hrule height 1pt}
\end{tabular}
\label{tab:exoplanet_detection_methods}
\end{table}

\section{State of the art in stellar multiplicity}
\label{sec:state_art_stellar_multiplicity}

In this section, we will present what the scientific community knows about the various aspects of multiplicity that this thesis focuses on, as well as the motivations for conducting this research. In the next section, we will outline the objectives for each described area.

\subsection{About the first astronomer in history to list binaries}
\label{sec:first_astronomer}

The interest in the study of stellar multiplicity has existed for millennia. Ptolemy, in his 2nd-century \textit{Almagest}, already showed interest in double stars. It has generally been accepted that the pioneers of double star astronomy were Christian Mayer and William Herschel. However, a little-known Italian astronomer Giovanni Battista Hodierna, published a list of twelve binary stars in his book \textit{De systemate orbis cometici} \citep{hodierna1654} more than a century before Mayer and Herschel, and they have never been deeply studied. Before Hodierna, others contributed to the knowledge of multiple stars, such as Giovanni Riciachini, James Bradley, Jean Cassini, and Gottfried Kirch. Aditionally, the discoveries of Castelli and Galileo in 1616--1617 and the publication of Huygens' book in 1659 also preceded Mayer and Herschel.

The only study that mentions the relationship between Hodierna and multiple stars is the Ph.D. thesis by \citet{longhitano11}. Other studies discuss various aspects of his life and different areas of research, leaving apart his studies about binary stars.

This gap in knowledge prompted our investigation into the list of binary stars published by the Sicilian astronomer to determine whether Hodierna's role in the history of astronomy should be reconsidered.

\subsection{M-L subdwarfs multiplicity}
\label{sec:m-l_subdwarfs_mult}

Subdwarfs are objects that lie significantly below the main sequence in the Hertzsprung-Russell diagram and were first discovered by \citet{kuiper39}. They are classified as luminosity class VI under the Yerkes spectral classification system \citep{morgan43} and appear less luminous than solar-metallicity dwarfs of similar spectral types due to their low abundances of elements heavier than helium. Subdwarfs belong to Population II and originate from the galactic thick disk or halo \citep{gizis99} and there is an interest in investigating the impact of metallicity on the binary fraction of low-mass, metal-poor systems and improving metallicity determination using the more massive binary companion.

We can find studies on the multiplicity of metal-poor populations, including works by \citet{chaname04}, \citet{zapatero04a}, \citet{riaz08a}, \citet{jao09a}, \citet{badenes18a}, \citet{moe19}, and \citet{elbadry19a}. Some facts that compose our knowledge up to now are:
\begin{itemize}
    \item Early M subdwarfs seem to have a significantly lower binary fraction \citep[3.3 $\pm$ 3.3\%,][]{riaz08a,lodieu09c}, compared to solar-metallicity M dwarfs \citep[23.5--42\%,][]{ward15,cortescontreras17b} at similar separation ranges.
    \item Hydrodynamic simulations predict a multiplicity fraction of 15--25\% for M dwarfs and 12\% for L dwarfs with subsolar metallicity \citep[0.1\,Z$_\odot$,][]{bate14b,bate19a}. The impact of metallicity appears limited, with fractions of 15--40\% for metal-poor M dwarfs and $\sim$10\% for metal-poor L dwarfs at $Z =$ 0.01\,Z$_\odot$. However, these theoretical studies consider separations below 10\,000\,au, mostly under 1000\,au.
    \item A previous study by \citep{lodieu17a} identified 100 subdwarfs using VO tools, later extended to 193 sources in the SVO late-type subdwarf archive.
\end{itemize}
The research motivation in M-L subdwarfs arises from the need to identify more massive companions to a sample of spectroscopically confirmed metal-poor M and L dwarfs and its multiplicity rate of M and L subdwarfs.
It is also interesting the investigation of the impact of metallicity on the binary fraction of low-mass, metal-poor systems. The obtained results through \textit{Gaia} can be compared with multiplicity studies focused on metal-poor populations. 

\subsection{Ultrawide binaries}
\label{sec:ultrawide_binaries}

A binary system consists of two stars orbiting a common center of gravity \citep{batten73}, while multiple systems involve three or more stars with hierarchical structures \citep{tokovinin97,tokovinin08,eggleton08,duchene13}. Wide multiple systems have large separations and low gravitational binding energies. Traditionally, the maximum separation rarely exceeds 0.1\,pc due to stellar formation and dynamical evolution \citep{tolbert64,kraicheva85,abt88,caballero09}, but recent studies extend this limit to 1\,pc \citep{jiang10,caballero10} or even 1--8 pc \citep{shaya11,kirkpatrick16,gonzalezpayo21}. At such distances, pairs are less likely to remain bound over time \citep{retterer82,weinberg87,dhital10}.

The available knowledge about binaries with separations greater than 1000\,arcsec before this research is:
\begin{itemize}
    \item At the time of this analysis, the WDS catalogue contained 504 pairs with separations greater than 1000\,arcsec. In comparison, a previous study \citep{caballero09} investigated about 105\,000 WDS pairs, of which only 35 had separations greater than 1000\,arcsec. This study is mentioned as the first in this series, focusing on WDS with the widest angular separations.
    \item The advent of the third \textit{Gaia} data release, providing precise parallaxes for many more stars, represented a significant advancement for the study of wide binary systems. This allowed for a more accurate validation of whether wide pairs were real physical systems or optical doubles.
    \item Previous searches for parallax and proper motion in the past decade \citep[e.g.][]{caballero10,shaya11,tokovinin12,kirkpatrick16} had already contributed to the knowledge of wide systems.
\end{itemize}
What motivates this research is to take advantage of the large amount and high precision of \textit{Gaia} DR3 data to conduct a comprehensive study of the widest binary systems catalogued in the WDS, specifically those with angular separations $\rho \ge$ 1000\,arcsec. With \textit{Gaia} DR3, the conditions are created to increase the sample size and to obtain more precise results compared to previous studies. If we are able to characterise these WDS wide pairs, and demonstrate their physical bound, we will expand the knowledge of the ultrawide binaries.

\subsection{Multiplicity of stars with planets}
\label{sec:multiplicity_stars_planets}

The study of stellar multiplicity has a long history, but the discovery of exoplanets is more recent. The first discoveries are just about 30 years ago \citep{campbell88,wolszczan92,mayor95}, and from that moment nearly 6000 candidates have been reported. Since a significant fraction of stars in the Galaxy are part of multiple star systems, many of the exoplanets discovered also belong to multiple star systems. However, there is an observational bias where many exoplanet searches tend to discard close binaries from their input catalogues.

What we know about this type of systems at less than 100\,pc of the Sun is:
\begin{itemize}
    \item In our immediate neighborhood (at 10\,pc radius centered on the Sun), \citet{reyle21} measured an overall multiplicity fraction of 27.4 $\pm$ 2.3\%. But this MF depends on the spectral type.
    \item Thousands of planetary systems were expected to be found among multiple star systems, given that most exoplanets orbit FGKM-type stars.
    \item Currently, there are a few hundred known multiple star systems with exoplanets. Some works include \citet{eggenberger04}, \citet{konacki05}, \citet{furlan17}, \citet{martin18}, and \citet{bonavita20}.
    \item Numerous observational studies had been conducted on the multiplicity of stellar systems with planets with different methodologies and maximum distances investigated, remarking the studies from \citet{fontanive21}, \citet{hirsch21}, and \citet{mugrauer19}, among many others.
    \item These previous studies had proposed various, sometimes inconsistent, results about the relationship between stellar multiplicity and planetary properties, such as orbital eccentricity and the presence of massive planets. For example, some studies suggested that massive planets with short periods in binaries tend to have low eccentricities \citep{eggenberger04,udry04}, while others found the opposite \citep{moutou17}.
    \item The challenge of searching for planets in multiple star systems compared to single stars was well known \citep{eggenberger04}.
\end{itemize}

The main motivation for this work is to revisit the topic of stellar multiplicity with planets within 100\,pc, using the latest astrometric data from \textit{Gaia} DR3 and advanced statistical methodologies. There are aspects that are still pending to know in depth, as the impact of stellar multiplicity on the presence and properties of exoplanets. With \textit{Gaia} DR3 we will be able to find new companions in star systems with known planets, and characterise them. The statistical tools can help to establish comparisons between planets in multiple systems and single stars. They can help us to determine tendencies for high-mass planets to be in closer orbits in multiple systems, or the orbital eccentricity of planets in multiple systems. The obtention of multiplicity fraction of stars with planets in the solar neighborhood can be compared to the multiplicity fraction of field stars in general.

In summary, the availability of the new and precise \textit{Gaia} DR3 data drives this research to provide a more comprehensive and statistically robust picture of the impact of stellar multiplicity on nearby planetary systems, building on and expanding the results of numerous previous studies.

\subsection{Multiplicity of stars within 10\,pc}
\label{sec:multiplicity_10pc}

As shown in former sections, multiplicity is a subject of constant research in astronomy. The opportunity of having a detailed map of the solar neighbourhood is now within reach using the data from \textit{Gaia} DR3, and the knowledge of multiple systems in that area can be completed soon.

The current knowledge about these multiples systems within 10\,pc is:
\begin{itemize}
    \item Previous studies used common proper motion and parallax criteria to identify stellar companions \citep[e.g.][]{gonzalezpayo23,gonzalezpayo24} using \textit{Gaia} DR3 data.

    \item There are studies that characterise the stars in the solar neighbourhood in high detail, such as the work from \citet{reyle21} at $d <$\,10\,pc, and the work from \citet{kirkpatrick21} at $d <$\,20\,pc.

    \item The challenges of identifying companions in very wide systems due to proper motion projection effects and in very close systems due to anomalies in proper motion and parallax are acknowledged \citep{wertheimer06,kervella19,brandt21}.

    \item There are still around 12 unresolved multiple pairs within 10\,pc, showing that some nearby multiple systems were already known.
\end{itemize}

The main motivation of this research is the analysis of the multiplicity of all stars and brown dwarfs up to a heliocentric distance of 10\,pc using \textit{Gaia} DR3 data and the WDS catalogue. The characterisation of all the multiple systems for different star types, especially M dwarfs, given their abundance in the sample, will fill the gaps in previous studies.

\section{Structure and objectives of this work}
\label{sec:thesis_structure}

The main astrophysical objectives of the whole doctoral thesis are improving of the general understanding of the properties of multiple star systems, and the investigation of the connection between stellar multiplicity and the presence and characteristics of planetary systems, stellar metallicity, and galactic dynamics.

The partial objectives established for each of the chapters are formulated with the intention of deepening the understanding of multiple stellar systems and their relationship with various astrophysical phenomena, taking advantage of advancements in astronomical data, especially those provided by the \textit{Gaia} mission. The following chapters detail the specific objectives for each of the studies, as well as the common objectives sought through their collective contribution.

Chapter~\ref{ch:hodierna} retrieves and analyses the list of double stars published by Giovanni Battista Hodierna in his 1654 work, that should be considered the first catalogue dedicated to binary stellar systems.
\begin{itemize}
    \item The main goal is to rescue from oblivion a significant historical contribution.
    \item Precise identify the twelve stellar pairs mentioned by Hodierna, using his descriptions of constellation positions, approximate ecliptic coordinates, estimated angular separations, and the aid of modern astronomical catalogues (WDS, \textit{Hipparcos}, \textit{Gaia)}.
    \item Assess the physical nature (bound binary or optical alignment) of the identified stellar systems using currently available astronomical data, including proper motions, parallaxes, and distances.
    \item Evaluate the accuracy of Hodierna's observations with his 17th-century instruments by comparing his estimates with modern measurements.
    \item Review and contextualise the history of binary star discoveries, recognising Hodierna’s early contribution to this field of astronomy.
\end{itemize}

Chapter~\ref{ch:multiplicity_of_ultra-cool_subdwarfs} conducts a systematic search for common proper motion stellar companions with wide separations around a well-defined sample of metal-poor M and L dwarfs. 
\begin{itemize}
    \item The main objective is to increase the number of known widely separated subdwarf binary systems.
    \item Investigate the potential correlation between subdwarf metallicity and the frequency of wide common proper motion companions, aiming to determine whether low metallicity influences the formation or survival of wide low-mass binary systems.
    \item Use the presence of more massive binary companions with better-determined metallicities to refine the metallicity estimates of M and L subdwarfs.
    \item Estimate the frequency of wide binary systems for different spectral subtypes of M and L subdwarfs.
    \item Perform a photometric and, if possible, preliminary spectroscopic characterisation of the newly identified wide companion systems to determine their spectral types and other relevant properties.
\end{itemize}

Chapter~\ref{ch:widest_binaries} analyses in detail the widest stellar pairs contained in the WDS catalogue with angular separations equal to or greater than 1000\,arcsec, using the most recent astrometric data from \textit{Gaia} DR3. 
\begin{itemize}
    \item The principal objective is to review and validate the physical nature of these ultra-wide systems.
    \item Confirm the physical association of these pairs within stellar systems by evaluating the consistency of their proper motions, parallaxes, and, where available, radial velocities.
    \item Search for and characterise additional companions at shorter separations within these ultra-wide systems to determine their full multiplicity order and hierarchy.
    \item Estimate the component masses and, based on them and their projected separations, calculate the gravitational potential energies of the systems.
    \item Explore the diffuse boundary between very wide binary systems and unbound members of young stellar kinematic groups, seeking criteria to distinguish them based on their dynamical and kinematic properties.
    \item Investigate the prevalence of high-multiplicity systems in the ultra-wide separation regime.
\end{itemize}

Chapter~\ref{ch:systems_with_planets} quantifies the influence of stellar multiplicity on the presence and properties of exoplanets discovered around stars located within 100\,pc of the Sun. 
\begin{itemize}
    \item The main goal is to obtain a statistical perspective on how companion stars affect planetary systems.
    \item Conduct a thorough search for new stellar companions in known exoplanet-hosting systems using \textit{Gaia} DR3 data and complementing it with other catalogue sources and the available literature.
    \item Statistically analyse the distributions of planetary orbital parameters, such as eccentricity, semi-major axis, and mass, in single and multiple planetary systems, looking for significant differences.
    \item Investigate the relationship between the projected physical separation between stars in multiple systems and the eccentricity of planetary orbits, searching for evidence of dynamic perturbations induced by companion stars.
    \item Determine the multiplicity fraction of stars hosting planets within 100\,pc and compare it with the overall field star multiplicity fraction.
\end{itemize}

Chapter~\ref{ch:characterisation_10pc} compiles and comprehensively characterises all multiple stellar systems located within a 10\,pc radius of the Sun.
\begin{itemize}
    \item The principal aim is to create a reference catalogue for the immediate solar neighbourhood in terms of stellar multiplicity.
    \item Accurately evaluate the fundamental physical properties of these systems, including their angular and physical separations, estimated component masses, and known or estimated orbital periods.
    \item Analyse the distribution of stellar multiplicity within this limited volume to obtain reliable statistics on the prevalence of binary, triple, and higher-order systems.
    \item Determine the multiplicity fraction (MF) and companion star fraction (CSF) for the local stellar population, with special attention to M dwarfs, the most abundant type of star in our Galaxy.
    \item Investigate the orbital period distribution of multiple systems and compare it with theoretical and empirical distribution functions, such as Öpik's law, to infer possible formation and evolutionary mechanisms.
    \item Identify multiple stellar systems that could be priority targets for future investigations of habitable exoplanets, considering the influence of companion stars on planetary system formation and stability.
\end{itemize}

From these works, the common objectives of the doctoral thesis are:
\begin{itemize}
    \item Improve the overall understanding of the prevalence and properties of multiple stellar systems in the solar neighbourhood.
    \item Demonstrate the usefulness and potential of high-precision astrometric data provided by the \textit{Gaia} mission for the detailed study of stellar multiplicity.
    \item Investigate the connection between stellar multiplicity and the presence and characteristics of planetary systems.
    \item Contribute to the creation of more complete and accurate catalogues of multiple stellar systems, including the identification of new companions and the reassessment of previously known systems.
    \item Apply rigorous analysis methodologies based on modern data to address fundamental questions about the formation, evolution, and dynamics of multiple stellar systems.
    \item Establish an observational reference framework for future theoretical and observational research in the field of stellar multiplicity and planetary systems.
    \item Conduct a comprehensive literature review to contextualise the findings and highlight the original contributions of this research.
    \item Promote collaboration and the use of Virtual Observatory tools for the analysis of large astronomical datasets.
    \item Contribute to the history of astronomy by recognising and valuing the contributions of past astronomers.
    \item Provide valuable information for future space and ground-based missions dedicated to the search for and characterisation of exoplanets, especially in multiple stellar systems.
\end{itemize}

\newpage

%%%%% CAPITULO 2 %%%%%

\chapter{The very first list of binaries in history} 
\label{ch:hodierna}
\vspace{2cm}
\pagestyle{fancy}
\fancyhf{}
\lhead[\small{\textbf{\thepage}}]{\textbf{Section \nouppercase{\rightmark}}}
\rhead[\small{\textbf{Chapter~\nouppercase{\leftmark}}}]{\small{\textbf{\thepage}}}

\begin{flushright}
\small{\textit{``...And many other twin stars, whose catalogue}}

\vspace{-2mm}
\small{\textit{would grow to infinity if individually noted''}}

\small{\textit{De systemate orbis cometici}}

\vspace{-2mm}
\small{\textit{-- Giovanni Battista Hodierna}} 
\end{flushright}
\bigskip

\begin{adjustwidth}{70pt}{70pt}
\tR{\small{The content of this chapter has been adapted from the article \citet{gonzalezpayo24b}: \textit{Discovery of stellar binaries by Giovanni Battista Hodierna in 1654}, published in Monthly Notices of the Royal Astronomical Society, \href{https://doi.org/10.1093/mnras/stae2010}{MNRAS 533, 3379 (2024)}.}}
\end{adjustwidth}
\bigskip

\lettrine[lines=3, lraise=0, nindent=0.1em, slope=0em]{F}{or centuries, but specially} during the last decades, stellar multiplicity has helped astronomers to learn about the formation, evolution, and parameters of stars \citep{duquennoy91,eggleton06,raghavan10,chabrier03,duchene13,tokovinin14b}.
Although we now use powerful telescopes to investigate them, double stars have been observed since ancient times.
Ptolemy (c. 100--c. 170 AD) was the first astronomer to assign the term ``diplous'' ($\delta\iota\pi\lambda \textit{o}\hat{\upsilon}\varsigma$) to a double star, specifically $\nu^{1}$~Sgr and $\nu^{2}$~Sgr. 
These stars, separated by about 14\,arcmin, were observed with the naked eye, recorded in Ptolemy's Almagest, \textit{``H\=e Megal\=e (Math\=ematik\=e) Syntaxis''}\footnote{``The Great (Mathematical) Treatise''.}, and are now identified as an optical double (i.e., non physically bound). 
In old Arab astronomy, numerous star names collectively denoted two or more adjacent stars easily distinguishable to the naked eye -- actually, $\nu^{1}$~Sgr and $\nu^{2}$~Sgr were \textit{Ain al Rami}, ``the eye of the Archer''.
Conversely, the widely recognised naked-eye duo \textit{Mizar} ($\zeta$~UMa) and \textit{Alcor} (g~UMa) within the Big Dipper garnered distinct names, likely due to their differing brightness levels, which did not portray them as a discernible ``pair'' of stars \citep{heintz78}.
\textit{Mizar} and \textit{Alcor} are not gravitationally bound either (but see \citealt{mamajek10}), although both of them are coeval members or the Ursa Major moving group \citep{riedel17,gagne18a}. 

It was not until the seventeenth century, with the invention of the telescope, that interest in double stars was rekindled. 
This led to the discovery of a new visual binary, \textit{Mizar} A and B, attributed to the Italian mathematician Benedetto Castelli. 
Castelli had requested Galileo to observe it in 1616, although the discovery was later inaccurately credited to Giovanni Battista Riccioli in 1650 \citep{ondra04};
\textit{Mizar} A and B form indeed a physical pair \citep{vogel1901}.
Galileo also resolved, on 4 February 1617, the three brightest stars of $\theta^{1}$~Ori, the central star of Orion's sword \citep{fedele49}. 
Besides, in his ``\textit{Almagestum Novum}'', \citet{riccioli1651} proposed another two pairs of stars separated by a few arcminutes in Capricorn and the Hyades (Sect.~\ref{sec:almagestum_novum}). 

Previously to our work, the next registered reference to multiple stars in the history was from the Dutch scientist Christiaan Huygens, who published in 1659 the book \textit{``Systema Saturnium''} \citep{huygens1659}.
There, Huygens presented an engraving announcing that $\theta^{1}$~Ori was actually a group of three very close stars surrounded by an extended nebulous region. 
Galileo had also noted the trio 42 years earlier, although he had not recognised the Orion Nebula due to unknown reasons.
The trio of stars was lately extended to a quartet in 1673 by Jean Picard \citep{bond1848,wesley1900}.

Huygens opened a new way in the search of new double stars to other astronomers such as Robert Hooke, who detected \textit{Mesarthim} ($\gamma$\,Ari) in 1664 \citep{argyle19}, or the two French Jesuit Fathers Jean de Fontaney and Jean Richaud, who discovered the duplicity of the stars \textit{Acrux} ($\alpha$ Cru) in the Southern Cross, in 1685 from Cape Town (South Africa), and $\alpha$\,Cen, in 1689 from Pondicherry (India), respectively \citep{henroteau28,kameswararao84,kochhar91}. 
Other astronomers contributed to the knowledge of multiple stars, such as 
Giovanni Biachini, 
James Bradley, 
Jean Cassini, 
Gottfried Kirch, 
Nevil Maskelyne, 
Charles Messier, 
John Michell, 
and Nathaniel Pigott, 
until the arrival of the first catalogue of double stars.
Published in 1779 by Christian Mayer, court astronomer at Mannheim, his book \textit{``De novis in coelo sidereo phaenomenis in miris stellarum fixarum comitibus''}\footnote{``On new phenomena in the starry sky among the amazing companions of the fixed stars.''} contained in its last pages the Tabula Nova Stellarum Duplicium, that is, the new table of double stars \citep{schlimmer07}. 
The first version of the catalogue by \citet{mayer1779} contained 72 double stars, which increased to 80 with the new version by \citet{bode1781} two years later.
Mayer's Tabula Nova Stellarum Duplicium is considered to be the first catalogue of double stars.

The objects within a double system were not initially considered as connected as the prevailing belief was that it was merely a coincidental occurrence. 
According to \citet{schlimmer07}, William Herschel heard about Mayer’s work and started his own observations of double stars.
Herschel published three catalogues of possible double stars: ``The Catalogue of Double Stars'' \citep{herschel1782} with 269 entries, ``Catalogue of Double Stars'' \citep{herschel1785} with 484 entries, and ``On the Places of 145 New Double Stars'' \citep{herschel1822}. 
When Herschel reexamined his first catalogue in 1802, after claims of common proper motion and orbital motion by \citet{mayer1786}, he confirmed that several double stars were gravitationally bound, which sparked modern astronomy \citep{williams14}.

In his PhD thesis, \citet{longhitano11} noticed that, between the discoveries of Castelli and Galileo in 1616--1617 and the publication of the book by Huygens in 1659, and over a century before the double star catalogues of Mayer and Herschel, a poorly known Sicilian astronomer, Giovanni Battista Hodierna, published a short list of possible binary systems that did not leave much of a mark on history and have gone since unnoticed. 
We focus on the seminal work of Hodierna by uncovering documentation of some pairs of stars deemed binary, a revelation potentially predating many celebrated observations. 
These celestial twins, concealed within Hodierna's principal opus, offer a compelling narrative of early astronomical inquiry. 
Our investigation aims to shed light on these overlooked observations, highlighting their significance in the annals of scientific discovery, and underscoring their potential to reshape our understanding of early modern astronomy.

\section{A historical review}
\label{sec:historical_review}

\subsection{Giovanni Battista Hodierna}
\label{sec:GB_Hodierna}

Giovanni Battista Hodierna was born on 13 April 1597 in Ragusa, Italy. 
Little is known about his early life and education, but that he studied at the University of Palermo, where he likely developed his interest in astronomy and mathematics. 
He became a catholic priest and served as such in the then Kingdom of Sicily. 
During his religious life, Hodierna dedicated much of his time to observing the night sky. His main focus was on cataloguing stars and creating celestial maps. 

Hodierna made his observations using rudimentary instruments, such as refracting telescopes of his own design. 
Despite the limitations of his tools, he managed to identify and catalogue numerous stars and celestial objects, some of which had not been documented before. 
Hodierna is particularly known for his star catalogue, entitled ``\textit{De systemate orbis cometici deque admirandis coeli caracteribus}\footnote{``On the system of the cometary universe and on the admirable characteristics of the sky.''}'' \citep{hodierna1654}. 
The work anticipated Messier's work, but had little impact, and neither Messier nor any European astronomer seem to have known of it \citep{serio85}. 
Therefore, Hodierna never reached the same fame as other contemporary astronomers, such as Galileo Galilei or Johannes Kepler. 
After his death on 6 April 1660 in Palma di Montechiaro, his work was largely forgotten and rediscovered only in later centuries when modern astronomers recognised the importance of his observations \citep{serio85,jones86}.

\subsection{\textit{De systemate orbis cometici}}
\label{sec:systemate}

\begin{figure}[H]
  \centering
  \includegraphics[width=0.5\linewidth, angle=0]{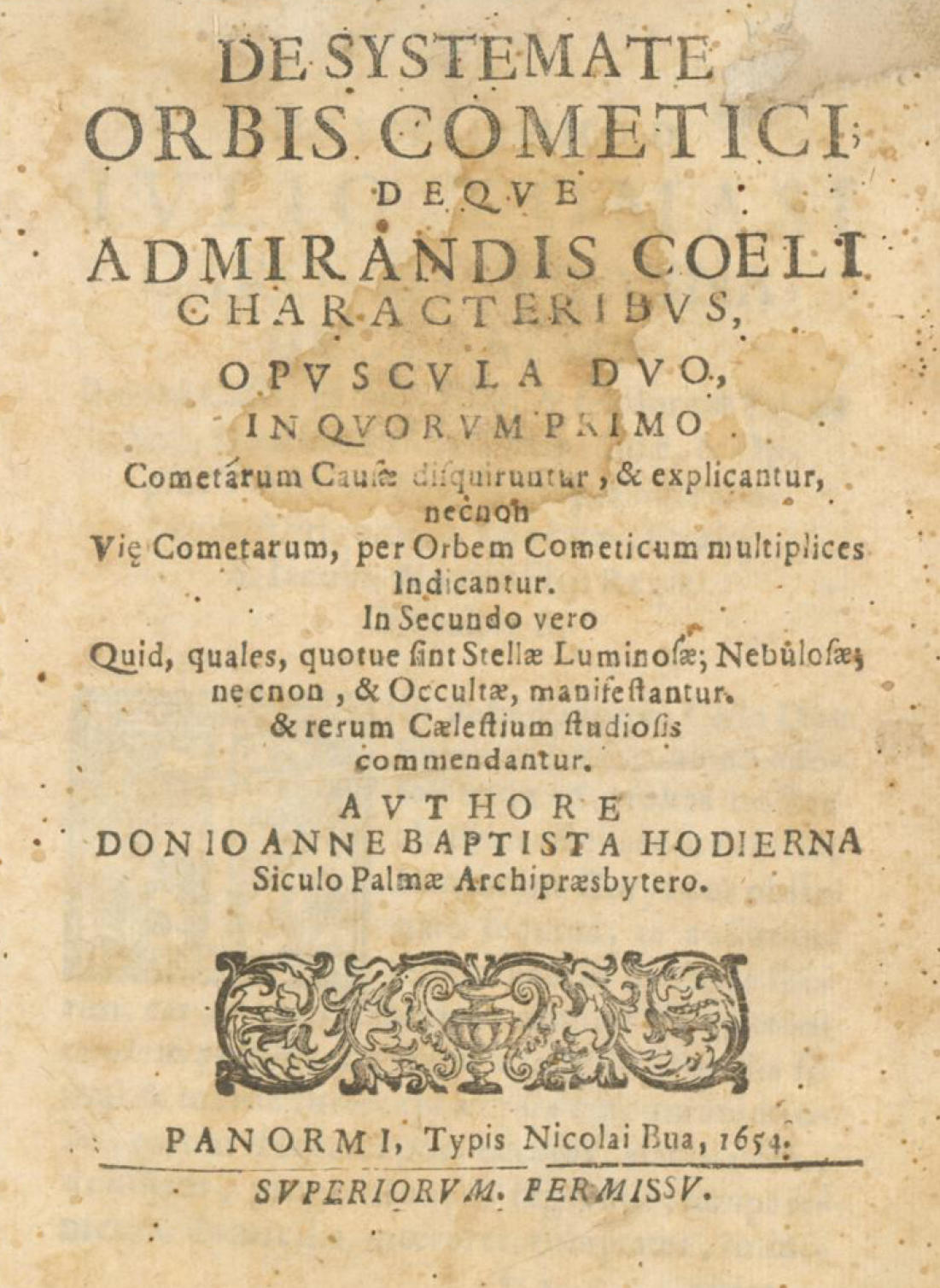}
 \caption[Front page of \textit{De systemate orbis cometici}, edition from Nicolai Bua (Palermo) of 1654.]{Front page of \textit{De systemate orbis cometici}, edition from Nicolai Bua (Palermo) of 1654.}
  \label{fig:desystemati}
\end{figure}

For our investigation, we used the ETH-Bibliothek Z\"urich copy of ``\textit{De systemate orbis cometici}'' (Rar 2876, see Fig.~\ref{fig:desystemati}).
The book is structured in two parts divided in their turn into four sections. 
The first part refers to the theory of comets. Following the path outlined by Galileo, Hodierna distinguished the nature of comets from that of nebulae, attributing an earthly nature to the former and recognising only a celestial, stellar nature to the latter \citep{pavone86}. 
In the second part, ``\textit{De admirandis coeli caracteribvs}''\footnote{``On the admirable characteristics of the sky''.}, he classified and catalogued deep sky objects, which he denominated nebulae \citep{jones91}.
There, he tabulated and provided finding charts for a number of prominent objects, including the $\alpha$~Persei open cluster (M20), the Butterfly Cluster (M6), or the Lagoon Nebula (M8), over a century before their independent discovery by Jean-Philippe Loys de Cheseaux, Guillaume Le Gentil, Charles Messier, or Caroline Herschel, just to mention some names \citep{serio85,williams14}.
The fourth section of the second part, entitled ``\textit{In qua de stellis contiguis duplicibus seu Geminis deque Mundani Systematys Coperniceorum implicantia ratiocinandum venit}''\footnote{``Wherein the discourse concerning close double stars, or twins, and the implications for the Copernican system of the Universe, is to be considered.''} includes the list of binary stars that we investigated here, and for which he used the word ``geminae'' (twins).

\subsection{``\textit{Stellae geminae}''}
\label{sec:description_binaries_text}

At the beginning of the section ``\textit{In qua de stellis contiguis duplicibus}'', Hodierna stated some sentences that were very advanced for their epoch: 
``everywhere throughout the ether, certain binary stars shine, which, although double in themselves, are so closely bound by a bond of contiguity that hardly two are distinguished from each other, by the bond of contiguity with which they appear to be connected, and thus are not considered double, but wholly single [...] but many, or countless, such double stars are found throughout the vastness of the sphere, as scarcely a constellation in the sky shines in which there is not at least one or two binaries, especially in the luminous nebulous regions and in the dark stretches of the sky, among which there are some very prominent ones that adhere to the ecliptic in the Zodiac''. 

\begin{figure}[h]
  \centering
  \includegraphics[width=0.9\linewidth, angle=0]{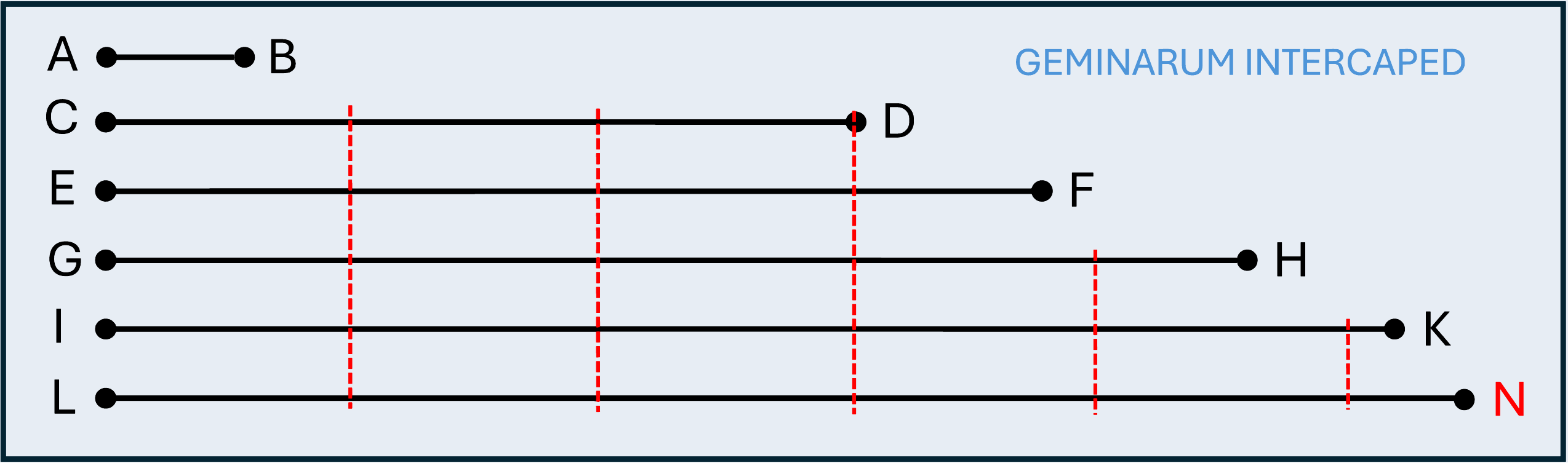}
 \caption[Adaptation of Hodierna's separations drawings.]{Adaptation of Hodierna's separations drawings. 
 Each segment corresponds to an angular separation.
 The vertical lines are 1\,arcmin ticks. 
 The N denotes a typographical error (the segment LN is referred in Hodierna's book as LM).}
  \label{fig:geminarum}
\end{figure}

Following, the author included a table containing five binary stars that he called ``\textit{Stellae geminae iuxta eclypticam}''\footnote{``Binary stars near the ecliptic''.}.
We show an adaptation of Hodierna's table in the Table~\ref{tab:hodierna_table_with_coordinates}.
For each of the five stars, he provided a Latin name, which is often a very brief description of its location within its constellation (for example, \textit{``Oculus Boreus Tauri''} is ``the northern eye of the bull''), and the ecliptic coordinates in degrees and minutes of latitude and longitude (see below).
The longitude includes the ``signum'', which is the zodiacal area of the sky (not necessarily coinciding with the current constellation definition), and the latitude includes an additional character for the ecliptic hemisphere, namely ``B'' for boreal (north) and ``A'' for austral (south).
We gave the identification numbers \#1 to \#5 to the five stars.

\begin{table}
 \centering
 \caption[Hodierna's star pairs with coordinates.]{Hodierna's star pairs with coordinates.}
 \footnotesize
 \scalebox{1}[1]{
 \begin{tabular}{@{\hspace{1mm}}c@{\hspace{1mm}}l@{\hspace{1mm}}l@{\hspace{1mm}}l@{\hspace{1mm}}c@{\hspace{1mm}}c@{\hspace{1mm}}c@{\hspace{1mm}}c@{\hspace{1mm}}c@{\hspace{1mm}}l@{\hspace{1mm}}c@{\hspace{1mm}}c@{\hspace{1mm}}}
 \noalign{\hrule height 1pt}
 \noalign{\smallskip}
 \noalign{\smallskip}
Id. & Star name  & Translation &  \multicolumn{3}{c}{Ecliptic longitude} & \multicolumn{3}{c}{Ecliptic latitude} & $\rho_{\rm text}$ & Segment & $\rho_{\rm drawing}$ \\
  &  & to English & Signum  & (deg) & (min) & (deg) & (min) & Hemis. & (arcmin) &  & (arcmin) \\
 \noalign{\smallskip}
 \hline
 \noalign{\smallskip}
\multirow{2}{*}{1} & \multirow{2}{*}{Orientalissima Pleiadum} & The easternmost & \multirow{2}{*}{Tauri} & \multirow{2}{*}{25} & \multirow{2}{*}{32} & \multirow{2}{*}{3} & \multirow{2}{*}{52} & \multirow{2}{*}{B} & \multirow{2}{*}{Three} & \multirow{2}{*}{EF} & \multirow{2}{*}{3.8} \\
 & &  of the Pleiades  & & & & & & & & & \\
 \noalign{\smallskip}
\multirow{2}{*}{2} & \multirow{2}{*}{Oculus Boreus Tauri} & The northern eye & \multirow{2}{*}{Gemin.} & \multirow{2}{*}{3} & \multirow{2}{*}{39} & \multirow{2}{*}{2} & \multirow{2}{*}{36} & \multirow{2}{*}{A} & \multirow{2}{*}{Almost five} & \multirow{2}{*}{GH} & \multirow{2}{*}{4.6} \\
 & & of the bull & & & & & & & & & \\
 \noalign{\smallskip}
\multirow{2}{*}{3} & \multirow{2}{*}{Lanx Austrina Libr\ae} & Southern pan & \multirow{2}{*}{Scorpionis} & \multirow{2}{*}{10} & \multirow{2}{*}{14} & \multirow{2}{*}{0} & \multirow{2}{*}{26} & \multirow{2}{*}{B} & \multirow{2}{*}{Three} & \multirow{2}{*}{CD} & \multirow{2}{*}{3.0} \\
 & & of the balance  & & & & & & & & & \\
 \noalign{\smallskip}
\multirow{3}{*}{4} & Cornu Occidentale Capric. & \multirow{2}{*}{The western horn} & \multirow{3}{*}{Capric.} & \multirow{3}{*}{29} & \multirow{3}{*}{4} & \multirow{3}{*}{7} & \multirow{3}{*}{2} & \multirow{3}{*}{B} & \multirow{2}{*}{Almost five} & \multirow{3}{*}{IK} & \multirow{3}{*}{5.2} \\
 & (In praecedenti Cornu & \multirow{2}{*}{of the capricorn} & & & & & & & \multirow{2}{*}{and a half} & & \\
  & (Capricorni) & & & & & & & & & & \\
 \noalign{\smallskip}
\multirow{2}{*}{5} & \multirow{2}{*}{Trium in frontem Occid.} & Three towards & \multirow{2}{*}{Scorpionis} & \multirow{2}{*}{28} & \multirow{2}{*}{20} & \multirow{2}{*}{1} & \multirow{2}{*}{40} & \multirow{2}{*}{B} & \multirow{2}{*}{...} & \multirow{2}{*}{...} & \multirow{2}{*}{...} \\
 & & the western front & & & & & & & & & \\
 \noalign{\smallskip}
 \noalign{\hrule height 1pt}
 \end{tabular}
} 
\label{tab:hodierna_table_with_coordinates}
\end{table}

After the table, Hodierna enumerated in the text seven binary stars for which he did not initially provide any quantitative datum, but only their Latin names.
We gave the Ids. \#6--\#12 to the seven stars.
Next, Hodierna displayed a drawing labelled ``\textit{Geminarum intercaped}'' (``Binaries separation'') with a pictographical representation of the angular separation between components of another seven binary systems.
We display in Fig.~\ref{fig:geminarum} our own adaptation of Hodierna's drawing. 
The ticks on the six segments indicate arcminutes.
There are six segments in Fig.~\ref{fig:geminarum} because the shortest one, of less than one arcminute, corresponds to two close binary stars that had not been mentioned before.
The seven binary systems for which Hodierna was able to measure separations with his rudimentary telescope were enumerated below the drawing, together with the approximate value of the angular separation in Latin. 
They have an entry in the last three columns of Table~\ref{tab:hodierna_table_without_coordinates}.
We provide both the translation of the angular separation in the text, $\rho_{\rm text}$, and our interpretation of the angular separation in the drawing, $\rho_{\rm drawing}$, with an uncertainty of 0.2\,arcmin, as they did not coincide exactly.
At least four of the seven stars were already in Hodierna's table, and the names of the other three had been enumerated in the text.
Finally, although Hodierna did not identify any more multiple systems, he emphasised the multitude of binary stars that were left out of his description: ``[...] and many others, whose catalogue would grow to infinity if they were individually noted''.

\begin{table*}[h]
 \centering
 \caption[Hodierna's star pairs without coordinates.]{Hodierna's star pairs without coordinates.}
 \footnotesize
 \scalebox{1}[1]{
 \begin{tabular}{c@{\hspace{2mm}}l@{\hspace{4mm}}l@{\hspace{4mm}}lcc}
 \noalign{\hrule height 1pt}
 \noalign{\smallskip}
Id. & Star name &  Translation to English & $\rho_{\rm text}$ & Segment & $\rho_{\rm drawing}$ \\
  &  &   & (arcmin) &  & (arcmin) \\
 \noalign{\smallskip}
 \hline
 \noalign{\smallskip}
 \multirow{2}{*}{6} & In pede sinistro & On the left foot of &  \multirow{2}{*}{...} &  \multirow{2}{*}{...} &  \multirow{2}{*}{...} \\
  & pr\ae cedentis Geminorum & the preceding twin & & & \\
 \noalign{\smallskip}  
 7 & Ceruice Leonis  & At the tail of the lion & ... & ... & ...  \\
  \noalign{\smallskip}
 8 & In ancone al\ae\, dextr\ae\, Cygni  & On the right wing of the swan & ... & ... & ... \\
  \noalign{\smallskip}  
 9 & In pede posteriori sinistro Leporis & On the left hind leg of the hare & ... & ... & ... \\
  \noalign{\smallskip}
 \multirow{2}{*}{10} & \multirow{2}{*}{Secunda spondili Scorpionis} & In the second vertebra  & \multirow{2}{*}{Five and a half} & \multirow{2}{*}{LM}  & \multirow{2}{*}{5.4} \\ 
   &  & of the scorpion & & & \\
  \noalign{\smallskip}  
 \multirow{3}{*}{11} & Capitis Draconis, quatuor &  On the head of the dragon, of the & \multirow{3}{*}{Less than one} & \multirow{3}{*}{AB} & \multirow{3}{*}{0.6} \\
   & rombum constituentium,  & four that constitute the diamond, & & & \\
   &  qu\ae\, sub Oculo exigua & the one that is slightly below the eye   & & & \\
  \noalign{\smallskip}  
 12 & Medie ensis Orionis & In the middle of Orion's sword & Less than one & AB & 0.6 \\ 
 \noalign{\smallskip}
 \noalign{\hrule height 1pt}
 \end{tabular}
} \label{tab:hodierna_table_without_coordinates}
\end{table*}

\section{Analysis and results}
\label{sec:hodierna_text_analysis}

We proceeded to the identification of the dozen stars mentioned by Hodierna and their companions using the coordinates, properties, and clues provided in his book.
First, for the five stars in Hodierna's table for which he provided ecliptic coordinates (Table~\ref{tab:hodierna_table_with_coordinates}), we computed their current equatorial coordinates.
For that, we first translated Hodierna's coordinates to a modern format using the relations of \citet{verbunt11}:
\begin{equation}
    \lambda=(Z-\text{1})\cdot \text{30} + G_{lon} + \frac{M_{lon}}{\text{60}}
\end{equation}
\noindent and
\begin{equation}
    \beta=\pm \,\, G_{lat} + \frac{M_{lat}}{\text{60}},
\end{equation}
\noindent where $\lambda$ and $\beta$ are the ecliptic coordinates, $G_{lon}$ and $M_{lon}$ the degrees and minutes of longitude, $G_{lat}$ and $M_{lat}$ the degrees and minutes of latitude, $Z$ is the number of order for the zodiacal ``signum'' (starting with 1 for Aries and finishing with 12 for Pisces), and the sign of $\beta$ is positive for northern longitudes (borealis) or negative for southern longitudes (australis).
Second, we transformed ecliptic to equatorial coordinates with the NASA/IPAC Extragalactic Database Coordinate Calculator tool\footnote{\url{https://ned.ipac.caltech.edu/coordinate_calculator}}.
Third, we searched for the closest naked-eye stars that fit to the Latin name description.
Fourth, we measured the angular separation between the computed and current J2000 coordinates.
The mean absolute separation of the five identified stars is 4.838$\pm$0.063\,deg, which is identical to the expected offset by the equinox precession since 1654 of 4.837\,deg. 
For the computation of the equinox precession we used the equation of \citet{capitaine03}:
\begin{equation}
    p_A = \text{5\,028.796195}\cdot T + \text{1.1054348} \cdot T^\text{2} 
\end{equation}
\noindent where $T$ is time in Julian centuries and $p_A$ is the absolute value of equinox precession in arcsec. 
The offset vectors in right ascension and declination were also very similar in the five cases.
Given the uncertainty in Hodierna's input coordinates, of arcminutes, and the relatively low proper motion of the stars, we did not take into account any proper motion correction.
The five identified stars are so bright that they have proper names: \textit{Atlas} (\#1), \textit{Ain} (\#2), \textit{Zubenelgenubi} (\#3), \textit{Algedi} (\#4), and \textit{Acrab} (\#5). 
Furthermore, the location of the five stars correspond to their descriptive Latin names 
(e.g., \textit{Atlas} is ``the easternmost of the Pleaides'').
They are listed in the top part of Table~\ref{tab:identified_pairs}.

\begin{figure}
  \centering
  \includegraphics[width=0.7\linewidth, angle=0]{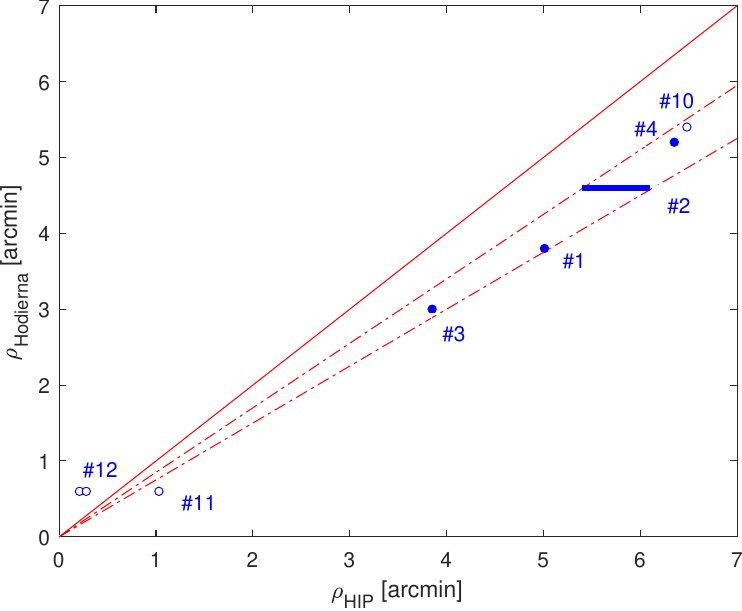}
 \caption[Binary angular separations $\rho$ supplied by Hodierna (when available) vs. $\rho$ calculated by us from \textit{Hipparcos} data.]{Binary angular separations $\rho$ supplied by Hodierna (when available) vs. $\rho$ calculated by us from \textit{Hipparcos} data. 
 Filled and empty dots are for stars with and without coordinates, respectively. The solid line has a slope equal to 1, while the dashed lines have slopes of 0.75 and 0.85. 
 The horizontal solid line represents the possible real values of separation that the companion of \textit{Ain} should have in case that it existed.} 
  \label{fig:rho}
\end{figure}

Next, we used the Aladin sky atlas \citep{bonnarel00} and the Washington Double Star catalogue \citep[WDS;][]{mason01} to search for bright stellar companions to the five stars.
Since Hodierna's telescope was rudimentary, such companions should have a magnitude comparable to those of their primaries (a modern telescope with an aperture of 35\,mm, such as one for children, can detect stars up to  9.9\,mag with a 5\,mm exit pupil; \citealt{north14}).
Four of the five stars had two angular separations measured by Hodierna, namely one from the text and the other from the segment drawing; hereafter, we used only the angular separation from the drawing, dubbed $\rho_{\rm drawing}$ (Fig.~\ref{fig:geminarum} and last column of Table~\ref{tab:hodierna_table_with_coordinates}).
We were able to identify naked-eye companions to three of the four stars.
The three pairs are \textit{Atlas} and \textit{Pleione} (\#1, the easternmost of the Pleiades),
\textit{Zubenelgenubi} and $\alpha^{\text{1}}$ Lib (\#3, the southern weighing pan of Libra), 
and \textit{Algedi} and \textit{Prima Giedi}\footnote{As mentioned in the introduction, there was a potential previous discovery of the \textit{Algedi} and \textit{Prima Giedi} pair by Martinus Hortensius as described in Riccioli's \textit{``Almagestum Novum''}. Prima Giedi should not be confused with Giedi Prime.} (\#4, the western horn of the goat). 
Fig.~\ref{fig:rho} shows a comparison of Hodierna's $\rho_{\rm drawing}$ and our own measurements using \textit{Hipparcos} data \citep{perryman97} computed with spherical astronomy equations as \citet{gonzalezpayo23}.
From the comparison of the three pairs, Hodierna measured smaller angular separations by a $\sim \text{0.80}$ scale factor.
The $V$-band magnitudes of the primaries and secondaries range in the 2.8--3.6\,mag and 4.3--5.2\,mag intervals, well within the naked eye limit with typical dark sky conditions.  

We were not able to identify any suitable companion to the fourth star with ecliptic coordinates and $\rho_{\rm drawing}$, namely \textit{Ain} (\#2, $\epsilon$~Tau).
From Fig.~\ref{fig:rho}, we expected such a suitable companion at an actual angular separation of about 5.4--6.1\,arcmin, from Hodierna's value of 4.6\,arcmin and $\sim$ 0.80 scale factor.
However, there are no stars brighter than 13.5\,mag in the wider 5--7\,arcmin annulus centred on \textit{Ain}. 
Trying to ascertain whether Hodierna's claim of a companion to the bright star in Taurus, not far from the ecliptic ($\beta \approx -$2.6\,deg), was the result of an observational mistake (e.g. an optical ghost) or an actual measurement (e.g. a bright Solar System body passing near the ecliptic) would be just speculating.
Nevertheless, we were confident of our identification, because \textit{Ain} is indeed ``the northern eye of the bull'' in many classical sky descriptions (e.g. ``\textit{Reliqua quae est in oculo boreali}'' in Ptolemy's Almagest, ``\textit{Oculus boreus}'' in \citealt{flamsteed1725}), and the offset between precession-corrected Hodierna's coordinates and ours in J2000 is only slightly less of 1\,arcmin.

The fifth star in Table~\ref{tab:hodierna_table_with_coordinates}, \textit{Acrab} (\#5), does not have a Hodierna's $\rho_{\rm drawing}$ but is the only naked eye star in the trio of the western front (claws) of the Scorpion that has a relatively bright companion at an angular separation comparable to those of the pairs \#1, \#3, and \#4. 
The most probable companion is HD~144273, which has a visual magnitude of 7.54\,mag and is separated to \textit{Acrab} by 8.65\,arcmin.

\begin{table}[H]
 \centering
 \caption[All star pairs described by Hodierna.]{All star pairs described by Hodierna.}
 \footnotesize
 \scalebox{1}[1]{
 \begin{tabular}{@{\hspace{1mm}}c@{\hspace{1mm}}l@{\hspace{1mm}}c@{\hspace{1mm}}c@{\hspace{1mm}}c@{\hspace{1mm}}c@{\hspace{1mm}}c@{\hspace{1mm}}c@{\hspace{1mm}}c@{\hspace{1mm}}c@{\hspace{1mm}}c@{\hspace{1mm}}}
 \noalign{\hrule height 1pt}
 \noalign{\smallskip}
Id & Star name & $\alpha$ (J2000) & $\delta$ (J2000) & $\mu_{\alpha}\cos\delta$ & $\mu_\delta$ & $d$ & $V$ & WDS & Disc. code & $\rho$ \\
  &  &  (hh:mm:ss.ss) & (dd:mm:ss.s) & (mas\,a$^{-\text{1}}$) & (mas\,a$^{-\text{1}}$) & (pc) & (mag) &   &  & (arcmin) \\
 \noalign{\smallskip}
 \hline
 \noalign{\smallskip}
 1 & \textit{Atlas} (27 Tau) & 03:49:09.74 & +24:03:12.3 & 21.761 & --46.241 & 118.6 & 3.63 & \multirow{2}{*}{...} & ... & \\
   & \textit{Pleione} (28 Tau) & 03:49:11.22 & +24:08:12.2 & 19.496 & --47.650 & 138.1 & 5.09 &  & ... & 5.01 \\  
 \noalign{\smallskip}
 \hline
 \noalign{\smallskip}
 2 & \textit{Ain} ($\epsilon$ Tau) & 04:28:37.00 & +19:10:49.6 & 107.53 & --36.200 & 44.71 & 3.53 & \multirow{2}{*}{...} & ... & \\  
   & Not identified & ... & ... & ... & ... & ... & ... &  & ... & ... \\  
 \noalign{\smallskip}
 \hline
 \noalign{\smallskip}
 3 & \textit{Zubenelgenubi} ($\alpha^{\text{2}}$ Lib) & 14:50:52.71 & --16:02:30.4 & --105.68 & --68.400 & 23.24 & 2.75 & \multirow{2}{*}{14509--1603} & SHJ 186 A &  \\
   & $\alpha^{\text{1}}$ Lib & 14:50:41.17 & --15:59:50.0 & --73.505 & --67.871 & 23.94 & 5.16 &   & SHJ 186 B & 3.85 \\  
 \noalign{\smallskip}
 \hline
 \noalign{\smallskip}
 4 & \textit{Algedi} ($\alpha^{\text{2}}$ Cap)  & 20:18:03.26 & --12:32:41.5  & 61.212 & 2.4120 & 33.43 & 3.58 & \multirow{2}{*}{20181--1233} & STFA 51 A &  \\
   & \textit{Prima Giedi} ($\alpha^{\text{1}}$ Cap) & 20:17:38.87 & --12:30:29.6 & 21.709 & 1.6430 & 249.0 & 4.27 &  & STFA 51 E & 6.35 \\  
 \noalign{\smallskip}
 \hline
 \noalign{\smallskip}   
 5 & \textit{Acrab} ($\beta$ Sco) & 16:05:26.23 & --19:48:19.4 &  --5.200 & --24.040  & 123.9 & 2.50 & \multirow{2}{*}{...} & ... &  \\  
  & HD 144273 &  16:05:44.84 & --19:40:51.8  & --10.160 & --21.824  & 152.7 & 7.54 &  & ... & 8.65 \\
 \noalign{\smallskip}
 \hline
 \noalign{\smallskip}  
6 & $\nu$ Gem & 06:28:57.79 & +20:12:43.7 & --4.669 & --14.402& 224.8 & 4.14 & \multirow{2}{*}{06290+2013} & STTA 77 A & \\
   & HD 257937 & 06:28:53.77 & +20:14:21.2 & --3.121 & --7.624& 122.8 & 8.97 &  & STTA 77 B & 1.88 \\  
 \noalign{\smallskip}
 \hline
 \noalign{\smallskip}
7 & \textit{Denebola} ($\beta$ Leo) & 11:49:03.58 & +14:34:19.4 & --497.68 & --114.67 & 11.00 & 2.13 & \multirow{2}{*}{11491+1434} & BU 604 A &  \\
 & BD+15 2382 & 11:48:59.18 & +14:30:27.2 & --45.481 & 31.050 & 117.9 & 8.46 &  & BU 604 D & 4.01 \\  
 \noalign{\smallskip}
 \hline
 \noalign{\smallskip} 
 8 & $\theta$ Cyg & 19:36:26.53 & +50:13:16.0 & --5.854 & 263.66 & 18.43 & 4.48 & \multirow{3}{*}{...} & ... &  \\
  & HD 185264$^{\text{(a)}}$ & 19:35:55.95 & +50:14:19.0 & --2.805 & 36.659 & 161.3 & 6.46 &  & ... & 4.99 \\
  & R Cyg$^{\text{(a)}}$ & 19:36:49.36 & +50:11:59.7 & --4.194 & --6.2730& 597.1 & 6.10 &  & ... & 3.89 \\
 \noalign{\smallskip}
 \hline
 \noalign{\smallskip}
 9 & $\delta$ Lep & 05:51:19.30 & --20:52:44.7 & 229.03 & --647.93 & 35.18 & 3.85 & \multirow{2}{*}{...} & ... &  \\  
   & HD 39405 & 05:51:36.77 & --20:50:13.0 & --4.186 & --21.897 & 315.5 & 8.25 &  & ... & 4.84 \\ 
 \noalign{\smallskip}
 \hline
 \noalign{\smallskip}  
  10 & $\zeta^{\text{2}}$ Sco & 16:54:35.01 & --42:21:40.7 &  --126.72 & --228.84 & 41.26 & 3.62 & \multirow{2}{*}{...} & ... &  \\ 
  & $\zeta^{\text{1}}$ Sco & 16:53:59.73 & --42:21:43.3 & --0.0940 & --3.3680 & 1\,708 & 4.79 &  & ... & 6.48 \\   
 \noalign{\smallskip}
 \hline
 \noalign{\smallskip} 
 11 & $\nu^{\text{2}}$ Dra & 17:32:16.04 & +55:10:22.5 & 144.34 & 59.585 & 29.92 & 4.87 & \multirow{2}{*}{17322+5511} & STFA 35 A & \\
   & $\nu^{\text{1}}$ Dra & 17:32:10.57 & +55:11:03.3 & 148.04 & 54.194 & 30.11 & 4.90 &  & STFA 35 B & 1.03 \\  
 \noalign{\smallskip}
 \hline
 \noalign{\smallskip} 
 12 & $\theta^{\text{1}}$ Ori C & 05:35:16.50 & --05:23:22.9 & 2.262 & 0.994 & 399.8 & 5.13 & \multirow{3}{*}{05353--0523} & STF 748 C & \\  
   & $\theta^{\text{1}}$ Ori A$^{\text{(b)}}$ & 05:35:15.83 & --05:23:14.3 & 1.355 & 0.250 & 378.4 & 6.73 &  & STF 748 A & 0.21 \\  
   & $\theta^{\text{1}}$ Ori D$^{\text{(b)}}$ & 05:35:17.26 & --05:23:16.6 & 1.822 & 0.393 & 438.21 & 6.70 &   & STF 748 D & 0.22 \\ 
 \noalign{\smallskip} 
 \noalign{\hrule height 1pt}
 \end{tabular}
}
 \label{tab:identified_pairs}
\begin{justify}
\scriptsize{\textbf{\textit{Notes. }}}
\scriptsize{$^{\text{(a)}}$ Both stars are possible companions of $\theta$ Cyg.}
\scriptsize{$^{\text{(b)}}$ Both stars are possible companions of $\theta^{\text{1}}$ Ori C. The three stars A, C, and D conform the Orion's Trapezium.}
\end{justify}
\end{table}

Next, we went on identifying the rest of stars without ecliptic coordinates.
First, we searched for bright stars that match the Latin descriptions of pairs \#10, \#11, and \#12 (see the translation to English in Table~\ref{tab:hodierna_table_without_coordinates}) and have relatively bright companions at angular separations similar to those estimated by Hodierna.
The two closest pairs, with $\rho_{\rm drawing}$ of ``less than one [arcminute]'' according to Hodierna, are \textit{Kuma}$^{\text{2}}$ ($\nu^{\text{2}}$ Dra) and \textit{Kuma}$^{\text{1}}$ ($\nu^{\text{1}}$ Dra) (\#11), whose components are actually separated by 1.03\,armin, and $\theta^{\text{1}}$~Ori (\#12).
The latter is the famous asterism of the Trapezium \citep{herbig86,mccaughrean94,simondiaz06}.
Hodierna may be able to see the three brightest components of the Trapezium, namely $\theta^{\text{1}}$ Ori A, C, and D, which had been already resolved by Galileo in 1617.
The separation between the eastern- and westernmost OB-type components of the Trapezium is about 0.2\,arcmin, probably close to the resolution limit of Hodierna's instrumentation and site.
The last pair with $\rho_{\rm drawing}$ is very likely $\zeta^{\text{2}}$~Sco and $\zeta^{\text{1}}$~Sco (\#10), which is made of two stars brighter than $V$ = 5\,mag separated by 6.48\,arcmin.
Hodierna's measurement of their separation was also affected by the same scale factor $\sim \text{0.80}$ as for the pairs \#1, \#3, and \#4 (Fig.~\ref{fig:rho}), which strengthen our identification.

Finally, also guided by the Latin description and existence of relatively bright companions at a few arcminutes to naked-eye stars, we tentatively identified the four remaining pairs (\#6, \#7, \#8, and \#9).
Save in one case, the components are separated by 3.9--6.5\,arcmin.
The exception is the pair $\nu$ Gem and HD~257937 (\#6), which are separated by only 1.9\,arcmin.
Similarly, the companions are probably at the faintest limit of Hodierna's instrumentation, at $V \approx$ 8.2--9.0\,mag, save again for one exception: the $\theta$ Cyg system (\#8).
This system has in common with the Orion's Trapezium that Hodierna may have seen three components, since R~Cyg and HD~185264 are more or less of the same visual magnitude, $V \approx$ 6.1--6.5\,mag, and at the same angular separation to $\theta$~Cyg.

\begin{figure}[H]
  \centering
  \includegraphics[width=0.85\linewidth, angle=0]{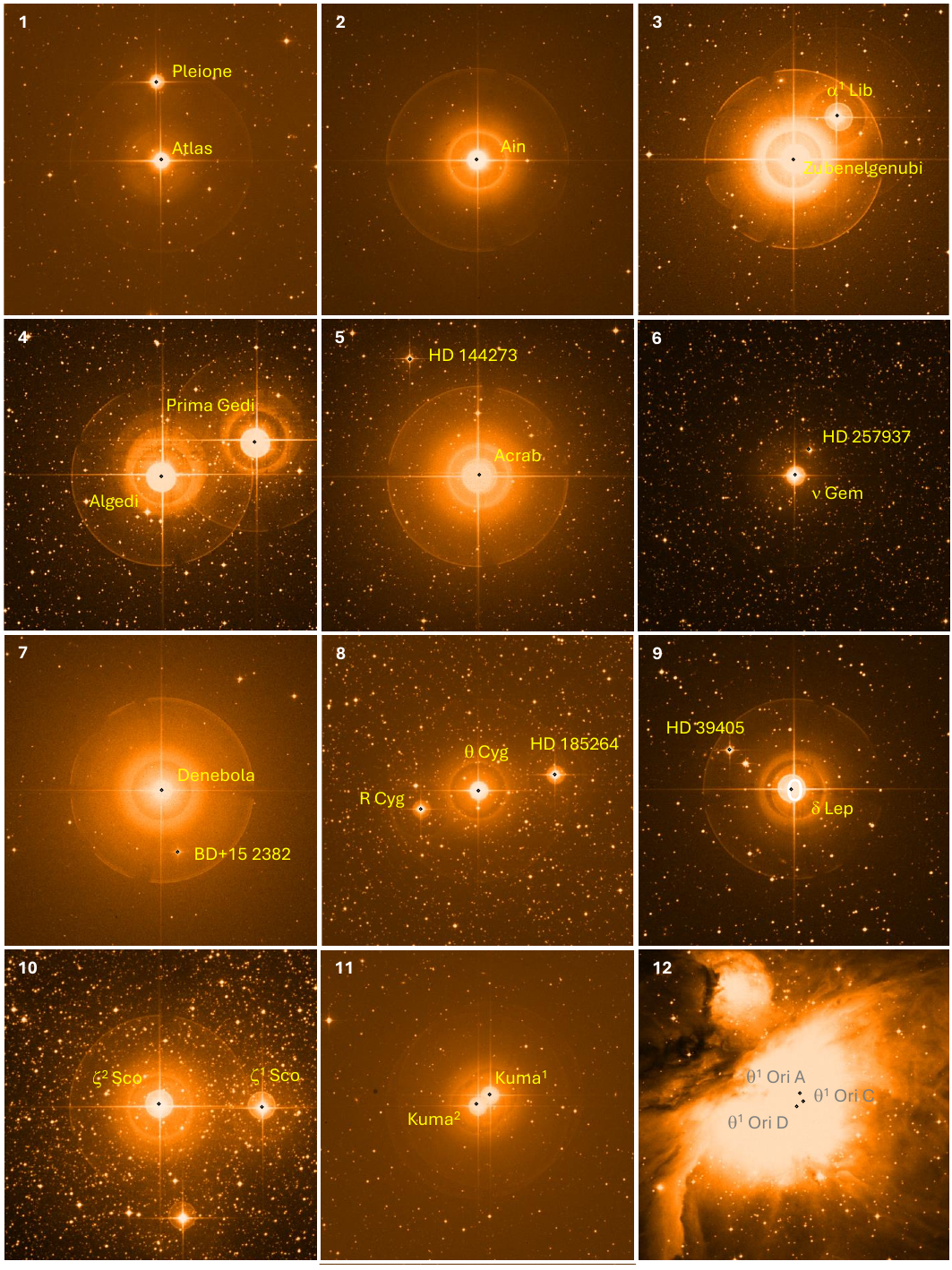}
 \caption[ESO DSS2 thumbnails of the Hodierna's binaries.]{ESO DSS2 images of the Hodierna's binaries. All images are 20 $\times$ 20\,arcmin and DSS2-red except image number 12 that is DSS2-infrared for a better visualization. (This set of images were published in the original article \citealt{gonzalezpayo24b} as a separate appendix.)}
  \label{fig:binary_stars_images}
\end{figure}

Table~\ref{tab:identified_pairs} collects the results and main properties of the twelve systems, including equatorial coordinates, proper motions, distances from \textit{Hipparcos} and \textit{Gaia}, and visual magnitudes from Simbad.
There are actually two triples and nine binaries, for which we provide our angular separations in the last column, and one single (\textit{Ain}).
The table also shows the WDS identifiers and discovery codes of six systems, assigned to F.\,G.\,W.~Struve (STF, STFA), J.~South \& J.~Herschel (SHJ), S.\,W.~Burnham (BU), and O.~Struve (STTA) instead of to Hodierna (or even to Galileo and Castelli in the case of the Orion's Trapezium, or to Hortensius and Riccioli for \textit{Algedi} and \textit{Prima Giedi}).
The twelve primary stars and their companions are also illustrated by the thumbnails in Fig.~\ref{fig:binary_stars_images}.

Not all of the eleven identified systems are physically bound.
Furthermore, there are optical systems with components at very different heliocentric distances.
For example, the system \#8 is made of an F3 dwarf at 18.4\,pc ($\theta$ Cyg), a G9\,III eruptive variable at $\sim$160\,pc (HD~185264), and a S-type star at $\sim$600\,pc (R~Cyg).
We applied the conditions outlined by \citet{gonzalezpayo23} for common proper motion and parallax systems, but accounting for the greater uncertainties in parallactic distances of the brightest stars.
Since they are too bright for \textit{Gaia}, we took their parallax values from \textit{Hipparcos}.
After that, there remain four systems that are actually bound or are part of the same astrophysical association.
Two of them are in young open clusters, namely \textit{Atlas} and \textit{Pleione} (\#1) in the Pleaides and $\theta^{\text{1}}$~Ori (\#12) in the Orion Nebula Cluster.
A third bound system, $\alpha$~Lib (\#3) is also young and has been proposed to belong to the contested $\sim$300\,Ma-old Castor stellar kinematic group \citep{barrado98,montes01a,mamajek13,zuckerman13} 
Regardless of the actual membership of $\alpha$~Lib in the group, the stellar system is indeed young based on a dusty disc and, possibly, mid-infrared emission features around the primary \citep{chen05}.
The system is actually quadruple, since both $\alpha^{\text{2}}$~Lib and $\alpha^{\text{1}}$~Lib are double themselves ($\alpha^{\text{2}}$~Lib: \citealt{slipher1904,lee1914,young1917,wilson53} -- $\alpha^{\text{1}}$~Lib: \citealt{duquennoy91,beuzit04,makarov05}).

Furthermore, \citet{caballero10} proposed a fifth component in the system at an extremely wide separation, KU~Lib, which also displays youth features such as high lithium abundance, strong X-ray emission, fast rotation, and photospheric dark spots \citep{gaidos98,gaidos00}.
The fourth and last bound system, namely \textit{Kuma}$^{\text{2}}$ and \textit{Kuma}$^{\text{1}}$ (\#11), is also triple (\textit{Kuma}$^{\text{2}}$ is a spectroscopic binary -- \citealt{batten78,pourbaix04}), but we did not find any reliable reference in the literature for their membership in any stellar kinematic group.
Accounting for the reported masses of all the stellar components, all four systems have binding energies greater than 10$^{\text{35}}$\,J and up to 10$^{\text{38}}$\,J, between three and six orders of magnitude greater than those of the most fragile multiple systems \citep{caballero09,gonzalezpayo23}.  
The other seven identified pairs (namely \textit{Algedi} [\#4], \textit{Acrab} [\#5], $\nu$~Gem [\#6], \textit{Denebola} [\#7], $\theta$~Cyg [\#8], $\delta$~Lep [\#9], and $\zeta^{\text{2}}$~Sco [\#10]) are optical and, therefore, are not bound.

\section{Conclusions}
\label{sec:hodierna_conclusions}

It is established that Christian Mayer published the first catalogue of possible double stars in 1779, just a few years ahead of the three catalogues of William Herschel. 
Hence, both Mayer and Herschel inaugurated the double star astronomy.
However, here we demonstrate that the Italian astronomer Giovanni Battista Hodierna identified twelve double (and triple) stars in his ``\textit{De systemate orbis cometici}'', which dates back to 1654.
Hodierna measured angular separations for a few of them and remarked that there were many more multiple systems.
Of the twelve systems, we were not able to identify only one, and of the remaining eleven systems, only four are gravitationally bound.
One of the four, namely $\theta^{\text{1}}$~Ori, had already been discovered by Castelli, Galileo, and Huygens.
This fact does not diminish Hodierna's status as the first astronomer in history to publish a list of multiple systems, over a century earlier than previously accepted.

\section{Appendix: Three binaries mentioned in the ``\textit{Almagestum Novum}''}
\label{sec:almagestum_novum}

Giovanni Battista Riccioli (1598--1671) was an Italian astronomer and Jesuit priest, famous for his experiments with pendulums and falling bodies, his analysis of 126 arguments about the Earth's motion, and for introducing the modern system of lunar nomenclature.
In his encyclopedic book \textit{``\textit{Almagestum novum astronomiam veterem novamque complectens observationibus aliorum et propriis novisque theorematibus, problematibus ac tabulis promotam in tres tomos distributam}''}\footnote{``Almagestum Novum, encompassing ancient and modern astronomy, advanced with observations of others and his own, as well as with new theorems, problems, and tables, distributed in three volumes''.}, \citet{riccioli1651} described three binaries hypothetically discovered by the Dutch astronomer and mathematician Martinus Hortensius (Martin van den Hove, 1605--1639).
In particular, when \citet{riccioli1651} portrayed the capacity of Hortensius' telescope to measure the diameter of stars, the former wrote:
``(...) and comparing this capacity with other distances, he found that the two contiguous stars in Capricorn are separated by one-eighth of the length of the tube, that is, 5 or 5\,{\textonehalf} arcminutes, and the two contiguous stars in the Hyades are nearly 5 or 4\,{\textonehalf} arcminutes'' (First volume, Sixth book, Chapter IX, Paragraph 4; translated from the original in Latin). 
We were not able to identify the original claim by Hortensius, nor a third double star in the Pleiades separated by astonishing 31\,arcmin.

Following the same procedure as described for Hodierna's double stars, we propose that the two first pairs of Hortensius and Riccioli were \textit{Algedi} ($\alpha^{\text{2}}$~Cap) and \textit{Prima Giedi} ($\alpha^{\text{1}}$~Cap) in Capricorn, which are separated by about 6.4\,arcmin in the sky and make up our system \#4, and $\theta^{2}$~Tau and $\theta^{\text{1}}$~Tau in the Hyades, which are separated by about 5.6\,arcmin.
Curiously, if these assignations are correct, Riccioli's estimations of angular separations were also affected by the same $\sim$ 0.80 scale factor as Hodierna's.

\newpage

%%%%% CAPITULO 3 %%%%%

\chapter{Multiplicity of ultra-cool subdwarfs} 
\label{ch:multiplicity_of_ultra-cool_subdwarfs}
\vspace{2cm}
\pagestyle{fancy}
\fancyhf{}
\lhead[\small{\textbf{\thepage}}]{\textbf{Section \nouppercase{\rightmark}}}
\rhead[\small{\textbf{Chapter~\nouppercase{\leftmark}}}]{\small{\textbf{\thepage}}}

\begin{flushright}
\small{\textit{``And in his eyes}}

\vspace{-2mm}
\small{\textit{The cold stars lighting, very old and bleak,}}

\vspace{-2mm}
\small{\textit{In different skies''}}

\small{\textit{The War Poems}}

\vspace{-2mm}
\small{\textit{-- Wilfred Owen}} 
\end{flushright}
\bigskip

\begin{adjustwidth}{70pt}{70pt}
\tGray{\small{The content of this chapter has been adapted from the article \citet{gonzalezpayo21}: \textit{Wide companions to M and L subdwarfs with Gaia and the Virtual Observatory}, published in Astronomy \& Astrophysics, \href{https://doi.org/10.1051/0004-6361/202140493}{A\&A 650, A190 (2021)}.}}
\end{adjustwidth}
\bigskip

\lettrine[lines=3, lraise=0, nindent=0.1em, slope=0em]{S}{ubdwarfs are objects} that lie appreciably below the main sequence on the Hertzsprung-Russell diagram and were first discovered by \citet{kuiper39}. They have a luminosity class VI under the Yerkes spectral classification system \citep{morgan43}, and appear less luminous than solar metallicity dwarfs with similar spectral types because of the low abundances in elements heavier than helium. Subdwarfs belong to population II and are stars from the galactic thick disk or halo \citep{gizis99}. 
Cool subdwarfs have spectral types G, K, and M, and are typically found to have thick disk or halo kinematics \citep{gizis97}. They are presumably relics of the early Galaxy, with ages of 10$-$12\,Gyr \citep{jofre11}, and are therefore excellent tracers of galactic chemical history, because they were formed at the early stage of the Milky Way. 
There are different metallicity classes of M subdwarfs based on a spectral index that measures the ratio of hydrides and oxides present in their atmospheres. The original classification for M subdwarfs (sdMs) and extreme subdwarfs (esdMs) developed by \citet{gizis97} was revised and extended by \citet{lepine07c}. A new class of subdwarfs, the ultra subdwarfs (usdMs), has been added to the sdMs and esdMs. Currently, dwarfs are classified as having solar metallicity, subdwarfs as having  moderately low metallicity, extreme subdwarfs as having very low metallicity, and ultra subdwarfs as having ultra low metallicity \citep{kirkpatrick05} with metallicity estimates of approximately [M/H]\,=\,0.0, -0.5, -1.0, and -2.0\,dex, respectively, with a typical dispersion of 0.5\,dex \citep{lodieu17a,zhang17a}.

Star-like objects with spectral types later than M5 and effective temperatures of less than $\sim$2900\,K \citep{kirkpatrick97b} are usually referred to as ultra-cool dwarfs. This heterogeneous group includes stars of extremely low mass as well as brown dwarfs, and represents about 15\% of the population of astronomical objects near the Sun \citep{gillon16}. The M dwarfs represent about two-thirds of the stars in the Milky Way and constitute around the 40\% of the total stellar mass in the Galaxy \citep{gould96,bochanski10}.

M dwarfs are now of popular interest in the search for extrasolar planets with complementary techniques, leading independent groups to define new methods to estimate their metallicities. As of now, most studies look at slightly metal-poor M dwarfs (typically $>$\,$-$0.5 dex) with an accuracy in their metallicity determination of the order of 0.15 dex, either photometrically \citep{bonfils05a,johnson09a,schlaufman10a,neves12,hejazi15,dittmann16}, or spectroscopically \citep{woolf05,woolf06,bean06a,woolf09,rojas_ayala10,rojas_ayala12,muirhead12,terrien12a,onehag12,neves13,neves14,hejazi15,newton15a,lindgren16a}. The sample of M dwarfs with metallicities of less than $-$0.5 dex is small, and very few M dwarfs in that sample have companions to more massive primaries with well-determined physical parameters to help to determine the metallicity scale of sdMs.

There is a direct correlation between the metallicity of stars and the occurrence of giant gaseous exoplanets \citep{papaloizou06,williams11}. In the core accretion model of planet formation \citep{pollack96,papaloizou06,udry07,boley09,janson11}, there are particular processes that depend on metallicity to form a planet from a dusty circumstellar disk \citep{weidenschilling80,armitage10}. Those processes tend to occur in greater numbers in metal-rich disks than in metal-poor disks \citep{johnson12}. Therefore, it is of prime importance to determine the metallicity of M dwarfs with the best accuracy possible in order to better characterise the properties of the planets in their vicinity.

In this chapter, we present a dedicated search for wide companions to known M and L subdwarfs reported in the literature \citep{lodieu12b,lodieu17a,zhang17b,zhang17a,zhang18b,zhang18a}. The objective of this chapter is two-fold: (i) to determine the multiplicity rate of the sample, and (ii) to improve estimates of the metallicity of M-L subdwarfs and their distances from their more massive primary. We looked for wide common proper motion companions in \textit{Gaia} DR2, and compared our results with multiplicity studies focusing on metal-poor populations such as those of \cite{chaname04}, \cite{zapatero04a}, \cite{riaz08a}, \cite{jao09a}, \cite{badenes18a}, \cite{moe19}, and \cite{elbadry19a}.

The chapter is structured as follows. First, we describe the sample of M and L subdwarfs and their properties. Later, we describe the methodology to identify wide common proper motion companions on the basis of astrometric and photometric criteria. The third step is the analysis of all potential companion candidates based on their proper motions, distances and photometry. Then, we present the results of the search and the most promising wide systems with additional photometric and spectroscopic characterisation. Finally, we provide the spectral type of each component of the multiple systems identified in the search, and discuss the frequency of M and L subdwarf systems with previous observational studies and theoretical predictions. 

\section{Sample selection}
\label{sec:sample_selection}

To achieve the scientific objectives, we worked with a sample of 219 known ultra-cool subdwarfs, 185 of which were taken from the SVO late-type subdwarf archive\footnote{The SVO late-type subdwarf archive, \url{http://svo2.cab.inta-csic.es/vocats/ltsa/index.php}} maintained by the Spanish Virtual Observatory\footnote{SVO, \url{https://svo.cab.inta-csic.es/main/index.php}}. The sample of subdwarfs contained in the archive is an extension of the list of 100 subdwarfs identified by \citet{lodieu17a} using Virtual Observatory (VO) tools, now containing 193 sources. This sample includes most of the known ultra-cool subdwarfs confirmed spectroscopically at the time of writing. we rejected eight of them because of the lack of spectral types derived from optical spectroscopy. For each object, the archive contains coordinates, identifiers, effective temperatures, proper motions, spectral types, and magnitudes in different passbands, which can be accessed through a very simple search interface that permits queries by coordinates or radius and/or a range of magnitudes, colours, and effective temperatures. All of these data come from the Two Microns All Sky Survey \citep[2MASS;][]{skrutskie06}, the United Kingdom InfraRed Telescope (UKIRT) Deep Sky Survey \citep[UKIDSS;][]{lawrence07}, the Visible and Infrared Survey Telescope for Astronomy (VISTA) Hemisphere Survey \citep{mcmahon12}, the Sloan Digital Sky Survey \citep[SDSS;][]{york00} and Wide Field Infrared Survey Explorer \citep[WISE;][]{wright10} public catalogues. Thirty-four additional objects were taken from more recent works \citep{zhang17b,zhang17a,zhang18b,zhang18a,zhang19g}, bringing the total sample used here to 219 ultra-cool subdwarfs. This sample contains all known subdwarfs with spectral types between M5 and L8\footnote{Note added during external revision: Detected a typo in the original article where L7 was indicated.} confirmed spectroscopically.

The 219 sources considered in this work cover spectral types from M5 to L8, and belong to different metallicity classes: subdwarfs, extreme subdwarfs, and ultra subdwarfs. The numeric identifiers (Id) of the 219 sources, common names, coordinates, source identification from \textit{Gaia} DR2 \citep{gaiacollaboration18} when existing, spectral types, and references are listed in Table~\ref{tab:coordinates_subdwarfs}, available in \textit{VizieR} \citep{ochsenbein00}. Coordinates were obtained from the catalogue that gives the name to the source: SDSS \citep[Sloan Digital Sky Survey;][]{adelman_mccarthy09,adelman_mccarthy12,alam15a}, ULAS \citep[UKIDSS Large Area Survey;][]{lawrence07}, 2MASS \citep[Two Micron All-Sky Survey;][]{skrutskie06}, LSR \citep[L\'epine, Shara, Rich;][]{lepine02}, LHS \citep[Luyten Half Second catalogue;][]{luyten79}, SSSPM \citep[SuperCOSMOS Sky Survey Proper Motion;][]{hambly01a}, and APMPM \citep[Automatic Plate Measuring Proper Motion;][]{kibblewhite71}.

\section{Methodology}
\label{sec:methodology}

In this section, we describe the methodology using proper motion, distance, and binding energy criteria to identify wide companions to the sample of M and L subdwarfs. We based all of the data on catalogue \textit{Gaia} DR2, and did not use the latest release of \textit{Gaia} because we started this work well ahead of the \textit{Gaia} EDR3 and DR3\@.

\subsection{Proper motions}
\label{sub:proper_motions}

To search for wide companions, we need the most accurate proper motions possible. To collect them, we looked for proper motions in the \textit{Gaia} DR2 catalogue. Whenever proper motions were not available, we computed them through a linear regression of the positions and epochs provided by different astrometric catalogues: 2MASS, SDSS DR7 \citep{adelman_mccarthy09}, SDSS DR9 \citep{adelman_mccarthy12}, SDSS DR12 \citep{alam15a}, UKIDSS DR9 \citep{lawrence07}, WISE \citep{wright10}, and Pan-STARRS1 \citep{chambers16a}. We used the \textit{Aladin} Sky Atlas \citep{bonnarel00} VO tool, \textit{Simbad} \citep{wenger00}, and \textit{VizieR} VO services to avoid mismatches and identify the correct detections of the sources in our sample from each catalogue. We checked whether or not the method agrees with \textit{Gaia} DR2 values when these latter were available, and we found variations of less than 4\% in right ascension and declination components and less than 10\% in the total proper motion.

Using the linear regression method, for most of the sources we found between 4 and 15 positions with different epochs, even in the same catalogue. The mean coverage is around 10 years, giving acceptable error bars. For the 219 M and L subdwarfs in this work, there are 149 with proper motions from \textit{Gaia} DR2 and 70 computed by linear regression using public catalogues. The median motion of the objects in the sample is around 255 mas a$^{-\text{1}}$. The mean uncertainty on the proper motions is about 1 mas a$^{-\text{1}}$ and 12 mas a$^{-\text{1}}$ for those in \textit{Gaia} DR2 and those computed by linear regression, respectively. We calculated the proper motion with just two values for two sources, yielding large errors in those specific cases (Id 29 and Id 134). We did not find any companions to these two sources even after applying such error margins. We list the proper motions and their references in the second, third, and fourth columns of Table~\ref{tab:astrometric_subdwarfs}. The data contained in this table can be retrieved from \textit{VizieR}.

\subsection{Distances}
\label{sub:distances}

We also need distances that are as accurate as possible. Therefore, we first considered the parallaxes from \textit{Gaia} DR2. If not available or if the parallax error was higher than 20\% of the parallax, we estimated spectrophotometric distances as in \citet{lodieu17a}. Although \citet{bailerjones18a} set a limiting value of the relative error of the parallax at 10\%, we allow a larger margin (20\%) to avoid rejecting possible candidates (e.g. unresolved binaries). Because of the weakness of the sources, the uncertainties on \textit{Gaia} DR2 parallaxes are similar to the uncertainties on the calculated distances. We found 126 M and L subdwarfs with \textit{Gaia} DR2 parallaxes and estimated spectro-photometric distances for another 93 objects in the sample. In some cases, the \textit{Gaia} distance with errors on the parallax higher than 20\% turns out to be more accurate than that estimated from the spectral type--magnitude relation. We therefore opted for the \textit{Gaia} distances in those specific cases.

To calculate the spectrophotometric distances, we used the \textit{J} band photometry in table~5 in \citet{lodieu17a} when available, and the photometry in the \textit{i} and \textit{J} bands in table~2 in \citet{zhang13} otherwise. The error on the spectrophotometric distances takes into account the 0.5 uncertainty on the spectral type and the associated error on the magnitudes, and other parameters used in the relation given in \citet{zhang13}.

\begin{figure}[h]
  \centering
  \includegraphics[width=0.6\linewidth, angle=0]{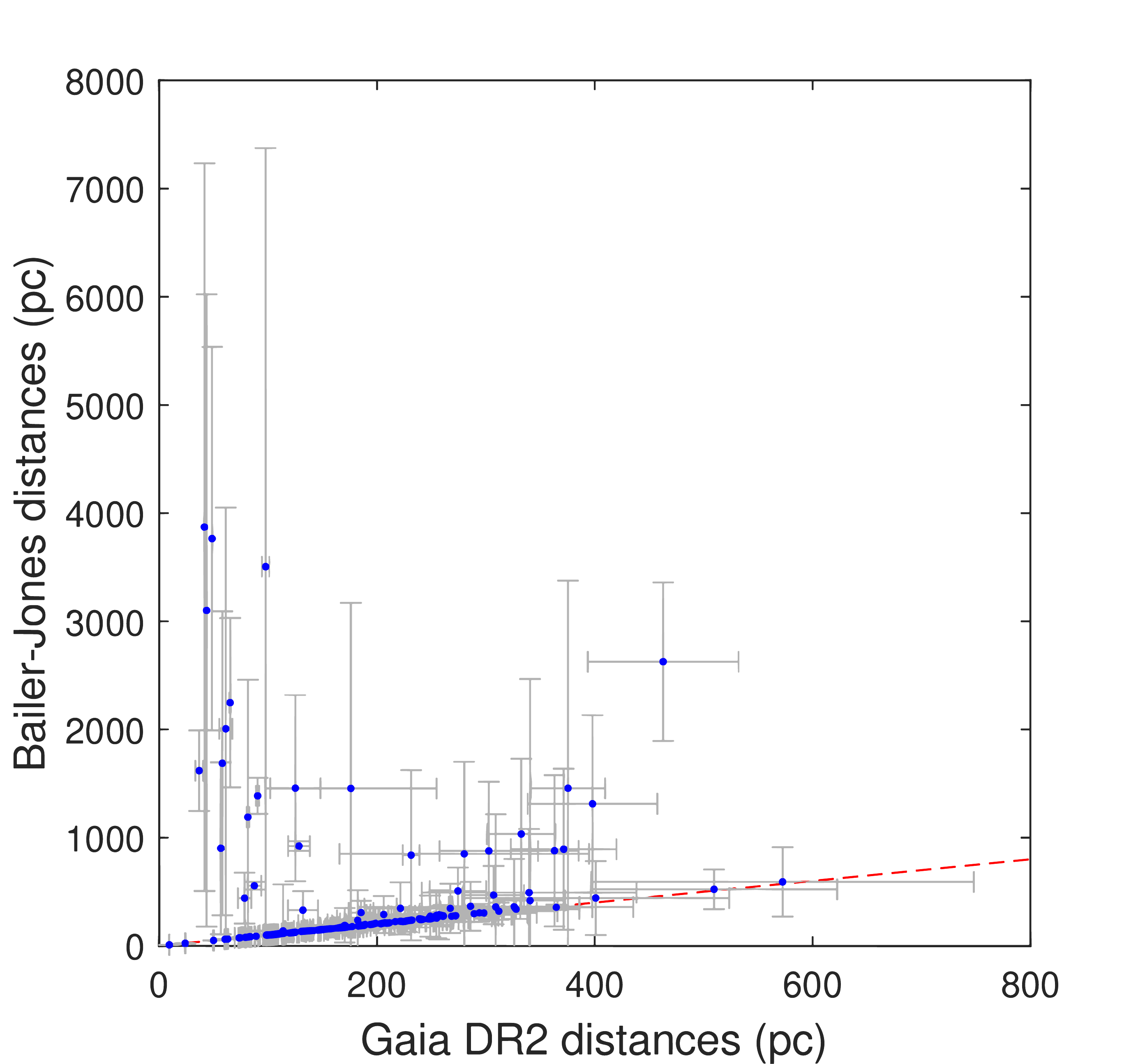}
     \caption[Comparison between \textit{Gaia} DR2 distances and Bailer-Jones distances.]{Comparison between \textit{Gaia} DR2 distances and the ones in \citet{bailerjones18a}. The red dashed line represents the 1:1 relation. 
}
   \label{fig:Bailer_Jones}
\end{figure}

The simple approach of inverting the parallax to estimate the distance can sometimes lead to potentially strong biases, especially (but not only) when the relative uncertainties are large and objects lie at large distances, as is often the case for members of the halo. A proper statistical treatment of the data, its uncertainties, and correlations may be required as advised by \citet{luri18a} and \citet{bailerjones18a}. For this reason, we compared the adopted \textit{Gaia} DR2 distances in this work with the values of \citet{bailerjones18a}, who recently proposed an alternative methodology to get reliable distances taking into account the non-linearity of the transformation and the asymmetry of the resulting probability distribution. In our case, although the \textit{Gaia} DR2 distances are not reliable when the error on the parallax is larger than or equal to 20\%, the comparison with the distances from \citet{bailerjones18a} shows that these distances have larger errors than the \textit{Gaia} DR2 ones (see Fig.~\ref{fig:Bailer_Jones}). We observe that 5 sources out of 123 lie noticeably away from the 1:1 relation, most of them showing unreliable distances and very large errors. We also note consistency between the calculated spectrophotometric distances and the ones in \textit{Gaia} DR2, when available. Therefore, the use of the distances from \citet{bailerjones18a} does not provide any significant advantage with respect to the use of \textit{Gaia} DR2 or spectrophotometric distances. Finally, We were able to compile and compute distances for all the sources in the sample, which are presented in the fifth column in Table~\ref{tab:astrometric_subdwarfs}, (in Appendix~\hyperref[ch:Appendix_C]{C}, and available too in \textit{VizieR}). The mean distance of the sample is 187\,pc with a mean error of 18.1\,pc in \textit{Gaia} distances and 33.7\,pc in the spectrophotometric distances.

\subsection{Radius of the search}
\label{sub:radius_search}

To perform the search for companions, We define a search radius for M and L subdwarfs equal to the maximum separation the system may have if gravitationally bound. Beyond that separation, the gravitational binding energy is too low to keep the system tight and pairs are no longer physically bound. This search radius is determined as a function of the binding energy ($W$) and the masses involved in the system:
\begin{equation}
W=G\frac{M_\text{1} M_\text{2}}{r}
\label{eqn:binding_subdwarfs}
\end{equation}
\noindent where $G$ is the gravitational constant and has a value of 6.674 $\cdot$ 10$^{-\text{11}}$\,Nm$^\text{2}$kg$^{-\text{2}}$ \citep{carroll07}, $M_\text{1}$ and $M_\text{2}$ are the masses in kilograms of the two components of the system, and $r$ is the projected physical separation between them in metres. The most commonly accepted value of the minimum energy required for two celestial bodies to be bound is 10$^{\text{33}}$\,J \citep{caballero09,dhital10b}. Accounting also for the maximum mass that a physical companion could have, we can obtain the maximum separation between components.

We set A0 to be the upper limit in the spectral type (i.e. in the mass) of the companion. Although lifetimes of A0 stars are typically shorter than lifetimes of subdwarfs, we select them in order to account for the extra mass involved in close binaries, that is, in the case of triple or multiple systems, that could yield larger projected physical separations. This maximum separation is calculated for each source in the sample using the average mass of an A0 star and the mass of the subdwarf. We estimated a mass of 2.36$\pm$0.035\,M$_\odot$ for an A0 star based on the data from \citet{popper80}, \citet{harmanec88}, and \citet{gray05}. We adopted a mass of 2.40\,M$_\odot$ accounting for the estimated error, which is in agreement with \citet{adelman04}. The masses of the low-mass stars vary depending on their spectral types and metallicity \citep{burrows89,kroupa97}. Because of the lack of dynamical masses for metal-poor ultra-cool dwarfs, we adopted the masses of main sequence solar-type M dwarfs from \citet{reid05b} as a first approximation. For stars with spectral types later than M9, we used the mass of 0.075\,M$_\odot$ as an upper limit, corresponding to the stellar--substellar boundary at solar-metallicity \citep{chabrier97}.

\subsection{Search criteria}
\label{sub:search_criteria}

Once we inferred the proper motion, distance, and search radius for each source in the sample, we looked for candidate companions in \textit{Gaia} DR2 through \texttt{Topcat} \citep{taylor05} and a code in Astronomical Data Query Language (ADQL) specifically written for our purposes \citep{yasuda04}. We imposed the following conditions, where \textit{Obj} refers to the subdwarf in the sample and \textit{Comp} to the companion candidate:

\begin{itemize}

\item [$\bullet$] The companion candidates must share the same proper motion as the M and L subdwarfs in the sample in each direction within 3$\sigma$: $\mu_{Obj}-3\sigma_{\mu.Obj}\leq \mu_{Comp} \leq \mu_{Obj}+3\sigma_{\mu.Obj}$.

\item [$\bullet$] The companion candidates must share the same distance as the object in the sample within 3$\sigma$: $d_{Obj}-3\sigma_{d.Obj}\leq d_{Comp} \leq d_{Obj}+3\sigma_{d.Obj}$. Here, we add a restriction into the search to avoid too many spurious candidates, restricting the possible candidates to those with a maximum relative error in their distances of 20\%.

\item [$\bullet$] The companion candidates must lie within the search radius previously defined for every source in the sample. The search radius used for each subdwarf in the sample is presented in the last column of Table~\ref{tab:astrometric_subdwarfs}, available in \textit{VizieR}.
\end{itemize}

\section{Analysis}
\label{sec:analysis_subdwarf}

\subsection{Performed search}
\label{sub:search}

We looked for common proper motion companions to the 219 M and L subdwarfs in the sample with a search radius defined for each of them varying from 10.8\,arcmin to 9.4\,deg, corresponding to an interval of projected physical separations of 1.5 to 3.7\,pc. Despite the wide range of radii, the typical search radius is 1\,deg with 90\% of the sources covered within 2.28\,deg. The lower limit of detection is given by the angular resolution of \textit{Gaia} DR2, and is 0.4\,arcsec. The detection of pairs with \textit{Gaia} DR2 is complete beyond 2.2\,arcsec \citep{gaiacollaboration18}, which is the lower limit that we adopted for completeness. This translates to minimum projected physical separations of between 20 and 1260\,au in the sample.

\subsection{Companion candidates}
\label{sub:companion_candidates}

We found 62 companion candidates around 12 M and L subdwarfs in the sample. Table~\ref{tab:number_wide_comp} shows the numerical identifiers of the sources with companion candidates, their coordinates, spectral types, search radii, and number of candidates.

We checked the re-normalised unit weight error (\texttt{RUWE}) in the \textit{Gaia} catalogue. This is an astrometric quality parameter that is high when the source has poor astrometric solutions \citep{lindegren18a}, and is sometimes influenced by the presence of another source. Hence, it can be used as an indication as to whether or not there could be an additional close companion. The additional documentation of \textit{Gaia} DR2 \citep{lindegren18b} shows that a value under 1.4 generally indicates a good solution because approximately 70\% of the sources have such a value. We noticed that 9 of the 62 candidates present \texttt{RUWE} values of between 1.4 and 6.6\@. Given that \textit{Gaia} DR2 exhibits \texttt{RUWE} values higher than 40, we do not consider 6.6 as a large value (i.e.\ meaning a poor astrometric solution). Therefore, we keep all 62 candidates for subsequent analysis. Table~\ref{tab:astrometric_subdwarfs_companions}, available in \textit{VizieR}, lists the numerical identifier of each candidate companion together with their \textit{Gaia} DR2 source identifier, coordinates, proper motions, distances, and angular separations. As is the case for the rest of the tables contained in the Appendices, these data can be found in \textit{VizieR}.

\begin{table}[H]
 \centering
 \caption[Number of companion candidates identified in this work.]{Number of companion candidates identified in this work.}
 \footnotesize
 \scalebox{1}[1]{
 \begin{tabular}{ccccccc}
 \noalign{\hrule height 1pt}
 \noalign{\smallskip}
 Id & $\alpha$ (J2000) & $\delta$ (J2000) & Spectral  & \textit{J} & Radius & Num. \\
 & (hh:mm:ss.ss) & (dd:mm:ss.s) & type & (mag) & (deg) &  \\
 \noalign{\smallskip}
 \hline
 \noalign{\smallskip}
11  & 01:18:24.90 & $+$03:41:30.4 & sdL0.0 & 18.2 & 0.50 & 32 \\
25  & 04:52:45.69 & $-$36:08:41.3 & esdL0.0 & 16.3 & 0.61 & 1 \\
73  & 10:46:57.93 & $-$01:37:46.4 & dM4.5/sdM5.0 & 16.5 & 0.33 & 1 \\
89  & 11:19:29.20 & $+$67:21:04.1 & sdM5.0$-$5.5 & 16.8 & 0.60 & 3 \\ 
107 & 12:41:04.75 & $-$00:05:31.6 & sdL0.0 & 18.5 & 0.45 & 2 \\
126 & 13:07:10.22 & $+$15:11:03.5 & sdL8.0 & 18.1 & 1.80 & 6 \\
128 & 13:09:59.60 & $+$05:29:38.7 & sdM6.5 & 18.4 & 0.22 & 1 \\
149 & 13:53:59.58 & $+$01:18:56.8 & sdL0.0 & 17.4 & 0.73 & 1 \\
150 & 13:55:28.24 & $+$06:51:14.6 & sdM5.5 & 17.6 & 0.39 & 1 \\
190 & 15:46:38.34 & $-$01:12:13.1 & sdL3.0 & 17.5 & 1.44 & 1 \\
213 & 22:59:02.15 & $+$11:56:02.1 & sdL0.0 & 17.0 & 0.84 & 1 \\
215 & 23:04:43.31 & $+$09:34:24.0 & sdL0.0 & 17.2 & 0.78 & 12 \\
 \noalign{\smallskip}
 \noalign{\hrule height 1pt}
 \end{tabular}
}
 \label{tab:number_wide_comp}
\end{table}

We assess the validity of the candidates through visual inspection of the proper motion diagrams (PMDs), colour--magnitude diagrams (CMDs), and tangential velocity--distance diagrams. When possible, we also use radial velocities from the literature to compute galactocentric velocities, which serve us as membership indicators of the different galactic populations, as explained below. Table~\ref{tab:proposed_candidates} summarises whe\-ther each candidate agrees (\textit{Yes}), disagrees (\textit{No}), or is not conclusive (\textit{?}) with the position of the subdwarfs in the diagrams or whether it exhibits thick disc or halo kinematics. The last column of the table indicates whether the candidates are likely (\textit{Yes}) or doubtful (\textit{Yes?}) companions, or have been rejected (\textit{No}) as bound companions.

We analysed two optical and one infrared CMDs using the \textit{Gaia} $G$ and $RP$ passbands, the $i,z$ filters from SDSS, and the $J,K$ filters from 2MASS when available, or UKIDSS otherwise (when the 2MASS quality flags differ from ``A'' or ``B''). For each CMD, we used the \textit{BT-Settl}\footnote{\url{https://phoenix.ens-lyon.fr/Grids/BT-Settl/CIFIST2011/ISOCHRONES/}} isochrones with metallicities [M/H] of $-$2.0 dex, $-$0.5 dex, and solar for comparison \citep{allard14}. 

We used the 10\,Gyr isochrones because ultra-cool subdwarfs are old objects. We complemented the \textit{Gaia} DR2 CMD with the observational H-R diagram in the same bands, plotting \textit{Gaia} sources with parallaxes larger than 10\,mas as a reference. All diagrams used for the analysis are displayed in Appendix~\hyperref[ch:Appendix_C]{C}, except from Id 150, which is shown as an example in Fig.~\ref{fig:plot_CMD_PM_150}.

Radial velocities of each component of a physically bound system should be similar because of their presumably common origin. Therefore, we looked for radial velocities of every subdwarf with candidate companions using the SVO Discovery tool\footnote{SVO Discovery tool, \url{http://sdc.cab.inta-csic.es/SVODiscoveryTool/jsp/searchform.jsp}} developed and maintained by the Spanish Virtual Observatory, which performs a search through the \textit{VizieR} VO service.

\begin{table}[H]
 \centering
 \caption[Summary of the companion candidates assessment.]{Summary of the companion candidates assessment.}
 \footnotesize
 \scalebox{0.85}[0.85]{
 \begin{tabular}{lcccccccc}
\noalign{\hrule height 1pt}
\noalign{\smallskip}
 Id & PMD & CMD & CMD & CMD & H-R & Kinematics & $V_{Tan}$ & Candidate \\
  & & $M_j/J-K$ & $M_G/G-RP$ & $M_i/i-z$ & diagram & & & \\ 
 \noalign{\smallskip}  
\hline
\noalign{\smallskip}
 11-1 & No & - & - & - & - & No & - & No \\
 11-2 & No & - & - & - & - & - & - & No \\ 
 11-3 & No & - & - & - & - & - & - & No \\ 
 11-4 & No & - & - & - & - & - & - & No \\ 
 11-5 & No & - & - & - & - & - & - & No \\ 
 11-6 & No & - & - & - & - & - & - & No \\
 11-7 & No & - & - & - & - & No & - & No \\ 
 11-8 & No & - & - & - & - & No & - & No \\ 
 11-9 & No & - & - & - & - & No & - & No \\ 
 11-10 & No & - & - & - & - & No & - & No \\
 11-11 & No & - & - & - & - & - & - & No \\ 
 11-12 & No & - & - & - & - & - & - & No \\ 
 11-13 & No & - & - & - & - & - & - & No \\ 
 11-14 & No & - & - & - & - & No & - & No \\ 
 11-15 & No & - & - & - & - & No & - & No \\ 
 11-16 & No & - & - & - & - & - & - & No \\ 
 11-17 & Yes & Yes & ? & Yes & Yes & - & Yes & Yes \\
 11-18 & No & - & - & - & - & - & - & No \\ 
 11-19 & Yes & Yes & ? & Yes & No & No & No & No \\
 11-20 & No & - & - & - & - & No & - & No \\ 
 11-21 & No & - & - & - & - & - & - & No \\ 
 11-22 & No & - & - & - & - & - & - & No \\ 
 11-23 & No & - & - & - & - & No & - & No \\ 
 11-24 & No & - & - & - & - & - & - & No \\ 
 11-25 & No & - & - & - & - & - & - & No \\ 
 11-26 & No & - & - & - & - & No & - & No \\ 
 11-27 & No & - & - & - & - & - & - & No \\ 
 11-28 & No & - & - & - & - & - & - & No \\ 
 11-29 & No & - & - & - & - & - & - & No \\ 
 11-30 & No & - & - & - & - & - & - & No \\ 
 11-31 & No & - & - & - & - & - & - & No \\ 
 11-32 & No & - & - & - & - & - & - & No \\ 
 25-1 & Ýes & Yes & Yes & - & Yes & - & Yes & Yes \\
 73-1 & Yes & Yes & Yes & Yes & Yes & - & Yes & Yes \\
 89-1 & ? & Yes & No & No & No & - & No & No \\
 89-2 & ? & Yes & No & No & No & - & No & No \\
 89-3 & ? & Yes & No & No & No & - & No & No \\
 107-1 & Yes & Yes & ? & Yes & Yes & - & Yes & Yes \\
 107-2 & ? & No & ? & Yes & No & - & No & No \\ 
 126-1 & No & ? & Yes & No & Yes & No & No & No \\ 
 126-2 & No & ? & Yes & Yes & Yes & No & Yes & No \\ 
 126-3 & No & ? & Yes & No & Yes & - & No & No \\ 
 126-4 & No & ? & No & No & No & - & No & No \\ 
 126-5 & No & No & No & No & Yes & No & No & No \\ 
 126-6 & No & ? & Yes & Yes & Yes & - & No & No \\ 
 128-1 & ? & No & ? & No & Yes & - & No & No \\
 149-1 & ? & No & ? & No & Yes & No & No & No \\
 150-1 & Yes & Yes & Yes & Yes & Yes & - & Yes & Yes \\
 190-1 & Yes & No & ? & Yes & Yes & - & Yes & Yes ? \\
 213-1 & No & ? & ? & Yes & Yes & No & No & No \\
 215-1 & No & Yes & ? & Yes & Yes & - & No & No \\ 
 215-2 & Yes & Yes & ? & ? & No & - & Yes & No \\ 
 215-3 & No & Yes & ? & No & Yes & - & No & No \\ 
 215-4 & No & No & ? & No & No & - & No & No \\ 
 215-5 & No & Yes & ? & No & Yes & - & No & No \\ 
 215-6 & No & ? & ? & No & No & - & No & No \\ 
 215-7 & No & Yes & ? & No & Yes & - & No & No \\ 
 215-8 & Yes & Yes & ? & No & Yes & No & Yes & No \\
 215-9 & Yes & Yes & ? & No & Yes & - & Yes & No \\ 
 215-10 & No & No & ? & No & No & - & No & No \\ 
 215-11 & No & No & ? & No & No & No & No & No \\ 
 215-12 & No & Yes & ? & Yes & Yes & - & No & No \\ 
\noalign{\smallskip}
\noalign{\hrule height 1pt}
\end{tabular}
 }
 \label{tab:proposed_candidates}
\end{table}

None of the 12 subdwarfs in Table~\ref{tab:number_wide_comp} have radial velocity measurements in the literature and, therefore, they could not be used for comparison. Nevertheless, we found values of radial velocities for 18 companion candidates \citep{gaiacollaboration18, zhong19, anguiano17, kervella19}, and with these data we are able to compute their $UVW$ galactic space velocities from their coordinates, proper motions, and distances \citep{johnson87}. We can assign the companions to the different populations in the Galaxy (thin and thick discs, transition between thin and thick discs, and halo) according to the increasing galactocentric velocities of these populations towards the galactic halo \citep{bensby03,bensby05}. This procedure is described in \citet{montes01a} and the updated version of the code used in this work was published in \citet{cortescontreras24}. As stated before, subdwarfs belong to the old galactic population, and therefore any candidate companion showing thick disc or halo kinematics would be a suitable companion. 

For all the subdwarfs and candidates, we can also use the tangencial velocities to assess old population kinematics, as in \citet{zhang18a}. 

\begin{itemize}

 \item [$\bullet$] \textbf{Id 11}: This subdwarf has a sdL0.0 spectral type and is located at about 176\,pc. We detected 32 possible companion candidates separated by between 1.7 and 29.9\,arcmin. The large number of spurious candidates is due to the high uncertainty in the proper motion computed in this work. As we are not able to reliably keep or discard any candidates, we reject all candidates above 1$\sigma$ in proper motion, leaving two companion candidates. Candidate Id number 17 shows higher proper motion than the subdwarf, lies between the $-$2.0 dex and $-$0.5 dex isochrones in the infrared CMD, and slightly above the solar-metallicity isochrone in the CMD with SDSS filters. Its position in the CMD with \textit{Gaia} photometry is slightly lower than the solar-metallicity isochrone, consistent with its position in the lower edge of the main sequence in the H-R diagram. Moreover, its tangential velocity, which is higher than that of the field population, and within 1$\sigma$ of the velocity of the subdwarf, makes this source a potential companion candidate. The position of candidate Id number 19 in the H-R diagram is above the solar-metallicity isochrone and within the main sequence, and its low tangential velocity and radial velocity suggest a location for this source of inside the thin disc; we therefore reject candidate number 19\@.
 
 \item [$\bullet$] \textbf{Id 25}: This subdwarf is an esdL0.0 source located at about 140\,pc with just one candidate separated by 1.9\,arcmin. It is interesting to remark that the companion candidate we find (denominated 25-1 hereafter) is Id 24 of the sample. The search program was not able to find a companion to Id 24 because of the very small margins of error on its proper motion provided by \textit{Gaia} DR2, despite the proximity of the values of both elements in the possible pair as shown in the PMD. The CMDs also show aligned positions for the subdwarf and for the candidate companion, except for the CMD with photometry from SDSS, which is not provided. The positions in the H-R diagram are below the main sequence, and their tangential velocities are similar in terms of distance. We fully support this pair as likely companions.
 
 \item [$\bullet$] \textbf{Id 73}: This dM4.5/sdM5.0 source located at about 572\,pc has one companion candidate, 15.9 arcmin away. The PMD shows similar proper motions. We favour the UKIDSS photometry for this potential companion. This candidate lies at the same distance to the isochrones in all CMDs. From the positions of each component in the H-R diagram, we suggest that both sources are probably solar metallicity rather than metal-poor, with a spectral type of dM2.0--dM2.5 as the spectroscopic follow-up suggests (Sect.~\ref{sub:NOT_ALFOSC_optical_spectroscopy}). Moreover, their tangential velocities show similarities in terms of their distance. All in all, we suggest both objects form a bound solar-metallicity pair. 
 
 \item [$\bullet$] \textbf{Id 89}: This is an sdM5.0--5.5 \citep{lodieu17a} object located at about 274\,pc with three companion candidates separated by 18.4, 34.7, and 40\,arcmin. The PMD shows very small values for the proper motions in all of the candidates and the subdwarf, and their positions in the CMDs are aligned. The $J-K$ colour of the ultra-cool subdwarf exhibits large error bars due to the poor quality flags in 2MASS photometry (Qflag\,=\,``CCD''). Nonetheless, this subdwarf lies below the main sequence in the H-R diagram, as expected. On the contrary, the three companion candidates lie within or slightly above the main sequence, perhaps reflecting different metallicities. The tangential velocity diagram shows that none of the candidates are valid. However, all tangential velocities are below the mean value for field stars, suggesting that these objects do not have thick disk or halo kinematics. Consequently, we do not support companionship of any of them.
 
 \item [$\bullet$] \textbf{Id 107}: This source, with spectral type sdL0.0, located at 197\,pc has two companion candidates separated by 3.5 and 10.9\,arcmin. According to the PMD, candidate number 2 has proper motions that are not agreement with those of the subdwarf but still remain within $3\sigma$. Because of the faintness of the subdwarf, even the UKIDSS photometry suffers from large error bars in the $J-K$ colour diagram. For the candidates, the photometry comes from 2MASS. We reject candidate number 2 as a bound companion because of its position in the CMDs. The position of candidate number 1 in the H-R diagram, below the main sequence, suggests subsolar metallicity, similar to the source. The subdwarf is not in \textit{Gaia} DR2 and therefore we cannot plot it on the H-R diagram, but we are able to plot the positions of the companion candidate number 1, compatible with a low-metallicity source. Finally, its tangential velocity is within the range of halo objects. Therefore, we consider this pair physically associated.
 
 \item [$\bullet$] \textbf{Id 126}: This sdL8.0 source is located at 49\,pc. We identify six possible companion candidates with separations of between 34 and 98.5\,arcmin. The large uncertainty in the proper motion of the subdwarf prevents us from obtaining any reliable candidate from the PMD. Additionally, none of the candidates show photometric criteria consistent with metal-poor isochrones in the CMD and H-R diagram\@. In particular, the radial velocity of candidate number 2 places it in the galactic thin disc. In conclusion, we reject all of these candidates.

 \item [$\bullet$] \textbf{Id 128}: This subdwarf is a sdM6.5 source located at 547\,pc, with a single companion candidate at 8.5 arcmin. According to the PMD, the candidate has a similar motion to the subdwarf. However, their positions in the CMDs suggest inconsistent metallicities, corroborated by their location in the H-R diagram where the companion follows with the main sequence solar-metallicity track. Additionally, their tangential velocities differ. Therefore, we discard the system. 
 
 \item [$\bullet$] \textbf{Id 149}: This source has a sdL0.0 spectral type and is located at 121\,pc. We detect a companion candidate at 36.1\,arcmin. The PMD shows that both objects have very similar proper motions. There is no photometry available in \textit{Gaia} for the subdwarf and the SDSS $i,z$ photometry of the candidate is clearly saturated, and so these CMDs do not provide useful information. Both objects show similar positions in the other CMD\@. The companion candidate is HD 120981, a G2/3\,V star \citep{houk88} with metallicity in the range 0.09--0.27 \citep{ammons06,stevens17}. Combining tangential velocity and radial velocity, we find that the companion shows galactocentric velocities typical of the thin disk. We discard this system as a physical pair.

  \begin{figure}
  \caption*{\textbf{Id 150}}
    \begin{subfigure}{.5\textwidth}
    \centering
    \includegraphics[width=0.91\linewidth]{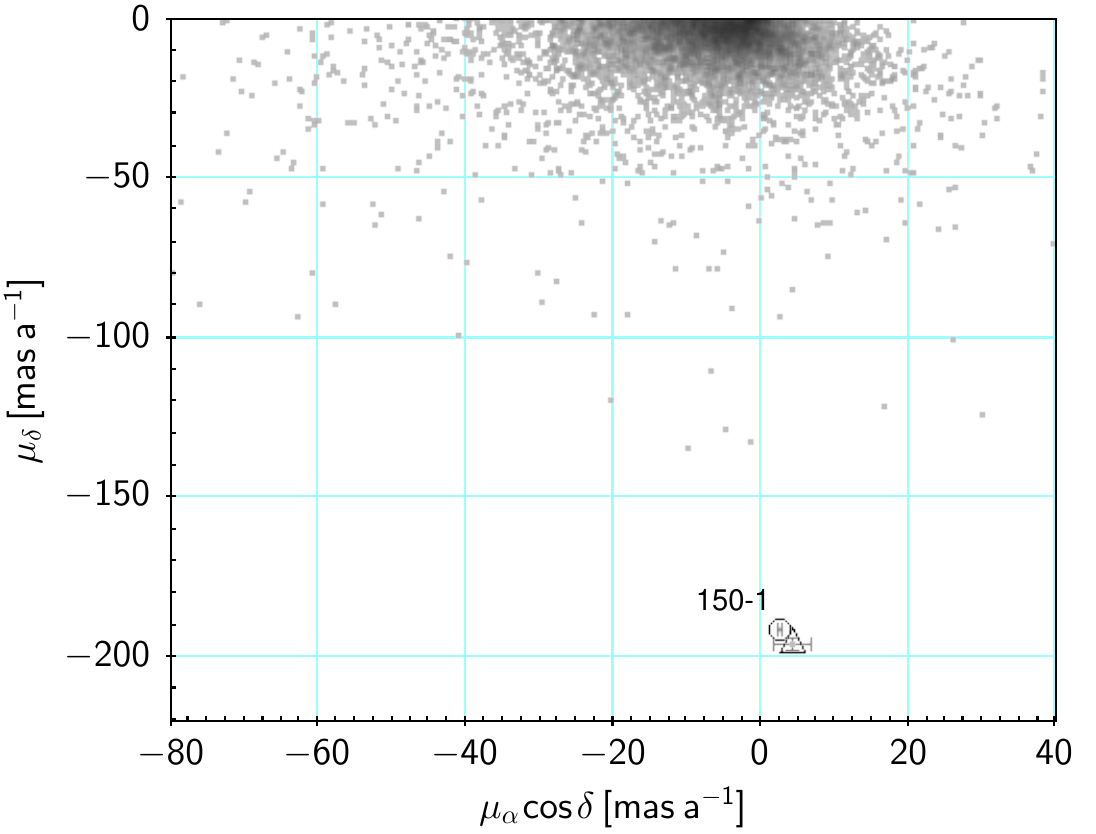}
    \end{subfigure}
    \begin{subfigure}{.5\textwidth}
     \centering
      \includegraphics[width=0.91\linewidth]{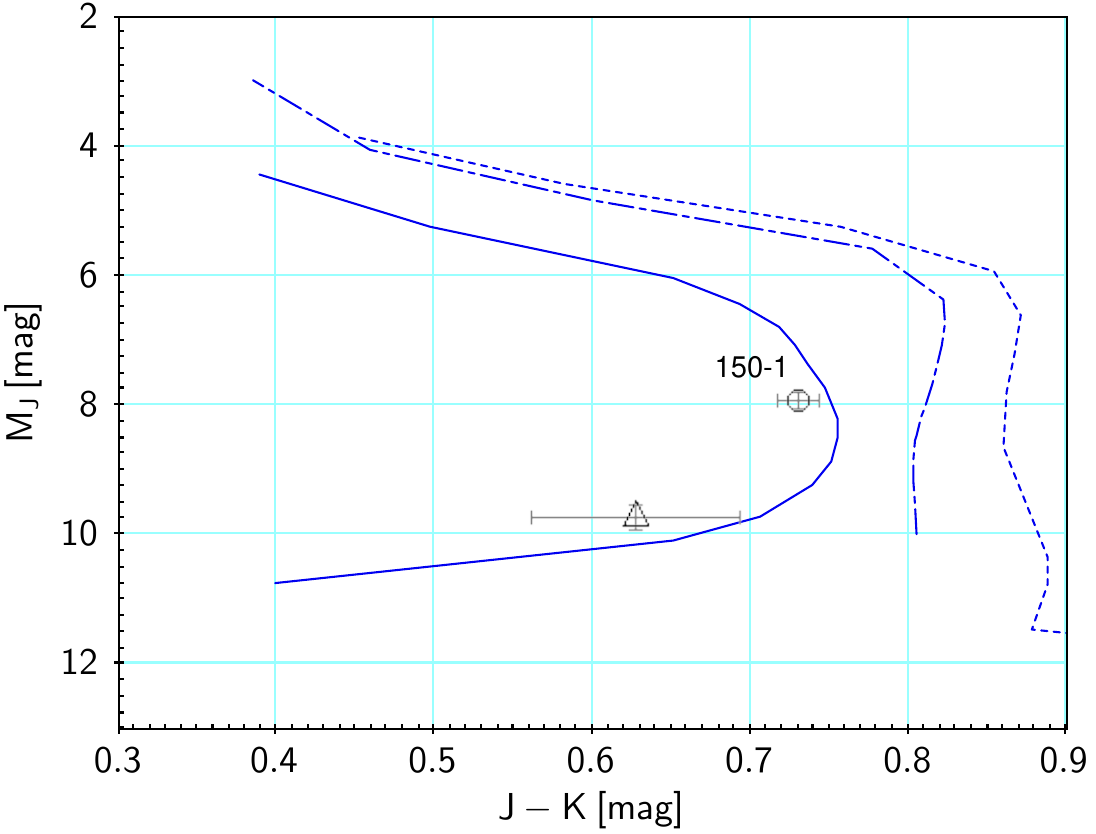}
    \end{subfigure}
  \vspace{2mm}
  
     \begin{subfigure}{.5\textwidth}
    \centering
    \includegraphics[width=0.91\linewidth]{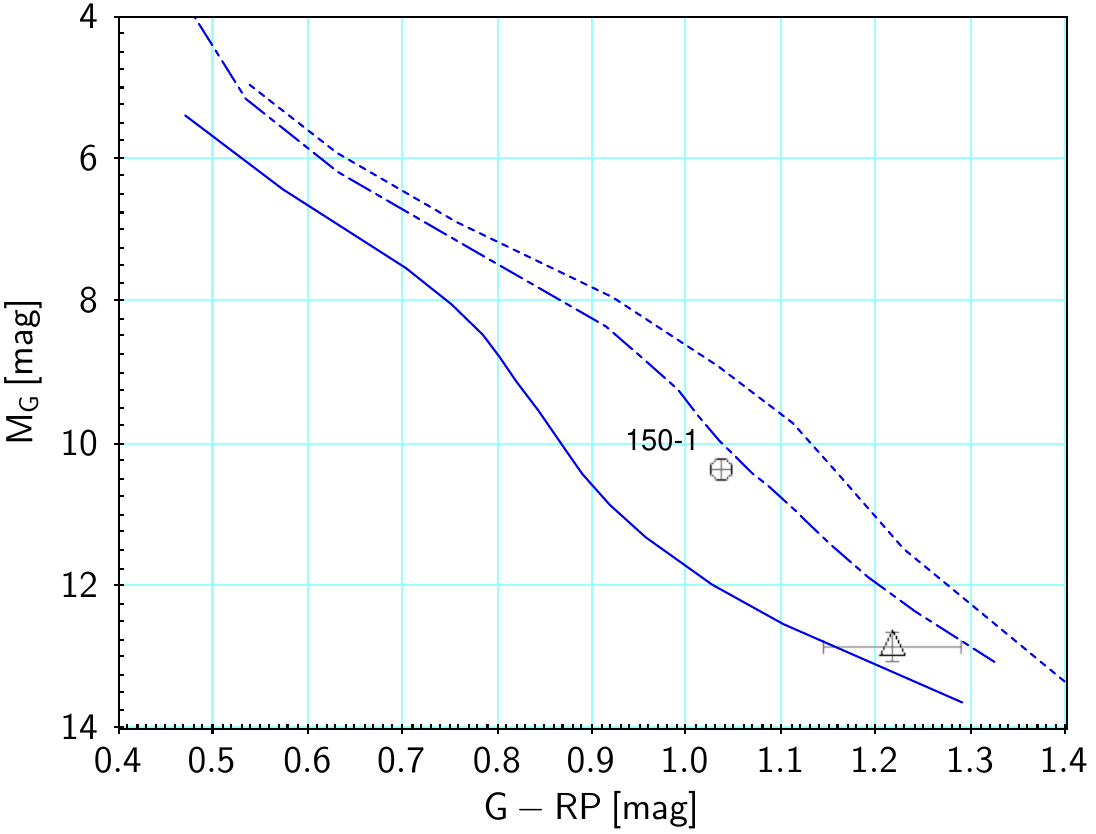}
    \end{subfigure}
    \begin{subfigure}{.5\textwidth}
     \centering
      \includegraphics[width=0.91\linewidth]{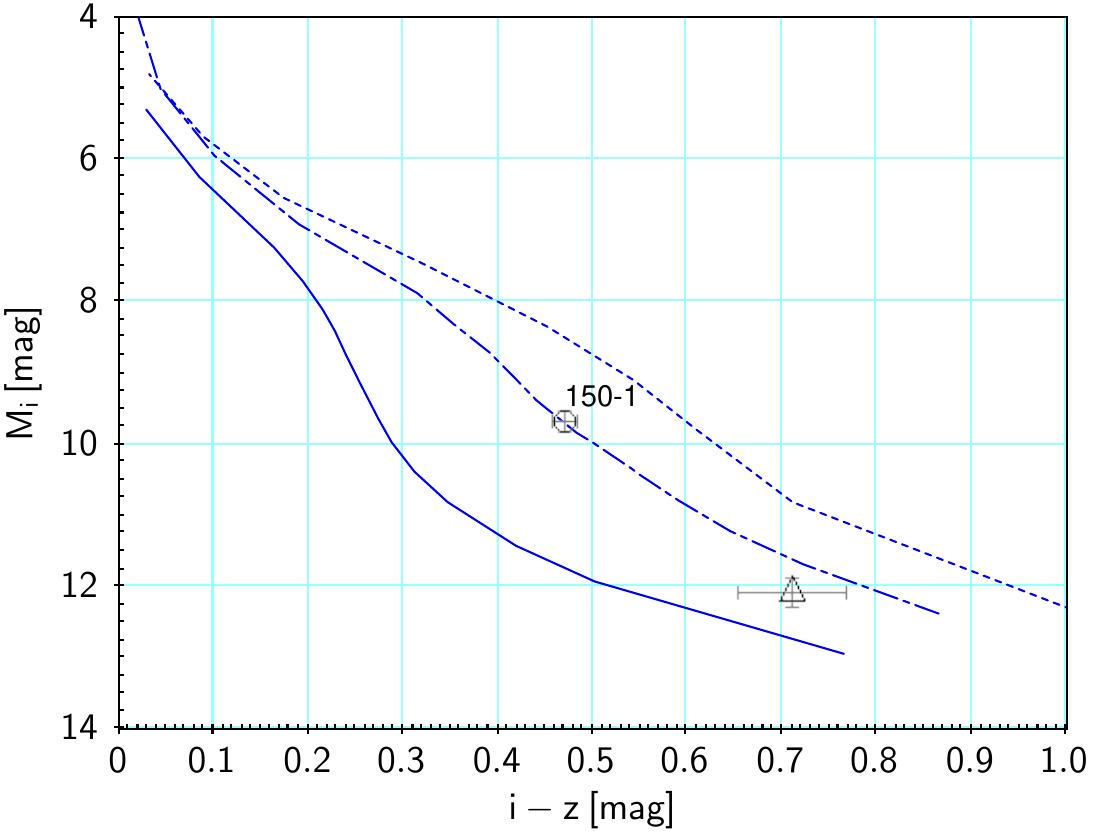}
    \end{subfigure}
    
  \vspace{2mm}  
     \begin{subfigure}{.5\textwidth}
    \centering
    \includegraphics[width=0.91\linewidth]{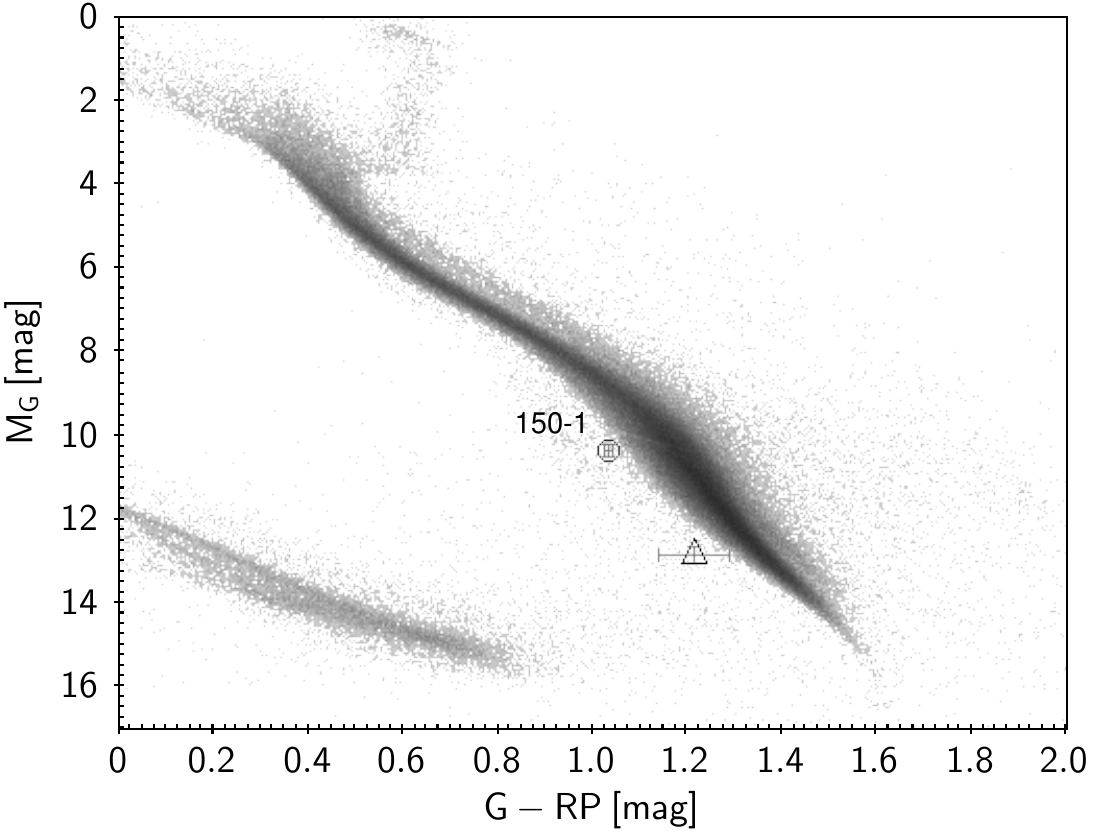}
    \end{subfigure}
    \begin{subfigure}{.5\textwidth}
     \centering
      \includegraphics[width=0.91\linewidth]{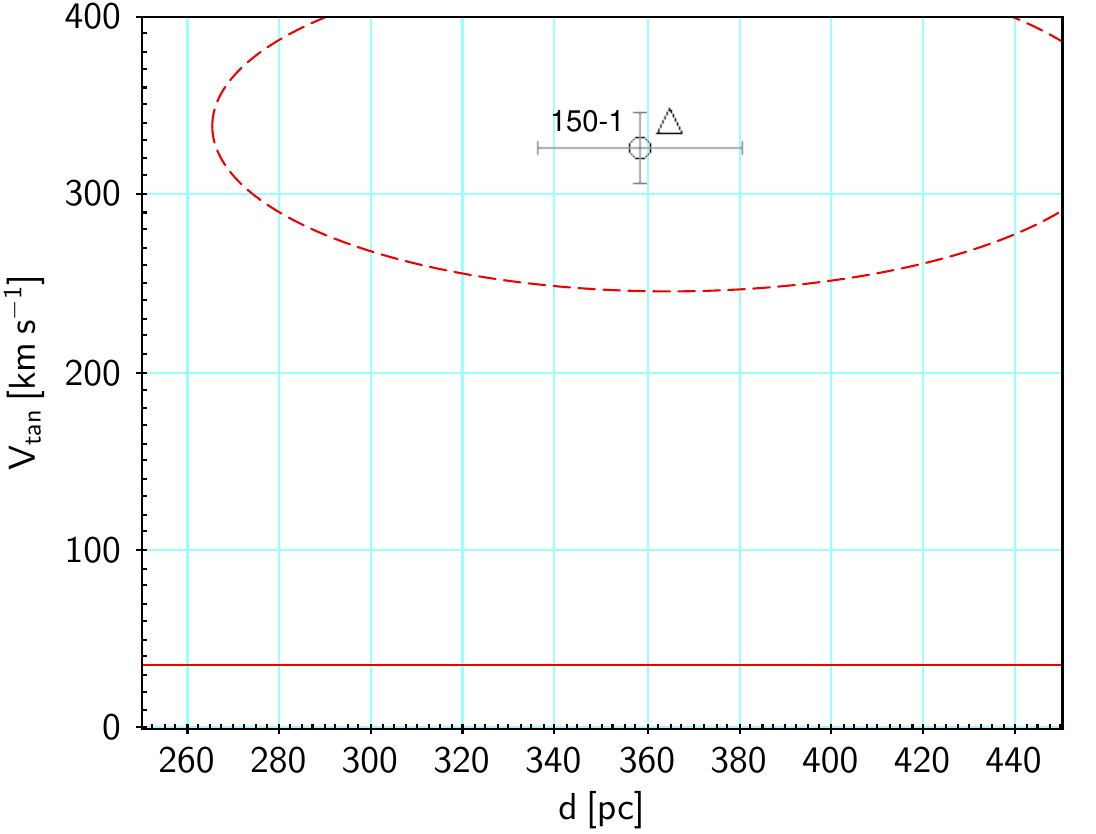}
    \end{subfigure}
  \vspace{2mm}
  \caption[PMD, CMDs, H-R diagram and tangential velocity vs distance diagram for the target Id 150 and its candidate companion.]{PMD (top left panel), CMDs (top right and mid panels), H-R diagram (bottom left panel), and tangential velocity--distance diagram (bottom right) for the target Id 150 and its candidate companion. The black triangle represents the source under study, and the numbered black circle represents the companion candidate, both with their corresponding error bars. Grey dots in the PMD represent field stars, and in the H-R diagram they are \textit{Gaia} DR2 sources with parallaxes $>$\,10\,mas used as a reference. The blue solid, dashed, and dotted lines stand for [M/H]\,=\,$-$2.0, $-$0.5, and 0.0 \textit{BT-Settl} isochrones in the CMDs, respectively. The red solid line in the tangential velocity plot marks the value $V_{tan}=$\,36\,km\,s$^{-\text{1}}$ which is the mean value for field stars \citep{zhang18a}, and the red dashed ellipse around Id 150 indicates its values of $V_{tan}\pm \text{3}\sigma$ and $d \pm \text{3}\sigma$.}
\label{fig:plot_CMD_PM_150}
\end{figure}
 
 \item [$\bullet$] \textbf{Id 150}: This source has a sdM5.5 spectral type and is located at about 365\,pc. It has just one companion candidate separated by 3.8\,arcsec, which is visually confirmed as a common proper motion pair in \textit{Aladin}. The positions in the PMD and CMDs support their companionship. Both stars are under the main sequence in the H-R diagram, which supports their metal-poor nature. The tangential velocities of both sources are very similar and well above the mean value of field sources of 36\,km\,s$^{-\text{1}}$ \citep{zhang18a}, in agreement with an old population member. The companion is in the LSPM catalogue \citep{lepine05d} with the identifier J1355$+$0651\@. We propose that this system is a bound pair.
 
 \item [$\bullet$] \textbf{Id 190}: This is a sdL3.0 source at $\sim$61\,pc with one candidate companion at 8.9\,arcmin. We do not have enough information to discard or confirm the candidate. The PMD shows very similar proper motions and the positions in one of the CMDs agree. The tangential velocity is above the mean value of field sources, outside the error limits of the value of the subdwarf, but with very close values. The current information is not conclusive but we decided to include this possible pair in a subsequent analysis.
 
 \item [$\bullet$] \textbf{Id 213}: This is a sdL0.0 located at about 105\,pc with a single candidate companion at 36.8\,arcmin. The large uncertainties in the input parameters of the subdwarf provide a unique candidate companion with very different proper motions. Its position on the CMDs and H-R diagram suggest a likely pair, but its tangential velocity diagram does not, and the kinematic analysis using its radial velocity suggests the companion as a thin disk star. We reject this system as a probable pair.
 
 \item [$\bullet$] \textbf{Id 215}: This subdwarf is a source with sdL0.0 spectral type located at 114\,pc with 12 candidates separated by 22.4 to 46.3\,arcmin. The large error bars in the proper motions of the subdwarf lead to a large number of spurious candidates. As in the case of Id 11, we refute all candidates with proper motions above 1$\sigma$ of the proper motion of the subdwarf. Only candidate numbers 2, 8, and 9 remain. There is no available photometry for the subdwarf in \textit{Gaia}, and these three candidates are aligned in the rest of the CMDs, except number 2 which is not aligned in the optical CMD. Based on the H-R diagram, we infer that the metallicity classes of candidates 2 and 9 probably differ from the subdwarf. Using the radial velocity of candidate 8, we place it in the thin disk. The tangential velocity of the three candidates is lower than the mean velocity of field stars. Consequently, we reject all of them.

\end{itemize}

\section{Spectroscopic follow-up}
\label{sec:spectroscopic_followup}

\subsection{WHT ACAM optical spectroscopy}
\label{sub:WHT_ACAM_optical_spectroscopy}

We collected a low-resolution optical spectrum of the candidate companion Id 150-1 (i.e.\ the primary) with the auxiliary-port camera (ACAM) mounted on the Cassegrain focus of the 4.2-m William Herschel telescope at Roque de los Muchachos observatory in La Palma (Table~\ref{tab:log_obs_spectro}). We carried out the observations in service mode on the night of 1 February 2019\@. The night of 1 February 2019 was clear with variable seeing between 1.0 and 1.6\,arcsec after UT\,$\sim$\,3\,h when the object was observed. 

ACAM is permanently mounted on the telescope as an optical imager and spectrograph. We used the VPH grism with a slit of 1.0\,arcsec to cover the 350--940\,nm wavelength range at a resolution of 430 and 570 at 565\,nm and 750\,nm, respectively. We did not use a second-order blocking filter, resulting in light contamination beyond 660\,nm{}. We use single on-source integrations of 900\,s (Table~\ref{tab:log_obs_spectro}).

\begin{figure}
  \centering
  \includegraphics[width=0.6\linewidth, angle=0]{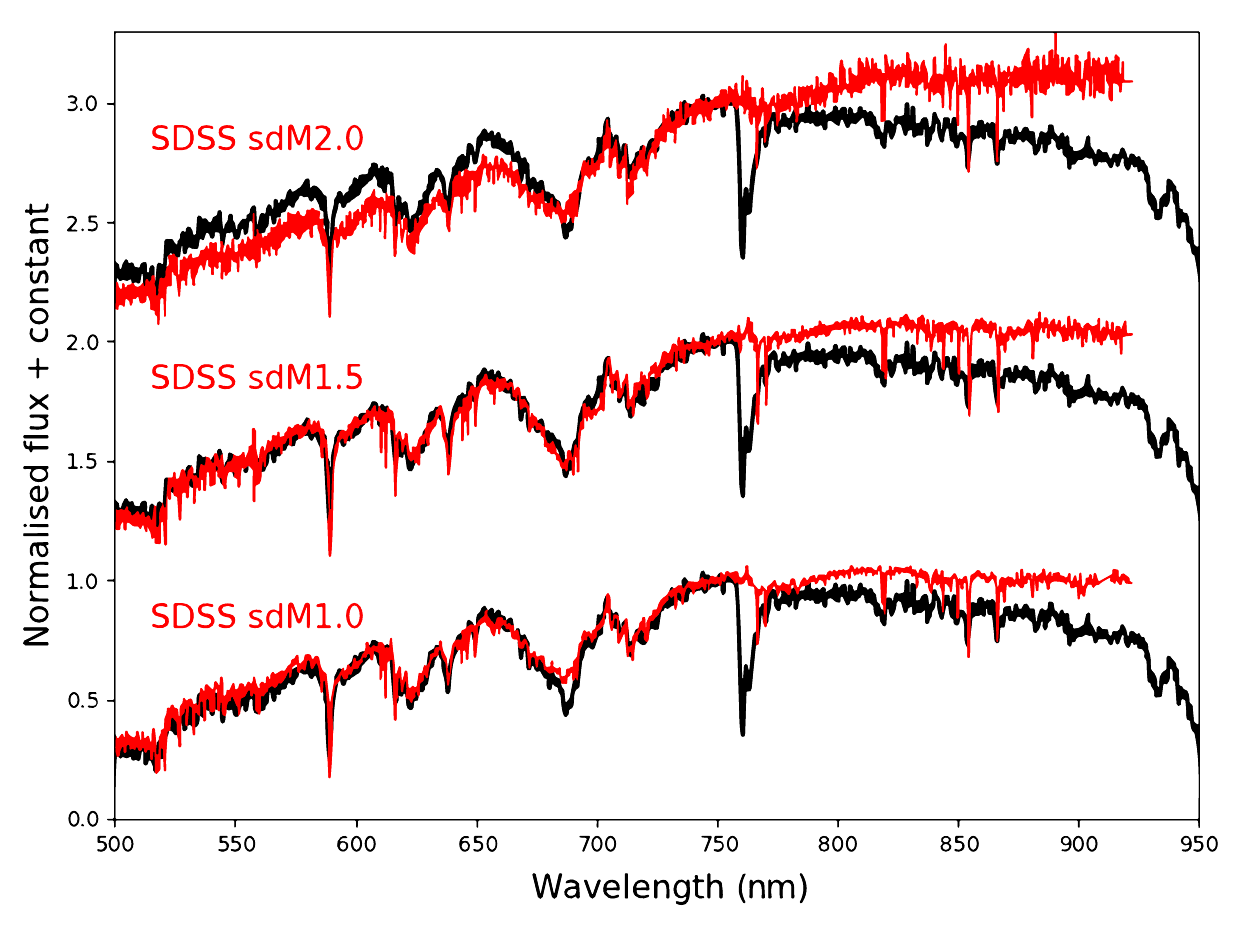}
  \includegraphics[width=0.6\linewidth, angle=0]{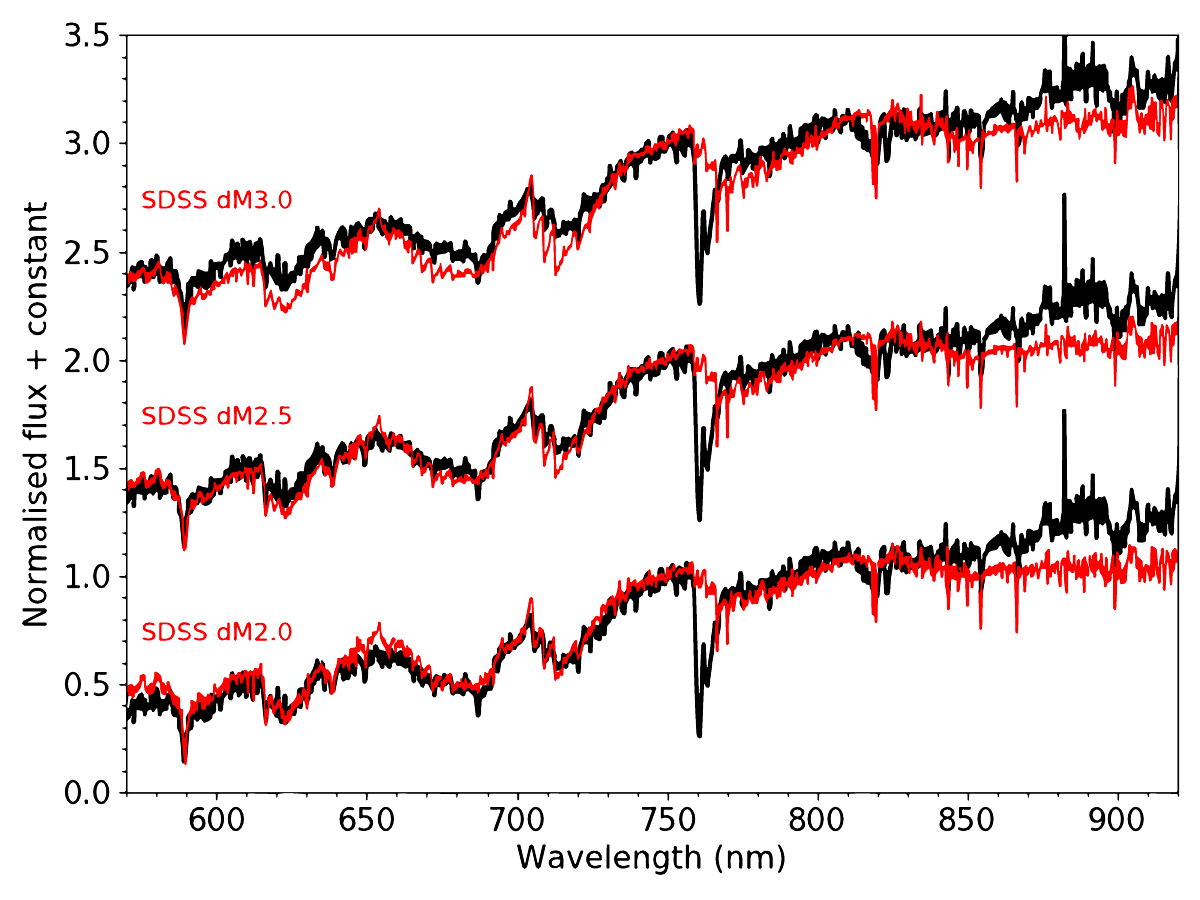}
  \includegraphics[width=0.6\linewidth, angle=0]{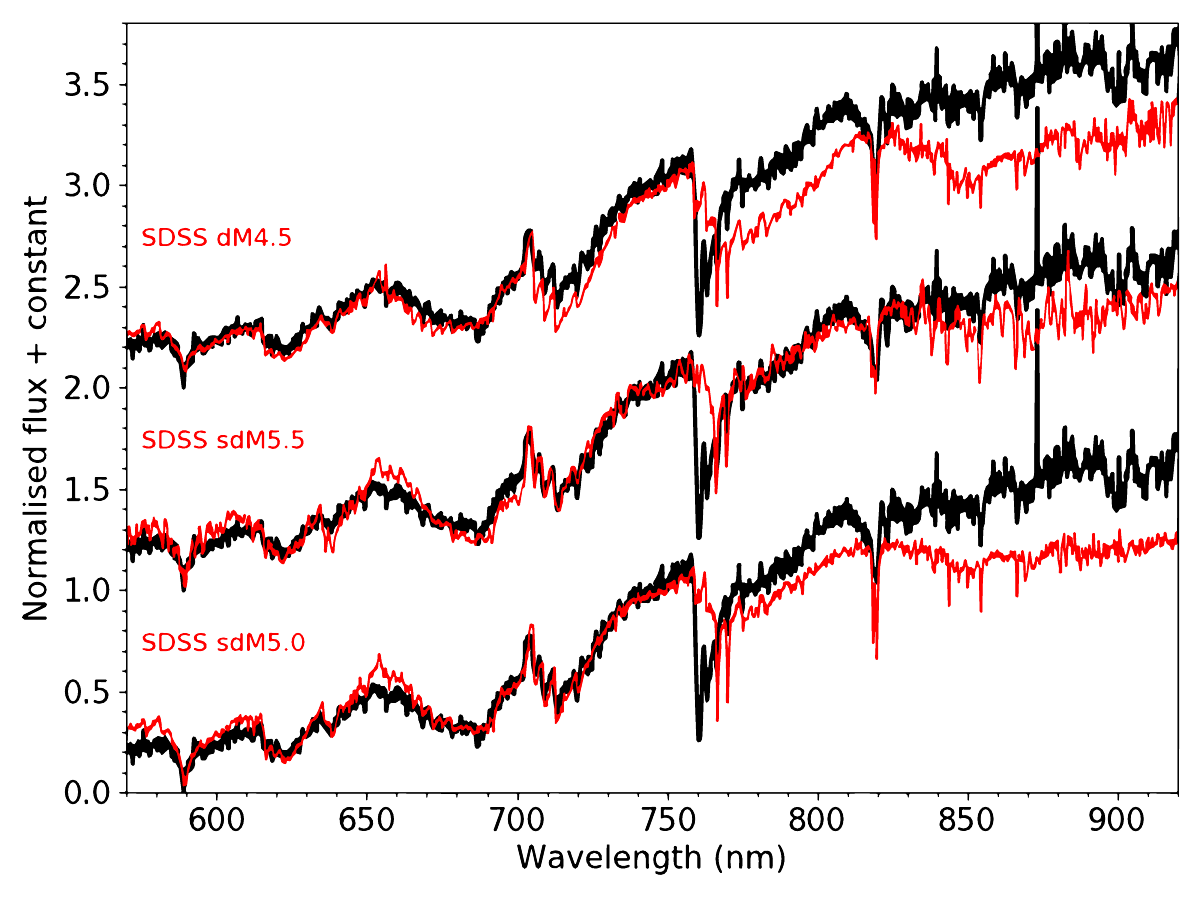}
   \caption[WHT/ACAM optical spectrum of the companion of subdwarf Id 150, NOT/ALFOSC spectra of Id 74 and Id 107 compared to Sloan solar-type and subdwarf templates.]{WHT/ACAM optical spectrum (black) of the companion of subdwarf Id 150 (top) and NOT/ALFOSC spectra (black) of Id 74 (middle) and Id 107 (bottom) compared to Sloan solar-type and subdwarf templates (red).}
   \label{fig:plot_spectra_WHT_ACAM}
\end{figure}

We reduced the data in a standard manner under \texttt{IRAF} \citep{tody86,tody93}, and subtracted a median-combined bias frame to the target's frame and later divided by the median flat-field. We extracted the one-dimensional spectrum in an optimal way by selecting the size of the aperture and the background regions on the left and right side of the target. We calibrated the spectra in wavelength with CuNe arc lamps taken just after the target, yielding rms better than 0.3\,nm. Finally, we calibrated the targets in flux with a spectro-photometric standard observed with the same setup during the night \citep[Ross640; DAZ5.5][]{greenstein67,monet03}. However, the flux calibration is uncertain beyond 660\,nm because the second-order blocking filter was not in place. The spectrum is displayed in Fig.~\ref{fig:plot_spectra_WHT_ACAM} along with Sloan metal-poor templates \citep{savcheva14}. From the spectral fits, we derive a spectral type of sdM1.5$\pm$0.5\@.

\subsection{NOT ALFOSC optical spectroscopy}
\label{sub:NOT_ALFOSC_optical_spectroscopy}

We obtained low-resolution optical spectra of two additional candidate companions (Id 73-1 and Id 107-1) with the Alhambra Faint Object Spectrograph and Camera (ALFOSC) on the 2.5-m Nordic Optical telescope (NOT) at Roque de los Muchachos observatory in La Palma (Table~\ref{tab:log_obs_spectro}). We collected the spectra as part of a service programme (SST2020-506; PI Cort\'es-Contreras) on the night of 12 December 2020, which presented cirrus and seeing between 1 and 1.5 arcsec during the observations between UT\,=\,5h--6h.

\begin{table}[h]
 \centering
 \caption[Logs of spectroscopic observations of three companion candidates]{Logs of spectroscopic observations of three companion candidates}
 \footnotesize
 \scalebox{1}[1]{
 \begin{tabular}{l@{\hspace{2mm}}c@{\hspace{2mm}}c@{\hspace{2mm}}c@{\hspace{2mm}}c@{\hspace{2mm}}c@{\hspace{2mm}}c@{\hspace{2mm}}c@{\hspace{2mm}}c@{\hspace{2mm}}c}
\noalign{\hrule height 1pt}
 \noalign{\smallskip}
Name  & $\alpha$ (J2000) & $\delta$ (J2000) &  Instr. & $i'$  & Date  & UT-MID & Airm & Exp.T & Spectral type\\
 & (hh:mm:ss.ss)   & (dd:mm:ss.s)  &  & (mag)   & (dd/mm/yyyy) & (hh:mm:ss.ss)  & & (s)  & \\
 \noalign{\smallskip}
 \hline 
 \noalign{\smallskip}
Id 150-1  & 13:55:28.38 & $+$06:51:18.5 & ACAM & 17.46  & 02/02/2019 & 06:15:03.48 & 1.08 &  900   & sdM1.5$\pm$0.5 \\
Id 73-1  & 10:47:13.20 & $-$01:22:21.4 & ALFOSC & 17.60  & 13/12/2020 & 05:15:07.02 & 1.22 &  2100   & dM2.0--dM2.5 \\
Id 107-1  & 12:40:35.18 & $-$00:13:28.7 & ALFOSC & 16.71  & 13/12/2020 & 06:00:48.00 & 1.41 &  900   & dM/sdM5.0$\pm$0.5 \\
 \noalign{\smallskip}
\noalign{\hrule height 1pt}
 \label{tab:log_obs_spectro}
 \end{tabular}
 }
\end{table}

ALFOSC is equipped with a 2048\,$\times$\,2048 CCD231-42-g-F61 back-illuminated, deep-depletion detector sensitive to optical wavelengths. The pixel size is 0.2138 arcsec and the field of view is 6.4\,$\times$\,6.4 arcmin across. We employed the VPH grism \#20 and a slit of 1.3 arcsec covering the 565--1015\,nm at a resolution of about 500\@. We used an on-source integration of 2100\,s and 900\,s for Id 73-1 and Id 107-1, respectively (Table~\ref{tab:log_obs_spectro}). A spectro-photometric standard star, Feige 66 \citep[sdB1; $V$\,=\,10.59 mag;][]{berger63} was observed to characterise the response of the detector.

We reduced the data with \texttt{IRAF} following standard procedures. We combined the bias and flat frames before subtracting each science spectrum by the median-combined bias and subtracting the normalised flat field. We optimally extracted the 2D spectra choosing the aperture and background regions interactively. We calibrated the spectra in wavelength with ThAr lamps taken at the end of the night, and corrected the science spectra with the response function derived from Feige 66\@. The NOT ALFOSC optical spectra, normalised at 750\,nm, are displayed in Fig.~\ref{fig:plot_spectra_WHT_ACAM} along with Sloan spectral M-type templates at solar and subsolar metallicities \citep{bochanski07a,savcheva14}. From the direct comparison with Sloan spectral templates, we classify Id 73-1 as a solar-metallicity M2.0--M2.5 dwarf. For Id 107-1, we find that this source shows features intermediate between a solar-metallicity and subdwarf with a mid-M spectral type, and we adopt a classification of dM/sdM5.0$\pm$0.5\@.

\section{Results of the search}
\label{sec:results_search}

Following the search and analysis performed in previous sections,we identified six M and L subdwarfs with potential candidate companions, some of them with high probabilities.
To further characterise these stars and their candidates companions, we built their spectral energy distribution (SED) using the VO SED Analyser \citep[VOSA;][]{bayo08}, which provides estimates of stellar physical parameters from the SED fitting to different collections of theoretical models. In this work, we used the \textit{BT-Settl} theoretical models \citep{baraffe15}, and limited the range of gravities to $4\leq log(g)\leq 6$, which are the normal values for old dwarfs \citep{cifuentes20}. We prove that metallicity does not have a significant impact on the determination of effective temperatures in VOSA, because it takes values in the range of $\pm$100\,K when metallicity varies from -3 to 0\,dex. Finally, we took the effective temperature proposed as the best one by VOSA with an error of $\pm$100\,K ---instead of $\pm$50\,K which would correspond to the grid error established by VOSA--- to provide a more realistic margin. The obtained effective temperatures of the candidate companions will give us information on their spectral types. 

In the calculation of the masses, we used different methods to estimate them. In the case of sdL sources, we took the values of table\,3 and fig.\,5 from \cite{zhang18a} in the range of 0.08--0.09\,M$_\odot$. For sdM sources, we estimate a range of masses with the isochrone model\footnote{BCAH97 isochrone models, \url{http://perso.ens-lyon.fr/isabelle.baraffe/BCAH97_models}} from \citet{baraffe97}, using as input the effective temperature obtained by VOSA and the J magnitude from the UKIDSS catalogue (or 2MASS if not available in  UKIDSS ). For solar metallicity sources, we used the isochrone model\footnote{BHAC15 isochrone models, \url{http://perso.ens-lyon.fr/isabelle.baraffe/BHAC15dir/}} from \citet{baraffe15}.

In the estimation of the spectral type, for the case of low-metallicity dwarfs, we checked table\,2 from \citet{lodieu19b}, and fig.\,4 from \citet{zhang18a}. For solar metallicities, the spectral type can be estimated following \citet{reid05b}.

Additionally, we intensively browsed the literature for any additional relevant information related to the companion candidates (spectral type, metallicity, age), including references to any known physically bound or unrelated companions.

\begin{itemize}
    \item [$\bullet$] \textbf{Id 11}: This subdwarf has a spectral type of sdL0.0 with one remaining candidate companion (Id 11-17). We obtained their SEDs with VOSA, from which we derived effective temperatures of 2600 $\pm$100\,K and 2900$\pm$100\,K for Id 11 and Id 11-17, respectively. Using \citet{lodieu19b} and \citet{zhang18a}, we estimated its spectral type as sdM8.0$\pm$0.5. We also calculated its binding energy, which is very low (about ten times lower than the accepted minimum energy to be considered a bound pair). We considered the possibility of both objects being part of a multiple system, but the small \texttt{RUWE} value of the candidate (1.11) points towards a single star and therefore a low chance of such a multiple system having a larger binding energy.
    
    \item [$\bullet$] \textbf{Id 25}: This subdwarf and its candidate companion are extreme subdwarfs, and are reported as a known pair by \citet{zhang19g}. Their spectral types are esdL0.0 and esdM1.0 respectively, which we support from the temperatures derived by VOSA (2800$\pm$100\,K and 3700$\pm$100\,K respectively). All the obtained data are in line with the reported ones by \citet{zhang19g}, with small differences due to the fact that Zhang estimated the properties with a lower metallicity than that used here (he used [M/H]\,=\,-1.4\,dex). The binding energy of the system is consistent with a physically bound pair.
    
    \item [$\bullet$] \textbf{Id 73}: This object in the sample is a solar metallicity dwarf with spectral type dM4.5, and an effective temperature of 3100$\pm$100\,K derived from its SED. The possible companion with Id 73-1 is a solar metallicity dwarf with a temperature of 3500$\pm$100\,K as NOT ALFOSC optical spectrum suggests, corresponding to a spectral type of dM2.0--2.5. With this new spectral type we recalculated the distance of the companion candidate through spectroscopic methods \citep{zhang13} and obtained 607.2$\pm$98.2\,pc, a value that is closer to the distance of Id 73 (instead of 860.9$\pm$159.7\,pc, the value obtained from the parallax provided by \textit{Gaia} DR2, as shown in Table~\ref{tab:astrometric_subdwarfs_companions}, available in \textit{VizieR}). This pair has a low binding energy, about 70\% of the minimum to be a bound pair. The \texttt{RUWE} value of the brightest source of the pair is low (0.95), indicating that it is unlikely to be part of a multiple system. The analysis of the spectrum at this resolution shows all lines to be single. At this stage, the pair seems to be bound.
    
    \item [$\bullet$] \textbf{Id 107}: The subdwarf has a spectral type of sdL0.0, and we suggest that its candidate companion is also metal poor in light of its position on the H-R diagram. There are not enough photometric points to build the SED of the subdwarf, but we estimate its temperature to about 2600$\pm$100\,K from its spectral type using fig.~4 of \citet{zhang18a}. VOSA provides the value of 3300$\pm$100\,K for the companion candidate, for which we estimate a spectral type of dM/sdM5.0$\pm0.5$, in agreement with the spectral classification of the NOT ALFOSC spectrum. The optical spectrum of the wide companion suggests that this system might have an intermediate metallicity between solar and $-$0.5 dex (typical of subdwarfs), and its lines do not appear deblended. In light of the possibility that this companion source is a dwarf, we calculated its range of mass using isochrone models from \citet{baraffe97} and \citet{baraffe15}. Here, the value of the binding energy is also low, three times lower than the minimum, and the companion candidate has a \texttt{RUWE} of 0.98. We conclude that the pair seems to be a bound system.
    
    \item [$\bullet$] \textbf{Id 150}: This subdwarf has a spectral type of sdM5.5, and the companion candidate is also a subdwarf, as suggested from its H-R diagram. The effective temperatures derived by VOSA (see Fig.~\ref{fig:vosa150}) are 3000$\pm$100\,K for the subdwarf and 3600$\pm$100\,K for the companion candidate. The temperature of the companion is in agreement with the spectral type of sdM2.0$\pm$0.5 from the performed spectroscopic follow-up (Sect.~\ref{sec:spectroscopic_followup}) based on table~2 of \citet{lodieu19c}. The two sources are very close (about 1360\,au), and we visually confirm them to be a co-moving pair in \textit{Aladin} as shown in Fig.~\ref{fig:visual150}. Their binding energy is more than 40 times the minimum energy of gravitationally bound pairs, reinforcing the system as a true pair. We conclude that this pair is the most secure in the sample.
        
    \begin{figure}[h]
    \centering
    \includegraphics[width=0.6\textwidth]{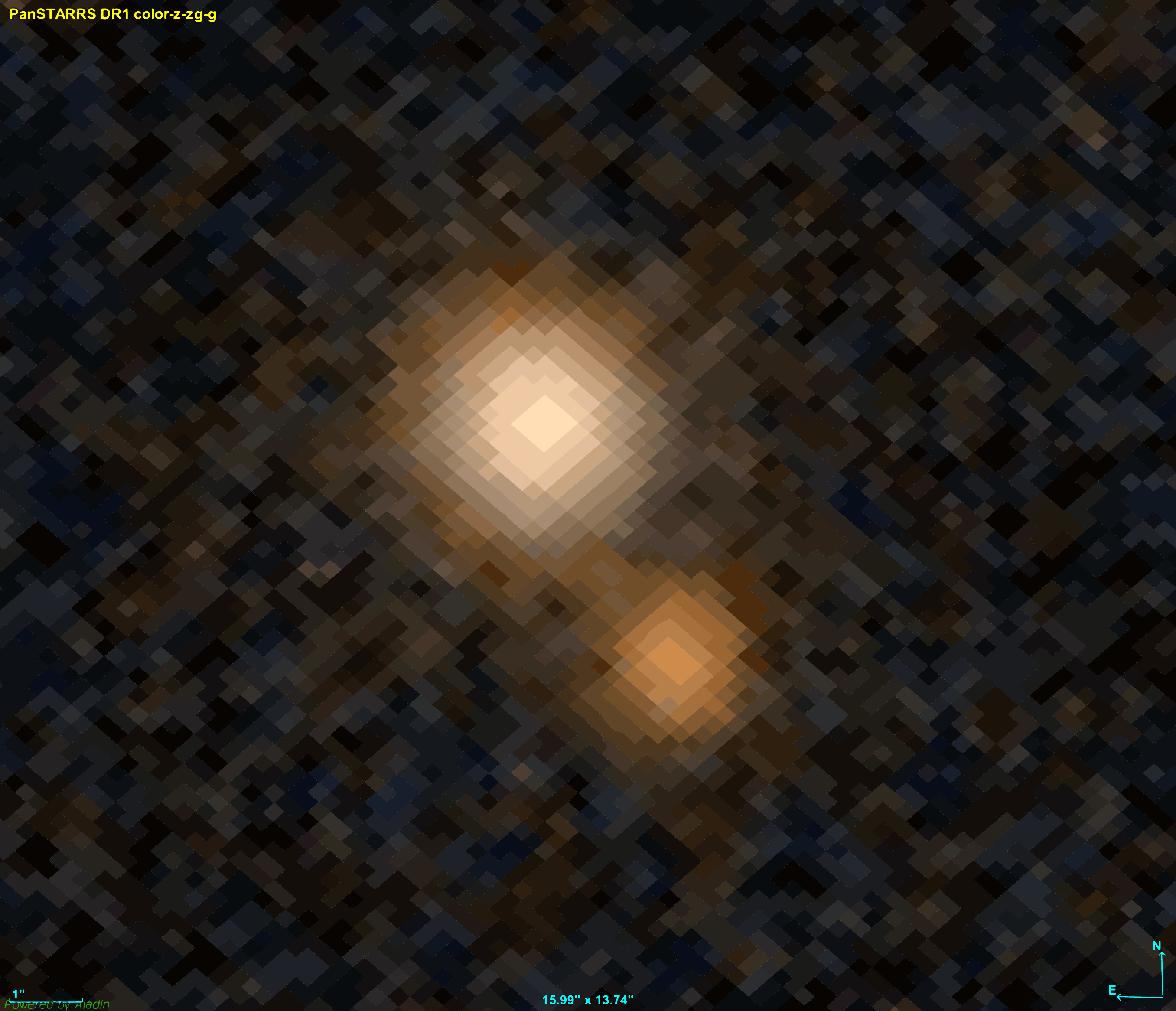}
    \caption[Visual image of pair Id 150.]{Visual image of pair Id 150 showing the primary star at the bottom of the figure, and the secondary star at the top, separated about 3.8\,arcsec. Image from PanSTARRS DR1 color z-zg-g \citep{chambers16a}. }
    \label{fig:visual150}
    \end{figure}
   
    \item [$\bullet$] \textbf{Id 190}: The subdwarf Id 190 has a spectral type of sdL3.0. The effective temperatures provided by VOSA for the subdwarf and the companion candidate are 2200$\pm$100\,K (consistent with the spectral type) and 3000$\pm$100\,K, respectively. As we did for other subdwarfs, we estimated the spectral type of the companion to sdM7.0$\pm$0.5 from its temperature \citep{lodieu19b}. Their masses and distances provide a value for the binding energy of 88\% of the minimum; with a low \texttt{RUWE} of 1.17 for the companion candidate, we conclude that this pair has a low chance of being a real system, and therefore, that this is the most doubtful pair in the sample.
    
    \end{itemize}

  \begin{figure}
  \centering
  \includegraphics[width=0.6\textwidth]{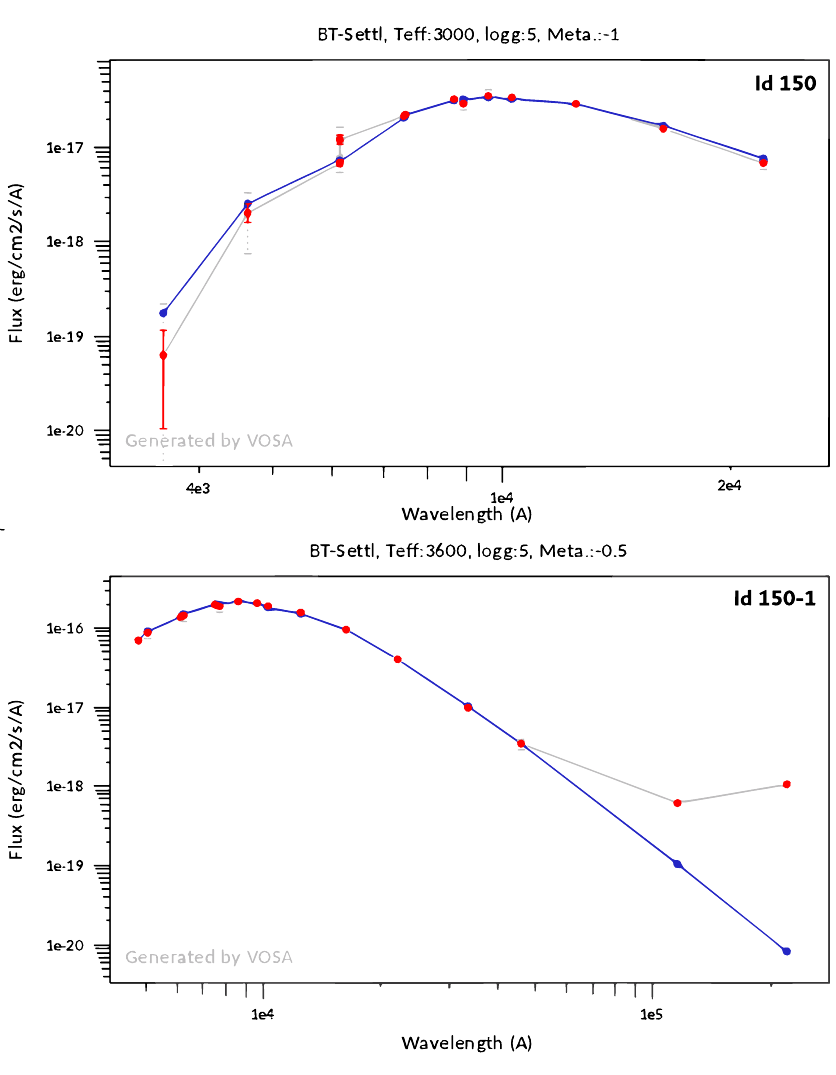}
   \caption[VOSA diagrams for the source Id 150 and its candidate companion.]{VOSA diagrams for the source Id 150 (top) and its candidate companion (bottom). The shown elements are: Observed flux + 3-sigma points (pale grey line + error bars), fitted flux (red dots + error bars), no fitted points (orange dots + error bars), and \textit{BT-Settl} model (blue line).}
   \label{fig:vosa150}
  \end{figure}

From the previous analysis, we propose one confirmed binary system (Id 150), four likely pairs (Id 11, Id 25, Id 73, and Id 107), and a doubtful pair (Id 190). Their identifiers, names, spectral types, effective temperatures, masses, proper motions, distances, physical and projected angular separations, and binding energies are provided in Table~\ref{tab:proposed_pairs}. 

As mentioned above, the most commonly accepted value of the minimum gravitational binding energy is 10$^{\text{33}}$\,J\@. In this case, only two pairs with the closest separations, Id 25/25-1 (0.08\,pc) and Id 150/150-1 (0.007\,pc), fulfil this criterion. The other four systems have much lower gravitational binding energies. However, this study does not account for hidden mass in the systems in the form of spectroscopic binaries or fainter sources not detected with \textit{Gaia} DR2. In terms of projected physical separations, the widest pair is separated by 2.64\,pc (Id 73/73-1). The classical maximum separation between components in wide systems rarely exceeds 0.1\,pc, driven by the dynamic processes of the stars formation and evolution \citep{tolbert64,kraicheva85,abt88,weinberg88,close90,latham91,wasserman91,garnavich93,allen98,caballero09}, and strongly depends on their mass (i.e. spectral type), age, and kinematics \citep{duquennoy91,jensen93,patience02,zapatero04a,kraus09}. 
There are newer studies that increase this maximum separation up to 1\,pc \citep{jiang10,caballero10}, or even to 1--8\,pc \citep{shaya11,kirkpatrick16,gonzalezpayo21}. 
At these separations, the pairs are less likely bound for extended lifetimes \citep{retterer82,weinberg87,dhital10}, and the sources of the sample are very old. This is not the case for the proposed subdwarfs, whose separation from the companions are lower than 0.7\,pc. In the case of the solar-type system Id 73, although it has a separation above the classical limits, the component sources are field dwarfs with a mean age of less than the metal-poor population.

\begin{table}[H]
 \centering
 \caption[Physical parameters of the proposed systems.]{Physical parameters of the proposed systems.}
 \footnotesize
 \scalebox{0.8}[0.8]{
 \begin{tabular}{@{\hspace{0mm}}l@{\hspace{1mm}}l@{\hspace{1mm}}c@{\hspace{1mm}}c@{\hspace{0mm}}c@{\hspace{-1mm}}c@{\hspace{1mm}}c@{\hspace{1mm}}c@{\hspace{1mm}}c@{\hspace{1mm}}c@{\hspace{1mm}}c@{\hspace{0mm}}c@{\hspace{0mm}}c@{\hspace{0mm}}}
\noalign{\hrule height 1pt}
\noalign{\smallskip}
 Id & Name & SpT & [M/H]$^{\text{(a)}}$ & T$_{eff}$ & M & \textit{J} & $\mu_{\alpha}\cos\delta$ & $\mu_\delta$ & \textit{d} & \multicolumn{2}{c}{\textit{s}} & W\\
    &      &     & (dex) &     $\pm$100 (K)   & (M$_\odot$) & (mag) & (mas\,a$^{-\text{1}}$) & (mas\,a$^{-\text{1}}$) & (pc) & (pc) & (arcmin)  & (10$^{\text{33}}$\,J)  \\
\noalign{\smallskip}
\hline
\noalign{\smallskip}
\multicolumn{13}{c}{\textit{Confirmed pairs}} \\
\noalign{\smallskip}
\hline
\noalign{\smallskip}
150   & Gaia DR2 3720832015084722304 & sdM5.5$\pm$0.5 & -0.5 & 3000 & 0.092--0.125  & 17.55$\pm$0.04 & 4.4$\pm$2.5 & -195.9$\pm$2.0 & 364.6$\pm$33.1 &  &  &  \\
150-1 & Gaia DR2 3720832010789680000 & sdM1.5$\pm$0.5 & -0.5 & 3600 & 0.149--0.279  & 15.71$\pm$0.01 & 2.5$\pm$0.4 & -191.7$\pm$0.3 & 358.2$\pm$22.2 &  0.007 & 3.8$^{\text{(b)}}$ & 42.6 \\
\noalign{\smallskip}
\hline
\noalign{\smallskip}
\multicolumn{13}{c}{\textit{Likely pairs}} \\
\noalign{\smallskip}
\hline
\noalign{\smallskip}
11    & ULAS J011824.89$+$034130.4     & sdL0.0$\pm$0.5  & -0.5 & 2600 & 0.080--0.090  & 18.18$\pm$0.05 & 19.9$\pm$29.0 & -33.8$\pm$27.0  & 176.2$\pm$13.7 &  &  &  \\
11-17  & Gaia DR2 2562996857437494528 & sdM8.0$\pm$0.5  & -0.5 & 2900 & 0.089--0.135  & 15.72$\pm$0.06 & 41.2$\pm$1.2 & -45.1$\pm$0.6 & 177.1$\pm$17.7 & 0.69  & 13.4 & 0.10 \\
\noalign{\smallskip}
\hline
\noalign{\smallskip}
25    & 2MASS J04524567--3608412   & esdL0.0$\pm$0.5  & -1.0 & 2800 & 0.080--0.090  & 16.26$\pm$0.10 & 147.5$\pm$0.8 & -168.1$\pm$1.0  & 140.2$\pm$10.0 &  &  &  \\
25-1  & Gaia DR2 4818823636756117504 & esdM1.0$\pm$0.5  & -1.0 & 3700 &  0.178--0.275 & 13.77$\pm$0.03 & 148.5$\pm$0.1 & -168.7$\pm$0.1 & 137.3$\pm$0.7 & 0.08  & 1.92 &  2.65 \\
\noalign{\smallskip}
\hline
\noalign{\smallskip}
73    & SDSS J10465793--0137464       & dM4.5$\pm$0.5  & 0.0 & 3100 & 0.119--0.413 & 16.54$\pm$0.01 & -25.2$\pm$1.0 & -9.8$\pm$0.8  & 572.5$\pm$175.6 &  &  &  \\
73-1  & Gaia DR2 3802720750608315392 & dM2.0--2.5  & 0.0 & 3500 & 0.289--0.524 & 15.88$\pm$0.01 & -22.4$\pm$0.4 & -10.1$\pm$0.3 & 607.2$\pm$98.2 &  2.64 & 15.9 &  0.70 \\
\noalign{\smallskip}
\hline
\noalign{\smallskip}
107   & ULAS J124104.75--000531.4     & sdL0.0$\pm$0.5 & -0.5 & 2600 &  0.080--0.090  & 18.46$\pm$0.10 & -42.9$\pm$8.0 & -26.2$\pm$6.0 & 196.5$\pm$15.3 &  &  & \\
107-1 & Gaia DR2 3695978963488707072 & dM/sdM5.0$\pm$0.5 & -0.5--0.0 & 3300 & 0.107--0.289 & 14.73$\pm$0.04 & -36.9$\pm$0.3 & -20.7$\pm$0.1 & 177.8$\pm$3.7 &  0.62 & 10.9 & 0.36 \\
\noalign{\smallskip}
\hline
\noalign{\smallskip}
\multicolumn{13}{c}{\textit{Doubtful pairs}} \\
\noalign{\smallskip}
\hline
\noalign{\smallskip}
190   & ULAS J154638.34-011213.0 & sdL3.0$\pm$0.5 & -0.5 & 2200 & 0.080--0.090 & 17.51$\pm$0.04 & -49.9$\pm$8.6 &  -107.1$\pm$7.6 & 61.1$\pm$6.0 &  &  &  \\
190-1 & Gaia DR2 4404197733205321216 & sdM7.0$\pm$0.5 & -0.5 & 3000 & 0.092--0.183 & 12.85$\pm$0.02 & -52.8$\pm$0.2 & -122.7$\pm$0.1 & 69.8$\pm$0.4 & 0.16 & 8.9 & 0.88 \\
\noalign{\smallskip} 
\noalign{\hrule height 1pt}
 \end{tabular}
}
\label{tab:proposed_pairs}
\begin{justify}
\scriptsize{\textbf{\textit{Notes. }}}
\scriptsize{$^{\text{(a)}}$ Not calculated but derived from the metallicity class definition. Uncertainty of 0.5 dex. $^{\text{(b)}}$ This value in arcsec.}
\end{justify}
\end{table}

\section{Discussion on multiplicity}
\label{sec:discussion_multiplicity}

\subsection{General considerations about multiplicity}
\label{sub:general_considerations_multiplicity}

The binary frequency decreases with decreasing spectral type \citep{fontanive18}. Over 70\% of massive B and A-type stars are part of binary or hierarchical systems \citep{kouwenhoven07,peter12}. The incidence of multiplicity is about 50--60\% for solar-type stars \citep{duquennoy91,raghavan10}, and about 30--40\% for M dwarfs \citep{fischer92,delfosse04,janson12}. Later surveys found that the multiplicity fraction of M-dwarfs drops to 23.5--42\% \citep{ward15,cortescontreras17b}. Thus, the overall trend is that the multiplicity rate of main sequence stars decreases with mass \citep{jao09a}.

The total multiplicity frequency of population II stars with spectral types from F6 to K5 (masses between 0.7\,M$_\odot$ and 1.3\,M$_\odot$ is 39$\pm$3\% \citep{jao09a}, dropping to 26$\pm$6\% for stars with spectral types between K6 and M7 \citep[0.1--0.6\,M$_\odot$;][]{rastegaev10}. Spectroscopic binaries with separations of less than a few astronomical units among population II stars seem to be equally frequent to younger, higher metallicity stars \citep{stryker85,latham02,goldberg02}. This fact appears to hold for separations of a few tens of au \citep{koehler00b,zinnecker04}, and at very wide separations for GKM stars \citep{allen00,zapatero04a}, suggesting that metallicity might not have a strong impact on the formation of wide double and multiple systems. At lower masses, the frequency of early-M subdwarfs appears significantly lower \citep[3.3$\pm$3.3\%;][]{riaz08a,lodieu09c} than the multiplicity of solar-metallicity M dwarfs \citep[between 23.5 and 42\%;][]{ward15,cortescontreras17b} over similar separation ranges\footnote{Note added during external revision: Besides, \citet{winters19} measured a multiplicity frequency of M dwarfs of 26.8$\pm$1.4\%.}.

\subsection{Multiplicity}
\label{sub:multiplicity}

The search for companions in the sample of 219 low-metallicity dwarfs found one solar-metallicity M-M pair (Id 73/73--1), which was confirmed by analysis of its optical spectrum. We exclude this system from the subsequent discussion because we focus on metal-poor systems. For the final sample of 218 M5-L8 subdwarfs, the search revealed five candidate companions: one clear metal-poor M-M pair (Id 150/150--1), and four M-L pairs. The physical projected separations of the companion candidates lie between 3.8\,arcsec and 13.4\,arcmin, equivalent to 1360\,au (0.007\,pc) and 142\,400\,au (0.69\,pc). In all cases, the candidate companions are warmer than the subdwarfs. We infer a binary fraction of 2.29$\pm$2.01\% from the five metal-poor pairs among the 218 sources, assuming Poisson statistics and a Wald 95\% confidence interval\footnote{Wald interval\,$=(\lambda - \text{1.96}\sqrt{\lambda/n}, \lambda + \text{1.96}\sqrt{\lambda/n})$, where $\lambda$ is the number of successes in $n$ trials.}. In any case, we find one clear co-moving low-metallicity pair (Id 150), placing the minimum probability of finding a metal-poor M5--L8 dwarf in a binary system at 0.46\small $_{-\text{0.46}}^{\text{+0.90}}$\% \normalsize. We discuss the binary fractions as a function of spectral type and/or metallicity class below, and summarise the results in Table~\ref{tab:binarity}.

\begin{table}[H]
 \centering
 \caption[Binary fractions obtained in this work.]{Binary fractions obtained in this work.}
 \footnotesize
 \scalebox{1}[1]{
 \begin{tabular}{ccccc}
\noalign{\hrule height 1pt}
\noalign{\smallskip} 
 Sample & [M/H] & Binaries & Sample size & Binary fraction \\
        &   (dex)     &          &             &       (\%)      \\
\noalign{\smallskip}        
 \hline
\noalign{\smallskip} 
sdM5--L8 & $\leq$\,--0.5 & 5 & 218 & 2.29$\pm$2.01 \\
\noalign{\smallskip}
sdM5--9.5 & $\leq$\,--0.5 & 2 & 167 & 0.60$_{-\text{0.60}}^{+\text{1.17}}$ \\ 
\noalign{\smallskip}
sdL & $\leq$\,--0.5 & 3 & 49 & 6.12$_{-\text{6.12}}^{+\text{6.93}}$ $^{\text{(a)}}$ \\
\noalign{\smallskip} 
 \hline
\noalign{\smallskip} 
sdM & --0.5 & 1 & 97 & 1.03$_{-\text{1.03}}^{+\text{2.02}}$ \\ 
\noalign{\smallskip}
esdM & --1 & 1 & 53 & 1.89$_{-\text{1.89}}^{+\text{3.70}}$ \\
\noalign{\smallskip}
usdM & --2 & 0 & 19 & $\leq$\,5.3 \\
\noalign{\smallskip} 
\noalign{\hrule height 1pt}
 \end{tabular}
}
 \label{tab:binarity}
\begin{center}
  \footnotesize{\textbf{\textit{Notes. }}}
  \footnotesize{$^{\text{(a)}}$ With doubtful pair Id 190/190--1, it can reach 8.16$\pm$8.00\%.}
\end{center}
\end{table}

We identify one M-M subdwarf pair, Id 150, with a separation of 1360 au. We have a total of 167\,M5--M9.5 metal-poor dwarfs in the sample and found only one M-M pair, yielding a frequency of 0.60\small $_{-\text{0.60}}^{+\text{1.17}}$\% \normalsize. In terms of metallicity, Id 150 is a subdwarf, and in the sample there are 97 M subdwarfs (sdMs), yielding a binarity of 1.03\small $_{-\text{1.03}}^{+\text{2.02}}$\% \normalsize. Id 25 has a lower metallicity because it is an extreme M subdwarf system, and so the frequency of esdM systems is 1.89\small $_{-\text{1.89}}^{+\text{3.70}}$\% \normalsize (1 out of 53), assuming Poisson statistics. We did not find any companion to the 19 usdM in the sample, implying an upper limit of 5.3\% on the binary fraction of ultra subdwarfs.

For the M-L pairs, the subdwarf of the sample is the L-type while the companion is the M-type. Following the general convention, the primary is the most massive of the pair, and hence the L subdwarfs are the secondaries. Therefore, We find three L-type secondaries around M subdwarfs (plus one doubtful) out of 49\,L subdwarfs in the sample, yielding a binarity of 6.12\small $_{-\text{6.12}}^{+\text{6.93}}$\% \normalsize, that can increase to 8.16$\pm$8.00\% \normalsize if we include the doubtful pair (Table \ref{tab:proposed_pairs}). This means that 6.12\% of the L subdwarfs of the sample are part of a multiple system with a maximal physical projected separation of 0.69\,pc (13.4\,arcmin) while the search is sensitive to separations up to 1.5\,pc. In this case, We are not strictly talking about a binary fraction because the search in \textit{Gaia} DR2 is not sensitive to lower mass companions to the L subdwarfs.

We should mention that two sources in the sample have known companions in the WDS catalogue \citep{mason01}. On the one hand, Id\,3 (WDS\,00259--0748) has a companion at 2.1\,arcsec with a magnitude of 22.6 in the F775W optical filter \citep{riaz08a}. On the other hand, Id\,154 (WDS\,14164+1348, also known as SDSS J1416+13AB) is an sdL7 source with a T5 companion separated by 9.3\,arcsec with $J$\,=\,17.26\,mag \citep{scholz10a,schmidt10a,bowler10a,burningham10a}. Both companions are beyond the reach of the survey because of their faintness. We do not identify new more massive companions to both objects. 

For the same separation range, the found frequency is much lower than the multiplicity of F6--K3 stars \citep[44$\pm$3\%;][]{fischer92} and K7$-$M6 \citep[23.5$\pm$3.2\%;][]{ward15}; see also table~1 in \citet{cortescontreras17b}. Our result is more in line with the frequency of early-M subdwarfs based on high-resolution \textit{Hubble} Space Telescope and Lucky Imaging \citep{riaz08a,lodieu09c}. The found multiplicity is much lower than the frequency of GKM subdwarfs \citep[13--15\%;][]{zapatero04a}.
\citet{moe19} state that there is no difference between the frequencies of solar or subsolar metallicity samples among wide ($a$\,$>$\,1000\,au) binaries. Similarly, \citet{elbadry19a} conclude that the wide binary fraction is almost constant with metallicity at large separations ($a\geq$\,250\,au), but decreases quickly with metallicity at smaller separations. These statements do not seem to hold for ultra-cool subdwarfs with spectral types later than M5. The binary frequency of M subdwarfs still remains unclear because of poor statistics. \citet{jao09a} studied a sample of 32 K and 37 M subdwarfs and derived a multiplicity of 26$\pm$6\% for separations larger than 110\,au and 6\% for lower separations.

To provide a more complete picture, we combined the samples of \citet{gizis00c}, \citet{zapatero04a}, \citet{riaz08a}, and \citet{lodieu09c} to perform a search for companions around metal-poor GKM dwarfs using CCD and Lucky Imaging. We selected all M subdwarfs in this compilation and built a new sample adding all M subdwarfs from the sample, regardless of their metallicity class. Only the work of \citet{zapatero04a} does not provide spectral types or effective temperatures and therefore we recovered M type stars from their colours using infrared and optical photometry and the updated version of table~5 in \citet{pecaut13}\footnote{\url{https://www.pas.rochester.edu/~emamajek/EEM_dwarf_UBVIJHK_colors_Teff.txt}}. These latter authors searched for companions around subdwarfs with separations of between 0.1--0.2 and 25\,arcsec. As \textit{Gaia} is able to resolve pairs at 2.2\,arcsec, we neglect all binaries found at closer separations and do not consider any companion at separations above 25\,arcsec in order to obtain a coherent binary fraction. This new sample contains 279 objects with 264 low-metallicity M and L dwarfs (215 M, 49 L), including 12 binaries of which only 6 lie in the 2.2--25 arcsec separation range. Two of those six binaries are in the sample, and five are M low-metallicity dwarfs, yielding a binary fraction of 2.33$\pm$2.04\%. This exercise supports our finding that ultra-cool subdwarfs have a much lower multiplicity fraction than higher mass subdwarfs.

As for theoretical predictions, hydrodynamical simulations predict a multiplicity fraction of 15--25\% and 12\% for subsolar metallicity (0.1\,$Z_{\odot}$) M and L dwarfs, respectively \citep{bate14b,bate19a}. The impact of metallicity on the multiplicity appears very limited with fractions of 15--40\% for metal-poor M dwarfs, and $\sim$10\% for metal-poor L dwarfs, when $Z$\,=\,0.01 $Z_{\odot}$. These studies show that our findings are far from these numbers, but those simulations only consider separations below 10\,000\,au, with most of them lower than 1000\,au, while the pairs of the sample are beyond the upper limits (16\,500--142\,400\,au), except for pair Id 150/150--1 at 1360\,au. Therefore, our study is not sensitive to such short separations and we are not able to directly compare or test theoretical predictions.

\section{Conclusions}
\label{sec:conclusions_subdwarfs}

We present a dedicated search for wide companions to a sample of spectroscopically confirmed M and L subdwarfs. We identified several candidates around six subdwarfs. Based on these findings, we come to the following conclusions:

\begin{enumerate}

\item We did not find low-mass companions to any of the 219 sources of the sample.

\item We did not detect companions colder than L-type sources because of the sensitivity limit of \textit{Gaia} DR2\@. With these data, We are not able to determine the multiplicity fraction of L subdwarfs.

\item We found a metal-poor M-M system, which has been confirmed spectroscopically, composed of \textit{Gaia} DR2 3720832015084722304 (with spectral type sdM5.5 and effective temperature of 3000\,K) and \textit{Gaia} DR2 3720832010789680000 (sdM1.5$\pm$0.5 and 3600\,K).

\item We found another M-M system, but of solar metallicity, whose spectroscopy leads to consider it as a bound system. As its metallicity is higher than that of all the other sources in the sample, we did not include it in the analysis of binarity of subdwarfs.

\item We identified four possible M-L systems, and the spectroscopy seems to confirm one of them as bound. This system is composed of ULAS J124104.75-000531.4 and \textit{Gaia} DR2 3695978963488707072 (whose spectral types and effective temperatures are sdL0$\pm$0.5 and 2600\,K, and sdM5$\pm$0.5 and 3300\,K, respectively), plus one more system confirmed by \cite{zhang19g} composed of \textit{Gaia} DR2 4818823636756117504 and 2MASS J04524567-3608412 (esdM$\pm$0.5 and esdL0$\pm$0.5, and 3700\,K and 2800\,K respectively). The remaining two systems should be confirmed spectroscopically in the future. This is an interesting result because we extend the known M-L systems from one to two, and probably four. These new systems are important targets to infer the metallicities of the L subdwarfs with higher precision.

\item We infer a frequency of wide systems among sdM5--sdM9.5 of 0.6\small $_{-\text{0.6}}^{+\text{1.2}}$\% \normalsize for projected physical separations larger than 1360\,au (up to 142\,400\,au).

\item We derive a binarity of 1.03\small $_{-\text{1.03}}^{+\text{2.02}}$\% \normalsize in M subdwarfs (sdM), while the multiplicity of M extreme subdwarfs (esdM) is 1.89\small $_{-\text{1.89}}^{+\text{3.70}}$\% \normalsize.

\item We did not find any companion to the M ultra-cool subdwarfs (usdM) in the sample, placing an upper limit on binarity of 5.3\%.

\end{enumerate}

This study reveals new wide companions around the largest sample of ultra-cool subdwarfs known to date but is limited in depth to higher mass companions. We plan to look for less massive companions with future multi-epoch deep surveys like Vera Rubin Large Synoptic Survey telescope \citep{ivezic08a} or in the infrared with upcoming space missions like Euclid \citep{laureijs11,amiaux12,mellier16a} or the Wide Field Infrared Survey Telescope \citep[WFIRST;][]{spergel15}.

\newpage

%%%%% CAPITULO 4 %%%%%

\chapter{The widest binaries} 
\label{ch:widest_binaries}
\vspace{2cm}
\pagestyle{fancy}
\fancyhf{}
\lhead[\small{\textbf{\thepage}}]{\textbf{Section \nouppercase{\rightmark}}}
\rhead[\small{\textbf{Chapter~\nouppercase{\leftmark}}}]{\small{\textbf{\thepage}}}

\begin{flushright}
\small{\textit{``Let nothing unite us, so that nothing can separate us''}}

\small{\textit{One hundred love sonnets}}\\
\small{\textit{-- Pablo Neruda}}
\end{flushright}
\bigskip

\begin{adjustwidth}{70pt}{70pt}
\tR{\small{The content of this chapter has been adapted from the article \citet{gonzalezpayo23}: \textit{Reaching the boundary between stellar kinematic groups and very wide binaries — IV. The widest Washington Double Star systems with $\rho\geq$\,1000 arcsec in Gaia DR3}, published in Astronomy \& Astrophysics, \href{https://doi.org/10.1051/0004-6361/202245476}{A\&A 670, A102 (2023)}.}}
\end{adjustwidth}
\bigskip

\lettrine[lines=3, lraise=0, nindent=0.1em, slope=0em]{W}{ashington Double Star catalogue} (WDS) is the reference for multiple stars systems. We selected from WDS the pairs with longest angular separation to try to find out in this chapter whether those pairs are really linked objects or not. It is difficult to establish a limit to determine which pairs are wide. Some authors put the limit in physical separations bigger than 20\,000\,au (about 0.01\,pc) because they are beyond the typical size of a collapsing cloud core ($\sim$5000--10\,000\,au) \citep{jimenezesteban19}. Other authors rise the limit to between 0.1 and 1\,pc \citep{tian20}, or even from 1\,pc on \citep{caballero09}. And, as we already mentioned in Sec.~\ref{sec:results_search}, \citet{shaya11} affirmed that the pairs separated up to 1--8\,pc are also possible. In the present work, the wide binaries will be related to their angular separation, independently of their distance, so we will not care about the physical separation. In particular, we will focus on those objects with an angular separation bigger than 1000\,arcsec (16.7\,arcmin). The characterisation of the proposed pairs will determine if they are so wide binaries.

In this work, we perform a detailed characterisation of the widest pairs in the Washington Double Star (WDS) catalogue \citep{mason01} by making use of the latest \textit{Gaia} DR3 data \citep{gaiacollaboration23b}.
The WDS, which is maintained by the United States Naval Observatory, is the world's principal database of astrometric double and multiple star information.
For each system, we ascertain their actual gravitational binding and search for additional companions.
Since we investigate pairs with angular separations, $\rho$, greater than 1000\,arcsec, this work can be understood as a \textit{Gaia} update of that by \citet{caballero09}, who also used $\rho =$ 1000\,arcsec as the minimum separation between the widest WDS pairs at that time, but had only {\it Hipparcos} \citep{perryman97} parallaxes for a few bright stars and relatively poor proper motions for the faintest components.
Furthermore, this work is the fourth item of the series initiated by \citet{caballero09}, which aims to shed light from an observational perspective on the formation and evolution of the most separated and fragile multiple stellar systems in the Milky Way.
Although young systems play an important role in our analysis, here we focus on field systems that are relatively evolved and old, at the brink of disruption by the galactic gravitational potential.

This chapter is structured as follows: We describe the stellar sample in Sect.~\ref{sec:sample_WDS}. 
Section~\ref{sec:analysis_WDS} shows the analysis that we followed to filter, classify, and characterise WDS pairs, and to search for other possible members of the multiple systems. 
We present the results obtained after our analysis, together with a discussion, in Sect.~\ref{sec:results_discussion_widest}.
Finally, Sect.~\ref{sec:summary_WDS} summarises our work. 

\section{Sample}
\label{sec:sample_WDS}

We built our sample from the latest version of the WDS\footnote{WDS Catalog with precise last only, \url{http://www.astro.gsu.edu/wds/Webtextfiles/wds_precise.txt}, accessed on 12 November 2022.}.
For each of the 155\,159 resolved pairs, WDS tabulates the WDS identifier (based on J2000 position), 
discoverer code and number, 
number of observations and of components (when there are more than two),
date, position angle ($\theta$, i.e. orientation on the celestial plane of the companion with respect to the primary), and $\rho$ of the first and last observations,
magnitudes, and proper motions of the two components,
along with the equatorial coordinates of the primary of the pair 
and notes about the pair.
In a few cases, WDS also tabulates the Durchmusterung number (Bonn, C\'ordoba, Cape -- \citealt{schonfeld1886}; \citealt{argelander1903}) and the spectral type of the primary or companion (or both).
There are numerous pairs that take part of multiple systems with, usually, the same primary star; in general, they share the same WDS identifier, but not always.

\begin{figure}[]
 \centering
 \includegraphics[width=0.5\linewidth, angle=0]{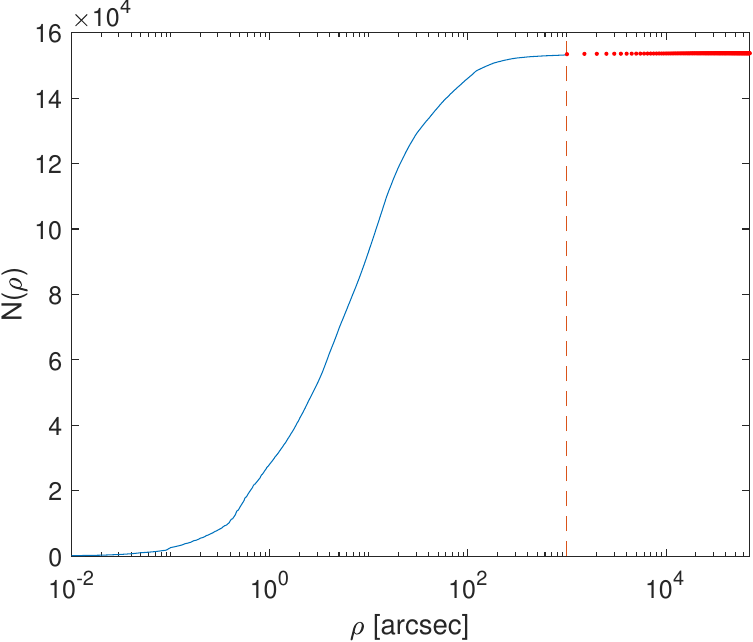}
 \caption[Cumulative number of WDS pairs as a function of $\rho$.]{Cumulative number of WDS pairs as a function of $\rho$. 
 Red data points with $\rho>$\,1000\,arcsec (to the right of the orange vertical dashed line) mark the 504 WDS pair candidates investigated by us. 
This figure can be compared with fig.~1 in \cite{caballero09}.
}
 \label{fig:acumhist}
\end{figure}

In Fig.\,\ref{fig:acumhist}, the cumulative number of WDS pair angular separations increases with a power law between $\rho\sim$\,0.4\,arcsec and $\rho\sim$\,100\,arcsec.
This distribution follows \"Opik's law \citep{opik24} for binaries with projected physical separations greater than 25\,au \citep{allen97}.
Outside the $\rho\sim$\,0.4--100\,arcsec range, the distribution flattens at both sides.
This flattening is an observational bias at short angular separations, as micrometer, speckle, lucky imaging, adaptive optics, and even imaging from space are limited by the atmospheric seeing, telescope size, or optical quality (but there seems to be a slight overabundance of close pairs of $\rho\sim$\,0.4--4.0\,arcsec with respect to more separated ones).

\begin{figure}[]
 \centering
 \includegraphics[width=0.65\linewidth, angle=0]{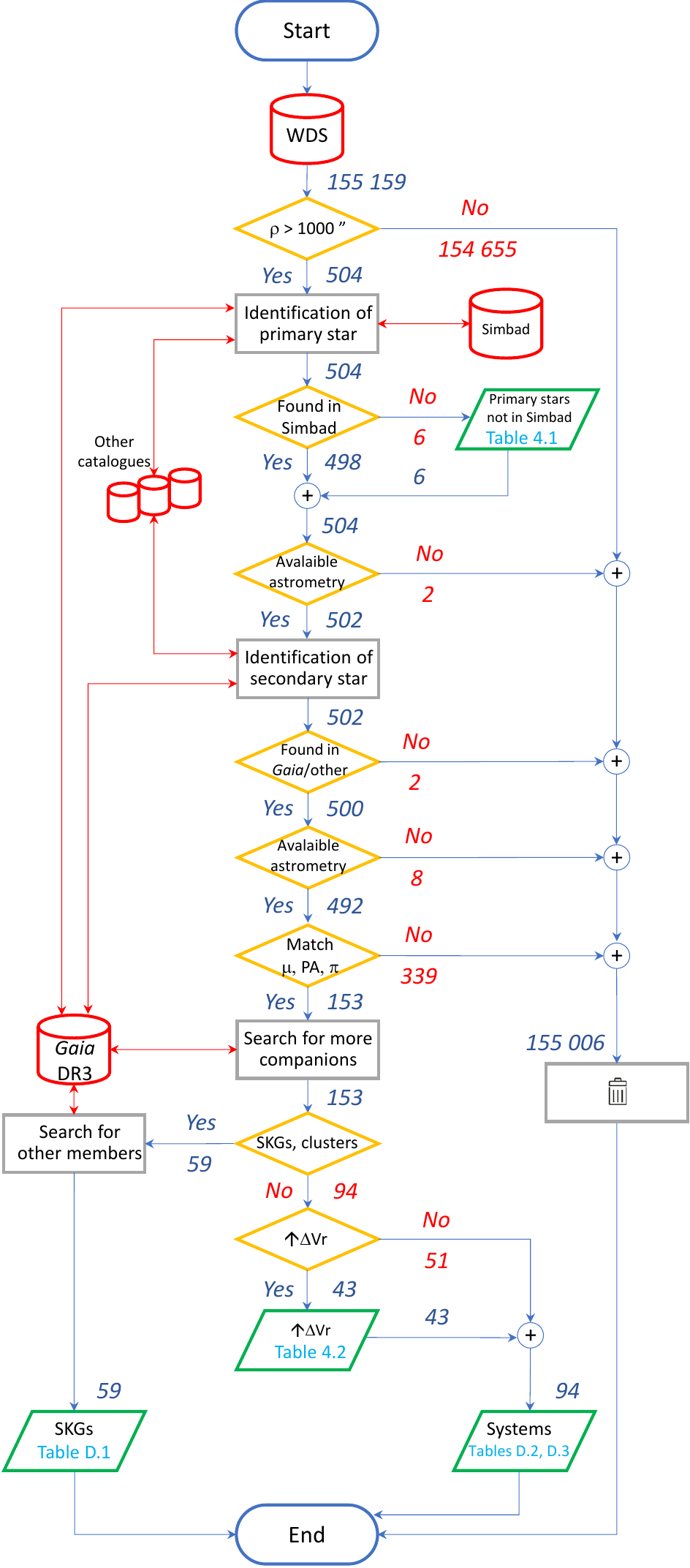}
 \caption{Flowchart describing the analysis. }
 \label{fig:flowchart1}
\end{figure}

The flattening of the $\rho$ distribution at wide separations, especially at $\rho\sim$\,200\,arcsec, is mainly due to the actual formation and evolution of multiple stellar systems, although there may also be a contribution from another observational bias: until the advent of \textit{Gaia} \citep{gaiacollaboration16b,gaiacollaboration18,gaiacollaboration21a}, accurate proper motion and parallax measurements were available only for a tiny fraction of stars, while most wide WDS pairs come from pre-\textit{Gaia} common proper motion surveys (e.g. \citealt{allen00}; \citealt{chaname04}; \citealt{lepine07a}; \citealt{dhital10}; \citealt{raghavan10}; \citealt{tokovinin12}; and references therein\footnote{There are also relevant unpublished contributions to the WDS, such as that of the Observatori Astron\`omic del Garraf \citep{caballero13}. See further details at \url{http://www.astro.gsu.edu/wds/wdstext.html\#intro}.}).
In spite of numerous common proper motion surveys, the observational bias remains at $\rho \gtrsim$ 200\,arcsec because most of them looked for companions at angular separations of up to a few arcminutes only, mostly due to past computational limitations. 
However, this difficulty is starting to be alleviated thanks to new \textit{Gaia} surveys (e.g. \citealt{kervella22,sarro23}).
Due to the observational bias or the actual difficulty in forming wide binaries \citep{kouwenhoven10,reipurth12,lee17,tokovinin17}, the distribution of ultrawide WDS pairs with $\rho \gtrsim$ 1000\,arcsec becomes extremely flat (Fig.~\ref{fig:acumhist}).

At the time of our analysis, WDS contained 504 pairs separated by more than 1000\,arcsec.
For comparison, \citet{caballero09} investigated about 105\,000 WDS pairs, of which only 35 had $\rho >$ 1000\,arcsec.
Together with the \textit{Gaia} DR3 data, a sample about 15 times larger represents a qualitative and quantitative leap with respect to the first item of this paper series.

\section{Analysis}
\label{sec:analysis_WDS}

Our analysis procedure is illustrated by the flowchart in Fig.~\ref{fig:flowchart1}, with the following elements: ovals represent the initial and final status of the process, diamonds are Boolean questions for which only possible answers are ``Yes'' or ``No'', 
rectangles are specific actions, trapezia are partial or final obtained results, and cylinders are databases (or catalogues) from where the data are collected or consulted. 
In Fig.~\ref{fig:flowchart1}, the incoming numbers in every block represent the number of processed WDS pairs in that block, and the outgoing numbers represent the number of pairs that match the condition or pass through to the next block of the flowchart.

\subsection{Primary stars in \textit{Gaia} DR3}
\label{sec:primary_stars}

For each primary star of the 504 WDS ultrawide pairs, we collected its main identifier in the Simbad astronomical database \citep{wenger00}. 
Of them, only six do not have a Simbad entry; they are shown in Table~\ref{tab:no_simbad}.
Using \texttt{Topcat} \citep[][]{taylor05} and the equatorial coordinates tabulated by WDS, we automatically cross-matched every primary with \textit{Gaia} DR3.
Next, we visually inspected all cross-matches with the help of the Aladin sky atlas \citep{bonnarel00}, and \textit{VizieR} \citep{ochsenbein00}.
When there were more than a single \textit{Gaia} counterpart per primary within our $\sim$5\,arcsec cross-match radius, we chose the right star by comparing WDS and \textit{Gaia} proper motions, magnitudes, and spectral types.

\begin{table}[H]
 \centering
 \caption[Primary stars without a Simbad entry.]{Primary stars without a Simbad entry.}
 \footnotesize
  \scalebox{1}[1]{
 \begin{tabular}{cllcccc}
\noalign{\hrule height 1pt}
 \noalign{\smallskip}
 WDS & Discoverer & Primary star &  $\alpha$ (J2000) & $\delta$  (J2000) & \textit{G} & \textit{d} \\
  & code & (\textit{Gaia} DR3 Id) & (hh:mm:ss.ss) & (dd:mm:ss.s) & (mag) & (pc) \\
 \noalign{\smallskip}
 \hline
 \noalign{\smallskip}
00474--7345 & OGL 84 & 4685766099704705280 & 00:47:25.09 & $-$73:44:42.5 & 17.5 & 1700$\pm$200 \\
00489--7434 & OGL 87 & 4685482661910629248 & 00:48:55.94 & $-$74:33:46.7 & 19.4 & 504$\pm$54 \\
01121--7400 & OGL 161 & 4686310976416294528 & 01:12:11.20 & $-$73:59:42.9 & 16.6 & 1870$\pm$150 \\
01235--7356 & OGL 188 & 4686225867342046848 & 01:23:33.07 & $-$73:55:34.7 & 14.3 & 468.0$\pm$3.1 \\
03074--4655 & TSN 110 & 4750712533547201920 & 03:07:26.03 & $-$46:54:44.8 & 16.6 & 136.0$\pm$1.0 \\
10181--0130 & TSN 113 & 3830436797339858176 & 10:18:03.35 & $-$01:30:11.6 & 17.8 & 397$\pm$21 \\
\noalign{\hrule height 1pt}
 \end{tabular}
 }
 \label{tab:no_simbad}
\end{table}

Of the 504 primaries, 44 are redundant (they belong to two or more pairs). 
Only 6 out of the 460 non-redundant primaries had no \textit{Gaia} DR3 entry because of their extreme brightness: \textit{Capella} ($\alpha$~Aur), \textit{Canopus} ($\alpha$~Car), \textit{Fomalhaut} ($\alpha$~PsA), $\alpha$~Cen, $\gamma$~Cen, and $\delta$~Vel, for which we took proper motions and parallaxes from \textit{Hipparcos}.
In addition, another 22 primaries are in \textit{Gaia} DR3, but do not have a five-parameter astrometric solution.
For 11 of them, namely, those of moderate brightness ($G$ = 2.3--9.5\,mag), we again took proper motions and parallaxes from \textit{Hipparcos}, while for 9 of them\footnote{The 9 primary stars with \textit{Gaia} DR2 data are: LP~295--49, G~202--45, 36~And~A, 4~Sex, HD~111456, HD~125354, HD~340345, BD-12~6174, and HD~213987.}, we took the data from \textit{Gaia} DR2 \citep{gaiacollaboration18}.
The 2 remaining stars are LSPM~J2323+6559 and 2MASS~J00202956--1535280, for which there are only proper motions available from \citet{lepine05d} and \citet{cutri14}, respectively.
Since the 2 later primaries do not have published parallaxes, we discarded the corresponding pairs from the analysis.
Accounting for these 2 discards, we retained 502 pairs for the next step of the analysis.

\subsection{Search for WDS companions}
\label{sec:search_for_wds_companions}

The WDS catalogue provides the relative positions of the companion stars of the pairs with respect to the primaries through $\rho$ and $\theta$. 
To manually locate the companions to the primaries, we used Aladin.
We loaded different catalogues and services, namely \textit{Gaia} DR3, 2MASS \citep{skrutskie06}, Simbad, and WDS, and we used the \texttt{dist} tool. 
For the correct identification of the companion, we proceeded to carry out a visual confirmation of the primary cross-match and we chose the \textit{Gaia} DR3 candidate companion within 10\,arcsec around the expected location that matched the WDS values of proper motion, magnitude, and spectral type. 
If the companion had not been identified, especially in the widest systems ($\rho >$ 10\,000\,arcsec), we enlarged the search radius in steps up to 120\,arcsec. 
For these outliers, we used all the available information, namely, the WDS remarks and the original publications.
There were only two cases where the companion star was not found by us with the $\rho$ and $\theta$ provided by WDS, even after enlarging the search radius and scouring the literature\footnote{The 2 WDS pairs with unidentified companion stars are 
WDS~03074--5655 (TSN~110) and 
WDS~03353--4020 (TSN~111).
In a preliminary analysis, there was a third unidentified system, namely
WDS~05463+5627 (LDS~3673), but it suffered from a typographical error in WDS that was corrected afterwards (\citealt{carro21}; B.\,D.~Mason, priv. comm.).
We revise its relative astrometry to $\rho$ = 57.1\,arcsec, $\theta$ = 262.5\,deg, and epoch = J2016.0.}.

As for the primaries, we retrieved Simbad identifiers and \textit{Gaia} DR3 for the corresponding companions. 
Only eight of the companions had no parallaxes (or even proper motions) available in any catalogue, and we also discarded them from the analysis\footnote{The 8 companions without parallax are: LSPM J1536+2856, SCR J1900--3939, UCAC3 208--200112, 2MASS J13543510--0607333, 2MASS J14313545--0313117, Gaia DR3 276070675205077632, Gaia DR3 4655216993788228480, and Gaia DR3 601133385210548736.}.

We computed our own $\rho$ and $\theta$ parameters for the 492 remaining pairs using the standard equations of spherical trigonometry (e.g. \citealt{smolinski06}):
\begin{equation}
\rho= \arccos{[\cos{(\Delta \alpha \cos{\delta_\text{1}})}\cos{(\Delta \delta)}]}
\label{eqn:rho_obtention}
,\end{equation}
\noindent and
\begin{equation}
\theta = \frac{\pi}{2} - \arctan\left[\frac{\sin{(\Delta \delta)}}{\cos{(\Delta \delta)}\sin{(\Delta \alpha \cos{\delta_\text{1}})}}\right],
\label{eqn:theta_obtention}
\end{equation}
\noindent where $\Delta \alpha$ = $\alpha_\text{2} - \alpha_\text{1}$,
$\Delta \delta$ = $\delta_\text{2} - \delta_\text{1}$,
and $\alpha_\text{1},\delta_\text{1}$ and $\alpha_\text{2},\delta_\text{2}$ are the equatorial coordinates of the primary and companion stars, respectively.

We compared the $\rho$ and $\theta$ values we measured  with those tabulated by WDS (in particular, with the latest measurements, i.e. \texttt{sep2} and \texttt{pa2}).
For the position angle, the standard deviation of the differences between our measurements and those from WDS is 0.84\,deg.
The distribution of the differences in $\theta$ is not Gaussian, with a narrow peak centred at 0\,deg and wide, but shallow, wings at both sides.
Of the 492 identified pairs with parallaxes, only 23 have absolute differences in $\theta$ greater than 1\,deg (and up to 4.2\,deg). Most of the kinds of differences ascribed to uncertainties propagated from inaccurate pre-\textit{Gaia} coordinates, especially for the widest systems, such as WDS~23127+6317 (e.g. with Eq.~\ref{eqn:theta_obtention} being highly non-linear).
The distribution of the differences in $\rho$ is similar to that of $\theta$, with a narrow peak centred at 0\,arcsec and a relatively large standard deviation of the differences of 21.0\,arcsec.
This large amount is originated by the difficulty in previous works to measure $\rho$ or even to identify the companion of the widest systems, such as the ``outliers'' described above and found at more than 10\,arcsec from their expected locations\footnote{For example, \citet{tokovinin12} and we ourselves measured $\rho$ = 1684.2\,arcsec and 1684.49\,arcsec, respectively, for the outlier system HD~45875 + Gaia DR3 1115649542191409664 (TOK~503, WDS~06387+7542~AD), but WDS instead tabulates 1898.64\,arcsec collected in 2015.}.

The distribution of our new values of $\rho$ are plotted with red data points in Fig.~\ref{fig:acumhist}.
Of the 492 identified pairs with parallax, 298 have $\rho$ = 1000--2000\,arcsec, 117 have $\rho$ = 2000--10\,000\,arcsec, and 77 have $\rho>$ 10\,000\,arcsec.
The latter ultrawide pairs come mostly from the works by \cite{probst83} and \cite{shaya11}.
The widest pair has $\rho$ = 66\,094\,arcsec (WDS 02157+6740, SHY~10; \citealt{shaya11}).
As described below, not all of them are physically bound.

\subsection{Pair validation}
\label{sec:pair_validation}

To validate the 492 pairs, we used the criteria established by \cite{montes18a} to distinguish between physical (bound) and optical (unbound) systems.
For that purpose, we computed two astrometric parameters that quantify the similarity of the proper motions of two stars:
\begin{equation}
\mu\,\text{ratio}=\sqrt{\frac{(\mu_{\alpha} \cos{\delta_\text{1}}-\mu_{\alpha} \cos{\delta_\text{2}})^\text{2}+(\mu_{\delta_\text{1}}-\mu_{\delta \text{2}})^\text{2}}{(\mu_{\alpha} \cos{\delta_\text{1}})^\text{2}+(\mu_{\delta_\text{1}})^\text{2}}}<\text{0.15},
\label{eqn:crit1a}
\end{equation}
\noindent and
\begin{equation}
\Delta PA =\lvert PA_\text{1} - PA_\text{2} \rvert<\text{15}\,\text{deg},
\label{eqn:crit2a}
\end{equation}
\noindent where PA$_i$ are the angles of the proper motion vectors, with $i=$\,1 for the primary star and $i=$\,2 for the companion.
We added an extra buffer in the $\mu$ ratio of up to 0.25 to account for projection effects on the celestial sphere of nearby ultrawide systems, as in the case of $\alpha$~Cen~AB + Proxima \citep{innes1915,wertheimer06,caballero09}.

\begin{figure}[H]
     \centering
     \includegraphics[width=0.55\linewidth]{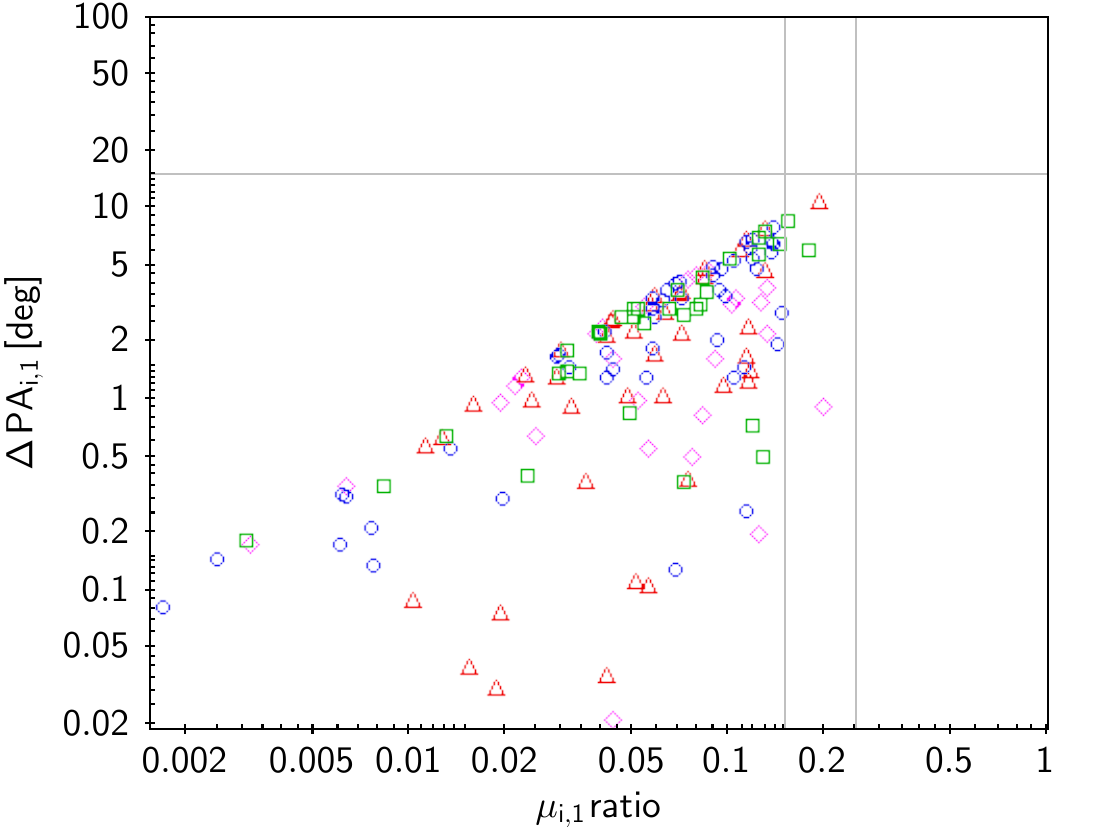}
     \centering
     \includegraphics[width=0.55\linewidth]{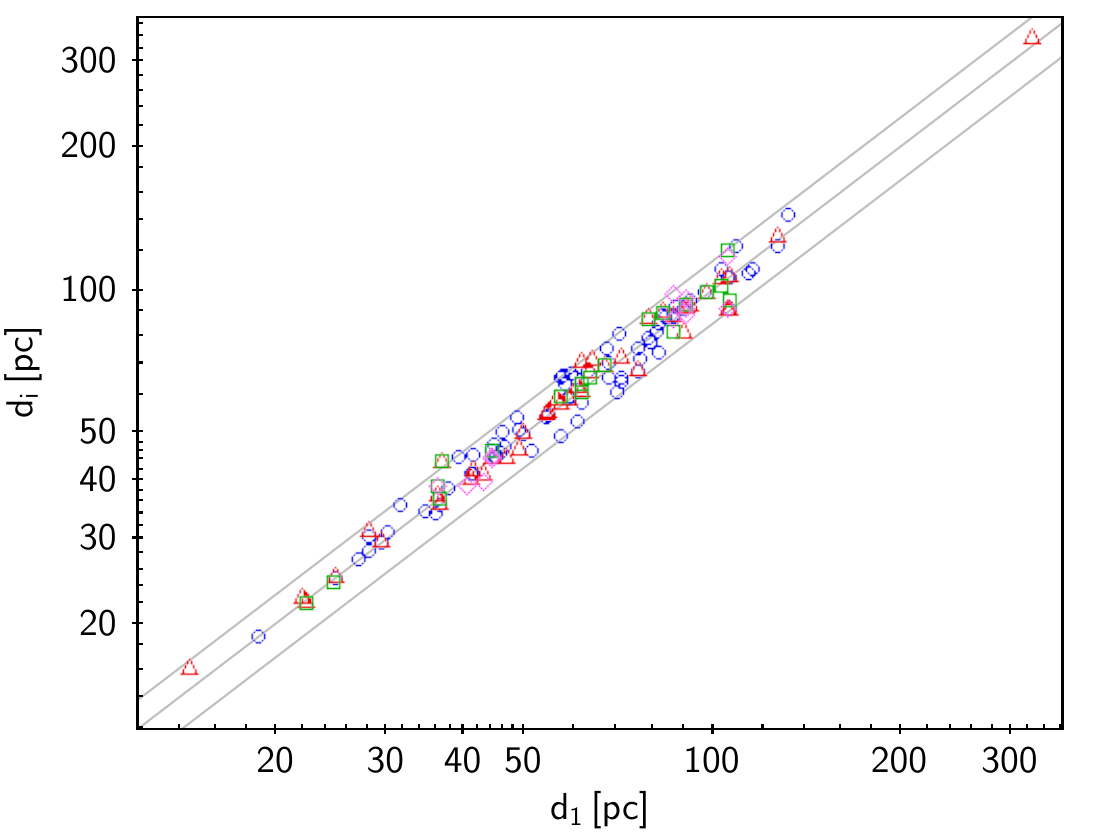}
  \vspace{2mm}
\caption[Astrometric criteria for pair validation.]{
 Astrometric criteria for pair validation.
 In both diagrams, we plot pairs between primaries and secondaries with blue circles, tertiaries with red triangles, quaternaries with green squares, and higher order companions with purple diamonds. 
 \textit{Top:} $\Delta PA_{1,i}$ vs. $\mu_{1,i}$ ratio diagram.
 Vertical and horizontal grey lines mark the $\mu$ ratios of 0.15 and 0.25, and $\Delta PA$ of 15\,deg, respectively. 
 Compare with fig.~2 in \citet{montes18a}.
 \textit{Bottom:} 
 Distance of companions vs. distance of primaries. 
 Diagonal lines indicate the 1.15:1, 1:1, and 0.85:1 distance relationships.
 The $\alpha$~Cen~AB + Proxima system, at $d$\,$\sim$\,1.3\,pc, is not shown.
}
\label{fig:dpamu_ddd}
\end{figure}

At the time of publication by \cite{montes18a}, \textit{Gaia} parallaxes were not available except the for 2.5 million stars of the Tycho-\textit{Gaia} Astrometric Solution \citep{michalik15}.
With the advent of the third \textit{Gaia} data release with precise parallaxes for $\sim$700 times more stars, we added one additional condition to our validation. 
\citet{cifuentes21} imposed parallactic distances to agree within 10\%, while \citet{gonzalezpayo21} did it within 15\%, which is the value we chose to impose. 
In short, our third astrometric criterion was:
\begin{equation}
\Biggl\lvert \frac{\pi_\text{1}^{-\text{1}}-\pi_\text{2}^{-\text{1}}}{\pi_\text{1}^{-\text{1}}} \Biggr\rvert < \text{0.15},
\label{eqn:crit3a}
\end{equation}
\noindent with $\pi_\text{1}$ and $\pi_\text{2}$ as the parallaxes of both components of the pair (we did not apply any colour correction for computing distances -- \citealt{bailerjones18a,lindegren21}).
In Fig.~\ref{fig:dpamu_ddd}, we show the relations between $\mu$ ratio and $\Delta$PA and between distances of the two components of the 153 pairs that satisfy the three imposed criteria simultaneously (and additionally, the multiple companions obtained in Sect.~\ref{sec:search_additional}). 
Although we refer to them as pairs, in many cases they are actually part of hierarchical multiple (triple, quadruple, quintuple...) systems made of stars at very different angular separations to their primaries.
This is described in detail below.

Except for 2 of them\footnote{The two systems at $d >$ 150\,pc are $\gamma$\,Cas + HD~5408 (188\,pc) and G~143--33 + G~143--27 (324\,pc).}, all the 153 pairs are located at heliocentric distances shorter than 150\,pc, with a distribution peaking at 40--50\,pc.
The distribution of total proper motions is, however, flatter, with only one pair\footnote{The high proper motion pair with $\mu >$\,700\,mas\,a$^{-\text{1}}$ is $\alpha$~Cen~AB + Proxima (3710\,mas\,a$^{-\text{1}}$).} with a $\mu$ greater than 700\,mas\,a$^{-\text{1}}$ and none with a $\mu$ less than 25\,mas\,a$^{-\text{1}}$.

We did not keep in our final list of validated pairs an ultrawide system candidate at about 2400\,pc towards the Magellanic Clouds, namely OGL~54 \citep{poleski12}.
It is made of OGLE~SMC-SC1~161-162 and Gaia DR3 4685747717242739328 (``SMC128.7.9551''), which are separated by about 12\,pc.
If truly linked, the pair would be much further and wider than any other system considered here.
Last but not least, we revised the system $\rho$ from 1017\,arcsec to 977\,arcsec, below our boundary at 1000\,arcsec.

\subsection{Additional companions and stellar kinematic groups in \textit{Gaia} DR3}
\label{sec:search_additional}

We looked for additional proper motion and parallax companions within 1\,pc around both the primary and the companion of the 153 validated pairs.
We followed the methodology described in Sect.~3 of \cite{gonzalezpayo21}; however, in our work, apart from \texttt{Topcat} and a customised code in astronomic data query language \citep{yasuda04}, we used \textit{Gaia} DR3 and the criteria imposed by Eqs.~\ref{eqn:crit1a}--\ref{eqn:crit3a}.
For a few cases of ultrawide WDS pairs with projected physical separations greater than 1\,pc, we extended the search radius up to the maximum separation between known components.

In our \textit{Gaia} search, we identified 349 additional common proper motion and parallax companions that satisfy the astrometric criteria of Eqs.~\ref{eqn:crit1a}, \ref{eqn:crit2a}, and~\ref{eqn:crit3a}.
Of these, 111 additional companions are catalogued by WDS and 239 are not.
The large multiplicity order of some system candidates, made of over a dozen pairs each (i.e. higher than dodecuple), together with the presence of debris discs in some of the components (e.g. $\alpha^{\text{01}}$~Lib, AU~Mic -- \citealt{kalas04,chen05,mizusawa12,gaspar13,mittal15,plavchan20}), has led us to investigate the membership of all our targets in young stellar kinematic groups (SKGs -- \citealt{eggen65,montes01a,zuckerman04}), stellar associations \citep{ambartsumian49,blaaw91,dezeeuw99}, and even open clusters.

\begin{figure}[h]
 \centering
 \includegraphics[width=1\linewidth]{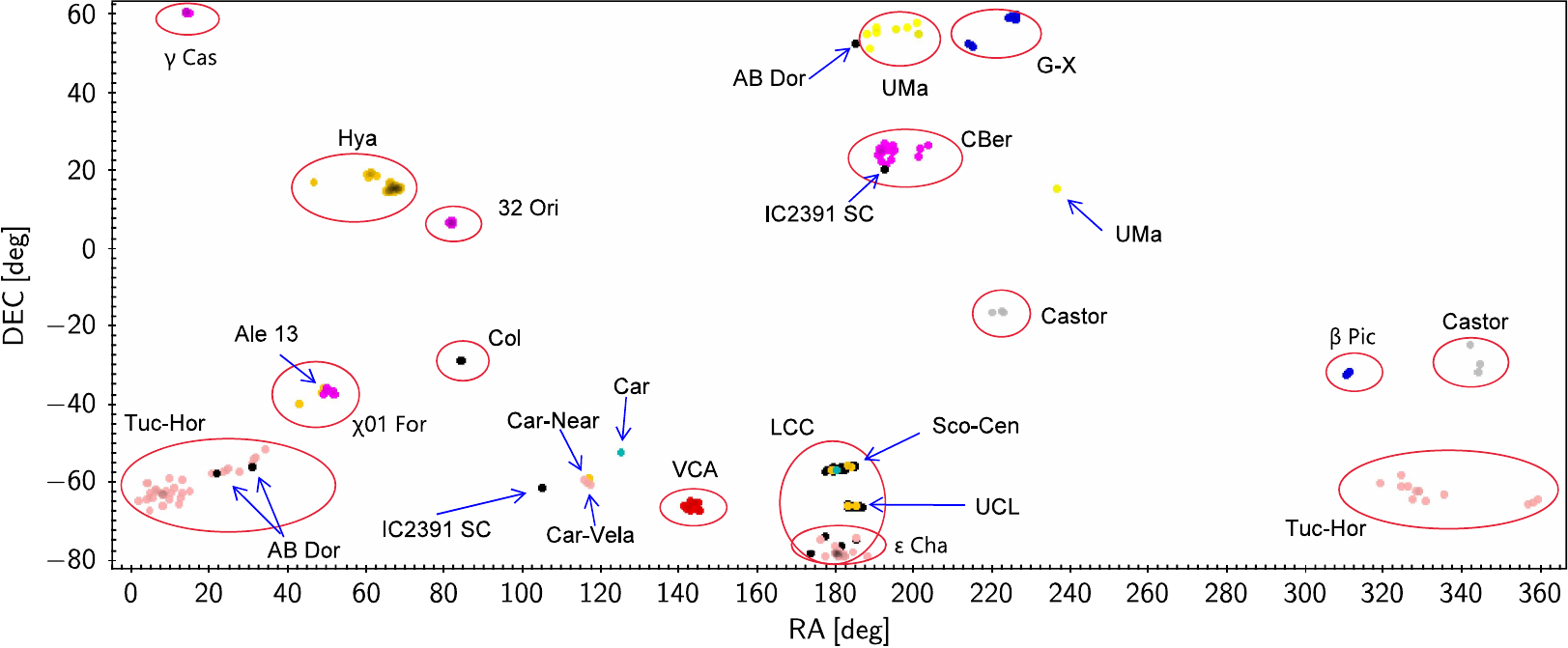}
 \caption[Spatial distribution of the 309 identified young stars in SKGs, associations, and open clusters.]{Spatial distribution of the 309 identified young stars in SKGs, associations, and open clusters.
}
\label{fig:spatialdistr}
\end{figure}

Of the 153 validated pairs in Sect.~\ref{sec:pair_validation}, there are 59 with at least one component (primary, companion, or both) that had previously been considered part of young SKGs, associations, and clusters such as the Tucana-Horologium and Coma Berenices moving groups, the $\epsilon$~Chamaeleontis association, or the Hyades open cluster (e.g.
\citealt{perryman98,murphy13,kraus14,pecaut16,riedel17,gagne18a,tang19}).
Furthermore, of the 239 additional astrometric companions not catalogued by WDS, a total of 199 share proper motion and parallax companions with these young pairs.
Table~\ref{tab:stars_young_skg} in Appendix~\hyperref[ch:Appendix_D]{D}, available in \textit{VizieR}, shows the name, equatorial coordinates, and $G$-band magnitude of 349 young stars and candidates, together with the corresponding group (SKG, association, or cluster) when available (309 cases), and references.
The full names and acronyms of the 22 considered groups, with ages ranging from 4--8\,Ma (of the Chamaeleon-Scorpius-Centaurus-Crux complex) to 600--800\,Ma (of the Hyades and [TPY2019]~Group-X), are provided in the table notes.
Discoverer codes are given for all components tabulated by WDS (some stars that belong to different WDS systems can have different entries\footnote{For example, HD~1466 is SHY~113~G, SHY~114~G, and CVN~33~G.}), while the 199 additional companions have the string ``...'' in the discoverer code column. 
The spatial distribution of the 309 young stars and candidates in SKGs, associations, and open clusters is shown in Fig.~\ref{fig:spatialdistr}.

About 80\% of the 199 additional companions had also been ascribed to young groups, but not all.
We report 40 stars, marked with ``...'' in the group column in Table~\ref{tab:stars_young_skg}, that are new candidate members in young SKGs, associations, and clusters.
Eight of them had actually been considered as previous members, but the most recent works have classified them as ``improbable members'' (e.g. HD~207377~AB in Tucana-Horologium; \citealt{zuckerman01}). 
In any case, some of the 40 stars, because of either their brightness (e.g. HD~71043, which is probably an A0\,V, $G\approx$\,5.9\,mag, 200--300\,Ma-old star in Carina) or faintness (e.g. 2MASS~J12145318--5519494, which probably is a young brown dwarf in Lower Centaurus-Crux; \citealt{folkes12}), may be interesting to confirm in future works.

One more wide pair, composed by the bright stars $\gamma$~Cas and HD~5408, resulted in a nonuple system after our initial astrometric analysis and bibliographic search.
Since $\gamma$~Cas is an extremely young classical Be star (\citealt{poeckert78,white82,henrichs83,stee95}), we also tabulated the resolved pair components in Table~\ref{tab:stars_young_skg}, although it has never been ascribed to any group in particular (but see \citealt{mamajek17}). 
The $\gamma$~Cas system is discussed in further detail in Sect.~\ref{sec:gam_cas_association}.

Despite the far-reaching title of this series of papers, in this particular work, we focus on relatively evolved and old systems in the galactic field.
Common proper motion (and parallax) surveys of resolved companions to bona fide SKG members is indeed a widely used and successful technique for discovering new young stars and brown dwarfs \citep[][and references therein]{alonsofloriano15}.
However, a dedicated work on disentangling actual very wide binaries from unbound components in SKGs with similar galactocentric space velocities is planned.

After removing the 59 pairs with stars in young SKGs, associations, and clusters, we kept 94 wide pairs in the galactic field.
To them, we added the 40 additional astrometric companions found in our \textit{Gaia} DR3 search and not catalogued by WDS plus 39 already reported by WDS and separated by less than 1000\,arcsec.
As a result, there were 266 stars\footnote{HD~79392 is catalogued by WDS as the primary of two different systems (WDS 09150+3837/TOK 525 and WDS 09150+3837/DAM1575).} in 243 resolved \textit{Gaia} sources and in 94 systems that passed to the next step of our analysis.
All the systems and resolved \textit{Gaia} sources are listed in Tables~\ref{tab:galactic_field_system} and~\ref{tab:masses_rho_theta_u_systems}.

\subsection[The $\gamma$ Cas association]{The $\gamma$ Cas association\footnote{This section was originally published in the article \citet{gonzalezpayo23} as a separate appendix.}}
\label{sec:gam_cas_association}

\begin{figure}[h]
 \includegraphics[width=1\linewidth,angle=0]{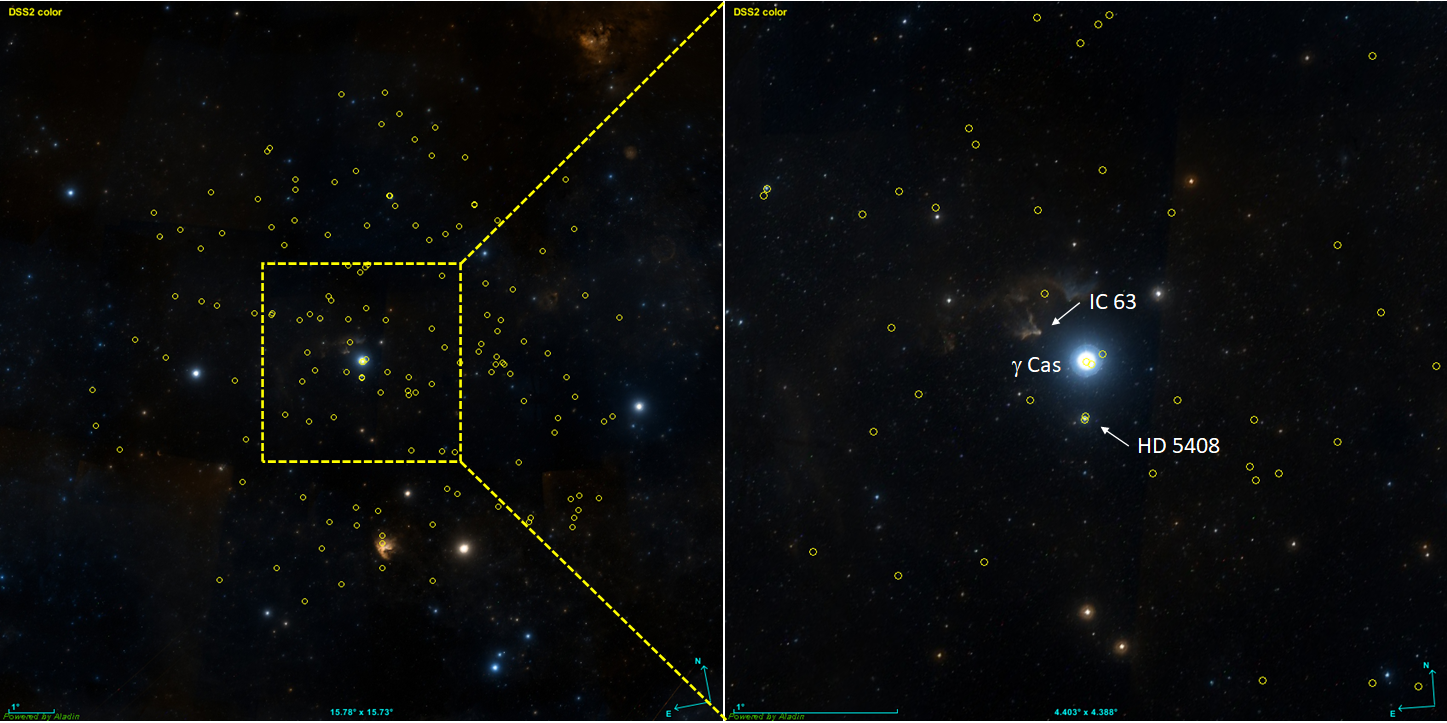}
 \caption[Spatial distribution of candidate young stars around $\gamma$~Cas.]{Spatial distribution of candidate young stars (open yellow circles) at less than 6\,deg ({\it left}) and 2\,deg ({\it right}) to $\gamma$~Cas.
 In the right panel we also highlight HD~5408 (the most massive star after $\gamma$~Cas) and the IC~63 emission nebula. 
 The images were created with the Aladin sky atlas and blue, red, and infrared Digitised Sky Survey data.}
 \label{fig:spatial_dist}
 \end{figure}

Figure~\ref{fig:spatial_dist} shows the spatial distribution around $\gamma$ Cas in two panels. The left panel shows the 145 stars in Table~\ref{tab:gamCas_candidates}, available from \textit{VizieR}, forming a circle area with a radius of 6\,degrees. The right panel reduce the radius to about 2\,degrees to have only 30 stars including $\gamma$ Cas. The IC 63 nebula emission, also named the ``Phantom nebula'', is easily observed in the right panel.

The stars $\gamma$~Cas and HD~5408, separated by 1274.5\,arcsec, constitute the wide pair MAM~20~AD \citep{mamajek17}.
They actually form a quintuple system, as they are reported to be close double (B0.5\,IV + F6\,V)\footnote{BU~499~AC is an optical pair, with ``ADS 782 C'' at 53\,arcsec to $\gamma$~Cas being a background star.} and triple (B7\,V + B9\,V + A1\,V) stars, respectively \citep{morgan43,osterbrock57,christy69,fekel79,nemravova12,hutter21}.

With such an early spectral type and an age of only about 8\,Ma \citep{zorec05}, $\gamma$~Cas is the ionising source of the nearby ($\rho$\,$\sim$\,1200\,arcsec) reflection nebulae IC~63 (The Ghost of Cassiopeia) and IC~59
\citep{hubble1922,sharpless59,jansen94}, as well as of an irregular, $\sim$3\,deg-diameter, H~{\sc ii} region \citep{karr05}.
With a mass of about 19\,M$_\odot$ and a surrounding disc, it is also the prototype of the $\gamma$~Cas type of stars (\citealt{poeckert78,stee95,naze22}).

Our astrometric search for common proper motion and parallax with the criteria in Sect.~\ref{sec:pair_validation} resulted in four additional stars not tabulated by WDS.
Of them, only one, namely UCAC4~752--011208 (M4\,Ve), had been catalogued in the literature \citep{nesci18}.
Together with the five components of the $\gamma$~Cas + HD~5408 system, they made an agglomerate of nine stars of 8\,Ma at about 188\,pc.
Since 19\,M$_\odot$-mass stars do not form in isolation \citep{kroupa01,chabrier03,penaramirez12}, we extended our astrometric search and found additional stars that satisfy our criteria.
In particular, we enlarged our search radius centred on $\gamma$~Cas in consecutive steps and, besides $\gamma$~Cas and HD~5408, we found 10, 30, 51, 85, 115, and 143 \textit{Gaia} DR3 stars with a 2MASS counterpart, and that satisfy our astrometric criteria, up to 1, 2, 3, 4, 5, and 6\,deg, respectively (Fig.~\ref{fig:spatial_dist}).
Some of these stars are in turn spectroscopic binaries, so their total number is larger.
Most selected stars follow the 10\,Ma theoretical isochrones of  PARSEC\footnote{CMD 3.7. A web interface dealing with stellar isochrones and their derivatives, \url{http://stev.oapd.inaf.it/cgi-bin/cmd}} \citep[][version 1.2S with the default values]{bressan12} at 188\,pc in \textit{Gaia}-2MASS colour-magnitude diagrams.
Besides, the mass function computed from masses derived from the $J$-band absolute magnitude and PARSEC models do not deviate too much from Salpeter's.
However, we did not find a clustering of stars towards the most massive stars, as it is usually observed in open clusters of similar age (e.g. \citealt{caballero08}).
The hypothetical stars of an open cluster or, more likely, a stellar association around $\gamma$~Cas \citep{mamajek17} overlaps with the extended and also young population of the Cas-Tau OB1 association \citep{blaaw91,dezeeuw99}.
As a result, additional work is necessary to disentangle the stars that were born together with $\gamma$~Cas and HD~5408.

\subsection{New close binary candidates from \textit{Gaia} data}
\label{sec:ruwe_vr}

We carried out a cross-matching with WDS and scoured the literature in search of additional companions not identified in our \textit{Gaia} DR3 search.
We did not find any additional WDS companion at $\rho <$\,1000 arcsec that were resolvable by \textit{Gaia} and that had not been recovered in our search.
However, WDS also tabulates very close systems ($\rho \lesssim$\,1.3\,arcsec) that were discovered and characterised with micrometers, speckle, lucky imaging, or adaptive optics, and which are unresolvable by \textit{Gaia} thanks to the close separation or relatively large magnitude difference between components (e.g. HD~6101, HD~102590, HD~186957).
In addition, there is a number of pair components that are spectroscopic binaries (e.g. HD~120510; \citealt{pourbaix04}) or triples (e.g. $\delta$~Vel, which is also an eclipsing binary with a close astrometric companion; \citealt{kervella13}), or very close binaries from proper motion anomalies (e.g. HD~125354; \citealt{kervella19}).
We further consider all this information in Sect.~\ref{sec:results_discussion_widest}.

Three pairs in multiple systems are in the 0.15--0.25 $\mu$ ratio buffer interval in the $\Delta$PA vs. $\mu$ ratio diagram (Fig.~\ref{fig:dpamu_ddd}), namely WDS~09487--2625, WDS~16278--0822, and WDS~23309--5807.
The origin of their large $\mu$ ratio lies on wide amplitude orbital (i.e. proper motion) variations induced by additional components in the systems at 1.83\,arcsec (HD 85043, 
I~205), $\sim$1.0\,arcsec ($\upsilon$~Oph, RST~3949), and 1.27\,arcsec (HD 221252, I~145) to the primaries or companions.
As a result, we also validated the three systems (one triple, one quadruple, and one quintuple) in spite of not satisfying our original $\mu$ ratio criterion.

Next, we cross-matched our 243 \textit{Gaia} sources in 94 systems with the {\it Hipparcos}-\textit{Gaia} catalogues of accelerations of \citet{kervella19} and \citet{brandt21}.
Of them, 54 have a measurable proper motion anomaly (Boolean variable set to unit in \textit{Gaia} DR2, proper motion anomaly binary flag \texttt{BinG2} --\citealt{kervella19}--, or \texttt{chi2} $>$ 11.8 --\citealt{brandt21}--) that are probably induced by unseen companions.
They are marked with a footnote in Tables~\ref{tab:galactic_field_system} and~\ref{tab:masses_rho_theta_u_systems}.

In addition, we looked for new very close binary candidates among the 94 systems.
First we used the \textit{Gaia} re-normalised unit weight error (\texttt{RUWE}), which is a robust indicator of the goodness of a star's astrometric solution \citep{arenou18,lindegren18a}.
Large \texttt{RUWE} values correspond to stars with angular separations small enough not to be resolved by \textit{Gaia}, but large enough to perturb the astrometric solution.
\textit{Gaia} DR3 provides \texttt{RUWE} values for 234 \textit{Gaia} entries.
The nine sources without \texttt{RUWE} values are either very bright stars ($\delta$~Vel, $\alpha$~Cen~A and B, $\upsilon$~Oph) or known close binaries with angular separations $\rho\sim$\,0.2--0.9\,arcsec (e.g. HD~6101).
There are 14 stars with \texttt{RUWE} $>$ 10.
They are also marked with a footnote in Tables~\ref{tab:galactic_field_system} and ~\ref{tab:masses_rho_theta_u_systems}.
The five \textit{Gaia} sources with the greatest \texttt{RUWE}, of about 20--40, are either already known close binaries below the \textit{Gaia} resolution limit (e.g. G~210--44, $\rho\sim$\,0.1\,arcsec -- HDS~2989 in the \textit{Hipparcos} Double Stars catalogue) or strong, relatively faint, new binary candidates (e.g. 2MASS~J02022892--3849021, UCAC3~109--11370, LSPM~J0956+0441, and HD~59438\,C). 
Of the other nine \textit{Gaia} sources with moderate \texttt{RUWE} of about 10--20, some have also been tabulated as candidate binaries, such as HD~75514 and HD~139696, which were listed in the {\it Hipparcos}–\textit{Gaia} catalogue of accelerations \citep{kervella19}, HD~210111, which is a $\lambda$~Bootis-type spectroscopic binary \citep{paunzen12}, and HD~215243, which is subgiant spectroscopic binary \citep{gorynya18}. 
The rest of \textit{Gaia} sources with \texttt{RUWE} $>$ 10 would need an independent confirmation of binarity.  
Being less conservative, we could have extended our analysis down to \texttt{RUWE} = 5, which is about three times greater than the critical value of 1.41 of \citet{arenou18}, \citet{lindegren18a}, or \cite{cifuentes20}.
There are only five \textit{Gaia} sources (in double or multiple systems) with 5\,$\leq$\,\texttt{RUWE}\,$\leq$\,10. However, as some careful studies of nearby stars indicate, \texttt{RUWE} values slightly larger than 1.4 do not necessarily translate into close binarity (\citealt{ramsay22,ribas23}. Since the confirmation of actual close binarity requires a radial-velocity or high-resolution imaging follow-up, we imposed a very conservative \texttt{RUWE} limit.

\begin{table}
 \centering
 \caption[Radial-velocity outlier candidates.]{Radial-velocity outlier candidates.}
 \footnotesize
 \label{tab:outliers}
  \scalebox{1}[1]{
\begin{tabular}{llllc}
    \noalign{\hrule height 1pt}
    \noalign{\smallskip}
WDS & Discoverer & Primary star & Companion star & $\lvert\Delta V_r\rvert$ \\
 & code &  &  & (km\,s$^{-\text{1}}$) \\
    \noalign{\smallskip}
    \hline
    \noalign{\smallskip}
01066+1353 & SHY 396 & HD 6566 & HD 5433$^{\text{(a)}}$ & 25.5$\pm$0.4 \\
02022--4550 & SHY 410 & HD 12586 & HD 12808 & 30.8$\pm$0.2 \\
02310+0823 & GIC  32 & G 4--24 & G 73--59$^{\text{(b)}}$ & 63.1$\pm$3.8 \\
02315+0106 & SHY 422 & BD+00 415B & HD 17000 & 4.4$\pm$0.3 \\
02462+0536 & TOK 651 & HD 17250 & HD 17163 & 18.1$\pm$0.6 \\
03503--0131 & SHY 164 & HD 24098A & HD 22584$^{\text{(a)}}$ & 5.9$\pm$0.2 \\
04346--3539 & TOK 488 & HD 29231 & L 447--2 & 25.5$\pm$0.4 \\
05222+0524 & TOK 497 & HD 35066[A]$^{\text{(a)}}$ & TYC 109--530--1 & 0.8$\pm$0.3 \\
08211+4021 & TOK 516 & BD+40 2030$^{\text{(a)}}$ & G 111--70 & 36.9$\pm$0.3 \\
08237--5519 & SHY 526 & HD 71257 & HD 72143$^{\text{(a)}}$ & 4.9$\pm$0.3 \\
08388--1315 & SHY 201 & HD 73583 & BD--09 2535 & 18.6$\pm$0.3 \\
08480--3115 & SHY 529 & HD 75269$^{\text{(a)}}$ & HD 75514$^{\text{(a,b)}}$ & 8.5$\pm$1.5 \\
09467+1632 & TOK 531 & BD+17 2130$^{\text{(a)}}$ & LP 428--36 & 56.6$\pm$2.7 \\
09487--2625 & TOK 532 & HD 85043A$^{\text{(a)}}$ & PM J09486--2644 & 1.8$\pm$0.3 \\
09568+0415 & TOK 533 & HD 86147 & LSPM J0956+0441$^{\text{(b)}}$ & 10.7$\pm$0.9 \\
10289+3453 & SHY 215 & HD 90681 & HD 92194 & 4.8$\pm$0.2 \\
10532--3006 & SHY 563 & HD 94375$^{\text{(a)}}$ & HD 94542$^{\text{(a)}}$ & 23.6$\pm$0.2 \\
11214+0638 & TOK 544 & HD 98697 & LP 552--34 & 19.4$\pm$3.5 \\
11455+4740 & LEP  45 & HD 102158 & G 122--46 & 19.6$\pm$0.6 \\
13305+2231 & SHY 626 & HD 117528 & BD+22 2587 & 93.2$\pm$0.2 \\
13470+3833 & SHY 633 & HD 120164 & HD 119767 & 21.6$\pm$0.2 \\
15120+0245 & WIS 281 & LP 562--9 & LP 562--10 & 24.5$\pm$3.7 \\
15208+3129 & LEP  74 & HD 136654 & AX CrB & 0.8$\pm$0.2 \\
15318--0204 & SHY 677 & HD 138370 & HD 138159 & 17.3$\pm$0.3 \\
15330--0111 & SHY 678 & 11 Ser & HD 142011 & 12.0$\pm$0.2 \\
15356+7726 & WIS 288 & LSPM J1535+7725 & LP 22--358 & 24.3$\pm$0.3 \\
15408--3252 & SHY 278 & HD 139696$^{\text{(a,b)}}$ & CD--32 10820 & 42.8$\pm$2.9 \\
15590+1820 & SHY 691 & HD 143292$^{\text{(a)}}$ & HD 142899 & 42.7$\pm$0.3 \\
16278--0822 & SHY 287 & $\upsilon$ Oph$^{\text{(a,c)}}$ & HD 144660 & 11.7$\pm$0.2 \\
17166+0325 & SHY 715 & HD 156287$^{\text{(a)}}$ & HD 159243 & 8.5$\pm$0.2 \\
18143--4309 & SHY 740 & HD 166793 & HD 166533 & 31.9$\pm$0.4 \\
18496+1313 & SHY 309 & HD 229635 & HD 229830 & 31.0$\pm$0.3 \\
18571+5143 & SHY 749 & HD 176341 & BD+49 2879 & 14.6$\pm$0.2 \\
18597+1615 & TOK 622 & HD 176441$^{\text{(a)}}$ & LSPM J1858+1613$^{\text{(b)}}$ & 19.3$\pm$0.5 \\
19290--4952 & SHY 319 & HD 182857 & HD 185112$^{\text{(a)}}$ & 14.0$\pm$0.3 \\
20084+1503 & LDS1033 & G 143--33 & G 143--27$^{\text{(c)}}$ & 66.3$\pm$1.7 \\
20371+6122 & SHY 780 & HD 196903 & HD 198662 & 12.4$\pm$0.2 \\
20404--3251 & SHY 781 & HD 196746$^{\text{(a)}}$ & HD 196189 & 26.2$\pm$0.2 \\
20489--6847 & SHY 782 & HD 197569 & HD 199760 & 7.6$\pm$0.2 \\
21105+2227 & SHY 793 & HD 201670 & HD 198759 & 54.4$\pm$17.8 \\
22175+2335 & GIC 179 & G 127--13$^{\text{(b)}}$ & G 127--14 & 21.7$\pm$0.9 \\
22220--3431 & SHY 802 & HD 212035 & HD 210111$^{\text{(c)}}$ & 11.4$\pm$0.5 \\
23506+5412 & SHY 840 & HD 223582$^{\text{(a)}}$ & HD 223788 & 1.7$\pm$0.2 \\
    \noalign{\smallskip}
    \noalign{\hrule height 1pt}
    \end{tabular}
 }
\begin{justify}
    \footnotesize{\textbf{\textit{Notes. }}}
    \footnotesize{$^{\text{(a)}}$ Stars with proper motion anomaly \citep{kervella19,brandt21}. $^{\text{(b)}}$ Stars with \texttt{RUWE}\,>\,10. $^{\text{(c)}}$ Known spectroscopic binaries.}
\end{justify}

\end{table}

Next, we used the standard deviation of the radial velocities, $V_r$, measured with the \textit{Gaia} Radial Velocity Spectrometer, which receives the misleading label \texttt{radial\_velocity\_error} \citep{gaiacollaboration23b,katz23}.
Of the 243 \textit{Gaia} entries, 182 have $V_r$ and its standard deviation, $\sigma_{Vr}$.
The median formal precision of the velocities for the brightest, most stable \textit{Gaia} stars lies at about 0.12\,km\,s$^{-\text{1}}$ to 0.15\,km\,s$^{-\text{1}}$ and smoothly increases for fainter stars \citep{katz23}. 
However, we identified at least six \textit{Gaia} sources that have significantly greater $\sigma_{Vr}$ than expected given their magnitudes. 
Being all stars of intermediate ages and spectral types in the main sequence (i.e. no pulsating giants or subgiants, nor very active T~Tauri stars), we adscribed the large $\sigma_{Vr}$ to spectroscopic binarity.
Actually, two of them had already been reported as spectroscopic binaries, namely HD~200077 (\citealt{konacki10,montes18a} and references therein) and HD~215243 \citep[][which also has a large \texttt{RUWE}]{gorynya18}.
A third one, namely HD~75514 \citep{kervella19,brandt21}, has a significant proper motion anomaly.
The other three new spectroscopic binary candidates have a large \texttt{RUWE} (8.1; BD+32~2868), 
moderate \texttt{RUWE} and $\sigma_{Vr}$ (2.48, 2.86\,km\,s$^{-\text{1}}$), or a small \texttt{RUWE} but a huge $\sigma_{Vr}$ for a bright single star ($G\approx$\,7.7\,mag, 17.76\,km\,s$^{-\text{1}}$; HD~201670).

At this stage, we may wonder why common parallax and proper motion criteria alone were used for system validation, instead of common radial velocities as well, at least for the 99 pairs with data for the two components.
We note that a large difference in radial velocities may be a symptom of long-period spectroscopic binarity of one of the components (by ``long period'', we mean longer than or of the same order of the 34 months of the \textit{Gaia} DR3 radial-velocity coverage).
Nevertheless, the above-mentioned properties of high \texttt{RUWE}, $\sigma_{Vr}$, or, especially, proper motion anomaly do not always indicate unknown close companions, but can also be produced by the already detected close companions. 
Some examples of known pairs with astrometric accelerations and orbital periods of tens to hundreds years are HD~6101, HD~59438, and HD~85043.

In Table~\ref{tab:outliers}, we list 43 pairs of primaries and companions with radial velocity differences larger than three times the quadratic sum of the respective $\sigma_{Vr}$.
Among them, we can find known spectroscopic binaries, components with large \texttt{RUWE} values, proper motion anomalies, or a combination of them.
Some of the pairs in Table~\ref{tab:outliers} may be false positives, that is, two unrelated stars with very similar proper motions and parallaxes but very different radial velocities.
However, with the data available to us, it is impossible to disentangle between them and true wide physical systems with one radial-velocity outlier component due to currently unknown long-period spectroscopic binarity.

\subsection{Colour-magnitude diagram}
\label{sec:HR_diagram}

\begin{figure}[H]
 \centering
 \includegraphics[width=0.7\linewidth, angle=0]{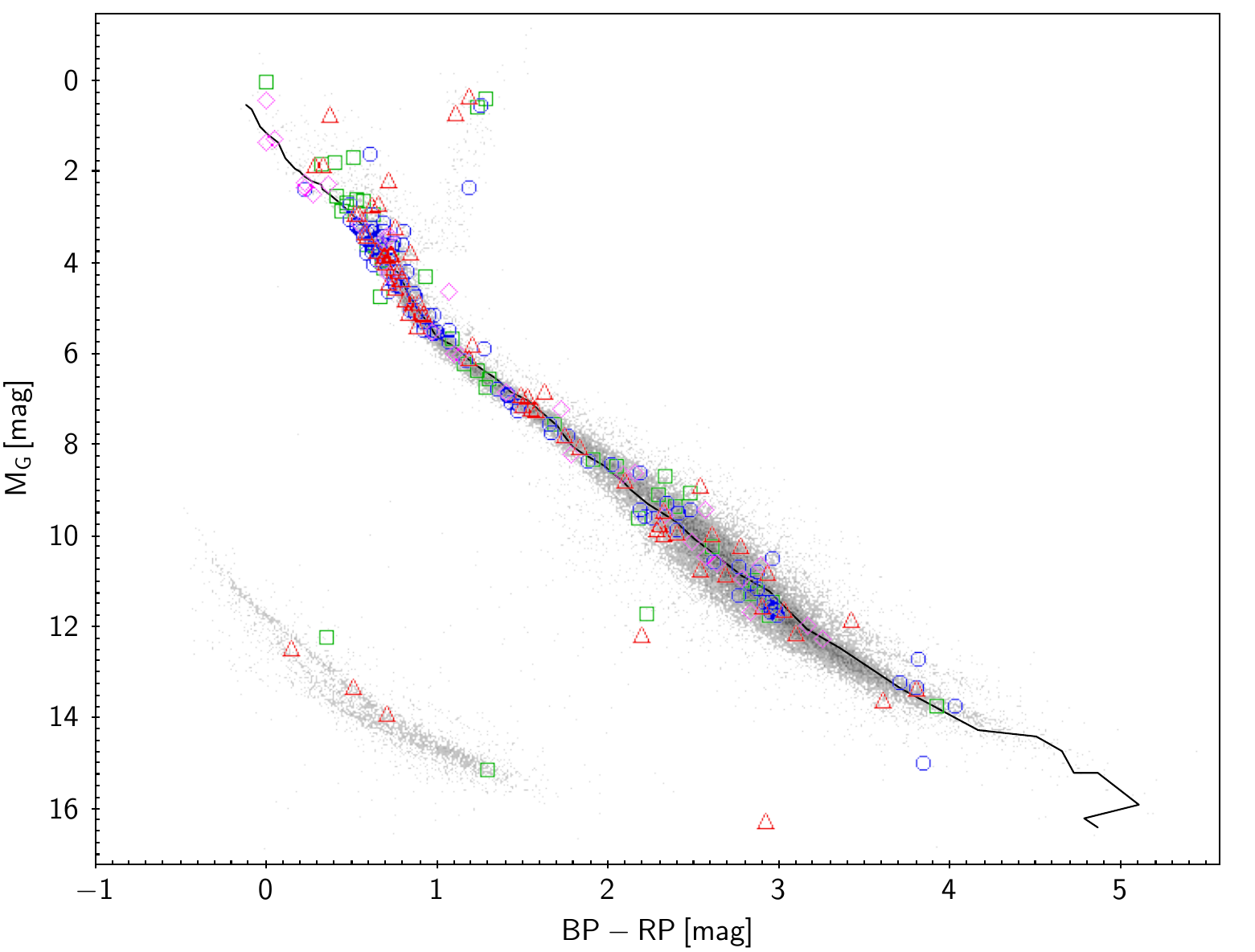}
 \caption[H-R diagram of all investigated stars.]{H-R diagram of all investigated stars.
 Coloured open symbols stand for stars in double (blue circles), triple (red triangles), quadruple (green squares), and higher-order multiple systems (purple diamonds).
 Grey dots represent selected field stars from \textit{Gaia}.
 The black solid line is the updated main sequence of \citet{pecaut13}.
 The stars outside the main sequence are discussed in the text.
}
 \label{fig:hrdplot}
\end{figure}

Stellar masses are needed to compute gravitational binding energies, while luminosity classes are needed to estimate stellar masses.
Estimated stellar ages are also needed to investigate the evolution of fragile multiple systems, while luminosity classes also shed light on stellar ages, especially outside the main sequence.
The luminosity class of the stars in the 243 \textit{Gaia} sources is illustrated by the Hertzprung-Russell (H-R) diagram of Fig.\,\ref{fig:hrdplot}.
We took parallaxes and $G$, $G_{BP}$, and $G_{RP}$ magnitudes from \textit{Gaia} DR3, except for four very bright stars ($\delta$~Vel, $\alpha$~Cen A and B, $\upsilon$~Oph), for which we estimated their magnitudes from their well-determined spectral types, published Johnson $B$, $V$, $R$ photometry, and the main-sequence colour-spectral type relation of \citet{pecaut13}.
We also plot this relation in the diagram, although the main sequence, together with the loci of white dwarfs and giants and subgiants beyond the turnoff point, is clearly marked by 57\,345 field stars with good \textit{Gaia} astrometry and photometry following the H-R example of \citet{taylor21}, but with the DR3 data set.

From their position in the H-R diagram, we identified eight giants and subgiants and five white dwarfs (listed in Table\,\ref{tab:giants}).
We confirmed their classification with a comprehensive bibliographic study.
Among the 13 stars, only one is part of a close pair unresolved by \textit{Gaia}, namely $\delta$~Vel.
Furthermore, all but one of the giants are so bright that were listed already by \citet{bayer1603} and \citet{flamsteed1725}.

Four of the five white dwarfs have a spectral type determination, with only one presented as a white dwarf candidate by \citet{gentilefusillo19}.
However, all of them are part of multiple systems (i.e. triple or higher).
For example, WDS~01024+0504 is made of two spectroscopic binaries, namely the double, early K dwarf HD~6101 and the double, DA5.9 white dwarf EGGR~7 \citep{giclas59,maxted00,lajoie07,caballero09,gianninas11,toonen17},
while WDS~06536-3956 is made of the early M dwarf L~454--11 \citep{lepine11} and two white dwarfs, WT~201 (DA8.0) and WT~202 (DA7.0) \citep{subasavage08}.
The system may also be quadruple because L~454--11 has a \texttt{RUWE} value of 18.0.
The other two white dwarfs are in triple and quadruple systems.

\begin{table}
 \centering
 \caption[Giants and white dwarfs in wide double and multiple systems.]{Giants and white dwarfs in wide double and multiple systems.}
 \footnotesize
 \scalebox{1}[1]{
 \begin{tabular}{@{\hspace{1mm}}l@{\hspace{2mm}}l@{\hspace{1mm}}l@{\hspace{1mm}}l@{\hspace{1mm}}c@{\hspace{1mm}}c@{\hspace{1mm}}c@{\hspace{1mm}}c@{\hspace{1mm}}}
 \noalign{\hrule height 1pt}
 \noalign{\smallskip}
 Star & Spectral & WDS & Discoverer & $\alpha$\,(J2000) & $\delta$\,(J2000) & $M$ & Age \\
 & type &  & code & (hh:mm:ss.ss) & (dd:mm:ss.s) & (M$_\odot$) & (Ga) \\
 \noalign{\smallskip}
 \hline
 \noalign{\smallskip}
 \multicolumn{8}{c}{\it Giants}\\
 \noalign{\smallskip}
 \hline
 \noalign{\smallskip}
$\delta$ Vel Aa & A2\,IV & 08447--5443 & SHY 49 & 08:44:42.23 & $-$54:42:31.7 & 3.19$\pm$0.03$^{\text{(a)}}$ & $\sim$0.431$^{\text{(a)}}$ \\
HD 120164 & K0\,III & 13470+3833 & SHY 633 & 13:46:59.77 & +38:32:33.7 & 2.42$\pm$0.24$^{\text{(b)}}$ & $\sim$0.7$^{\text{(b)}}$ \\
$\iota$ Vir & F7\,III & 14190--0636 & SHY 71 & 14:16:00.87 &$-$06:00:02.0 & $\sim$1.81$^{\text{(c)}}$ & 1.809$\pm$0.001$^{\text{(d)}}$ \\
11 Ser & K0\,III & 15330--0111 & SHY 678 & 15:32:57.94 & $-$01:11:11.0 & 1.27$\pm$0.35$^{\text{(e)}}$ & 2.75$^{+\text{0.88}}_{-\text{0.66}}$\,$^{\text{(e)}}$ \\
64 Aql & K1\,III--IV & 20080--0041 & SHY 325 & 20:08:01.82 & $-$00:40:41.5 & 1.00$\pm$0.27$^{\text{(e)}}$ & 9.33$\pm$4.17$^{\text{(e)}}$ \\
$\nu$ Aqr & K0\,III & 21096--1122 & TOK 633 & 21:09:35.64 & $-$11:22:18.1 & 2.01$_{-\text{0.11}}^{+\text{0.04}}$\,$^{\text{(f)}}$ & 1.26$_{-\text{0.19}}^{+\text{0.22}}$\,$^{\text{(f)}}$ \\
$\kappa$ Aqr & K1.5\,III & 22378--0414 & TOK 640 & 22:37:45.38 & $-$04:13:41.0 & 2.55$\pm$0.13$^{\text{(g)}}$ & 2.79$\pm$1.16$^{\text{(h)}}$ \\
$\iota$ Cep & K1\,III & 22497+6612 & SHY 359 & 22:49:40.81 & +66:12:01.4 & 1.55$_{-\text{0.20}}^{+\text{0.05}}$\,$^{\text{(f)}}$ & 2.57$_{-\text{0.38}}^{+\text{0.18}}$\,$^{\text{(f)}}$ \\
 \noalign{\smallskip}
 \hline
 \noalign{\smallskip}
 \multicolumn{8}{c}{\it White dwarfs}\\
 \noalign{\smallskip}
 \hline
 \noalign{\smallskip}
EGGR 7 & DA5.9 & 01024+0504 & WNO 50 & 01:03:49.92 & +05:04:30.6 & $\sim$ 0.77$^{\text{(i)}}$ &  ... \\
WT 202 & DA7.0 & 06536--3956 & SUB 2 & 06:53:35.44 & $-$39:55:34.8 & 0.64$\pm$0.02$^{\text{(j)}}$ & 2.4$_{-\text{0.1}}^{+\text{1.0}}$\,$^{\text{(j)}}$ \\
WT 201 & DA8.0 & 06536--3956 & SUB 2 & 06:53:30.21 & $-$39:54:29.1 & 0.64$\pm$0.02$^{\text{(j)}}$ & 3.2$_{-\text{0.1}}^{+\text{1.1}}$\,$^{\text{(j)}}$ \\
Gaia DR3 812109085097488768 & ...$^{\text{(k)}}$ & 09150+3837 & DAM1575 & 09:14:58.95 & +38:36:58.3 & 0.5$\pm$0.1$^{\text{(l)}}$ & ... \\
SDSS J230056.41+640815.5 & DC & 22497+6612 & ... & 23:00:56.46 & +64:08:16.0 & 0.5$\pm$0.1$^{\text{(l)}}$ & ... \\
 \noalign{\smallskip}
 \noalign{\hrule height 1pt}
 \end{tabular}
 }
 \label{tab:giants}
 \begin{justify}
 \footnotesize{\textbf{\textit{Notes. }}
    $^{\text{(a)}}$ \citet{david15};
    $^{\text{(b)}}$ \citet{dasilva15};
    $^{\text{(c)}}$ \citet{gontcharov10};
    $^{\text{(d)}}$ \citet{eker18};
    $^{\text{(e)}}$ \citet{feuillet16};
    $^{\text{(f)}}$ \citet{stock18};
    $^{\text{(g)}}$ \citet{kervella19};
    $^{\text{(h)}}$ \citet{soubiran08};
    $^{\text{(i)}}$ \citet{lajoie07};
    $^{\text{(j)}}$ Rebassa- Mansergas (priv. comm.);
    $^{\text{(k)}}$ \citet{gentilefusillo19};
    $^{\text{(l)}}$ This work.}
 \end{justify}
\end{table}

Apart from the 13 giants, subgiants, and white dwarfs in Table\,\ref{tab:giants}, there are still some objects lying outside the main sequence in the colour-magnitude diagram of Fig.~\ref{fig:hrdplot}.
In particular, there are 4 sources apparently below the main sequence.
The origin of this discrepancy lies in the four cases on wrong photometry:
2MASS J02004917--3848535 in the double system WDS~02025--3849 is an $\sim$M7--8 ultra-cool dwarf with $G_{BP}$ fainter than the \textit{Gaia} limit \citep{smart19}; 
Gaia DR3 749786356557791744 in the triple system WDS 10289+3453 is another ultra-cool dwarf with $G_{BP}$ fainter than the \textit{Gaia} limit, but with a spectral type at the M-L boundary; 
Gaia DR3 3923191426460144896 in the quadruple system WDS~11486+1417 is an $\sim$M4--5 late-type dwarf at $\rho\approx$\,10.1\,arcsec of the very bright ($G\approx$\,5.9\,mag) A8+G2 binary HD~102590;
and Gaia DR3 1367008242580377216 in the triple system WDS~17415+4924 is an $\sim$M4--5 late-type dwarf with a relatively high value of $G_{BP}/G_{RP}$ excess factor, \texttt{E(BP/RP)}, which is an indicator of systematic errors in photometry \citep{riello18}.
Remarkably, 2 of the 4 \textit{Gaia} sources with the wrong photometry are the least massive stars in our sample (Sect.~\ref{sec:masses_gravitational_bindings}).
The rest of the \textit{Gaia} sources, which are especially redder than the subgiant turnoff point, are reasonably  matched to the main sequence.

\subsection{Masses and gravitational binding energies}
\label{sec:masses_gravitational_bindings}

For stars in the main sequence, we determined stellar masses, M, from the $G$-band absolute magnitude, \textit{Gaia} and 2MASS colours, spectral types, and the updated version of table~5 in \citet{pecaut13}\footnote{\url{https://www.pas.rochester.edu/~emamajek/EEM_dwarf_UBVIJHK_colors_Teff.txt}}.
The match between spectral types derived by us from colours and absolute magnitudes and compiled from the bibliography is excellent (although we estimated spectral types for some \textit{Gaia} sources that had previously gone unreported in the literature).
In the case of unresolved (spectroscopic binaries and close WDS astrometric binaries), very bright stars (e.g. $\alpha$~Cen A and B), giants, subgiants, and white dwarfs (Table\,\ref{tab:giants}), we compiled mass values from the bibliography \citep[e.g.][]{lajoie07,soubiran08,feuillet16,eker18,stock18,gentilefusillo19}.
If unavailable, we determined their mass from colours and absolute magnitudes by assuming either two equal-mass stars in double-lined spectroscopic binaries or that the mass of the companion, M$_\text{2}$ is much less than the mass of the primary, M$_\text{1}$, in single-lined spectroscopic binaries \citep{latham02}.
Because of this na\"ive approach, we established an uncertainty of 10\% for our mass values \citep{mann19,schweitzer99}, which may actually be larger in poorly investigated, single-lined spectroscopic binaries. 
In only two cases, namely, of white dwarfs without a public mass determination, we estimated their mass as in \citet{rebassamansergas21}.
For the giants, subgiants, and white dwarfs we also compiled ages from the literature, as summarised in the last column of Table\,\ref{tab:giants}; such ages can be extrapolated to their wide companions.
While the masses of the white dwarfs vary between about 0.5 and 0.8\,M$_\odot$ and of the giants between 1.0 and 3.2\,M$_\odot$, the masses of the stars on or near the main sequence vary from about 0.08 to 2.8\,M$_\odot$.
The latter extremes correspond to the new ultra-cool dwarf Gaia DR3 749786356557791744 at the M-L boundary, which is at 13.7\,arcsec to the solar-like HD~90681 star and the B9.5\,IV HD~188162, which is the most massive star of a septuple system candidate (Sect.~\ref{sec:results_discussion_widest}).

Next, we computed the projected physical separation, $s$, between every two \textit{Gaia}-resolved components in each pair from the angular separation, $\rho$, and the distance, $d$, to the primary.
Given the wide separations considered, instead of using the $s \approx d ~ \rho$ approximation, we used instead the exact definition from the trigonometry:
\begin{equation}
\label{Eq_s}
s = d ~ \sin{\rho}.
\end{equation}
We considered the distance to the primary star (which usually has the smallest parallax uncertainty) as the distance to the whole system.
The determined $s$ vary from $\sim$11\,au in the case of nearby, close astrometric binaries (e.g. HD~6101), to $\sim$2.3 $\cdot$ 10$^\text{6}$\,au (about 11\,pc) in the case of the very widest companions (see below).
The uncertainty in $s$ is underestimated for primaries whose parallaxes may be affected by close binarity.

Finally, we determined reduced binding energies of the wide systems as in \citet{caballero09}:
\begin{equation}
|U^*_g| = G\frac{M_\text{1} \cdot M_\text{2}}{s},
\label{eqn:binding_WDS}
\end{equation}
They are ``reduced'' because we used the projected physical separation for computing $|U^*_g|$ instead of the actual separation or the semi-major axis, $a$, which is unknown.
We did not apply a most probable conversion factor between $a$ and $s$ for easier computation and, especially, comparisons with previous works \citep{close03,burgasser07a,radigan09,caballero10,faherty10}.
This conversion factor, resulting from a uniform distribution of tridimensional vectors projected on a bidimensional plane \citep{abt76,fischer92}, would lead to about 26\% larger actual separations and, therefore, 26\% smaller (non-reduced) binding energies\footnote{\citet{fischer92} determined the statistical correction $\overline{a} \approx \text{1.26} ~ \overline{s}$ between projected separation ($s$)
and true separation ($a$) from Monte Carlo
simulations over a full suite of binary parameters.} \citep{dhital10,oelkers17}.

The resulting $M_\text{1}$, $M_\text{2}$, $\rho$, $\theta$, $s$, and $|U^*_g|$ are listed in Table~\ref{tab:masses_rho_theta_u_systems}. 
We computed $|U^*_g|$ only for systems with double-like hierarchy, that is, actual doubles and multiple systems with $\rho_{\text{1}, \rm wide}$\,$\gg$\,$\rho_{\text{1},i}$.
Here, ``wide'' indicates resolved companions at $\rho_{\text{1}, \rm wide}$\,>\,1000\,arcsec and ``$i$'' other components.
As a result, we did not compute $|U^*_g|$ of 14 multiple systems with $\rho_{\rm wide}$\,$\sim$\,$\rho_{\text{1},i}$, which we called trapezoidal systems or trapezia.

\section{Results and discussion}
\label{sec:results_discussion_widest}

Among the 155\,159 pairs contained in the WDS catalogue at the time of our analysis, 153 pairs with com\-mon-parallax, common-proper motion, ultrawide components at $\rho >$ 1000\,arcsec passed our astrometric criteria in Sect.~\ref{sec:pair_validation}, of which 59 (38.6$\pm$9.8\%) are part of young SKGs, associations, or open clusters (Table~\ref{tab:stars_young_skg}), and 95 (61.4$\pm$12.4\%) are ultrawide pairs in 94 galactic systems --one triple is made of two pairs with $\rho >$ 1000\,arcsec and different WDS entries (see Sect.~\ref{sec:search_additional}), which makes 95 WDS pairs.
Because of the small sample size, we used the Wald interval \citep{agresti98} with a 95\% of confidence to calculate the ratio uncertainties\footnote{Wald 95\% confidence interval is $(\lambda - \text{1.96}\sqrt{\lambda/n}, \lambda + \text{1.96}\sqrt{\lambda/n})$, where $\lambda$ is the number of successes in $n$ trials.}. 
To the galactic systems, we added 39 companions from the literature and separated by $\rho <$ 1000\,arcsec.
In our \textit{Gaia} DR3 search, we also found 39 additional astrometric companions not catalogued by WDS; that is, we found new companions in about a quarter of the investigated systems. 
In contrast, WDS tabulated a number of additional companion candidates with accurate \textit{Gaia} DR3 data that did not pass our conservative astrometric criteria (Sect.\,\ref{sec:pair_validation}).
Most, but not all, of them are flagged by WDS with ``\texttt{U}'' (proper motion or other technique indicates that this pair is non-physical).

The 94 galactic field systems and their components are listed in Tables\,\ref{tab:galactic_field_system} (basic astrometry and photometry) and~\ref{tab:masses_rho_theta_u_systems} (stellar masses, angular and projected physical separations, position angles, and binding energies).
We remark that we reclassified the stars in Tables\,\ref{tab:galactic_field_system} and~\ref{tab:masses_rho_theta_u_systems} as primaries, secondaries, tertiaries, and so on, according to their $G$-band magnitudes. As a result, the WDS nomenclature ``A'', ``B'', ``C'' (etc.) does not always match our re-ordering.

Among the 94 galactic field systems, there are 48 double, 24 triple, 14 quadruple, 2 quintuple, 3 sextuples, and 2 septuples.
The corresponding minimum multiplicity order rates are displayed in Table~\ref{tab:multiplicity}.
The estimated multiplicity order rates and, therefore, the number of multiple systems increase significantly at the expense of the number of doubles if the new candidate companions with large \texttt{RUWE}, $\sigma_{Vr}$, or proper motion anomaly are included (Sect.~\ref{sec:ruwe_vr}).
When these close binary candidates are taken into account, most of the ultrawide systems (67.8\%) become multiple: 23.7\% are triple, 21.5\% are quadruple, and 22.6\% have a higher multiplicity order.
These rates are far greater than what is found in less separated multiple systems in the field \citep{tokovinin97,chaname04,duchene13}.
Such a higher-than-usual multiplicity order implies a larger total mass, which, in turn, implies a larger binding energy.

\begin{table}[H]
 \centering
 \caption[Multiplicity order rates of ultrawide galactic systems.]{Multiplicity order rates of ultrawide galactic systems.}
 \footnotesize
 \scalebox{1}[1]{
 \begin{tabular}{lcc}
 \noalign{\hrule height 1pt}
 \noalign{\smallskip}
System type & Minimum rate$^{\text{(a)}}$ & Estimated rate$^{\text{(b)}}$ \\
 & (\%) & (\%) \\
 \noalign{\smallskip}
 \hline
 \noalign{\smallskip}
 Double & 51.6 $\pm$ 14.6 &  32.3 $\pm$ 11.5 \\
 Triple & 25.8 $\pm$ 10.3 &  23.7 $\pm$ 9.9 \\
 Quadruple & 15.1 $\pm$ 7.9 &  21.5 $\pm$ 9.4 \\
 Quintuple or higher & 7.5 $\pm$ 5.6 &  22.6 $\pm$ 9.7 \\
 \noalign{\smallskip}
 \noalign{\hrule height 1pt}
 \end{tabular}
 }
 \label{tab:multiplicity}
 \begin{justify}
 \footnotesize{\textbf{\textit{Notes. }}
    $^{\text{(a)}}$ Multiplicity order rate including only systems resolved by \textit{Gaia} or tabulated by WDS.
   $^{\text{(b)}}$ Multiplicity order rate including also close binary candidates with large \texttt{RUWE}, $\sigma_{Vr}$, or proper motion anomaly.}
  \end{justify}  
\end{table}

\begin{figure}[H]
    \centering
    \includegraphics[width=0.64\linewidth]{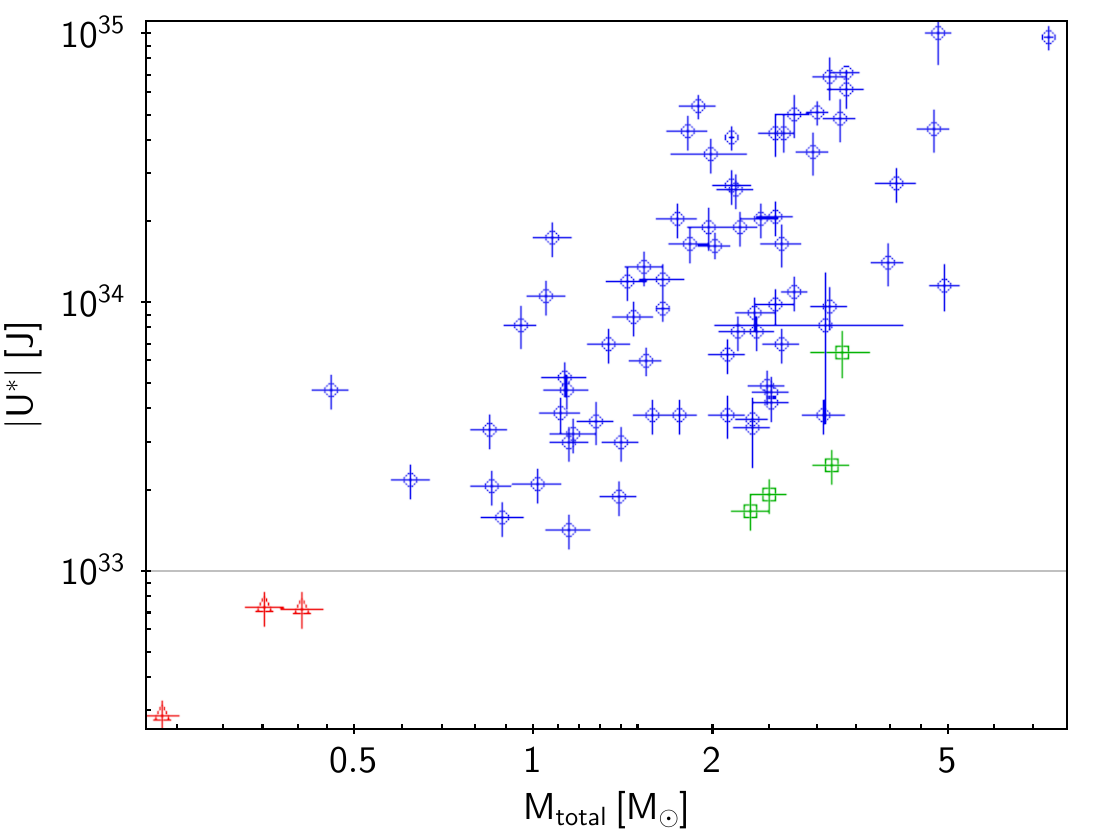}
    \includegraphics[width=0.64\linewidth]{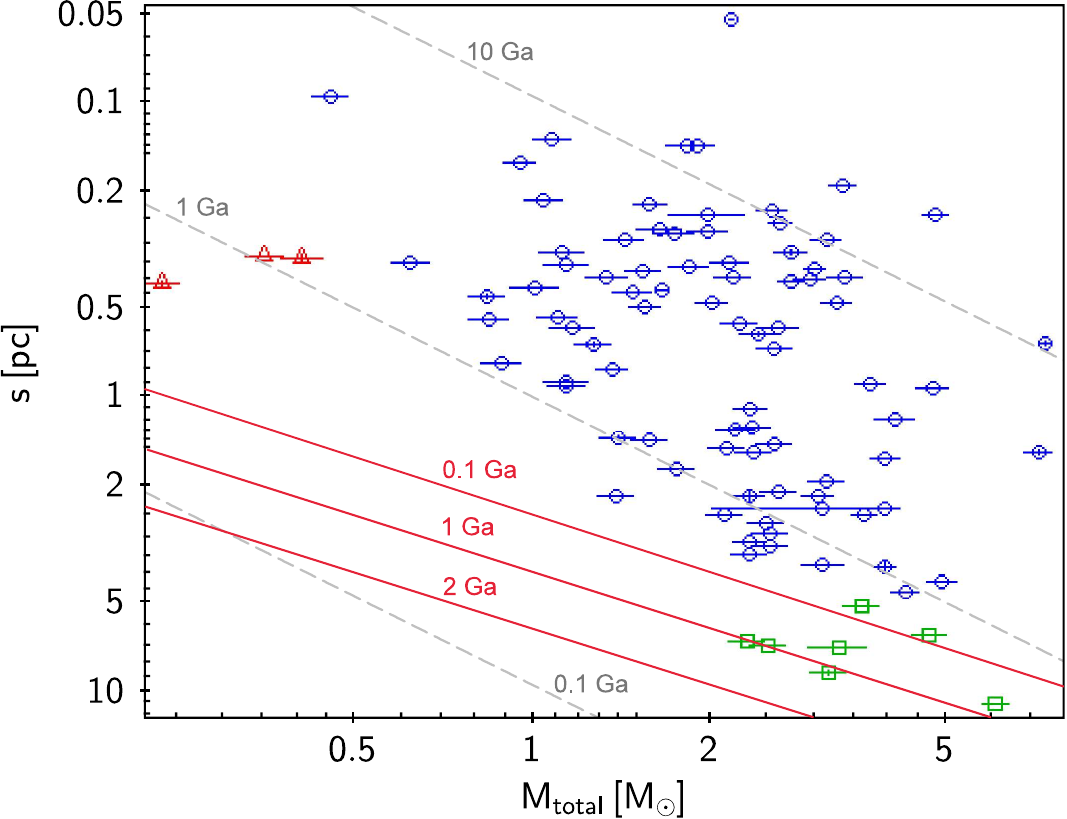}
  \vspace{2mm}
 \caption[Reduced binding energy and projected physical separation as functions of ultrawide system total mass.]{Reduced binding energy ({\it top}) and projected physical separation ({\it bottom}) as functions of ultrawide system total mass.
 In both panels, the three most fragile systems (top of Table~\ref{tab:fragile}) and the most separated systems (bottom of Table~\ref{tab:fragile}) are plotted in red triangles and green squares, respectively, while the rest of investigated ultrawide systems are plotted in blue circles.
 The error bars in $s$ are smaller than the used symbols.
 In the {\it top} panel, the horizontal line marks the limit of $|U^*_g|$ at $10^{\text{33}}$\,J.
 In the {\it bottom} panel, the grey dashed diagonal lines mark the statistical maximum ages of 0.1, 1, and 10\,Ga at which the systems are likely bound (computed with Eq.~\ref{eqn:seplimit}), while the red solid diagonal lines mark the corresponding orbital periods of 0.1, 1, and 2\,Ga (computed with Kepler's third law). 
 In both cases we used the correction $a\approx$\,1.26\,$s$ \citep{fischer92}.
}
\label{fig:massenergysep}
\end{figure}

In the left panel of Fig.~\ref{fig:massenergysep}, we display the minimum reduced gravitational binding energy of the 80 systems for which we were able to compute their $|U^*_g|$ as a function of the total mass in the system, M$_{\rm total} = \sum \text{M}_i$, with $i=$ 1 : 7 (i.e. all except for the 14 trapezia).
This diagram would be complete only by adding systems with angular separations $\rho<$\, 1000\,arcsec but with very low masses \citep[e.g.][]{caballero07a,caballero07b,artigau07,rica12}.
It is complete, however, at the highest total masses and lowest binding energies.
Actual total masses and binding energies, when close binary candidates are taken into account, are larger.

\begin{table}[H]
 \centering
 \caption[The most fragile and the most separated systems.]{The most fragile ($|U^*_g|<$\,10$^{\text{33}}$\,J) and the most separated ($s\geq$\,5\,pc) systems.}
 \footnotesize
 \scalebox{1}[1]{
 \begin{tabular}{@{\hspace{2mm}}c@{\hspace{3mm}}l@{\hspace{2mm}}l@{\hspace{2mm}}c@{\hspace{2mm}}c@{\hspace{2mm}}c@{\hspace{2mm}}c@{\hspace{2mm}}}
 \noalign{\hrule height 1pt}
 \noalign{\smallskip}
WDS & Discoverer & Star & $M$ & \textit{d}$^{\text{(a)}}$ & \textit{s}$^{\text{(b)}}$ & $|U^*_g|$ \\
 & code &  & (M$_\odot$) & (pc) & (10$^\text{3}$\,au) & (10$^{\text{33}}$\,J) \\
 \noalign{\smallskip}
 \hline
 \noalign{\smallskip}
 \multicolumn{7}{c}{\textit{The most fragile systems}}\\
 \noalign{\smallskip}
 \hline
 \noalign{\smallskip}
00016--0102 &  & 2MASS J00013688-0101441 & 0.22$\pm$0.02 & \multirow{2}{*}{60.33$\pm$0.13} &  \multirow{2}{*}{68.5$\pm$0.2} & \multirow{2}{*}{0.74$\pm$0.10}  \\
 & WIS 1 & SIPS J0000-0112 & 0.13$\pm$0.01 & & & \\
 \noalign{\smallskip}  
 \hline
 \noalign{\smallskip}
02025--3849 &  & 2MASS J02022892-3849021$^{\text{(c)}}$ & 0.32$\pm$0.03 & \multirow{2}{*}{59.46$\pm$2.29} & \multirow{2}{*}{69.3$\pm$2.7} & \multirow{2}{*}{0.72$\pm$0.11} \\
 & WIS 248 & 2MASS J02004917-3848535 & 0.09$\pm$0.01 & & & \\
 \noalign{\smallskip}
  \hline
 \noalign{\smallskip}
15488+4929 & & LSPM J1548+4928 & 0.12$\pm$0.01 & \multirow{2}{*}{76.81$\pm$0.63} & \multirow{2}{*}{85.0$\pm$0.7} & \multirow{2}{*}{0.29$\pm$0.04} \\
 & WIS 295 & LSPM J1550+4921 & 0.11$\pm$0.01 & & & \\
 \noalign{\smallskip}
 \hline
 \noalign{\smallskip}
 \multicolumn{7}{c}{\textit{The most separated systems}}\\  
 \noalign{\smallskip} 
 \hline
 \noalign{\smallskip} 
02315+0106 &  & HD 15695 & 1.75$\pm$0.18 & \multirow{6}{*}{105.52$\pm$0.33} & \multirow{6}{*}{2284.7$\pm$7.2} & \multirow{6}{*}{...} \\
 & STF 274 & BD+00 415B & 1.63$\pm$0.16 &  &  &  \\
 & SHY 422 & HD 17000 & 1.14$\pm$0.11 &  &  &  \\
 & ... & HD 16985 & 1.07$\pm$0.11 &  &  &  \\
 & ... & Gaia DR3 2497835645142616192$^{\text{(d)}}$ & 0.30$\pm$0.03 &  &  &  \\
 & ... & Gaia DR3 2514005200579732608 & 0.20$\pm$0.02 &  &  &  \\
 \noalign{\smallskip} 
 \hline
 \noalign{\smallskip}
07166--2319 &  & HD 56578$^{\text{(e)}}$ & 2.42$\pm$0.24 & \multirow{3}{*}{106.32$\pm$0.47} & \multirow{3}{*}{1340.9$\pm$6.0} & \multirow{3}{*}{...} \\
 & SHY 508 & HD 57527$^{\text{(e)}}$ & 1.92$\pm$0.19 &  &  &  \\
 & ... & Gaia DR3 5613164850183516544 & 0.35$\pm$0.03 &  &  &  \\
 \noalign{\smallskip} 
 \hline
 \noalign{\smallskip}
10532--3006 &  & HD 94375$^{\text{(e)}}$ & 1.32$\pm$0.13 & \multirow{2}{*}{82.17$\pm$0.16} & \multirow{2}{*}{1439.6$\pm$2.8} & \multirow{2}{*}{1.91$\pm$0.27} \\
 & SHY 563 & HD 94542$^{\text{(e)}}$ & 1.19$\pm$0.12 &  &  &  \\
 \noalign{\smallskip} 
 \hline
 \noalign{\smallskip}
15330--0111 &  & 11 Ser & 1.27$\pm$0.35 & \multirow{4}{*}{83.61$\pm$0.42} & \multirow{4}{*}{1483.0$\pm$7.4} & \multirow{4}{*}{3.08$\pm$0.62} \\
 & SHY 678 & HD 142011 & 1.21$\pm$0.12 &  &  &  \\
 & ... & Gaia DR3 4403070145373483392 & 0.43$\pm$0.04 &  &  &  \\
 & ... & Gaia DR3 4403070149671286272$^{\text{(d)}}$ & 0.40$\pm$0.04 &  &  &  \\
 \noalign{\smallskip} 
 \hline
 \noalign{\smallskip}
17166+0325 &  & HD 156287$^{\text{(e)}}$ & 1.24$\pm$0.12 & \multirow{2}{*}{82.15$\pm$0.17} & \multirow{2}{*}{1407.6$\pm$2.8} & \multirow{2}{*}{1.68$\pm$0.24} \\
 & SHY 715 & HD 159243 & 1.08$\pm$0.11 &  &  &  \\
 \noalign{\smallskip} 
 \hline
 \noalign{\smallskip}
21105+2227 &  & HD 201670$^{\text{(d)}}$ & 1.74$\pm$0.17 & \multirow{2}{*}{113.81$\pm$0.97} & \multirow{2}{*}{1783.3$\pm$15.2} & \multirow{2}{*}{2.49$\pm$0.35} \\
 & SHY 793 & HD 198759 & 1.45$\pm$0.15 &  &  &  \\
 \noalign{\smallskip} 
 \hline
 \noalign{\smallskip}
22497+6612 &  & $\iota$ Cep & 1.55$\pm$0.20 & \multirow{4}{*}{36.65$\pm$0.18} & \multirow{4}{*}{1061.3$\pm$5.1} & \multirow{4}{*}{...} \\
 & SHY 359 & HD 215588 & 1.23$\pm$0.12 &  &  &  \\
 & ... & UCAC3 297-187960 & 0.35$\pm$0.03 & & & \\
 & ... & SDSS J230056.41+640815.5 & 0.50$\pm$0.10 &  &  &  \\
 \noalign{\smallskip}
 \noalign{\hrule height 1pt}
 \noalign{\smallskip}
 \end{tabular}
 }
 \label{tab:fragile}
\begin{justify}
    \footnotesize{\textbf{\textit{Notes. }}
    $^{\text{(a)}}$ Distance of the primary star.
    $^{\text{(b)}}$ Maximum separation between stars inside the system.
    $^{\text{(c)}}$ \texttt{RUWE}$>$\,10.
    $^{\text{(d)}}$ Large $\sigma_{Vr}$ for its $G$ magnitude.
    $^{\text{(e)}}$ Proper motion anomaly measured by \citet{kervella19}, \citet{brandt21} or both.}
\end{justify}
\end{table}

There are three systems with $|U^*_g|<$ 10$^{\text{33}}$\,J, significantly lower that those of the other 77 systems.
They are listed at the top of Table~\ref{tab:fragile} with their WDS identifiers, discoverer codes (i.e. Wide-field Infrared Survey Explorer, WIS, \citealt{kirkpatrick16}), Simbad names, stellar masses, distances, projected physical separations, and gravitational binding energies.
The three systems are doubles composed of M3--6\,V primaries and M5--9\,V secondaries.
These spectral types were estimated by us from $M_G$ from the relations of \citet{pecaut13} and \citet{cifuentes20}, except for the secondary star in WDS~15488+4929, namely, LSPM~J1550+4921, whose spectral type M7.0\,V was determined by \citet{west11} from low-resolution spectroscopy.
 
With a mass of about 0.09\,M$_\odot$, the secondary in the system WDS~02025--3849, namely, 2MASS J02004917--3848535 ($\sim$M7--8\,V), is the second-least massive star in our whole sample.
The low masses of the system components and the wide separations, of about 68--85 $\cdot$ 10$^\text{3}$\,au (six to eight times wider than $\alpha$~Cen + Proxima), explain the very low $|U^*_g|$.
Actual binding energies may be larger, as the primary in WDS~02025--3849 has a \texttt{RUWE} = 40.7; assuming an equal-mass binary, the corrected binding energy would double.
None of the three systems have radial-velocity determinations (from \textit{Gaia} DR3 and \citealt{west11}) for the two resolved components. 
Given their relatively large $\mu$ ratios and $\Delta$PA (but within our boundary conditions), a dedicated radial-velocity follow-up would be necessary to ascertain whether the three fragile binaries are actually triples. 

Even if each of the three systems had an additional component and, therefore, higher total masses and binding energies than estimated above, there seems to be a lower boundary of $|U^*_g|$ for the most fragile systems at about 10$^{\text{33}}$\,J (first mentioned by \citealt{caballero10}).
This lower limit may be a consequence of the tidal disruption of wide systems by the galactic gravitational potential, via energy and momentum exchange in encounters with other stars or even the interstellar medium \citep{heggie75,draine80,bahcall81,dhital10,jiang10}.
Actually, during the lifetime of a stellar system, the continuous small and dissipative encounters with other stars are much more disruptive than occasional single catastrophic encounters \citep{retterer82,weinberg87}.
As a result of these interactions, the initial distribution of separations of stellar systems change (increase) over time until eventual disruption.
Using the Fokker–Planck coefficients to describe the effects produced on the orbital binding energies due to those small encounters over time, \citet{weinberg87} estimated the average lifetime of a binary as:
\begin{equation}
\begin{aligned}
t_*(a) \simeq \text{18}\,\text{Ga} ~
\left(\frac{n_*}{\text{0.05}\,\text{pc}^{-\text{3}}}\right)^{-\text{1}}
\left(\frac{M_*}{M_\odot}\right)^{-\text{2}}
\left(\frac{M_{\rm tot}}{M_\odot}\right)\\
\left(\frac{V_{\rm rel}}{\text{20}\,\text{km\,s}^{-\text{1}}}\right)
\left(\frac{a}{\text{0.1}\,\text{pc}}\right)^{-\text{1}}
\text{ln}^{-\text{1}}\Lambda,
\end{aligned}
\label{eqn:weinberg}
\end{equation}
\noindent where $n_*$ and M$_*$ are the number density and average mass of the perturber objects, $V_{\rm rel}$ is the relative velocity between the binary system and the perturber, M$_{\rm tot}$ and $a$ are the total mass and semi-major axis of the binary system, and ln\,$\Lambda$ is the Coulomb logarithm.
The calculation was simplified by \citet{dhital10} by setting the values $n_*=$\,0.1\,M$_\odot$\,pc$^{-\text{3}}$, $M_*=$\,0.7\,M$_\odot$, 
V$_{\rm rel}=$\,20\,km\,s$^{-\text{1}}$, and $\ln{\Lambda}=$\,1 \citep{close07}, and produced an equation that describes in a statistical way the maximum separation of a surviving stellar system at a given age:
\begin{equation}
a \simeq \text{1.212}\,\frac{M_{\text{total}}}{t_*},
\label{eqn:seplimit}
\end{equation}
\noindent where the total mass is in M$_\odot$, the average lifetime in Ga, and the semi-major axis in~pc. 

We plot the projected physical separation $s$ as a function of the total mass M$_\odot$ of the 94 ultrawide systems in the right panel of Fig.~\ref{fig:massenergysep}.
Overplotted on them, we display the physical separations corresponding to 0.1, 1.0, and 10\,Ga and the orbital periods for 0.1, 1.0, and 2\,Ga.
The three most fragile systems may have survived in their current configuration by about 1\,Ga or slightly less in the case of WDS~15488+4929.
However, there are other systems that are less fragile (i.e. have higher reduced binding energies, comparable to those of well-recognised systems) but that can be disrupted in a few hundred million years.
As we may expect, they are among the most separated systems.

There are seven system candidates with $s$ = 1.1--2.3 $\cdot$ 10$^\text{6}$\,au (5.1--11.1\,pc), listed at the bottom of Table~\ref{tab:fragile}. 
These refer to the kinds of systems that lend their name to the topic of this work (Reaching the boundary between stellar kinematic groups and very wide binaries).
In the spherical volume of radius 10\,pc centred on the Sun, according to the exhaustive compendium by \citet{reyle21}, there are 339 systems containing stars, brown dwarfs, and exoplanets.
As a result, regardless of their (unknown) age, the ultrawide binary and multiple systems may be at the last stages of disruption and follow the formation-evolution-dissolution sequence described by \citet{close03}, who predicted an overabundance of very low-mass binaries far from the centre of the original ``minicluster''.
This is the case of the most separated components in the systems WDS~02315+0106 and WDS~15330--0111, which have low masses and large $\sigma_{Vr}$ for their $G$ magnitudes.

The seven system candidates, all of them identified by \citet{shaya11}, may be the remnants of previous SKGs that are being dissolved in the Milky Way and that are older than the ones identified in Sect.~\ref{sec:search_additional}.
Three system candidates, including the sextuple (perhaps septuple) WDS~02315+0106 system, are trapezia and, therefore, their reduced binding energies were not computed.
The other four systems consist of three very wide binaries made of two bright Henry Draper stars \citep{cannon1918} and one double-like hierarchical quadruple (perhaps quintuple) system.
The latter, namely WDS~15330--01110, is made of the K0 giant 11~Ser (Table~\ref{tab:giants}), the G1 dwarf HD~142011, and two anonymous early M dwarfs, one of which has a large $\sigma_{Vr}$ for their $G$ magnitude.
These two (or three) M dwarfs are very close to each other ($s \sim$ 92\,au) and to the Sun-like star ($s \sim$ 630--690\,au), which allowed us to compute $|U^*_g|$.
However, the K0 giant and the G1+M+M triple are separated by about 1.5 $\cdot$ 10$^\text{6}$\,au (7.2\,pc).
Between them,  dozens of unrelated stars with similar parallaxes must exist, but with very different proper motions that exert smooth ``continuous small and dissipative'' gravitational thrusts. 

\begin{figure}[H]
 \centering
 \includegraphics[width=0.9\linewidth, angle=0]{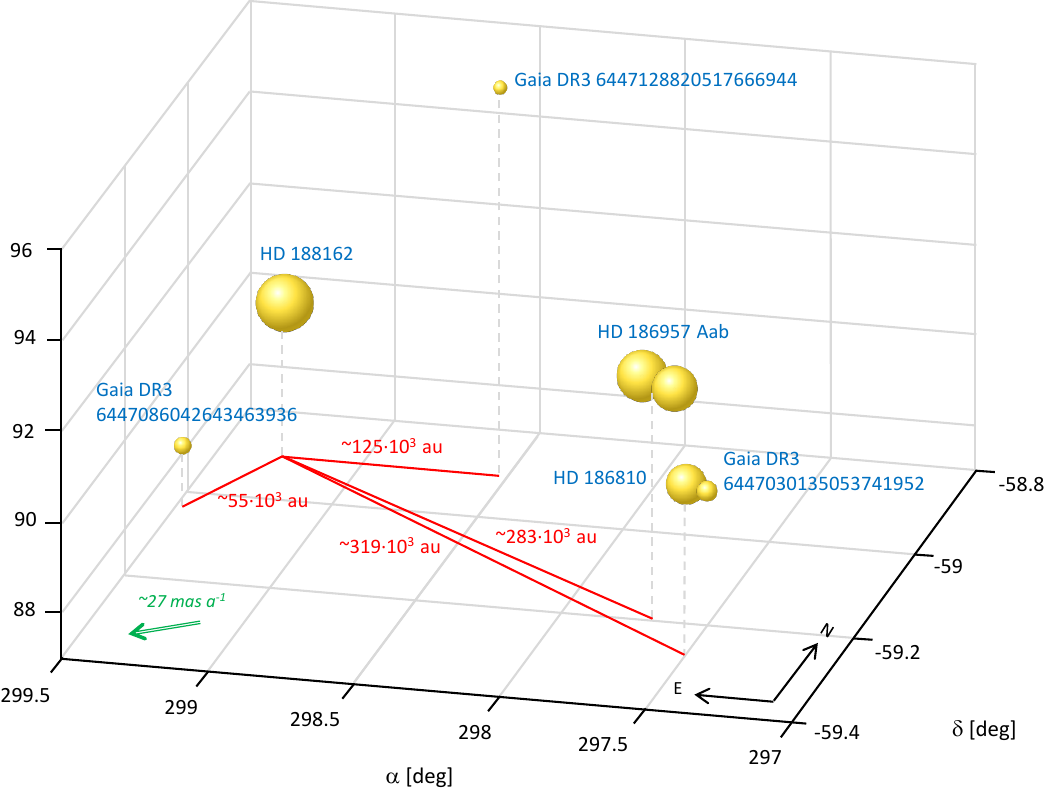}
 \caption[Spatial distribution of the septuple system WDS 19507--5912.]{Spatial distribution of the septuple system WDS 19507--5912. 
 The size of the spheres representing every star, labelled in blue, are proportional to their brightness. 
 The projected physical separations from the primary star, HD 188162, are in red, and 
 the overall proper motion is in green.}
 \label{fig:septuple1}
\end{figure}

\begin{figure}[H]
 \centering
 \includegraphics[width=1\linewidth, angle=0]{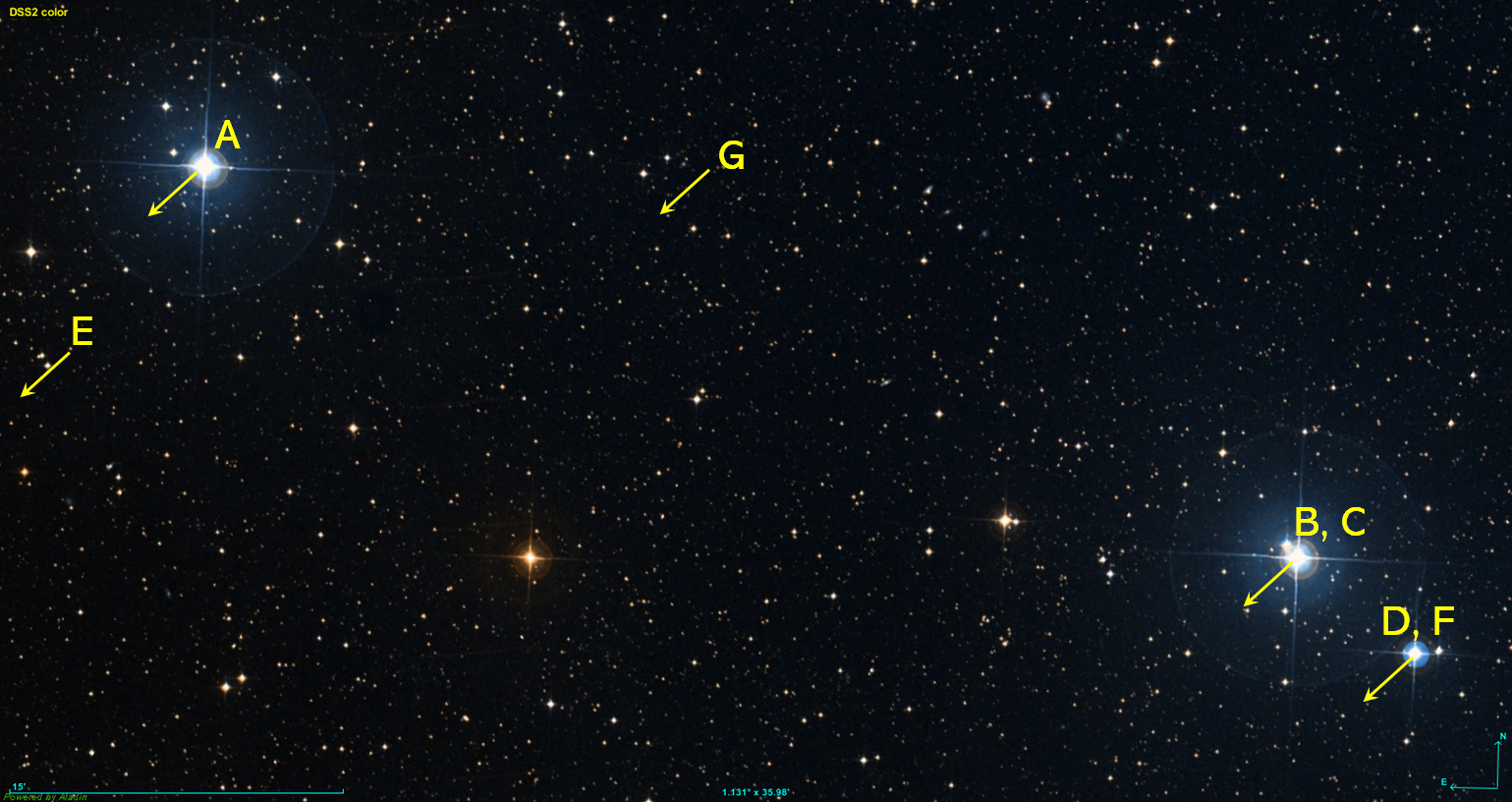}
 \caption[Visual image of the septuple system WDS 19507--5912.]{Visual image of the septuple system WDS 19507--5912 composed of stars A:~HD 188162, B:~HD 186957[Aa], C:~HD 186957[Ab], D:~HD 186810, E:~\textit{Gaia} DR3 6447086042643463936, F:~\textit{Gaia} DR3 6447030135053741952, and G:~\textit{Gaia} DR3 6447128820517666944. The yellow arrows represent the direction of movement of the stars. Image from Digitised Sky Survey II \citep{lasker96}.}
 \label{fig:septuple2}
\end{figure}

Something similar may occur in the case of the 14 ultrawide trapezia, which tend to have a high multiplicity hierarchy: 1 of the 2 septuple systems, the 3 sextuples, 7 of the 15 quadruples, and 3 of the 25 triples are trapezia.
In particular, the trapezoidal septuple system WDS~19507--5912 is sketched in Fig.~\ref{fig:septuple1} and Fig.~\ref{fig:septuple2}.
It consists of three late-B-to-early-A stars (one is a spectroscopic binary) and three intermediate-to-late M dwarfs (one is close to an A star).
Given the early spectral types of the most massive stars, this system may approximately be the age of the Hyades  (600--700\,Ma; \citealt{perryman98,gossage18,martin18}) and, therefore, stand as an unidentified sparse young SKG\footnote{If confirmed in the future, we propose naming the SKG following the discovery name of the brightest, earliest star: HD~188162.}.
All these results are in accordance with the suggestions by \citet{basri06} and \citet{caballero07a}, who proposed a major prevalence of wide triples over wide binaries.
We also confirm that the individual components of systems at very wide separations are often multiple systems themselves, as stated by
\citet{cifuentes21}.

Five of the seven ultrawidest systems have orbital periods of the order of 1\,Ga, even older than the Hyades.
Such long orbital periods stand as a challenge to the ``binary'' definition itself, namely: a system of two stars that are gravitationally bound to and in orbit around each other. 
As a result, the ultrawidest pairs may have not completed one revolution either because of their young age or because they were recently disrupted by passing stars and, therefore, should not be called ``binaries''.
This statement should also be extrapolated to the system candidates in young stellar kinematic groups (Section~\ref{sec:search_additional}), including less separated but also less massive, pairs.
Furthermore, \citet{caballero09} already claimed that the AU~Mic + AT~Mic system in the $\beta$~Pictoris moving group has only completed at most two orbital periods since its formation.
All of this makes the boundary between stellar kinematic groups and very wide binaries blurrier and blurrier.

Finally, we used the accurate \textit{Gaia} astrometry to measure the relative transverse velocity, $\Delta V$, as a function of the projected physical separation of the 48 widest systems with $s >$ 0.1\,pc, and compared it with the maximum velocity allowed for a bound binary, as done by some other authors (e.g. \citealt{elbadry19b,elbadry21}).
This comparison may interpret several observational studies that have reported that the difference in the proper motions or radial velocities of the components of nearby wide binaries appear larger than predicted by Kepler’s laws, indicating a potential breakdown of general relativity at low accelerations.
However, our data, which are relatively scarce compared to extensive simulations (cf. \citealt{elbadry19b}), do not even show projection effects.
Furthermore, inner subsystems, which are frequent, disturb our $\Delta V$ estimates.
As a result, the actual bound nature of our most fragile system is hardly verifiable.

To sum up, wide pairs with very low-mass components and $|U^*_g|$\,$\sim$\,10$^{\text{33}}$\,J (e.g. ultra-cool dwarf binaries with late-M and early-L spectral types and projected physical separations of a few thousand astronomical units -- \citealt{caballero07a,caballero07b,artigau07}) are perhaps more relevant for investigating the disruption of fragile systems by the galactic gravitational potential rather than ultrawide systems of gargantuan projected physical separations, much larger than those proposed by \citet{tolbert64}, \citet{bahcall81}, or \citet{retterer82}, caught in the act of destruction and, that probably are the leftover of past SKGs.

\section{Summary}
\label{sec:summary_WDS}

Thanks to the \textit{Gaia} DR3 \citep{gaiacollaboration23b} and a number of parallax and proper motion searches in the previous decade \citep[e.g.][]{caballero10, shaya11, tokovinin12, kirkpatrick16}, we present a leap forward with respect to the first item of this series of papers, on the Washington double stars with the widest angular separations \citep{caballero09}.
Accordingly, we increase by over an order of magnitude the sample size and the astrometric precision of systems with angular separations $\rho >$ 1000\,arcsec.
Among other results of our analysis, we present:
($i$) 40 additional astrometric companions not catalogued yet by WDS, including several ultra-cool dwarfs at the M-L boundary and one hot white dwarf, not counting several dozens close binary candidates from large \textit{Gaia} DR3 \texttt{RUWE} and $\sigma_{\rm RV}$; 
($ii$) a general confusion in the literature between actual, physically bound, ultrawide pairs and components in young SKGs, associations, and even clusters with identical galactocentric space velocities;
($iii$) three very fragile systems discovered by \citet{kirkpatrick16} that are made of intermediate and late M dwarfs with large projected physical separations of 0.33--0.41\,pc and small reduced binding energies $|U_g^*|\lesssim$\,10$^{\text{33}}$\,J, which is probably the smallest value found among gravitationally bound systems;
($iv$) the individual components of systems at very wide separations are often multiple systems themselves \citep{cifuentes21}, which implies an overabundance of high-order multiples (triples, quadruples, quintuples, and more) among the widest systems, and larger binding energies and total masses than systems with comparable separations but lower multiplicity order;  
and ($v$) an additional observational confirmation of classical theoretical predictions \citep[e.g.][]{weinberg87} of disruption of binary systems by the galactic gravitational potential, which destroys the ultrawidest systems with total masses below 10\,M$_\odot$ in less that 600--700\,Ma (the age of the Hyades cluster).
As incidental results, we report 40 new candidate stars in known young SKGs and at least one new young stellar association around the bright star $\gamma$~Cas, although some of our highest-order multiple systems, such as the septuple around the B9.5\,IV star HD~188162, may also be association remnants.

In conclusion, the total mass, binding energy, and probability that an ultrawide system is actually bound increase as the stellar multiplicity order also increases.
Many systems reported here have overwhelmingly large projected physical separations, but have instead total masses large enough for the binding energies being comparable to those of less separated, less massive systems that are widely accepted as physical.
However, none of the ultrawide systems will survive for another few hundred million years.
In a sense, the widest multiple systems of today, which are now being torn apart by the Galaxy, will be the single stars of tomorrow.

\newpage

%%%%% CAPITULO 5 %%%%%

\chapter{Multiplicity of stars with planets} 
\label{ch:systems_with_planets}
\vspace{2cm}
\pagestyle{fancy}
\fancyhf{}
\lhead[\small{\textbf{\thepage}}]{\textbf{Section \nouppercase{\rightmark}}}
\rhead[\small{\textbf{Chapter~\nouppercase{\leftmark}}}]{\small{\textbf{\thepage}}}

\begin{flushright}
\small{\textit{``If the Lord Almighty had consulted me before embarking upon\\
his creation, I should have recommended something simpler''}}

\small{\textit{-- Alfonso X of Castile}}
\end{flushright}
\bigskip

\begin{adjustwidth}{70pt}{70pt}
\tR{\small{The content of this chapter has been adapted from the article \citet{gonzalezpayo24}: \textit{Multiplicity of stars with planets in the solar neighbourhood}, published in Astronomy \& Astrophysics, \href{https://doi.org/10.1051/0004-6361/202450048}{A\&A, 689, A302 (2024)}.}}
\end{adjustwidth}
\bigskip

\lettrine[lines=3, lraise=0, nindent=0.1em, slope=0em]{A}{fter more than thirty years} since the discovery of the first exoplanets \citep{campbell88,wolszczan92,mayor95}, almost 6000 exoplanet candidates have been reported. 
Except for a few ``rogue'' planets \citep[and references therein]{caballero18} and bodies of unknown nature found around compact objects such as white dwarfs and neutron stars \citep{bailes11,vanderburg20}, the majority of these exoplanet candidates orbit around main-sequence and giant stars.
Planet occurrence rates determined with different techniques indicate that, on average, there are slightly more than one planet per star \citep{cassan12,dressing15,baron19,savel20,sabotta21,ribas23}.
Additionally, a significant fraction of the stars in the Galaxy are part of stellar multiple systems, including double, triple, or higher-order systems \citep[][and many others]{abt76,duchene13,tokovinin08}.
As a consequence, many of the discovered exoplanet candidates are also part of stellar multiple systems.

Stellar multiplicity has been studied for centuries \citep{mayer1778,herschel1802,duquennoy91,jao09a,duchene13}.
In our immediate vicinity, that is, in the 10\,pc-radius volume centred on the Sun, \citet{reyle21} measured an overall multiplicity fraction (MF) of 27.4 $\pm$ 2.3\,\%.
Actually, the MF decreases from the earliest to the latest spectral types:
It varies from virtually 100\% for O stars (except for a few runaway stars; they are located in dense star-forming regions), to more than 70\% for B and A stars \citep{kouwenhoven07,mason09,sana13,caballero14,maizapellaniz19},
44--67\% for solar-type stars \citep{duquennoy91,raghavan10,duchene13}, 26--40\% for M dwarfs \citep{fischer92,reid97c,delfosse04,janson12,ward15,cortescontreras17b,winters19}, and to below 20\% for L, T, and Y ultra-cool dwarfs \citep{burgasser03c,burgasser05,burgasser07b,fontanive18}.
Since most of the reported exoplanet candidates orbit FGKM-type stars, several thousands of planetary systems would also be expected among multiple stellar systems.
This fact is actually modulated by an observational bias in which many exoplanet surveys tend to discard close binaries in their input catalogues (e.g. \citealt{raghavan06,winn15,sebastian21,ribas23}).
Currently, there are a few hundred multiple stellar systems with known exoplanet candidates \citep[][and many others]{eggenberger04b,konacki05,furlan17,martin18,bonavita20}.
 
\begin{table}[H]
 \centering
 \caption[Observational works on multiplicity of stellar systems with planets.]{Observational works on multiplicity of stellar systems with planets.}
 \footnotesize
 \scalebox{1}[1]{
 \begin{tabular}{@{\hspace{2mm}}l@{\hspace{2mm}}l@{\hspace{1mm}}c@{\hspace{1mm}}c@{\hspace{1mm}}}
 \noalign{\hrule height 1pt}
 \noalign{\smallskip}
 Title & Reference & $N^{\text{(a)}}$ & Methodology$^{\text{(b)}}$ \\
 \noalign{\smallskip}
 \hline
 \noalign{\smallskip} 
 Multiplicity of stars with planets in the solar neighbourhood & This work & 215  &  Misc. \\
 An early catalog of planet-hosting multiple-star systems of...  & \citet{cuntz22} & 40 & Misc. \\
 Speckle observations of TESS exoplanet host stars. II...  & \citet{lester21} & 102 &  SI \\   
 The census of exoplanets in visual binaries: population...  & \citet{fontanive21} & 218 & Misc. \\ 
 How many suns are in the sky? A SPHERE multiplicity...  & \citet{ginski21} & 4 & AO \\ 
 Frequency of planets in binaries & \citet{bonavita20} & 313 & RV \\
 Understanding the impacts of stellar companions on planet... & \citet{hirsch21} & 109 & AO, RV \\
 Search for stellar companions of exoplanet host stars by... & \citet{mugrauer19} & 204 & WFI \\
 The SEEDS high-contrast imaging survey of exoplanets... & \citet{uyama17} & 68 & AO \\
 A lucky imaging multiplicity study of exoplanet host stars...  & \citet{ginski16} & 60 & LI \\
 High-contrast imaging search for stellar and substellar... & \citet{mugrauer15} & 33 & AO \\
 A Lucky Imaging search for stellar companions to transiting... & \citet{wollert15} & 49 & LI \\
 Binary frequency of planet-host stars at wide separations...  & \citet{lodieu14} & 37 & WFI \\
 The multiplicity status of three exoplanet host stars & \citet{ginski13} & 3 & LI \\
 Stellar companions to exoplanet host stars: Lucky imaging... & \citet{bergfors13} & 21 & LI  \\
 Extrasolar planets in stellar multiple systems  & \citet{roell12} & 57 &  AO \\
 Know the star, know the planet. I. Adaptive optics of...  & \citet{roberts11} & 62 & AO  \\
 Know the star, know the planet. II. Speckle interferometry...  & \citet{mason11} & 118 &  SI \\
 Binarity of transit host stars. Implications for planetary... & \citet{daemgen09} & 14 & LI \\
 The frequency of planets in multiple systems  & \citet{bonavita07} & 202 & Misc. \\
 The multiplicity of exoplanet host stars. New low-mass... & \citet{mugrauer09} & 43 & WFI \\
 The multiplicity of planet host stars - new low-mass...  & \citet{mugrauer07} & 3 &  IS \\
 The impact of stellar duplicity on planet occurrence and...  & \citet{eggenberger07} & 57 & AO \\   
 A search for wide visual companions of exoplanet host... & \citet{mugrauer06} & 44 & WFI \\
 Two suns in the sky: Stellar multiplicity in exoplanet systems & \citet{raghavan06} & 36 & Misc.  \\
 Detection and properties of extrasolar planets in double... & \citet{eggenberger04} & 15 & AO \\
 Planets in multiple-star systems: properties and detections & \citet{udry04} & 15 & Misc. \\
Stellar companions to stars with planets & \citet{patience02} & 11 & Misc. \\
 \noalign{\smallskip}
 \noalign{\hrule height 1pt}
 \end{tabular}
 }
 \label{tab:studies} 
 \begin{justify}
    \footnotesize{\textbf{\textit{Notes. }}
    $^{\text{(a)}}$ $N$: number of investigated multiple stellar systems with exoplanets.}
    $^{\text{(b)}}$ AO: Adaptive optics; IS: Infrared spectroscopy; LI: Lucky imaging; RV: Radial velocity; SI: Speckle imaging; WFI: Wide field imaging; Misc.: Miscellanea.
 \end{justify}   
\end{table}

The consequence of a star being in a multiple system is not only relevant for the formation and evolution of the star itself but also in the formation and evolution of planets, especially if the stellar companions are close to each other.
Previous theoretical works predicted that nearby companion stars can significantly disrupt circumstellar discs and hinder the process of planetary formation 
(\citealt{lissauer87,gladman93,jensen96,artymowicz96,pichardo05,quintana06,hamers21}; see also \citealt{thebault15} and the series of papers initiated by \citealt{wang14a}),
while other theoretical works have focused on the long-term stability of planets in binary systems. 
For example, \citet{holman99} were among the first to study such stability in great detail.
A number of similar works on the long-term stability of S- (around one star) and P-type (around the two stars) systems were published afterwards by, for example, \citet{pilatlohinger03}, \citet{musielak05}, \citet{mudryk06}, and \citet{doolin11}.
There has also been theoretical work on the long-term stability of triple systems with planets \citep[e.g.][]{busetti18}.
Most of these publications, however, have focused on the eccentricity of stellar orbits rather than on the eccentricity of planet orbits.
The relation between stellar multiplicity and planet eccentricity was first investigated by \citet{mazeh97}, who analysed the stability of the 16~Cyg system.
The system contains three stars (A: G1.5\,V; B: G3\,V; C: mid-M) and a 1.8\,M$_{\rm Jup}$-mass planet in an eccentric orbit ($e \approx$ 0.68) around the secondary \citep{cochran97,turner01,rosenthal21}.

Likewise, there are a number of observational works on the multiplicity of stellar systems with planets.
Table~\ref{tab:studies} enumerates many of the relevant observational publications on this topic.
The authors have used a diversity of target stars with planets, methodologies, and even maximum distances (e.g. \citealt{hirsch21}: 25\,pc; \citealt{eggenberger07}: 50\,pc; \citealt{fontanive21}: 200\,pc; \citealt{mugrauer19,lester21}: 500\,pc). 
A wealth of results have been proposed after these observations.
For example, \citet{eggenberger04} and \citet{udry04} found that massive planets ($M \sin{i} >$ 2\,M$_{\rm Jup}$) with moderately short periods ($P \le$ 40--100\,d) in binaries tend to have low eccentricities, while \citet{moutou17}, with a more powerful facility (SPHERE at the Very Large Telescope), also concluded that the majority of high-eccentricity planets are ``not embedded'' in multiple stellar systems.
These results are rather inconsistent with those of \citet{raghavan06}, who showed that planets in systems with confirmed stellar companions instead generally have higher eccentricities.
The authors reasoned that companion stars would have a greater gravitational influence on the planets' orbits and could shorten the periods of Kozai cycles \citep{innanen97,wu03,tamuz08,naoz16}.
We finish this introduction with a sentence written by \citet{eggenberger04} exactly two decades ago and that is still valid:
``{The studies [enumerated above] emphasise the importance of searching for planets in multiple star systems, even though it is more challenging to carry out than the search for planets around individual stars}''.

In this work, we revisit the topic of multiplicity of stars with planets at less than 100\,pc.
After this introduction, we describe our target sample in Sect.~\ref{sec:sample_hosts}. 
Next, we detail in Sect.~\ref{sec:analysis_hosts} our analysis, which consists of an individualised search for common proper-motion and parallax companions to exoplanet host stars with \textit{Gaia} DR3 complemented with data from the Washington Double Star catalogue and the literature.
The results are presented in Sect.~\ref{sec:results_hosts}, where we report on new stellar systems with planets; the relationships between star-star separations, star-planet semi-major axes, and planet eccentricities; and the number and masses of planets in single and multiple systems, and we conclude with the properties of multiple star systems with planets.
Finally, in Sects.~\ref{sec:discussion_hosts} and~\ref{sec:summary_hosts} we discuss and summarise our results.

\section{Sample}
\label{sec:sample_hosts}

Our sample was built on the basis of the two most widely used exoplanet databases. We took all the host stars tabulated by either the Extrasolar Planets Encyclopaedia\footnote{Now Encyclopaedia of Exoplanetary Systems, \url{http://exoplanet.eu/}} \citep{schneider11} or the NASA Exoplanet Archive\footnote{\url{https://exoplanetarchive.ipac.caltech.edu/}} \citep{akeson13}. At the moment of downloading (3 January 2024), the first database contained 5576 planets in 4114 planetary systems, and the second one had 5566 planets in 4145 systems. 
We discarded all of the duplicate host stars in both databases. Sometimes the same star was tabulated with different names in each database, and did not always have the same exact coordinates. 
We finally obtained a set of 4612 non-duplicate host stars, all of which have an entry in the Simbad astronomical database \citep{wenger00}.

Secondly, we looked for the \textit{Gaia} DR3 \citep{gaiacollaboration23b} counterpart of every host star. 
For some cases that required a visual inspection, we used the Aladin Sky Atlas \citep{bonnarel00}. 
Of the 4612 host stars, 339 do not have a \textit{Gaia} DR3 entry. 
Of them, eight are just too bright for \textit{Gaia}\footnote{The eight very bright exoplanet host stars are: $\alpha$~PsA (Fomalhaut), $\alpha$~Ari (Hamal), $\alpha$~Tau (Aldebaran), $\beta$~And (Mirach), $\beta$~Gem (Pollux), $\beta$~UMi (Kochab), $\gamma$~Lib (Zubenelhakrabi), and $\gamma^{\text{01}}$~Leo.}, while the other 331 stars without a \textit{Gaia} DR3 entry are distant microlensing objects from the OGLE\footnote{Optical Gravitational Lensing Experiment, \url{https://ogle.astrouw.edu.pl/}} \citep{udalski92}, KMT\footnote{Korea Microlensing Telescope Network, \url{https://kmtnet.kasi.re.kr/kmtnet/}} \citep{henderson14}, and MOA\footnote{Microlensing Observations in Astrophysics, \url{http://www2.phys.canterbury.ac.nz/moa/}} \citep{alcock95,alcock97} surveys, and some pulsars. 

\begin{figure}[H]
 \centering
   \includegraphics[width=0.6\linewidth, angle=0]{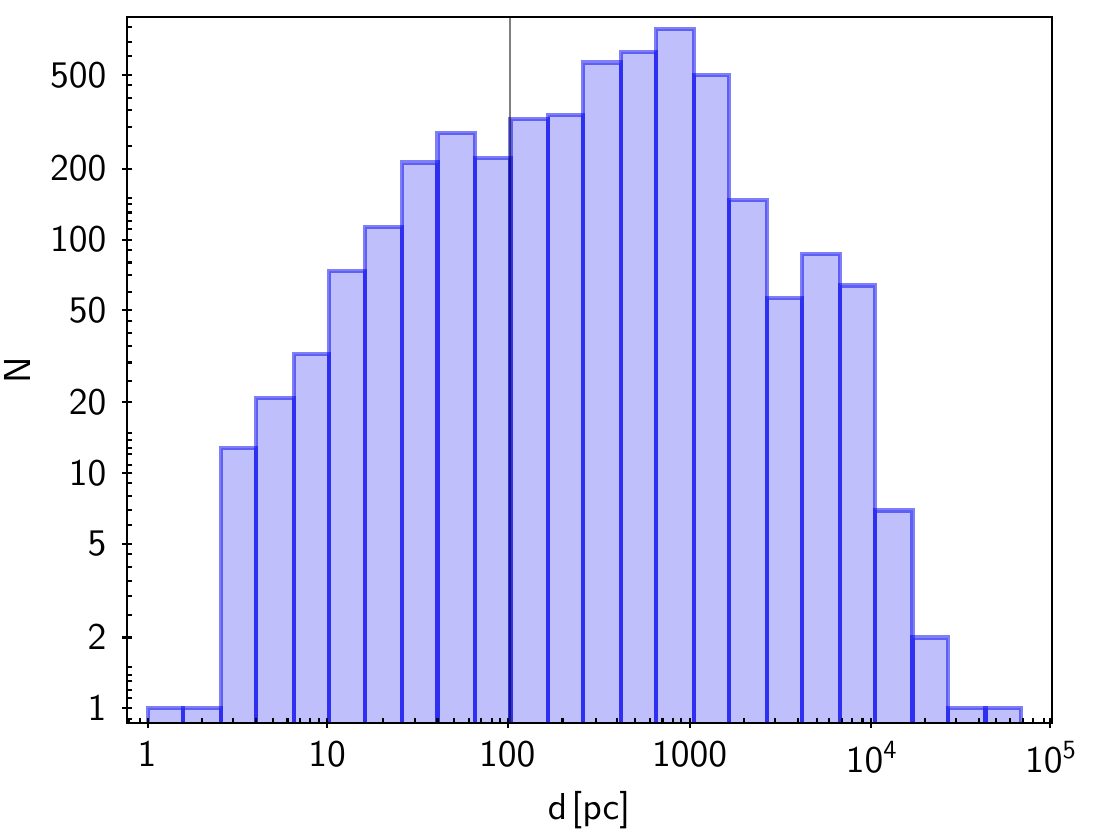}
   \includegraphics[width=0.6\linewidth, angle=0]{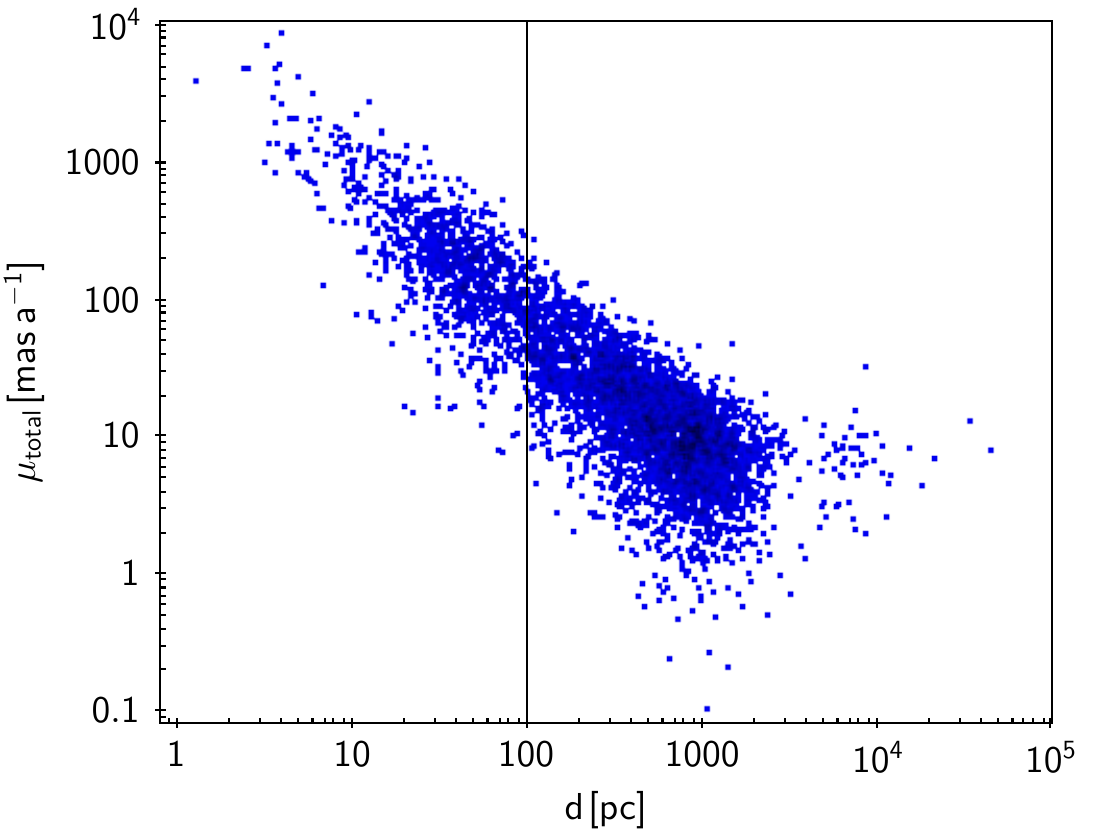}
 \caption[Histogram of distances of non-duplicated exoplanet host stars in the initial sample; and total proper motion as a function of distance for the same stars.]{Diagrams to determine the maximum distance of our analysis.\textit{Top panel}: Histogram of distances of the 4583 non-duplicated exoplanet host stars in the initial sample.
 \textit{Bottom panel}: Total proper motion as a function of distance for the same stars.
 The vertical line at $d$ = 100\,pc in both panels indicate the search limit.}
 \label{fig:distances}
\end{figure}

Of the 4273 stars with an entry in \textit{Gaia} DR3, 4232 have a positive parallax value. 
For the remaining 41 stars (with a negative parallax or no parallax at all) plus the former 339 with no counterpart in \textit{Gaia} DR3, we looked for their parallaxes or spectro-photometric distances in the \textit{Hipparcos} \citep{perryman97}, \textit{Gaia} DR2 \citep{gaiacollaboration18}, \textit{Gaia} EDR3 \citep[Third Early Data Release --][]{gaiacollaboration21a},
and UCAC4 \citep[Fourth United States Naval Observatory CCD Astrograph Catalogues --][]{zacharias13}
catalogues and in the literature (e.g. \citealt{shkolnik12,gagne15,leggett17,finch18,winters19}). 
We were able to find a parallax or a distance for 351 of the mentioned 380 stars. 
To sum up, a total of 4583 exoplanet host stars in our sample have a tabulated distance and only 29 (0.63\% of the 4612 non-duplicate host stars) do not.
The distribution of distances of the 4583 stars is shown in the top panel of Fig.~\ref{fig:distances}.

Next, we restricted the analysis to stars with distances less than 100\,pc, which is the limit of the solar neighbourhood \citep{gaiacollaboration21b}. 
There are many practical reasons behind this search limit selection, such as manageability of the final sample for detailed analysis, increase of both the incompleteness of the exoplanet searches and of the astrometric uncertainty for the search for common proper motion companions at longer distances, as illustrated by the bottom panel of Fig.~\ref{fig:distances}; and reliability of the parameters of the planetary systems, which have been mostly detected with the radial velocity and transit methods.
Finally, 998 non-duplicate exoplanet host stars are located at less than 100\,pc.
They are listed in Table~\ref{tab:sample_host_stars}.

\section{Analysis}
\label{sec:analysis_hosts}

\begin{figure}
 \centering
 \includegraphics[width=0.55\linewidth, angle=0]{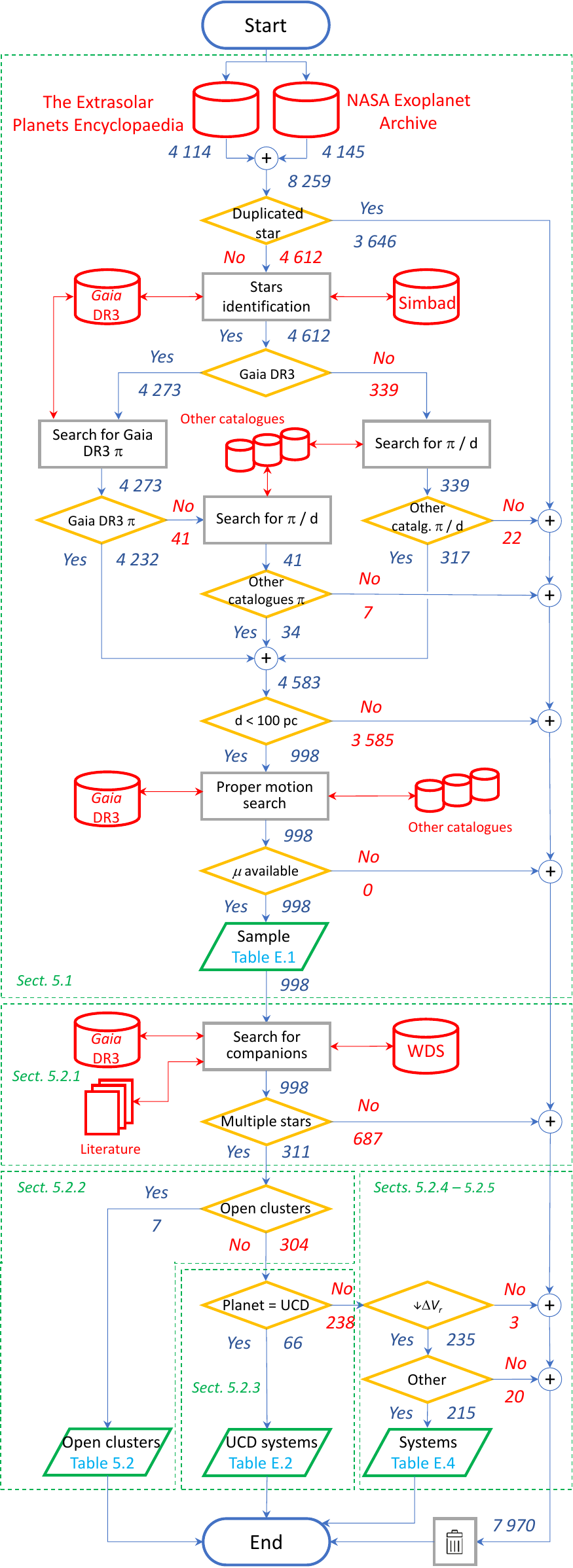}
 \caption{Flowchart describing the sample preparation and following analysis.}
 \label{fig:flowchart2}
\end{figure}

The flowchart in Fig.~\ref{fig:flowchart2} summarises the sample preparation and the following analysis.
In the diagram, the stadia indicate the beginning and ending, cylinders the databases, small circles the connectors, diamonds the decisions, rectangles the processes, stacks of rectangles the literature documents, rhomboids the inputting and outputting data, and flow lines (arrowheads) the processes' order of operation.
Every flowline is labelled with the corresponding number of host stars.

\subsection{Search for stellar companions}
\label{sec:search_for_stellar_companions}

The first step of the analysis was to look for companions of common \textit{Gaia} DR3 proper motion and parallax to our 998 non-duplicate exoplanet host stars at less than 100\,pc. 
We used the same methodology as \cite{gonzalezpayo21} (their Sect.~3).
In particular, we used \texttt{Topcat} \citep[Tool for OPerations on Catalogues And Tables;][]{taylor05} with a customised code in ADQL \citep[Astronomic Data Query Language;][]{yasuda04} to search for companions at projected physical separations, $s = \rho \cdot d$, of up to 1\,pc that satisfy the following criteria to distinguish between physical (bound) and optical (unbound) systems: 
\begin{equation}
\mu\,\mathrm{ratio}=\sqrt{\frac{(\mu_{\alpha} \cos{\delta_\text{1}}-\mu_{\alpha} \cos{\delta_\text{2}})^\text{2}+(\mu_{\delta_\text{1}}-\mu_{\delta_\text{2}})^\text{2}}{(\mu_{\alpha} \cos{\delta_\text{1}})^\text{2}+(\mu_{\delta_\text{1}})^\text{2}}}<\text{0.15},
\label{eqn:crit1b}
\end{equation}
\begin{equation}
\Delta \text{PA}=\lvert  \text{PA}_\text{1} - \text{PA}_\text{2} \rvert<\text{15}\,\mathrm{\text{deg}},
\label{eqn:crit2b}
\end{equation}
\noindent and
\begin{equation}
\Biggl\lvert \frac{\pi_\text{1}^{-\text{1}}-\pi_\text{2}^{-\text{1}}}{\pi_\text{1}^{-\text{1}}} \Biggr\rvert < \text{0.15},
\label{eqn:crit3b}
\end{equation}
\noindent where PA$_\text{i}$ is the angle between the proper motion vectors, with i\,=\,1 for the primary star and i\,=\,1 for the companion.
The inverse of the parallax, $\pi_\text{i}^{-\text{1}}$, is the distance, which for $d <$ 100\,pc in general does not need any further correction (e.g. \citealt{bailerjones18a,luri18a}). 
The motivation of the 0.15 and 15\,deg values by \cite{gonzalezpayo21} was justified by the dissimilarity of proper motions and parallaxes of bona fide physically bound stars of different \textit{Gaia} colours in close resolved systems with astrometric solutions (i.e., we neither applied a colour-term parallax correction nor subtracted relative motions in systems with orbital periods of a few years).
We found 230 pairs of stars that satisfy these criteria. 
While resolved binary systems are made of one pair of stars, resolved multiple systems are made of two (triple) or three (quadruple) pairs. 
See \citet{gonzalezpayo21} for further details on the search methodology.

Next, we complemented our \textit{Gaia} DR3 search for companions with a cross match with data in the Washington Double Star catalogue \citep[WDS;][]{mason01}.
Currently, they tabulate angular separations and position angles for the first and last epochs of observation of about 156\,000 multiple systems.
While they also tabulate other parameters (e.g. equatorial coordinates, magnitudes, notes on systems), additional information can be obtained from the WDS team upon request.

There are 687 WDS pairs in 341 systems containing at least one of the 998 input stars or the 230 \textit{Gaia} DR3 companions from our previous search.
Since there are ultra-wide multiple systems with extremely large projected separations \citep{caballero09,shaya11,gonzalezpayo23}, we kept WDS systems with $s >$ 1\,pc if they satisfy the three astrometric criteria (Eqs.~\ref{eqn:crit1b}, \ref{eqn:crit2b}, and \ref{eqn:crit3b}).
The remaining stars, either host stars or \textit{Gaia} companions, are not part of any WDS system. Of the 687 WDS pairs, we rejected 417 because their stars have very different \textit{Gaia} DR3 distances and proper motion moduli and direction (i.e. do not satisfy at least one of Eqs. \ref{eqn:crit1b}, \ref{eqn:crit2b}, and~\ref{eqn:crit3b}).
Generally, the discarded companions are further in the background and have lower proper motions than the planet host stars.
Many of them have the WDS ``U'' flag: ``Proper motion or other technique indicates that this pair is non-physical''.
The other 270 WDS pairs passed to the next step of our analysis.

\begin{figure}
 \centering
 \includegraphics[width=0.65\linewidth, angle=0]{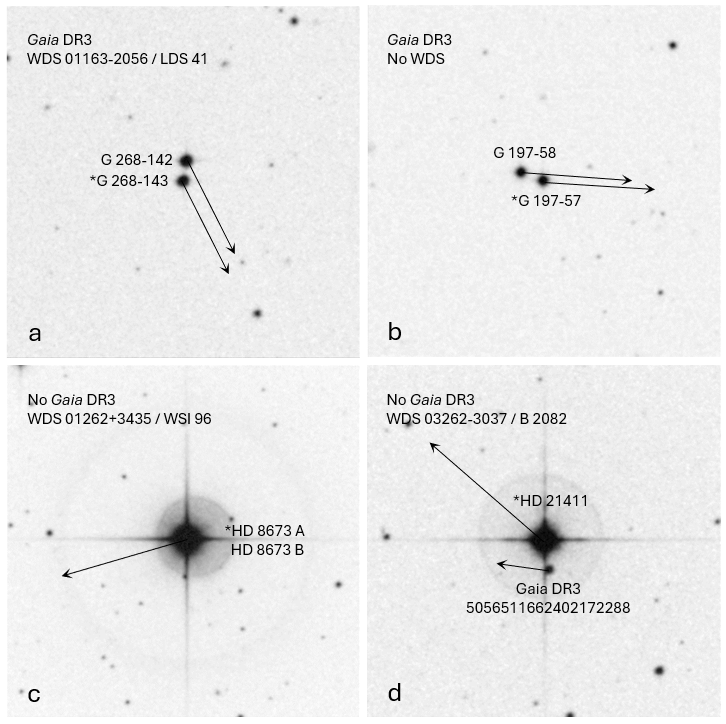}
 \caption[European Southern Observatory Digital Sky Survey DDS2 Red images centred on four representative examples of analysed systems.]{European Southern Observatory Digital Sky Survey DDS2 Red images centred on four representative examples of analysed systems.
 The field of view is $5 \times 5$\,arcmin, north is up and east is left.
 Stars are labelled, and exoplanet host names are preceded by an asterisk.
 Arrows indicate modulus and direction of \textit{Gaia} proper motions.
 The four representative types are:
 (a) a resolved physical double identified in our \textit{Gaia} search and with a WDS entry,
 (b) a resolved physical double found with \textit{Gaia} and not in WDS (but reported by \citealt{gaiacollaboration21b}),
 (c) a close physical double with a WDS entry ($\rho =$ 0.31\,arcsec) but not resolved by \textit{Gaia},
 and (d) an optical double in WDS but not identified in our \textit{Gaia} search.
 }
 \label{fig:pairs_examples}
\end{figure}

\begin{figure}
 \centering
 \includegraphics[width=0.65\linewidth, angle=0]{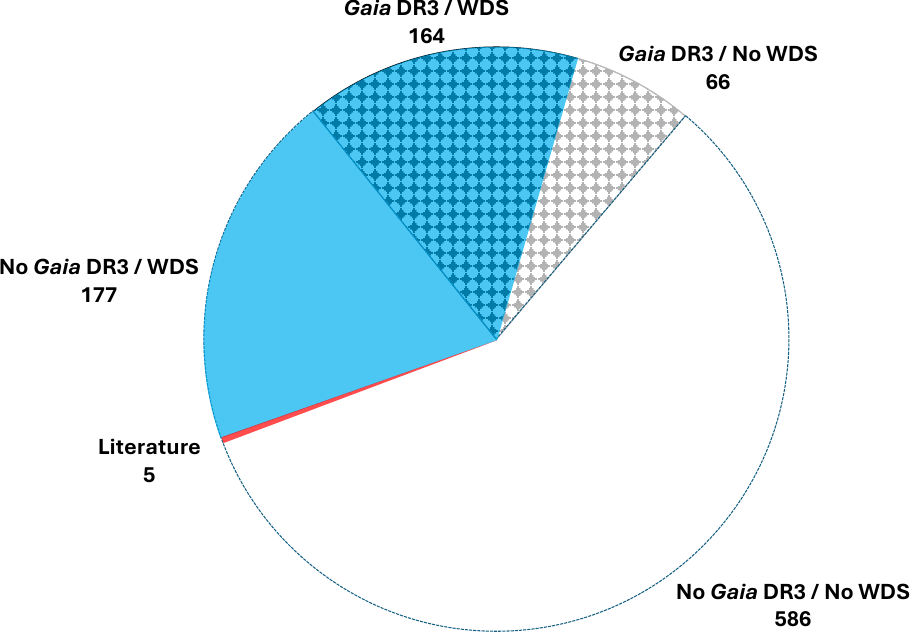}
 \caption[Pie chart of \textit{Gaia}, WDS, and literature systems.]{Pie chart of \textit{Gaia} (rhombus checkered), WDS (blue), and literature (red) systems.
 Slices are labelled with the number of systems of each type.}
 \label{fig:pie_chart}
\end{figure}

Not all the WDS pairs were identified in the \textit{Gaia} DR3 search, nor all the \textit{Gaia} DR3 common $\mu$ and $\pi$ companions were tabulated by WDS.
Most, if not all, WDS companions not found in our \textit{Gaia} DR3 search are in very close systems unresolvable by \textit{Gaia}, have moderate angular separations but accompany very bright stars, or are very faint ultra-cool dwarfs beyond the \textit{Gaia} limit at about $G \approx$ 20.3\,mag \citep{smart17}. 
Likewise, there are known WDS systems with additional \textit{Gaia} DR3 common $\mu$ and $\pi$ companions reported here for the first time (see below). 
As a result of our searches, we found companions to 311 of the 998 stars in the sample, including those in the 270 WDS pairs. Four representative examples of analysed systems are shown in Fig.~\ref{fig:pairs_examples}. 

We categorised the remaining 687 exoplanet host stars as single, either because WDS and the literature do not report any companions for them or because we were not able to find any common parallax and proper motion companion (by chance, there were also 687 WDS optical and physical pairs).
The list of single stars may actually be shorter, as they may have faint companions beyond the \textit{Gaia} magnitude limit, perhaps in the substellar domain \citep{smart17, smart19, marocco20}, or may be unidentified very close spectroscopic or astrometric binaries.

Finally, we also added five additional multiple systems with extremely close separations: 
the 3 circumbinary systems at less than 100\,pc (RR~Cae~AB, Kepler-16~AB, and HD~202206~AB -- \citealt{qian12,doyle11,benedict17}, respectively),
1 spectroscopic binary (HD~42936~AB -- \citealt{barnes20}),
and 1 very close pair (LP~413--32~AB -- \citealt{feinstein19}). 
The five of them have been reported in the literature but not tabulated by WDS.
We recovered them from an exhaustive object-by-object search through the literature. 
Fig.~\ref{fig:pie_chart} summarises the casuistry of the different types of systems in the \textit{Gaia}, WDS, and literature searches.

\subsection{Wide pairs in open clusters and associations}
\label{sec:wide_pairs_in_open_clusters_and_associations}

There are reasonable doubts about the true binding of ultra-wide pairs of young stars, as they may actually be part of stellar kinematic groups and associations \citep[][and references therein]{caballero10}.
The doubts are more reasoned when the pairs belong to nearby open clusters and OB associations, and especially when the measured separation resembles the typical separation between cluster members.
This is the case of seven of the 311 host stars with \textit{Gaia} DR3 or WDS companions, shown in Table~\ref{tab:clusters}, which we excluded from our analysis.
They belong to the nearby Hyades open cluster \citep{reid93,perryman98} and Lower Centaurus Crux OB association \citep{dezeeuw99}. 
As expected, all of them have a large number of common $\mu$ and $\pi$ companions (and companions to companions) at wide projected physical separations, of up to 37 in the case of one of the Hyades (being the 37 companions known Hyades, too).
The stars in Table~\ref{tab:clusters} compose a complete volume-limited sample of exoplanet host stars in open clusters and dense OB associations, and could be used for chemical tagging studies linked to exoplanets \citep[e.g.][]{desilva06,liu19}.

\begin{table}[H]
 \centering
 \caption[Host stars in open clusters and OB associations with discarded wide common $\mu$ and $\pi$ companion candidates.]{Host stars in open clusters and OB associations with discarded wide common $\mu$ and $\pi$ companion candidates.}
 \footnotesize
 \scalebox{1}[1]{
 \begin{tabular}{lcccc}
 \noalign{\hrule height 1pt}
 \noalign{\smallskip}
 Star$^{\text{(a)}}$ & $\alpha$\,(J2000) & $\delta$\,(J2000) & $d$ & Open \\
 & (hh:mm:ss.ss) & (dd:mm:ss.s) & (pc) & cluster \\
 \noalign{\smallskip}
 \hline
 \noalign{\smallskip} 
  HD 285507 & 04:07:01.23 & +15:20:06.1 & 45.0 & Hya  \\
  K2--25  & 04:13:05.61 & +15:14:52.0 & 44.7 & Hya \\
  K2--136 & 04:29:38.99 & +22:52:57.8 & 58.9 & Hya \\
  HD 28736 & 04:32:04.81 & +05:24:36.1 & 43.3 & Hya \\
  HD 283869 & 04:47:41.80 & +26:09:00.8 & 47.4 & Hya \\
  HD 95086 & 10:57:03.02 & --68:40:02.4 & 86.5 & LCC \\
  HD 114082 & 13:09:16.19 & --60:18:30.1 & 95.1 & LCC \\ 
 \noalign{\smallskip}
 \noalign{\hrule height 1pt}
 \end{tabular}
 }
 \label{tab:clusters} 
 \begin{justify}
  \footnotesize{\textbf{\textit{Notes. }}
    $^{\text{(a)}}$ The list does not contain $\epsilon$ Tau in the Hyades open cluster and b~Cen~(AB) in the Upper Centaurus-Lupus OB association.}
 \end{justify}
\end{table}

Table~\ref{tab:clusters} does not include two close double candidates.
We kept $\epsilon$~Tau as a single host star in the Hyades although WDS tabulates a very close companion (WSI~53~Ab) undetected by \textit{Gaia}.
For this pair, \citet{mason11} measured $\rho$ = 0.237\,arcsec and $\theta$ = 108.3\,deg on only one epoch (J2005.8687), and estimated $\Delta$ = 2.4\,mag from their speckle interferometric measurements in the optical.
However, the companion candidate was not detected in an earlier speckle survey of the Hyades by \citet{mason93}, nor has been detected yet by any other team.
In particular, there are no hints for any close companion beyond the 7.6\,M$_{\rm Jup}$-minimum-mass planet around $\epsilon$~Tau \citep{sato07}. 
The other close double is b~Cen~(AB), which is made of a B2.5\,V star in the Upper Centaurus-Lupus OB association with a very close, later companion of uncertain properties and a wide-orbit substellar object detected in direct imaging \citep[][]{janson21}; this system was, however, filtered in the following~step.

\subsection{Ultra-cool dwarfs}
\label{sec:ultra-cool_dwarfs}

In their desire to be as complete as possible, either the Extrasolar Planets Encyclopaedia, the NASA Exoplanet Archive, or both often tabulate exoplanet candidates that are far from being actual exoplanets according to the International Astronomical Union definition of a planet in the Solar System\footnote{\url{https://www.iau.org/static/resolutions/Resolution_GA26-5-6.pdf}}.
For example, although much has been written about the differences between brown dwarfs and substellar objects below the deuterium burning mass limit at about 13\,M$_{\rm Jup}$ \citep[][and references therein]{caballero18}, the Extrasolar Planets Encyclopaedia tabulates the first unambiguous brown dwarfs discovered, namely Teide~1 \citep[$\sim$65\,M$_{\rm Jup}$;][]{rebolo95} and GJ~229\,B \citep[$\sim$35\,M$_{\rm Jup}$;][]{nakajima95}, counting as exoplanets\footnote{\url{https://exoplanet.eu/catalog/\#inclusion-criteria-section}}, but not hundreds of other ultra-cool dwarfs (UCDs) with spectral type M7\,V or later, \textit{Gaia} parallax, and mass at or below the hydrogen burning limit (e.g. \citealt{basri00,kirkpatrick05,smart19}).
Something similar happens to a few M-type companions to young stars in stellar kinematic groups and star-forming regions, and LTY-type companions to very nearby stars and brown dwarfs detected via direct imaging. 
These UCD companions, which are counted as exoplanets in one or the two catalogues, orbit at a few arcseconds from their primaries, comparable to the separations between stars in multiple systems, have been resolved from their stars and characterised photometrically and even spectroscopically, or have high-uncertainty model-dependent masses at or above the deuterium burning limit. 
Because of their extreme heterogeneity, we discarded a total of 66 stars in systems with UCDs discovered by direct imaging, which left us only exoplanet candidates in compact orbits detected with the RV and transit methods. 
The 66 stars with imaged UCD companions are collected in Table~\ref{tab:UCD_systems} with their respective references.
Actually, there are 65 systems because both young late M dwarfs 2MASS J14504216--7841413 and 2MASS J14504113--7841383, which form a common $\mu$ and $\pi$ double, are each counted as exoplanet candidates.
Among the 65 discarded systems, one can find cornerstone star-brown dwarf systems such as 
GJ~229 (HD 42581) itself \citep{nakajima95},
G~196--3 \citep{rebolo98},
$\eta$~Tel \citep{lowrance00},
TWA~5 (CD--33~7795) \citep{macintosh01},
or CD--52~381 \citep{neuhauser03}.
We included here one of the three found systems with possible circumbinary planets, HD~202206, because the planet is also a brown dwarf \citep{benedict17}. 
Most of the 66 discarded stars in 65 systems in Table~\ref{tab:UCD_systems} have UCDs companions that are too faint (and, sometimes, too close) for \textit{Gaia}.

\subsection{Close binaries without \textit{Gaia} astrometric solution}
\label{close_binaries_without_Gaia}

At this stage, we were back to the 238 multiple systems that remained from the original sample of 311 stars with companions after discarding the 7 stars in Table~\ref{tab:clusters} and the 66 stars in Table~\ref{tab:UCD_systems}.
All the 238 exoplanet host stars have both $\mu$ and $\pi$ from \textit{Gaia} DR3 except for 3 very bright stars (Aldebaran, Fomalhaut, and $\gamma^{\text{01}}$~Leo), and LP~413--32~B, which was recovered (see above) from \citet{feinstein19}.
However, there are 79 companions without \textit{Gaia} DR3 $\mu$ and $\pi$.
Of them, 20 have a \textit{Gaia} DR3 entry but not a five-parameter astrometric solution:
one star that is the early K companion at 1.2\,arcsec to the late-F planet-host star HD~176051;
six fainter close companions with similar or slightly greater angular separations $\rho$\,$\sim$\,1.2--5.0\,arcsec but much larger magnitude differences $\Delta G$\,$\sim$\,5.5--11.5\,mag to their primaries (Gaia DR3 959971546241605120, Gaia DR3 1220404653532465536, Gaia DR3 5917231492302779776, Gaia DR3 4122133830617000320, Gaia DR3 4296208099223616640, and Gaia DR3 6407428994689762560); 
12 stars that are wide companions to planet host stars and that are in turn part of close binaries tabulated by WDS ($\rho \approx$ 0.06--1.3\,arcsec), namely $\psi^{\text{01}}$~Aqr~[BC]\footnote{The square brackets indicate that we use these designations for the first time.}, 94~Cet~[BC] (both with very large uncertainties in \textit{Gaia} equatorial coordinates for its magnitude), HD~178911~[AB], HD~222582~[BC] (LP~703--44~AB), BD--17~588[BC], and LSPM~J1301+6337~[AB] (a wide companion to HD~113337 with a bimodal distribution in its $G$-band light curve that depends on the scan direction across the source); and 1 wide companion that is also a close binary, namely L~72--1~[AB], but is presented here for the first time (WDS 15154--7032; Appendix~\hyperref[ch:Appendix_E]{E}).
The \textit{Gaia} DR3 $G$-band light curve of L~72--1~[AB] displays the same bimodal distribution as LSPM~J1301+6337~[AB], which, together with its missing \textit{Gaia} astrometric solution, points towards a close binarity of the order of 0.2\,arcsec.

The other 59 companions do not even have a \textit{Gaia} DR3 entry.
They include the three companions reported in the literature but not tabulated by WDS that were mentioned above (in two circumbinary systems and one spectroscopic binary) and  56 companions in very close WDS systems, usually resolved with adaptive optics and beyond \textit{Gaia}'s capabilities (e.g. $\gamma$~Cep~AB; \citealt{neuhauser07}), that are close to naked-eye stars with very bright haloes ($\tau$~Boo~B, 26~Dra~B, 54~Psc~B) or that are very bright themselves (e.g. $\alpha$~Cen~A and~B, $\gamma^{\text{02}}$~Leo). 

Although we could recover proper motions, parallaxes, or proper motions different from \textit{Gaia} DR3 only for a small fraction of them, given the robust multiplicity classification in most cases, we kept all the known close companions in our analysis.
Thanks to this analysis, though, we were able to add a new component to a system that was considered to be double and is instead triple (WDS 15154--7032; Appendix~\hyperref[ch:Appendix_E]{E}).

\subsection{Final cut}
\label{sec:final_cut}

We went on with the revision of known exoplanet candidates, and discarded 18 additional systems from our analysis:
one system, namely GJ~682, with a bright companion candidate at 0.17\,arcsec reported by \citet{ward15} and discarded with deep imaging by \citet{desgrange23};
two systems, namely 14~Her and 70~Vir, with faint companion candidates ($\Delta I=$ 10.9--11.4\,mag, $\rho=$ 2.9--4.3\,arcsec and 42.8\,arcsec) detected in imaging searches by \citet{pinfield06} and \citet{roberts11} on only one epoch that are background sources according to, for example, \citet{patience02}, \citet{grether06}, \citet{carson09}, \citet{leconte10}, \citet{ginski12}, \citet{durkan16}, and \citet{fontanive21} (see also \citealt{rodigas11} and \citealt{dalba21} for searches at even closer angular separations);
two systems, namely HD~9578 and TOI--717, with planet candidates that have not undergone publication within a peer-reviewed journal yet\footnote{HD~9578 was only announced in a press release in October 2009 during an international conference.};
one system, namely Fomalhaut ($\alpha$~PsA), with a planet candidate proposed by \citet{kalas05, kalas08}, considered to be the first candidate imaged at visible wavelengths, which is actually an expanding blob of debris from a massive planetesimal collision in the disc \citep{gaspar20, gaspar23};
one triple system, namely LP~563--38, with two M dwarfs and an eclipsing brown dwarf \citep[][]{irwin10}; 
six systems, namely 11~Com,
HD~26161,
HD~77065,
HD~109988,
HD~127506,
and BD+24~4697, with radial-velocity companions with minimum masses between 15.5\,M$_{\rm Jup}$ and 53.0\,M$_{\rm Jup}$ and therefore actual masses well above the deuterium burning mass limit; 
and five systems with exoplanet candidates with astrometric masses above the hydrogen burning mass limit, that is, in the stellar domain.
The five discarded stellar companions are
HD~184601\,B \citep[$M = \text{117}^{+\text{36}}_{-\text{32}}\,{\rm M}_{\rm Jup}$;][]{xiao23}, 
HD~211847\,B \citep[$M = \text{148}\pm\text{5}\,{\rm M}_{\rm Jup}$;][]{philipot23}, 
HD~283668\,B \citep[$M = \text{319}\pm\text{19}\,{\rm M}_{\rm Jup}$;][]{xiao23}, 
HD~114762\,B \citep[$M = \text{210}\pm\text{10}\,{\rm M}_{\rm Jup}$;][]{gaiacollaboration23a}, 
and BD--02~2198\,B ($M = \text{196.9}^{+\text{5.0}}_{-\text{4.9}}\,{\rm M}_{\rm Jup}$; \citealt{baroch21} -- see also \citealt{biller22}).
Of them, HD~114762\,B has received much attraction in the last decades (\citealt{latham89,patience02,bowler09,kane11,kiefer19} -- See \citealt{latham12} for a historical review).
On the contrary to the 66 discarded stars in Sect.~\ref{sec:ultra-cool_dwarfs}, these 18 additional systems have companion candidates that have never been resolved from their host stars.

Including the new companion in the close binary with missing \textit{Gaia} DR3 astrometric solution and bimodal distribution of $G$-band light curve (L~72--1~[AB]), we had a total of 64 companions found in our common $\mu$ and $\pi$ search but not reported by WDS.
We carefully investigated the literature associated with these 64 stars and found that 44 of them had already been proposed as companions to the exoplanet host stars by other authors \cite[e.g.][]{fontanive21,gaiacollaboration21b}.

\begin{table}[H]
 \centering
 \caption[Systems with new common $\mu$ and $\pi$ companions.]{Systems with new common $\mu$ and $\pi$ companions.}
 \footnotesize
 \scalebox{0.87}[0.87]{
 \begin{tabular}{l@{\hspace{1mm}}l@{\hspace{2mm}}l@{\hspace{2mm}}c@{\hspace{1mm}}c@{\hspace{2mm}}c@{\hspace{2mm}}c@{\hspace{1mm}}l@{\hspace{2mm}}l@{\hspace{2mm}}c@{\hspace{2mm}}c@{\hspace{1mm}}}
 \noalign{\hrule height 1pt}
 \noalign{\smallskip}
 Star & \multicolumn{2}{c}{Spectral type} & \multicolumn{2}{c}{$M$} & $V_r$ & $\rho$ & WDS & Disc. code / & System & |U$_g^*$| \\
  & Value & Ref.$^{\text{(a)}}$ & (M$_{\odot}$) & Ref.$^{\text{(a)}}$ & (km~s$^{-\text{1}}$) & (arcsec)  & & Reference$^{\text{(a,b)}}$ & schema$^{\text{(c)}}$ & (10$^{\text{33}}$\,J) \\
 \noalign{\smallskip}
 \hline
 \noalign{\smallskip} 
* HD 1466 & F8\,V & Tor06 & 1.16 & Des21 & +6.53$\pm$0.16  &  & ... &  & \multirow{4}{*}{\includegraphics[width=7.5mm]{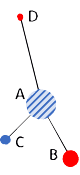}} & \multirow{4}{*}{...} \\
UCAC3 52--533 & M1.8\,(V) & Kra14 & 0.45 & Pec13 &  +5.97$\pm$24.0 & 3023 &  & New &  & \\
UCAC3 53--724 & M5.5\,V & Kir11 & 0.12 & Pec13 & ...  & 1817 &  & Gai21 &  & \\
2MASS J00191296--6226005 & L1 & Gag15 & 0.08 & Pec13 & ... & 3772 &  & New &  & \\
 \noalign{\smallskip}
 \hline
 \noalign{\smallskip} 
* HD 27196 & K1--2\,V & Ker22 & 0.90 & Ker22 & +0.41$\pm$0.14 &  & ... &  & \multirow{2}{*}{\includegraphics[width=6mm]{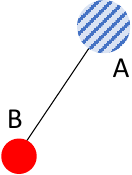}} & \multirow{2}{*}{176} \\
2MASS J04164536--2646205 & $\sim$M4.5\,V & This work & 0.18 & Pec13 & ... & 28 &  & New &  & \\
\noalign{\smallskip}
\hline
\noalign{\smallskip}
* HD 30177 & G8\,V & Hou75 & 0.98 & Fen22 & +62.60$\pm$0.12 &  & ... &  & \multirow{2}{*}{\includegraphics[width=4.5mm]{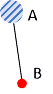}} & \multirow{2}{*}{4.0} \\
Gaia DR3 4774292312023378816 & M7\,(V) & Rey18 & 0.10 & Pec13 & ... & 780 &  & New & \\
 \noalign{\smallskip}
 \hline
 \noalign{\smallskip} 
* HD 88072[A] & G3\,V & Gra03 & 1.04 & Fen22 & --18.00$\pm$0.13 &  & ... &  &  \multirow{3}{*}{\includegraphics[width=7.5mm]{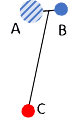}} & \multirow{3}{*}{1.7}  \\
HD 88072B & $\sim$M5\,V  & This work & 0.14 & Pec13 & ... & 3.7 &  &  Gai21 &  & \\
LP 609--39 & $\sim$M5.5\,V & This work & 0.12 & Pec13 & ... &  4772 &  &  New &  & \\
 \noalign{\smallskip}
 \hline
 \noalign{\smallskip} 
* HD 93396 & G8/K0\,IV & Hou99 & 1.46 & Tur15 & +35.00$\pm$0.12 &  & ... &  & \multirow{2}{*}{\includegraphics[width=10mm]{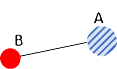}} & \multirow{2}{*}{14} \\
UCAC3 162--116679 & $\sim$K6\,V & This work & 0.69 & Pec13 & +34.60$\pm$0.71 & 1266 &  & New &  &  \\
 \noalign{\smallskip}
 \hline
 \noalign{\smallskip} 
* HD 94834 & K1\,IV & Egg60 & 1.11 & Luh19 & +2.79$\pm$0.12 &  & ... &  &  \multirow{2}{*}{\includegraphics[width=4.5mm]{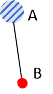}} & \multirow{2}{*}{1178} \\
Gaia DR3 3995918691799173376 & $\sim$M4\,V  & This work & 0.23 & Pec13 & ... & 3.9 &  & New &  &  \\
 \noalign{\smallskip}
 \hline
 \noalign{\smallskip} 
* HD 96700 & G0\,V & Gra06 & 0.99 & Tur15 & +12.80$\pm$0.12 &  & ... &  & \multirow{2}{*}{\includegraphics[width=4.5mm]{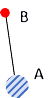}} & \multirow{2}{*}{7.2} \\
CD--27 7881 & K6\,V & Gra06 & 0.72 & Pec13 & ... & 6867 &  & New &  & \\
 \noalign{\smallskip}
 \hline
 \noalign{\smallskip} 
* CD--39 7945 & $\sim$G8\,V & This work & 0.92 & Fri20 & --14.00$\pm$0.16 &  & ... & & \multirow{2}{*}{\includegraphics[width=8mm]{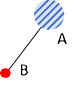}} & \multirow{2}{*}{2.7} \\
Gaia DR3 6140359952471910272 & $\sim$M5.5\,V & This work & 0.12 & Pec13 & ... & 726 &  & New & &  \\
 \noalign{\smallskip}
 \hline
 \noalign{\smallskip} 
* UPM J1349--4603 & $\sim$M3\,V & This work & 0.39 & Hob23 & --15.00$\pm$2.72 &  & ... &  & \multirow{2}{*}{\includegraphics[width=12mm]{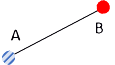}} & \multirow{2}{*}{4.5} \\
Gaia DR3 6107347154502415744 & D: & Gen19 & 0.65 & Gen19 & ... & 1361 &  & New &  &  \\
 \noalign{\smallskip}
 \hline
 \noalign{\smallskip} 
* HD 134606 & G6\,IV & Gra06 & 1.06 & Luh19 & +1.94$\pm$0.12 &  & 15154--7032 &  & \multirow{2}{*}{\includegraphics[width=8mm]{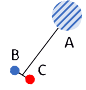}} & \multirow{2}{*}{729} \\
L 72--1[AB] & $\sim$M3+$\sim$M3 & This work & 0.60 & Pec13 & +8.54$\pm$6.11 & 57 &  & FMR 173/New &  & \\
 \noalign{\smallskip}
 \hline
 \noalign{\smallskip} 
* HD 135872 & G5\,IV & Hou99 & 1.65 & Fen22 & --20.00$\pm$0.13 &  & ... &  & \multirow{2}{*}{\includegraphics[width=5.5mm]{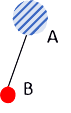}} & \multirow{2}{*}{61} \\
Gaia DR3 6334935890969037696 & $\sim$K4\,V & This work & 0.73 & Pec13 & --20.00$\pm$0.48 & 423 &  & New & \\
 \noalign{\smallskip}
 \hline
 \noalign{\smallskip} 
* CD--24 12030 & $\sim$K2.5\,V & This work & 0.80 & Chr22 & +5.17$\pm$0.29 &  & ... & & \multirow{3}{*}{\includegraphics[width=16mm]{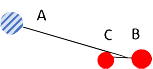}} & \multirow{3}{*}{29} \\
Gaia DR3 6214879765359934592 & $\sim$M2\,V & This work & 0.45 & Pec13 & +8.78$\pm$3.18 & 231 &  & New &  &  \\
Gaia DR3 6214879769657777536 & $\sim$M4\,V & This work & 0.25 & Pec13 & +42.30$\pm$4.30 & 226 &  & New &  &  \\
 \noalign{\smallskip}
 \hline
 \noalign{\smallskip} 
* HD 143361 & G6\,V & Hou78 & 0.95 & Min09 & --0.60$\pm$0.16 &  & ... &  & \multirow{2}{*}{\includegraphics[width=15mm]{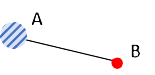}} & \multirow{2}{*}{2.3} \\
Gaia DR3 5994315469412318848 & $\sim$M4.5\,V & This work & 0.19 & Pec13 & ... & 1964 &  & New & & \\
 \noalign{\smallskip}
 \hline
 \noalign{\smallskip} 
* TOI--1410 & $\sim$K4\,V & This work & 0.71 & Sta19 & +2.46$\pm$0.10 &  & ... & & \multirow{2}{*}{\includegraphics[width=12mm]{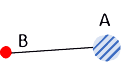}} & \multirow{2}{*}{69} \\
Gaia DR3 1958584427911285120 & $\sim$M3\,V & This work & 0.36 & Pec13 & +3.38$\pm$1.87 & 91 &  & New & & \\
 \noalign{\smallskip}
 \hline
 \noalign{\smallskip} 
* HD 222259A & G6\,V & Tor06 & 0.96 & Ben19 & +7.13$\pm$0.28 &  & 23397--6912 &  & \multirow{3}{*}{\includegraphics[width=16mm]{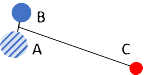}}  & \multirow{3}{*}{4.6} \\
HD 222259B & K3\,V & Tor06 & 0.78 & Pec13 & +5.17$\pm$0.42 & 5.4 &  & R   348 &  & \\
2MASS J23321028--6926537 & M5.3\,V & Ujj20 & 0.17 & Pec13 & ... & 2560 &  & New &  & \\
 \noalign{\smallskip}
 \noalign{\hrule height 1pt}
 \end{tabular}
 } % \scalebox
 \label{tab:new_detected_systems}  
 \footnotesize
 \begin{justify}
    \textbf{\textit{Notes. }}
    $^{\text{(a)}}$ References: 
      Alo15: \citet{alonsofloriano15}; 
      Ben19: \citet{benatti19}; 
      Chr22: \citet{christiansen22}; 
      Des21: \citet{desidera21};
      Egg60: \citet{eggen60}; 
      Fen22: \citet{feng22}; 
      Fri20: \citet{fridlund20}; 
      Gag15: \citet{gagne15}; 
      Gai21: \citet{gaiacollaboration21b}; 
      Gen19: \citet{gentilefusillo19}; 
      Gra03: \citet{gray03}; 
      Gra06: \citet{gray06}; 
      Hob23: \citet{hobson23}; 
      Hou75: \citet{houk75}; 
      Hou78: \citet{houk78}; 
      Hou99: \citet{houk99}; 
      Ker22: \citet{kervella22};
      Kir11: \citet{kirkpatrick11}; 
      Kra14: \citet{kraus14}; 
      Lep13: \citet{lepine13}; 
      Low00: \citet{lowrance00};
      Luh19: \citet{luhn19}; 
      Min09: \citet{minniti09}; 
      Pec13: \citet{pecaut13}; 
      Rey18: \citet{reyle18}; 
      Rie17: \citet{riedel17b}; 
      Smi18: \citet{smith18}; 
      Sta19: \citet{stassun19}; 
      Tor06: \citet{torres06}; 
      Tur15: \citet{turnbull15};
      Ujj20: \citet{ujjwal20}.
    $^{\text{(b)}}$ WDS discoverer codes are written in uppercase, and literature references in lowercase.
    $^{\text{(c)}}$ Blue circles: already reported stars. Red circles: new stars in system. Stripped circles: exoplanet host stars. Symbol sizes are proportional to stellar masses.
 \end{justify}
 \normalsize
\end{table}

We looked for available \textit{Gaia} (mean) absolute radial velocities for the remaining 20 stars (in 18 systems).
Because of the typical faintness of the new companions reported here and the \textit{Gaia} limit at $G \sim$ 16\,mag for radial velocities \citep{katz23}, there are measurements for both the exoplanet host star and companion candidates for only eight systems.
RAVE and other radial-velocity surveys \citep{steinmetz06,kunder17} did not provide with additional measurements of companions.
Of the eight systems, three binary systems, namely those with exoplanet host stars HD~168009, HD~210193, and K2--137, have radial velocities that differ by 80.90 $\pm$ 0.27, 32.55 $\pm$ 0.19, and 34.4 $\pm$ 2.9\,km\,s$^{-\text{1}}$, respectively.
Such differences may be difficult to ascribe to unknown close companions of the secondaries, as they may actually be happening to a lesser degree than to the M-dwarf companion to CD--24~12030, or among the G--K components of HD~222259.
Therefore, the stars in the three pairs do not have the same galactocentric velocities and are optical pairs.
Hence, we discarded them from our analysis.
This deletion left us with 17 genuinely new companions candidates, which are discussed in Sect.~\ref{sec:new_stellar_systems} and shown in Tables~\ref{tab:new_detected_systems} and \ref{tab:new_detected_systems_astrometry}.
The former tabulates main derived data (stellar mass $M_\star$, systemic radial velocity $V_r$, angular separation $\rho$, WDS identifier and discoverer code when available, a pictographical system schema, and reduced binding energy $|U_g^*|$ -- see below), while the later tabulates the astrometric data (parallax, proper motion in right ascension and declination, $\mu_{\rm ratio}$, and $\Delta$PA).

\begin{table*}
 \centering
 \caption[Binary systems with detected planets around both stars.]{Binary systems with detected planets around both stars.}
 \footnotesize
  \scalebox{1}[1]{
 \begin{tabular}{lc@{\hspace{1mm}}ccccc@{\hspace{1mm}}cc}
 \noalign{\hrule height 1pt}
 \noalign{\smallskip}
 Star & \multicolumn{2}{c}{Spectral type} & $\alpha$\,(J2000) & $\delta$\,(J2000) & $d$ & \multicolumn{2}{c}{Planets}  & \textit{s}\\
  & Value & Reference$^{\text{(a)}}$ & (hh:mm:ss.ss) & (dd:mm:ss.s) & (pc) & Number & Reference$^{\text{(a)}}$ & (au) \\
 \noalign{\smallskip}
 \hline
 \noalign{\smallskip} 
 HD 20782 & G1.5\,V & Gra06 & 03:20:03.58 & --28:51:14.7 & 35.9 & 1 & Jon06 & \multirow{2}{*}{9\,075} \\
 HD 20781 & G9.5\,V & Gra06 & 03:20:02.94 & --28:47:01.8 & 36.0 & 4 &  May11, Udr19 & \\
 \noalign{\smallskip}
 UCAC3 234--106607 & M1.3\,V & Bir20 & 12:58:23.27 & +26:30:09.1 & 58.7 & 1 & Lew22 & \multirow{2}{*}{909} \\
 Sand 178$^{\text{(b)}}$ & M2.4\,V & Bir20 & 12:58:22.24 & +26:30:16.1 & 58.8 & 1 & Lew22 & \\
 \noalign{\smallskip}
 HD\,133131A & G2.0\,V & Sto72 & 15:03:35.45 & --27:50:33.2 & 51.5& 2 & Tes16 & \multirow{2}{*}{379} \\
 HD\,133131B & G2.0\,V & Sto72 & 15:03:35.81& --27:50:27.6 & 51.5 & 1 & Tes16 & \\
 \noalign{\smallskip}
 \noalign{\hrule height 1pt}
 \end{tabular}
 }
 \label{tab:both_stars_with_planets}
 \small
 \begin{justify}
    \textbf{\textit{Notes. }}
    $^{\text{(a)}}$
    Bir20: \citealt{birky20}; 
    Gra06: \citealt{gray06}; 
    Jon06: \citealt{jones06}; 
    Lew22: \citealt{lewis22}; 
    May11: \citealt{mayor11}; 
    Sto72: \citealt{stock72}; 
    Tes16: \citealt{teske16}; 
    Udr19: \citealt{udry19}.
    $^{\text{(b)}}$ \citet{lewis22} stated that ``depending on the true period and eccentricity of the system, the minimum companion mass may fall in a range from 2\,M$_{\rm Jup}$ to 50\,M$_{\rm Jup}$, making it a possible brown dwarf candidate''.
 \end{justify}
 \normalsize
\end{table*}

After all these considerations, we identified 215 exoplanet host stars in 212 multiple systems, of which 173 are binary, 39 are triple, and three are quadruple. 
The list of these multiple systems is shown in Table~\ref{tab:basic_data_identified_hosts}.
We tabulate only 212 entries because there are three binary systems with planets discovered around both stars.
The three systems, accounting for six stars and ten planets, are also displayed in Table~\ref{tab:both_stars_with_planets}. 

\subsection{Planetary systems}
\label{sec:results_planetary_systems}

\begin{figure*}
 \centering
 \includegraphics[width=1\linewidth, angle=0]{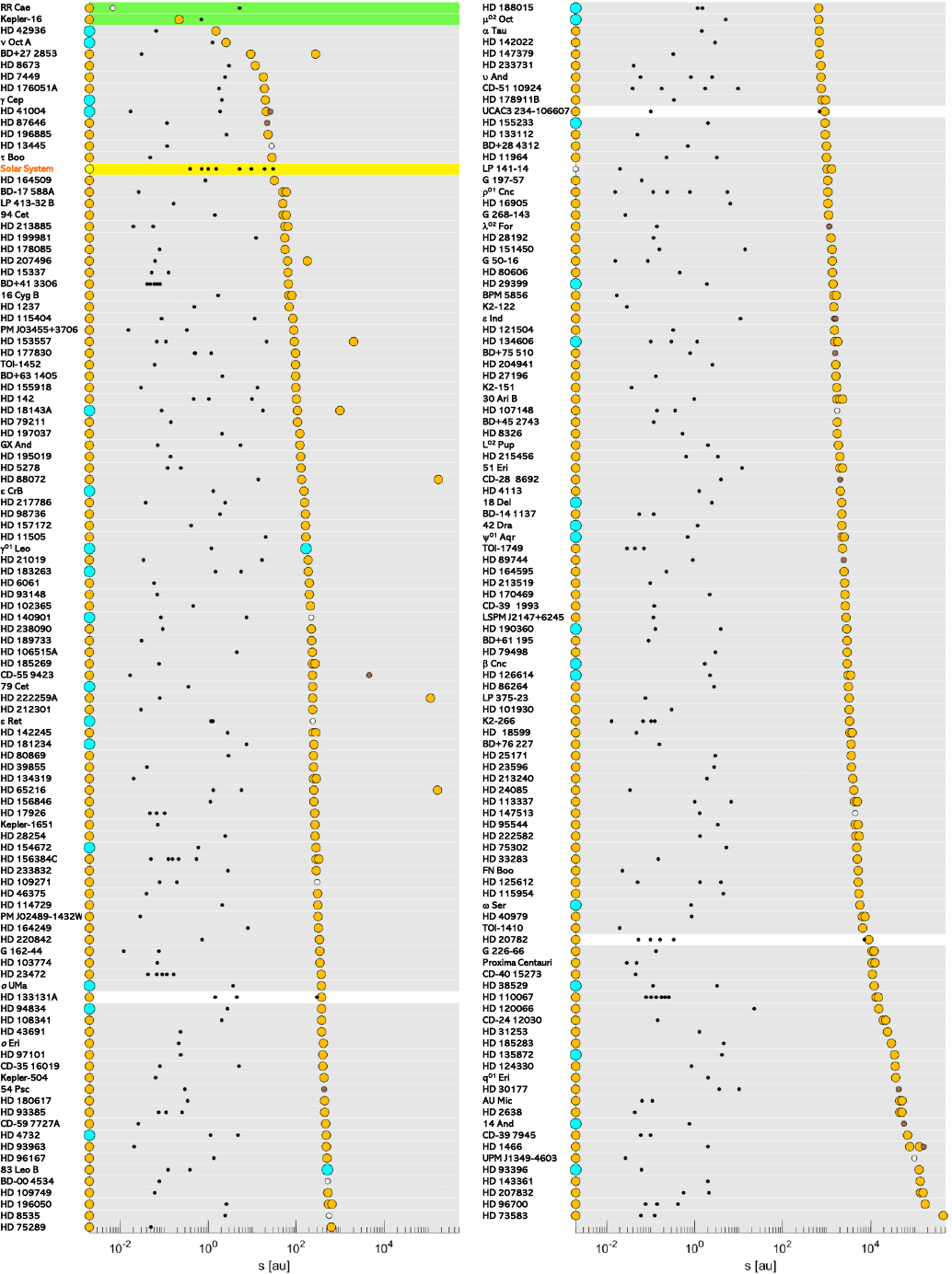}
 \caption[Schematic configurations of multiple stellar systems with exoplanets.]{Schematic configurations of multiple stellar systems with exoplanets.
 Orange circles are main-sequence stars, cyan circles are subgiant and giant stars, white circles are white dwarfs, small brown circles are brown dwarfs, and black dots are planets.
 We display our 212 systems with grey background, except for the two systems with circumbinary planets, namely RR~Cae and Kepler-16, with green background, and the three systems from Table~\ref{tab:both_stars_with_planets} with planets around both stars, with white background.
 The systems are sorted by increasing separation from the planet host star to the closest companion star.
 The abscissa is in logarithmic scale.
 We also display the Solar System in yellow as a comparison.}
 \label{fig:planets}
\end{figure*}

For the 215 stars in 212 multiple systems, we compiled the main parameters of 276 of the 302 planets orbiting them from the Extrasolar Planets Encyclopaedia and NASA Exoplanet Archive. 
In particular, we retrieved their orbital periods, $P$, semi-major axes, $a$, eccentricities, $e$, and masses, $M_{\rm pl}$, or minimum masses, $M_{\rm pl} \sin{i}$, for transiting or radial-velocity planets, respectively.
When exoplanets are common to both databases, we took the most recent parameters from the NASA Exoplanet Archive, except for very few cases with significantly smaller uncertainties that we took from the Extrasolar Planets Encyclopaedia.
The remaining 26 planets miss at least one datum.
The schematic configurations of the 212 systems with all their members (stars, white dwarfs, brown dwarfs, planets) are represented in Fig.~\ref{fig:planets}.
In occasions, the planet host star is not the brightest primary in the system (e.g. $\alpha$~Cen, 26~Dra, 83~Leo, GJ~667).
A total of 32 of the investigated exoplanets orbit giant or subgiant stars and only one around a white dwarf (LP~141--14\,b; Appendix~\hyperref[ch:Appendix_E]{E}). 
The rest of the host stars are main-sequence stars (see Sect.~\ref{sec:discussion_hosts} and Fig.~\ref{fig:HRD_new_detected_systems}).

Most exoplanet searches have focused on stars without close stellar companions.
For example, the CARMENES survey for exoplanets around nearby M dwarfs \citep{quirrenbach14} discarded from their guaranteed and legacy time observations target stars that have companions at less than 5\,arcsec, including astrometric and spectroscopic binaries, but kept resolved stellar systems with wider separations \citep{caballero16,cortescontreras17b,jeffers18,reiners18b}.
Since the CARMENES survey is a prototypical radial-velocity exoplanet search and our sample is built upon a collection of results of searches, our target list of 212 systems is naturally biased against very close binary systems with planets. 
Acknowledging this bias prior to commencing any discussion is therefore mandatory.
Furthermore, this deficit of close binary systems is also found in transit exoplanet searches, which are also partly affected by this bias \citep{ziegler21}. 

For comparison purposes, we defined a control sample of single stars with exoplanets, but no known stellar or brown dwarf companion at any separation.
For that, we took a few steps back in our sample definition and discarded multiple systems with planets (Table~\ref{tab:basic_data_identified_hosts}), or with one or more ultra-cool dwarfs (Table~\ref{tab:UCD_systems}), and retained only single stars at $d <$ 100\,pc.
For the 687 single stars with planets, we searched for $P$, $a$, $e$, and $M_{\rm pl}$ or $M_{\rm pl} \sin{i}$ of their 1029 planets exactly as we did for the multiple systems with planets. 
From now on, we called this set of stars the ``single star sample''.

\section{Results}
\label{sec:results_hosts}

\subsection{New stellar systems}
\label{sec:new_stellar_systems}

Of the 17 genuinely new companions in Table~\ref{tab:new_detected_systems}, 12 are part of completely new systems, while the other five are additional companions to four systems tabulated by WDS (two) or reported in the literature (two).
In these four systems, the K- and M-dwarf companions had never been described in the literature and therefore, we tabulated their \textit{Gaia} DR3 identifiers (the companion of HD~143361 was catalogued by \citealt{gaiacollaboration21b}).
On the other hand, the 17 companions are members of 15 systems, four of which are triple, and one is quadruple.
The four known systems with new additional companions and the new triple system are described in Appendix~\hyperref[ch:Appendix_E]{E}.
The remaining ten systems are new binaries, that were thought to be single.

As already noticed by \citet{shaya11}, \citet{newton19}, or \citet{gaiacollaboration21b}, the very large projected physical separations between primary stars and companions casts doubts on the actual gravitational binding of some systems.
In our sample, a few of the systems are very wide. 
In particular, there are at least two systems in young associations and separations greater than 1000\,arcsec (HD~1466 with its M-L companions, and WDS 23397--6912), which make the multiple system classification even blurrier \citep{caballero10,tokovinin12,duchene13}.

To assess whether the 17 genuinely new companions in Table~\ref{tab:new_detected_systems} are indeed bound or not, we calculated the reduced binding energy value by following the methodology outlined by \citet{caballero09}:
\begin{equation}
|U^*_g| = G\frac{M_{\star,\text{1}} M_{\star,\text{2}}}{s},
\label{eqn:binding}
\end{equation}
\noindent where $G$ is the gravitational constant, $M_{\star,\text{1}}$ and $M_{\star,\text{2}}$ are the masses of both stars, and $s$ their projected physical separation.
The asterisk in $|U^*_g|$ indicates that the absolute values of the ``true'' potential energies $U_g$ using the physical separation $r$ must be lower than in Table~\ref{tab:new_detected_systems} \citep{caballero09}.
For the three triple systems, we used the combined masses of the close pairs in the calculus (e.g. at a large separation, LP~609--39 feels the gravitational attraction of HD~88072AB as if it were a single, more massive star).
We refrained from using the $\overline{r} \approx$\,1.26\,$~ \overline{s}$ relation between projected and true separations determined by \citet{fischer92} to facilitate direct comparisons with certain other studies \citep{close03, burgasser07a, radigan09, caballero10, faherty10}.

\begin{figure}
 \centering \includegraphics[width=0.6\linewidth, angle=0]{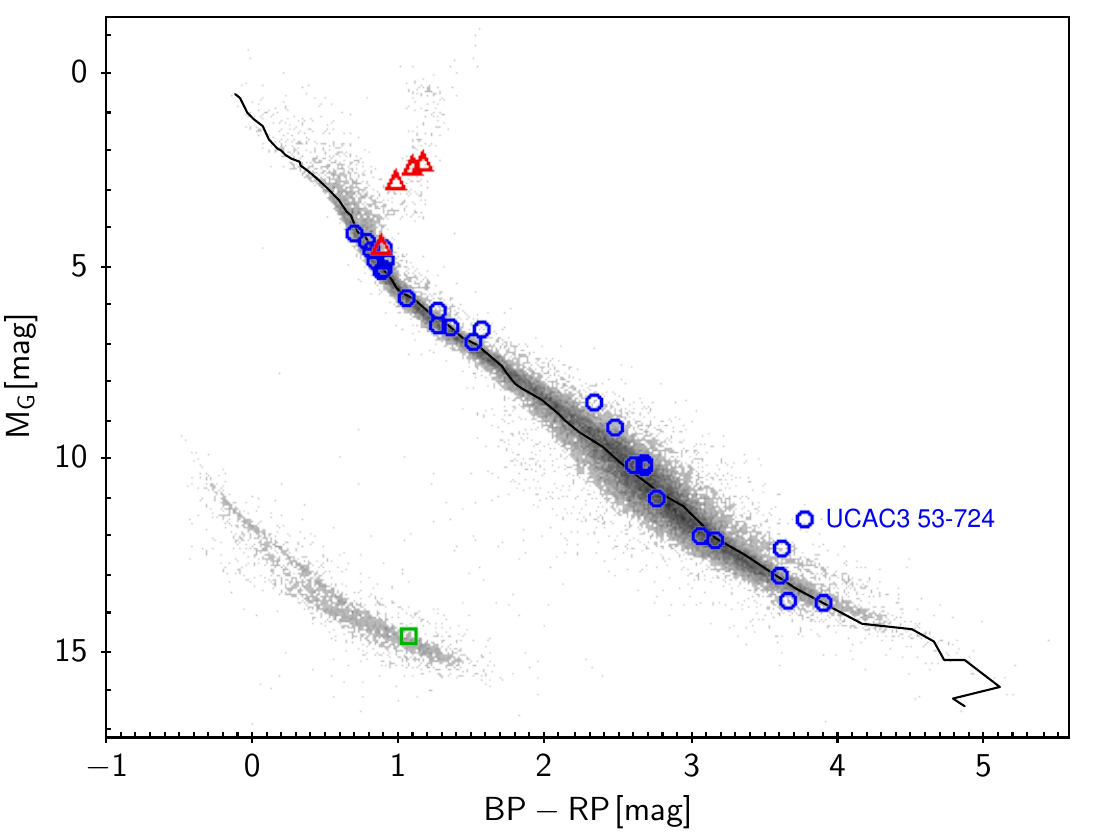}
 \caption[\textit{Gaia} colour-magnitude diagram of the systems with new common $\mu$ and $\pi$ companions in Table~\ref{tab:new_detected_systems}.]{\textit{Gaia} colour-magnitude diagram of the systems with new common $\mu$ and $\pi$ companions in Table~\ref{tab:new_detected_systems}.
 Blue circles, red triangles, and the green square stand for the main-sequence, subgiant, and white dwarf stars, while the small grey points are the over 50\,000 random \textit{Gaia} sources retrieved as \citet{taylor21} did.
 The young overluminous star UCAC3\,53--724 in Tucana-Horologium is labelled.
 The black solid line is the updated main sequence of \citet{pecaut13}.
 We did not apply any colour or magnitude correction for reddening.}
 \label{fig:HRD_new_detected_systems}
\end{figure}

We computed the projected physical separation from the angular separation and the (parallactic) distance to the primary and used the colour-magnitude diagram in Fig.~\ref{fig:HRD_new_detected_systems} as a guide to determine stellar masses.
For resolved stars in the main sequence, we used the conversion from \textit{Gaia} $G$-band absolute magnitudes and $B_P-R_P$ colours to masses outlined in table~5 of \citet{pecaut13}, which is available in an enhanced and updated version on line\footnote{\url{https://www.pas.rochester.edu/~emamajek/EEM_dwarf_UBVIJHK_colors_Teff.txt}}.
For six stars out of the main sequence (e.g., four subgiants, one white dwarf, and one young overluminous star), we took their masses from the literature \citep{turnbull15, livingston18, gentilefusillo19, luhn19, feng22}.
For completeness, we also took spectral types from Simbad; when not available, we estimated them from photometry by relying on \citet{pecaut13} and \citet{cifuentes20}.

The names, spectral types (and references), stellar masses (and references), radial velocities for the 35 resolved stars, and angular separation, WDS identifier, pair discovery code or reference, and a schema of the 15 systems with new proper motion and parallax companions are displayed in  Table~\ref{tab:new_detected_systems}, and their astrometric properties are shown in Table~\ref{tab:new_detected_systems_astrometry}.
The asterisks denote the planet-host stars, which in all cases are the system primaries.
We were able to compute $|U^*_g|$ values for all but one trapezoidal system, of which the host star is HD~1466 in Tucana-Horologium \citep{gonzalezpayo23}.  
The reduced binding energies vary from over 10$^{\text{36}}$\,J for the HD~94834 binary (K1\,IV + $\sim$M4\,V) to merely a few 10$^{\text{33}}$\,J for the widest systems, with separations of up to 6\,900\,arcsec.
These small $|U^*_g|$ values are, however, slightly greater than 10$^{\text{33}}$\,J, which may represent the minimum threshold for binding gravitational energy before disruption by the galactic potential \citep{caballero10}, as well as the minimum value of reduced binding energy of moderately separated binary systems of very low mass \citep{chauvin04,artigau07,caballero07a,radigan09}.
Even for the system HD~88072\,AB and LP~609--39, the lowest $|U^*_g|$ in Table~\ref{tab:new_detected_systems} do not deviate from what is expected for gravitationally bound systems in the Milky Way (see fig.~6 of \citealt{gonzalezpayo23}).
Besides, the actual binding of HD~1466 and its companions in Tucana-Horologium and similar systems will be the subject of a forthcoming work, and it is out of the scope of this paper.
Furthermore, we kept in Table~\ref{tab:basic_data_identified_hosts} two systems with ultra-wide separations separations greater than 6900\,arcsec: WDS~14396--6050 ($\alpha$~Cen~AB and Proxima, $\rho\sim$\,7960\,arcsec; \citealt{innes1915}) and WDS~08388--1315 (HD~73583 and BD--09~2535, $\rho\sim$\,14\,240\,arcsec; 
\citealt{shaya11}).
As a result, we maintained the 15 systems for the following analysis.

\subsection{Semi-major axes and separations}
\label{sec:separations}

\begin{figure}
 \centering
 \includegraphics[width=0.6\linewidth, angle=0]{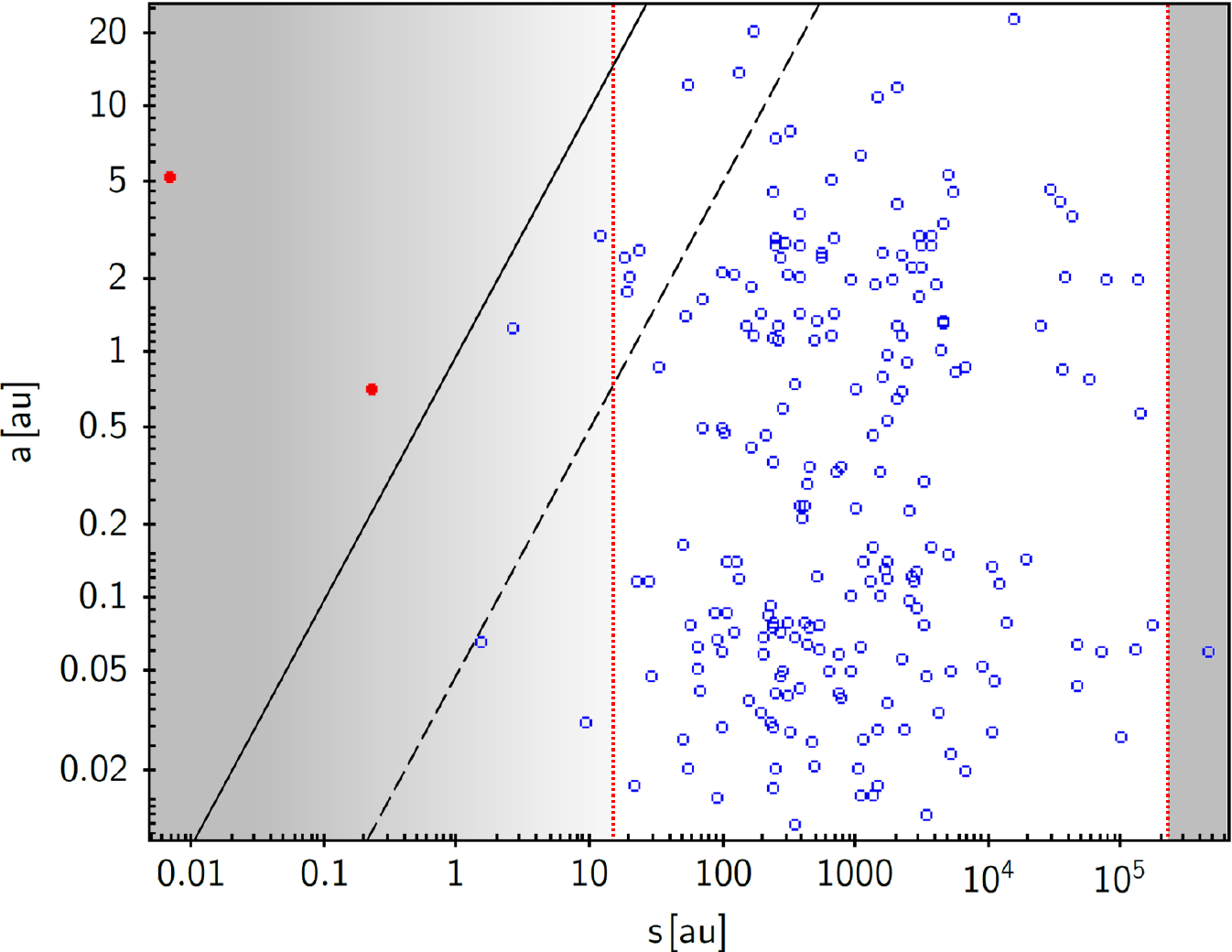}
 \caption[Exoplanet semi-major axis versus star-projected physical separation.]{Exoplanet semi-major axis versus star-projected physical separation.
 Solid and dashed black lines mark the $s$:$a$ relationships at 1:1 and 20:1, respectively.
 The two circumbinary systems are marked with red filled symbols.
 Grey regions delimited by red dotted lines (16.5\,au and 1\,pc = 648\,000/$\pi \approx$ 206\,264.81\,au) indicate the approximate incompleteness areas (see main text). 
 }
 \label{fig:s-a-plot}
\end{figure}

As a first step of the analysis, we compared the projected physical separations between stars in multiple systems, $s$, and the exoplanet semi-major axes, $a$. 
In particular we compared the $s$ between the exoplanet host star and the closest stellar (or white dwarf, or brown dwarf) companion in triple and quadruple systems, and the $a$ of the most separated planet in multi-planet systems.
This comparison is illustrated by Fig.~\ref{fig:s-a-plot}.
We mark the incompleteness areas in shadows of grey.
The outer limit of the completeness area is defined by our maximum search $s$ at 1\,pc, while the approximate inner limit is set by $\overline{s} = \rho_{Gaia} \cdot \overline{d}$, where $\overline{s}$ and $\overline{d}$ are the median of the individual $s$ and $d$, and $\rho_{Gaia}$ is the critical value at 0.4\,arcsec for spatial resolution of close binaries by \textit{Gaia} \citep{lindegren18a,lindegren21}.
There have been, though, adaptive optics and speckle imaging searches that have explored inner regions ($\rho <$ 0.4\,arcsec).
Besides, the investigated range of projected physical separations is directly proportional to the $d$, so the inner regions of nearby stars are in general better studied, and vice versa. 
Therefore, the darker the region, the more incomplete it is.

In Fig.~\ref{fig:s-a-plot}, besides the two close binaries with circumbinary planets (RR~Cae and Kepler--16), there is a number of remarkable systems that stand out because of their relatively small $s/a$ ratios.
They are relatively close binaries with planetary systems, some of which might challenge current exoplanet formation scenarios in truncated protoplanetary discs.
About 25\% of the displayed multiple systems have $s/a <$\,188 (first quartile), and about 10\% have $s/a <$\,34 (first decile).
We list in Table~\ref{tab:systems_with_sa20} the only nine systems with $s/a <$\,20, together with their corresponding references.
The 9 systems represent about 8\% of the systems (there is a gap in the $s/a$ distribution at $\sim$13--22).
Most of the references, especially those presenting systems with the smallest $s/a$ ratios, already made extensive discussion on the different formation, evolution, and stability mechanisms that gave rise to such peculiar systems \citep[e.g.][]{neuhauser07,ramm21,feng22}.
Some stellar parameter values may be biased due to companion blend (e.g. HD~176051\,AB: \citealt{muterspaugh10} did not know around which component the planet b, detected by astrometry, is orbiting).
Actually, some stellar systems secondaries, marked with ``[B]'' in Table~\ref{tab:systems_with_sa20}, are so close to their primaries that they do not have an entry in Simbad and have only been resolved with high-resolution imagers.
The non-tabulated system that stands out in Fig.~\ref{fig:s-a-plot} with $s \approx$ 1.5\,au and $a \approx$ 0.07\,au ($s/a \approx$ 22) is HD~42936, which is made of a K0\,IV primary orbited by a super-Earth at $P \approx$ 6.67\,d and a very low mass star at $P \approx$ 507\,d \citep{barnes20}.

\begin{table*}[h]
 \footnotesize
 \centering
 \caption[Multiple systems with planets and $s/a<$\,20.]{Multiple systems with planets and $s/a<$\,20.}
  \scalebox{1}[1]{
 \begin{tabular}{@{\hspace{2mm}}l@{\hspace{2mm}}l@{\hspace{2mm}}l@{\hspace{0mm}}c@{\hspace{2mm}}c@{\hspace{2mm}}c@{\hspace{2mm}}l@{\hspace{2mm}}}
 \noalign{\hrule height 1pt}
 \noalign{\smallskip}
 Host star & Companion & Planet & $s$ & $a$ & $e$ & References \\  
  & star &  & (au) & (au) &  &  \\ 
 \noalign{\smallskip}
 \hline
 \noalign{\smallskip} 
 HD 11505[A] & HD 11505B & b & 167.19$\pm$0.20 & 20.2$^\text{+4.8}_\text{-3.4}$ & 0.144$^\text{+0.011}_\text{-0.056}$  & \citet{feng22} \\ 
 \noalign{\smallskip}
 HD 88072[A] & HD 88072B & b & 132.77$\pm$0.19 & 13.9$^\text{+4.1}_\text{-2.0}$ & 0.16$^\text{+0.14}_\text{-0.10}$ & \citet{feng22} \\ 
 \noalign{\smallskip}
 HD 8673[A] & HD 8673[B]  & b & 11.8 & 2.97$^\text{+0.15}_\text{-0.17}$  & 0.730$^\text{+0.042}_\text{-0.026}$  & \citet{hartmann10,feng22} \\ 
 \noalign{\smallskip}
 HD 196885 & HD 196885B & b & 23.0 & 2.6$\pm$0.1 & 0.48$\pm$0.02  & \citet{correia08,chauvin11}  \\ 
 \noalign{\smallskip}
 HD 7449[A] & HD 7449[B] & b & 20.8$\pm$0.3 & 2.438$^\text{+0.062}_\text{-0.063}$ & 0.752$^\text{+0.035}_\text{-0.032}$  & \citet{dumusque11,feng22}  \\ 
 \noalign{\smallskip}
 \multirow{2}{*}{$\gamma$ Cep [A]} & \multirow{2}{*}{$\gamma$ Cep [B]} & \multirow{2}{*}{b} & \multirow{2}{*}{20.18$\pm$0.66} & \multirow{2}{*}{2.05$\pm$0.06} & \multirow{2}{*}{0.049$\pm$0.034} & \citet{hatzes03,neuhauser07} \\ 
  &  &  &  &  &  & \citet{endl11} \\ 
 \noalign{\smallskip}
 HD 176051A & HD 176051B & b & 19.14$\pm$0.04 & 1.76  & 0 (fixed)  & \citet{muterspaugh10} \\ % Some stellar parameter values may be biased due to companion blend (WDS Catalog)
 \noalign{\smallskip}
 HD 199981[A] & HD 199981B & b  & 55.16$\pm$0.03 & 12.3$^\text{+3.6}_\text{-2.2}$  & 0.155$^\text{+0.073}_\text{-0.041}$ & \citet{feng22} \\ 
 \noalign{\smallskip}
 $\nu$ Oct [A]  & $\nu$ Oct [B] & b & 2.6 & 1.25$\pm$0.05 & 0.11$\pm$0.02 & \citet{ramm16,ramm21} \\ 
 \noalign{\smallskip}
 \noalign{\hrule height 1pt}
 \end{tabular}
 }
 \label{tab:systems_with_sa20}
\end{table*}

According to \citet{winn15}, ``the rough rule-of-thumb [for system stability is] that the planet’s period should differ by at least a factor of three from the binary’s period, even for a mass ratio as low as 0.1''.
The three stars in Table~\ref{tab:systems_with_sa20} with the smallest $s/a$ ratios are: 
HD~199981 \citep[][$s/a \approx$ 4.48]{feng22}, % P ratio about 8
HD~8673 \citep[][$s/a \approx$ 3.97]{hartmann10,feng22}, % P ratio about 7
and $\nu$~Oct \citep[][$s/a \approx$ 2.04]{ramm21}. % P ratio about 2.6
Their corresponding $P_{\text{star}} / P_{\text{planet}}$ ratios, after retrieving the stellar masses from the literature \citep[e.g.][]{roberts15} and applying Third Kepler's Law, are  about 8, 7, and 2.6, respectively.
HD~199981 (a late-K primary and a mid-M secondary separated by 2.7\,arcsec) and HD~8673 (a late-F primary and an early M secondary separated by 0.31\,arcsec), both with massive substellar companions at the planet-brown dwarf boundary, seem to be stable systems in spite of their small $s/a$ and $P_{\text{star}} / P_{\text{planet}}$ ratios (see again \citealt{feng22}).
However, the $\nu$~Oct system (an early K giant with a $\sim$0.58\,M$_\odot$ companion separated by 0.11\,arcsec and a planetary candidate in a retrograde orbit -- \citealt{ramm09,ramm16,ramm21,ramm15}), is catalogued in the Extrasolar Planets Encyclopaedia as ``Unconfirmed'' and is not catalogued at all by the NASA Exoplanet Archive.
As a result, the hypothetical challenge for formation and stability scenarios may not apply in this particular case, nor in the other confirmed systems with larger $s/a$ and $P_{\text{star}} / P_{\text{planet}}$ ratios.

\subsection{Eccentricities}
\label{sec:eccentricities}

\begin{figure}[H]
 \centering \includegraphics[width=0.6\linewidth, angle=0]{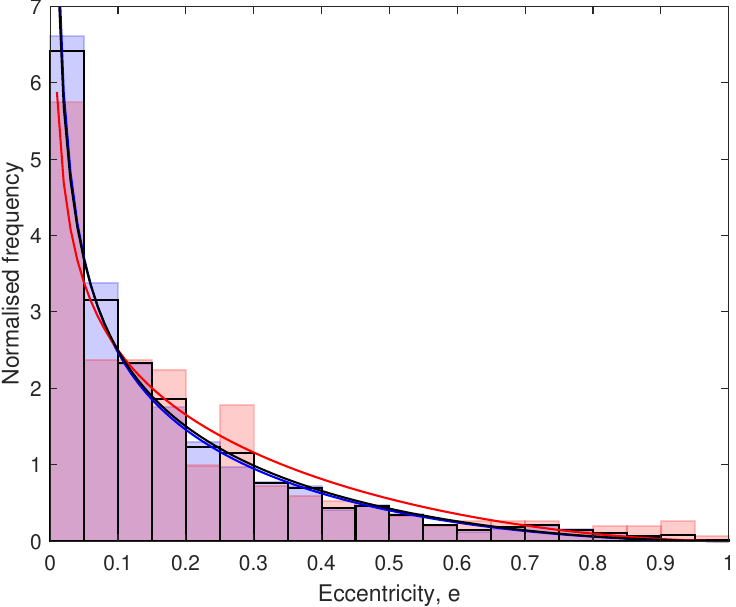}
 \caption[Distributions of eccentricities of planetary orbits and fitted beta distributions.]{Distributions of eccentricities of planetary orbits and fitted beta distributions.  Black open bins: joint sample (1332 planets in single plus multiple stars); blue filled bins: single-star sample (1029 planets); red filled bins: multiple-star system sample (303 planets). The histograms are normalised to have a unit area. The three curve fittings (black for the joint sample, blue for single-star systems, and red for multiple-star systems) have beta distributions with different $\alpha$ and $\beta$ parameters. See text for a description.}
 \label{fig:ecc_global}
\end{figure}

The previous section prepared the ground for this one where the inter-comparison of the planet eccentricity distributions gets more complicated and uses new tools introduced here.
Except for the unconfirmed astrometric exoplanet candidate around one of the two stars in HD~176051AB and for $\gamma$~Cep~Ab (which was originally discovered by \citealt{campbell88} but confirmed 15 years later by \citealt{hatzes03}), all the planets in Table~\ref{tab:systems_with_sa20} have eccentricities $e$ that are significantly different from null.
Furthermore there are two planets with $e >$ 0.7 and a third planet with $e \approx$ 0.48.
Therefore, we addressed the question of whether the multiplicity of the host stellar system has any effect on the eccentricities of the planetary orbits, as previously suggested by other authors \citep[e.g.][]{eggenberger04b,raghavan06,lester21}. 
In Fig.~\ref{fig:ecc_global} we show the distribution of eccentricities of our combined sample of planets around single and multiple stars. 
Apparently, there are more planets with low eccentricity in single systems and more planets with high eccentricities in multiple systems.

To assess the reliability of this possible effect, we first modelled the observed distribution of planet eccentricities in both single and multiple star samples using a unique probability distribution.
We found that the beta distribution, the simplest distribution representing a continuum variable taking values between 0 and 1, can indeed be used to model such a distribution.
The likelihood can therefore be written:
\begin{equation}
P(e|\alpha,\beta)=\frac{e^{\alpha-\text{1}}(\text{1}-e)^{\beta-\text{1}}}{B(\alpha, \beta)},
 \label{eqn:beta}
\end{equation}
\noindent where $B(\alpha, \beta)$ is the beta function as a function of gamma functions:
\begin{equation}
B(\alpha, \beta) = \frac{\Gamma(\alpha) \Gamma(\beta)}{\Gamma(\alpha+\beta)}.
 \label{eqn:gamma}
\end{equation}
In order to fit a beta distribution to these data, we constructed a Bayesian Markov chain Monte Carlo (MCMC) model written in Stan\footnote{\url{https://mc-stan.org}} \citep{carpenter17}. 
Stan is a programming language that makes use of the Hamiltonian Monte Carlo no U-turn sampler (HMC+NUTS) algorithm \citep[][]{neal11,hoffman11} to perform Bayesian modelling and inference. 
This programming technique allowed us to introduce in the calculations the errors in the measured eccentricity, which are typically asymmetrical.
In some cases, only an upper value for the eccentricity is available. 
This lack was solved by incorporating asymmetric Gaussian distributions in our Stan program. 
For the few planetary orbits without an estimation of the eccentricity error we assumed a typical value of 0.1.

Instead of working with the usual $\alpha$ and $\beta$ parameters of the beta distribution, we parameterised it using the more intuitive parameters mean, $p=\alpha/(\alpha+\beta)$, and concentration, $\theta=\alpha+\beta$. 
First, for the joint sample of 1332 planetary orbits in 902 single and multiple systems (1029 planets around 687 single stars, 303 planets in 215 multiple systems, including the three binary systems with detected planets around both stars), we derived the  
posterior probability distributions for the parameters and obtained $p=$\,0.1719\,$\pm$\,0.0050 and $\theta=$\,3.42\,$\pm$\,0.18. 
The quoted errors are the standard deviation of these distributions, although they are not necessarily Gaussian. 
We did not fit the curve to the represented binned data, but to the original unbinned eccentricities using their measurement errors.
The resulting beta distribution is plotted in black in Fig.~\ref{fig:ecc_global}. 

To be able to assess the possible effect of the multiplicity of the host system on the eccentricity we fitted two additional models. 
In the second one we fitted the $p$ and $\theta$ parameters separately for planets in single and multiple systems. 
The result was that there is no statistically significant difference in the $\theta$ values between both samples, with a mean difference of $\Delta\theta = |\theta_{\rm single} - \theta_{\rm multiple}| =$\,0.22\,$\pm$\,0.21. 

We finally fitted a third model in which the $\theta$ value was the same for both samples, whereas the $p$ parameter was allowed to vary between them. 
The posterior parameters for this model were $\theta =$\,3.45\,$\pm$\,0.18, $p_{\rm single}=$\,0.1627\,$\pm$\,0.0058, and $p_{\rm multiple}=$\,0.203\,$\pm$\,0.012. 
Fig.~\ref{fig:ecc_global} also displays the distributions of eccentricities of planets in single and multiple systems and their corresponding beta fits (with $\theta_{\rm single} = \theta_{\rm multiple}$), in blue and red respectively.

\begin{figure}
 \centering \includegraphics[width=0.6\linewidth, angle=0]{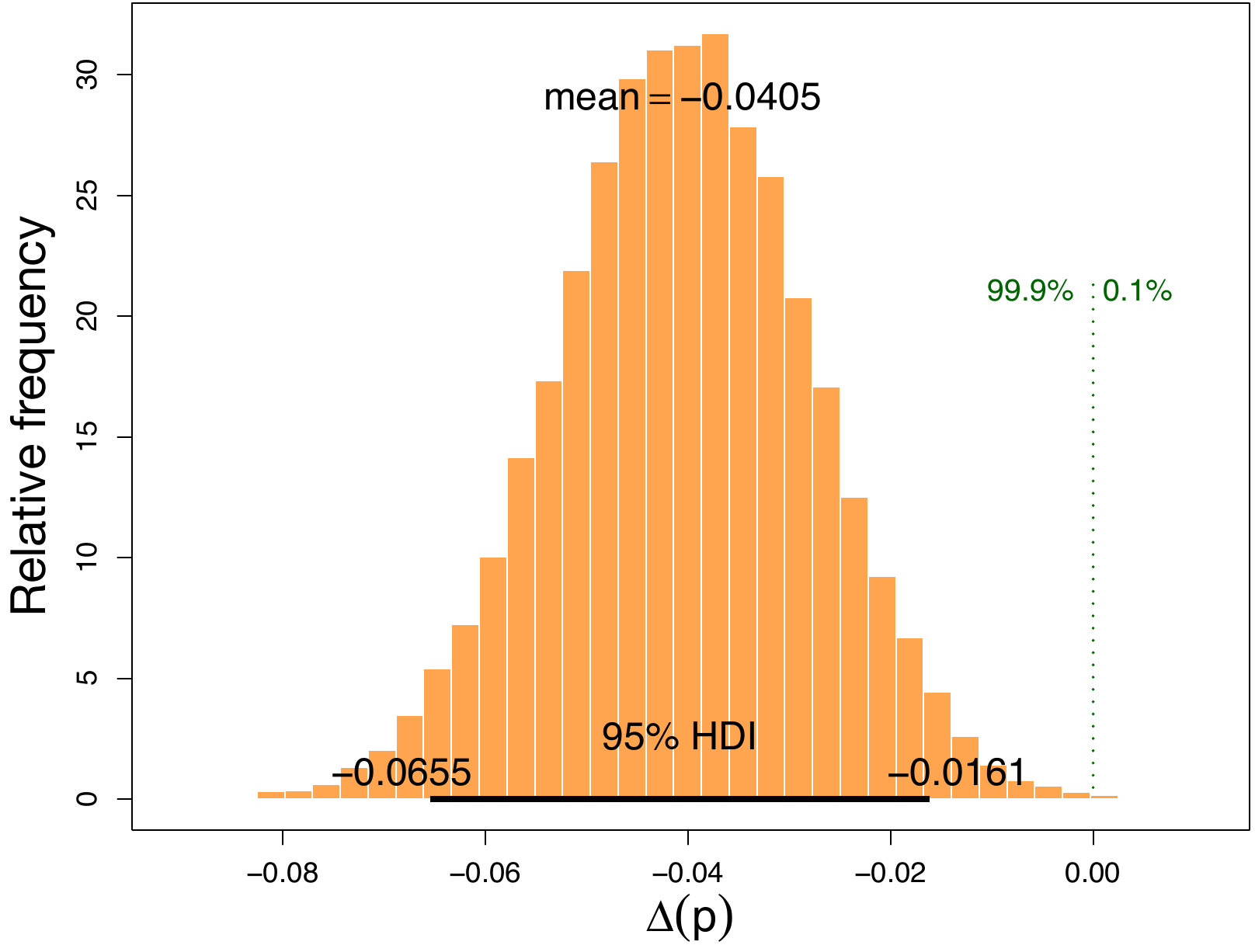}
 \caption[Posterior probability distribution for the difference in the mean $p$ parameter of the beta distributions fitted to the eccentricity distributions of planetary orbits in simple and multiple stars.]{Posterior probability distribution for the difference in the mean $p$ parameter of the beta distributions fitted to the eccentricity distributions of planetary orbits in simple and multiple stars. Here $\Delta p  = p_{\rm single} - p_{\rm multiple}$, and $\theta_{\rm single} = \theta_{\rm multiple}$. The thick horizontal black line represents the 95\% HDI (see text), while the black labels indicate the upper and lower values of the 95\% HDI and the mean $\Delta p$. 
 The vertical green dotted line indicates the location of the null difference, while the percentages show %the magnitudes of 
 the relative areas of the distribution at both sides of this line. 
 This figure was created using a modified version of the software provided by \citet{kruschke15}.}
 \label{fig:ecc_deltap}
\end{figure}

To carry out an objective comparison among the three models, we applied an information criterion, namely the Watanabe-Akaike Information Criterion \citep[WAIC;][]{watanabe13}. 
The particular values obtained for the first (same $p$ and $\theta$ parameters), second (different $p$ and $\theta$), and third (fixed $\theta$, different $p$) were $-$2891.8, $-$2908.2, and $-$2909.3, respectively.
We obtained identical results with the Leave-One-Out Cross-Validation criterion \citep[LOO-CV;][]{vehtari17}.
Therefore, the third model, with shared $\theta$ and different $p$ parameters, was indeed the best model representing the data according to this criterion. 
The analysis with the MCMC technique provides an estimate of the unknown parameters of the fitted model for each step of the Markov chains. In this case, we ran four separate chains with 12\,500 points each and therefore obtained 50\,000 estimates for the difference $\Delta p=p_{\rm single} - p_{\rm multiple}$. 
A histogram of these values is represented in Fig~\ref{fig:ecc_deltap}. 
According to \citet{gelman14}, who thoroughly discussed the Bayesian data analysis methods, this distribution is an unbiased true representation of the actual posterior distribution of the parameters. 
The thick horizontal line in the figure shows a 95\% highest density interval (HDI). The HDI is the Bayesian counterpart of the classical confidence interval, and it is calculated as the narrowest interval that contains a certain percentage of the probability density. In other words, it encompasses the most credible values of the distribution though not necessarily symmetrical nor centred on the arithmetic mean (see, for example, \citealt{kruschke15} for a definition and discussion of the HDI).

In our case we obtained a posterior distribution that implies $\Delta p = -$0.041\,$\pm$\,0.013. 
In Fig.~\ref{fig:ecc_deltap} we compare the extent of the 95\% HDI bar with the position of the zero value.  
Only 0.1\% of the posterior estimates are positive, and therefore our conclusion is that $\Delta p < 0$ with a significance level of 0.001 in conventional statistics.  
Thus, we concluded that planetary orbits in multiple stellar systems do exhibit significantly larger eccentricities than those around single stars.
This result is expanded in the next section.

\subsection{Semi-major axes, separations, and eccentricities}
\label{sec:semi-major_sep_eccen}

Given the result obtained in Sect.~\ref{sec:eccentricities}, one would expect that the physical separation between the stars in the closest multiple systems may have an effect on the eccentricities of planetary orbits, though this effect is negligible for the widest pairs. 
To check and quantify this hypothesis, we fitted a model in which we allowed the mean $p$ of the beta distribution of the eccentricities $e$ to vary with the physical separation $s$ between stars.
For this purpose we constructed a generalised lineal model. 
This kind of models is useful to explore regressions with dependent variables not following a Gaussian distribution, such as the orbit eccentricities in our case, although its use in astrophysical research is limited.
We refer the reader to \citet{desouza15a,desouza15b}, \citet{elliot15}, and \citet{hilbe17} for discussion of this statistical technique in the context of astrophysical problems, including some useful examples.

\begin{figure}[H]
 \centering \includegraphics[width=0.7\linewidth, angle=0]{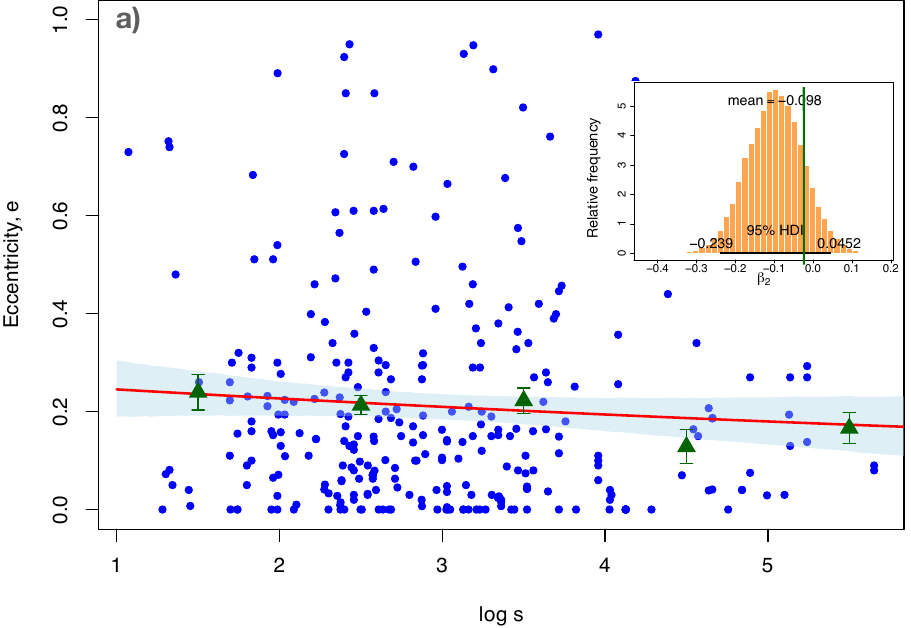}
 \includegraphics[width=0.7\linewidth, angle=0]{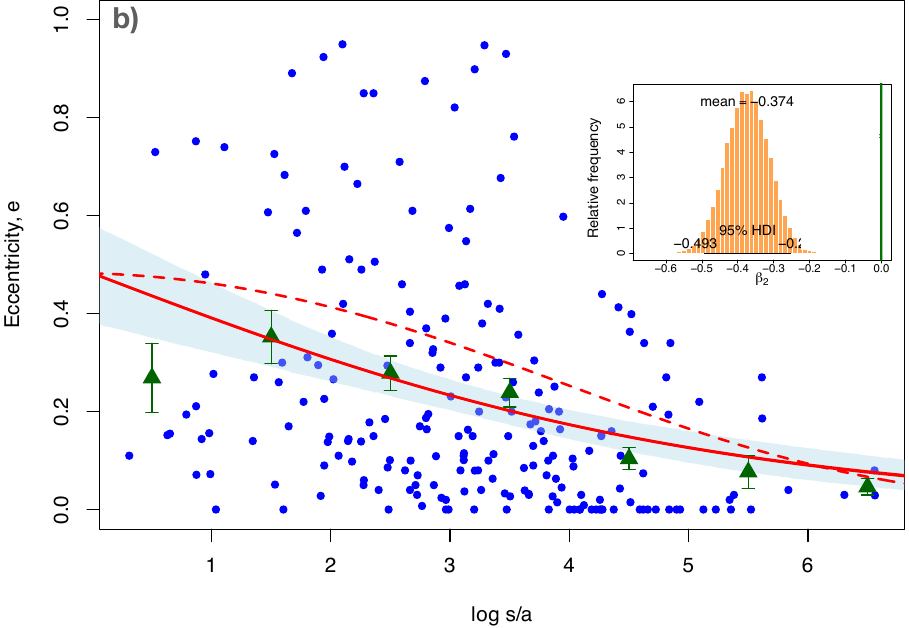}
 \caption[Planet eccentricity as a function of star-star separation in au, and of ratio between star-star separation and semi-major axis.]{Planet eccentricity as a function of star-star separation (panel a, \textit{top}) in au, and of ratio between star-star separation and semi-major axis (panel b, \textit{bottom}).
 The red lines show the predicted mean value $p$ of the beta distributions as a function of the dependent variable, in logarithmic scale. 
 The blue shades around these lines correspond to their 95\% HDI. 
 The insets show the corresponding posterior probability distribution of the $\beta_2$ parameters, and the vertical green lines mark $\beta_2=$ 0. 
 Green triangles mark the mean eccentricities in bins of 1\,dex, together with error bars for their formal errors. The dashed red line in panel b shows the result of a generalised model using a quadratic fit.}
 \label{fig:panelsep}
\end{figure}

In our case we fitted a generalised model in which the possible effect of physical separation in logarithmic units, $\log{s}$, was introduced by computing a linear predictor $\eta$:
\begin{equation}
\eta=\beta_\text{1} + \beta_\text{2}\log{s},
\label{eqn:eta}
\end{equation}
\noindent which was related to the mean $p$ of the beta distribution by a logit link function \citep{berkson44,mcfadden73}:
\begin{equation}
\eta \equiv {\rm logit}~p=\log\frac{p}{\text{1}-p}
\label{eqn:logit}
\end{equation}
We introduced the link function to transform from the unbounded scale of $\eta$ to the bounded range of $p$ between zero and one. 
The $\theta$ parameter of the beta distribution was not allowed to vary with the stellar separations, though. 
To perform the fit, we created another MCMC model in Stan to derive the posterior probability distributions for the three parameters $\theta$, $\beta_\text{1}$, and $\beta_\text{2}$. 
These distributions can be summarised as $\theta=$\,1.981\,$\pm$\,0.160,  $\beta_\text{1}=-$1.034\,$\pm$\,0.226, and $\beta_\text{2}=-$0.098\,$\pm$\,0.072. It is already apparent from the estimate for $\beta_2$ that the effect of the stellar separation on the eccentricities is not statistically significant. This is more clearly seen in the top panel of Fig.~\ref{fig:panelsep}, where we represent the prediction for the $p$ parameter, and its 95\% HDI in the $e-\log{s}$ plane. 
To guide the eye, and although they were not used in the fitting procedure, we included in Fig.~\ref{fig:panelsep} the mean eccentricities in bins of 1\,dex.

The above result is not surprising since one would expect that the effect of star separation $s$ on the orbit eccentricities should not depend just on the absolute value of $s$, but on its relative value compared to the star-planet separation. 
To confirm this hypothesis, we repeated the above analysis but replaced $s$ by the ratio $s/a$ between the separation between stars and the semi-major axis of the planetary orbit. 
As before, in the case of multiple planetary systems we used the $a$ value of the outermost planet, and in the case of triple and quadruple systems we used the $s$ value of the innermost star (or white dwarf or brown dwarf). 
The results are illustrated by the bottom panel of Fig.~\ref{fig:panelsep} and summarised by the following parameters: $\theta=$\,2.09\,$\pm$\,0.20,  $\beta_\text{1}=-$0.07\,$\pm$\,0.21, and $\beta_\text{2}=-$0.374\,$\pm$\,0.062. 
In the $e-\log{s/a}$ plane, the $\beta_\text{2}$ parameter is significantly different from zero; the equivalent significance level in conventional statistics is $<$\,2\,$\cdot$\,10$^{-\text{5}}$ (i.e. beyond a 4$\sigma$ effect). 
As a result, for a fixed semi-major axis $a$, planets in multiple systems with shorter star-star separations $s$ tend to exhibit larger eccentricities $e$.

The mean binned eccentricities plotted in the bottom panel of Fig.~\ref{fig:panelsep} suggests a flattening of the relation for the lowest values of $\log(s/a)$. 
To assess the reliability of this effect we added a quadratic term to the relation between $\eta$ and $\log(s/a)$, thus fitting the following relation:
\begin{equation}
\eta = \beta_\text{1} + \beta_\text{2} \log(s/a) + \beta_\text{3}  \log^\text{2}(s/a)    
\end{equation}
\noindent The result, which is also shown in the right panel of Fig.~\ref{fig:panelsep}, confirms the suspected flattening. However, according to the values of the WAIC information criterion, this expanded model does not improve the linear one, which we kept afterwards.

\subsection{Number of planets in single and multiple systems}
\label{sec:planet_frequency}

We also investigated the planet rate in single (with one star) and multiple systems (with two, three, or four stars).
The arithmetic mean numbers of planets around the 687 single and 215 multiple stars in our two samples are 1.51 and 1.41, respectively, which suggest that single stars tend to host slightly more planets. 
To test whether this measured difference is statistically significant, we applied the MCMC technique, programmed in Stan, to fit Poisson distributions to the detected number of planets in each sample. 
Since the two samples do not include cases with no planets orbiting the host star, we used a zero-truncated Poisson (ZTP) distribution \citep{hilbe17}, which corrects the classical Poisson model to exclude the possibility of observations with zero counts. In particular, 
if \textit{N} denotes the random variable representing the number of planets, and $\lambda$ is the parameter of the Poisson distribution, the ZTP probability distribution function takes the form:
\begin{equation}
P(N=k)=\frac{\lambda^k e^{-\lambda}}{k!\,(\text{1}-e^{-\lambda})},
\label{eqn:ztp}
\end{equation}
\noindent where the corrected mean number of planets $\mu$ is:
\begin{equation}
\mu = \frac{\lambda}{\text{1}-e^{-\lambda}}.
\label{eqn:mu_ztp}
\end{equation}
\noindent Here $e^{-\lambda}$ is the expected number of zeroes in a Poisson distribution with parameter $\lambda$.

\begin{figure}
 \centering \includegraphics[width=0.6\linewidth, angle=0]{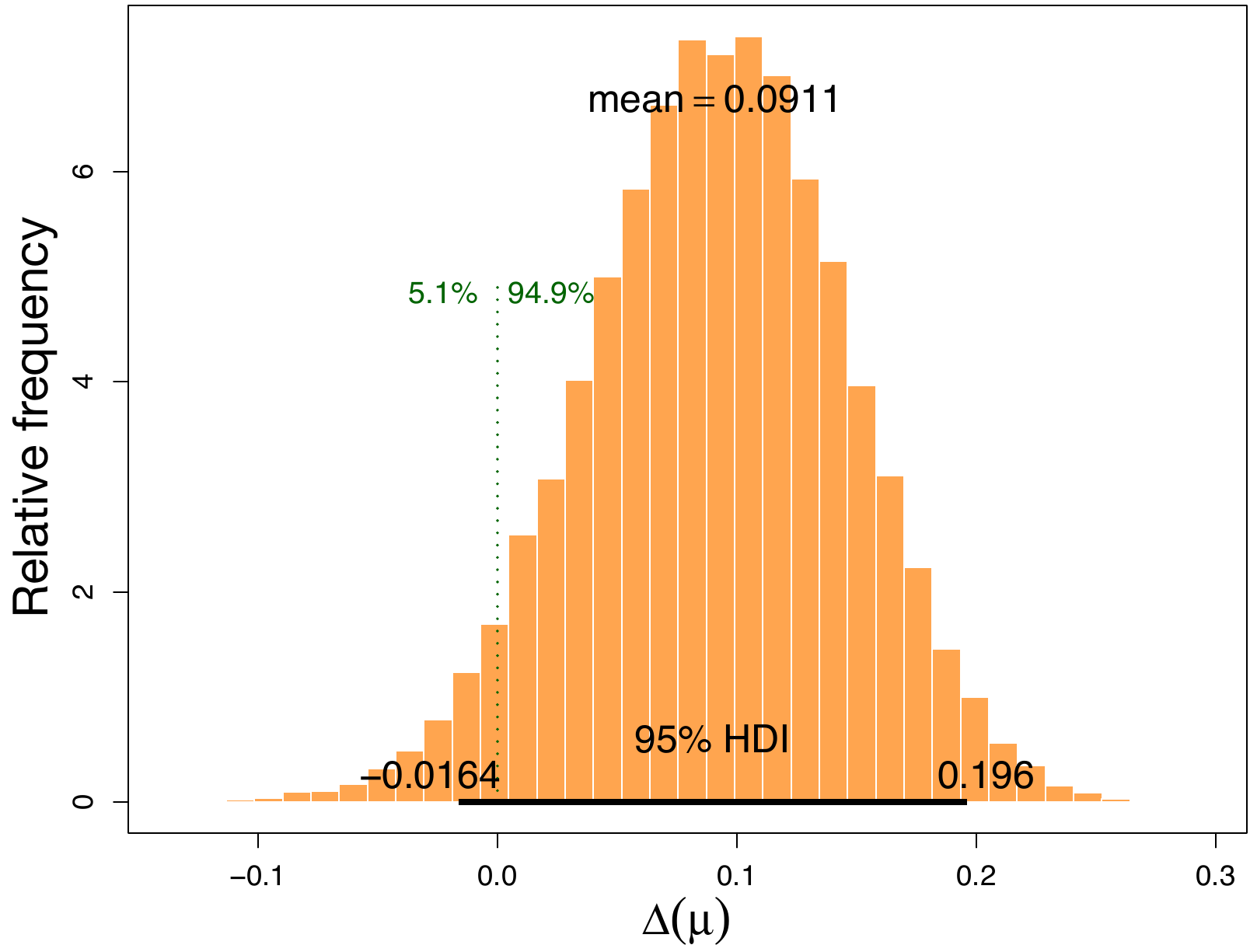}
 \caption[Same as Fig.~\ref{fig:ecc_deltap} but for the difference in the number of planets in single and multiple systems.]{Same as Fig.~\ref{fig:ecc_deltap} but for the difference in the number of planets in single and multiple systems, $\Delta \mu = \mu_{\rm single} - \mu_{\rm multiple}$.}
 \label{fig:difmeanp}
\end{figure}

The corresponding means of planets per host star are $\mu_{\rm single}$ = 1.508 $\pm$ 0.029 and $\mu_{\rm multiple}$ = 1.416 $\pm$ 0.046, virtually identical to the arithmetic means. 
The posterior probability distribution of the difference of the two means is represented in Fig.~\ref{fig:difmeanp}, which displays a mean offset of $\Delta\mu=$\,0.091\,$\pm$\,0.054. 
This difference is only marginally significant and hovers at the edge of the 95\% confidence interval. 
As a conclusion, although the data suggest a larger number of planets around single stars when compared to multiple systems, the statistical significance of this effect is weak and we could not derive a firm conclusion.

One caveat of the previous analysis is that the applied ZTP distribution is not completely appropriate since, as it is often the case in real observational data, the observed dispersion in $N$ is larger than the one expected from a Poisson distribution (i.e. $\sigma>\sqrt{\lambda}$). 
To solve this problem, we used an alternative probability distribution, namely the negative binomial, which is similar to the Poisson distribution but with one more parameter to account for a possible overdispersion. 
We again refer the reader to \citet{desouza15b} and \citet{hilbe17} for descriptions of the use of the negative binomial distribution in astrophysical contexts. 
To analyse the possible effects of this more reliable distribution, we constructed an MCMC model introducing a zero-truncated negative binomial distribution and applied it to our data. 
The result is that the mean offset in the number of planets between both samples becomes $\Delta\mu=$\,0.088\,$\pm$\,0.066. 
This difference from zero is even less significant than our estimate with the Poisson distribution and therefore we reinforce our previous conclusion of a non-significant effect of the stellar multiplicity on the number of planets per star.

Given the results of Sect.~\ref{sec:semi-major_sep_eccen}, in which we showed how the increase of eccentricities is more apparent in low-$s/a$ systems (i.e. systems in which the star-star separation is not much wider than the size of the planetary orbit), a possible effect of a lower number of planets for multiple systems could be hindered by the inclusion of high-$s/a$ systems, which could be more similar to single systems. 
To check this hypothesis, we constructed a generalised linear model in which the mean parameter of the Poisson distribution was allowed to vary with $\log(s/a)$. 
As in Sect.~\ref{sec:separations}, in the case of multi-planet systems, we used the $a$ value of the outermost planet and $s$ of the closest stellar companion, so $s/a$ should be read as ${\rm min}(s/a)$.
In particular, we used an exponential link function to go from the unbounded scale of a linear relation to the positive scale of the parameter of the Poisson distribution, that is:
\begin{equation}
   \lambda = e^{\beta_\text{1} + \beta_\text{2}\log(s/a)} 
\end{equation}
The result for the slope parameter of the linear relation is $\beta_\text{2}=-$0.015\,$\pm$\,0.080. Therefore, it is not significantly different from zero, and we concluded that there is no apparent effect of $s/a$ on the observed number of planets.
A computation of the corresponding WAICs also indicates that this model is worst than the constant $\lambda$ model.

\subsection{Planet masses in single and multiple systems}
\label{planet_masses}

\begin{figure}[h]
 \centering \includegraphics[width=0.6\linewidth, angle=0]{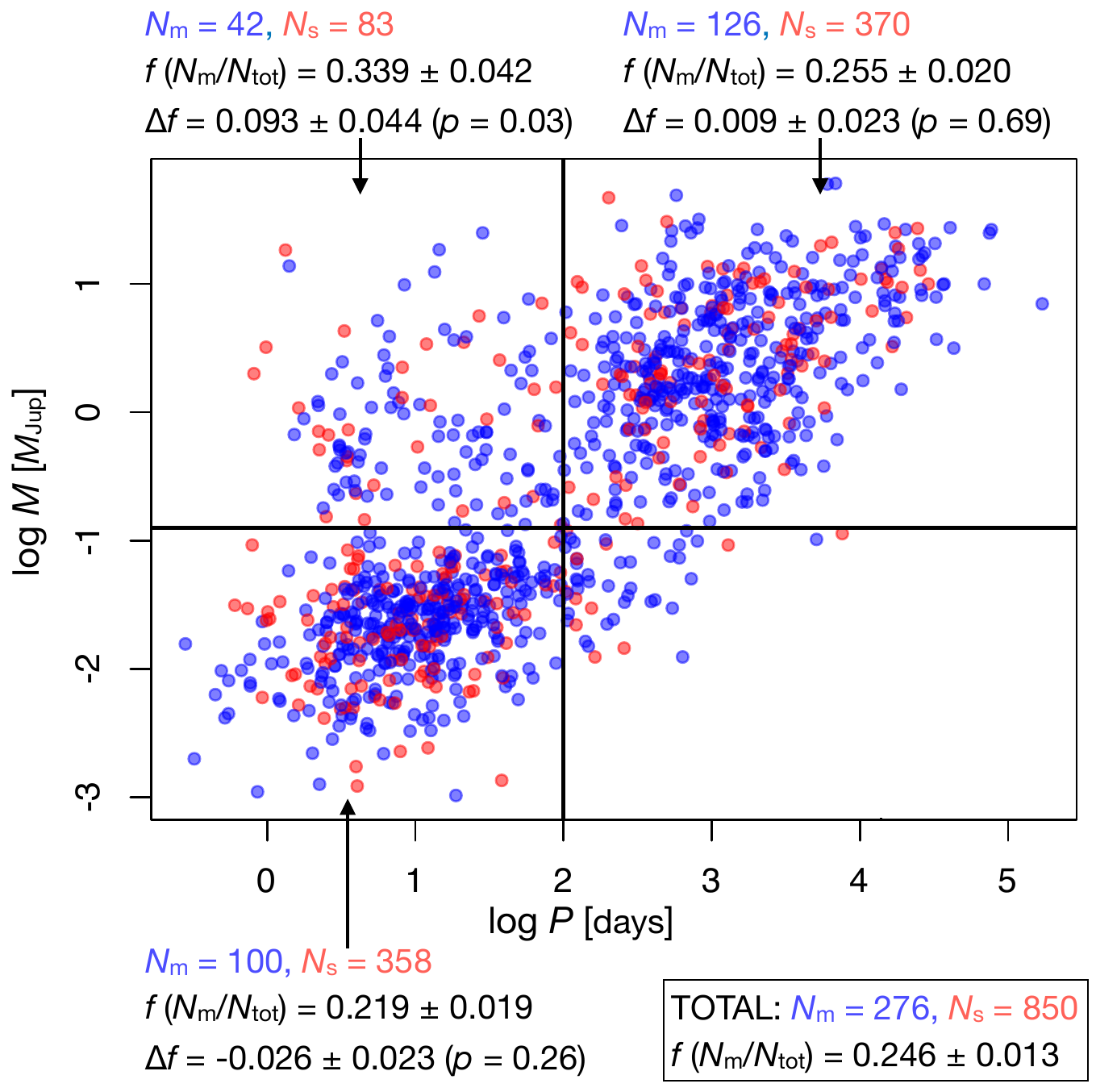}
 \caption[Period-mass diagram for planets in single and multiple systems.]{Period-mass diagram for planets in single (red circles) and multiple systems (blue circles). 
 The horizontal black line divides the sample into high- and low-mass regimes using a cutoff of 0.1258\,M$_{\rm Jup}$ (40\,M$_\oplus$), while the vertical black line at $P$\,=\,100\,d marks the position of the arithmetic mean of the distribution in orbital periods. 
 The information showed for each quadrant is the number of planets in multiple, $N_{\rm m}$, and single systems, $N_{\rm s}$, the fraction of planets in multiple systems with respect to the whole sample, $f=N_{\rm m}/(N_{\rm m}+N_{\rm s})$, the difference $\Delta f$ between this ratio and the value computed for the total sample ($f_{\rm tot}=$\,0.246), and, in parenthesis, the statistical significance $p$ of the differences. 
 No information is offered for the bottom-right quadrant due to the insufficient number of planets. 
 }
 \label{fig:period_masses}
\end{figure}

In this section, we explore whether stellar multiplicity has an impact on the locus of planets in the orbital period-planetary mass diagram.
For example, \citet{fontanive19} and \citet{fontanive21}, while not offering specific statistics, suggested that the separation of high-mass planets is influenced by stellar multiplicity. 
With the expanded and enhanced sample presented in our work, we can address these concerns with greater confidence.
In Fig.~\ref{fig:period_masses} we display the period--mass diagram where we discriminate between planets in single and multiple systems. 
As shown in the plot, our sample comprises 850 planets around single stars and 276 in multiple systems with all data, which account for a 25\% of the total sample. 
At a first glance, Fig.~\ref{fig:period_masses} suggests that high-mass planets with short periods may be relatively more frequent in multiple systems. 
To verify this hypothesis, we divided the plane into four quadrants, as depicted in Fig.~\ref{fig:period_masses}, and determined a number of parameters, together with their errors, by fitting binomial distributions using the MCMC technique. 
We found a marginally significant difference only in the upper-left quadrant, which suggested that high-mass planets ($M >$\,40\,{\rm M}$_\oplus$) in close orbits ($P<$\,100\,d) appear to be relatively more frequent in multiple systems than around single stars.
Other authors, such as \citet{eggenberger04} and \citet{fontanive21}, have also concluded that the most massive short-period planets (with masses greater than 2\,M$_{\rm Jup}$) also tend to orbit in multiple star systems.

\begin{figure}[H]
 \centering \includegraphics[width=0.55\linewidth, angle=0]{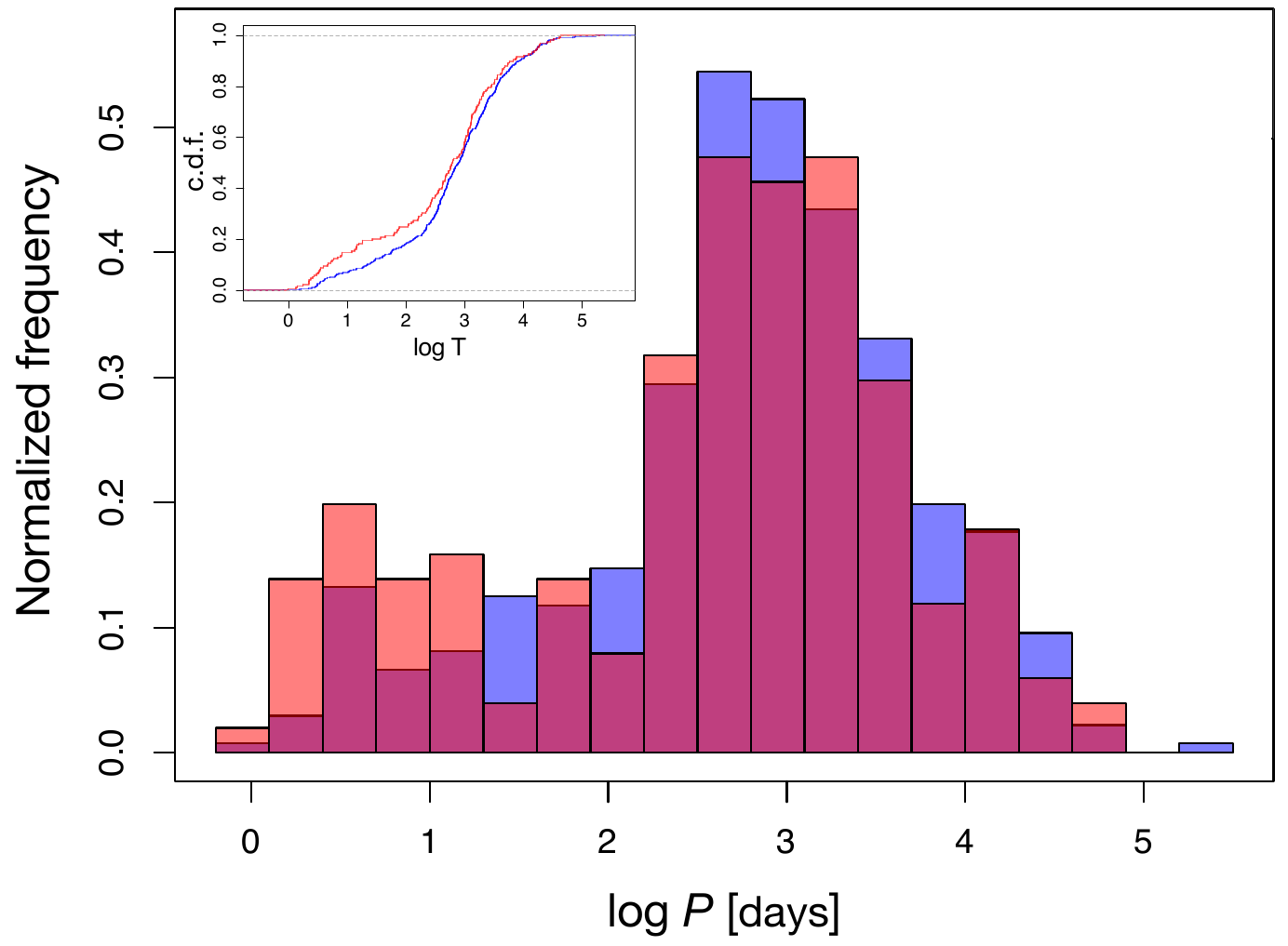}
 \caption[Histogram of the distributions of orbital periods for high-mass planets in single and multiple systems.]{Histogram of the distributions of orbital periods for high-mass ($M>$\,40\,M$_\oplus$) planets in single (blue bars) and multiple (orange bars) systems. The inset of the figure compares the cumulative distribution functions of both subsamples.}
 \label{fig:hist_period}
\end{figure}

To delve deeper into these potential findings, in Fig.~\ref{fig:hist_period} we present histograms illustrating the distribution of orbital periods for high-mass ($M$\,>\,40\,M$_\oplus$) planets in both single and multiple systems. 
The relative increase in the proportion of planets with low orbital periods in the latter systems becomes apparent, as suggested by the previous quadrant analysis.
To assess the reliability of this observation, we compared the cumulative distribution functions of both subsamples, as depicted in the inset of Fig.~\ref{fig:hist_period}. For this purpose, we conducted an Anderson-Darling test \citep{anderson52}, which is a modification of the Kolmogorov-Smirnov test, but more suitable for situations where greater emphasis must be placed on the tails of the distribution\footnote{\url{https://asaip.psu.edu/Articles/beware-the-kolmogorov-smirnov-test}}, as it is the case with our data.
The outcome of this test indicates that the statistical significance of the difference between both distributions is $p=$\,0.054. Consequently, we cannot draw a definitive conclusion, and larger samples are evidently required to evaluate the reliability of this trend.

\section{Discussion}
\label{sec:discussion_hosts}

\subsection{Missing multiple systems}
\label{sec:missing_multiple_systems}

We evaluated the existence of unidentified multiple systems in our sample.
First, our common proper motion and parallax search was limited by the \textit{Gaia} completeness at $G \sim$ 20.3\,mag.
According to table~7 of \citet{smart19}, our \textit{Gaia} search was complete down to spectral type M8$\pm$1\,V up to 100\,pc, and L8$\pm$1 up to 10\,pc.
As a result, we were expected not to be able to detect ultra-cool dwarfs M9\,V and later in the whole surveyed volume, nor T- and Y-type brown dwarfs just beyond 10\,pc.
Actually, we identified an L1 ultra-cool dwarf as a wide companion to HD~1466 in Tucana-Horologium, but it is overluminous because of its youth (Appendix~\hyperref[ch:Appendix_E]{E}).
While we might have conservatively stated that our \textit{Gaia} search for resolved companions was virtually complete in the stellar domain down to about 0.1\,M$_\odot$ \citep{cifuentes20}, it was far from being complete for the least massive stars and the whole substellar domain.
Furthermore, most of the brown dwarf companions analysed here came from WDS or literature works that employed deep adaptive optics imaging (e.g. GJ~229\,B; \citealt{nakajima95}).
Additional wide stellar companions with spectral types earlier than M8\,V might have also been uncatalogued by \textit{Gaia} in crowded regions at low galactic latitudes \citep{reyle21}.

Second, there may be more unidentified multiple systems, although hidden in plain sight.
Exoplanet-host stars have naturally been monitored and validated with high-resolution spectroscopy in search for close unresolved binaries.
Except for a few cases with long-term radial-velocity trends superimposed on short-term exoplanetary signals (e.g. 51~Peg itself), most host stars are known to have only exoplanetary companions at close separations.
However, their resolved companions, usually fainter and with later spectral types, have in general not been analysed so thoroughly, so it may happen that some of the identified double systems actually are triple systems (and triple systems are quadruple systems, and so on).
Furthermore, the proverb ``these things always come in threes'' seems to apply well to multiple stellar systems through a major prevalence of wide triples over wide binaries \citep{basri06,tokovinin06,caballero07a,cifuentes21}.

There have been only a few papers with exhaustive searches for companions of very different masses and at very different projected separations from the primary stars.
For example, \citet{caballero22} compiled CHARA interferometry, NICMOS/\textit{Hubble} Space Telescope and PUEO/Canada-France-Hawai'i Telescope high-resolution imaging, CARMENES/3.5\,m Calar Alto and MAROON-X/Gemini high-resolution spectroscopy, and \textit{Gaia} spectrophotometry, and ruled out the presence of stellar and high-mass brown-dwarf companions to the exoplanet-host star Gliese~486 from the limit of the \textit{Hubble} observations at 24--32\,au up to 100\,000\,au and of planets with minimum masses of $\sim$30\,M$_\oplus$ with periods up to 20\,000 days ($\sim$11\,au).
Other planetary systems have not been the subject of such a massive observational effort, but reasonable upper limits to the presence of additional companions exist for many of them.
As a result, some of the 687 single stars with planets may not be single, but there are high chances that the great majority of them, especially those at the closest distances, do not have stellar companions at close or wide separations.

\subsection{Eccentricity: Nature or nurture}
\label{sec:eccentricity}

One of the main results of our work is that the orbital eccentricity, $e$, of known planets in multiple stellar systems does not depend on the projected physical separation between component stars, $s$, but on the ratio between this separation and the planet semi-major axis, $a$.
Therefore, the smaller the $s/a$ ratio, the more probable the planet $e$ is significantly larger than zero.
However, in spite of this significant result after so many previous searches and analyses, we are still far from understanding the nature-nurture dilemma of eccentric planets in close multiple stellar systems or the role of the modulation of the orbital eccentricity, which strongly depends on the relative inclination between the plane of motion of the planet and that of the binary \citep{mazeh97}.
This relative inclination is known for very few systems, if any, as the planet must (in general) transit and the stellar binary orbit must have an astrometric solution.
The collection of multiple stellar systems with exoplanets in Table~\ref{tab:basic_data_identified_hosts}, especially those with short projected angular separations, is an excellent input for forthcoming determinations of astrometric orbit parameters with periods of a few years.
Such determinations will open the door to further dynamical analyses of systems with stellar and planetary companions orbiting primary stars at different angles, from which to extract conclusions on their formation, evolution, and stability.

We also made a visual inspection of planet $e$ as a function of total mass of the system, and did not find any hint of relation that could be statistically analysed.
It happened the same when we inspected the exoplanets orbital periods and the number of planets in system as a function of total mass.
However, trying to analyse the detectability of exoplanets and therefore the number and (minimum) mass of exoplanets per system as a function of parameters of only the planet-host stars, such as their mass, spectral type, and degree of activity, is beyond the scope of this work \citep[e.g.][]{sabotta21}.

\subsection{Multiplicity function and future exoplanet surveys}
\label{sec:multiplicity_function_future_exop_surveys}

We went on by comparing our results with those from the literature.
The stellar MF of exoplanet systems (i.e. the fraction of exoplanet systems with stellar companions) has been determined since the very beginning of exoplanetology at about 20\% \citep{bakos06,bonavita07,mugrauer07}. 
More precise determinations have varied around the narrow interval of 23\% (30 multiple systems in 131 planetary systems; \citealt{raghavan06}) and 23.2 $\pm$ 1.6\% (218 multiple systems in 938 planetary systems; \citealt{fontanive21}), compatible with the value of at least 12\% proposed by \citet{roell12}.
In our case, after discarding pairs in open clusters and associations and with ultra-cool dwarf companions, we determined an MF of exoplanet systems of 21.6 $\pm$ 2.9\% (212 multiple systems in 899 planetary systems).
We stress on three issues in the comparison with the recent value of \citet{fontanive21}: 
($i$) the actual MF of exoplanet systems must be slightly larger because all searches, including ours, are incomplete at the latest spectral types (Sect.~\ref{sec:missing_multiple_systems});
($ii$) the identical value but the larger uncertainty of our measurement in spite of the similar numbers.
In particular, we used the Wald interval \citep{agresti98} assuming Poisson statistics for computing our uncertainties with a Wald 95\% confidence interval (Wald interval\,$=(\lambda - \text{1.96}\sqrt{\lambda/n}, \lambda + \text{1.96}\sqrt{\lambda/n})$, where $\lambda$ is the number of successes in $n$ trials);
and ($iii$) they explored a volume eight times larger than us, since they studied systems up to 200\,pc, and we did it up to 100\,pc.
Our value of 21.6 $\pm$ 2.9\% must necessarily be compared with MFs of nearby stars regardless planet presence.
Since the MF depends on spectral type, the comparison should be done with that of field stars of spectral types between late F and mid M, as most radial-velocity and transit surveys during the last three decades have focused on them.
All reference MF values enumerated in the beginning of this chapter are larger than our MF.
Although there have been claims establishing that the influence of a companion star results in the system's stars being unable to host planets, whether due to formation or stellar evolution,\citep[e.g.][]{wang14a,kraus16}, we instead ascribe our lower multiplicity fraction to the aforementioned exoplanet survey bias (\citealt{winn15} and the CARMENES examples; Sect.~\ref{sec:results_planetary_systems}).

There is no exoplanet survey free of multiplicity bias, unless especially designed to overcome it.
For example, one could envisage a search for smaller-amplitude signals in radial-velocity and light curves of known spectroscopic and eclipsing binaries, which are generally discarded in the first stages of exoplanet surveys \citep{garciamelendo11,gillon17,jeffers18,tal-or18,parviainen19,kristiansen22}\footnote{Note added during external revision: Other works on the topic that were not compared with by us are \citet{hirsch21} and \citet{michel24}.}.
Although several major transiting surveys are or will be greatly affected by multiplicity because of their poor spatial resolution, such as TESS \citep{ricker15} and PLATO \citep{rauer25},
wide multiplicity barely affects radial-velocity exoplanet searches.
Actually, some of the stars in wide multiple systems monitored by CARMENES, with companions at more than 5\,arcsec, eventually turned out to host new exoplanets \citep[e.g.][]{trifonov18,reiners18a,kaminski18,gonzalezalvarez20,kossakowski21}.
The CARMENES limit at 5\,arcsec was defined from the size of the spectrograph's optical fibre projected on the sky, 1.5\,arcsec, the typical seeing at the Calar Alto observatory, and the maximum allowed spectral contamination by a close companion \citep{quirrenbach14,cortescontreras17b}.
Similar minimum separations have been defined for other radial velocity searches with HARPS \citep{bonfils13} or SPIRou \citep{moutou17b,fouque18}, just too put two comparable examples.
A separation of 5\,arcsec is, however, too short for TESS and PLATO, which have pixel sizes of 15.6--21.0\,arcsec, and therefore they need compulsory \textit{Gaia}, lucky imaging, or adaptive optics data in order to validate transiting targets or even pre-launch scientific prioritising targets \citep{collins18,clark22,nascimbeni22}.
As a result, the crux of the matter is perhaps the definition of a boundary between ``wide'' and ``close'' multiples, which depends on the exoplanet survey.
For example, one of the key exoplanet surveys of the next decades will be the Habitable Worlds Observatory\footnote{\url{https://habitableworldsobservatory.org}} \citep[HWO;][]{clery23},
for which preliminary target lists are currently starting to be defined \citep{tuchow24,harada24}.
Although the HWO scientific requirements and therefore the mission specifications are far from being fixed, a wide multiple stellar system (i.e. one in which a stellar companion does not affect observations of host stars) must necessarily be less than 1\,arcsec for HWO\footnote{Note added during external revision: The actual value of minimum separation between stellar pairs to be observed by HWO is to be defined and will probably be larger than 1\,arcsec. It will depend on the
magnitude difference of the stars, the final optical design of the facility, the science case, and many more factors.}.
The same shall happen to the Large Interferometer For Exoplanets\footnote{\url{https://life-space-mission.com}} \citep[LIFE;][]{quanz22}.
The bulk of the resolved stellar systems have angular separations greater than that value (there is a recent \"Opik diagram in fig.~1 of \citealt{gonzalezpayo23}).
However, the astronomical community must make a joint effort to complement the forthcoming \textit{Gaia} DR4 data in order to analyse in great detail the innermost arcsecond of the nearest stars regardless their age, using high-resolution spectroscopy, searching for new spectroscopic multiples, or obtaining deep imaging (e.g. \citealt{oppenheimer01,mccarthy04,dieterich12,gauza21}).

\subsection{Spectral types and hierarchy}

\begin{figure}[h]
 \centering \includegraphics[width=0.6\linewidth, angle=0]{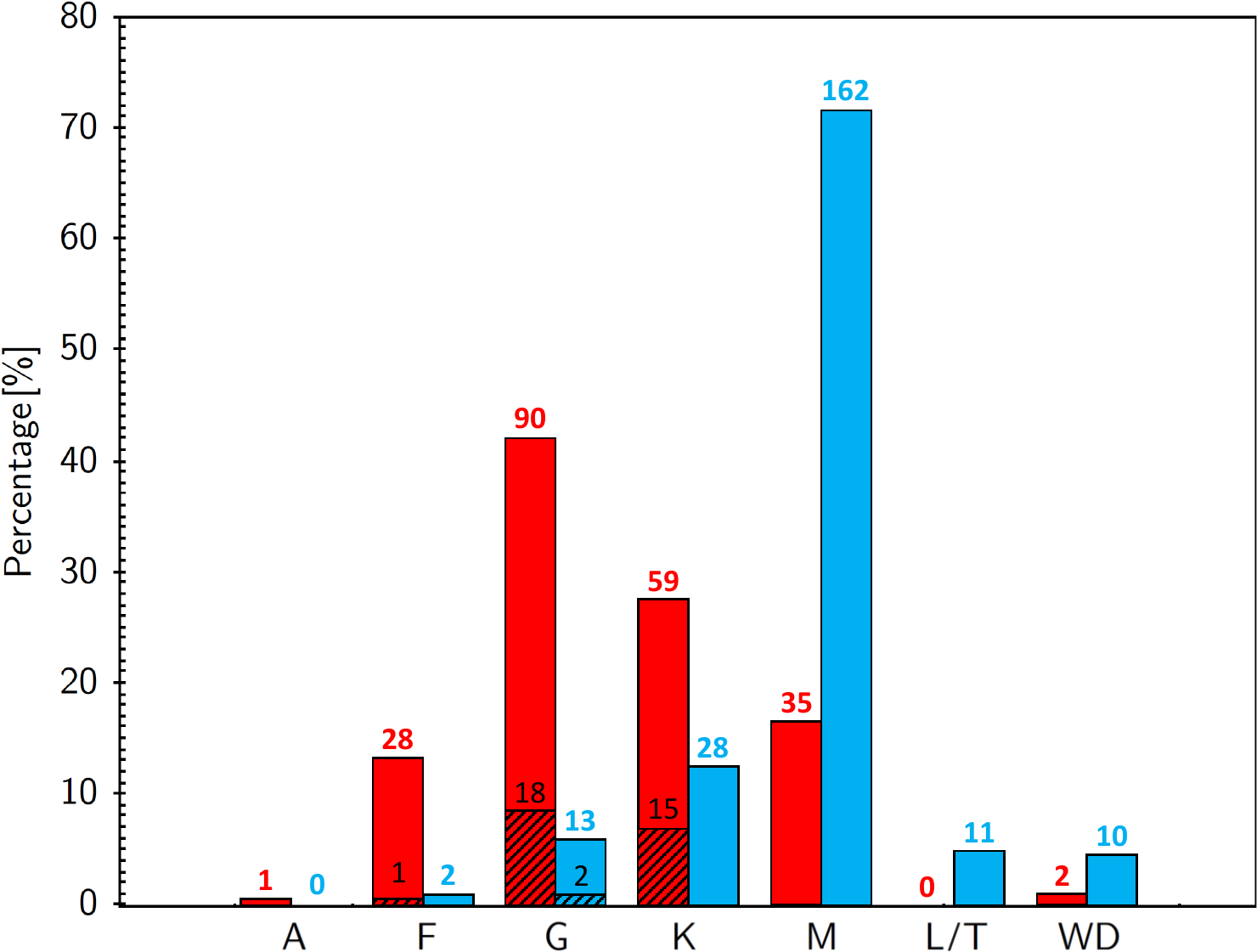}
 \caption[Relative distribution of host stars and companion stars per spectral type.]{Relative distribution of host stars (red) and companion stars (blue) per spectral type.
 Each bar is labelled with the absolute number of stars per bin.
 The black stripped area represents the number of subgiant and giant stars in each spectral type.
 Compare with fig.~3 of \citet{fontanive21}.}
 \label{fig:spectral_types}
\end{figure}

Following the comparison with previous works, we also studied the relative distribution of spectral types of all the stars considered in our sample, distinguishing between host and companion stars, as illustrated by Fig.~\ref{fig:spectral_types}. 
Our sample contains stars (and brown dwarfs) from types A, F, G, M, L, T, and white dwarfs (WD). 
As already noticed by \citet{fontanive21}, most host stars are obviously G- and K-type stars, while most companions are M-type stars (see also \citealt{mugrauer07}).
However, we found a larger proportion of late spectral types  as companions, with higher proportion of M stars (71.7\%), and of L and T ultra-cool dwarfs (4.9\%). 
Many of them were missing in \citet{fontanive21}'s work.
Here the effect of the different maximum survey distances and Malmquist bias played a role. 

Finally, we devote the last part of our discussion to systems hierarchy.
As already mentioned, we found 40 triple and three quadruple stellar systems with exoplanets. 
Accounting for the 172 binaries with exoplanets, this means that 18.9\% and 1.4\% of the multiple systems are triple and quadruple, respectively.
The values reasonably match those reported by \citet{duchene13} and \citet{reyle21} for field stars in general, regardless they have planets or not.
For example, \citet{reyle21} found that $\sim$20\,\% and $\sim$5\,\% of the multiple systems in the 10\,pc neighbourhood are triple and quadruple+quintuple, respectively. % 69+19+3+2=93
In the case of quadruple systems, \citet{cuntz22} concluded that they are in general made of two groups of binary stars.
This is indeed the case for two of our three quadruples, namely 30\,Ari and HD~18599, but not for the third one centred on HD~1466 (Table~\ref{tab:new_detected_systems}).
This is, however, a young system in Tucana-Horologium \citep{riedel17,gagne18a} that will likely be disrupted in a few tens of millions of years \citep{caballero09}.
In the case of the triple systems, the 40 of them are made of a close pair plus a separated companion (i.e. AB--C or A--BC), in line with what was already presented by \citet{mugrauer07b}, but with a much poorer statistics of just five triple systems.
As in the case of \citet{cuntz22}, we did not find any system of higher order (i.e. quintuple or higher).

\subsection[Remarkable multiple systems with planets]{Remarkable multiple systems with planets\footnote{This section was originally published in the article \citet{gonzalezpayo24} as a separate appendix.}}
\label{sec:remarkable_multiple_systems_planets}

This section presents some of the systems we consider the as most remarkable, that is, the most interesting within the multiple planetary systems. Among them, we find very wide systems, systems where we have found an additional stellar component, systems with multiple M-type dwarfs. In short, systems that offer distinguishing characteristics compared to more well-known multiple planetary systems. Among them, the following systems stand out:

\begin{itemize}
    \item \textit{HD~1466, two M dwarfs, and an early L ultra-cool dwarf:} Before our analysis, this system was made of the young late F-type star HD~1466, which hosts an astrometrically inferred Jovian exoplanet \citep{mesa22}, in the Tucana-Horologium association and a wide M5.5\,V companion at about 30\,arcmin \citep{gaiacollaboration21b}. We added to the system, which became quadruple, an early M dwarf and and L1 ultra-cool dwarf already reported as members of the Tucana-Horologium association \citep{kraus14,gagne15} at about twice the projected angular separation of the wide M5.5\,V companion. The ultra-cool dwarf is the latest, least massive system component reported in this work.
    
    \item \textit{HD~88072~AB and LP~609--39:}  The bright, close ($\rho$\,$\sim$\,3.7\,arcsec) pair HD~88072~AB was reported by \citet{gaiacollaboration21b}, while the planet ($\rho_{\rm planet}$\,$\sim$\,0.39\,arcsec calculated from the semi-major axis $a$ and distance $d$) was reported by \citet{feng22}. To this compact, relatively new system we added another new component, namely LP~609--39, which is a poorly investigated $\sim$M5.5\,V companion at an extremely wide separation of about 80\,arcmin.

    \item \textit{HD~134606 and L~72--1~[AB] (WDS~15154--7032):} It was made of the three-exoplanet host star HD~134606 \citep{mayor11} and an $\sim$M3.5\,V companion at about 1\,arcmin, namely L~72--1. Interestingly, the wide pair was discovered by an amateur astronomer \citep{rica12}. As discussed in Sect.~\ref{close_binaries_without_Gaia}, L~72--1~[AB] displays a bimodal distribution of the \textit{Gaia} DR3 $G$-band light curve, which is indicative of a close binarity of the order of 0.2\,arcsec. Furthermore, in his personal notes, F.\,M.~Rica also annotated a possible close binarity of L~72--1 based on the difference between its $V$-band absolute and apparent magnitudes. Furthermore, the \textit{Gaia} DR3 $d$ value of L~72--1~[AB] by \citet{finch18} in Table~\ref{tab:basic_data_identified_hosts} is affected by close multiplicity.

    \item \textit{CD--24~12030 and two M-dwarf companions:} The K2.5\,V-type primary star has received little attention since the tabulation by \citet{thome1892} in his Cordoba Durchmusterung (e.g. \citealt{degeus90,cruzalebes19,trifonov20}). It hosts a transiting planet discovered by K2 \citep{zink21,christiansen22}. The two co-moving companions at about 3.7\,arcmin, which are separated by 10.2\,arcsec between them (and 46.6\,deg of position angle), have not been previously discussed in the literature. The abnormal radial velocity of the faintest component is probably due to the orbital effect of the close binary.

    \item \textit{HD~222259\,AB and 2MASS J23321028--6926537 (WDS~23397--6912):} The system was originally made of the bright, close binary system HD~222259\,A and B, which are late G and early K dwarfs, respectively. The pair, separated by about 5\,arcmin, was resolved for the first time by H.\,C.~Russell at the end of the 19th century, and is best known in the literaure as DS~Tuc~AB. The primary, DS~Tuc~A, hosts an inflated transiting planet \citep{newton19,benatti19}. We added a $\sim$M5\,V wide companion at about 43\,arcmin, namely 2MASS J23321028--6926537, which is also a member of Tucana-Horologium \citep{gagne15,ujjwal20}. Actually, \citet{newton19} noticed the co-moving, co-distant candidate companion, but given its very wide separation they classified it as ``likely [...] unbound member of Tuc-Hor and not a bound companion of DS~Tuc''. Its binding energy is $|U_g| \sim$ 4.6 $\cdot$ 10$^{\text{33}}$\,J, comparable to other wide binaries reported in the literature.

    \item \textit{LP~141--14 and two M-dwarf companions:} LP~141--14 is a white dwarf \citep[spectral type DC\,D,][]{dupuis94} located in a triple system, along with stars G 229--20A and G 229--20B, which are two M dwarfs with very similar individual masses and a combined mass of about 0.65\,M$_{\odot}$ \citep{stephan21}. The separation of LP~141--14 with the two close M dwarfs is 43.7\,arcsec (1082\,au), while the separation between the planet \citep[LP~141--14\,b,][]{vanderburg20} and the white dwarf it is orbiting around is 0.0204\,au. Together with the circumbinary planet RR\,Cae\,b, LP~141--14\,b is the only planet orbiting a white dwarf in our sample.
\end{itemize}

\section{Summary}
\label{sec:summary_hosts}

In this study, we have compiled all stars with reported planets from the Extrasolar Planets Encyclopaedia and the NASA Exoplanet Archive at less than 100\,pc.
Employing primarily \textit{Gaia} DR3 data supplemented by searches in the Washington Double Star catalogue and the available literature,
we identified 215 exoplanet host stars in 212 multiple-star systems with common parallax and proper motion.
During the analysis, we rejected stellar systems that are part of open clusters (Hyades) or OB associations (Lower Centaurus Crux), and planetary systems that have controversial planet candidates (e.g. with minimum masses above the hydrogen burning limit). 
Of the 212 multiple-star systems, 173 are binary (including two systems with circumbinary planets), 39 are triple, and three are quadruple, as expected from the rate of triple or higher-order multiples of field systems.
There are three binary systems with detected planets around both stars.

We identified 17 new companions in 15 systems (ten binaries, four triples, one quadruple) with known planet host stars.
All the new physically bound companions have spectral types, either determined by other teams or estimated by us, between K4\,V and L1, except for a white dwarf candidate companion.
Some of the components in the widest systems have been reported to belong to very young stellar kinematic groups, such as Tucana-Horologium, which may indicate that they are currently in the process of disruption by the galactic gravitational potential.

We were able to measure projected physical separations between stars with \textit{Gaia} data in most cases, and to compile key parameters for 276 planets in multiple systems, with which we statistically analysed a series of hypotheses.
First, we computed the ratio between separations and semi-major axes for all multiple systems and identified the nine of them with the smallest ratio, $s/a<$\,20.
Only one controversial planet candidate, namely $\nu$~Oct~b, poses a challenge to current planet formation and evolution models. 
Next, we constructed a single planet-host star sample for comparison purposes and found a significant difference between the eccentricities of planets in multiple systems and those of planets around single stars.
In particular, planets in multiples systems with small $s/a$ ratios have eccentricities that are significantly larger than those of planets with larger $s/a$ ratios.
This result is in line with a number of theoretical predictions; however, our sample is more complete, and our statistical tools are more advanced than most previous observational studies.
Actually, we have a higher proportion of faint M, L, and T companions than previous searches, which provides more statistical robustness to our findings.

Concerning the comparison in the number of planets around multiple and single systems, we  found a possible trend in the sense that the former tend to host a smaller number of planets, although the significance level of this result (around 0.05) precludes us from reaching a firm conclusion. 
Also, we detected a greater frequency of high-mass planets in close orbits in multiple systems compared to single ones but again, this result is on the verge of being statistically significant.
Finally, we estimated a multiplicity fraction of stars with planets of 21.6\,$\pm$\,2.9\%, a value close to what has been reported by similar studies but also smaller than the multiplicity fraction of field stars, regardless of whether they have planets. 
We ascribed this difference not to a different formation or evolution mechanism, but to a well-known observational bias in exoplanet searches, especially in radial-velocity surveys.

One may speculate on extending the survey limit up to 200\,pc or beyond as a future work.
Nonetheless, we must acknowledge the considerable time invested in the individualised analysis of thousands of planetary system candidates, some of which we had to discard.
We instead propose prioritising the continued investigation of stellar multiplicity with planets with a previous work aimed at confirming and characterising the myriad of controversial planetary systems that have polluted past and current analyses. Eventually, we can leverage \textit{Gaia} DR4 for this purpose.

\newpage

%%%%% CAPITULO 6 %%%%%

\chapter{Multiple systems within 10\,pc} 
\label{ch:characterisation_10pc}
\vspace{2cm}
\pagestyle{fancy}
\fancyhf{}
\lhead[\small{\textbf{\thepage}}]{\textbf{Section \nouppercase{\rightmark}}}
\rhead[\small{\textbf{Chapter~\nouppercase{\leftmark}}}]{\small{\textbf{\thepage}}}

\begin{flushright}
\small{\textit{``Look up at the sky and count the stars, if indeed you can count them''}}

\small{\textit{-- Genesis 15:5}}
\end{flushright}

\bigskip

\begin{adjustwidth}{70pt}{70pt}
\tR{\small{The content of this chapter constitutes the first version of the article \textit{Characterisation of all known multiple stellar systems within 10\,pc} submitted to Astronomy \& Astrophysics on 27 February 2025, and currently under referee review.}}
\end{adjustwidth}
\bigskip

\lettrine[lines=3, lraise=0, nindent=0.1em, slope=0em]{T}{he study of stellar multiplicity} has interested astronomers for millennia. From Ptolemy in his Almagest in the 2nd century \citep{peters1915,grasshoff90,argyle19}, through Hodierna in the first catalogue of binary stars in 1654 \citep{gonzalezpayo24b}, to modern astronomers who have compiled the most recent and comprehensive lists of stars and brown dwarfs in the solar neighbourhood at $d <$ 10\,pc \citep{reyle21}, 20\,pc \citep{kirkpatrick24}, and 100\,pc \citep{gaiacollaboration21b}.
Understanding stars and brown dwarfs in the solar neighbourhood %and further 
provides insights into galactic structure, stellar formation and evolution, planetary system dynamics, and, as developed below, even the search for habitable worlds. % life beyond Earth. 

The study of resolved stellar multiplicity has evolved with observing techniques (naked eye, position micrometers, seeing-limited imaging from the ground with photographic plates and digital detectors, eyepieces, speckle and multi-aperture interferometry, lucky imaging, adaptive optics, high-resolution imaging from space), as well as with the explored angular separations, at both the closest and widest separation ends \citep{batten73,caballero09,duchene13,tokovinin14b,tokovinin18,dhital15}.
The interest of astronomers in stellar multiplicity has also evolved, and it seems to have reached a plateau just when \textit{Gaia} is releasing the greatest and most accurate astro-photometric datasets in history \citep{gaiacollaboration23b}. 
There is, however, a renewed interest in stellar multiplicity studies, especially in the solar neighbourhood. 
This fact comes from current searches for Earth-mass exoplanets around nearby M dwarfs and future searches for biosignatures on such exoplanets, but around solar-like stars.
On the one hand, many close binaries are discarded from extreme-precision radial-velocity searches \citep{roell12,cortescontreras17b,fouque18,gonzalezpayo24,cifuentes25}, while wide binaries help in determining M-dwarf planet-host parameters that are challenging to determine in single, cool stars.
Examples of those parameters are stellar masses (e.g. dynamical masses in a spectro-astrometric binary -- \citealt{gonzalezalvarez20} for GJ~338) and element abundances that are critical for the formation and composition of rocky planets (e.g. Mg and Si in multiple systems with FGK-type primaries and M-type secondaries -- \citealt{tabernero24}).

On the other hand, both NASA and ESA are starting to design their flagship missions for the characterisation of habitable planets, which are set to launch in the 2040--2050 time frame.
NASA's Habitable Worlds Observatory\footnote{\url{https://science.nasa.gov/astrophysics/programs/habitable-worlds-observatory/}} (HWO -- \citealt{clery23,dressing24}) will likely be in orbit first, and will explore exoplanet atmospheres in the ultraviolet, visible, and near infrared with a $\sim$6-m segmented primary mirror.
On the other side of the Atlantic, one of the final recommendations from the ESA's Voyage 2050 Senior Committee was to develop a mission specifically focusing on the ``Characterisation of Temperate Exoplanets'' science theme\footnote{\url{https://www.cosmos.esa.int/web/voyage-2050}}.
The Large Interferometer For Exoplanets\footnote{\url{https://life-space-mission.com/}} (LIFE -- \citealt{quanz22}), which adequately fits this theme, might be in orbit just afterwards, and complement HWO with interferometric observations in the mid-infrared \citep{alei24}.
In order to optimise exoplanet yield simulations and inform the spacecraft optomechanical designs with accurate scientific requirements, both HWO and LIFE need preliminary target lists \citep{mamajek23,tuchow24,harada24,menti24}.
Both mission teams emphasise the study of stellar multiplicity, as angular separation and magnitude difference between components are key parameters for current planet yield calculations and the next generation of atmospheric retrieval simulations \citep{konrad22,carriongonzalez23,morgan24}.
Similarly to radial-velocity searches, close binaries will be discarded from future HWO and LIFE programs, while wide multiples will help in determining precise parameters of planet host stars (the close-wide boundary an arbitrary angular separation that depends on the used observing facility; \citealt{cifuentes25}). 
In any case, studying Earth-like exoplanets is not the only reason for the renewed interest of astronomers in stellar multiplicity. 
A few recent examples of astrophysical themes that require stellar multiplicity as a key input are the determination of the local (initial) mass function \citep{kirkpatrick24}, the formation and evolution of ultra-cool dwarfs \citep{baig24}, and the characterisation of white dwarfs \citep{obrien24}. 

This is the last item of a trio of papers on a thorough investigation of the widest multiple systems with angular separations $\rho \ge$ 1000\,arcsec \citep{gonzalezpayo23},
the multiplicity of stars with planets at $d <$ 100\,pc \citep{gonzalezpayo24}, 
and of all multiple systems within 10\,pc (this work).
The three of them share an identical methodology and the extensive use of data from \textit{Gaia} DR3 \citep{gaiacollaboration23b} and the Washington Double Star catalogue (WDS -- \citealt{mason01}), along with a comprehensive analysis and data compilation from the literature.
The current work should pave the way for new multiplicity surveys \citep{littlefield24,leblanc24,henry24}, 
updates of stellar and substellar catalogues up to 10--20\,pc \citep{reyle21,kirkpatrick24}, and ongoing work on stellar multiplicity for HWO and LIFE target lists (C.\,K.~Harada, E.\,E.~Mamajek, F.~Menti, Z.~Hartman, priv. comm.).

\section{Sample}
\label{sec:sample_10pc}

The list of 541 stars, brown dwarfs, and exoplanets in 336 systems within 10\,pc from the Sun compiled by \citet{reyle22} represents the most complete volume-limited sample based on current knowledge.
This list is the first update to the ``The 10~parsec sample in the \textit{Gaia} era'' \citep{reyle21}.
The completeness of their compilation has been confirmed later by several authors \citep[e.g.][]{golovin23,kirkpatrick24}, as well as by cross-match with contemporaneous catalogues \citep{hirsch21}.
The latest version of the list of \citet{reyle22} can be retrieved from VizieR and its dedicated website\footnote{\url{https://gruze.org/10pc/}}.
We only had to delete two entries, corresponding to the L2 and L4 components of the binary 2MASS J06174191+1945135~AB (CWISE J061741.79+194512.8~AB), which have a new spectrophotometric distance estimate of 28.2 $\pm$ 5.7\,pc, far beyond the 10\,pc limit \citep{humphreys23}.
There are also a few targets with distances slightly farther than $d =$ 10\,pc, but consistent with that value within uncertainties, which we kept in our analysis.

Table~\ref{tab:sample_10pc} displays the default name in the Simbad database, Gliese-Jahreiss identifier when available, J2000 equatorial coordinates, and proper motions of 425 stars and brown dwarfs at less than 10\,pc. 
The \textit{Gaia} early third data release (EDR3; \citealt{gaiacollaboration21a}) was the main provider of astrometric data (identical to that from DR3). 
When EDR3 data are missing, \citet{reyle21,reyle22} compiled them from other sources (e.g. \citealt{vanleeuwen07} for bright stars or close binaries without \textit{Gaia} astrometric solution, but also \citealt{dupuy12} or \citealt{kirkpatrick19,kirkpatrick21} for faint low-mass stars and brown dwarfs).

There are fewer objects catalogued by Simbad than by \citet{reyle22}, which explains the difference in the number of objects in Table~\ref{tab:sample_10pc}.
All of the apparently missing objects are in close multiple systems.
We added square brackets in the names for clarity and easy cross-match with Simbad.
As an example, \citet{reyle22} tabulated $\chi$~Dra~A and $\chi$~Dra~B separately, we show $\chi$~Dra~[AB] in a single row in Table~\ref{tab:sample_10pc}, and Simbad only catalogues $\chi$~Dra.

\section{Analysis}
\label{sec:analysis}

As mentioned in the introduction, we followed the same methodology as \citet{gonzalezpayo23,gonzalezpayo24}.
First we looked for common proper motion and common parallax \textit{Gaia} DR3 companions to the {425} stars and brown dwarfs in Table~\ref{tab:sample_10pc}.
We imposed maximum relative proper motion modulus and parallax differences of 15\,\% and maximum proper motion angle difference of 15\,deg.
These conservatively large criteria limits were empirically justified by \citet{gonzalezpayo23,gonzalezpayo24} to account for high proper motion projection effects in nearby, very wide multiple systems \citep{wertheimer06} and %for 
proper motion and parallax anomalies in very close systems \citep{kervella19,brandt21}.
We set a maximum projected physical separation of our search of 1\,pc ($s \sim$ 2 $\cdot \text{10}^\text{5}$\,au). 

Next, we complemented our \textit{Gaia} DR3 search with WDS, discarding from the analysis all background and spurious sources (usually with the ``U'' flag).
All our companions found in the \textit{Gaia} search are already tabulated by WDS.
However, WDS tabulates a number of systems not found in the \textit{Gaia} search, due to either very close separation between components (e.g. G~158--50~A,B, WDS~00155--1608, HEI~299: $\rho\,\approx$\,0.3037\,arcsec as measured by \citealt{ment23}), or faintness of the companion, with a $G$ magnitude beyond the \textit{Gaia} completeness (e.g. HD~42581~A,Ba,Bb, WDS~06106--2152, NAJ~1: the companion is the famous ``GJ~229\,B'' T-type brown dwarf discovered by \citealt{nakajima95}).
We discarded only one WDS system, namely Wolf~358 (J10509+0648, RAO~256), which is described in detail in Appendix~\hyperref[ch:Appendix_F]{F}.

We complemented our \textit{Gaia} and WDS searches with a detailed review of the literature.
We identified multiple systems that were absent in our \textit{Gaia} search or tabulated by WDS.
Some of them are unresolved close pairs, including spectroscopic binaries, or with ultra-cool companions resolved recently (e.g. GJ~229\,Ba,Bb, \citealt{xuan24}).
Most of them were already identified by \citet{reyle22} or \citet{kirkpatrick24}.

\begin{table*}[h]
 \footnotesize
 \centering
 \caption{Unresolved multiple pairs at d $<$ 10\,pc.}
 \begin{tabular}{l@{\hspace{3mm}}l@{\hspace{3mm}}l@{\hspace{3mm}}l@{\hspace{3mm}}c@{\hspace{3mm}}l@{\hspace{3mm}}}
 \noalign{\hrule height 1pt}
 \noalign{\smallskip}
 Name & WDS & Spectral & Class & $P$ & Ref.$^{\text{(b)}}$ \\
  &  & type$^{\text{(a)}}$ & & (d) & \\
 \noalign{\smallskip}
 \hline
 \noalign{\smallskip} 
 \noalign{\smallskip}
Wolf 227 [AB] & ... & M4.5\,V+T: & SB1+Astrom. & $\text{10.59472}^{+\text{0.000020}}_{-\text{0.000016}}$ & Fit24 \\ 
 \noalign{\smallskip}
QY Aur [AB] & ... & M4.5\,V+M: & SB2+Astrom. & 10.42672 $\pm$ 0.00006 & Bar12  \\ 
 \noalign{\smallskip}
Ross 619 [AB]$^{\text{(c)}}$ & ... & M4.5\,V+? & \texttt{RUWE} & ... & ... \\ %Cif25 
 \noalign{\smallskip}
G 41--14 A[ab]$^{\text{(c)}}$ & 08589+0829 & M3.5\,V+M4.0:\,V & SB2 & 7.5555 $\pm$ 0.0002 & Del99  \\
 \noalign{\smallskip}
$\xi$ UMa B[ab]$^{\text{(c)}}$ & 11182+3112 & G2\,V+M: & SB1 & 3.980507 $\pm$ 6 $\cdot \text{10}^{-\text{6}}$ & Gri98 \\ 
 \noalign{\smallskip}
L 768--119 [AB] & ... & M3.0\,V+M: & SB1+Astrom. & 62.6277 $\pm$ 0.0001 & Nid02 \\ 
 \noalign{\smallskip}
HD 152751 [Bab]$^{\text{(c)}}$ & 16555--0820 & M3.5\,V+M: & SB2 & 2.965522 $\pm$ 0.000014 & Maz01 \\ 
 \noalign{\smallskip}
G 203--47 [AB]$^{\text{(c)}}$ & ... & M3.5\,V+D: & SB1 & 14.7136 $\pm$ 0.0005 & Del99 \\
 \noalign{\smallskip}
41 Ara B[ab]$^{\text{(c)}}$ & 17191--4638 & M0.0\,V+? &  Astrom. & 87.9 $\pm$ 3.2 & \textit{Gaia} DR3 \\
  \noalign{\smallskip}
2MASS J17502484--0016151 [AB] & ... & L5.5+L--T & Astrom. & 2460 $\pm$ 140 & Hen18 \\
 \noalign{\smallskip}
FK Aqr [AB]$^{\text{(c)}}$ & 22388--2037 & M1.5\,V+M: & SB2+Astrom. & 4.08319598 $\pm$ 9.5 $\cdot \text{10}^{-\text{7}}$ & Tsv24 \\
 \noalign{\smallskip}
FL Aqr [AB]$^{\text{(c)}}$ & 22388--2037 & M3.5\,V+M: & SB1 & 1.795 $\pm$ 0.017 & Dav14 \\ 
\noalign{\smallskip}
 \noalign{\hrule height 1pt}
 \end{tabular}
 \label{tab:unresolved_systems_10pc} 
 \footnotesize
 \begin{justify}
    \textbf{\textit{Notes. }}
    $^{\text{(a)}}$ Colons indicate spectral-type estimations done by us.
    Two companions, marked with a question mark, have no estimations.
    $^{\text{(b)}}$ References for $P$:
    Bar12: \citet{barry12}; 
    Dav14: \citet{davison14};
    Del99: \citep{delfosse99a};
    Fit24: \citet{fitzmaurice24};
    \textit{Gaia} DR3: \citet{gaiacollaboration23a};
    Gri98: \citet{griffin98};
    Han02: \citet{han02}; 
    Hen18: \citet{henry18}; 
    Maz01: \citet{mazeh01};
    Nid02: \citet{nidever02};
    Tsv24: \citet{tsvetkova24}.
    $^{\text{(c)}}$ Systems described in Appendix~\hyperref[ch:Appendix_F]{F}.
 \end{justify}
 \normalsize
\end{table*}

Finally, we used \textit{Gaia} data to enhance the information of our sample of multiple stars.
Specifically, we applied the criteria to identify unresolved sources based on \textit{Gaia} DR3 statistical indicators summarised in table~3 of \citet{cifuentes25}.
These indicators include the renormalised unit weight error (\texttt{RUWE}), which quantifies the goodness of fit between the observed astrometric data and a single-star model \citep{arenou18,lindegren18a}, \texttt{ipd\_gof\_harmonic\_amplitude}, which is a flag for spurious solutions in resolved doubles \citep{fabricius21}, and several measures of radial-velocity variability among all the \textit{Gaia} measurement epochs \citep{katz23}.
There are 16 single stars and resolved binary components that satisfy at least one of the close-multiplicity criteria.
Of them, fifteen are very bright stars with high \texttt{RUWE} (e.g. $\tau$~Cet, $\epsilon$~Eri, $\pi^{\text{03}}$~Ori), M dwarfs with astrometric signals produced by giant planets (e.g. GJ~876), white dwarfs with spurious radial-velocity or astrometric perturbations (e.g. EGGR~290, L~88--59), or well-investigated stars with wrong duplicated-source labels (e.g. GJ~486).
The only remaining star with a reliable \textit{Gaia} multiplicity indicator is Ross~619, which has a \texttt{RUWE} value of 2.1.
This moderate \texttt{RUWE} excess is difficult to reconcile with a truly single nature.

\section{Results}
\label{sec:results}

We identified 217 stars and brown dwarfs in 112 resolved pairs in 87 multiple systems (Table~\ref{tab:resolved_systems_10pc}) and 24 stars and brown dwarfs in 12 unresolved pairs in 11 multiple systems (Table~\ref{tab:unresolved_systems_10pc}) at $d <$ 10\,pc.
The remaining stars and brown dwarfs in Table~\ref{tab:sample_10pc} not displayed in Tables~\ref{tab:resolved_systems_10pc} or~\ref{tab:unresolved_systems_10pc} are considered to be single with current data (including 2MASS J09393548--2448279; \citealt{burgasser08}).
A large fraction of them have been extensively studied with extreme precision spectrographs \citep[e.g.][]{mayor11,howard16,wittenmyer16,ribas23,mignon24,harada24b} and with high-resolution, high-contrast imagers \citep[e.g.][]{oppenheimer01, chauvin10, vigan21, gauza21}, besides with \textit{Gaia}.
The confirmation of the hypothetical single stars that have not been yet the subject of in-depth multiplicity studies, such as the M2.5\,V star TYC~3980--1081--1 at $d\,\approx$\,8.1\,pc (\citealt{caballero24}; \'E.~Artigau, priv.~comm.), will be the topic of forthcoming work.

For all star and brown dwarf components of each of the 87 systems with resolved pairs at $d <$ 10\,pc, Table~\ref{tab:resolved_systems_10pc} displays the default Simbad name of the stellar or brown-dwarf component (in parenthesis, common name), WDS system identifier, pair discoverer code (``...'' when unavailable), heliocentric distance from \citet{reyle22}, angular separation from primary to companion $\rho$, position angle $\theta$ (measured from North to East; ``...'' when $\rho <$ 1\,arcsec), projected physical separation ($s \approx \rho \cdot d$; see exact formula in eq.~6 of \citealt{gonzalezpayo23}), reference for the later astrometric data (``This work'' for 55 of the 112 resolved pairs), spectral type, \textit{Gaia} magnitude $G$, mass $M$, and reference of the resolved components, orbital period $P$ and reference of the pairs, reduced binding energy modulus of the system $|U_g^*| = G M_i M_j / s_{ij}$, where $i$ and $j$ denote two components in a pair \citep{caballero07a}, and, for completeness, reported planet candidates, for which we updated the results of \citet{gonzalezpayo24} with the Encyclopaedia of exoplanetary systems\footnote{\url{https://exoplanet.eu/}} \citep{schneider11} and NASA Exoplanet Archive\footnote{\url{https://exoplanetarchive.ipac.caltech.edu/}} \citep{akeson13}.
Finally, the last column displays a diagram sketch of each system.

New $\rho$ and $\theta$ values were measured by us with % from 
\textit{Gaia} DR3 astrometric data (epoch J2016.0).
In triple and higher order systems with hierarchical arrangements and at least two pairs of components in Table~\ref{tab:resolved_systems_10pc}, we show $\rho$ between the primary and the wide companion and between the components in close pairs (e.g. $\epsilon$~Ind~A,Ba,Bb, WDS~22034--5647, SOZ~1: $\rho_{\rm A-Ba,Bb}$ = 402.439\,arcsec and $\rho_{\rm Ba-Bb}$ = 0.66158\,arcsec).

In general, spectral types, $G$ magnitudes, and $M$ were taken from Simbad, \textit{Gaia} DR3, and the literature (especially \citealt{kirkpatrick24}), respectively.
In %For 
many cases, we estimated their magnitudes and spectral types from available data, which are marked with a colon in Table~\ref{tab:resolved_systems_10pc}.
When missing or dubious, we also estimated masses from $G$-band absolute magnitudes and the relations of \citet{pecaut13} and \citet{cifuentes20}.

We made extensive use of the Sixth Catalog of Orbits of Visual Binary Stars\footnote{\url{https://crf.usno.navy.mil/wdsorb6}} (ORB6) of the WDS to complement our literature search for $P$ of pairs from astrometric solutions.
Only when not available, we estimated $P$ from $M_i$, $M_j$, $s$ in Table~\ref{tab:resolved_systems_10pc}, and Kepler's third law, $P^\text{2} = (M_i + M_j) ~  a^\text{3}$ in Solar System units, where $a$ = 1.26 $\cdot s$ to account for random orbital inclinations  \citep{fischer92,close03,burgasser07a,radigan09,caballero10,faherty10}.
While all $|U_g^*|$ values were computed homogeneously with $s$ and can therefore be compared among themselves and with published works \citep{close03,burgasser07a,caballero10,gonzalezpayo21,gonzalezpayo23,gonzalezpayo24,cifuentes25}, our estimated $P$ from $a$ needed the 1.26 correcting factor to be compared with the rest of ``true'' $P$ from orbital astrometric fits.
For triple and higher-order multiples we used the same definition for $P$ as for $\rho$ (e.g. we tabulate two periods in the $\epsilon$~Ind system: $P_{\rm A-Ba,Bb}$ and $P_{\rm Ba-Bb}$).
However, we computed only one $|U_g^*|$ per resolved system, corresponding to the most separated pair, except for trapezium-like systems that do not have hierarchical arrangements \citep{ambartsumian55,gonzalezpayo23}.

Not all nearby multiple systems have been resolved yet.
In Table~\ref{tab:unresolved_systems_10pc}, we list 12 very close unresolved pairs at $d <$ 10\,pc.
Six of them are in five hierarchical systems and are therefore listed in Table~\ref{tab:resolved_systems_10pc} as one resolved component.
Of the 12 pairs, four have very precise orbital periods from simultaneous spectroscopic and astrometric data, five are spectroscopic binaries (three single- and two double-line spectroscopic binaries), two are astrometric binaries, and %the remaining 
one is the candidate binary from a moderately large \texttt{RUWE} presented in Sect.~\ref{sec:analysis}, namely Ross~619. 
The hierarchical systems, the candidate binary with large \texttt{RUWE}, and one of the spectroscopic binaries that has a poorly characterised white-dwarf candidate are described, with the corresponding references, in Appendix~\hyperref[ch:Appendix_F]{F}.

Six pairs in Tables~\ref{tab:resolved_systems_10pc} and~\ref{tab:unresolved_systems_10pc} are compiled in the Ninth Catalogue of Spectroscopic Binary Orbits (SB9)\footnote{\url{https://sb9.astro.ulb.ac.be/}} of \citet{pourbaix04}.
In all cases, the SB9 periods and those compiled by us from ORB6 or the literature match within uncertainties, except for BD--18~359\,AB.
\citet{nidever02} determined $P$ = 18.7 $\pm$ 6.8\,a from spectroscopic observations, but we kept the most precise value of $P$ = 13.328 $\pm$ 0.037\,a from astrometric observations by \citet{mann19}.

Taking into account the 87 systems with resolved pairs in Table~\ref{tab:resolved_systems_10pc} and the 6 unresolved pairs without additional companions in Table~\ref{tab:unresolved_systems_10pc}, there are 93 multiple systems known at $d <$ 10\,pc.
Of them, 69 are double, 19 triple, 3 quadruple, and 2 quintuple.
We compared our multiple system classification with those of \citet{reyle22} and \citet{kirkpatrick24}, and found a few differences:
WDS~J00363+1821 was single for \citet{reyle22}, but double for \citet{kirkpatrick24} and for us;
WDS J06106--2152 was double for both \citet{reyle22} and \citet{kirkpatrick24}, but triple for us;
WDS~J06523--0510 was triple for \citet{reyle22}, but double for \citet{kirkpatrick24} and for us;
and J08115757+0846220 (Ross~619) was single for both \citet{reyle22} and \citet{kirkpatrick24}, but double for us.
All of them are described again in Appendix~\hyperref[ch:Appendix_F]{F}.

\section{Discussion}
\label{sec:discussion}

\subsection{Multiplicity fraction and companion star fraction}
\label{sec:MSF_CSF_10pc}

With the numbers of single stars and binary and high-order multiples, we determined the multiplicity fraction (MF) and companion star fraction (CSF) as defined by \citet{batten73} and \citet{reipurth93}, and recently summarised by \citet{cifuentes25}.
The MF of a given stellar population or volume represents the fraction of systems in which the primary star belongs to that population or volume, while the CSF is a measure of the average number of companions per system.
We measured MF = 30.0$^{\text{+5.3}}_{-\text{4.8}}$\,\% and CSF = 40.0$^{+\text{5.5}}_{-\text{5.3}}$\,\%, including Ross~619 as an unresolved binary, where the uncertainties are 95\,\% Wilson's confidence levels (ibid. \citealt{cifuentes25}).
The values above correspond to global values of MF and CSF, as they do not take into account the mass of the primary star, but rather include all masses.
Without proper primary identification, direct comparisons between MF and CSF values and the overall stellar population may lead to misleading interpretations.
When restricted to systems with M-dwarf primaries only, the revised MF$_{\rm M}$ and CSF$_{\rm M}$ values become 20.8$^{+\text{5.2}}_{-\text{4.4}}$\,\% and 26.6$^{+\text{5.5}}_{-\text{4.9}}$\,\%, respectively (we did not count systems with white-dwarf components because their stellar progenitors were earlier than M).
Because of the completeness and comprehensiveness of our analysis, these MF$_{\rm M}$ and CSF$_{\rm M}$ values may be the most accurate determined to date for M dwarfs.

Since M dwarfs are by far the most abundant type of stars in the solar neighbourhood, and in the 10\,pc sample in particular \citep{henry06,reyle21}, the MF$_{\rm M}$ and CSF$_{\rm M}$ values are in line with other analyses of multiplicity of low-mass stars \citep{henry91,reid97c,janson12,jodar13,cortescontreras17b,winters19,susemiehl22,clark24}.
Including new binary candidates identified in a similar way to this work, \citet{cifuentes25} reported higher MF and CSF for M dwarfs, which validates their hypothesis that their sample was biased by previously unknown unresolved binaries.
These results also support the need to perform multiplicity statistics on well-defined volume-limited samples and to monitor new \textit{Gaia} DR3 binary candidates with high-resolution imaging and spectroscopic facilities.

\subsection{Magnitude differences and orbital periods}
\label{sec:magnitud_differences_10pc}

\begin{figure}[H]
    \centering
    \includegraphics[width=0.6\linewidth]{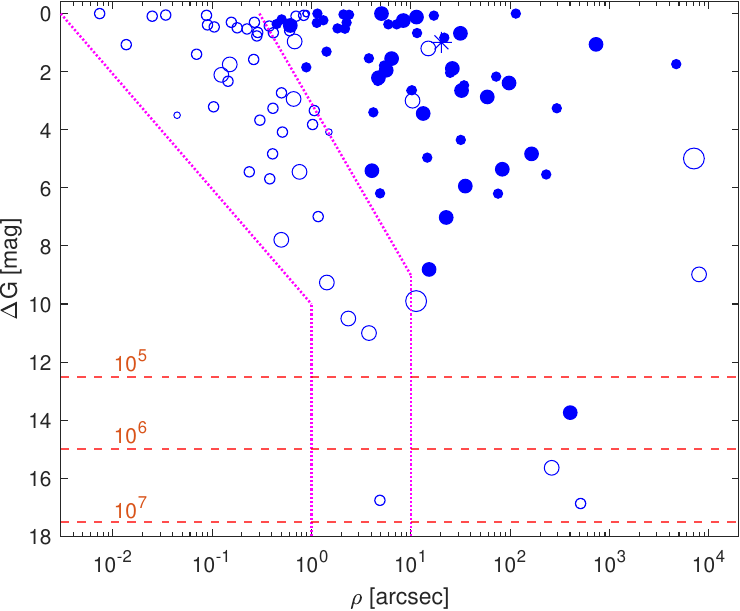}
    \caption[Absolute magnitude difference in $G$ band as a function of angular separation for each resolved pair.]{Absolute magnitude difference in $G$ band as a function of angular separation for each resolved pair. Filled circles represent pairs with measured magnitudes of both components, while empty circles represent pairs with estimated magnitudes for one or both components. Symbol size is approximately proportional to the mass of the primary (the only double white-dwarf binary is marked with an asterisk). Horizontal lines indicate primary-to-companion contrast ratios of 10$^\text{5}$, 10$^\text{6}$, and 10$^\text{7}$ from top to bottom.
    The diagonal and vertical lines indicate the inaccessible regions with current technology for photometry of both pair components (right) and only one component (left).}
     \label{fig:10pc_rhoG}
\end{figure}

Here we review the principal parameters of the investigated pairs.
Fig.~\ref{fig:10pc_rhoG} shows the magnitude difference in the \textit{Gaia} $G$ band between components in resolved pairs as a function of their angular separation, $\rho$.
As expected, there is a lack of very close pairs with high contrast ratios between components.
This contrast is independent of the primary spectral type.
\textit{Gaia} is able to measure magnitudes of two equal-brightness components separated by $\sim$0.4\,arcsec, while some ground facilities (especially interferometers) can do that down to $\sim$0.01\,arcsec \citep[][and references therein]{gaiacollaboration18,lindegren18a,lindegren18b,ziegler18,brandeker19,cifuentes25}.
At angular separations greater than 1\,arcsec, \textit{Gaia} is limited mostly by its magnitude limit at $G\,\approx$\,20.4\,mag, except for naked-eye primaries.

The earliest and, therefore, most massive stars in multiple systems within 10\,pc are $\gamma$~Lep (F6\,V), Sirius~[A] (A0\,V), and Procyon~[A] (F5\,IV--V), all with masses greater than 1.2\,M$_\odot$.
Conversely, there are 11 late T dwarfs and one early Y dwarf in binary (WISE J045853.89+643452.5 [AB], Scholz’s Star [B], WISE J121756.90+162640.8 [AB], ULAS J141623.94+134836.3, SCR J1845--6357B), triple (GJ~229B[ab], $\epsilon$~Ind B[b]), quadruple (GJ~570D), and quintuple systems ($\xi$~UMa~C).
In all cases except for GJ~229B[ab], their masses between the hydrogen and deuterium mass limits are poorly constrained \citep{caballero18}.
The values of $|U_g^*|$ of all systems are greater than the minimum value required to avoid disruption by the galactic potential \citep{bahcall81,weinberg87,jiang10}, as well as than the lowest reduced binding energy observed in moderately separated, very low-mass binary systems \citep{chauvin04,artigau07,caballero07a,radigan09}, which is about 10$^{\text{33}}$\,J \citep{caballero10}. 

Angular separations, $\rho$, of resolved systems range from less than 10\,mas (G~184--19~AC, GJ~229B[ab], EZ~Aqr~BC), to more than 1000\,arcsec (AU~Mic, Fomalhaut, and $\alpha$~Cen systems), and corresponding projected physical separations from less than 0.1\,au to over 100\,000\,au.
Likewise, the compiled and estimated orbital periods range more than ten orders of magnitude from $P$ = 1.795 $\pm$ 0.017\,d for the single-lined spectroscopic binary FL~Aqr~[AB] \citep{davison14} to tens of million years for AU~Mic and AT~Mic and Fomalhaut [A] and C, which reach the blurred boundary between the widest multiple systems and chance alignments in young stellar kinematic groups \citep{caballero10,shaya11,mamajek13}. 
The orbital period is a manifestation of the masses of the system components and their semi-major axis, $a$, which itself depends on the projected angular separation, distance, and orbital elements.
Observables $\rho$ and $d$ are available for all resolved systems in Table~\ref{tab:resolved_systems_10pc}, while orbital elements besides $P$ and $a$ (namely eccentricity $e$, inclination $i$, longitude of ascending node $\Omega$, argument of periastron $\omega$, time of periastron passage $T_\text{0}$) are available only for a few, compiled mostly by ORB6.
$P$ is also available for all unresolved pairs in Table~\ref{tab:unresolved_systems_10pc}.

\subsection{\"Opik's law in the \textit{Gaia} era}
\label{sec:opiks_law_10pc}

\citet{kuiper55} suggested that the distribution of semi-major axes is equivalent to ``the frequency distribution of total angular momentum present in the original protostars, after allowance is made for components ejected from multiple stars and for those wide binaries that were subsequently dissolved by stellar encounters''.
As a result, the distribution function of $a$ provides a unique input to stellar formation models.
The distribution function of $a$ and, more often, projected physical separations $s$ of multiple stellar systems in the field have been approximated by various functions in the last century \citep{opik24,kuiper35,heintz69,zinnecker84,heacox96}.
First of all, \citet{opik24} suggested that the distribution of binaries with semi-major axes larger than $\sim$60\,au follows the form $f (a) \propto a^{-\text{1}}$.
Next, \citet{kuiper42, kuiper55} proposed that the overall distribution could be represented instead by a Gaussian.
Afterwards, authors have used power-law \citep{poveda04,longhitano10}, log-normal \citep{duquennoy91,reipurth07}, or both functions \citep{kouwenhoven07} to study $a$ distributions.
\"Opik's law suggests a scale-free process, while a log-normal function might imply a preferred spatial scale for companion formation.
Besides, the system's separation can be significantly altered by dynamical evolution, which complicates the interpretation \citep[][and references therein]{duchene13}.

\begin{figure}[H]
    \centering
    \includegraphics[width=0.6\linewidth]{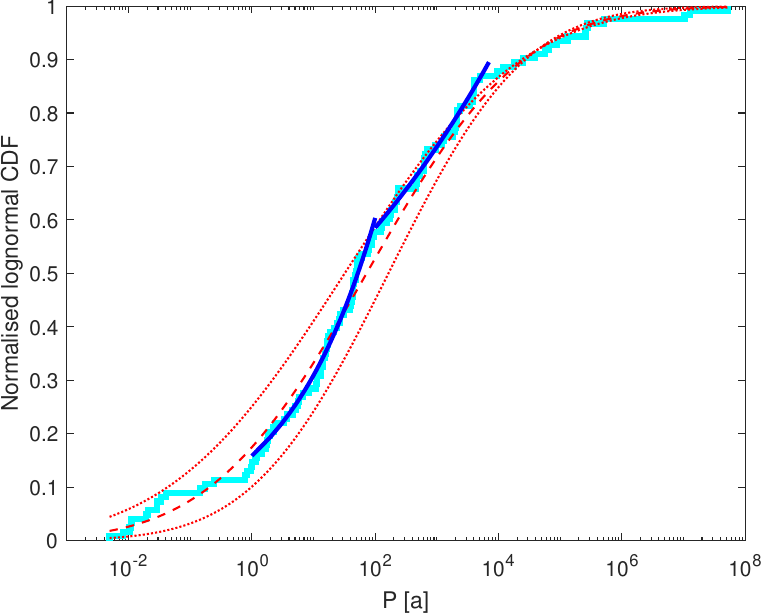}
    \caption[CDF of $P$ of all pairs at $d <$ 10\,pc.]{CDF of $P$ of all pairs at $d <$ 10\,pc (cyan thick staired line). The red dashed and dotted lines are the log-normal fit with its lower and upper limits, while the navy blue solid lines are the two power-law fits in the 1--100\,a and 100--7000\,a intervals.}
     \label{fig:10pc_periods}
\end{figure}

Our sample of multiple stellar systems is complete in volume ($d <$ 10\,pc), in physical separations (from a few solar radii to half a parsec), and in mass of primary and companion stars and brown dwarfs (from 2\,M$_\odot$ to almost the deuterium burning mass limit at about 0.01\,M$_\odot$).
Previous multiplicity surveys have been incomplete in volume, separation, mass, or, more likely, all of them simultaneously.
As a result, our sample allows investigating for the first time the full range of  separations.
Of all the compiled orbital elements, $P$ is the most precise and available one for all pairs except one (Ross~619~[AB]), and is a proxy for $a$.
As a result, we computed the cumulative distribution function (CDF) of $P$ of all our pairs (i.e. in both Tables~\ref{tab:resolved_systems_10pc} or~\ref{tab:unresolved_systems_10pc}), as shown in Fig.~\ref{fig:10pc_periods}.

There are several advantages of using CDF over non-cumulative distribution functions (e.g. histograms), such as easier comparison of different data sets, detection of outliers, clusters, and, especially, distribution types, as well as being safer against misinterpretation and manipulation.
Besides, the CDF can be defined for any kind of random variable (discrete, continuous, and mixed) and facilitates direct quantitative reading of essential key values (minimum, maximum, median, percentiles).

We fitted the data to two different CDFs.
First, we fitted them to a log-normal CDF of the form:
\begin{equation}
D (P) = \frac{k}{\text{2}} \left[ \text{1}+ {\rm erf} \left( \frac{\ln{P}-\mu}{\sigma \sqrt{\text{2}}} \right) \right],
\end{equation}
\noindent where erf is the error function, $\mu$ and $\sigma$ are the expected value (mean) and standard deviation of the natural logarithm of a variable in a log-normal distribution, and $k$ is the normalisation factor, which in our case we fixed to 123, i.e. the number of pairs with known periods.
We used the \texttt{logncdf} function in the MATLAB\footnote{\url{https://www.mathworks.com}} programming language \citep{moler80} for the fit in the whole range of $P$, and got $\mu$ = 4.28 $\pm$ 0.82 and $\sigma$ = 4.58 $\pm$ 0.58  with confidence intervals of 95\%.
Next, we fitted the data to a power law of the form:
\begin{equation}
D (P) = \beta P^\alpha.
\end{equation}
A single power law is not able to properly fit the data in a wide $P$ range, so we actually fitted two power laws in the ranges $P_{\rm short}$ = 1--10\,a and $P_{\rm long}$ = 10--7000\,a.
We obtained the fitting parameters $\alpha_{\rm short}$ = 0.2906 $\pm$ 0.0039 and $\beta_{\rm short}$ = 0.1583 $\pm$ 0.0018, and  $\alpha_{\rm long}$ = 0.1000 $\pm$ 0.0022 and $\beta_{\rm long}$ = 0.3693 $\pm$ 0.0054, respectively.
Any of the different power-law indices $\alpha_i$ do not necessarily match the original \"Opik law index, as we fitted the CDF instead of the distribution function itself and there is a factor 3/2 from the $P$ to $a$ transformation convolved with the sum of the masses of individual components in each pair from the Kepler third law.
In contrast to the power-law fit, the log-normal fit can be successfully applied in the whole $P$ interval from days to millions of years, which supports the use of a Gaussian distribution function by \citet{duquennoy91} and \citet{reipurth07}, among other authors.
This result leaves the window open for further theoretical work that can interpret the physics behind the fitted $\mu$ and $\sigma$ parameters ($\mu$ is of the same order of magnitude as the natural logarithm of the median $P$).

\subsection[Remarkable systems within 10\,pc]{Remarkable systems within 10\,pc\footnote{This section was written in the original article González-Payo et al. 2025, (under referee review) as a separate appendix.}}
\label{sec:remarkable_systems_10pc}

Here we highlight some remarkable stellar systems in the solar neighbourhood:

\begin{itemize}
    \item \textit{WDS~J00363+1821:} LSPM~J0036+1821 (2MASS J00361617+1821104) was first discovered by \citet{reid00} and later classified by \cite{reid06} as an L3.5+L6: unresolved ultra-cool dwarf pair. The system was subsequently resolved by \citet{bernat10} with aperture masking interferometry and Palomar laser-guide-star adaptive optics in the $K_s$ band. The reported angular separation was only $\rho=$ 89.5\,mas. This discovery was later confirmed by \cite{pope13} in the $J$ and $H$ bands with a kernel phase interferometry reanalysis of archival NICMOS/{\it Hubble}.

    \item \textit{WDS~J06106--2152:} HD~42581 (GJ\,229) is a triple system consisting of an M dwarf and two brown dwarfs. Considered single for several decades, GJ\,229\,B was the first T dwarf to be detected \citep{nakajima95}, with confirmed methane absorption features \citep{oppenheimer95, oppenheimer98}. It orbits the primary at $a=$ 33.3\,au with a period of 237.9\,a \citep{brandt21}. The T dwarf was further resolved into a 12.1-day binary with $a=$ 0.042\,au \citep{xuan24}. Additionally, two circumstellar planetary companions to GJ\,229\,A have been proposed (\citealt{tuomi14,feng20,feng22}).

    \item \textit{WDS~J06523--0510:} HD~50281~AB consists of an early-K primary and a early-M secondary. The pair has been known since the middle of the 20th century \citep[e.g.][]{eggen56}. WDS tabulates two more pairs in the system: A 9.6\,arcsec companion candidate reported by \citet{tanner10} is a background star according to \textit{Gaia} DR3, while the hypothetical close binarity of the secondary star (WSI~BaBb, $\rho =$ 0.04\,arcsec on a single epoch in 2010, $\Delta V \sim$ 0.5\,mag) has been repeatedly questioned \citep{tokovinin15, tokovinin20, tokovinin18, mason18}. 

    \item \textit{J08115757+0846220:} Ross~619 \citep{ross1927} was considered to be a single M4.5\,V star after observations with high-resolution imagers \citep{oppenheimer01,hinz02,janson14} and spectrographs \citep{bonfils13,ribas23,mignon24}. However, it is an unresolved binary candidate based on astrometric data from \textit{Gaia} DR2 \citep{vrijmoet20} and DR3 \citep{clark24,cifuentes25}, especially because of its \texttt{RUWE} and \textit{Gaia} \texttt{ipd\_frac
    \_multi\_peak} values.

    \item \textit{WDS~J08589+0829:} G~41--14 \citep{giclas59} is a triple system made of roughly similar M-type dwarfs \citep{stephenson86,kirkpatrick12}. The primary is a 7.6\,d-period double-lined spectroscopic binary \citep{reid97b}, while the third component was discovered by \citet{delfosse99a} with adaptive optics at an angular separation of 0.620\,arcsec. They estimated an orbital period of approximately 10 years. After two decades collecting more astrometric data \cite[e.g.][]{hartkopf12,janson14}, the actual orbital period is about half that value \citep{tokovinin23}.

    \item \textit{WDS~J10509+0648:} \citet{lamman20} reported a close companion for Wolf~358 (EE~Leo) with a separation of 0.098 $\pm$ 0.070\,arcsec, detected in a single Robo-AO observation from 2016, although no further confirmation of multiplicity has been reported. In fact, it appears to be single based on extremely precise astrometry from \textit{Gaia} \citep[e.g.,][]{kervella19, vrijmoet20} and, in particular, radial velocity measurements from CARMENES \citep{ribas23}.

    \item \textit{WDS~J11182+3112:} $\xi$~UMa (Alula~Australis) is a quintuple system. According to \citet{mason95}, it was one of the first binaries reported \citep{herschel1803}, with precise relative astrometry ($\Sigma$~1523: \citealt{struve1827}), calculated orbit \citep{savary1827}, and also a ``definitive'' orbit \citep{norlund1905,vandenbos1928,heintz67,worley83}. The two main components of $\xi$~UMa are main-sequence, solar-like stars with angular separation of about 1–3\,arcsec, with an orbital period of 59.89\,a. Each component of this double star is itself a close binary. The 1.834\,a eccentric orbit of the A pair has been determined from spectroscopy and speckle interferometry \citep{heintz67,mason95}, while $\xi$~UMa~B is a single-lined spectroscopic binary with a circular 3.98-day orbit \citep{berman31,griffin98}. Using Wide-field Infrared Survey Explorer (WISE) data, \citet{wright13} discovered the fifth component of the system, WISE~J111838.70+312537.9 ($\xi$~UMa~C), a T8.5 brown dwarf located 8.5\,arcmin away.

    \item \textit{WDS~J16555--0820:} This complex hierarchical system \citep[or mini-cluster, e.g.][]{eggen65} is quintuple and the closest such system to the Sun \citep{wolf1917,kuiper34}. The primary component, HD~152751, has an M3\,V spectral type \citep{reid95} and is a triple-lined spectroscopic binary \citep{joy47,delfosse99a}, with inner and outer orbital periods of 2.965 and 627 days, respectively, and a combined mass of roughly one solar mass \citep{mazeh01}. Wolf~629, an M3.5\,V star \citep{reid95}, is the fourth component, located 72.3\,arcsec away from the primary. The fifth component, vB~8, is an M7.0\,V star located 220\,arcsec from the primary \citep{vanbiesbroeck61, alonsofloriano15}. Early infrared speckle interferometry by \citet{mccarthy85} suggested the presence of a cool companion at 1\,arcsec to vB~8, consistent with a brown dwarf. However, this claim was later disproven and vB~8 was confirmed as a single star within the AstraLux sensitivity range \citep{janson14}. 

    \item \textit{G~203--47:} This is a single-lined spectroscopic binary discovered by \citet{reid97b}, composed by an M3.5\,V star and a white dwarf, with the first (and only) orbit determination by \citet{delfosse99a}. A large radial velocity scatter has been observed in several surveys, including \textit{Gaia} DR3, but no re-determination of the orbit has been made since.

    \item \textit{WDS~J22388--2037:} FK~Aqr and FL~Aqr form a visual pair ($\rho\simeq$\,25\,arcsec) of magnetically-active cool dwarfs \citep[][and references therein]{vyssotsky52,herbig65,tsvetkova24}. Both components have been further identified as close binaries, therefore constituting a quadruple star system. FK~Aqr is a double-lined spectroscopic binary in a 4.083-day orbit with astrometric signatures consisting of two early-M dwarfs of nearly equal mass \citep{herbig65,delfosse99a,kervella19}. FL~Aqr is a single-lined spectroscopic binary with an ultra-cool dwarf companion in a 1.795-day orbit \citep[][who reported a minimum mass of the companion of 67\,M$_{\rm Jup}$]{davison14}. This is the only quadruple system within 10\,pc of the Sun comprising just M-type (or M- and L-type) components.
\end{itemize}

\section{Summary and future work}
\label{sec:summary}

In this study, we analysed the multiplicity of all stars and brown dwarfs at a maximum heliocentric distance of 10\,pc, originally compiled by \citet{reyle22}. 
Our analysis focused on identifying, confirming, and characterising the multiple systems in the sample, compiling and determining their astrometric properties and masses, and deriving their orbital periods and reduced binding energies. 
For this purpose, we used data from \textit{Gaia} DR3 and the WDS catalogues, together with the available literature. 
As a result, we identified 217 stars and brown dwarfs in 93 systems at all physical separations from less than 0.1\,au to over half a parsec. 
Of these, 69 systems are double, 19 are triple, 3 are quadruple, and 2 are quintuple.
Only eleven systems contain at least one unresolved pair.
In terms of the mass distribution of stars in multiple systems, there are only three stars more massive than the Sun, and 11 late T dwarfs and one early Y binary at the opposite mass end. 
Since M dwarfs represent about 70\% of the investigated sample, we measured the multiplicity fraction and companion star fraction of systems with M-dwarf primaries with an unprecedented accuracy from the most complete volume limited sample near the Sun. 
The measured values of MF$_{\rm M}$ = 20.8$^{+\text{5.2}}_{-\text{4.4}}$\,\% and CSF$_{\rm M}$ = 26.6$^{+\text{5.5}}_{-\text{4.9}}$\,\% align well with multiplicity fractions of low-mass stars reported by previous studies.
The computed (reduced) binding energies indicate that all systems are stable enough to survive over time against disruptive galactic forces.
Finally, we compiled or estimated orbital periods for 123 of the 124 identified pairs, ranging from 1.795\,d to tens of million years. 
We fitted its cumulative distribution function to a log-normal function at all period ranges, and by extension, all physical separations, and compared it to power-law fits.
In a sense, the derived $\mu$ and $\sigma$ values from the log-normal fit are a 21st-century revision of the \"Opik's law.

We finish this work by enumerating a list of forthcoming work necessary to complete the results presented here.
Most of them are related to new observations for improving the characterisation of companions in a few poorly investigated pairs.
We itemise below the potential targets, necessary observations, and expected outcome:
\begin{itemize}

\item Ross~619~[AB]: both high-resolution imaging and spectroscopy for resolving the binary, computing a preliminary orbit, and determining the spectral-type and mass of the secondary;

\item 41~Ara~B[ab]: intermediate- or high-resolution spectroscopy for ascertaining the nature of the astrometric companion and improving the orbital parameters;

\item 2MASS J17502484--0016151 [AB]: additional astrometry and intermediate-resolution spectroscopy for characterising the L--T companion;

\item G 203--47[AB]: intermediate- or high-resolution spectroscopy for confirming the nature of the hypothetical white-dwarf companion;

\item LSPM J0036+1821 AB, G~100--28 [AB], G~119--36 [AB], WT~460 [AB], SCR~J1546--5334 [AB], $\mu^{\text{02}}$~Her AB, G~184--19 AC, SCR~J1845--6357 AB: additional sub-arcsecond adaptive optics imaging for computing orbits and, therefore, system total masses, periods (between weeks and a few decades), and semi-major axes for the first time;

\item BD--17~588 AB, BD+16~2708 [A,Bab], CD--48~11837 AB, HD~191408 AB (all other resolved pairs with $P \sim$ 100--600\,a from this work): seeing-limited imaging and analysis of old public data for determining preliminary orbital elements;

\item Fifteen pairs with previous astrometric solutions but no uncertainties in parameters (all other resolved pairs with $P\lesssim$ 1000\,a from the literature): seeing-limited and high-resolution imaging and analysis of old public data for improving orbital elements.
This list include Scholz's Star [AB], $\xi$~UMa~A[ab], and Ross~775 [AB], which have periods of only 0.15--8.3\,a from the literature;

\item All single stars at $d <$ 10\,pc not monitored with high-resolution imagers and spectrographs: confirmation of their single nature. 

\end{itemize}
A number of the systems enumerated above, plus others better studied and shown in Table~\ref{tab:resolved_systems_10pc}, have white-dwarf, ultra-cool-dwarf, and exoplanet companions, which also need further investigation (e.g. precise magnitude differences and spectral types of close resolved companions, stellar and planet mass improvement, ratio of orbital separations in multiple systems, orbits in 3D).
Data from \textit{Gaia} DR4, to be released in the second half of 2026, will certainly help in knowing better our closest multiple systems.
However, there must be a concerted effort worldwide to characterise them in detail beyond \textit{Gaia}'s capabilities.
Actually, some of the systems reported here are in the lists of potential targets of HWO and LIFE, so this work also does its little part in the search for habitable exoplanets.

%%%%% CAPITULO 7 %%%%%

\chapter{Conclusions and future work} 
\label{ch:conclusions_future}
\vspace{2cm}
\pagestyle{fancy}
\fancyhf{}
\lhead[\small{\textbf{\thepage}}]{\textbf{Section \nouppercase{\rightmark}}}
\rhead[\small{\textbf{Chapter~\nouppercase{\leftmark}}}]{\small{\textbf{\thepage}}}

\begin{flushright}
\small{\textit{``For my part, I know nothing with any certainty\\but the sight of the stars makes me dream''}}

\small{\textit{-- Vincent Van Gogh}}
\end{flushright}
\bigskip

\lettrine[lines=3, lraise=0, nindent=0.1em, slope=0em]{T}{his doctoral thesis has explored} various aspects of multiple stellar systems, focusing on their characterisation in the solar neighborhood, the search for wide companions in specific stellar populations, the analysis of ultra-wide systems, the influence of multiplicity on planetary systems, and the recovery of an early historical contribution to double star astronomy. Through the analysis of extensive modern astronomical datasets, primarily those provided by the \textit{Gaia} mission, and the review of existing literature, a series of objectives have been achieved that significantly expand our understanding of these dynamic systems. This final chapter consolidates the key conclusions drawn from this research, emphasising how these findings refine our knowledge of multiple stellar systems across various spatial and evolutionary scales. Moreover, it outlines promising avenues for future investigations, including continued observational follow-up of newly identified systems, high-resolution spectroscopy and imaging campaigns, theoretical modeling of complex hierarchical systems, and statistical studies of stellar and planetary multiplicity using forthcoming \textit{Gaia} data releases. 

The work presented in this thesis lays the foundation for further discoveries and advancements in the study of multiple stellar systems, reinforcing the importance of high-precision astrometric data in unveiling the intricate nature of stellar relationships in our Galaxy.

\section{Conclusions}
\label{sec:conclusions}

Throughout this work, several significant conclusions have been reached that shed light on the nature and behavior of multiple stellar systems.

The contents of Chapter~\ref{ch:hodierna} show the recovery and analysis of Giovanni Battista Hodierna’s 1654 catalogue of double stars vindicated his pioneering contribution to the history of binary star astronomy, successfully identifying most of the stellar pairs he described and assessing their physical nature with modern astronomical data. The main findings are:

\begin{itemize}
    \item \textit{Hodierna's work about binaries:} 
    \begin{itemize}
        \item The memory of the Italian astronomer Giovanni Battista Hodierna has been recovered; the work demonstrates that he published the first catalogue of binary stars in 1654.
        \item This discovery predates by over a century the contributions of Christian Mayer and William Herschel, who have traditionally been considered the pioneers of double-star astronomy.
    \end{itemize}
    \item \textit{First list of binaries:} 
    \begin{itemize}
        \item The fourth section of Hodierna’s 1654 book, \textit{De systemate orbis cometici}, has been analysed, where he listed a dozen star pairs, referred by him as \textit{stellae geminae} (twin stars). For five of them Hodierna provided ecliptic coordinates, and for the rest he included a drawing representing the angular separation between their components pictographically. Additionally, the astronomer provided a description in Latin for all of them.
    \end{itemize}
    \item \textit{Identification of the listed stars:} 
    \begin{itemize}
        \item The article presents an attempt to identify the twelve double stars mentioned by Hodierna using modern astronomical catalogues such as WDS catalogue, \textit{Gaia} DR3 and \textit{Hipparcos} data.
        \item It was determined that Hodierna measured smaller angular separations compared to modern \textit{Hipparcos} measurements, with an approximate scaling factor of $\sim$0.80.
        \item Of the twelve systems described by Hodierna, the work plausibly identified eleven. Among them, four appear to be gravitationally bound systems (\textit{Atlas}/\textit{Pleione},  $\alpha^{\text{1,2}}$ Lib, \textit{Kuma}$^\text{1,2}$, and $\theta^\text{1}$ Ori A/D/C).
        \item It is noteworthy that Hodierna observed the $\theta^\text{1}$ Ori system (the \textit{Trapezium}) before Huygens published his findings on its triple nature.
        \item The remaining seven identified pairs were found to be optical alignments of bright stars at different distances.
    \end{itemize}
\end{itemize}

The impact of the findings in this article on previous research is:
\begin{itemize}
    \item \textit{Rewriting history:} This discovery rewrites the history of double-star astronomy, recognising Hodierna as the first astronomer to publish a list of multiple-star systems. Previously, the birth of double-star astronomy was primarily attributed to the observations and catalogues of Christian Mayer (late 18th century) and William Herschel (late 18th and early 19th centuries). Hodierna’s work predates these contributions by over a hundred years. This finding highlights the importance of reviewing and analysing lesser-known historical astronomical documents, which may contain significant scientific discoveries that were overlooked at the time.
\end{itemize}

In Chapter~\ref{ch:multiplicity_of_ultra-cool_subdwarfs}, a systematic search for wide companions around M and L subdwarfs led to the identification of new systems and the estimation of their frequency. Preliminary photometric and spectroscopic characterisations allowed for the inference of multiplicity rates for different subdwarf types. Key results include:

\begin{itemize}
    \item \textit{Multiplicity rates by subdwarf type:}
    \begin{itemize}
        \item A confirmed wide companion was identified for an M subdwarf, leading to a multiplicity estimate of 1.03\small $_{-\text{1.03}}^{+\text{2.02}}$\% \normalsize at projected physical separations of up to 743\,000 au.
        \item Among sdM5--sdM9.5 subdwarfs, the inferred frequency of wide systems is 0.6\small $_{-\text{0.6}}^{+\text{1.2}}$\% \normalsize for separations larger than 1360 au (up to 142\,400 au).
        \item M extreme subdwarfs (esdM) show a multiplicity rate of 1.89\small $_{-\text{1.89}}^{+\text{3.70}}$\% \normalsize.
        \item No companions were found for the M ultra-cool subdwarfs (usdM) in the sample, setting an upper limit on binarity at 5.3\%.
    \end{itemize}
    
    \item \textit{Identification of new systems:}
    \begin{itemize}
        \item Four M-L systems were detected, three of which are new. These systems are valuable for improving the determination of L subdwarf metallicities using their M-type companions.
        \item No low-mass or colder-than-L-type companions were identified among the 219 M and L subdwarfs studied, likely due to detection limitations in \textit{Gaia} DR2.
        \item One confirmed binary system was proposed, along with four probable pairs and one uncertain pair.
    \end{itemize}
    
    \item \textit{Spectroscopic follow-up:}
    \begin{itemize}
        \item Spectral classification was obtained for three candidate companions through dedicated spectroscopic follow-up observations.
    \end{itemize}
        
    \item\textit{Comparison with solar-metallicity M dwarfs:}
    \begin{itemize}
        \item The binary fraction of metal-poor M dwarfs appears significantly lower than that of solar-metallicity M dwarfs within similar separation ranges.
    \end{itemize}
    
\end{itemize}
These findings contribute to the understanding of how metallicity influences the binary fraction in low-mass systems and improve metallicity estimates by leveraging the properties of the more massive component in binary systems. Compared to previous studies, this work introduces several key distinctions:
\begin{itemize}
    \item \textit{Focus on metallicity:} Unlike most multiplicity studies, which focus on main-sequence stars or nearby M dwarfs, this research specifically investigates a low-metallicity population. This approach provides insights into how varying chemical compositions affect star and multiple system formation. While other studies have also examined metal-poor stars, this work presents unique results for a well-defined sample of M and L subdwarfs.
    \item \textit{Emphasis on wide separations:} The study focuses on the search for wide companions using \textit{Gaia} DR2, with projected separations predominantly exceeding 1360 au. This contrasts with previous research that has examined a broader range of separations, including closer systems.
    \item \textit{Acknowledgment of detection limits:} The study highlights \textit{Gaia} DR2’s limitations in detecting very low-mass companions, particularly those cooler than spectral type L. This limitation should be considered when comparing results with studies using deeper imaging techniques, such AO.
\end{itemize}
These results refine our understanding of low-metallicity multiple systems and highlight the need for deeper observational studies to uncover faint companions and further characterise subdwarf populations. In summary, this chapter is important because it provides new statistics on the multiplicity of a specific stellar population composed of M and L subdwarfs, highlighting the possible influence of metallicity and pointing out the limitations of current observations.

In Chapter~\ref{ch:widest_binaries}, a detailed analysis of the widest pairs contained in the WDS catalogue, using \textit{Gaia} DR3 data, enabled the validation of their physical association and the identification of additional companions. This study revealed a significant prevalence of high-multiplicity systems in the ultra-wide regime and explored the boundary between wide binaries and members of young kinematic groups. The most important findings were:
\begin{itemize}
    \item \textit{Identification and classification of ultra-wide systems:} 
    \begin{itemize}
        \item Out of 155\,159 pairs in the WDS catalogue, 504 were identified with angular separations $\rho >$ 1000 arcsec.
        \item After a deep analysis using \textit{Gaia} DR3, 94 ultra-wide pairs were confirmed as gravitationally bound in 94 galactic field systems.
        \item 59 pairs (38.6 $\pm$ 9.8\%) were associated with young kinematic groups, open clusters, or stellar associations and were excluded from the field system analysis.
    \end{itemize}
    \item \textit{Discovery of new companions:}
    \begin{itemize}
        \item A total of 79 new stars were found within these 94 ultra-wide systems: 39 were previously reported in the WDS at separations lower than 1000 arcsec, and 40 were uncatalogued in the WDS, including ultra-cool dwarfs at the M-L boundary and a hot white dwarf.
        \item Several dozen candidate close binaries were identified based on high \texttt{RUWE} and $\sigma_{Rv}$ values from \textit{Gaia} DR3.
    \end{itemize}
    \item \textit{Multiplicity and system structure:} 
    \begin{itemize}
        \item The study identified 14 quadruple, 2 quintuple, 3 sextuple, and 2 septuple systems, highlighting a high fraction of hierarchical and non-hierarchical configurations.
        \item A trend toward higher multiplicity was observed in ultra-wide systems, with a relative decrease in purely binary systems when accounting for close binary candidates.
        \item When considering these candidate close binaries, 67.8\% of ultra-wide systems were classified as multiple.
    \end{itemize}
    \item \textit{Fragility and disruption of ultra-wide systems:}
    \begin{itemize}
        \item The minimum calculated binding energies align with theoretical predictions for tidal disruption by the galactic gravitational potential.
        \item The most fragile and massive systems exhibit projected physical separations larger than 1\,pc, suggesting they may be in the process of disruption or belong to yet-unidentified young kinematic groups.
        \item Seven candidate systems were found with separations of between 1.1 $\cdot$ 10$^\text{6}$ and 2.3 $\cdot$ 10$^\text{6}$\,au (5.1--11.1\,pc), possibly remnants of dissolving kinematic groups.
        \item Systems with very low-mass components and binding energies near 10$^{\text{33}}$\,J may be more relevant for studying galactic potential disruption than ultra-wide systems with extremely large separations.
    \end{itemize}
\end{itemize}
This study represents a major advancement in understanding ultra-wide binary systems ($\rho >$ 1000 arcsec) by the use of a larger sample and higher-precision astrometric data from \textit{Gaia} DR3. Compared to previous research, it introduces several key findings:
\begin{itemize}
    \item \textit{Expansion of sample size:} The study analyses a significantly larger sample of ultra-wide pairs than previous works, enabling more robust statistical insights into these systems.
    \item \textit{Clarification of binary and kinematic group membership:} It addresses confusion in the literature between physically bound ultra-wide pairs and unbound components in young kinematic groups, stellar associations, and clusters with similar galactocentric velocities. This distinction is crucial for determining the true binary fraction at large separations.
    \item \textit{Discovery of extremely fragile systems:} Three highly fragile systems, first reported by \citet{kirkpatrick16}, were confirmed, consisting of mid-to-late M dwarfs with projected separations of 0.33--0.41\,pc and extremely low binding energies ($|U_g^*| \lessapprox \text{10}^{\text{33}}$\,J). A lower limit of $\text{10}^{\text{33}}$\,J is suggested for the most fragile systems, likely due to tidal disruption by the galactic gravitational potential.
    \item \textit{Confirmation of the abundance of high-order multiple systems:} The study reinforces the idea that many ultra-wide systems contain multiple subsystems, leading to an overabundance of high-order multiples (triples, quadruples, quintuples, etc.). These higher-order systems tend to have greater total mass and binding energy than systems of similar separations with fewer components.
    \item \textit{Observational evidence for galactic disruption:} This work provides observational confirmation of classic theoretical predictions \citep[e.g.][]{weinberg87} regarding the disruption of ultra-wide binaries by the galactic gravitational potential. The findings suggest that binary systems with total masses $<$10\,M$_\odot$ are disrupted within 600--700 Myr, comparable to the age of the Hyades cluster.
\end{itemize}
All of these results refine our understanding of wide binary formation, stability, and disruption, emphasising the role of the galactic gravitational field in shaping the structure of multiple-star systems.

Chapter~\ref{ch:systems_with_planets} provides critical insights into the impact of stellar multiplicity on exoplanetary systems by analysing a large sample of exoplanet host stars within 100\,pc. Using \textit{Gaia} DR3 data, complemented by the WDS catalogue and previous literature, the study confirms a strong correlation between planetary eccentricity and stellar separation. It also identifies new stellar companions, updates the estimated multiplicity fraction of exoplanet host stars, and refines our understanding of planetary system architectures in multiple-star environments. Among its key contributions:
\begin{itemize}
    \item \textit{Identification of systems:}
    \begin{itemize}
        \item A total of 215 exoplanet-host stars were identified in 212 multiple stellar systems within 100\,pc. Among these, 173 are binaries, 39 are triples, and 3 are quadruples, showing consistency with general field star multiplicity rates. Three binary systems have planets detected around both stars.
        \item  Most triple systems have a hierarchical configuration (AB--C or A--BC).
        \item Up of 17 new stellar companions in 15 known exoplanet host systems were discovered (we report one new companion in 13 systems and two new companions in two systems). These companions either belong to newly identified systems or are additions to previously catalogued ones. These discoverings expand the sample of known multiple systems with exoplanets. 
        \item The reduced binding energy, |$U^*_g$|, was calculated for the newly identified companions, indicating that most are gravitationally bound, even in very wide systems.
    \end{itemize}
    
    \item \textit{Planetary eccentricities and stellar separation relationship:} 
    \begin{itemize}
        \item It was confirmed that planets in multiple systems tend to have significantly higher orbital eccentricities than those orbiting single stars, particularly in systems with small stellar separations relative to the planetary orbit size (low $s/a$ ratio).
        \item Significant confirmation ($>$4$\sigma$) of a correlation between planetary eccentricity and stellar separation, suggesting that nearby stellar companions influence planetary orbit eccentricity.
    \end{itemize}
    
    \item \textit{Multiplicity fraction of exoplanet host stars:} 
    \begin{itemize}
        \item A multiplicity fraction of 21.6 $\pm$ 2.9\% was determined for exoplanet-hosting systems. This value is slightly lower than the overall field star multiplicity fraction, which is attributed to an observational bias in exoplanet searches that tend to exclude or fail to detect close binary systems.
        \item The investigation of the multiplicity of stars hosting planets within 100\,pc quantified the influence of stellar companions on planetary systems. The identification and characterisation of new stellar companions, along with a statistical analysis of planetary orbital properties in single and multiple systems, revealed a significant correlation between stellar separation relative to planetary orbit size and planetary eccentricity. 
    \end{itemize}

    \item \textit{Statistical analysis:} 
    \begin{itemize}
        \item No statistically significant difference was found in the average number of planets between single and multiple stellar systems.
        \item A tendency was observed for high-mass planets ($M >$ 40\,M$_\oplus$) to be relatively more frequent in close orbits ($P <$ 10\, $d$) in multiple systems compared to single stars, although this difference was only marginally significant.
        \item The spectral type distribution was analysed, showing that most host stars are of spectral types G and K, while most companions are M dwarfs, with a notable proportion of ultra-cool L and T dwarfs compared to previous studies.
    \end{itemize}
    
\end{itemize}
These findings refine previous studies by leveraging a significantly larger and more precise dataset. Comparing to previous research, these are the key findings provided in this study:

\begin{itemize}
    \item \textit{Expanded and updated sample:} The study examines 215 exoplanet host stars in 212 multiple-star systems, alongside a comparison group of 687 single exoplanet host stars. This dataset is considerably larger than those used in previous studies.
    \item \textit{Completeness:} \textit{Gaia} DR3’s astrometric precision enables the detection of new stellar companions, improving sample completeness.
    \item \textit{Identification of new companions:} The mentioned 17 unknown stellar companions that were discovered, has increased the number of known multiple-star systems with exoplanets. This expands the dataset for studying how stellar companions affect planetary system architecture.
    \item \textit{Statistical analysis of planetary properties:} The study rigorously analyses how exoplanet characteristics differ between multiple and single-star systems. It finds that high-mass planets in multiple-star systems tend to have closer orbits than their counterparts in single-star systems.
    \item \textit{Planetary eccentricity and stellar separation:} A strong correlation ($>$ 4$\sigma$) is confirmed between planetary eccentricity and the ratio of stellar separation to planetary semi-major axis. This suggests that nearby stellar companions influence planetary orbits, refining prior studies that reported inconsistent results.
    \item \textit{Multiplicity fraction of exoplanetary systems:} The fraction of multiple-star systems among exoplanet hosts is consistent with past estimates. However, the study suggests that the actual fraction could be higher due to detection biases, especially for late-type stars.
    \item \textit{Hierarchy of planetary systems:} The study identifies 43 multiple systems (double and triple systems) with exoplanets. No higher-order systems (quintuples or beyond) were found.
\end{itemize}

The last work, described in Chapter~\ref{ch:characterisation_10pc} is a comprehensive characterisation of multiple stellar systems within 10\,pc of the Sun. A detailed catalogue has been compiled, precisely evaluating their physical properties, including separations, estimated masses, and orbital periods. The main results of this sudy are:

\begin{itemize}
    \item \textit{Identification of all multiple systems:}
    \begin{itemize}
        \item A total of 217 stars and brown dwarfs were identified in 93 multiple stellar systems within a 10 pc radius. These systems are distributed as follows: 69 binaries, 19 triples, 3 quadruples, and 2 quintuples.
        \item Among these systems, only 10 contain at least one unresolved pair.
    \end{itemize}
    \item \textit{Multiplicity:}
    \begin{itemize}
        \item The multiplicity fraction (MFM) and companion star fraction (CSFM) were precisely measured for systems with M-dwarf primaries, yielding values of 20.8$^{+\text{5.2}}_{-\text{4.4}}$\,\% and 26.6$^{+\text{5.5}}_{-\text{4.9}}$\,\%, respectively, excluding systems with white dwarfs as companions.
        \item The analysis of the MFM and the CSF for the solar neighborhood, with a special focus on M dwarfs, has provided valuable statistical insights into the local stellar population.
    \end{itemize}
    \item \textit{Systems properties:}
    \begin{itemize}
        \item Orbital periods of multiple systems span ten orders of magnitude, ranging from approximately one day to millions of years, and can be fitted with a log-normal cumulative distribution function across the entire range.
        \item An evaluation of the masses of the stars and brown dwarfs that make up the identified multiple systems has been conducted. It is emphasized that this sample is volume-complete ($d <$\,10\,pc) and covers the mass range of both primary and companion stars, from approximately 2\,M$_\odot$ down to the deuterium-burning mass limit (around 0.01\,M$_\odot$). This has allowed, for the first time, an investigation of multiplicity across a complete mass range.
    \end{itemize}
\end{itemize}

This analysis was conducted using data from \textit{Gaia} DR3 and the WDS catalogue, along with an extensive literature review. and these are the innovations compared to previous studies:

\begin{itemize}
    \item \textit{Update of last studies:} This study focuses on the immediate solar neighborhood (within 10 pc) to obtain complete and precise data for reliable stellar multiplicity statistics in relation to the last performed studies. An exhaustive compilation of WDS data and literature was carried out for all known multiple systems across all separation ranges, supplemented with a search for common proper motion and parallax using \textit{Gaia} DR3 data.

    \item \textit{Identification of unsolved sources:} Criteria were applied to identify unresolved sources based on \textit{Gaia} DR3 statistical indicators, such as the renormalized unit weight error (\texttt{RUWE}).

    \item \textit{Accuracy in multiplicity estimation:} The accuracy in measuring the MFM and CSFM for M dwarfs is considered unprecedented due to the comprehensiveness of the analysis and the most complete volume-limited sample near the Sun.

    \item \textit{Ongoing analysis:} Some differences in the classification of multiple systems were found compared to previous studies by \citet{reyle22} and \citet{kirkpatrick24}, highlighting the need for ongoing analyses.
    
\end{itemize}

Taken together, these conclusions highlight the complexity and diversity of multiple stellar systems in our Galaxy, demonstrating the unprecedented power of \textit{Gaia} DR3 data in advancing this field. The research also underscores the interconnectedness between stellar multiplicity and planetary system properties while recovering a forgotten chapter in the history of astronomy.

\section{Future work}
\label{sec:future_work}

The research lines initiated in this thesis open numerous promising avenues for future work, with the potential to yield new discoveries and further deepen our understanding of multiple stellar systems. Some anticipated future research directions include:

\subsection{Future work about Hodierna's investigations}
\label{sec:future_work_hodierna}
\begin{itemize}
    \item \textit{Deeper historical analysis of lesser known astronomers' work:} While the article provides an initial analysis of the twelve star pairs identified by Hodierna, a more exhaustive investigation into the historical and scientific context of his observations could be conducted. But not only for the Sicilian astronomer, but for many astronomers whose work probably has not been deeply studied, such as Al-Zarqali (1029--1087), Bhaskara II (1114--1185), Levi ben Gerson (1288--1344), Georg von Peuerbach (1423--1461), or Christopher Clavius (1538--1612), among others.  This may include a detailed examination of their observational methods with rudimentary instruments, the influence of other astronomers of their time, and the reasons why their work was forgotten.
    \item \textit{Comparison of Hodierna’s methods with those of Mayer and Herschel:} The article highlights that Hodierna published the first catalogue of binary stars more than a century before Mayer and Herschel. A comparative study of the methodologies used by these three astronomers to identify and catalogue double stars could be highly insightful. This could reveal the evolution of observational techniques and the changing understanding of the nature of double stars over time.
    \item \textit{Exploration of other possible discoveries by Hodierna:} The article focused on the double star section of Hodierna’s book. Future research could analyse other parts of his work \textit{De systemate orbis cometici} to identify other possible astronomical discoveries that may have gone unnoticed.
    \item \textit{Impact of this rediscovery on the history of astronomy:} Confirming Hodierna as the first cataloguer of double stars redefines the early history of this field in astronomy. A future study could explore the implications of this finding for the historical narrative of astronomy and how it influences our understanding of the development of binary star astrophysics.
    \item \textit{Deeper investigation about the not identified companion:} Try to figure out which is the undetected companion. There are reasons to think it could be an astronomer's misidentification, detecting a glow inside the instrument, a distant nova or supernova, an asteroid, or a dwarf planet. It is certain that Ceres passed through that region of space on a date that could have been close to the observation (E. Mamajek, priv. comm.).
\end{itemize}

\subsection{Future work for M-L subdwarf companions}
\label{sec:future_work_M-L_subdwarfs}

\begin{itemize}
    \item \textit{Spectroscopic confirmation of candidate systems:} The study identified several probable binary systems. The study explicitly mentions the need for spectroscopic confirmation of some of these systems to ensure they are truly gravitationally bound.
    \item \textit{Search for lower-mass companions:} The current study was limited in depth to detect lower-mass companions. The search for these lower-mass companions can be performed using future deep multi-epoch surveys such as the \textit{Vera Rubin} Large Synoptic Survey Telescope (LSST) and infrared observations from upcoming space missions like \textit{Euclid} \citep{laureijs11,amiaux12,mellier16a} or the Wide Field Infrared Survey Telescope \citep[WFIRST,][]{spergel15}. These surveys will enable a more complete detection of companions, including brown dwarfs and planetary-mass objects.
    \item \textit{Further characterisation of identified systems:} For confirmed systems, a more detailed characterisation is required. This could include obtaining high-resolution spectroscopy to more precisely determine spectral types, metallicities, and radial velocities of the components. Additionally, a deeper analysis of light curves could be conducted to search for variability, which might indicate the presence of additional companions or stellar activity.
    \item \textit{Investigating the impact of metallicity on binary fraction:} The main goal of the project was to study the effect of metallicity on the binary fraction of low-mass metal-poor systems. With a larger and better-characterised sample of M and L subdwarf binary systems, more precise multiplicity statistics can be obtained, leading to a better understanding of how metallicity influences the formation and evolution of these systems.
    \item \textit{Improving metallicity and distance determinations:} The study aimed to improve the metallicity determination of M-L subdwarfs based on their more massive primaries. Confirming more binary systems and refining the characterisation of their components will enhance the relationships between the binary components' properties, ultimately leading to improved metallicity and distance estimates for these types of stars.
    \item \textit{Analysis of gravitational binding energy:} The work calculated the binding energy of candidate systems. Future studies could refine these calculations by considering possible stellar orbits and searching for hidden mass within the systems (e.g., through the detection of spectroscopic binaries) to gain a more complete picture of their gravitational stability.
    \item \textit{Testing theoretical models:} The obtained multiplicity rates can be used to compare with theoretical models of binary star and multiple system formation in low-metallicity environments. The authors point out that their findings differ from the predictions of hydrodynamic simulations that consider smaller separations.
    \item \textit{Exploration of planetary formation:} Although this work does not focus on exoplanets, the identified multiple subdwarf systems could be targets for future planet searches. Understanding the presence and properties of planets around low-mass, low-metallicity stars in binary systems could provide valuable insights into planetary formation processes in different stellar environments.
\end{itemize}

\subsection{Future work for ultrawide binaries ($\rho >$ 1000 arcsec)}
\label{sec:future_work_ultrawide_binaries}

\begin{itemize}
    \item \textit{Distinction between wide-separation binaries and SKGs:} A dedicated study is planned to distinguish genuine very wide binaries from unbound components in young Stellar Kinematic Groups (SKGs) that share similar galactocentric space velocities. This would involve developing methodologies to differentiate gravitationally bound systems from stellar associations that move together but are not gravitationally linked.
    \item \textit{Analysis of multiple systems:} Since the individual components of very wide systems are often themselves multiple systems, future research could focus on a deeper analysis of the hierarchy of these higher-order systems. This might include studying the multiplicity distribution within wide systems and how it affects their stability and properties.
    \item \textit{Disruption of binary systems:} The study observationally confirms classical theoretical predictions regarding the disruption of binary systems by the galactic gravitational potential, particularly for the widest systems with total masses below 10\,M$_\odot$. This could lead to future research exploring in greater detail the mechanisms and timescales of such disruption, possibly using N-body simulations with even more precise \textit{Gaia} data, such as DR4 or DR5.
    \item \textit{More detailed characterisation of new discovered companions:} The discovered 40 astrometric companions not yet catalogued by the WDS, including several near the M-L boundary and a hot white dwarf open the door to detailed characterisations of these new companions, including obtaining spectra to determine their spectral types and masses more precisely, as well as investigating their possible orbital properties.
    \item \textit{Deeper study of fragile systems:} The discovered three extremely fragile systems composed of mid-to-late M dwarfs with large projected physical separations and small reduced binding energies could be subject to long-term observational follow-up to attempt to detect orbital motion and better understand the stability limits of binary systems with such low binding energies.
    \item \textit{Further detection of weaker companions:} Since the search for companions with \textit{Gaia} DR3 was limited by its magnitude threshold, future research could focus on searching for lower-mass companions (brown dwarfs and substellar objects) or companions at greater distances using deeper survey data or high-contrast imaging techniques such as AO.
    \item \textit{Wide binaries census completeness:} The study notes that there may be wide stellar companions (with spectral types earlier than M8\,V) not catalogued by \textit{Gaia} in dense stellar regions at low galactic latitudes. A targeted search in these regions using other datasets or methodologies could be conducted to complete the census of wide binaries.
    \item \textit{More detailed study of $\gamma$ Cas:} The discovery of a group of nine stars around $\gamma$ Cas and HD 5408 suggests the need for further investigation into the possible physical association of these stars. This could include using radial velocities and a more detailed analysis of proper motions to confirm whether they form an open cluster or a stellar association to understand their formation history.
    \item \textit{Galactic disruption mechanisms new models:} Based on the equations presented for estimating the lifetime and maximum separation of stellar systems, a more in-depth theoretical investigation could be conducted on the statistical distribution of wide binary separations in the context of galactic disruption mechanisms, using the new observational data provided in the article to refine the models.
\end{itemize}

\subsection{Future work about multiple systems with planets in the solar neighbourhood}
\label{sec:future_work_multiple_systems_planets_solar_neighbourhood}

\begin{itemize}
    \item \textit{Larger samples studies:} Since some of the results from the statistical analysis (such as the comparison of the average number of planets and the tendency of high-mass planets to be in closer orbits in multiple systems) only reached a statistical significance of 2$\sigma$, the need for studies with larger samples to confirm these trends could be an interesting future work.
    \item \textit{Continuation of studies on multiplicity in stellar systems with planets:} The article emphasises the importance of continued research on stellar multiplicity in systems with planets, highlighting previous work aimed at confirming and characterising controversial planetary systems. In this regard, the analysis of data from the future \textit{Gaia} DR4 catalogue is presented as a crucial tool for advancing this field.
    \item \textit{Further detection of weaker companions:} The same as proposed in the former study, since the search for companions in this study was limited by \textit{Gaia}'s completeness at faint magnitudes, deeper searches for lower-mass companions and in dense stellar regions should be performed using high-contrast imaging techniques.
    \item \textit{Detailed characterisation of nearby stars:} The discussion on defining the boundary between ``wide'' and ``close'' multiple systems in the context of future missions like HWO and LIFE suggests the need for a detailed characterisation of the immediate environment (up to one arcsec) of nearby stars. This could be achieved through high-resolution spectroscopy and deep imaging to inform target lists for these missions.
    \item \textit{Understanding the stability of wide and fragile systems:} For the 17 newly discovered genuine companions in this work, reduced binding energies were calculated. A long-term observational follow-up to determine the orbital parameters of these systems and better understand the stability of wide and fragile systems would be a natural research direction.
    \item \textit{Search for all spectral types:} Although the study compares the multiplicity function of exoplanetary systems with previous works, it acknowledges the incompleteness of all searches for the latest spectral types. Future studies aimed at obtaining a more complete census of companions, especially ultra-cool dwarfs, would be highly valuable.
    \item \textit{Deeper study about hierarchical systems:} The analysis of the hierarchy of multiple systems with exoplanets revealed triple and quadruple proportions consistent with those of field stars. Further investigation into the formation and evolution of these hierarchical systems hosting planets could be pursued.
    \item \textit{Detailed analysis of HD 1466 system:} The work already mentions that a detailed analysis of the HD 1466 system and its companions in the Tucana-Horologium association will be the subject of future work. 
    \item \textit{Study of multiplicity of stars inside dense areas:} Since the study excluded systems belonging to open clusters and stellar associations, future research could specifically focus on the multiplicity of stars with planets within these dense environments. This would allow for a comparison of the multiplicity rate and planetary system properties across different star-forming environments.
    \item \textit{Analysis of UCD companions:} The study also excluded systems with UCDs discovered through direct imaging. A dedicated study analysing the influence of UCD companions on planetary systems, considering their separations, masses, and potential dynamical interactions, would be an interesting research direction.
    \item \textit{Improving the statistical significance of results:} As more precise and complete \textit{Gaia} data become available (such as \textit{Gaia} DR4), the statistical results of the article could be revisited and refined, including the search for new companions and the improved characterisation of known systems. This could confirm or refute the observed trends with greater statistical significance.
    \item \textit{Planetary formation simulations:} Theoretical simulations of planetary formation and evolution could be carried out for the specific types of multiple systems found in the sample, to better understand how the presence of one or more companion stars affects planetary accretion and migration processes.
    \item \textit{Comparison with systems without planets:} Comparing the multiplicity function of stars with planets to that of field stars without detected planets in the same solar neighborhood would be valuable. This could help determine whether stellar multiplicity has a significant impact on the presence or detection of exoplanets.
    \item \textit{New companions follow-up:} Given the discovery of 17 new companions, a detailed photometric and spectroscopic follow-up of these systems could be performed to refine their spectral types, estimate their masses more precisely, and search for possible signs of activity or variability that may be related to the presence of a planet in the primary system.
\end{itemize}

\subsection{Future work about multiple systems within 10\,pc}
\label{sec:future_work_10pc}

\begin{itemize}
    \item \textit{Further observations:} New observations are needed to improve the characterisation of companions in some understudied pairs, including very high-resolution imaging to resolve binaries such as Ross 619 [AB], and extreme adaptive optics to measure the magnitudes of closely bound components in systems like $\xi$ UMa, G 184--19, and EZ Aqr.

    \item \textit{Deeper study of single stars:} Confirmation of the unique nature of single stars within 10\,pc that have not yet been monitored with high-resolution imaging and spectroscopy is proposed. The confirmation of hypothetical single stars that have not yet been studied in-depth for multiplicity will be the subject of future research.

    \item \textit{Properties accuracy improving:} Additional investigation is required for systems containing ultra-cool dwarfs, white dwarfs, and exoplanets as companions to improve the accuracy of magnitude differences, spectral types, and orbital parameters.

    \item \textit{Gaia DR4 use:} The upcoming release of \textit{Gaia} DR4 in the second half of 2026 is expected to enhance our understanding of the closest multiple systems. It will offer improved astrometric precision and a longer time baseline, enabling more accurate orbital solutions. This, in turn, will allow the detection of subtle perturbations indicative of additional, unseen companions. Furthermore, enhanced photometric and spectroscopic data will aid in characterising the component stars in terms of their masses, luminosities, and evolutionary states.

    \item \textit{Search for habitable planets:} A global concerted effort is emphasized to characterise these systems in detail beyond \textit{Gaia}’s capabilities, particularly because some reported systems are potential targets for the Habitable Worlds Observatory (HWO) and the Large Interferometer For Exoplanets (LIFE) missions, contributing to the search for habitable exoplanets.

\end{itemize}

\subsection{Future work summary}
\label{sec:future_work_summary}

This thesis has unveiled numerous opportunities for future research in multiple stellar systems. Expanding historical investigations, particularly on overlooked astronomers, could provide new insights into early observational techniques. Spectroscopic and astrometric follow-ups will be crucial for confirming and characterising new binary and multiple systems. The upcoming \textit{Gaia} DR4, LSST, and space missions like \textit{Euclid} will enhance the detection of faint companions and improve metallicity and distance estimations. Refining theoretical models on binary formation and galactic disruption will provide a deeper understanding of wide binaries. Investigating planetary formation in multiple systems will help assess their stability and evolution. Additionally, targeted searches in dense stellar regions may reveal new low-mass companions. Continued observational efforts will be key to completing the census of multiple systems, understanding their impact on exoplanetary systems, contributing to the search for habitable exoplanets, and refining stellar system models.

\newpage

%%%%% BIBLIOGRAFIA %%%%%

\thispagestyle{plain}
\vspace{9cm}
\lhead[\small{\textbf{\thepage}}]{\textbf{Bibliography}}
\rhead[\small{\textbf{Bibliography}}]{\small{\textbf{\thepage}}}
\phantomsection %Para conseguir salto correcto desde el indice
\addcontentsline{toc}{chapter}{Bibliography}
\bibliography{biblio}

@ARTICLE{abbot16,
       author = {{Abbott}, B.~P. and {Abbott}, R. and {Abbott}, T.~D. and {Abernathy}, M.~R. and {Acernese}, F. and {Ackley}, K. and {Adams}, C. and {Adams}, T. and {Addesso}, P. and {Adhikari}, R.~X. and {Adya}, V.~B. and {Affeldt}, C. and {Agathos}, M. and {Agatsuma}, K. and {Aggarwal}, N. and {Aguiar}, O.~D. and {Aiello}, L. and {Ain}, A. and {Ajith}, P. and {Allen}, B. and {Allocca}, A. and {Altin}, P.~A. and {Anderson}, S.~B. and {Anderson}, W.~G. and {Arai}, K. and {Arain}, M.~A. and {Araya}, M.~C. and {Arceneaux}, C.~C. and {Areeda}, J.~S. and {Arnaud}, N. and {Arun}, K.~G. and {Ascenzi}, S. and {Ashton}, G. and {Ast}, M. and {Aston}, S.~M. and {Astone}, P. and {Aufmuth}, P. and {Aulbert}, C. and {Babak}, S. and {Bacon}, P. and {Bader}, M.~K.~M. and {Baker}, P.~T. and {Baldaccini}, F. and {Ballardin}, G. and {Ballmer}, S.~W. and {Barayoga}, J.~C. and {Barclay}, S.~E. and {Barish}, B.~C. and {Barker}, D. and {Barone}, F. and {Barr}, B. and {Barsotti}, L. and {Barsuglia}, M. and {Barta}, D. and {Bartlett}, J. and {Barton}, M.~A. and {Bartos}, I. and {Bassiri}, R. and {Basti}, A. and {Batch}, J.~C. and {Baune}, C. and {Bavigadda}, V. and {Bazzan}, M. and {Behnke}, B. and {Bejger}, M. and {Belczynski}, C. and {Bell}, A.~S. and {Bell}, C.~J. and {Berger}, B.~K. and {Bergman}, J. and {Bergmann}, G. and {Berry}, C.~P.~L. and {Bersanetti}, D. and {Bertolini}, A. and {Betzwieser}, J. and {Bhagwat}, S. and {Bhandare}, R. and {Bilenko}, I.~A. and {Billingsley}, G. and {Birch}, J. and {Birney}, R. and {Birnholtz}, O. and {Biscans}, S. and {Bisht}, A. and {Bitossi}, M. and {Biwer}, C. and {Bizouard}, M.~A. and {Blackburn}, J.~K. and {Blair}, C.~D. and {Blair}, D.~G. and {Blair}, R.~M. and {Bloemen}, S. and {Bock}, O. and {Bodiya}, T.~P. and {Boer}, M. and {Bogaert}, G. and {Bogan}, C. and {Bohe}, A. and {Bojtos}, P. and {Bond}, C. and {Bondu}, F. and {Bonnand}, R. and {Boom}, B.~A. and {Bork}, R. and {Boschi}, V. and {Bose}, S. and {Bouffanais}, Y. and {Bozzi}, A. and {Bradaschia}, C. and {Brady}, P.~R. and {Braginsky}, V.~B. and {Branchesi}, M. and {Brau}, J.~E. and {Briant}, T. and {Brillet}, A. and {Brinkmann}, M. and {Brisson}, V. and {Brockill}, P. and {Brooks}, A.~F. and {Brown}, D.~A. and {Brown}, D.~D. and {Brown}, N.~M. and {Buchanan}, C.~C. and {Buikema}, A. and {Bulik}, T. and {Bulten}, H.~J. and {Buonanno}, A. and {Buskulic}, D. and {Buy}, C. and {Byer}, R.~L. and {Cabero}, M. and {Cadonati}, L. and {Cagnoli}, G. and {Cahillane}, C. and {Bustillo}, J. Calder{\'o}n and {Callister}, T. and {Calloni}, E. and {Camp}, J.~B. and {Cannon}, K.~C. and {Cao}, J. and {Capano}, C.~D. and {Capocasa}, E. and {Carbognani}, F. and {Caride}, S. and {Casanueva Diaz}, J. and {Casentini}, C. and {Caudill}, S. and {Cavagli{\`a}}, M. and {Cavalier}, F. and {Cavalieri}, R. and {Cella}, G. and {Cepeda}, C.~B. and {Baiardi}, L. Cerboni and {Cerretani}, G. and {Cesarini}, E. and {Chakraborty}, R. and {Chalermsongsak}, T. and {Chamberlin}, S.~J. and {Chan}, M. and {Chao}, S. and {Charlton}, P. and {Chassande-Mottin}, E. and {Chen}, H.~Y. and {Chen}, Y. and {Cheng}, C. and {Chincarini}, A. and {Chiummo}, A. and {Cho}, H.~S. and {Cho}, M. and {Chow}, J.~H. and {Christensen}, N. and {Chu}, Q. and {Chua}, S. and {Chung}, S. and {Ciani}, G. and {Clara}, F. and {Clark}, J.~A. and {Cleva}, F. and {Coccia}, E. and {Cohadon}, P. -F. and {Colla}, A. and {Collette}, C.~G. and {Cominsky}, L. and {Constancio}, M. and {Conte}, A. and {Conti}, L. and {Cook}, D. and {Corbitt}, T.~R. and {Cornish}, N. and {Corsi}, A. and {Cortese}, S. and {Costa}, C.~A. and {Coughlin}, M.~W. and {Coughlin}, S.~B. and {Coulon}, J. -P. and {Countryman}, S.~T. and {Couvares}, P. and {Cowan}, E.~E. and {Coward}, D.~M. and {Cowart}, M.~J. and {Coyne}, D.~C. and {Coyne}, R. and {Craig}, K. and {Creighton}, J.~D.~E. and {Creighton}, T.~D. and {Cripe}, J. and {Crowder}, S.~G. and {Cruise}, A.~M. and {Cumming}, A. and {Cunningham}, L. and {Cuoco}, E. and {Dal Canton}, T. and {Danilishin}, S.~L. and {D'Antonio}, S. and {Danzmann}, K. and {Darman}, N.~S. and {Da Silva Costa}, C.~F. and {Dattilo}, V. and {Dave}, I. and {Daveloza}, H.~P. and {Davier}, M. and {Davies}, G.~S. and {Daw}, E.~J. and {Day}, R. and {De}, S. and {DeBra}, D. and {Debreczeni}, G. and {Degallaix}, J. and {De Laurentis}, M. and {Del{\'e}glise}, S. and {Del Pozzo}, W. and {Denker}, T. and {Dent}, T. and {Dereli}, H. and {Dergachev}, V. and {DeRosa}, R.~T. and {De Rosa}, R. and {DeSalvo}, R. and {Dhurandhar}, S. and {D{\'\i}az}, M.~C. and {Di Fiore}, L. and {Di Giovanni}, M. and {Di Lieto}, A. and {Di Pace}, S. and {Di Palma}, I. and {Di Virgilio}, A. and {Dojcinoski}, G. and {Dolique}, V. and {Donovan}, F. and {Dooley}, K.~L. and {Doravari}, S. and {Douglas}, R. and {Downes}, T.~P. and {Drago}, M. and {Drever}, R.~W.~P. and {Driggers}, J.~C. and {Du}, Z. and {Ducrot}, M. and {Dwyer}, S.~E. and {Edo}, T.~B. and {Edwards}, M.~C. and {Effler}, A. and {Eggenstein}, H. -B. and {Ehrens}, P. and {Eichholz}, J. and {Eikenberry}, S.~S. and {Engels}, W. and {Essick}, R.~C. and {Etzel}, T. and {Evans}, M. and {Evans}, T.~M. and {Everett}, R. and {Factourovich}, M. and {Fafone}, V. and {Fair}, H. and {Fairhurst}, S. and {Fan}, X. and {Fang}, Q. and {Farinon}, S. and {Farr}, B. and {Farr}, W.~M. and {Favata}, M. and {Fays}, M. and {Fehrmann}, H. and {Fejer}, M.~M. and {Feldbaum}, D. and {Ferrante}, I. and {Ferreira}, E.~C. and {Ferrini}, F. and {Fidecaro}, F. and {Finn}, L.~S. and {Fiori}, I. and {Fiorucci}, D. and {Fisher}, R.~P. and {Flaminio}, R. and {Fletcher}, M. and {Fong}, H. and {Fournier}, J. -D. and {Franco}, S. and {Frasca}, S. and {Frasconi}, F. and {Frede}, M. and {Frei}, Z. and {Freise}, A. and {Frey}, R. and {Frey}, V. and {Fricke}, T.~T. and {Fritschel}, P. and {Frolov}, V.~V. and {Fulda}, P. and {Fyffe}, M. and {Gabbard}, H.~A.~G. and {Gair}, J.~R. and {Gammaitoni}, L. and {Gaonkar}, S.~G. and {Garufi}, F. and {Gatto}, A. and {Gaur}, G. and {Gehrels}, N. and {Gemme}, G. and {Gendre}, B. and {Genin}, E. and {Gennai}, A. and {George}, J. and {Gergely}, L. and {Germain}, V. and {Ghosh}, Abhirup and {Ghosh}, Archisman and {Ghosh}, S. and {Giaime}, J.~A. and {Giardina}, K.~D. and {Giazotto}, A. and {Gill}, K. and {Glaefke}, A. and {Gleason}, J.~R. and {Goetz}, E. and {Goetz}, R. and {Gondan}, L. and {Gonz{\'a}lez}, G. and {Castro}, J.~M. Gonzalez and {Gopakumar}, A. and {Gordon}, N.~A. and {Gorodetsky}, M.~L. and {Gossan}, S.~E. and {Gosselin}, M. and {Gouaty}, R. and {Graef}, C. and {Graff}, P.~B. and {Granata}, M. and {Grant}, A. and {Gras}, S. and {Gray}, C. and {Greco}, G. and {Green}, A.~C. and {Greenhalgh}, R.~J.~S. and {Groot}, P. and {Grote}, H. and {Grunewald}, S. and {Guidi}, G.~M. and {Guo}, X. and {Gupta}, A. and {Gupta}, M.~K. and {Gushwa}, K.~E. and {Gustafson}, E.~K. and {Gustafson}, R. and {Hacker}, J.~J. and {Hall}, B.~R. and {Hall}, E.~D. and {Hammond}, G. and {Haney}, M. and {Hanke}, M.~M. and {Hanks}, J. and {Hanna}, C. and {Hannam}, M.~D. and {Hanson}, J. and {Hardwick}, T. and {Harms}, J. and {Harry}, G.~M. and {Harry}, I.~W. and {Hart}, M.~J. and {Hartman}, M.~T. and {Haster}, C. -J. and {Haughian}, K. and {Healy}, J. and {Heefner}, J. and {Heidmann}, A. and {Heintze}, M.~C. and {Heinzel}, G. and {Heitmann}, H. and {Hello}, P. and {Hemming}, G. and {Hendry}, M. and {Heng}, I.~S. and {Hennig}, J. and {Heptonstall}, A.~W. and {Heurs}, M. and {Hild}, S. and {Hoak}, D. and {Hodge}, K.~A. and {Hofman}, D. and {Hollitt}, S.~E. and {Holt}, K. and {Holz}, D.~E. and {Hopkins}, P. and {Hosken}, D.~J. and {Hough}, J. and {Houston}, E.~A. and {Howell}, E.~J. and {Hu}, Y.~M. and {Huang}, S. and {Huerta}, E.~A. and {Huet}, D. and {Hughey}, B. and {Husa}, S. and {Huttner}, S.~H. and {Huynh-Dinh}, T. and {Idrisy}, A. and {Indik}, N. and {Ingram}, D.~R. and {Inta}, R. and {Isa}, H.~N. and {Isac}, J. -M. and {Isi}, M. and {Islas}, G. and {Isogai}, T. and {Iyer}, B.~R. and {Izumi}, K. and {Jacobson}, M.~B. and {Jacqmin}, T. and {Jang}, H. and {Jani}, K. and {Jaranowski}, P. and {Jawahar}, S. and {Jim{\'e}nez-Forteza}, F. and {Johnson}, W.~W. and {Johnson-McDaniel}, N.~K. and {Jones}, D.~I. and {Jones}, R. and {Jonker}, R.~J.~G. and {Ju}, L. and {Haris}, K. and {Kalaghatgi}, C.~V. and {Kalogera}, V. and {Kandhasamy}, S. and {Kang}, G. and {Kanner}, J.~B. and {Karki}, S. and {Kasprzack}, M. and {Katsavounidis}, E. and {Katzman}, W. and {Kaufer}, S. and {Kaur}, T. and {Kawabe}, K. and {Kawazoe}, F. and {K{\'e}f{\'e}lian}, F. and {Kehl}, M.~S. and {Keitel}, D. and {Kelley}, D.~B. and {Kells}, W. and {Kennedy}, R. and {Keppel}, D.~G. and {Key}, J.~S. and {Khalaidovski}, A. and {Khalili}, F.~Y. and {Khan}, I. and {Khan}, S. and {Khan}, Z. and {Khazanov}, E.~A. and {Kijbunchoo}, N. and {Kim}, C. and {Kim}, J. and {Kim}, K. and {Kim}, Nam-Gyu and {Kim}, Namjun and {Kim}, Y. -M. and {King}, E.~J. and {King}, P.~J. and {Kinzel}, D.~L. and {Kissel}, J.~S. and {Kleybolte}, L. and {Klimenko}, S. and {Koehlenbeck}, S.~M. and {Kokeyama}, K. and {Koley}, S. and {Kondrashov}, V. and {Kontos}, A. and {Koranda}, S. and {Korobko}, M. and {Korth}, W.~Z. and {Kowalska}, I. and {Kozak}, D.~B. and {Kringel}, V. and {Krishnan}, B. and {Kr{\'o}lak}, A. and {Krueger}, C. and {Kuehn}, G. and {Kumar}, P. and {Kumar}, R. and {Kuo}, L. and {Kutynia}, A. and {Kwee}, P. and {Lackey}, B.~D. and {Landry}, M. and {Lange}, J. and {Lantz}, B. and {Lasky}, P.~D. and {Lazzarini}, A. and {Lazzaro}, C. and {Leaci}, P. and {Leavey}, S. and {Lebigot}, E.~O. and {Lee}, C.~H. and {Lee}, H.~K. and {Lee}, H.~M. and {Lee}, K. and {Lenon}, A. and {Leonardi}, M. and {Leong}, J.~R. and {Leroy}, N. and {Letendre}, N. and {Levin}, Y. and {Levine}, B.~M. and {Li}, T.~G.~F. and {Libson}, A. and {Littenberg}, T.~B. and {Lockerbie}, N.~A. and {Logue}, J. and {Lombardi}, A.~L. and {London}, L.~T. and {Lord}, J.~E. and {Lorenzini}, M. and {Loriette}, V. and {Lormand}, M. and {Losurdo}, G. and {Lough}, J.~D. and {Lousto}, C.~O. and {Lovelace}, G. and {L{\"u}ck}, H. and {Lundgren}, A.~P. and {Luo}, J. and {Lynch}, R. and {Ma}, Y. and {MacDonald}, T. and {Machenschalk}, B. and {MacInnis}, M. and {Macleod}, D.~M. and {Maga{\~n}a-Sandoval}, F. and {Magee}, R.~M. and {Mageswaran}, M. and {Majorana}, E. and {Maksimovic}, I. and {Malvezzi}, V. and {Man}, N. and {Mandel}, I. and {Mandic}, V. and {Mangano}, V. and {Mansell}, G.~L. and {Manske}, M. and {Mantovani}, M. and {Marchesoni}, F. and {Marion}, F. and {M{\'a}rka}, S. and {M{\'a}rka}, Z. and {Markosyan}, A.~S. and {Maros}, E. and {Martelli}, F. and {Martellini}, L. and {Martin}, I.~W. and {Martin}, R.~M. and {Martynov}, D.~V. and {Marx}, J.~N. and {Mason}, K. and {Masserot}, A. and {Massinger}, T.~J. and {Masso-Reid}, M. and {Matichard}, F. and {Matone}, L. and {Mavalvala}, N. and {Mazumder}, N. and {Mazzolo}, G. and {McCarthy}, R. and {McClelland}, D.~E. and {McCormick}, S. and {McGuire}, S.~C. and {McIntyre}, G. and {McIver}, J. and {McManus}, D.~J. and {McWilliams}, S.~T. and {Meacher}, D. and {Meadors}, G.~D. and {Meidam}, J. and {Melatos}, A. and {Mendell}, G. and {Mendoza-Gandara}, D. and {Mercer}, R.~A. and {Merilh}, E. and {Merzougui}, M. and {Meshkov}, S. and {Messenger}, C. and {Messick}, C. and {Meyers}, P.~M. and {Mezzani}, F. and {Miao}, H. and {Michel}, C. and {Middleton}, H. and {Mikhailov}, E.~E. and {Milano}, L. and {Miller}, J. and {Millhouse}, M. and {Minenkov}, Y. and {Ming}, J. and {Mirshekari}, S. and {Mishra}, C. and {Mitra}, S. and {Mitrofanov}, V.~P. and {Mitselmakher}, G. and {Mittleman}, R. and {Moggi}, A. and {Mohan}, M. and {Mohapatra}, S.~R.~P. and {Montani}, M. and {Moore}, B.~C. and {Moore}, C.~J. and {Moraru}, D. and {Moreno}, G. and {Morriss}, S.~R. and {Mossavi}, K. and {Mours}, B. and {Mow-Lowry}, C.~M. and {Mueller}, C.~L. and {Mueller}, G. and {Muir}, A.~W. and {Mukherjee}, Arunava and {Mukherjee}, D. and {Mukherjee}, S. and {Mukund}, N. and {Mullavey}, A. and {Munch}, J. and {Murphy}, D.~J. and {Murray}, P.~G. and {Mytidis}, A. and {Nardecchia}, I. and {Naticchioni}, L. and {Nayak}, R.~K. and {Necula}, V. and {Nedkova}, K. and {Nelemans}, G. and {Neri}, M. and {Neunzert}, A. and {Newton}, G. and {Nguyen}, T.~T. and {Nielsen}, A.~B. and {Nissanke}, S. and {Nitz}, A. and {Nocera}, F. and {Nolting}, D. and {Normandin}, M.~E.~N. and {Nuttall}, L.~K. and {Oberling}, J. and {Ochsner}, E. and {O'Dell}, J. and {Oelker}, E. and {Ogin}, G.~H. and {Oh}, J.~J. and {Oh}, S.~H. and {Ohme}, F. and {Oliver}, M. and {Oppermann}, P. and {Oram}, Richard J. and {O'Reilly}, B. and {O'Shaughnessy}, R. and {Ott}, C.~D. and {Ottaway}, D.~J. and {Ottens}, R.~S. and {Overmier}, H. and {Owen}, B.~J. and {Pai}, A. and {Pai}, S.~A. and {Palamos}, J.~R. and {Palashov}, O. and {Palomba}, C. and {Pal-Singh}, A. and {Pan}, H. and {Pan}, Y. and {Pankow}, C. and {Pannarale}, F. and {Pant}, B.~C. and {Paoletti}, F. and {Paoli}, A. and {Papa}, M.~A. and {Paris}, H.~R. and {Parker}, W. and {Pascucci}, D. and {Pasqualetti}, A. and {Passaquieti}, R. and {Passuello}, D. and {Patricelli}, B. and {Patrick}, Z. and {Pearlstone}, B.~L. and {Pedraza}, M. and {Pedurand}, R. and {Pekowsky}, L. and {Pele}, A. and {Penn}, S. and {Perreca}, A. and {Pfeiffer}, H.~P. and {Phelps}, M. and {Piccinni}, O. and {Pichot}, M. and {Pickenpack}, M. and {Piergiovanni}, F. and {Pierro}, V. and {Pillant}, G. and {Pinard}, L. and {Pinto}, I.~M. and {Pitkin}, M. and {Poeld}, J.~H. and {Poggiani}, R. and {Popolizio}, P. and {Post}, A. and {Powell}, J. and {Prasad}, J. and {Predoi}, V. and {Premachandra}, S.~S. and {Prestegard}, T. and {Price}, L.~R. and {Prijatelj}, M. and {Principe}, M. and {Privitera}, S. and {Prix}, R. and {Prodi}, G.~A. and {Prokhorov}, L. and {Puncken}, O. and {Punturo}, M. and {Puppo}, P. and {P{\"u}rrer}, M. and {Qi}, H. and {Qin}, J. and {Quetschke}, V. and {Quintero}, E.~A. and {Quitzow-James}, R. and {Raab}, F.~J. and {Rabeling}, D.~S. and {Radkins}, H. and {Raffai}, P. and {Raja}, S. and {Rakhmanov}, M. and {Ramet}, C.~R. and {Rapagnani}, P. and {Raymond}, V. and {Razzano}, M. and {Re}, V. and {Read}, J. and {Reed}, C.~M. and {Regimbau}, T. and {Rei}, L. and {Reid}, S. and {Reitze}, D.~H. and {Rew}, H. and {Reyes}, S.~D. and {Ricci}, F. and {Riles}, K. and {Robertson}, N.~A. and {Robie}, R. and {Robinet}, F. and {Rocchi}, A. and {Rolland}, L. and {Rollins}, J.~G. and {Roma}, V.~J. and {Romano}, J.~D. and {Romano}, R. and {Romanov}, G. and {Romie}, J.~H. and {Rosi{\'n}ska}, D. and {Rowan}, S. and {R{\"u}diger}, A. and {Ruggi}, P. and {Ryan}, K. and {Sachdev}, S. and {Sadecki}, T. and {Sadeghian}, L. and {Salconi}, L. and {Saleem}, M. and {Salemi}, F. and {Samajdar}, A. and {Sammut}, L. and {Sampson}, L.~M. and {Sanchez}, E.~J. and {Sandberg}, V. and {Sandeen}, B. and {Sanders}, G.~H. and {Sanders}, J.~R. and {Sassolas}, B. and {Sathyaprakash}, B.~S. and {Saulson}, P.~R. and {Sauter}, O. and {Savage}, R.~L. and {Sawadsky}, A. and {Schale}, P. and {Schilling}, R. and {Schmidt}, J. and {Schmidt}, P. and {Schnabel}, R. and {Schofield}, R.~M.~S. and {Sch{\"o}nbeck}, A. and {Schreiber}, E. and {Schuette}, D. and {Schutz}, B.~F. and {Scott}, J. and {Scott}, S.~M. and {Sellers}, D. and {Sengupta}, A.~S. and {Sentenac}, D. and {Sequino}, V. and {Sergeev}, A. and {Serna}, G. and {Setyawati}, Y. and {Sevigny}, A. and {Shaddock}, D.~A. and {Shaffer}, T. and {Shah}, S. and {Shahriar}, M.~S. and {Shaltev}, M. and {Shao}, Z. and {Shapiro}, B. and {Shawhan}, P. and {Sheperd}, A. and {Shoemaker}, D.~H. and {Shoemaker}, D.~M. and {Siellez}, K. and {Siemens}, X. and {Sigg}, D. and {Silva}, A.~D. and {Simakov}, D. and {Singer}, A. and {Singer}, L.~P. and {Singh}, A. and {Singh}, R. and {Singhal}, A. and {Sintes}, A.~M. and {Slagmolen}, B.~J.~J. and {Smith}, J.~R. and {Smith}, M.~R. and {Smith}, N.~D. and {Smith}, R.~J.~E. and {Son}, E.~J. and {Sorazu}, B. and {Sorrentino}, F. and {Souradeep}, T. and {Srivastava}, A.~K. and {Staley}, A. and {Steinke}, M. and {Steinlechner}, J. and {Steinlechner}, S. and {Steinmeyer}, D. and {Stephens}, B.~C. and {Stevenson}, S.~P. and {Stone}, R. and {Strain}, K.~A. and {Straniero}, N. and {Stratta}, G. and {Strauss}, N.~A. and {Strigin}, S. and {Sturani}, R. and {Stuver}, A.~L. and {Summerscales}, T.~Z. and {Sun}, L. and {Sutton}, P.~J. and {Swinkels}, B.~L. and {Szczepa{\'n}czyk}, M.~J. and {Tacca}, M. and {Talukder}, D. and {Tanner}, D.~B. and {T{\'a}pai}, M. and {Tarabrin}, S.~P. and {Taracchini}, A. and {Taylor}, R. and {Theeg}, T. and {Thirugnanasambandam}, M.~P. and {Thomas}, E.~G. and {Thomas}, M. and {Thomas}, P. and {Thorne}, K.~A. and {Thorne}, K.~S. and {Thrane}, E. and {Tiwari}, S. and {Tiwari}, V. and {Tokmakov}, K.~V. and {Tomlinson}, C. and {Tonelli}, M. and {Torres}, C.~V. and {Torrie}, C.~I. and {T{\"o}yr{\"a}}, D. and {Travasso}, F. and {Traylor}, G. and {Trifir{\`o}}, D. and {Tringali}, M.~C. and {Trozzo}, L. and {Tse}, M. and {Turconi}, M. and {Tuyenbayev}, D. and {Ugolini}, D. and {Unnikrishnan}, C.~S. and {Urban}, A.~L. and {Usman}, S.~A. and {Vahlbruch}, H. and {Vajente}, G. and {Valdes}, G. and {Vallisneri}, M. and {van Bakel}, N. and {van Beuzekom}, M. and {van den Brand}, J.~F.~J. and {Van Den Broeck}, C. and {Vander-Hyde}, D.~C. and {van der Schaaf}, L. and {van Heijningen}, J.~V. and {van Veggel}, A.~A. and {Vardaro}, M. and {Vass}, S. and {Vas{\'u}th}, M. and {Vaulin}, R. and {Vecchio}, A. and {Vedovato}, G. and {Veitch}, J. and {Veitch}, P.~J. and {Venkateswara}, K. and {Verkindt}, D. and {Vetrano}, F. and {Vicer{\'e}}, A. and {Vinciguerra}, S. and {Vine}, D.~J. and {Vinet}, J. -Y. and {Vitale}, S. and {Vo}, T. and {Vocca}, H. and {Vorvick}, C. and {Voss}, D. and {Vousden}, W.~D. and {Vyatchanin}, S.~P. and {Wade}, A.~R. and {Wade}, L.~E. and {Wade}, M. and {Waldman}, S.~J. and {Walker}, M. and {Wallace}, L. and {Walsh}, S. and {Wang}, G. and {Wang}, H. and {Wang}, M. and {Wang}, X. and {Wang}, Y. and {Ward}, H. and {Ward}, R.~L. and {Warner}, J. and {Was}, M. and {Weaver}, B. and {Wei}, L. -W. and {Weinert}, M. and {Weinstein}, A.~J. and {Weiss}, R. and {Welborn}, T. and {Wen}, L. and {We{\ss}els}, P. and {Westphal}, T. and {Wette}, K. and {Whelan}, J.~T. and {Whitcomb}, S.~E. and {White}, D.~J. and {Whiting}, B.~F. and {Wiesner}, K. and {Wilkinson}, C. and {Willems}, P.~A. and {Williams}, L. and {Williams}, R.~D. and {Williamson}, A.~R. and {Willis}, J.~L. and {Willke}, B. and {Wimmer}, M.~H. and {Winkelmann}, L. and {Winkler}, W. and {Wipf}, C.~C. and {Wiseman}, A.~G. and {Wittel}, H. and {Woan}, G. and {Worden}, J. and {Wright}, J.~L. and {Wu}, G. and {Yablon}, J. and {Yakushin}, I. and {Yam}, W. and {Yamamoto}, H. and {Yancey}, C.~C. and {Yap}, M.~J. and {Yu}, H. and {Yvert}, M. and {Zadro{\.Z}ny}, A. and {Zangrando}, L. and {Zanolin}, M. and {Zendri}, J. -P. and {Zevin}, M. and {Zhang}, F. and {Zhang}, L. and {Zhang}, M. and {Zhang}, Y. and {Zhao}, C. and {Zhou}, M. and {Zhou}, Z. and {Zhu}, X.~J. and {Zucker}, M.~E. and {Zuraw}, S.~E. and {Zweizig}, J. and {LIGO Scientific Collaboration} and {Virgo Collaboration}},
        title = "{Observation of Gravitational Waves from a Binary Black Hole Merger}",
      journal = {PRL},
         year = 2016,
        month = feb,
       volume = {116},
       number = {6},
          eid = {061102},
        pages = {061102},
          doi = {10.1103/PhysRevLett.116.061102},
archivePrefix = {arXiv},
       eprint = {1602.03837},
 primaryClass = {gr-qc},
       adsurl = {https://ui.adsabs.harvard.edu/abs/2016PhRvL.116f1102A}
}

@ARTICLE{abbot17,
       author = {{Abbott}, B.~P. and {Abbott}, R. and {Abbott}, T.~D. and {Acernese}, F. and {Ackley}, K. and {Adams}, C. and {Adams}, T. and {Addesso}, P. and {Adhikari}, R.~X. and {Adya}, V.~B. and {Affeldt}, C. and {Afrough}, M. and {Agarwal}, B. and {Agathos}, M. and {Agatsuma}, K. and {Aggarwal}, N. and {Aguiar}, O.~D. and {Aiello}, L. and {Ain}, A. and {Ajith}, P. and {Allen}, B. and {Allen}, G. and {Allocca}, A. and {Altin}, P.~A. and {Amato}, A. and {Ananyeva}, A. and {Anderson}, S.~B. and {Anderson}, W.~G. and {Angelova}, S.~V. and {Antier}, S. and {Appert}, S. and {Arai}, K. and {Araya}, M.~C. and {Areeda}, J.~S. and {Arnaud}, N. and {Arun}, K.~G. and {Ascenzi}, S. and {Ashton}, G. and {Ast}, M. and {Aston}, S.~M. and {Astone}, P. and {Atallah}, D.~V. and {Aufmuth}, P. and {Aulbert}, C. and {AultONeal}, K. and {Austin}, C. and {Avila-Alvarez}, A. and {Babak}, S. and {Bacon}, P. and {Bader}, M.~K.~M. and {Bae}, S. and {Bailes}, M. and {Baker}, P.~T. and {Baldaccini}, F. and {Ballardin}, G. and {Ballmer}, S.~W. and {Banagiri}, S. and {Barayoga}, J.~C. and {Barclay}, S.~E. and {Barish}, B.~C. and {Barker}, D. and {Barkett}, K. and {Barone}, F. and {Barr}, B. and {Barsotti}, L. and {Barsuglia}, M. and {Barta}, D. and {Barthelmy}, S.~D. and {Bartlett}, J. and {Bartos}, I. and {Bassiri}, R. and {Basti}, A. and {Batch}, J.~C. and {Bawaj}, M. and {Bayley}, J.~C. and {Bazzan}, M. and {B{\'e}csy}, B. and {Beer}, C. and {Bejger}, M. and {Belahcene}, I. and {Bell}, A.~S. and {Berger}, B.~K. and {Bergmann}, G. and {Bernuzzi}, S. and {Bero}, J.~J. and {Berry}, C.~P.~L. and {Bersanetti}, D. and {Bertolini}, A. and {Betzwieser}, J. and {Bhagwat}, S. and {Bhandare}, R. and {Bilenko}, I.~A. and {Billingsley}, G. and {Billman}, C.~R. and {Birch}, J. and {Birney}, R. and {Birnholtz}, O. and {Biscans}, S. and {Biscoveanu}, S. and {Bisht}, A. and {Bitossi}, M. and {Biwer}, C. and {Bizouard}, M.~A. and {Blackburn}, J.~K. and {Blackman}, J. and {Blair}, C.~D. and {Blair}, D.~G. and {Blair}, R.~M. and {Bloemen}, S. and {Bock}, O. and {Bode}, N. and {Boer}, M. and {Bogaert}, G. and {Bohe}, A. and {Bondu}, F. and {Bonilla}, E. and {Bonnand}, R. and {Boom}, B.~A. and {Bork}, R. and {Boschi}, V. and {Bose}, S. and {Bossie}, K. and {Bouffanais}, Y. and {Bozzi}, A. and {Bradaschia}, C. and {Brady}, P.~R. and {Branchesi}, M. and {Brau}, J.~E. and {Briant}, T. and {Brillet}, A. and {Brinkmann}, M. and {Brisson}, V. and {Brockill}, P. and {Broida}, J.~E. and {Brooks}, A.~F. and {Brown}, D.~A. and {Brown}, D.~D. and {Brunett}, S. and {Buchanan}, C.~C. and {Buikema}, A. and {Bulik}, T. and {Bulten}, H.~J. and {Buonanno}, A. and {Buskulic}, D. and {Buy}, C. and {Byer}, R.~L. and {Cabero}, M. and {Cadonati}, L. and {Cagnoli}, G. and {Cahillane}, C. and {Calder{\'o}n Bustillo}, J. and {Callister}, T.~A. and {Calloni}, E. and {Camp}, J.~B. and {Canepa}, M. and {Canizares}, P. and {Cannon}, K.~C. and {Cao}, H. and {Cao}, J. and {Capano}, C.~D. and {Capocasa}, E. and {Carbognani}, F. and {Caride}, S. and {Carney}, M.~F. and {Carullo}, G. and {Casanueva Diaz}, J. and {Casentini}, C. and {Caudill}, S. and {Cavagli{\`a}}, M. and {Cavalier}, F. and {Cavalieri}, R. and {Cella}, G. and {Cepeda}, C.~B. and {Cerd{\'a}-Dur{\'a}n}, P. and {Cerretani}, G. and {Cesarini}, E. and {Chamberlin}, S.~J. and {Chan}, M. and {Chao}, S. and {Charlton}, P. and {Chase}, E. and {Chassande-Mottin}, E. and {Chatterjee}, D. and {Chatziioannou}, K. and {Cheeseboro}, B.~D. and {Chen}, H.~Y. and {Chen}, X. and {Chen}, Y. and {Cheng}, H. -P. and {Chia}, H. and {Chincarini}, A. and {Chiummo}, A. and {Chmiel}, T. and {Cho}, H.~S. and {Cho}, M. and {Chow}, J.~H. and {Christensen}, N. and {Chu}, Q. and {Chua}, A.~J.~K. and {Chua}, S. and {Chung}, A.~K.~W. and {Chung}, S. and {Ciani}, G. and {Ciolfi}, R. and {Cirelli}, C.~E. and {Cirone}, A. and {Clara}, F. and {Clark}, J.~A. and {Clearwater}, P. and {Cleva}, F. and {Cocchieri}, C. and {Coccia}, E. and {Cohadon}, P. -F. and {Cohen}, D. and {Colla}, A. and {Collette}, C.~G. and {Cominsky}, L.~R. and {Constancio}, M. and {Conti}, L. and {Cooper}, S.~J. and {Corban}, P. and {Corbitt}, T.~R. and {Cordero-Carri{\'o}n}, I. and {Corley}, K.~R. and {Cornish}, N. and {Corsi}, A. and {Cortese}, S. and {Costa}, C.~A. and {Coughlin}, M.~W. and {Coughlin}, S.~B. and {Coulon}, J. -P. and {Countryman}, S.~T. and {Couvares}, P. and {Covas}, P.~B. and {Cowan}, E.~E. and {Coward}, D.~M. and {Cowart}, M.~J. and {Coyne}, D.~C. and {Coyne}, R. and {Creighton}, J.~D.~E. and {Creighton}, T.~D. and {Cripe}, J. and {Crowder}, S.~G. and {Cullen}, T.~J. and {Cumming}, A. and {Cunningham}, L. and {Cuoco}, E. and {Dal Canton}, T. and {D{\'a}lya}, G. and {Danilishin}, S.~L. and {D'Antonio}, S. and {Danzmann}, K. and {Dasgupta}, A. and {Da Silva Costa}, C.~F. and {Dattilo}, V. and {Dave}, I. and {Davier}, M. and {Davis}, D. and {Daw}, E.~J. and {Day}, B. and {De}, S. and {DeBra}, D. and {Degallaix}, J. and {De Laurentis}, M. and {Del{\'e}glise}, S. and {Del Pozzo}, W. and {Demos}, N. and {Denker}, T. and {Dent}, T. and {De Pietri}, R. and {Dergachev}, V. and {De Rosa}, R. and {DeRosa}, R.~T. and {De Rossi}, C. and {DeSalvo}, R. and {de Varona}, O. and {Devenson}, J. and {Dhurandhar}, S. and {D{\'\i}az}, M.~C. and {Dietrich}, T. and {Di Fiore}, L. and {Di Giovanni}, M. and {Di Girolamo}, T. and {Di Lieto}, A. and {Di Pace}, S. and {Di Palma}, I. and {Di Renzo}, F. and {Doctor}, Z. and {Dolique}, V. and {Donovan}, F. and {Dooley}, K.~L. and {Doravari}, S. and {Dorrington}, I. and {Douglas}, R. and {Dovale {\'A}lvarez}, M. and {Downes}, T.~P. and {Drago}, M. and {Dreissigacker}, C. and {Driggers}, J.~C. and {Du}, Z. and {Ducrot}, M. and {Dudi}, R. and {Dupej}, P. and {Dwyer}, S.~E. and {Edo}, T.~B. and {Edwards}, M.~C. and {Effler}, A. and {Eggenstein}, H. -B. and {Ehrens}, P. and {Eichholz}, J. and {Eikenberry}, S.~S. and {Eisenstein}, R.~A. and {Essick}, R.~C. and {Estevez}, D. and {Etienne}, Z.~B. and {Etzel}, T. and {Evans}, M. and {Evans}, T.~M. and {Factourovich}, M. and {Fafone}, V. and {Fair}, H. and {Fairhurst}, S. and {Fan}, X. and {Farinon}, S. and {Farr}, B. and {Farr}, W.~M. and {Fauchon-Jones}, E.~J. and {Favata}, M. and {Fays}, M. and {Fee}, C. and {Fehrmann}, H. and {Feicht}, J. and {Fejer}, M.~M. and {Fernandez-Galiana}, A. and {Ferrante}, I. and {Ferreira}, E.~C. and {Ferrini}, F. and {Fidecaro}, F. and {Finstad}, D. and {Fiori}, I. and {Fiorucci}, D. and {Fishbach}, M. and {Fisher}, R.~P. and {Fitz-Axen}, M. and {Flaminio}, R. and {Fletcher}, M. and {Fong}, H. and {Font}, J.~A. and {Forsyth}, P.~W.~F. and {Forsyth}, S.~S. and {Fournier}, J. -D. and {Frasca}, S. and {Frasconi}, F. and {Frei}, Z. and {Freise}, A. and {Frey}, R. and {Frey}, V. and {Fries}, E.~M. and {Fritschel}, P. and {Frolov}, V.~V. and {Fulda}, P. and {Fyffe}, M. and {Gabbard}, H. and {Gadre}, B.~U. and {Gaebel}, S.~M. and {Gair}, J.~R. and {Gammaitoni}, L. and {Ganija}, M.~R. and {Gaonkar}, S.~G. and {Garcia-Quiros}, C. and {Garufi}, F. and {Gateley}, B. and {Gaudio}, S. and {Gaur}, G. and {Gayathri}, V. and {Gehrels}, N. and {Gemme}, G. and {Genin}, E. and {Gennai}, A. and {George}, D. and {George}, J. and {Gergely}, L. and {Germain}, V. and {Ghonge}, S. and {Ghosh}, Abhirup and {Ghosh}, Archisman and {Ghosh}, S. and {Giaime}, J.~A. and {Giardina}, K.~D. and {Giazotto}, A. and {Gill}, K. and {Glover}, L. and {Goetz}, E. and {Goetz}, R. and {Gomes}, S. and {Goncharov}, B. and {Gonz{\'a}lez}, G. and {Gonzalez Castro}, J.~M. and {Gopakumar}, A. and {Gorodetsky}, M.~L. and {Gossan}, S.~E. and {Gosselin}, M. and {Gouaty}, R. and {Grado}, A. and {Graef}, C. and {Granata}, M. and {Grant}, A. and {Gras}, S. and {Gray}, C. and {Greco}, G. and {Green}, A.~C. and {Gretarsson}, E.~M. and {Groot}, P. and {Grote}, H. and {Grunewald}, S. and {Gruning}, P. and {Guidi}, G.~M. and {Guo}, X. and {Gupta}, A. and {Gupta}, M.~K. and {Gushwa}, K.~E. and {Gustafson}, E.~K. and {Gustafson}, R. and {Halim}, O. and {Hall}, B.~R. and {Hall}, E.~D. and {Hamilton}, E.~Z. and {Hammond}, G. and {Haney}, M. and {Hanke}, M.~M. and {Hanks}, J. and {Hanna}, C. and {Hannam}, M.~D. and {Hannuksela}, O.~A. and {Hanson}, J. and {Hardwick}, T. and {Harms}, J. and {Harry}, G.~M. and {Harry}, I.~W. and {Hart}, M.~J. and {Haster}, C. -J. and {Haughian}, K. and {Healy}, J. and {Heidmann}, A. and {Heintze}, M.~C. and {Heitmann}, H. and {Hello}, P. and {Hemming}, G. and {Hendry}, M. and {Heng}, I.~S. and {Hennig}, J. and {Heptonstall}, A.~W. and {Heurs}, M. and {Hild}, S. and {Hinderer}, T. and {Ho}, W.~C.~G. and {Hoak}, D. and {Hofman}, D. and {Holt}, K. and {Holz}, D.~E. and {Hopkins}, P. and {Horst}, C. and {Hough}, J. and {Houston}, E.~A. and {Howell}, E.~J. and {Hreibi}, A. and {Hu}, Y.~M. and {Huerta}, E.~A. and {Huet}, D. and {Hughey}, B. and {Husa}, S. and {Huttner}, S.~H. and {Huynh-Dinh}, T. and {Indik}, N. and {Inta}, R. and {Intini}, G. and {Isa}, H.~N. and {Isac}, J. -M. and {Isi}, M. and {Iyer}, B.~R. and {Izumi}, K. and {Jacqmin}, T. and {Jani}, K. and {Jaranowski}, P. and {Jawahar}, S. and {Jim{\'e}nez-Forteza}, F. and {Johnson}, W.~W. and {Johnson-McDaniel}, N.~K. and {Jones}, D.~I. and {Jones}, R. and {Jonker}, R.~J.~G. and {Ju}, L. and {Junker}, J. and {Kalaghatgi}, C.~V. and {Kalogera}, V. and {Kamai}, B. and {Kandhasamy}, S. and {Kang}, G. and {Kanner}, J.~B. and {Kapadia}, S.~J. and {Karki}, S. and {Karvinen}, K.~S. and {Kasprzack}, M. and {Kastaun}, W. and {Katolik}, M. and {Katsavounidis}, E. and {Katzman}, W. and {Kaufer}, S. and {Kawabe}, K. and {K{\'e}f{\'e}lian}, F. and {Keitel}, D. and {Kemball}, A.~J. and {Kennedy}, R. and {Kent}, C. and {Key}, J.~S. and {Khalili}, F.~Y. and {Khan}, I. and {Khan}, S. and {Khan}, Z. and {Khazanov}, E.~A. and {Kijbunchoo}, N. and {Kim}, Chunglee and {Kim}, J.~C. and {Kim}, K. and {Kim}, W. and {Kim}, W.~S. and {Kim}, Y. -M. and {Kimbrell}, S.~J. and {King}, E.~J. and {King}, P.~J. and {Kinley-Hanlon}, M. and {Kirchhoff}, R. and {Kissel}, J.~S. and {Kleybolte}, L. and {Klimenko}, S. and {Knowles}, T.~D. and {Koch}, P. and {Koehlenbeck}, S.~M. and {Koley}, S. and {Kondrashov}, V. and {Kontos}, A. and {Korobko}, M. and {Korth}, W.~Z. and {Kowalska}, I. and {Kozak}, D.~B. and {Kr{\"a}mer}, C. and {Kringel}, V. and {Krishnan}, B. and {Kr{\'o}lak}, A. and {Kuehn}, G. and {Kumar}, P. and {Kumar}, R. and {Kumar}, S. and {Kuo}, L. and {Kutynia}, A. and {Kwang}, S. and {Lackey}, B.~D. and {Lai}, K.~H. and {Landry}, M. and {Lang}, R.~N. and {Lange}, J. and {Lantz}, B. and {Lanza}, R.~K. and {Larson}, S.~L. and {Lartaux-Vollard}, A. and {Lasky}, P.~D. and {Laxen}, M. and {Lazzarini}, A. and {Lazzaro}, C. and {Leaci}, P. and {Leavey}, S. and {Lee}, C.~H. and {Lee}, H.~K. and {Lee}, H.~M. and {Lee}, H.~W. and {Lee}, K. and {Lehmann}, J. and {Lenon}, A. and {Leon}, E. and {Leonardi}, M. and {Leroy}, N. and {Letendre}, N. and {Levin}, Y. and {Li}, T.~G.~F. and {Linker}, S.~D. and {Littenberg}, T.~B. and {Liu}, J. and {Liu}, X. and {Lo}, R.~K.~L. and {Lockerbie}, N.~A. and {London}, L.~T. and {Lord}, J.~E. and {Lorenzini}, M. and {Loriette}, V. and {Lormand}, M. and {Losurdo}, G. and {Lough}, J.~D. and {Lousto}, C.~O. and {Lovelace}, G. and {L{\"u}ck}, H. and {Lumaca}, D. and {Lundgren}, A.~P. and {Lynch}, R. and {Ma}, Y. and {Macas}, R. and {Macfoy}, S. and {Machenschalk}, B. and {MacInnis}, M. and {Macleod}, D.~M. and {Maga{\~n}a Hernandez}, I. and {Maga{\~n}a-Sandoval}, F. and {Maga{\~n}a Zertuche}, L. and {Magee}, R.~M. and {Majorana}, E. and {Maksimovic}, I. and {Man}, N. and {Mandic}, V. and {Mangano}, V. and {Mansell}, G.~L. and {Manske}, M. and {Mantovani}, M. and {Marchesoni}, F. and {Marion}, F. and {M{\'a}rka}, S. and {M{\'a}rka}, Z. and {Markakis}, C. and {Markosyan}, A.~S. and {Markowitz}, A. and {Maros}, E. and {Marquina}, A. and {Marsh}, P. and {Martelli}, F. and {Martellini}, L. and {Martin}, I.~W. and {Martin}, R.~M. and {Martynov}, D.~V. and {Marx}, J.~N. and {Mason}, K. and {Massera}, E. and {Masserot}, A. and {Massinger}, T.~J. and {Masso-Reid}, M. and {Mastrogiovanni}, S. and {Matas}, A. and {Matichard}, F. and {Matone}, L. and {Mavalvala}, N. and {Mazumder}, N. and {McCarthy}, R. and {McClelland}, D.~E. and {McCormick}, S. and {McCuller}, L. and {McGuire}, S.~C. and {McIntyre}, G. and {McIver}, J. and {McManus}, D.~J. and {McNeill}, L. and {McRae}, T. and {McWilliams}, S.~T. and {Meacher}, D. and {Meadors}, G.~D. and {Mehmet}, M. and {Meidam}, J. and {Mejuto-Villa}, E. and {Melatos}, A. and {Mendell}, G. and {Mercer}, R.~A. and {Merilh}, E.~L. and {Merzougui}, M. and {Meshkov}, S. and {Messenger}, C. and {Messick}, C. and {Metzdorff}, R. and {Meyers}, P.~M. and {Miao}, H. and {Michel}, C. and {Middleton}, H. and {Mikhailov}, E.~E. and {Milano}, L. and {Miller}, A.~L. and {Miller}, B.~B. and {Miller}, J. and {Millhouse}, M. and {Milovich-Goff}, M.~C. and {Minazzoli}, O. and {Minenkov}, Y. and {Ming}, J. and {Mishra}, C. and {Mitra}, S. and {Mitrofanov}, V.~P. and {Mitselmakher}, G. and {Mittleman}, R. and {Moffa}, D. and {Moggi}, A. and {Mogushi}, K. and {Mohan}, M. and {Mohapatra}, S.~R.~P. and {Molina}, I. and {Montani}, M. and {Moore}, C.~J. and {Moraru}, D. and {Moreno}, G. and {Morisaki}, S. and {Morriss}, S.~R. and {Mours}, B. and {Mow-Lowry}, C.~M. and {Mueller}, G. and {Muir}, A.~W. and {Mukherjee}, Arunava and {Mukherjee}, D. and {Mukherjee}, S. and {Mukund}, N. and {Mullavey}, A. and {Munch}, J. and {Mu{\~n}iz}, E.~A. and {Muratore}, M. and {Murray}, P.~G. and {Nagar}, A. and {Napier}, K. and {Nardecchia}, I. and {Naticchioni}, L. and {Nayak}, R.~K. and {Neilson}, J. and {Nelemans}, G. and {Nelson}, T.~J.~N. and {Nery}, M. and {Neunzert}, A. and {Nevin}, L. and {Newport}, J.~M. and {Newton}, G. and {Ng}, K.~K.~Y. and {Nguyen}, P. and {Nguyen}, T.~T. and {Nichols}, D. and {Nielsen}, A.~B. and {Nissanke}, S. and {Nitz}, A. and {Noack}, A. and {Nocera}, F. and {Nolting}, D. and {North}, C. and {Nuttall}, L.~K. and {Oberling}, J. and {O'Dea}, G.~D. and {Ogin}, G.~H. and {Oh}, J.~J. and {Oh}, S.~H. and {Ohme}, F. and {Okada}, M.~A. and {Oliver}, M. and {Oppermann}, P. and {Oram}, Richard J. and {O'Reilly}, B. and {Ormiston}, R. and {Ortega}, L.~F. and {O'Shaughnessy}, R. and {Ossokine}, S. and {Ottaway}, D.~J. and {Overmier}, H. and {Owen}, B.~J. and {Pace}, A.~E. and {Page}, J. and {Page}, M.~A. and {Pai}, A. and {Pai}, S.~A. and {Palamos}, J.~R. and {Palashov}, O. and {Palomba}, C. and {Pal-Singh}, A. and {Pan}, Howard and {Pan}, Huang-Wei and {Pang}, B. and {Pang}, P.~T.~H. and {Pankow}, C. and {Pannarale}, F. and {Pant}, B.~C. and {Paoletti}, F. and {Paoli}, A. and {Papa}, M.~A. and {Parida}, A. and {Parker}, W. and {Pascucci}, D. and {Pasqualetti}, A. and {Passaquieti}, R. and {Passuello}, D. and {Patil}, M. and {Patricelli}, B. and {Pearlstone}, B.~L. and {Pedraza}, M. and {Pedurand}, R. and {Pekowsky}, L. and {Pele}, A. and {Penn}, S. and {Perez}, C.~J. and {Perreca}, A. and {Perri}, L.~M. and {Pfeiffer}, H.~P. and {Phelps}, M. and {Piccinni}, O.~J. and {Pichot}, M. and {Piergiovanni}, F. and {Pierro}, V. and {Pillant}, G. and {Pinard}, L. and {Pinto}, I.~M. and {Pirello}, M. and {Pitkin}, M. and {Poe}, M. and {Poggiani}, R. and {Popolizio}, P. and {Porter}, E.~K. and {Post}, A. and {Powell}, J. and {Prasad}, J. and {Pratt}, J.~W.~W. and {Pratten}, G. and {Predoi}, V. and {Prestegard}, T. and {Prijatelj}, M. and {Principe}, M. and {Privitera}, S. and {Prix}, R. and {Prodi}, G.~A. and {Prokhorov}, L.~G. and {Puncken}, O. and {Punturo}, M. and {Puppo}, P. and {P{\"u}rrer}, M. and {Qi}, H. and {Quetschke}, V. and {Quintero}, E.~A. and {Quitzow-James}, R. and {Raab}, F.~J. and {Rabeling}, D.~S. and {Radkins}, H. and {Raffai}, P. and {Raja}, S. and {Rajan}, C. and {Rajbhandari}, B. and {Rakhmanov}, M. and {Ramirez}, K.~E. and {Ramos-Buades}, A. and {Rapagnani}, P. and {Raymond}, V. and {Razzano}, M. and {Read}, J. and {Regimbau}, T. and {Rei}, L. and {Reid}, S. and {Reitze}, D.~H. and {Ren}, W. and {Reyes}, S.~D. and {Ricci}, F. and {Ricker}, P.~M. and {Rieger}, S. and {Riles}, K. and {Rizzo}, M. and {Robertson}, N.~A. and {Robie}, R. and {Robinet}, F. and {Rocchi}, A. and {Rolland}, L. and {Rollins}, J.~G. and {Roma}, V.~J. and {Romano}, J.~D. and {Romano}, R. and {Romel}, C.~L. and {Romie}, J.~H. and {Rosi{\'n}ska}, D. and {Ross}, M.~P. and {Rowan}, S. and {R{\"u}diger}, A. and {Ruggi}, P. and {Rutins}, G. and {Ryan}, K. and {Sachdev}, S. and {Sadecki}, T. and {Sadeghian}, L. and {Sakellariadou}, M. and {Salconi}, L. and {Saleem}, M. and {Salemi}, F. and {Samajdar}, A. and {Sammut}, L. and {Sampson}, L.~M. and {Sanchez}, E.~J. and {Sanchez}, L.~E. and {Sanchis-Gual}, N. and {Sandberg}, V. and {Sanders}, J.~R. and {Sassolas}, B. and {Sathyaprakash}, B.~S. and {Saulson}, P.~R. and {Sauter}, O. and {Savage}, R.~L. and {Sawadsky}, A. and {Schale}, P. and {Scheel}, M. and {Scheuer}, J. and {Schmidt}, J. and {Schmidt}, P. and {Schnabel}, R. and {Schofield}, R.~M.~S. and {Sch{\"o}nbeck}, A. and {Schreiber}, E. and {Schuette}, D. and {Schulte}, B.~W. and {Schutz}, B.~F. and {Schwalbe}, S.~G. and {Scott}, J. and {Scott}, S.~M. and {Seidel}, E. and {Sellers}, D. and {Sengupta}, A.~S. and {Sentenac}, D. and {Sequino}, V. and {Sergeev}, A. and {Shaddock}, D.~A. and {Shaffer}, T.~J. and {Shah}, A.~A. and {Shahriar}, M.~S. and {Shaner}, M.~B. and {Shao}, L. and {Shapiro}, B. and {Shawhan}, P. and {Sheperd}, A. and {Shoemaker}, D.~H. and {Shoemaker}, D.~M. and {Siellez}, K. and {Siemens}, X. and {Sieniawska}, M. and {Sigg}, D. and {Silva}, A.~D. and {Singer}, L.~P. and {Singh}, A. and {Singhal}, A. and {Sintes}, A.~M. and {Slagmolen}, B.~J.~J. and {Smith}, B. and {Smith}, J.~R. and {Smith}, R.~J.~E. and {Somala}, S. and {Son}, E.~J. and {Sonnenberg}, J.~A. and {Sorazu}, B. and {Sorrentino}, F. and {Souradeep}, T. and {Spencer}, A.~P. and {Srivastava}, A.~K. and {Staats}, K. and {Staley}, A. and {Steinke}, M. and {Steinlechner}, J. and {Steinlechner}, S. and {Steinmeyer}, D. and {Stevenson}, S.~P. and {Stone}, R. and {Stops}, D.~J. and {Strain}, K.~A. and {Stratta}, G. and {Strigin}, S.~E. and {Strunk}, A. and {Sturani}, R. and {Stuver}, A.~L. and {Summerscales}, T.~Z. and {Sun}, L. and {Sunil}, S. and {Suresh}, J. and {Sutton}, P.~J. and {Swinkels}, B.~L. and {Szczepa{\'n}czyk}, M.~J. and {Tacca}, M. and {Tait}, S.~C. and {Talbot}, C. and {Talukder}, D. and {Tanner}, D.~B. and {T{\'a}pai}, M. and {Taracchini}, A. and {Tasson}, J.~D. and {Taylor}, J.~A. and {Taylor}, R. and {Tewari}, S.~V. and {Theeg}, T. and {Thies}, F. and {Thomas}, E.~G. and {Thomas}, M. and {Thomas}, P. and {Thorne}, K.~A. and {Thorne}, K.~S. and {Thrane}, E. and {Tiwari}, S. and {Tiwari}, V. and {Tokmakov}, K.~V. and {Toland}, K. and {Tonelli}, M. and {Tornasi}, Z. and {Torres-Forn{\'e}}, A. and {Torrie}, C.~I. and {T{\"o}yr{\"a}}, D. and {Travasso}, F. and {Traylor}, G. and {Trinastic}, J. and {Tringali}, M.~C. and {Trozzo}, L. and {Tsang}, K.~W. and {Tse}, M. and {Tso}, R. and {Tsukada}, L. and {Tsuna}, D. and {Tuyenbayev}, D. and {Ueno}, K. and {Ugolini}, D. and {Unnikrishnan}, C.~S. and {Urban}, A.~L. and {Usman}, S.~A. and {Vahlbruch}, H. and {Vajente}, G. and {Valdes}, G. and {Vallisneri}, M. and {van Bakel}, N. and {van Beuzekom}, M. and {van den Brand}, J.~F.~J. and {Van Den Broeck}, C. and {Vander-Hyde}, D.~C. and {van der Schaaf}, L. and {van Heijningen}, J.~V. and {van Veggel}, A.~A. and {Vardaro}, M. and {Varma}, V. and {Vass}, S. and {Vas{\'u}th}, M. and {Vecchio}, A. and {Vedovato}, G. and {Veitch}, J. and {Veitch}, P.~J. and {Venkateswara}, K. and {Venugopalan}, G. and {Verkindt}, D. and {Vetrano}, F. and {Vicer{\'e}}, A. and {Viets}, A.~D. and {Vinciguerra}, S. and {Vine}, D.~J. and {Vinet}, J. -Y. and {Vitale}, S. and {Vo}, T. and {Vocca}, H. and {Vorvick}, C. and {Vyatchanin}, S.~P. and {Wade}, A.~R. and {Wade}, L.~E. and {Wade}, M. and {Walet}, R. and {Walker}, M. and {Wallace}, L. and {Walsh}, S. and {Wang}, G. and {Wang}, H. and {Wang}, J.~Z. and {Wang}, W.~H. and {Wang}, Y.~F. and {Ward}, R.~L. and {Warner}, J. and {Was}, M. and {Watchi}, J. and {Weaver}, B. and {Wei}, L. -W. and {Weinert}, M. and {Weinstein}, A.~J. and {Weiss}, R. and {Wen}, L. and {Wessel}, E.~K. and {We{\ss}els}, P. and {Westerweck}, J. and {Westphal}, T. and {Wette}, K. and {Whelan}, J.~T. and {Whitcomb}, S.~E. and {Whiting}, B.~F. and {Whittle}, C. and {Wilken}, D. and {Williams}, D. and {Williams}, R.~D. and {Williamson}, A.~R. and {Willis}, J.~L. and {Willke}, B. and {Wimmer}, M.~H. and {Winkler}, W. and {Wipf}, C.~C. and {Wittel}, H. and {Woan}, G. and {Woehler}, J. and {Wofford}, J. and {Wong}, K.~W.~K. and {Worden}, J. and {Wright}, J.~L. and {Wu}, D.~S. and {Wysocki}, D.~M. and {Xiao}, S. and {Yamamoto}, H. and {Yancey}, C.~C. and {Yang}, L. and {Yap}, M.~J. and {Yazback}, M. and {Yu}, Hang and {Yu}, Haocun and {Yvert}, M. and {Zadro{\.Z}ny}, A. and {Zanolin}, M. and {Zelenova}, T. and {Zendri}, J. -P. and {Zevin}, M. and {Zhang}, L. and {Zhang}, M. and {Zhang}, T. and {Zhang}, Y. -H. and {Zhao}, C. and {Zhou}, M. and {Zhou}, Z. and {Zhu}, S.~J. and {Zhu}, X.~J. and {Zimmerman}, A.~B. and {Zucker}, M.~E. and {Zweizig}, J. and {LIGO Scientific Collaboration} and {Virgo Collaboration}},
        title = "{GW170817: Observation of Gravitational Waves from a Binary Neutron Star Inspiral}",
      journal = {PRL},
         year = 2017,
        month = oct,
       volume = {119},
       number = {16},
          eid = {161101},
        pages = {161101},
          doi = {10.1103/PhysRevLett.119.161101},
archivePrefix = {arXiv},
       eprint = {1710.05832},
 primaryClass = {gr-qc},
       adsurl = {https://ui.adsabs.harvard.edu/abs/2017PhRvL.119p1101A}
}

@ARTICLE{abt76,
       author = {{Abt}, H.~A. and {Levy}, S.~G.},
        title = "{Multiplicity among solar-type stars.}",
      journal = {ApJS},
         year = 1976,
        month = mar,
       volume = {30},
        pages = {273-306},
          doi = {10.1086/190363},
       adsurl = {https://ui.adsabs.harvard.edu/abs/1976ApJS...30..273A}
}

@ARTICLE{abt88,
       author = {{Abt}, Helmut A.},
        title = "{Maximum Separations among Cataloged Binaries}",
      journal = {ApJ},
         year = 1988,
        month = aug,
       volume = {331},
        pages = {922},
          doi = {10.1086/166609},
       adsurl = {https://ui.adsabs.harvard.edu/abs/1988ApJ...331..922A}
}

@ARTICLE{abt90,
       author = {{Abt}, Helmut A. and {Gomez}, Ana E. and {Levy}, Saul G.},
        title = "{The Frequency and Formation Mechanism of B2--B5 Main-Sequence Binaries}",
      journal = {ApJS},
         year = 1990,
        month = oct,
       volume = {74},
        pages = {551},
          doi = {10.1086/191508},
       adsurl = {https://ui.adsabs.harvard.edu/abs/1990ApJS...74..551A}
}

@ARTICLE{adams05,
       author = {{Adams}, F.~C. and {Bodenheimer}, P. and {Laughlin}, G.},
        title = "{M dwarfs: planet formation and long term evolution}",
      journal = {Astron. Nachr.},
         year = 2005,
        month = dec,
       volume = {326},
       number = {10},
        pages = {913-919},
          doi = {10.1002/asna.200510440},
       adsurl = {https://ui.adsabs.harvard.edu/abs/2005AN....326..913A}
}

@INPROCEEDINGS{adelman04,
       author = {{Adelman}, Saul J.},
        title = "{The physical properties of normal A stars}",
    booktitle = {The A-Star Puzzle},
       series = {Proc. Int. Astron. Union},
         year = 2004,
       editor = {{Zverko}, Juraj and {Ziznovsky}, Jozef and {Adelman}, Saul J. and {Weiss}, Werner W.},
       volume = {224},
        month = dec,
        pages = {1-11},
          doi = {10.1017/S1743921304004314},
       adsurl = {https://ui.adsabs.harvard.edu/abs/2004IAUS..224....1A}
}

@ARTICLE{adelman_mccarthy09,
   author = {{Adelman-McCarthy}, J.~K. and {et al.}},
    title = "{The SDSS Photometric Catalog, Release 7 (Adelman-McCarthy+, 2009)}",
  journal = {VizieR Online Data Catalog},
     year = 2009,
    month = jun,
   volume = 2294,
    pages = {0},
   adsurl = {http://cdsads.u-strasbg.fr/abs/2009yCat.2294....0A}
}

@ARTICLE{adelman_mccarthy12,
   author = {{Ahn}, C.~P. and {Alexandroff}, R. and {Allende Prieto}, C. and 
	{Anderson}, S.~F. and {Anderton}, T. and {Andrews}, B.~H. and 
	{Aubourg}, {\'E}. and {Bailey}, S. and {Balbinot}, E. and {Barnes}, R. and et al.},
    title = "{The Ninth Data Release of the Sloan Digital Sky Survey: First Spectroscopic Data from the SDSS-III Baryon Oscillation Spectroscopic Survey}",
  journal = {ApJS},
archivePrefix = "arXiv",
   eprint = {1207.7137},
 primaryClass = "astro-ph.IM",
     year = 2012,
    month = dec,
   volume = 203,
      eid = {21},
    pages = {21},
      doi = {10.1088/0067-0049/203/2/21},
   adsurl = {http://cdsads.u-strasbg.fr/abs/2012ApJS..203...21A}
}

@ARTICLE{agati15,
       author = {{Agati}, J. -L. and {Bonneau}, D. and {Jorissen}, A. and {Souli{\'e}}, E. and {Udry}, S. and {Verhas}, P. and {Dommanget}, J.},
        title = "{Are the orbital poles of binary stars in the solar neighbourhood anisotropically distributed?}",
      journal = {A\&A},
         year = 2015,
        month = feb,
       volume = {574},
          eid = {A6},
        pages = {A6},
          doi = {10.1051/0004-6361/201323056},
archivePrefix = {arXiv},
       eprint = {1411.4919},
 primaryClass = {astro-ph.SR},
       adsurl = {https://ui.adsabs.harvard.edu/abs/2015A&A...574A...6A}
}

@ARTICLE{agresti98,
     ISSN = {00031305},
      URL = {http://www.jstor.org/stable/2685469},
    author = {{Agresti},A. and {Coull}, B.A.},
   journal = {Am. Stat.},
    number = {2},
     pages = {119--126},
 publisher = {[American Statistical Association, Taylor & Francis, Ltd.]},
     title = {Approximate Is Better than "Exact" for Interval Estimation of Binomial Proportions},
   urldate = {2022-06-20},
    volume = {52},
      year = {1998}
}

@BOOK{aitken1932,
       author = {{Aitken}, Robert Grant and {Doolittle}, Eric},
        title = "{New General Catalogue of Double Stars within 120{\textdegree} of the North Pole}",
         year = 1932,
       adsurl = {https://ui.adsabs.harvard.edu/abs/1932ngcd.book.....A},
    publisher = {Carnegie Institution of Washington}
}

@BOOK{aitken64,
       author = {{Aitken}, Robert Grant},
        title = "{The Binary Stars}",
         year = 1964,
       adsurl = {https://ui.adsabs.harvard.edu/abs/1964bist.book.....A},
    publisher = {Dover Publications. University of Michigan} 
}

@ARTICLE{akana23,
       author = {{Akana Murphy}, Joseph M. and {Batalha}, Natalie M. and {Scarsdale}, Nicholas and {Isaacson}, Howard and {Ciardi}, David R. and {Gonzales}, Erica J. and {Giacalone}, Steven and {Twicken}, Joseph D. and {Dattilo}, Anne and {Fetherolf}, Tara and {Rubenzahl}, Ryan A. and {Crossfield}, Ian J.~M. and {Dressing}, Courtney D. and {Fulton}, Benjamin and {Howard}, Andrew W. and {Huber}, Daniel and {Kane}, Stephen R. and {Petigura}, Erik A. and {Robertson}, Paul and {Roy}, Arpita and {Weiss}, Lauren M. and {Beard}, Corey and {Chontos}, Ashley and {Dai}, Fei and {Rice}, Malena and {Van Zandt}, Judah and {Lubin}, Jack and {Blunt}, Sarah and {Polanski}, Alex S. and {Behmard}, Aida and {Dalba}, Paul A. and {Hill}, Michelle L. and {Rosenthal}, Lee J. and {Brinkman}, Casey L. and {Mayo}, Andrew W. and {Turtelboom}, Emma V. and {Angelo}, Isabel and {Mo{\v{c}}nik}, Teo and {MacDougall}, Mason G. and {Pidhorodetska}, Daria and {Tyler}, Dakotah and {Kosiarek}, Molly R. and {Holcomb}, Rae and {Louden}, Emma M. and {Hirsch}, Lea A. and {Gilbert}, Emily A. and {Anderson}, Jay and {Valenti}, Jeff A.},
        title = "{The TESS-Keck Survey. XVI. Mass Measurements for 12 Planets in Eight Systems}",
      journal = {AJ},
         year = 2023,
        month = oct,
       volume = {166},
       number = {4},
          eid = {153},
        pages = {153},
          doi = {10.3847/1538-3881/ace2ca},
archivePrefix = {arXiv},
       eprint = {2306.16587},
 primaryClass = {astro-ph.EP},
       adsurl = {https://ui.adsabs.harvard.edu/abs/2023AJ....166..153A}
}

@ARTICLE{akeson13,
       author = {{Akeson}, R.~L. and {Chen}, X. and {Ciardi}, D. and {Crane}, M. and {Good}, J. and {Harbut}, M. and {Jackson}, E. and {Kane}, S.~R. and {Laity}, A.~C. and {Leifer}, S. and {Lynn}, M. and {McElroy}, D.~L. and {Papin}, M. and {Plavchan}, P. and {Ram{\'\i}rez}, S.~V. and {Rey}, R. and {von Braun}, K. and {Wittman}, M. and {Abajian}, M. and {Ali}, B. and {Beichman}, C. and {Beekley}, A. and {Berriman}, G.~B. and {Berukoff}, S. and {Bryden}, G. and {Chan}, B. and {Groom}, S. and {Lau}, C. and {Payne}, A.~N. and {Regelson}, M. and {Saucedo}, M. and {Schmitz}, M. and {Stauffer}, J. and {Wyatt}, P. and {Zhang}, A.},
        title = "{The NASA Exoplanet Archive: Data and Tools for Exoplanet Research}",
      journal = {PASP},
         year = 2013,
        month = aug,
       volume = {125},
       number = {930},
        pages = {989},
          doi = {10.1086/672273},
archivePrefix = {arXiv},
       eprint = {1307.2944},
 primaryClass = {astro-ph.IM},
       adsurl = {https://ui.adsabs.harvard.edu/abs/2013PASP..125..989A}
}

@ARTICLE{akeson21,
       author = {{Akeson}, Rachel and {Beichman}, Charles and {Kervella}, Pierre and {Fomalont}, Edward and {Benedict}, G. Fritz},
        title = "{Precision Millimeter Astrometry of the {\ensuremath{\alpha}} Centauri AB System}",
      journal = {AJ},
         year = 2021,
        month = jul,
       volume = {162},
       number = {1},
          eid = {14},
        pages = {14},
          doi = {10.3847/1538-3881/abfaff},
archivePrefix = {arXiv},
       eprint = {2104.10086},
 primaryClass = {astro-ph.SR},
       adsurl = {https://ui.adsabs.harvard.edu/abs/2021AJ....162...14A}
}

@ARTICLE{al-wardat02,
       author = {{Al-Wardat}, M.~A.},
        title = "{Spectrophotometry of speckle binary stars}",
      journal = {BSAO},
         year = 2002,
        month = jun,
       volume = {53},
        pages = {58-77},
       adsurl = {https://ui.adsabs.harvard.edu/abs/2002BSAO...53...58A}
}

@ARTICLE{alam15a,
       author = {{Alam}, S. and {Albareti}, F.~D. and {Allende Prieto}, C. and 
	{Anders}, F. and {Anderson}, S.~F. and {Anderton}, T. and {Andrews}, B.~H. and 
	{Armengaud}, E. and {Aubourg}, {\'E}. and {Bailey}, S. and et al.},
        title = "{The Eleventh and Twelfth Data Releases of the Sloan Digital Sky Survey: Final Data from SDSS-III}",
      journal = {ApJS},
archivePrefix = "arXiv",
       eprint = {1501.00963},
 primaryClass = "astro-ph.IM",   
         year = 2015,
        month = jul,
       volume = 219,
          eid = {12},
        pages = {12},
          doi = {10.1088/0067-0049/219/1/12},
       adsurl = {http://cdsads.u-strasbg.fr/abs/2015ApJS..219...12A}
}

@ARTICLE{alcock95,
       author = {{Alcock}, C. and {Allsman}, R.~A. and {Alves}, D. and {Axelrod}, T.~S. and {Bennett}, D.~P. and {Cook}, K.~H. and {Freeman}, K.~C. and {Griest}, K. and {Guern}, J. and {Lehner}, M.~J. and {Marshall}, S.~L. and {Peterson}, B.~A. and {Pratt}, M.~R. and {Quinn}, P.~J. and {Rodgers}, A.~W. and {Stubbs}, C.~W. and {Sutherland}, W.},
        title = "{First Observation of Parallax in a Gravitational Microlensing Event}",
      journal = {ApJL},
         year = 1995,
        month = dec,
       volume = {454},
        pages = {L125},
          doi = {10.1086/309783},
archivePrefix = {arXiv},
       eprint = {astro-ph/9506114},
 primaryClass = {astro-ph},
       adsurl = {https://ui.adsabs.harvard.edu/abs/1995ApJ...454L.125A}
}

@ARTICLE{alcock97,
       author = {{Alcock}, C. and {Allen}, W.~H. and {Allsman}, R.~A. and {Alves}, D. and {Axelrod}, T.~S. and {Banks}, T.~S. and {Beaulieu}, S.~F. and {Becker}, A.~C. and {Becker}, R.~H. and {Bennett}, D.~P. and {Bond}, I.~A. and {Carter}, B.~S. and {Cook}, K.~H. and {Dodd}, R.~J. and {Freeman}, K.~C. and {Gregg}, M.~D. and {Griest}, K. and {Hearnshaw}, J.~B. and {Heller}, A. and {Honda}, M. and {Jugaku}, J. and {Kabe}, S. and {Kaspi}, S. and {Kilmartin}, P.~M. and {Kitamura}, A. and {Kovo}, O. and {Lehner}, M.~J. and {Love}, T.~E. and {Maoz}, D. and {Marshall}, S.~L. and {Matsubara}, Y. and {Minniti}, D. and {Miyamoto}, M. and {Morse}, J.~A. and {Muraki}, Y. and {Nakamura}, T. and {Peterson}, B.~A. and {Phillips}, M.~M. and {Pratt}, M.~R. and {Quinn}, P.~J. and {Reid}, I.~N. and {Reid}, M. and {Reiss}, D. and {Retter}, A. and {Rodgers}, A.~W. and {Sargent}, W.~L.~W. and {Sato}, H. and {Sekiguchi}, M. and {Stetson}, P.~B. and {Stubbs}, C.~W. and {Sullivan}, D.~J. and {Sutherland}, W. and {Tomaney}, A. and {Vandehei}, T. and {Watase}, Y. and {Welch}, D.~L. and {Yanagisawa}, T. and {Yoshizawa}, M. and {Yock}, P.~C.~M.},
        title = "{MACHO Alert 95-30: First Real-Time Observation of Extended Source Effects in Gravitational Microlensing}",
      journal = {ApJ},
         year = 1997,
        month = dec,
       volume = {491},
       number = {2},
        pages = {436-450},
          doi = {10.1086/304974},
archivePrefix = {arXiv},
       eprint = {astro-ph/9702199},
 primaryClass = {astro-ph},
       adsurl = {https://ui.adsabs.harvard.edu/abs/1997ApJ...491..436A}
}

@ARTICLE{alei24,
       author = {{Alei}, E. and {Quanz}, S.~P. and {Konrad}, B.~S. and {Garvin}, E.~O. and {Kofman}, V. and {Mandell}, A. and {Angerhausen}, D. and {Molli{\`e}re}, P. and {Meyer}, M.~R. and {Robinson}, T. and {Rugheimer}, S. and {the LIFE Collaboration}},
        title = "{Large Interferometer For Exoplanets (LIFE): XIII. The value of combining thermal emission and reflected light for the characterization of Earth twins}",
      journal = {A\&A},
         year = 2024,
        month = sep,
       volume = {689},
          eid = {A245},
        pages = {A245},
          doi = {10.1051/0004-6361/202450320},
archivePrefix = {arXiv},
       eprint = {2406.13037},
 primaryClass = {astro-ph.EP},
       adsurl = {https://ui.adsabs.harvard.edu/abs/2024A&A...689A.245A}
}

@ARTICLE{alfonsogarzon12,
       author = {{Alfonso-Garz{\'o}n}, J. and {Domingo}, A. and {Mas-Hesse}, J.~M. and {Gim{\'e}nez}, A.},
        title = "{The first INTEGRAL-OMC catalogue of optically variable sources}",
      journal = {A\&A},
         year = 2012,
        month = dec,
       volume = {548},
          eid = {A79},
        pages = {A79},
          doi = {10.1051/0004-6361/201220095},
archivePrefix = {arXiv},
       eprint = {1210.0821},
 primaryClass = {astro-ph.IM},
       adsurl = {https://ui.adsabs.harvard.edu/abs/2012A&A...548A..79A}
}

@INPROCEEDINGS{allard14,
       author = {{Allard}, F.},
        title = "{The BT-Settl Model Atmospheres for Stars, Brown Dwarfs and Planets}",
    booktitle = {Exploring the Formation and Evolution of Planetary Systems},
         year = 2014,
       editor = {{Booth}, Mark and {Matthews}, Brenda C. and {Graham}, James R.},
       volume = {299},
        month = jan,
        pages = {271-272},
          doi = {10.1017/S1743921313008545},
       adsurl = {https://ui.adsabs.harvard.edu/abs/2014IAUS..299..271A}
}

@INBOOK{allen97,
       author = {{Allen}, C. and {Poveda}, A. and {Herrera}, M.~A.},
        title = "{The Distribution of Separations of Wide Binaries}",
    booktitle = {Visual Double Stars : Formation, Dynamics and Evolutionary Tracks},
         year = 1997,
    publisher = {Springer},
       volume = {223},
        pages = {133},
          doi = {10.1007/978-94-009-1477-3\_18},
       adsurl = {https://ui.adsabs.harvard.edu/abs/1997ASSL..223..133A}
}

@INPROCEEDINGS{allen98,
       author = {{Allen}, C. and {Herrera}, M.~A. and {Poveda}, A.},
        title = "{Wide Binaries among High-Velocity and Metal-Poor Stars}",
    booktitle = {IX Latin American Regional IAU Meeting, ``Focal Points in Latin American Astronomy''},
         year = 1998,
       editor = {{Aguilar}, A. and {Carrami\~nana}, A.},
        month = nov,
          eid = {21},
        pages = {21},
       adsurl = {https://ui.adsabs.harvard.edu/abs/1998larm.confE..21A}
}

@ARTICLE{allen00,
       author = {{Allen}, C. and {Poveda}, A. and {Herrera}, M.~A.},
        title = "{Wide binaries among high-velocity and metal-poor stars}",
      journal = {A\&A},
         year = 2000,
        month = apr,
       volume = 356,
        pages = {529-540},
       adsurl = {http://cdsads.u-strasbg.fr/abs/2000A\&A...356..529A}
}

@ARTICLE{alonsofloriano15,
       author = {{Alonso-Floriano}, F.~J. and {Caballero}, J.~A. and {Cort{\'e}s-Contreras}, M. and {Solano}, E. and {Montes}, D.},
        title = "{Reaching the boundary between stellar kinematic groups and very wide binaries. III. Sixteen new stars and eight new wide systems in the {\ensuremath{\beta}} Pictoris moving group}",
      journal = {A\&A},
         year = 2015,
        month = nov,
       volume = {583},
          eid = {A85},
        pages = {A85},
          doi = {10.1051/0004-6361/201526795},
archivePrefix = {arXiv},
       eprint = {1508.06929},
 primaryClass = {astro-ph.SR},
       adsurl = {https://ui.adsabs.harvard.edu/abs/2015A&A...583A..85A}
}

@ARTICLE{althaus10,
       author = {{Althaus}, Leandro G. and {C{\'o}rsico}, Alejandro H. and {Isern}, Jordi and {Garc{\'\i}a-Berro}, Enrique},
        title = "{Evolutionary and pulsational properties of white dwarf stars}",
      journal = {A\&ARv},
         year = 2010,
        month = oct,
       volume = {18},
       number = {4},
        pages = {471-566},
          doi = {10.1007/s00159-010-0033-1},
archivePrefix = {arXiv},
       eprint = {1007.2659},
 primaryClass = {astro-ph.SR},
       adsurl = {https://ui.adsabs.harvard.edu/abs/2010A&ARv..18..471A}
}

@ARTICLE{ambartsumian49,
       author = {{Ambartsumian}, V.~A.},
        title = "{Stellar Associations}",
      journal = {Sov. Astron.},
         year = 1949,
        month = jan,
       volume = {26},
        pages = {3},
       adsurl = {https://ui.adsabs.harvard.edu/abs/1949AZh....26....3A}
}

@ARTICLE{ambartsumian55,
       author = {{Ambartsumian}, V.~A.},
        title = "{Stellar systems of positive total energy}",
      journal = {The Observatory},
         year = 1955,
        month = apr,
       volume = {75},
        pages = {72-78},
       adsurl = {https://ui.adsabs.harvard.edu/abs/1955Obs....75...72A}
}

@INPROCEEDINGS{amiaux12,
   author = {{Amiaux}, J. and {Scaramella}, R. and {Mellier}, Y. and {Altieri}, B. and 
	{Burigana}, C. and {Da Silva}, A. and {Gomez}, P. and {Hoar}, J. and 
	{Laureijs}, R. and {Maiorano}, E. and {Magalh{\~a}es Oliveira}, D. and 
	{Renk}, F. and {Saavedra Criado}, G. and {Tereno}, I. and {Augu{\`e}res}, J.~L. and 
	{Brinchmann}, J. and {Cropper}, M. and {Duvet}, L. and {Ealet}, A. and 
	{Franzetti}, P. and {Garilli}, B. and {Gondoin}, P. and {Guzzo}, L. and 
	{Hoekstra}, H. and {Holmes}, R. and {Jahnke}, K. and {Kitching}, T. and 
	{Meneghetti}, M. and {Percival}, W. and {Warren}, S.},
    title = "{Euclid mission: building of a reference survey}",
booktitle = {Proc. SPIE},
     year = 2012,
   series = {Posp},
   volume = 8442,
archivePrefix = "arXiv",
   eprint = {1209.2228},
 primaryClass = "astro-ph.IM",
    month = sep,
      eid = {84420Z},
    pages = {84420Z},
      doi = {10.1117/12.926513},
   adsurl = {http://cdsads.u-strasbg.fr/abs/2012SPIE.8442E..0ZA}
}

@ARTICLE{ammons06,
       author = {{Ammons}, S. Mark and {Robinson}, Sarah E. and {Strader}, Jay and {Laughlin}, Gregory and {Fischer}, Debra and {Wolf}, Aaron},
        title = "{The N2K Consortium. IV. New Temperatures and Metallicities for More than 100,000 FGK Dwarfs}",
      journal = {ApJ},
         year = 2006,
        month = feb,
       volume = {638},
       number = {2},
        pages = {1004-1017},
          doi = {10.1086/498490},
archivePrefix = {arXiv},
       eprint = {astro-ph/0510237},
 primaryClass = {astro-ph},
       adsurl = {https://ui.adsabs.harvard.edu/abs/2006ApJ...638.1004A}
}

@ARTICLE{anderson52,
       author = {{Anderson}, T.~W. and {Darling}, D.~A.},
        title = "{Asymptotic Theory of Certain ``Goodness of Fit'' Criteria Based on Stochastic Processes}",
      journal = {Ann. Math. Statist.},
         year = 1952,
        month = jun,
       volume = {469},
       number = {23.2},
        pages = {193-212},
          doi = {10.1214/aoms/1177729437},
          url = {https://projecteuclid.org/journals/annals-of-mathematical-statistics/volume-23/issue-2/Asymptotic-Theory-of-Certain-Goodness-of-Fit-Criteria-Based-on/10.1214/aoms/1177729437.full}
}

@ARTICLE{andrae10,
       author = {{Andrae}, Rene},
        title = "{Error estimation in astronomy: A guide}",
      journal = {arXiv e-prints},
         year = 2010,
        month = sep,
          eid = {arXiv:1009.2755},
        pages = {arXiv:1009.2755},
          doi = {10.48550/arXiv.1009.2755},
archivePrefix = {arXiv},
       eprint = {1009.2755},
 primaryClass = {astro-ph.IM},
       adsurl = {https://ui.adsabs.harvard.edu/abs/2010arXiv1009.2755A}
}

@ARTICLE{anguiano17,
       author = {{Anguiano}, B. and {Rebassa-Mansergas}, A. and {Garc{\'\i}a-Berro}, E. and
         {Torres}, S. and {Freeman}, K.~C. and {Zwitter}, T.},
        title = "{The kinematics of the white dwarf population from the SDSS DR12}",
      journal = {MNRAS},
         year = 2017,
        month = aug,
       volume = {469},
       number = {2},
        pages = {2102-2120},
          doi = {10.1093/mnras/stx796},
archivePrefix = {arXiv},
       eprint = {1703.09152},
 primaryClass = {astro-ph.GA},
       adsurl = {https://ui.adsabs.harvard.edu/abs/2017MNRAS.469.2102A}
}

@ARTICLE{apps23,
       author = {{Apps}, Kevin and {Luque}, Rafael},
        title = "{HD 110067 is a Wide Hierarchical Triple System}",
      journal = {RNAAS},
         year = 2023,
        month = dec,
       volume = {7},
       number = {12},
          eid = {264},
        pages = {264},
          doi = {10.3847/2515-5172/ad12d0},
archivePrefix = {arXiv},
       eprint = {2312.04599},
 primaryClass = {astro-ph.EP},
       adsurl = {https://ui.adsabs.harvard.edu/abs/2023RNAAS...7..264A}
}

@ARTICLE{arenou18,
       author = {{Arenou}, F. and {Luri}, X. and {Babusiaux}, C. and {Fabricius}, C. and {Helmi}, A. and {Muraveva}, T. and {Robin}, A.~C. and {Spoto}, F. and {Vallenari}, A. and {Antoja}, T. and {Cantat-Gaudin}, T. and {Jordi}, C. and {Leclerc}, N. and {Reyl{\'e}}, C. and {Romero-G{\'o}mez}, M. and {Shih}, I. -C. and {Soria}, S. and {Barache}, C. and {Bossini}, D. and {Bragaglia}, A. and {Breddels}, M.~A. and {Fabrizio}, M. and {Lambert}, S. and {Marrese}, P.~M. and {Massari}, D. and {Moitinho}, A. and {Robichon}, N. and {Ruiz-Dern}, L. and {Sordo}, R. and {Veljanoski}, J. and {Eyer}, L. and {Jasniewicz}, G. and {Pancino}, E. and {Soubiran}, C. and {Spagna}, A. and {Tanga}, P. and {Turon}, C. and {Zurbach}, C.},
        title = "{Gaia Data Release 2. Catalogue validation}",
      journal = {A\&A},
         year = 2018,
        month = aug,
       volume = {616},
          eid = {A17},
        pages = {A17},
          doi = {10.1051/0004-6361/201833234},
archivePrefix = {arXiv},
       eprint = {1804.09375},
 primaryClass = {astro-ph.GA},
       adsurl = {https://ui.adsabs.harvard.edu/abs/2018A&A...616A..17A}
}

@ARTICLE{argelander1903,
       author = {{Argelander}, Friedrich Wilhelm August},
        title = "{Bonner Durchmusterung des nordlichen Himmels.}",
      journal = {Eds Marcus and Weber's Verlag},
         year = 1903,
        month = jan,
        pages = {0},
       adsurl = {https://ui.adsabs.harvard.edu/abs/1903BD....C......0A}
}

@book{argyle12,
        title = {Observing and Measuring Visual Double Stars},
       author = {Argyle, R.W.},
         isbn = {9781461439455},
       series = {The Patrick Moore Practical Astronomy Series},
          url = {https://books.google.es/books?id=DR9j2PODQOwC},
         year = {2012},
    publisher = {Springer New York}
}

@BOOK{argyle19,
       author = {{Argyle}, Robert W. and {Swan}, Mike and {James}, Andrew},
        title = "{An anthology of visual double stars}",
    publisher = {Cambridge University Press},
         year = 2019,
       adsurl = {https://ui.adsabs.harvard.edu/abs/2019avds.book.....A}
}

@ARTICLE{armitage10,
       author = {{Armitage}, Philip J. and {Valencia}, Diana},
        title = "{Astrophysics of Planet Formation}",
      journal = {Phys. Today},
         year = 2010,
        month = jan,
       volume = {63},
       number = {12},
        pages = {63},
          doi = {10.1063/1.3529004},
       adsurl = {https://ui.adsabs.harvard.edu/abs/2010PhT....63l..63A}
}

@BOOK{arnett96,
        title = {Supernovae and nucleosynthesis: an investigation of the history of matter, from the big bang to the present},
       author = {{Arnett}, David},
       volume = {7},
         year = {1996},
    publisher = {Princeton University Press}
}

@ARTICLE{artigau07,
       author = {{Artigau}, {\'E}tienne and {Lafreni{\`e}re}, David and {Doyon}, Ren{\'e} and {Albert}, Lo{\"\i}c and {Nadeau}, Daniel and {Robert}, Jasmin},
        title = "{Discovery of the Widest Very Low Mass Binary}",
      journal ={ApJL},
         year = 2007,
        month = apr,
       volume = {659},
       number = {1},
        pages = {L49-L52},
          doi = {10.1086/516710},
archivePrefix = {arXiv},
       eprint = {astro-ph/0702647},
 primaryClass = {astro-ph},
       adsurl = {https://ui.adsabs.harvard.edu/abs/2007ApJ...659L..49A}
}

@ARTICLE{artymowicz96,
       author = {{Artymowicz}, Pawel and {Lubow}, Stephen H.},
        title = "{Mass Flow through Gaps in Circumbinary Disks}",
      journal = {ApJL},
         year = 1996,
        month = aug,
       volume = {467},
        pages = {L77},
          doi = {10.1086/310200},
       adsurl = {https://ui.adsabs.harvard.edu/abs/1996ApJ...467L..77A}
}

@INPROCEEDINGS{arzoumanian96,
       author = {{Arzoumanian}, Z. and {Joshi}, K. and {Rasio}, F.~A. and {Thorsett}, S.~E.},
        title = "{Orbital Parameters of the PSR B1620-26 Triple System}",
    booktitle = {IAU Colloq. 160: Pulsars: Problems and Progress},
         year = 1996,
       editor = {{Johnston}, S. and {Walker}, M.~A. and {Bailes}, M.},
       series = {ASPCS},
       volume = {105},
        month = jan,
        pages = {525-530},
archivePrefix = {arXiv},
       eprint = {astro-ph/9605141},
 primaryClass = {astro-ph},
       adsurl = {https://ui.adsabs.harvard.edu/abs/1996ASPC..105..525A}
}

@ARTICLE{asada04,
       author = {{Asada}, Hideki and {Akasaka}, Toshio and {Kasai}, Masumi},
        title = "{Inversion Formula for Determining the Parameters of an Astrometric Binary}",
      journal = {PASJ},
         year = 2004,
        month = dec,
       volume = {56},
        pages = {L35-L38},
          doi = {10.1093/pasj/56.6.L35},
archivePrefix = {arXiv},
       eprint = {astro-ph/0409613},
 primaryClass = {astro-ph},
       adsurl = {https://ui.adsabs.harvard.edu/abs/2004PASJ...56L..35A}
}

@ARTICLE{astronomicalregister1881,
        title = "{Reviews.}",
       author = {{Astronomical Register}},
      journal = {Astronomical Register Reviews},
         year = 1881,
        month = jan,
       volume = {19},
        pages = {253-256},
       adsurl = {https://ui.adsabs.harvard.edu/abs/1881AReg...19..253.}
}

@BOOK{astronomischen1781,
       author = {{Astronomischen Rechen-Institut}},
        title = {Astronomisches Jahrbuch: 1784},
          url = {https://books.google.com.fj/books?id=9PdEAAAAcAAJ},
         year = {1781},
    publisher = {D{\"u}mmler}
}

@ARTICLE{babusiaux23,
       author = {{Babusiaux}, C. and {Fabricius}, C. and {Khanna}, S. and {Muraveva}, T. and {Reyl{\'e}}, C. and {Spoto}, F. and {Vallenari}, A. and {Luri}, X. and {Arenou}, F. and {{\'A}lvarez}, M.~A. and {Anders}, F. and {Antoja}, T. and {Balbinot}, E. and {Barache}, C. and {Bauchet}, N. and {Bossini}, D. and {Busonero}, D. and {Cantat-Gaudin}, T. and {Carrasco}, J.~M. and {Dafonte}, C. and {Diakit{\'e}}, S. and {Figueras}, F. and {Garcia-Gutierrez}, A. and {Garofalo}, A. and {Helmi}, A. and {Jim{\'e}nez-Arranz}, {\'O}. and {Jordi}, C. and {Kervella}, P. and {Kostrzewa-Rutkowska}, Z. and {Leclerc}, N. and {Licata}, E. and {Manteiga}, M. and {Masip}, A. and {Mongui{\'o}}, M. and {Ramos}, P. and {Robichon}, N. and {Robin}, A.~C. and {Romero-G{\'o}mez}, M. and {S{\'a}ez}, A. and {Santove{\~n}a}, R. and {Spina}, L. and {Torralba Elipe}, G. and {Weiler}, M.},
        title = "{Gaia Data Release 3. Catalogue validation}",
      journal = {A\&A},
         year = 2023,
        month = jun,
       volume = {674},
          eid = {A32},
        pages = {A32},
          doi = {10.1051/0004-6361/202243790},
archivePrefix = {arXiv},
       eprint = {2206.05989},
 primaryClass = {astro-ph.SR},
       adsurl = {https://ui.adsabs.harvard.edu/abs/2023A&A...674A..32B}
}

@ARTICLE{badenes18a,
       author = {{Badenes}, Carles and {Mazzola}, Christine and {Thompson}, Todd A. and {Covey}, Kevin and {Freeman}, Peter E. and {Walker}, Matthew G. and {Moe}, Maxwell and {Troup}, Nicholas and {Nidever}, David and {Allende Prieto}, Carlos and {Andrews}, Brett and {Barb{\'a}}, Rodolfo H. and {Beers}, Timothy C. and {Bovy}, Jo and {Carlberg}, Joleen K. and {De Lee}, Nathan and {Johnson}, Jennifer and {Lewis}, Hannah and {Majewski}, Steven R. and {Pinsonneault}, Marc and {Sobeck}, Jennifer and {Stassun}, Keivan G. and {Stringfellow}, Guy S. and {Zasowski}, Gail},
        title = "{Stellar Multiplicity Meets Stellar Evolution and Metallicity: The APOGEE View}",
      journal = {ApJ},
         year = 2018,
        month = feb,
       volume = {854},
       number = {2},
          eid = {147},
        pages = {147},
          doi = {10.3847/1538-4357/aaa765},
archivePrefix = {arXiv},
       eprint = {1711.00660},
 primaryClass = {astro-ph.SR},
       adsurl = {https://ui.adsabs.harvard.edu/abs/2018ApJ...854..147B}
}

@INPROCEEDINGS{baglin06,
       author = {{Baglin}, A. and {Auvergne}, M. and {Barge}, P. and {Deleuil}, M. and {Catala}, C. and {Michel}, E. and {Weiss}, W. and {COROT Team}},
        title = "{Scientific Objectives for a Minisat: CoRoT}",
    booktitle = {The CoRoT Mission Pre-Launch Status - Stellar Seismology and Planet Finding},
         year = 2006,
       editor = {{Fridlund}, M. and {Baglin}, A. and {Lochard}, J. and {Conroy}, L.},
       series = {ESA Special Publication},
       volume = {1306},
        month = nov,
        pages = {33},
       adsurl = {https://ui.adsabs.harvard.edu/abs/2006ESASP1306...33B}
}

@ARTICLE{bahcall81,
       author = {{Bahcall}, J.~N. and {Soneira}, R.~M.},
        title = "{The distribution of stars to V = 16th magnitude near the north galactic pole - Normalization, clustering properties, and counts in various bands.}",
      journal ={ApJ},
         year = 1981,
        month = may,
       volume = {246},
        pages = {122-135},
          doi = {10.1086/158905},
       adsurl = {https://ui.adsabs.harvard.edu/abs/1981ApJ...246..122B}
}

@ARTICLE{baig24,
       author = {{Baig}, Sayan and {Smart}, R.~L. and {Jones}, Hugh R.~A. and {Gagn{\'e}}, Jonathan and {Pinfield}, D.~J. and {Cheng}, Gemma and {Moranta}, Leslie},
        title = "{The Gaia ultracool dwarf sample - V: the ultracool dwarf companion catalogue}",
      journal = {MNRAS},
         year = 2024,
        month = oct,
       volume = {533},
       number = {4},
        pages = {3784-3810},
          doi = {10.1093/mnras/stae2005},
archivePrefix = {arXiv},
       eprint = {2408.07024},
 primaryClass = {astro-ph.SR},
       adsurl = {https://ui.adsabs.harvard.edu/abs/2024MNRAS.533.3784B}
}

@ARTICLE{bailerjones18a,
       author = {{Bailer-Jones}, C.~A.~L. and {Rybizki}, J. and {Fouesneau}, M. and {Mantelet}, G. and {Andrae}, R.},
        title = "{Estimating Distance from Parallaxes. IV. Distances to 1.33 Billion Stars in Gaia Data Release 2}",
      journal = {AJ},
         year = 2018,
        month = aug,
       volume = {156},
       number = {2},
          eid = {58},
        pages = {58},
          doi = {10.3847/1538-3881/aacb21},
archivePrefix = {arXiv},
       eprint = {1804.10121},
 primaryClass = {astro-ph.SR},
       adsurl = {https://ui.adsabs.harvard.edu/abs/2018AJ....156...58B}
}

@ARTICLE{bailes11,
       author = {{Bailes}, M. and {Bates}, S.~D. and {Bhalerao}, V. and {Bhat}, N.~D.~R. and {Burgay}, M. and {Burke-Spolaor}, S. and {D'Amico}, N. and {Johnston}, S. and {Keith}, M.~J. and {Kramer}, M. and {Kulkarni}, S.~R. and {Levin}, L. and {Lyne}, A.~G. and {Milia}, S. and {Possenti}, A. and {Spitler}, L. and {Stappers}, B. and {van Straten}, W.},
        title = "{Transformation of a Star into a Planet in a Millisecond Pulsar Binary}",
      journal = {Science},
         year = 2011,
        month = sep,
       volume = {333},
       number = {6050},
        pages = {1717},
          doi = {10.1126/science.1208890},
archivePrefix = {arXiv},
       eprint = {1108.5201},
 primaryClass = {astro-ph.SR},
       adsurl = {https://ui.adsabs.harvard.edu/abs/2011Sci...333.1717B}
}

@ARTICLE{baize50,
       author = {{Baize}, P.},
        title = "{Second catalogue d'orbites d'Etoiles Doubles visuelles}",
      journal = {JO},
         year = 1950,
        month = jan,
       volume = {33},
        pages = {1},
       adsurl = {https://ui.adsabs.harvard.edu/abs/1950JO.....33....1B}
}

@ARTICLE{bakos06,
       author = {{Bakos}, G{\'a}sp{\'a}r {\'A}. and {P{\'a}l}, Andr{\'a}s and {Latham}, David W. and {Noyes}, Robert W. and {Stefanik}, Robert P.},
        title = "{A Stellar Companion in the HD 189733 System with a Known Transiting Extrasolar Planet}",
      journal = {ApJL},
         year = 2006,
        month = apr,
       volume = {641},
       number = {1},
        pages = {L57-L60},
          doi = {10.1086/503671},
archivePrefix = {arXiv},
       eprint = {astro-ph/0602136},
 primaryClass = {astro-ph},
       adsurl = {https://ui.adsabs.harvard.edu/abs/2006ApJ...641L..57B}
}

@PROCEEDINGS{banday01,
        title = "{Mining the Sky}",
    booktitle = {Mining the Sky},
         year = 2001,
       editor = {{Banday}, Anthony J. and {Zaroubi}, Saleem and {Bartelmann}, Matthias},
        month = jan,
          doi = {10.1007/b82674},
       adsurl = {https://ui.adsabs.harvard.edu/abs/2001misk.conf.....B}
}

@ARTICLE{baraffe97,
       author = {{Baraffe}, I. and {Chabrier}, G. and {Allard}, F. and {Hauschildt}, P.~H.
	},
        title = "{Evolutionary models for metal-poor low-mass stars. Lower main sequence of globular clusters and halo field stars}",
      journal = {A\&A},
         year = 1997,
        month = nov,
       volume = 327,
        pages = {1054-1069},
       adsurl = {http://cdsads.u-strasbg.fr/cgi-bin/nph-bib_query?bibcode=1997A\&A...327.1054B&amp;db_key=AST}
}

@ARTICLE{baraffe15,
   author = {{Baraffe}, I. and {Homeier}, D. and {Allard}, F. and {Chabrier}, G.
	},
    title = "{New evolutionary models for pre-main sequence and main sequence low-mass stars down to the hydrogen-burning limit}",
  journal = {A\&A},
archivePrefix = "arXiv",
   eprint = {1503.04107},
 primaryClass = "astro-ph.SR",
     year = 2015,
    month = may,
   volume = 577,
      eid = {A42},
    pages = {A42},
      doi = {10.1051/0004-6361/201425481},
   adsurl = {http://cdsads.u-strasbg.fr/abs/2015A&A...577A..42B}
}

@ARTICLE{bardalezgagliuffi14,
       author = {{Bardalez Gagliuffi}, Daniella C. and {Burgasser}, Adam J. and {Gelino}, Christopher R. and {Looper}, Dagny L. and {Nicholls}, Christine P. and {Schmidt}, Sarah J. and {Cruz}, Kelle and {West}, Andrew A. and {Gizis}, John E. and {Metchev}, Stanimir},
        title = "{SpeX Spectroscopy of Unresolved Very Low Mass Binaries. II. Identification of 14 Candidate Binaries with Late-M/Early-L and T Dwarf Components}",
      journal = {ApJ},
         year = 2014,
        month = oct,
       volume = {794},
       number = {2},
          eid = {143},
        pages = {143},
          doi = {10.1088/0004-637X/794/2/143},
archivePrefix = {arXiv},
       eprint = {1408.3089},
 primaryClass = {astro-ph.SR},
       adsurl = {https://ui.adsabs.harvard.edu/abs/2014ApJ...794..143B}
}

@ARTICLE{barnes20,
       author = {{Barnes}, John R. and {Haswell}, Carole A. and {Staab}, Daniel and {Anglada-Escud{\'e}}, Guillem and {Fossati}, Luca and {Doherty}, James P.~J. and {Cooper}, Joseph and {Jenkins}, James S. and {D{\'\i}az}, Mat{\'\i}as R. and {Soto}, Maritza G. and {Pe{\~n}a Rojas}, Pablo A.},
        title = "{An ablating 2.6 M$_{{\ensuremath{\oplus}}}$ planet in an eccentric binary from the Dispersed Matter Planet Project}",
      journal = {Nat. Astron.},
         year = 2020,
        month = jan,
       volume = {4},
        pages = {419-426},
          doi = {10.1038/s41550-019-0972-z},
archivePrefix = {arXiv},
       eprint = {1912.10793},
 primaryClass = {astro-ph.EP},
       adsurl = {https://ui.adsabs.harvard.edu/abs/2020NatAs...4..419B}
}

@ARTICLE{baroch21,
       author = {{Baroch}, D. and {Morales}, J.~C. and {Ribas}, I. and {B{\'e}jar}, V.~J.~S. and {Reffert}, S. and {Cardona Guill{\'e}n}, C. and {Reiners}, A. and {Caballero}, J.~A. and {Quirrenbach}, A. and {Amado}, P.~J. and {Anglada-Escud{\'e}}, G. and {Colom{\'e}}, J. and {Cort{\'e}s-Contreras}, M. and {Dreizler}, S. and {Galad{\'\i}-Enr{\'\i}quez}, D. and {Hatzes}, A.~P. and {Jeffers}, S.~V. and {Henning}, Th. and {Herrero}, E. and {Kaminski}, A. and {K{\"u}rster}, M. and {Lafarga}, M. and {Lodieu}, N. and {L{\'o}pez-Gonz{\'a}lez}, M.~J. and {Montes}, D. and {Pall{\'e}}, E. and {Perger}, M. and {Pollacco}, D. and {Rodr{\'\i}guez-L{\'o}pez}, C. and {Rodr{\'\i}guez}, E. and {Rosich}, A. and {Sch{\"o}fer}, P. and {Schweitzer}, A. and {Shan}, Y. and {Tal-Or}, L. and {Zechmeister}, M.},
        title = "{The CARMENES search for exoplanets around M dwarfs. Spectroscopic orbits of nine M-dwarf multiple systems, including two triples, two brown dwarf candidates, and one close M-dwarf-white dwarf binary}",
      journal = {A\&A},
         year = 2021,
        month = sep,
       volume = {653},
          eid = {A49},
        pages = {A49},
          doi = {10.1051/0004-6361/202141031},
archivePrefix = {arXiv},
       eprint = {2105.14770},
 primaryClass = {astro-ph.SR},
       adsurl = {https://ui.adsabs.harvard.edu/abs/2021A&A...653A..49B}
}

@ARTICLE{baron12,
       author = {{Baron}, F. and {Monnier}, J.~D. and {Pedretti}, E. and {Zhao}, M. and {Schaefer}, G. and {Parks}, R. and {Che}, X. and {Thureau}, N. and {ten Brummelaar}, T.~A. and {McAlister}, H.~A. and {Ridgway}, S.~T. and {Farrington}, C. and {Sturmann}, J. and {Sturmann}, L. and {Turner}, N.},
        title = "{Imaging the Algol Triple System in the H Band with the CHARA Interferometer}",
      journal = {ApJ},
         year = 2012,
        month = jun,
       volume = {752},
       number = {1},
          eid = {20},
        pages = {20},
          doi = {10.1088/0004-637X/752/1/20},
archivePrefix = {arXiv},
       eprint = {1205.0754},
 primaryClass = {astro-ph.SR},
       adsurl = {https://ui.adsabs.harvard.edu/abs/2012ApJ...752...20B}
}

@ARTICLE{baron19,
       author = {{Baron}, Fr{\'e}d{\'e}rique and {Lafreni{\`e}re}, David and {Artigau}, {\'E}tienne and {Gagn{\'e}}, Jonathan and {Rameau}, Julien and {Delorme}, Philippe and {Naud}, Marie-Eve},
        title = "{Constraints on the Occurrence and Distribution of 1-20 M $_{Jup}$ Companions to Stars at Separations of 5-5000 au from a Compilation of Direct Imaging Surveys}",
      journal = {AJ},
         year = 2019,
        month = nov,
       volume = {158},
       number = {5},
          eid = {187},
        pages = {187},
          doi = {10.3847/1538-3881/ab4130},
archivePrefix = {arXiv},
       eprint = {1909.06255},
 primaryClass = {astro-ph.EP},
       adsurl = {https://ui.adsabs.harvard.edu/abs/2019AJ....158..187B}
}

@ARTICLE{barrado98,
       author = {{Barrado y Navascu{\'e}s}, D.},
        title = "{The Castor moving group. The age of Fomalhaut and VEGA}",
      journal = {A\&A},
         year = 1998,
        month = nov,
       volume = {339},
        pages = {831-839},
archivePrefix = {arXiv},
       eprint = {astro-ph/9905243},
 primaryClass = {astro-ph},
       adsurl = {https://ui.adsabs.harvard.edu/abs/1998A&A...339..831B}
}

@ARTICLE{barry12,
       author = {{Barry}, R.~K. and {Demory}, B. -O. and {S{\'e}gransan}, D. and {Forveille}, T. and {Danchi}, W.~C. and {Di Folco}, E. and {Queloz}, D. and {Spooner}, H.~R. and {Torres}, G. and {Traub}, W.~A. and {Delfosse}, X. and {Mayor}, M. and {Perrier}, C. and {Udry}, S.},
        title = "{A Precise Physical Orbit for the M-dwarf Binary Gliese 268}",
      journal = {ApJ},
         year = 2012,
        month = nov,
       volume = {760},
       number = {1},
          eid = {55},
        pages = {55},
          doi = {10.1088/0004-637X/760/1/55},
       adsurl = {https://ui.adsabs.harvard.edu/abs/2012ApJ...760...55B}
}

@ARTICLE{basri00,
       author = {{Basri}, Gibor},
        title = "{Observations of Brown Dwarfs}",
      journal = {ARA\&A},
         year = 2000,
        month = jan,
       volume = {38},
        pages = {485-519},
          doi = {10.1146/annurev.astro.38.1.485},
       adsurl = {https://ui.adsabs.harvard.edu/abs/2000ARA&A..38..485B}
}

@ARTICLE{basri06,
       author = {{Basri}, Gibor and {Reiners}, Ansgar},
        title = "{A Survey for Spectroscopic Binaries among Very Low Mass Stars}",
      journal ={AJ},
         year = 2006,
        month = aug,
       volume = {132},
       number = {2},
        pages = {663-675},
          doi = {10.1086/505198},
archivePrefix = {arXiv},
       eprint = {astro-ph/0604259},
 primaryClass = {astro-ph},
       adsurl = {https://ui.adsabs.harvard.edu/abs/2006AJ....132..663B}
}

@ARTICLE{bate97,
       author = {{Bate}, Matthew R. and {Bonnell}, Ian A.},
        title = "{Accretion during binary star formation - II. Gaseous accretion and disc formation}",
      journal = {MNRAS},
         year = 1997,
        month = feb,
       volume = {285},
       number = {1},
        pages = {33-48},
          doi = {10.1093/mnras/285.1.33},
       adsurl = {https://ui.adsabs.harvard.edu/abs/1997MNRAS.285...33B}
}

@ARTICLE{bate14b,
   author = {{Bate}, M.~R.},
    title = "{The statistical properties of stars and their dependence on metallicity: the effects of opacity}",
  journal = {MNRAS},
archivePrefix = "arXiv",
   eprint = {1405.5583},
 primaryClass = "astro-ph.SR",
     year = 2014,
    month = jul,
   volume = 442,
    pages = {285-313},
      doi = {10.1093/mnras/stu795},
   adsurl = {http://cdsads.u-strasbg.fr/abs/2014MNRAS.442..285B}
}

@ARTICLE{bate19a,
   author = {{Bate}, M.~R.},
    title = "{The statistical properties of stars and their dependence on metallicity}",
  journal = {MNRAS},
archivePrefix = "arXiv",
   eprint = {1901.03713},
 primaryClass = "astro-ph.SR",
     year = 2019,
    month = apr,
   volume = 484,
    pages = {2341-2361},
      doi = {10.1093/mnras/stz103},
   adsurl = {http://cdsads.u-strasbg.fr/abs/2019MNRAS.484.2341B}
}

@ARTICLE{batten67,
       author = {{Batten}, Alan H.},
        title = "{On the Interpretation of Statistics of Double Stars}",
      journal = {ARAA},
         year = 1967,
        month = jan,
       volume = {5},
        pages = {25},
          doi = {10.1146/annurev.aa.05.090167.000325},
       adsurl = {https://ui.adsabs.harvard.edu/abs/1967ARA&A...5...25B}
}

@BOOK{batten73,
       author = {{Batten}, Alan H.},
        title = "{Binary and multiple systems of stars}",
    publisher = {Volume 51 in International Series in Natural Philosophy. Pergamon Press, New York},        
         year = 1973,
       adsurl = {https://ui.adsabs.harvard.edu/abs/1973bmss.book.....B}
}

@ARTICLE{batten77,
       author = {{Batten}, Alan H.},
        title = "{The Struves of Pulkovo- A Family of Astronomers}",
      journal = {J. R. Astron. Soc. Can.},
         year = 1977,
        month = oct,
       volume = {71},
        pages = {345},
       adsurl = {https://ui.adsabs.harvard.edu/abs/1977JRASC..71..345B}
}

@ARTICLE{batten78,
       author = {{Batten}, A.~H. and {Fletcher}, J.~M. and {Mann}, P.~J.},
        title = "{Seventh catalogue of the orbital elements of spectroscopic binary systems.}",
      journal = {DAOP},
         year = 1978,
        month = jan,
       volume = {15},
        pages = {121},
       adsurl = {https://ui.adsabs.harvard.edu/abs/1978PDAO...15..121B}
}

@ARTICLE{batten89,
       author = {{Batten}, Alan H. and {Fletcher}, J.~M. and {MacCarthy}, D.~G.},
        title = "{Catalogue of the orbital elements of spectroscopic binary systems : 8 : 1989}",
      journal = {DAOP},
         year = 1989,
        month = jan,
       volume = {17},
        pages = {1},
       adsurl = {https://ui.adsabs.harvard.edu/abs/1989PDAO...17....1B}
}

@ARTICLE{batten89b,
       author = {{Batten}, Alan H.},
        title = "{Two Centuries of Study of Algol Systems}",
      journal = {Space Sci. Rev.},
         year = 1989,
        month = jun,
       volume = {50},
       number = {1-2},
        pages = {1-8},
          doi = {10.1007/BF00215914},
       adsurl = {https://ui.adsabs.harvard.edu/abs/1989SSRv...50....1B}
}

@BOOK{bayer1603,
       author = {{Bayer}, Johann},
        title = "{Uranometria omnium asterismorum continens schemata, nova methodo delineata aereis laminis expressa}",
         year = 1603,
          doi = {10.3931/e-rara-309},
    publisher = {Augustae Vindelicorum: excudit Christophorus Mangus},
       adsurl = {https://ui.adsabs.harvard.edu/abs/1603uoac.book.....B}
}

@ARTICLE{bayo08,
   author = {{Bayo}, A. and {Rodrigo}, C. and {Barrado Y Navascu{\'e}s}, D. and 
	{Solano}, E. and {Guti{\'e}rrez}, R. and {Morales-Calder{\'o}n}, M. and 
	{Allard}, F.},
    title = "{VOSA: virtual observatory SED analyzer. An application to the Collinder 69 open cluster}",
  journal = {A\&A},
archivePrefix = "arXiv",
   eprint = {0808.0270},
     year = 2008,
    month = dec,
   volume = 492,
    pages = {277-287},
      doi = {10.1051/0004-6361:200810395},
   adsurl = {http://cdsads.u-strasbg.fr/abs/2008A&A...492..277B}
}

@ARTICLE{bean06a,
   author = {{Bean}, J.~L. and {Sneden}, C. and {Hauschildt}, P.~H. and {Johns-Krull}, C.~M. and 
	{Benedict}, G.~F.},
    title = "{Accurate M Dwarf Metallicities from Spectral Synthesis: A Critical Test of Model Atmospheres}",
  journal = {ApJ},
   eprint = {astro-ph/0608093},
     year = 2006,
    month = dec,
   volume = 652,
    pages = {1604-1616},
      doi = {10.1086/508321},
   adsurl = {http://cdsads.u-strasbg.fr/abs/2006ApJ...652.1604B}
}

@ARTICLE{beaulieu06,
       author = {{Beaulieu}, J. -P. and {Bennett}, D.~P. and {Fouqu{\'e}}, P. and {Williams}, A. and {Dominik}, M. and {J{\o}rgensen}, U.~G. and {Kubas}, D. and {Cassan}, A. and {Coutures}, C. and {Greenhill}, J. and {Hill}, K. and {Menzies}, J. and {Sackett}, P.~D. and {Albrow}, M. and {Brillant}, S. and {Caldwell}, J.~A.~R. and {Calitz}, J.~J. and {Cook}, K.~H. and {Corrales}, E. and {Desort}, M. and {Dieters}, S. and {Dominis}, D. and {Donatowicz}, J. and {Hoffman}, M. and {Kane}, S. and {Marquette}, J. -B. and {Martin}, R. and {Meintjes}, P. and {Pollard}, K. and {Sahu}, K. and {Vinter}, C. and {Wambsganss}, J. and {Woller}, K. and {Horne}, K. and {Steele}, I. and {Bramich}, D.~M. and {Burgdorf}, M. and {Snodgrass}, C. and {Bode}, M. and {Udalski}, A. and {Szyma{\'n}ski}, M.~K. and {Kubiak}, M. and {Wi{\c{e}}ckowski}, T. and {Pietrzy{\'n}ski}, G. and {Soszy{\'n}ski}, I. and {Szewczyk}, O. and {Wyrzykowski}, {\L}. and {Paczy{\'n}ski}, B. and {Abe}, F. and {Bond}, I.~A. and {Britton}, T.~R. and {Gilmore}, A.~C. and {Hearnshaw}, J.~B. and {Itow}, Y. and {Kamiya}, K. and {Kilmartin}, P.~M. and {Korpela}, A.~V. and {Masuda}, K. and {Matsubara}, Y. and {Motomura}, M. and {Muraki}, Y. and {Nakamura}, S. and {Okada}, C. and {Ohnishi}, K. and {Rattenbury}, N.~J. and {Sako}, T. and {Sato}, S. and {Sasaki}, M. and {Sekiguchi}, T. and {Sullivan}, D.~J. and {Tristram}, P.~J. and {Yock}, P.~C.~M. and {Yoshioka}, T.},
        title = "{Discovery of a cool planet of 5.5 Earth masses through gravitational microlensing}",
      journal = {Nature},
         year = 2006,
        month = jan,
       volume = {439},
       number = {7075},
        pages = {437-440},
          doi = {10.1038/nature04441},
archivePrefix = {arXiv},
       eprint = {astro-ph/0601563},
 primaryClass = {astro-ph},
       adsurl = {https://ui.adsabs.harvard.edu/abs/2006Natur.439..437B}
}

@ARTICLE{bedin24,
       author = {{Bedin}, L.~R. and {Dietrich}, J. and {Burgasser}, A.~J. and {Apai}, D. and {Libralato}, M. and {Griggio}, M. and {Fontanive}, C. and {Pourbaix}, D.},
        title = "{HST astrometry of the closest brown dwarfs-II. Improved parameters and constraints on a third body}",
      journal = {Astron. Nachr.},
         year = 2024,
        month = jan,
       volume = {345},
       number = {1},
          eid = {e20230158},
        pages = {e20230158},
          doi = {10.1002/asna.20230158},
archivePrefix = {arXiv},
       eprint = {2403.08865},
 primaryClass = {astro-ph.EP},
       adsurl = {https://ui.adsabs.harvard.edu/abs/2024AN....34530158B}
}

@ARTICLE{beichman14,
       author = {{Beichman}, Charles and {Benneke}, Bjoern and {Knutson}, Heather and {Smith}, Roger and {Lagage}, Pierre-Olivier and {Dressing}, Courtney and {Latham}, David and {Lunine}, Jonathan and {Birkmann}, Stephan and {Ferruit}, Pierre and {Giardino}, Giovanna and {Kempton}, Eliza and {Carey}, Sean and {Krick}, Jessica and {Deroo}, Pieter D. and {Mandell}, Avi and {Ressler}, Michael E. and {Shporer}, Avi and {Swain}, Mark and {Vasisht}, Gautam and {Ricker}, George and {Bouwman}, Jeroen and {Crossfield}, Ian and {Greene}, Tom and {Howell}, Steve and {Christiansen}, Jessie and {Ciardi}, David and {Clampin}, Mark and {Greenhouse}, Matt and {Sozzetti}, Alessandro and {Goudfrooij}, Paul and {Hines}, Dean and {Keyes}, Tony and {Lee}, Janice and {McCullough}, Peter and {Robberto}, Massimo and {Stansberry}, John and {Valenti}, Jeff and {Rieke}, Marcia and {Rieke}, George and {Fortney}, Jonathan and {Bean}, Jacob and {Kreidberg}, Laura and {Ehrenreich}, David and {Deming}, Drake and {Albert}, Lo{\"\i}c and {Doyon}, Ren{\'e} and {Sing}, David},
        title = "{Observations of Transiting Exoplanets with the James Webb Space Telescope (JWST)}",
      journal = {PASP},
         year = 2014,
        month = dec,
       volume = {126},
       number = {946},
        pages = {1134},
          doi = {10.1086/679566},
       adsurl = {https://ui.adsabs.harvard.edu/abs/2014PASP..126.1134B}
}

@INBOOK{beichman17,
       author = {{Beichman}, Charles A. and {Greene}, Thomas P.},
        title = {Observing Exoplanets with the James Webb Space Telescope},
    bookTitle = {Handbook of Exoplanets},
         year = {2017},
    publisher = {Springer International Publishing},
        pages = {1--26},
         isbn = {978-3-319-30648-3},
          doi = {10.1007/978-3-319-30648-3_85-1},
          url = {https://doi.org/10.1007/978-3-319-30648-3_85-1}
}

@ARTICLE{bell94,
       author = {{Bell}, J.~F.},
        title = "{PSR J0045-7319: a massive SMC binary.}",
      journal = {Proc. Astron. Soc. Australia},
         year = 1994,
        month = apr,
       volume = {11},
       number = {1},
        pages = {81},
       adsurl = {https://ui.adsabs.harvard.edu/abs/1994PASA...11...81B}
}

@ARTICLE{bell17,
       author = {{Bell}, Cameron P.~M. and {Murphy}, Simon J. and {Mamajek}, Eric E.},
        title = "{A stellar census of the nearby, young 32 Orionis group}",
      journal ={MNRAS},
         year = 2017,
        month = jun,
       volume = {468},
       number = {1},
        pages = {1198-1220},
          doi = {10.1093/mnras/stx535},
archivePrefix = {arXiv},
       eprint = {1703.00015},
 primaryClass = {astro-ph.SR},
       adsurl = {https://ui.adsabs.harvard.edu/abs/2017MNRAS.468.1198B}
}

@ARTICLE{belokurov20,
       author = {{Belokurov}, Vasily and {Penoyre}, Zephyr and {Oh}, Semyeong and {Iorio}, Giuliano and {Hodgkin}, Simon and {Evans}, N. Wyn and {Everall}, Andrew and {Koposov}, Sergey E. and {Tout}, Christopher A. and {Izzard}, Robert and {Clarke}, Cathie J. and {Brown}, Anthony G.~A.},
        title = "{Unresolved stellar companions with Gaia DR2 astrometry}",
      journal = {MNRAS},
         year = 2020,
        month = aug,
       volume = {496},
       number = {2},
        pages = {1922-1940},
          doi = {10.1093/mnras/staa1522},
archivePrefix = {arXiv},
       eprint = {2003.05467},
 primaryClass = {astro-ph.SR},
       adsurl = {https://ui.adsabs.harvard.edu/abs/2020MNRAS.496.1922B}
}

@ARTICLE{benatti19,
       author = {{Benatti}, S. and {Nardiello}, D. and {Malavolta}, L. and {Desidera}, S. and {Borsato}, L. and {Nascimbeni}, V. and {Damasso}, M. and {D'Orazi}, V. and {Mesa}, D. and {Messina}, S. and {Esposito}, M. and {Bignamini}, A. and {Claudi}, R. and {Covino}, E. and {Lovis}, C. and {Sabotta}, S.},
        title = "{A possibly inflated planet around the bright young star DS Tucanae A}",
      journal = {A\&A},
         year = 2019,
        month = oct,
       volume = {630},
          eid = {A81},
        pages = {A81},
          doi = {10.1051/0004-6361/201935598},
archivePrefix = {arXiv},
       eprint = {1904.01591},
 primaryClass = {astro-ph.SR},
       adsurl = {https://ui.adsabs.harvard.edu/abs/2019A&A...630A..81B}
}

@ARTICLE{benedict16,
       author = {{Benedict}, G.~F. and {Henry}, T.~J. and {Franz}, O.~G. and {McArthur}, B.~E. and {Wasserman}, L.~H. and {Jao}, Wei-Chun and {Cargile}, P.~A. and {Dieterich}, S.~B. and {Bradley}, A.~J. and {Nelan}, E.~P. and {Whipple}, A.~L.},
        title = "{The Solar Neighborhood. XXXVII: The Mass-Luminosity Relation for Main-sequence M Dwarfs}",
      journal = {AJ},
         year = 2016,
        month = nov,
       volume = {152},
       number = {5},
          eid = {141},
        pages = {141},
          doi = {10.3847/0004-6256/152/5/141},
archivePrefix = {arXiv},
       eprint = {1608.04775},
 primaryClass = {astro-ph.SR},
       adsurl = {https://ui.adsabs.harvard.edu/abs/2016AJ....152..141B}
}

@ARTICLE{benedict17,
       author = {{Benedict}, G. Fritz and {Harrison}, Thomas E.},
        title = "{HD 202206: A Circumbinary Brown Dwarf System}",
      journal = {AJ},
         year = 2017,
        month = jun,
       volume = {153},
       number = {6},
          eid = {258},
        pages = {258},
          doi = {10.3847/1538-3881/aa6d59},
archivePrefix = {arXiv},
       eprint = {1705.00659},
 primaryClass = {astro-ph.SR},
       adsurl = {https://ui.adsabs.harvard.edu/abs/2017AJ....153..258B}
}

@ARTICLE{benitez14,
       author = {{Benitez}, N. and {Dupke}, R. and {Moles}, M. and {Sodre}, L. and {Cenarro}, J. and {Marin-Franch}, A. and {Taylor}, K. and {Cristobal}, D. and {Fernandez-Soto}, A. and {Mendes de Oliveira}, C. and {Cepa-Nogue}, J. and {Abramo}, L.~R. and {Alcaniz}, J.~S. and {Overzier}, R. and {Hernandez-Monteagudo}, C. and {Alfaro}, E.~J. and {Kanaan}, A. and {Carvano}, J.~M. and {Reis}, R.~R.~R. and {Martinez Gonzalez}, E. and {Ascaso}, B. and {Ballesteros}, F. and {Xavier}, H.~S. and {Varela}, J. and {Ederoclite}, A. and {Vazquez Ramio}, H. and {Broadhurst}, T. and {Cypriano}, E. and {Angulo}, R. and {Diego}, J.~M. and {Zandivarez}, A. and {Diaz}, E. and {Melchior}, P. and {Umetsu}, K. and {Spinelli}, P.~F. and {Zitrin}, A. and {Coe}, D. and {Yepes}, G. and {Vielva}, P. and {Sahni}, V. and {Marcos-Caballero}, A. and {Kitaura}, F. -S. and {Maroto}, A.~L. and {Masip}, M. and {Tsujikawa}, S. and {Carneiro}, S. and {Gonzalez Nuevo}, J. and {Carvalho}, G.~C. and {Reboucas}, M.~J. and {Carvalho}, J.~C. and {Abdalla}, E. and {Bernui}, A. and {Pigozzo}, C. and {Ferreira}, E.~G.~M. and {Chandrachani Devi}, N. and {Bengaly}, Jr., C.~A.~P. and {Campista}, M. and {Amorim}, A. and {Asari}, N.~V. and {Bongiovanni}, A. and {Bonoli}, S. and {Bruzual}, G. and {Cardiel}, N. and {Cava}, A. and {Cid Fernandes}, R. and {Coelho}, P. and {Cortesi}, A. and {Delgado}, R.~G. and {Diaz Garcia}, L. and {Espinosa}, J.~M.~R. and {Galliano}, E. and {Gonzalez-Serrano}, J.~I. and {Falcon-Barroso}, J. and {Fritz}, J. and {Fernandes}, C. and {Gorgas}, J. and {Hoyos}, C. and {Jimenez-Teja}, Y. and {Lopez-Aguerri}, J.~A. and {Lopez-San Juan}, C. and {Mateus}, A. and {Molino}, A. and {Novais}, P. and {OMill}, A. and {Oteo}, I. and {Perez-Gonzalez}, P.~G. and {Poggianti}, B. and {Proctor}, R. and {Ricciardelli}, E. and {Sanchez-Blazquez}, P. and {Storchi-Bergmann}, T. and {Telles}, E. and {Schoennell}, W. and {Trujillo}, N. and {Vazdekis}, A. and {Viironen}, K. and {Daflon}, S. and {Aparicio-Villegas}, T. and {Rocha}, D. and {Ribeiro}, T. and {Borges}, M. and {Martins}, S.~L. and {Marcolino}, W. and {Martinez-Delgado}, D. and {Perez-Torres}, M.~A. and {Siffert}, B.~B. and {Calvao}, M.~O. and {Sako}, M. and {Kessler}, R. and {Alvarez-Candal}, A. and {De Pra}, M. and {Roig}, F. and {Lazzaro}, D. and {Gorosabel}, J. and {Lopes de Oliveira}, R. and {Lima-Neto}, G.~B. and {Irwin}, J. and {Liu}, J.~F. and {Alvarez}, E. and {Balmes}, I. and {Chueca}, S. and {Costa-Duarte}, M.~V. and {da Costa}, A.~A. and {Dantas}, M.~L.~L. and {Diaz}, A.~Y. and {Fabregat}, J. and {Ferrari}, F. and {Gavela}, B. and {Gracia}, S.~G. and {Gruel}, N. and {Gutierrez}, J.~L.~L. and {Guzman}, R. and {Hernandez-Fernandez}, J.~D. and {Herranz}, D. and {Hurtado-Gil}, L. and {Jablonsky}, F. and {Laporte}, R. and {Le Tiran}, L.~L. and {Licandro}, J and {Lima}, M. and {Martin}, E. and {Martinez}, V. and {Montero}, J.~J.~C. and {Penteado}, P. and {Pereira}, C.~B. and {Peris}, V. and {Quilis}, V. and {Sanchez-Portal}, M. and {Soja}, A.~C. and {Solano}, E. and {Torra}, J. and {Valdivielso}, L.},
        title = "{J-PAS: The Javalambre-Physics of the Accelerated Universe Astrophysical Survey}",
      journal = {arXiv e-prints},
         year = 2014,
        month = mar,
          eid = {arXiv:1403.5237},
        pages = {arXiv:1403.5237},
          doi = {10.48550/arXiv.1403.5237},
archivePrefix = {arXiv},
       eprint = {1403.5237},
 primaryClass = {astro-ph.CO},
       adsurl = {https://ui.adsabs.harvard.edu/abs/2014arXiv1403.5237B}
}

@ARTICLE{bensby03,
       author = {{Bensby}, T. and {Feltzing}, S. and {Lundstr{\"o}m}, I.},
        title = "{Elemental abundance trends in the Galactic thin and thick disks as traced by nearby F and G dwarf stars}",
      journal = {A\&A},
         year = "2003",
        month = "Nov",
       volume = {410},
        pages = {527-551},
          doi = {10.1051/0004-6361:20031213},
       adsurl = {https://ui.adsabs.harvard.edu/abs/2003A&A...410..527B}
}

@ARTICLE{bensby05,
       author = {{Bensby}, T. and {Feltzing}, S. and {Lundstr{\"o}m}, I. and {Ilyin}, I.},
        title = "{{\ensuremath{\alpha}}-, r-, and s-process element trends in the Galactic thin and thick disks}",
      journal = {A\&A},
         year = "2005",
        month = "Apr",
       volume = {433},
       number = {1},
        pages = {185-203},
          doi = {10.1051/0004-6361:20040332},
archivePrefix = {arXiv},
       eprint = {astro-ph/0412132},
 primaryClass = {astro-ph},
       adsurl = {https://ui.adsabs.harvard.edu/abs/2005A&A...433..185B}
}

@ARTICLE{berger63,
       author = {{Berger}, Jacques},
        title = "{Radial Velocities and Spectra of Faint Blue Stars at High Galactic Latitudes}",
      journal = {PASP},
         year = 1963,
        month = oct,
       volume = {75},
       number = {446},
        pages = {393},
          doi = {10.1086/127980},
       adsurl = {https://ui.adsabs.harvard.edu/abs/1963PASP...75..393B}
}

@ARTICLE{bergfors10,
       author = {{Bergfors}, C. and {Brandner}, W. and {Janson}, M. and {Daemgen}, S. and {Geissler}, K. and {Henning}, T. and {Hippler}, S. and {Hormuth}, F. and {Joergens}, V. and {K{\"o}hler}, R.},
        title = "{Lucky Imaging survey for southern M dwarf binaries}",
      journal = {A\&A},
         year = 2010,
        month = sep,
       volume = {520},
          eid = {A54},
        pages = {A54},
          doi = {10.1051/0004-6361/201014114},
archivePrefix = {arXiv},
       eprint = {1006.2377},
 primaryClass = {astro-ph.SR},
       adsurl = {https://ui.adsabs.harvard.edu/abs/2010A&A...520A..54B}
}

@ARTICLE{bergfors13,
       author = {{Bergfors}, C. and {Brandner}, W. and {Daemgen}, S. and {Biller}, B. and {Hippler}, S. and {Janson}, M. and {Kudryavtseva}, N. and {Gei{\ss}ler}, K. and {Henning}, T. and {K{\"o}hler}, R.},
        title = "{Stellar companions to exoplanet host stars: Lucky Imaging of transiting planet hosts}",
      journal = {MNRAS},
         year = 2013,
        month = jan,
       volume = {428},
       number = {1},
        pages = {182-189},
          doi = {10.1093/mnras/sts019},
archivePrefix = {arXiv},
       eprint = {1209.4087},
 primaryClass = {astro-ph.SR},
       adsurl = {https://ui.adsabs.harvard.edu/abs/2013MNRAS.428..182B}
}

@ARTICLE{berkson44,
         ISSN = {01621459},
          URL = {http://www.jstor.org/stable/2280041},
       author = {Joseph Berkson},
      journal = {J. Am. Stat. Assoc.},
       number = {227},
        pages = {357--365},
    publisher = {[American Statistical Association, Taylor & Francis, Ltd.]},
        title = {Application of the Logistic Function to Bio-Assay},
      urldate = {2024-02-06},
       volume = {39},
         year = {1944}
}

@ARTICLE{berman31,
       author = {{Berman}, Louis},
        title = "{The spectroscopic orbit of the fainter component in the system [xi] Ursae Majoris}",
      journal = {Lick Observatory Bulletin},
         year = 1931,
        month = jan,
       volume = {432},
        pages = {109-116},
          doi = {10.5479/ADS/bib/1931LicOB.15.109B},
       adsurl = {https://ui.adsabs.harvard.edu/abs/1931LicOB..15..109B}
}

@ARTICLE{bernat10,
       author = {{Bernat}, David and {Bouchez}, Antonin H. and {Ireland}, Michael and {Tuthill}, Peter and {Martinache}, Frantz and {Angione}, John and {Burruss}, Rick S. and {Cromer}, John L. and {Dekany}, Richard G. and {Guiwits}, Stephen R. and {Henning}, John R. and {Hickey}, Jeff and {Kibblewhite}, Edward and {McKenna}, Daniel L. and {Moore}, Anna M. and {Petrie}, Harold L. and {Roberts}, Jennifer and {Shelton}, J. Chris and {Thicksten}, Robert P. and {Trinh}, Thang and {Tripathi}, Renu and {Troy}, Mitchell and {Truong}, Tuan and {Velur}, Viswa and {Lloyd}, James P.},
        title = "{A Close Companion Search Around L Dwarfs Using Aperture Masking Interferometry and Palomar Laser Guide Star Adaptive Optics}",
      journal = {ApJ},
         year = 2010,
        month = jun,
       volume = {715},
       number = {2},
        pages = {724-735},
          doi = {10.1088/0004-637X/715/2/724},
archivePrefix = {arXiv},
       eprint = {1004.0223},
 primaryClass = {astro-ph.SR},
       adsurl = {https://ui.adsabs.harvard.edu/abs/2010ApJ...715..724B}
}

@ARTICLE{bertini23,
       author = {{Bertini}, L. and {Roccatagliata}, V. and {Kim}, M.},
        title = "{Flybys in debris disk systems with Gaia eDR3}",
      journal = {A\&A},
         year = 2023,
        month = mar,
       volume = {671},
          eid = {L2},
        pages = {L2},
          doi = {10.1051/0004-6361/202245415},
archivePrefix = {arXiv},
       eprint = {2302.06302},
 primaryClass = {astro-ph.EP},
       adsurl = {https://ui.adsabs.harvard.edu/abs/2023A&A...671L...2B}
}

@ARTICLE{beuzit04,
       author = {{Beuzit}, J. -L. and {S{\'e}gransan}, D. and {Forveille}, T. and {Udry}, S. and {Delfosse}, X. and {Mayor}, M. and {Perrier}, C. and {Hainaut}, M. -C. and {Roddier}, C. and {Roddier}, F. and {Mart{\'\i}n}, E.~L.},
        title = "{New neighbours. III. 21 new companions to nearby dwarfs, discovered with adaptive optics}",
      journal = {A\&A},
         year = 2004,
        month = oct,
       volume = {425},
        pages = {997-1008},
          doi = {10.1051/0004-6361:20048006},
archivePrefix = {arXiv},
       eprint = {astro-ph/0106277},
 primaryClass = {astro-ph},
       adsurl = {https://ui.adsabs.harvard.edu/abs/2004A&A...425..997B}
}

@ARTICLE{biller06,
       author = {{Biller}, B.~A. and {Kasper}, M. and {Close}, L.~M. and {Brandner}, W. and {Kellner}, S.},
        title = "{Discovery of a Brown Dwarf Very Close to the Sun: A Methane-rich Brown Dwarf Companion to the Low-Mass Star SCR 1845-6357}",
      journal = {ApJL},
         year = 2006,
        month = apr,
       volume = {641},
       number = {2},
        pages = {L141-L144},
          doi = {10.1086/504256},
archivePrefix = {arXiv},
       eprint = {astro-ph/0601440},
 primaryClass = {astro-ph},
       adsurl = {https://ui.adsabs.harvard.edu/abs/2006ApJ...641L.141B}
}

@ARTICLE{biller22,
       author = {{Biller}, B.~A. and {Grandjean}, A. and {Messina}, S. and {Desidera}, S. and {Delorme}, P. and {Lagrange}, A. -M. and {Hambsch}, F. -J. and {Mesa}, D. and {Janson}, M. and {Gratton}, R. and {D'Orazi}, V. and {Langlois}, M. and {Maire}, A. -L. and {Schlieder}, J. and {Henning}, T. and {Zurlo}, A. and {Hagelberg}, J. and {Brown-Sevilla}, S. and {Romero}, C. and {Bonnefoy}, M. and {Chauvin}, G. and {Feldt}, M. and {Meyer}, M. and {Vigan}, A. and {Pavlov}, A. and {Soenke}, C. and {LeMignant}, D. and {Roux}, A.},
        title = "{Dynamical masses for two M1 + mid-M dwarf binaries monitored during the SPHERE-SHINE survey}",
      journal = {A\&A},
         year = 2022,
        month = feb,
       volume = {658},
          eid = {A145},
        pages = {A145},
          doi = {10.1051/0004-6361/202142438},
archivePrefix = {arXiv},
       eprint = {2112.05457},
 primaryClass = {astro-ph.SR},
       adsurl = {https://ui.adsabs.harvard.edu/abs/2022A&A...658A.145B}
}

@ARTICLE{bird07,
       author = {{Bird}, A.~J. and {Malizia}, A. and {Bazzano}, A. and {Barlow}, E.~J. and {Bassani}, L. and {Hill}, A.~B. and {B{\'e}langer}, G. and {Capitanio}, F. and {Clark}, D.~J. and {Dean}, A.~J. and {Fiocchi}, M. and {G{\"o}tz}, D. and {Lebrun}, F. and {Molina}, M. and {Produit}, N. and {Renaud}, M. and {Sguera}, V. and {Stephen}, J.~B. and {Terrier}, R. and {Ubertini}, P. and {Walter}, R. and {Winkler}, C. and {Zurita}, J.},
        title = "{The Third IBIS/ISGRI Soft Gamma-Ray Survey Catalog}",
      journal = {ApJS},
         year = 2007,
        month = may,
       volume = {170},
       number = {1},
        pages = {175-186},
          doi = {10.1086/513148},
archivePrefix = {arXiv},
       eprint = {astro-ph/0611493},
 primaryClass = {astro-ph},
       adsurl = {https://ui.adsabs.harvard.edu/abs/2007ApJS..170..175B}
}

@ARTICLE{birky20,
       author = {{Birky}, Jessica and {Hogg}, David W. and {Mann}, Andrew W. and {Burgasser}, Adam},
        title = "{Temperatures and Metallicities of M Dwarfs in the APOGEE Survey}",
      journal = {ApJ},
         year = 2020,
        month = mar,
       volume = {892},
       number = {1},
          eid = {31},
        pages = {31},
          doi = {10.3847/1538-4357/ab7004},
archivePrefix = {arXiv},
       eprint = {2001.04962},
 primaryClass = {astro-ph.SR},
       adsurl = {https://ui.adsabs.harvard.edu/abs/2020ApJ...892...31B}
}

@ARTICLE{biro17,
   author = {{Biro}, Susana},
    title = "{La nebulosa de Orión: Una historia visual}",
  journal = {Elementos: Ciencia y Cultura 108},
     year = 2017,
    month = dec,
    pages = {31-37}
}

@INPROCEEDINGS{blaaw91,
       author = {{Blaauw}, Adriaan},
        title = "{OB Associations and the Fossil Record of Star Formation}",
    booktitle = {The Physics of Star Formation and Early Stellar Evolution},
         year = 1991,
       editor = {{Lada}, Charles J. and {Kylafis}, Nikolaos D.},
       series = {NATO Advanced Study Institute (ASI) Series C},
       volume = {342},
        month = jan,
        pages = {125},
       adsurl = {https://ui.adsabs.harvard.edu/abs/1991ASIC..342..125B}
}

@ARTICLE{bochanski07a,
   author = {{Bochanski}, J.~J. and {West}, A.~A. and {Hawley}, S.~L. and 
	{Covey}, K.~R.},
    title = "{Low-Mass Dwarf Template Spectra from the Sloan Digital Sky Survey}",
  journal = {AJ},
   eprint = {arXiv:astro-ph/0610639},
     year = 2007,
    month = feb,
   volume = 133,
    pages = {531-544},
      doi = {10.1086/510240},
   adsurl = {http://cdsads.u-strasbg.fr/abs/2007AJ....133..531B}
}

@ARTICLE{bochanski10,
   author = {{Bochanski}, J.~J. and {Hawley}, S.~L. and {Covey}, K.~R. and 
	{West}, A.~A. and {Reid}, I.~N. and {Golimowski}, D.~A. and 
	{Ivezi{\'c}}, {\v Z}.},
    title = "{The Luminosity and Mass Functions of Low-mass Stars in the Galactic Disk. II. The Field}",
  journal = {AJ},
archivePrefix = "arXiv",
   eprint = {1004.4002},
 primaryClass = "astro-ph.SR",
     year = 2010,
    month = jun,
   volume = 139,
    pages = {2679-2699},
      doi = {10.1088/0004-6256/139/6/2679},
   adsurl = {http://cdsads.u-strasbg.fr/abs/2010AJ....139.2679B}
}

@BOOK{bode1781,
       author = {{Bode}, Johann Elert},
        title = "{Astronomisches Jahrbuch für das Jahr 1784. Nebst einer Sammlung der neuesten in die astronomischen Wissenschaften einschlagenden Abhandlungen, Beobachtungen und Nachrichten}",
         year = 1781,
    publisher = {Gedruckt bey George Jacob Decker Königl. Hofbuchdrucker. Berlin},
}

@ARTICLE{boden05,
       author = {{Boden}, Andrew F. and {Sargent}, Anneila I. and {Akeson}, Rachel L. and {Carpenter}, John M. and {Torres}, Guillermo and {Latham}, David W. and {Soderblom}, David R. and {Nelan}, Ed and {Franz}, Otto G. and {Wasserman}, Lawrence H.},
        title = "{Dynamical Masses for Low-Mass Pre-Main-Sequence Stars: A Preliminary Physical Orbit for HD 98800 B}",
      journal = {ApJ},
         year = 2005,
        month = dec,
       volume = {635},
       number = {1},
        pages = {442-451},
          doi = {10.1086/497328},
archivePrefix = {arXiv},
       eprint = {astro-ph/0508331},
 primaryClass = {astro-ph},
       adsurl = {https://ui.adsabs.harvard.edu/abs/2005ApJ...635..442B}
}

@INPROCEEDINGS{boffin16,
       author = {{Boffin}, Henri M.~J.},
        title = "{An interferometric view of binary stars}",
    booktitle = {Optical and Infrared Interferometry and Imaging V},
         year = 2016,
       editor = {{Malbet}, Fabien and {Creech-Eakman}, Michelle J. and {Tuthill}, Peter G.},
       series = {Society of Photo-Optical Instrumentation Engineers (SPIE) Conference Series},
       volume = {9907},
        month = aug,
          eid = {99073P},
        pages = {99073P},
          doi = {10.1117/12.2232081},
archivePrefix = {arXiv},
       eprint = {1607.07231},
 primaryClass = {astro-ph.SR},
       adsurl = {https://ui.adsabs.harvard.edu/abs/2016SPIE.9907E..3PB}
}

@ARTICLE{bohn22,
       author = {{Bohn}, Alexander J. and {Ginski}, Christian and {Kenworthy}, Matthew A. and {Mamajek}, Eric E. and {Meshkat}, Tiffany and {Pecaut}, Mark J. and {Reggiani}, Maddalena and {Seay}, Christopher R. and {Brown}, Anthony G.~A. and {Cugno}, Gabriele and {Henning}, Thomas and {Launhardt}, Ralf and {Quirrenbach}, Andreas and {Rickman}, Emily L. and {S{\'e}gransan}, Damien},
        title = "{Unveiling wide-orbit companions to K-type stars in Sco-Cen with Gaia EDR3}",
      journal = {A\&A},
         year = 2022,
        month = jan,
       volume = {657},
          eid = {A53},
        pages = {A53},
          doi = {10.1051/0004-6361/202039917},
archivePrefix = {arXiv},
       eprint = {2109.09185},
 primaryClass = {astro-ph.EP},
       adsurl = {https://ui.adsabs.harvard.edu/abs/2022A&A...657A..53B}
}

@ARTICLE{boley09,
       author = {{Boley}, Aaron C.},
        title = "{The Two Modes of Gas Giant Planet Formation}",
      journal = {ApJL},
         year = 2009,
        month = apr,
       volume = {695},
       number = {1},
        pages = {L53-L57},
          doi = {10.1088/0004-637X/695/1/L53},
archivePrefix = {arXiv},
       eprint = {0902.3999},
 primaryClass = {astro-ph.EP},
       adsurl = {https://ui.adsabs.harvard.edu/abs/2009ApJ...695L..53B}
}

@ARTICLE{bolton72,
       author = {{Bolton}, C.~T.},
        title = "{Identification of Cygnus X-1 with HDE 226868}",
      journal = {Nature},
         year = 1972,
        month = feb,
       volume = {235},
       number = {5336},
        pages = {271-273},
          doi = {10.1038/235271b0},
       adsurl = {https://ui.adsabs.harvard.edu/abs/1972Natur.235..271B}
}

@ARTICLE{bonavita07,
       author = {{Bonavita}, M. and {Desidera}, S.},
        title = "{The frequency of planets in multiple systems}",
      journal = {A\&A},
         year = 2007,
        month = jun,
       volume = {468},
       number = {2},
        pages = {721-729},
          doi = {10.1051/0004-6361:20066671},
archivePrefix = {arXiv},
       eprint = {astro-ph/0703754},
 primaryClass = {astro-ph},
       adsurl = {https://ui.adsabs.harvard.edu/abs/2007A&A...468..721B}
}

@ARTICLE{bonavita20,
       author = {{Bonavita}, Mariangela and {Desidera}, Silvano},
        title = "{Frequency of Planets in Binaries}",
      journal = {Galaxies},
         year = 2020,
        month = feb,
       volume = {8},
       number = {1},
        pages = {16},
          doi = {10.3390/galaxies8010016},
archivePrefix = {arXiv},
       eprint = {2002.11734},
 primaryClass = {astro-ph.EP},
       adsurl = {https://ui.adsabs.harvard.edu/abs/2020Galax...8...16B}
}

@ARTICLE{bond1848,
       author = {{Bond}, W.~C.},
        title = "{Description of the Nebula about the Star {\ensuremath{\theta}} Orionis}",
      journal = {Mem. Am. Acad. Arts Sci.},
         year = 1848,
        month = jan,
       volume = {3},
        pages = {87},
          doi = {10.2307/25058145},
       adsurl = {https://ui.adsabs.harvard.edu/abs/1848MAAAS...3...87B}
}

@ARTICLE{bond15,
       author = {{Bond}, Howard E. and {Gilliland}, Ronald L. and {Schaefer}, Gail H. and {Demarque}, Pierre and {Girard}, Terrence M. and {Holberg}, Jay B. and {Gudehus}, Donald and {Mason}, Brian D. and {Kozhurina-Platais}, Vera and {Burleigh}, Matthew R. and {Barstow}, Martin A. and {Nelan}, Edmund P.},
        title = "{Hubble Space Telescope Astrometry of the Procyon System}",
      journal = {ApJ},
         year = 2015,
        month = nov,
       volume = {813},
       number = {2},
          eid = {106},
        pages = {106},
          doi = {10.1088/0004-637X/813/2/106},
archivePrefix = {arXiv},
       eprint = {1510.00485},
 primaryClass = {astro-ph.SR},
       adsurl = {https://ui.adsabs.harvard.edu/abs/2015ApJ...813..106B}
}

@ARTICLE{bond17,
       author = {{Bond}, Howard E. and {Schaefer}, Gail H. and {Gilliland}, Ronald L. and {Holberg}, Jay B. and {Mason}, Brian D. and {Lindenblad}, Irving W. and {Seitz-McLeese}, Miranda and {Arnett}, W. David and {Demarque}, Pierre and {Spada}, Federico and {Young}, Patrick A. and {Barstow}, Martin A. and {Burleigh}, Matthew R. and {Gudehus}, Donald},
        title = "{The Sirius System and Its Astrophysical Puzzles: Hubble Space Telescope and Ground-based Astrometry}",
      journal = {ApJ},
         year = 2017,
        month = may,
       volume = {840},
       number = {2},
          eid = {70},
        pages = {70},
          doi = {10.3847/1538-4357/aa6af8},
archivePrefix = {arXiv},
       eprint = {1703.10625},
 primaryClass = {astro-ph.SR},
       adsurl = {https://ui.adsabs.harvard.edu/abs/2017ApJ...840...70B}
}

@ARTICLE{bond20,
       author = {{Bond}, Howard E. and {Schaefer}, Gail H. and {Gilliland}, Ronald L. and {VandenBerg}, Don A.},
        title = "{Hubble Space Telescope Astrometry of the Metal-poor Visual Binary {\ensuremath{\mu}} Cassiopeiae: Dynamical Masses, Helium Content, and Age}",
      journal = {ApJ},
         year = 2020,
        month = dec,
       volume = {904},
       number = {2},
          eid = {112},
        pages = {112},
          doi = {10.3847/1538-4357/abc172},
archivePrefix = {arXiv},
       eprint = {2010.06609},
 primaryClass = {astro-ph.SR},
       adsurl = {https://ui.adsabs.harvard.edu/abs/2020ApJ...904..112B}
}

@ARTICLE{bonfils05a,
   author = {{Bonfils}, X. and {Delfosse}, X. and {Udry}, S. and {Santos}, N.~C. and 
	{Forveille}, T. and {S{\'e}gransan}, D.},
    title = "{Metallicity of M dwarfs. I. A photometric calibration and the impact on the mass-luminosity relation at the bottom of the main sequence}",
  journal = {A\&A},
   eprint = {astro-ph/0503260},
     year = 2005,
    month = nov,
   volume = 442,
    pages = {635-642},
      doi = {10.1051/0004-6361:20053046},
   adsurl = {http://cdsads.u-strasbg.fr/abs/2005A&A...442..635B}
}

@ARTICLE{bonfils13,
       author = {{Bonfils}, X. and {Delfosse}, X. and {Udry}, S. and {Forveille}, T. and {Mayor}, M. and {Perrier}, C. and {Bouchy}, F. and {Gillon}, M. and {Lovis}, C. and {Pepe}, F. and {Queloz}, D. and {Santos}, N.~C. and {S{\'e}gransan}, D. and {Bertaux}, J. -L.},
        title = "{The HARPS search for southern extra-solar planets. XXXI. The M-dwarf sample}",
      journal = {A\&A},
         year = 2013,
        month = jan,
       volume = {549},
          eid = {A109},
        pages = {A109},
          doi = {10.1051/0004-6361/201014704},
archivePrefix = {arXiv},
       eprint = {1111.5019},
 primaryClass = {astro-ph.EP},
       adsurl = {https://ui.adsabs.harvard.edu/abs/2013A&A...549A.109B}
}

@ARTICLE{bonnarel00,
   author = {{Bonnarel}, F. and {Fernique}, P. and {Bienaym{\'e}}, O. and 
	{Egret}, D. and {Genova}, F. and {Louys}, M. and {Ochsenbein}, F. and 
	{Wenger}, M. and {Bartlett}, J.~G.},
    title = "{The ALADIN interactive sky atlas. A reference tool for identification of astronomical sources}",
  journal = {A\&AS},
     year = 2000,
    month = apr,
   volume = 143,
    pages = {33-40},
      doi = {10.1051/aas:2000331},
   adsurl = {http://cdsads.u-strasbg.fr/abs/2000A&AS..143...33B}
}

@ARTICLE{bordier24,
       author = {{Bordier}, E. and {de Wit}, W. -J. and {Frost}, A.~J. and {Sana}, H. and {Pauwels}, T. and {Koumpia}, E.},
        title = "{The onset of stellar multiplicity in massive star formation: A search for low-mass companions of massive young stellar objects with L′-band adaptive optics imaging}",
      journal = {A\&A},
         year = 2024,
        month = jan,
       volume = {681},
          eid = {A85},
        pages = {A85},
          doi = {10.1051/0004-6361/202347548},
archivePrefix = {arXiv},
       eprint = {2311.06131},
 primaryClass = {astro-ph.SR},
       adsurl = {https://ui.adsabs.harvard.edu/abs/2024A&A...681A..85B}
}

@ARTICLE{borodina21,
       author = {{Borodina}, Olga I. and {Carraro}, Giovanni and {Seleznev}, Anton F. and {Danilov}, Vladimir M.},
        title = "{Unresolved Multiple Stars and Galactic Clusters' Mass Estimates}",
      journal = {ApJ},
         year = 2021,
        month = feb,
       volume = {908},
       number = {1},
          eid = {60},
        pages = {60},
          doi = {10.3847/1538-4357/abd562},
archivePrefix = {arXiv},
       eprint = {2101.01417},
 primaryClass = {astro-ph.SR},
       adsurl = {https://ui.adsabs.harvard.edu/abs/2021ApJ...908...60B}
}

@ARTICLE{borucki10,
       author = {{Borucki}, William J. and {Koch}, David and {Basri}, Gibor and {Batalha}, Natalie and {Brown}, Timothy and {Caldwell}, Douglas and {Caldwell}, John and {Christensen-Dalsgaard}, J{\o}rgen and {Cochran}, William D. and {DeVore}, Edna and {Dunham}, Edward W. and {Dupree}, Andrea K. and {Gautier}, Thomas N. and {Geary}, John C. and {Gilliland}, Ronald and {Gould}, Alan and {Howell}, Steve B. and {Jenkins}, Jon M. and {Kondo}, Yoji and {Latham}, David W. and {Marcy}, Geoffrey W. and {Meibom}, S{\o}ren and {Kjeldsen}, Hans and {Lissauer}, Jack J. and {Monet}, David G. and {Morrison}, David and {Sasselov}, Dimitar and {Tarter}, Jill and {Boss}, Alan and {Brownlee}, Don and {Owen}, Toby and {Buzasi}, Derek and {Charbonneau}, David and {Doyle}, Laurance and {Fortney}, Jonathan and {Ford}, Eric B. and {Holman}, Matthew J. and {Seager}, Sara and {Steffen}, Jason H. and {Welsh}, William F. and {Rowe}, Jason and {Anderson}, Howard and {Buchhave}, Lars and {Ciardi}, David and {Walkowicz}, Lucianne and {Sherry}, William and {Horch}, Elliott and {Isaacson}, Howard and {Everett}, Mark E. and {Fischer}, Debra and {Torres}, Guillermo and {Johnson}, John Asher and {Endl}, Michael and {MacQueen}, Phillip and {Bryson}, Stephen T. and {Dotson}, Jessie and {Haas}, Michael and {Kolodziejczak}, Jeffrey and {Van Cleve}, Jeffrey and {Chandrasekaran}, Hema and {Twicken}, Joseph D. and {Quintana}, Elisa V. and {Clarke}, Bruce D. and {Allen}, Christopher and {Li}, Jie and {Wu}, Haley and {Tenenbaum}, Peter and {Verner}, Ekaterina and {Bruhweiler}, Frederick and {Barnes}, Jason and {Prsa}, Andrej},
        title = "{Kepler Planet-Detection Mission: Introduction and First Results}",
      journal = {Science},
         year = 2010,
        month = feb,
       volume = {327},
       number = {5968},
        pages = {977},
          doi = {10.1126/science.1185402},
       adsurl = {https://ui.adsabs.harvard.edu/abs/2010Sci...327..977B}
}

@ARTICLE{bouvier08,
       author = {{Bouvier}, J. and {Kendall}, T. and {Meeus}, G. and {Testi}, L. and {Moraux}, E. and {Stauffer}, J.~R. and {James}, D. and {Cuillandre}, J. -C. and {Irwin}, J. and {McCaughrean}, M.~J. and {Baraffe}, I. and {Bertin}, E.},
        title = "{Brown dwarfs and very low mass stars in the Hyades cluster: a dynamically evolved mass function}",
      journal = {A\&A},
         year = 2008,
        month = apr,
       volume = {481},
       number = {3},
        pages = {661-672},
          doi = {10.1051/0004-6361:20079303},
archivePrefix = {arXiv},
       eprint = {0801.0670},
 primaryClass = {astro-ph},
       adsurl = {https://ui.adsabs.harvard.edu/abs/2008A&A...481..661B}
}

@ARTICLE{bowler09,
       author = {{Bowler}, Brendan P. and {Liu}, Michael C. and {Cushing}, Michael C.},
        title = "{The Benchmark Ultracool Subdwarf HD 114762B: A Test of Low-metallicity Atmospheric and Evolutionary Models}",
      journal = {ApJ},
         year = 2009,
        month = dec,
       volume = {706},
       number = {2},
        pages = {1114-1135},
          doi = {10.1088/0004-637X/706/2/1114},
archivePrefix = {arXiv},
       eprint = {0910.1604},
 primaryClass = {astro-ph.SR},
       adsurl = {https://ui.adsabs.harvard.edu/abs/2009ApJ...706.1114B}
}

@ARTICLE{bowler10a,
   author = {{Bowler}, B.~P. and {Liu}, M.~C. and {Dupuy}, T.~J.},
    title = "{SDSS J141624.08+134826.7: A Nearby Blue L Dwarf From the Sloan Digital Sky Survey}",
  journal = {ApJ},
archivePrefix = "arXiv",
   eprint = {0912.3796},
 primaryClass = "astro-ph.SR",
     year = 2010,
    month = feb,
   volume = 710,
    pages = {45-50},
      doi = {10.1088/0004-637X/710/1/45},
   adsurl = {http://cdsads.u-strasbg.fr/abs/2010ApJ...710...45B}
}

@ARTICLE{brandeker19,
       author = {{Brandeker}, Alexis and {Cataldi}, Gianni},
        title = "{Contrast sensitivities in the Gaia Data Release 2}",
      journal = {A\&A},
         year = 2019,
        month = jan,
       volume = {621},
          eid = {A86},
        pages = {A86},
          doi = {10.1051/0004-6361/201834321},
archivePrefix = {arXiv},
       eprint = {1811.05301},
 primaryClass = {astro-ph.IM},
       adsurl = {https://ui.adsabs.harvard.edu/abs/2019A&A...621A..86B}
}

@ARTICLE{brandt21,
       author = {{Brandt}, Timothy D.},
        title = "{The Hipparcos-Gaia Catalog of Accelerations: Gaia EDR3 Edition}",
      journal ={ApJS},
         year = 2021,
        month = jun,
       volume = {254},
       number = {2},
          eid = {42},
        pages = {42},
          doi = {10.3847/1538-4365/abf93c},
archivePrefix = {arXiv},
       eprint = {2105.11662},
 primaryClass = {astro-ph.GA},
       adsurl = {https://ui.adsabs.harvard.edu/abs/2021ApJS..254...42B}
}

@ARTICLE{bressan12,
       author = {{Bressan}, Alessandro and {Marigo}, Paola and {Girardi}, L{\'e}o. and {Salasnich}, Bernardo and {Dal Cero}, Claudia and {Rubele}, Stefano and {Nanni}, Ambra},
        title = "{PARSEC: stellar tracks and isochrones with the PAdova and TRieste Stellar Evolution Code}",
      journal ={MNRAS},
         year = 2012,
        month = nov,
       volume = {427},
       number = {1},
        pages = {127-145},
          doi = {10.1111/j.1365-2966.2012.21948.x},
archivePrefix = {arXiv},
       eprint = {1208.4498},
 primaryClass = {astro-ph.SR},
       adsurl = {https://ui.adsabs.harvard.edu/abs/2012MNRAS.427..127B}
}

@ARTICLE{bromley21,
       author = {{Bromley}, Benjamin C. and {Kenyon}, Scott J.},
        title = "{On the Estimation of Circumbinary Orbital Properties}",
      journal = {AJ},
         year = 2021,
        month = jan,
       volume = {161},
       number = {1},
          eid = {25},
        pages = {25},
          doi = {10.3847/1538-3881/abcbfb},
archivePrefix = {arXiv},
       eprint = {2011.13376},
 primaryClass = {astro-ph.EP},
       adsurl = {https://ui.adsabs.harvard.edu/abs/2021AJ....161...25B}
}

@INPROCEEDINGS{brooks15,
       author = {{Brooks}, Thomas and {Stahl}, H.~P. and {Arnold}, William R.},
        title = "{Advanced Mirror Technology Development (AMTD) thermal trade studies}",
    booktitle = {Optical Modeling and Performance Predictions VII},
         year = 2015,
       editor = {{Kahan}, Mark A. and {Levine-West}, Marie B.},
       series = {Society of Photo-Optical Instrumentation Engineers (SPIE) Conference Series},
       volume = {9577},
        month = sep,
        pages = {E3},
          doi = {10.1117/12.2188371},
       adsurl = {https://ui.adsabs.harvard.edu/abs/2015SPIE.9577E..03B}
}

@ARTICLE{bruch98,
       author = {{Bruch}, Albert and {Diaz}, Marcos P.},
        title = "{The Eclipsing Precataclysmic Binary RR Caeli}",
      journal = {AJ},
         year = 1998,
        month = aug,
       volume = {116},
       number = {2},
        pages = {908-916},
          doi = {10.1086/300471},
       adsurl = {https://ui.adsabs.harvard.edu/abs/1998AJ....116..908B}
}

@ARTICLE{burgasser00,
       author = {{Burgasser}, Adam J. and {Kirkpatrick}, J. Davy and {Cutri}, Roc M. and {McCallon}, Howard and {Kopan}, Gene and {Gizis}, John E. and {Liebert}, James and {Reid}, I. Neill and {Brown}, Michael E. and {Monet}, David G. and {Dahn}, Conard C. and {Beichman}, Charles A. and {Skrutskie}, Michael F.},
        title = "{Discovery of a Brown Dwarf Companion to Gliese 570ABC: A 2MASS T Dwarf Significantly Cooler than Gliese 229B}",
      journal = {ApJL},
         year = 2000,
        month = mar,
       volume = {531},
       number = {1},
        pages = {L57-L60},
          doi = {10.1086/312522},
archivePrefix = {arXiv},
       eprint = {astro-ph/0001194},
 primaryClass = {astro-ph},
       adsurl = {https://ui.adsabs.harvard.edu/abs/2000ApJ...531L..57B}
}

@ARTICLE{burgasser03b,
    author = {{Burgasser}, A.~J. and {Kirkpatrick}, J.~D. and {Burrows}, A. and 
	{Liebert}, J. and {Reid}, I.~N. and {Gizis}, J.~E. and {McGovern}, M.~R. and 
	{Prato}, L. and {McLean}, I.~S.},
    title = "{The First Substellar Subdwarf? Discovery of a Metal-poor L Dwarf with Halo Kinematics}",
    journal = {ApJ},
    year = 2003,
    month = aug,
    volume = 592,
    pages = {1186-1192},
    adsurl = {http://cdsads.u-strasbg.fr/cgi-bin/nph-bib_query?bibcode=2003ApJ...592.1186B&amp;db_key=AST}
}

@ARTICLE{burgasser03c,
       author = {{Burgasser}, Adam J. and {Kirkpatrick}, J. Davy and {Reid}, I. Neill and {Brown}, Michael E. and {Miskey}, Cherie L. and {Gizis}, John E.},
        title = "{Binarity in Brown Dwarfs: T Dwarf Binaries Discovered with the Hubble Space Telescope Wide Field Planetary Camera 2}",
      journal = {ApJ},
         year = 2003,
        month = mar,
       volume = {586},
       number = {1},
        pages = {512-526},
          doi = {10.1086/346263},
archivePrefix = {arXiv},
       eprint = {astro-ph/0211470},
 primaryClass = {astro-ph},
       adsurl = {https://ui.adsabs.harvard.edu/abs/2003ApJ...586..512B}
}

@ARTICLE{burgasser04b,
   author = {{Burgasser}, A.~J. and {Kirkpatrick}, J.~D. and {McGovern}, M.~R. and 
	{McLean}, I.~S. and {Prato}, L. and {Reid}, I.~N.},
    title = "{S Orionis 70: Just a Foreground Field Brown Dwarf?}",
  journal = {ApJ},
   eprint = {astro-ph/0312285},
     year = 2004,
    month = apr,
   volume = 604,
    pages = {827-831},
      doi = {10.1086/382129},
   adsurl = {http://cdsads.u-strasbg.fr/cgi-bin/nph-bib_query?bibcode=2004ApJ...604..827B&db_key=AST}
}

@ARTICLE{burgasser05,
       author = {{Burgasser}, Adam J. and {Kirkpatrick}, J. Davy and {Lowrance}, Patrick J.},
        title = "{Multiplicity among Widely Separated Brown Dwarf Companions to Nearby Stars: Gliese 337CD}",
      journal = {AJ},
         year = 2005,
        month = jun,
       volume = {129},
       number = {6},
        pages = {2849-2855},
          doi = {10.1086/430218},
archivePrefix = {arXiv},
       eprint = {astro-ph/0503379},
 primaryClass = {astro-ph},
       adsurl = {https://ui.adsabs.harvard.edu/abs/2005AJ....129.2849B}
}

@INPROCEEDINGS{burgasser07a,
       author = {{Burgasser}, A.~J. and {Reid}, I.~N. and {Siegler}, N. and {Close}, L. and {Allen}, P. and {Lowrance}, P. and {Gizis}, J.},
        title = "{Not Alone: Tracing the Origins of Very-Low-Mass Stars and Brown Dwarfs Through Multiplicity Studies}",
    booktitle = {Protostars and Planets V},
         year = 2007,
       editor = {{Reipurth}, Bo and {Jewitt}, David and {Keil}, Klaus},
        month = jan,
        pages = {427},
archivePrefix = {arXiv},
       eprint = {astro-ph/0602122},
 primaryClass = {astro-ph},
       adsurl = {https://ui.adsabs.harvard.edu/abs/2007prpl.conf..427B}
}

@ARTICLE{burgasser07b,
       author = {{Burgasser}, Adam J.},
        title = "{Binaries and the L Dwarf/T Dwarf Transition}",
      journal = {ApJ},
         year = 2007,
        month = apr,
       volume = {659},
       number = {1},
        pages = {655-674},
          doi = {10.1086/511027},
archivePrefix = {arXiv},
       eprint = {astro-ph/0611505},
 primaryClass = {astro-ph},
       adsurl = {https://ui.adsabs.harvard.edu/abs/2007ApJ...659..655B}
}

@ARTICLE{burgasser08,
       author = {{Burgasser}, Adam J. and {Tinney}, C.~G. and {Cushing}, Michael C. and {Saumon}, Didier and {Marley}, Mark S. and {Bennett}, Clara S. and {Kirkpatrick}, J. Davy},
        title = "{2MASS J09393548-2448279: The Coldest and Least Luminous Brown Dwarf Binary Known?}",
      journal = {ApJL},
         year = 2008,
        month = dec,
       volume = {689},
       number = {1},
        pages = {L53},
          doi = {10.1086/595747},
       adsurl = {https://ui.adsabs.harvard.edu/abs/2008ApJ...689L..53B}
}

@ARTICLE{burgasser08b,
       author = {{Burgasser}, Adam J.},
        title = "{Brown dwarfs: Failed stars, super Jupiters}",
      journal = {Phys. Today},
         year = 2008,
        month = jan,
       volume = {61},
       number = {6},
        pages = {70},
          doi = {10.1063/1.2947658},
       adsurl = {https://ui.adsabs.harvard.edu/abs/2008PhT....61f..70B}
}

@ARTICLE{burgasser15,
       author = {{Burgasser}, Adam J. and {Gillon}, Micha{\"e}l and {Melis}, Carl and {Bowler}, Brendan P. and {Michelsen}, Eric L. and {Bardalez Gagliuffi}, Daniella and {Gelino}, Christopher R. and {Jehin}, E. and {Delrez}, L. and {Manfroid}, J. and {Blake}, Cullen H.},
        title = "{WISE J072003.20-084651.2: an Old and Active M9.5 + T5 Spectral Binary 6 pc from the Sun}",
      journal = {AJ},
         year = 2015,
        month = mar,
       volume = {149},
       number = {3},
          eid = {104},
        pages = {104},
          doi = {10.1088/0004-6256/149/3/104},
archivePrefix = {arXiv},
       eprint = {1410.4288},
 primaryClass = {astro-ph.SR},
       adsurl = {https://ui.adsabs.harvard.edu/abs/2015AJ....149..104B}
}

@ARTICLE{burgay03,
       author = {{Burgay}, M. and {D'Amico}, N. and {Possenti}, A. and {Manchester}, R.~N. and {Lyne}, A.~G. and {Joshi}, B.~C. and {McLaughlin}, M.~A. and {Kramer}, M. and {Sarkissian}, J.~M. and {Camilo}, F. and {Kalogera}, V. and {Kim}, C. and {Lorimer}, D.~R.},
        title = "{An increased estimate of the merger rate of double neutron stars from observations of a highly relativistic system}",
      journal = {Nature},
         year = 2003,
        month = dec,
       volume = {426},
       number = {6966},
        pages = {531-533},
          doi = {10.1038/nature02124},
archivePrefix = {arXiv},
       eprint = {astro-ph/0312071},
 primaryClass = {astro-ph},
       adsurl = {https://ui.adsabs.harvard.edu/abs/2003Natur.426..531B}
}

@ARTICLE{burkart14,
       author = {{Burkart}, Joshua and {Quataert}, Eliot and {Arras}, Phil},
        title = "{Dynamical resonance locking in tidally interacting binary systems}",
      journal = {MNRAS},
         year = 2014,
        month = oct,
       volume = {443},
       number = {4},
        pages = {2957-2973},
          doi = {10.1093/mnras/stu1366},
archivePrefix = {arXiv},
       eprint = {1312.4966},
 primaryClass = {astro-ph.SR},
       adsurl = {https://ui.adsabs.harvard.edu/abs/2014MNRAS.443.2957B}
}

@BOOK{burnham1906,
       author = {{Burnham}, Sherburne Wesley},
        title = "{A General Catalogue of Double Stars within 121{\textdegree} of the North Pole}",
         year = 1906,
       adsurl = {https://ui.adsabs.harvard.edu/abs/1906gcds.book.....B},
    publisher = {Carnegie institution of Washington. University of Chicago Press}
}

@ARTICLE{burningham10a,
   author = {{Burningham}, B. and {Leggett}, S.~K. and {Lucas}, P.~W. and 
	{Pinfield}, D.~J. and {Smart}, R.~L. and {Day-Jones}, A.~C. and 
	{8 co-authors}},
    title = "{The discovery of a very cool binary system}",
  journal = {MNRAS},
archivePrefix = "arXiv",
   eprint = {1001.4393},
 primaryClass = "astro-ph.SR",
     year = 2010,
    month = jun,
   volume = 404,
    pages = {1952-1961},
      doi = {10.1111/j.1365-2966.2010.16411.x},
   adsurl = {http://cdsads.u-strasbg.fr/abs/2010MNRAS.404.1952B}
}

@ARTICLE{burrows89,
       author = {{Burrows}, Adam and {Hubbard}, W.~B. and {Lunine}, Jonathan I.},
        title = "{Theoretical Models of Very Low Mass Stars and Brown Dwarfs}",
      journal = {ApJ},
         year = 1989,
        month = oct,
       volume = {345},
        pages = {939},
          doi = {10.1086/167964},
       adsurl = {https://ui.adsabs.harvard.edu/abs/1989ApJ...345..939B}
}

@ARTICLE{burrows02,
       author = {{Burrows}, Adam and {Burgasser}, Adam J. and {Kirkpatrick}, J. Davy and {Liebert}, James and {Milsom}, J.~A. and {Sudarsky}, D. and {Hubeny}, I.},
        title = "{Theoretical Spectral Models of T Dwarfs at Short Wavelengths and Their Comparison with Data}",
      journal = {ApJ},
         year = 2002,
        month = jul,
       volume = {573},
       number = {1},
        pages = {394-417},
          doi = {10.1086/340584},
archivePrefix = {arXiv},
       eprint = {astro-ph/0109227},
 primaryClass = {astro-ph},
       adsurl = {https://ui.adsabs.harvard.edu/abs/2002ApJ...573..394B}
}

@ARTICLE{busetti18,
       author = {{Busetti}, F. and {Beust}, H. and {Harley}, C.},
        title = "{Stability of planets in triple star systems}",
      journal = {A\&A},
         year = 2018,
        month = nov,
       volume = {619},
          eid = {A91},
        pages = {A91},
          doi = {10.1051/0004-6361/201833097},
archivePrefix = {arXiv},
       eprint = {1811.08221},
 primaryClass = {astro-ph.EP},
       adsurl = {https://ui.adsabs.harvard.edu/abs/2018A&A...619A..91B}
}

@ARTICLE{byrne84,
       author = {{Byrne}, P.~B. and {Beesley}, D.~E. and {Dunsby}, P.},
        title = "{An Amateur Filar Micrometer (An Evaluation)}",
      journal = {Ir. Astron. J.},
         year = 1984,
        month = mar,
       volume = {16},
        pages = {181},
       adsurl = {https://ui.adsabs.harvard.edu/abs/1984IrAJ...16..181B}
}

@ARTICLE{caballero07a,
       author = {{Caballero}, J.~A.},
        title = "{The widest ultracool binary}",
      journal ={A\&A},
         year = 2007,
        month = feb,
       volume = {462},
       number = {3},
        pages = {L61-L64},
          doi = {10.1051/0004-6361:20066814},
archivePrefix = {arXiv},
       eprint = {astro-ph/0612234},
 primaryClass = {astro-ph},
       adsurl = {https://ui.adsabs.harvard.edu/abs/2007A&A...462L..61C}
}

@ARTICLE{caballero07b,
       author = {{Caballero}, Jos{\'e} Antonio},
        title = "{Southern Very Low Mass Stars and Brown Dwarfs in Wide Binary and Multiple Systems}",
      journal ={ApJ},
         year = 2007,
        month = sep,
       volume = {667},
       number = {1},
        pages = {520-526},
          doi = {10.1086/520873},
archivePrefix = {arXiv},
       eprint = {0706.1346},
 primaryClass = {astro-ph},
       adsurl = {https://ui.adsabs.harvard.edu/abs/2007ApJ...667..520C}
}

@ARTICLE{caballero08,
       author = {{Caballero}, Jos{\'e} A.},
        title = "{Spatial distribution of stars and brown dwarfs in {\ensuremath{\sigma}} Orionis}",
      journal ={MNRAS},
         year = 2008,
        month = jan,
       volume = {383},
       number = {1},
        pages = {375-382},
          doi = {10.1111/j.1365-2966.2007.12555.x},
archivePrefix = {arXiv},
       eprint = {0710.1255},
 primaryClass = {astro-ph},
       adsurl = {https://ui.adsabs.harvard.edu/abs/2008MNRAS.383..375C}
}

@ARTICLE{caballero09,
       author = {{Caballero}, J.~A.},
        title = "{Reaching the boundary between stellar kinematic groups and very wide binaries. The Washington double stars with the widest angular separations}",
      journal = {A\&A},
         year = 2009,
        month = nov,
       volume = {507},
       number = {1},
        pages = {251-259},
          doi = {10.1051/0004-6361/200912596},
archivePrefix = {arXiv},
       eprint = {0908.2761},
 primaryClass = {astro-ph.SR},
       adsurl = {https://ui.adsabs.harvard.edu/abs/2009A&A...507..251C}
}

@ARTICLE{caballero10,
       author = {{Caballero}, J.~A.},
        title = "{Reaching the boundary between stellar kinematic groups and very wide binaries . II. {\ensuremath{\alpha}} Librae + KU Librae: a common proper motion system in Castor separated by 1.0 pc}",
      journal = {A\&A},
         year = 2010,
        month = may,
       volume = {514},
          eid = {A98},
        pages = {A98},
          doi = {10.1051/0004-6361/200913986},
archivePrefix = {arXiv},
       eprint = {1001.5432},
 primaryClass = {astro-ph.SR},
       adsurl = {https://ui.adsabs.harvard.edu/abs/2010A&A...514A..98C}
}

@INPROCEEDINGS{caballero13,
       author = {{Caballero}, J.~A. and {Genebriera}, J. and {Tobal}, T. and {Miret}, F.~X. and {Rica}, F.~M. and {Cairol}, J. and {Miret}, N. and {Novalbos}, I. and {Montes}, D. and {Klutsch}, A.},
        title = "{Collaborating with ``professional'' amateurs: low-mass stars in fragile multiple systems}",
    booktitle = {Highlights of Spanish Astrophysics VII},
         year = 2013,
       editor = {{Guirado}, J.~C. and {Lara}, L.~M. and {Quilis}, V. and {Gorgas}, J.},
        month = may,
        pages = {971-976},
archivePrefix = {arXiv},
       eprint = {1211.2934},
 primaryClass = {astro-ph.SR},
       adsurl = {https://ui.adsabs.harvard.edu/abs/2013hsa7.conf..971C}
}

@ARTICLE{caballero14,
       author = {{Caballero}, J.~A.},
        title = "{Stellar multiplicity in the sigma Orionis cluster: a review}",
      journal = {The Observatory},
         year = 2014,
        month = oct,
       volume = {134},
        pages = {273-287},
          doi = {10.48550/arXiv.1408.2231},
archivePrefix = {arXiv},
       eprint = {1408.2231},
 primaryClass = {astro-ph.SR},
       adsurl = {https://ui.adsabs.harvard.edu/abs/2014Obs...134..273C}
}

@INPROCEEDINGS{caballero16,
       author = {{Caballero}, J.~A. and {Cort{\'e}s-Contreras}, M. and {Alonso-Floriano}, F.~J. and {Montes}, D. and {Quirrenbach}, A. and {Amado}, P.~J. and {Ribas}, I. and {Reiners}, A. and {Abellan}, F.~J. and {B{\'e}jar}, V.~J.~S. and {Brinkm{\"o}ller}, M. and {Czesla}, S. and {Dorda}, R. and {Gallardo}, I. and {Gonz{\'a}lez-{\'A}lvarez}, E. and {Hidalgo}, D. and {Holgado}, G. and {Jeffers}, S.~V. and {Kim}, M. and {Klutsch}, A. and {Lamert}, A. and {Llamas}, M. and {L{\'o}pez-Santiago}, J. and {Mart{\'\i}nez-Rodr{\'\i}guez}, H. and {Morales}, J.~C. and {Mundt}, R. and {Passegger}, V.~M. and {Sch{\"o}fer}, P. and {Seifert}, W. and {Zechmeister}, M.},
        title = "{Carmencita, The CARMENES Input Catalogue of Bright, Nearby M Dwarfs}",
    booktitle = {19th Cambridge Workshop on Cool Stars, Stellar Systems, and the Sun (CS19)},
         year = 2016,
        month = aug,
          eid = {148},
        pages = {148},
          doi = {10.5281/zenodo.60060},
       adsurl = {https://ui.adsabs.harvard.edu/abs/2016csss.confE.148C}
}

@ARTICLE{caballero18,
       author = {{Caballero}, Jos{\'e} A.},
        title = "{A Review on Substellar Objects below the Deuterium Burning Mass Limit: Planets, Brown Dwarfs or What?}",
      journal = {Geosciences},
         year = 2018,
        month = sep,
       volume = {8},
       number = {10},
        pages = {362},
          doi = {10.3390/geosciences8100362},
archivePrefix = {arXiv},
       eprint = {1808.07798},
 primaryClass = {astro-ph.SR},
       adsurl = {https://ui.adsabs.harvard.edu/abs/2018Geosc...8..362C}
}

@ARTICLE{caballero22,
       author = {{Caballero}, J.~A. and {Gonz{\'a}lez-{\'A}lvarez}, E. and {Brady}, M. and {Trifonov}, T. and {Ellis}, T.~G. and {Dorn}, C. and {Cifuentes}, C. and {Molaverdikhani}, K. and {Bean}, J.~L. and {Boyajian}, T. and {Rodr{\'\i}guez}, E. and {Sanz-Forcada}, J. and {Zapatero Osorio}, M.~R. and {Abia}, C. and {Amado}, P.~J. and {Anugu}, N. and {B{\'e}jar}, V.~J.~S. and {Davies}, C.~L. and {Dreizler}, S. and {Dubois}, F. and {Ennis}, J. and {Espinoza}, N. and {Farrington}, C.~D. and {L{\'o}pez}, A. Garc{\'\i}a and {Gardner}, T. and {Hatzes}, A.~P. and {Henning}, Th. and {Herrero}, E. and {Herrero-Cisneros}, E. and {Kaminski}, A. and {Kasper}, D. and {Klement}, R. and {Kraus}, S. and {Labdon}, A. and {Lanthermann}, C. and {Le Bouquin}, J. -B. and {L{\'o}pez Gonz{\'a}lez}, M.~J. and {Luque}, R. and {Mann}, A.~W. and {Marfil}, E. and {Monnier}, J.~D. and {Montes}, D. and {Morales}, J.~C. and {Pall{\'e}}, E. and {Pedraz}, S. and {Quirrenbach}, A. and {Reffert}, S. and {Reiners}, A. and {Ribas}, I. and {Rodr{\'\i}guez-L{\'o}pez}, C. and {Schaefer}, G. and {Schweitzer}, A. and {Seifahrt}, A. and {Setterholm}, B.~R. and {Shan}, Y. and {Shulyak}, D. and {Solano}, E. and {Sreenivas}, K.~R. and {Stef{\'a}nsson}, G. and {St{\"u}rmer}, J. and {Tabernero}, H.~M. and {Tal-Or}, L. and {ten Brummelaar}, T. and {Vanaverbeke}, S. and {von Braun}, K. and {Youngblood}, A. and {Zechmeister}, M.},
        title = "{A detailed analysis of the Gl 486 planetary system}",
      journal = {A\&A},
         year = 2022,
        month = sep,
       volume = {665},
          eid = {A120},
        pages = {A120},
          doi = {10.1051/0004-6361/202243548},
archivePrefix = {arXiv},
       eprint = {2206.09990},
 primaryClass = {astro-ph.EP},
       adsurl = {https://ui.adsabs.harvard.edu/abs/2022A&A...665A.120C}
}

@INPROCEEDINGS{caballero24,
       author = {{Caballero}, J. A. and {Ribas}, I. and {Reiners}, A. and {Quirrenbach}, A. and {Amado}, P.},
        title = "{CARMENES: The Importance of Being Complete}",
    booktitle = {The 22nd Cambridge Workshop on Cool Stars, Stellar Systems, and the Sun (San Diego, June 2024)},
         year = 2024,
       series = {},
       volume = {},
        month = jun,
          eid = {},
        pages = {},
       adsurl = {https://doi.org/10.5281/zenodo.13959771}
}

@ARTICLE{calamari24,
       author = {{Calamari}, Emily and {Faherty}, Jacqueline K. and {Visscher}, Channon and {Gemma}, Marina E. and {Burningham}, Ben and {Rothermich}, Austin},
        title = "{Predicting Cloud Conditions in Substellar Mass Objects Using Ultracool Dwarf Companions}",
      journal = {ApJ},
         year = 2024,
        month = mar,
       volume = {963},
       number = {1},
          eid = {67},
        pages = {67},
          doi = {10.3847/1538-4357/ad1f6d},
       adsurl = {https://ui.adsabs.harvard.edu/abs/2024ApJ...963...67C}
}

@ARTICLE{campante15,
       author = {{Campante}, T.~L. and {Barclay}, T. and {Swift}, J.~J. and {Huber}, D. and {Adibekyan}, V. Zh. and {Cochran}, W. and {Burke}, C.~J. and {Isaacson}, H. and {Quintana}, E.~V. and {Davies}, G.~R. and {Silva Aguirre}, V. and {Ragozzine}, D. and {Riddle}, R. and {Baranec}, C. and {Basu}, S. and {Chaplin}, W.~J. and {Christensen-Dalsgaard}, J. and {Metcalfe}, T.~S. and {Bedding}, T.~R. and {Handberg}, R. and {Stello}, D. and {Brewer}, J.~M. and {Hekker}, S. and {Karoff}, C. and {Kolbl}, R. and {Law}, N.~M. and {Lundkvist}, M. and {Miglio}, A. and {Rowe}, J.~F. and {Santos}, N.~C. and {Van Laerhoven}, C. and {Arentoft}, T. and {Elsworth}, Y.~P. and {Fischer}, D.~A. and {Kawaler}, S.~D. and {Kjeldsen}, H. and {Lund}, M.~N. and {Marcy}, G.~W. and {Sousa}, S.~G. and {Sozzetti}, A. and {White}, T.~R.},
        title = "{An Ancient Extrasolar System with Five Sub-Earth-size Planets}",
      journal = {ApJ},
         year = 2015,
        month = feb,
       volume = {799},
       number = {2},
          eid = {170},
        pages = {170},
          doi = {10.1088/0004-637X/799/2/170},
archivePrefix = {arXiv},
       eprint = {1501.06227},
 primaryClass = {astro-ph.EP},
       adsurl = {https://ui.adsabs.harvard.edu/abs/2015ApJ...799..170C}
}

@ARTICLE{campbell88,
       author = {{Campbell}, Bruce and {Walker}, G.~A.~H. and {Yang}, S.},
        title = "{A Search for Substellar Companions to Solar-type Stars}",
      journal = {ApJ},
         year = 1988,
        month = aug,
       volume = {331},
        pages = {902},
          doi = {10.1086/166608},
       adsurl = {https://ui.adsabs.harvard.edu/abs/1988ApJ...331..902C}
}

@ARTICLE{cannon1918,
       author = {{Cannon}, Annie J. and {Pickering}, Edward C.},
        title = "{The Henry Draper catalogue 0h, 1h, 2h, and 3h}",
      journal = {Annals of Harvard College Observatory},
         year = 1918,
        month = jan,
       volume = {91},
        pages = {1-290},
       adsurl = {https://ui.adsabs.harvard.edu/abs/1918AnHar..91....1C}
}

@ARTICLE{cantat-gaudin18,
       author = {{Cantat-Gaudin}, T. and {Jordi}, C. and {Vallenari}, A. and {Bragaglia}, A. and {Balaguer-N{\'u}{\~n}ez}, L. and {Soubiran}, C. and {Bossini}, D. and {Moitinho}, A. and {Castro-Ginard}, A. and {Krone-Martins}, A. and {Casamiquela}, L. and {Sordo}, R. and {Carrera}, R.},
        title = "{A Gaia DR2 view of the open cluster population in the Milky Way}",
      journal ={A\&A},
         year = 2018,
        month = oct,
       volume = {618},
          eid = {A93},
        pages = {A93},
          doi = {10.1051/0004-6361/201833476},
archivePrefix = {arXiv},
       eprint = {1805.08726},
 primaryClass = {astro-ph.GA},
       adsurl = {https://ui.adsabs.harvard.edu/abs/2018A&A...618A..93C}
}

@ARTICLE{cantrell13,
       author = {{Cantrell}, Justin R. and {Henry}, Todd J. and {White}, Russel J.},
        title = "{The Solar Neighborhood XXIX: The Habitable Real Estate of Our Nearest Stellar Neighbors}",
      journal = {AJ},
         year = 2013,
        month = oct,
       volume = {146},
       number = {4},
          eid = {99},
        pages = {99},
          doi = {10.1088/0004-6256/146/4/99},
archivePrefix = {arXiv},
       eprint = {1307.7038},
 primaryClass = {astro-ph.SR},
       adsurl = {https://ui.adsabs.harvard.edu/abs/2013AJ....146...99C}
}

@ARTICLE{capitaine03,
       author = {{Capitaine}, N. and {Wallace}, P.~T. and {Chapront}, J.},
        title = "{Expressions for IAU 2000 precession quantities}",
      journal = {A\&A},
         year = 2003,
        month = dec,
       volume = {412},
        pages = {567-586},
          doi = {10.1051/0004-6361:20031539},
       adsurl = {https://ui.adsabs.harvard.edu/abs/2003A&A...412..567C}
}

@ARTICLE{carleo20,
       author = {{Carleo}, Ilaria and {Gandolfi}, Davide and {Barrag{\'a}n}, Oscar and {Livingston}, John H. and {Persson}, Carina M. and {Lam}, Kristine W.~F. and {Vidotto}, Aline and {Lund}, Michael B. and {Villarreal D'Angelo}, Carolina and {Collins}, Karen A. and {Fossati}, Luca and {Howard}, Andrew W. and {Kubyshkina}, Daria and {Brahm}, Rafael and {Oklop{\v{c}}i{\'c}}, Antonija and {Molli{\`e}re}, Paul and {Redfield}, Seth and {Serrano}, Luisa Maria and {Dai}, Fei and {Fridlund}, Malcolm and {Borsa}, Francesco and {Korth}, Judith and {Esposito}, Massimiliano and {D{\'\i}az}, Mat{\'\i}as R. and {Dyregaard Nielsen}, Louise and {Hellier}, Coel and {Mathur}, Savita and {Deeg}, Hans J. and {Hatzes}, Artie P. and {Benatti}, Serena and {Rodler}, Florian and {Alarcon}, Javier and {Spina}, Lorenzo and {Santos}, {\^A}ngela R.~G. and {Georgieva}, Iskra and {Garc{\'\i}a}, Rafael A. and {Gonz{\'a}lez-Cuesta}, Luc{\'\i}a and {Ricker}, George R. and {Vanderspek}, Roland and {Latham}, David W. and {Seager}, Sara and {Winn}, Joshua N. and {Jenkins}, Jon M. and {Albrecht}, Simon and {Batalha}, Natalie M. and {Beard}, Corey and {Boyd}, Patricia T. and {Bouchy}, Fran{\c{c}}ois and {Burt}, Jennifer A. and {Butler}, R. Paul and {Cabrera}, Juan and {Chontos}, Ashley and {Ciardi}, David R. and {Cochran}, William D. and {Collins}, Kevin I. and {Crane}, Jeffrey D. and {Crossfield}, Ian and {Csizmadia}, Szilard and {Dragomir}, Diana and {Dressing}, Courtney and {Eigm{\"u}ller}, Philipp and {Endl}, Michael and {Erikson}, Anders and {Espinoza}, Nestor and {Fausnaugh}, Michael and {Feng}, Fabo and {Flowers}, Erin and {Fulton}, Benjamin and {Gonzales}, Erica J. and {Grieves}, Nolan and {Grziwa}, Sascha and {Guenther}, Eike W. and {Guerrero}, Natalia M. and {Henning}, Thomas and {Hidalgo}, Diego and {Hirano}, Teruyuki and {Hjorth}, Maria and {Huber}, Daniel and {Isaacson}, Howard and {Jones}, Matias and {Jord{\'a}n}, Andr{\'e}s and {Kab{\'a}th}, Petr and {Kane}, Stephen R. and {Knudstrup}, Emil and {Lubin}, Jack and {Luque}, Rafael and {Mireles}, Ismael and {Narita}, Norio and {Nespral}, David and {Niraula}, Prajwal and {Nowak}, Grzegorz and {Palle}, Enric and {P{\"a}tzold}, Martin and {Petigura}, Erik A. and {Prieto-Arranz}, Jorge and {Rauer}, Heike and {Robertson}, Paul and {Rose}, Mark E. and {Roy}, Arpita and {Sarkis}, Paula and {Schlieder}, Joshua E. and {S{\'e}gransan}, Damien and {Shectman}, Stephen and {Skarka}, Marek and {Smith}, Alexis M.~S. and {Smith}, Jeffrey C. and {Stassun}, Keivan and {Teske}, Johanna and {Twicken}, Joseph D. and {Van Eylen}, Vincent and {Wang}, Sharon and {Weiss}, Lauren M. and {Wyttenbach}, Aur{\'e}lien},
        title = "{The Multiplanet System TOI-421}",
      journal = {AJ},
         year = 2020,
        month = sep,
       volume = {160},
       number = {3},
          eid = {114},
        pages = {114},
          doi = {10.3847/1538-3881/aba124},
archivePrefix = {arXiv},
       eprint = {2004.10095},
 primaryClass = {astro-ph.EP},
       adsurl = {https://ui.adsabs.harvard.edu/abs/2020AJ....160..114C}
}

@ARTICLE{carpenter17,
       author = {{Carpenter}, Bob and {Gelman}, Andrew and {Hoffman}, Matthew D. and {Lee}, Daniel and {Goodrich}, Ben and {Betancourt}, Michael and {Brubaker}, Marcus and {Guo}, Jiqiang and {Li}, Peter and {Riddell}, Allen},
        title = "{Stan: A Probabilistic Programming Language}",
      journal = {J. Stat. Softw.},
         year = 2017,
        month = jan,
       volume = {76},
       number = {1},
        pages = {1},
       adsurl = {https://ui.adsabs.harvard.edu/abs/2017JSS....76....1C}
}

@ARTICLE{carriongonzalez23,
       author = {{Carri{\'o}n-Gonz{\'a}lez}, {\'O}scar and {Kammerer}, Jens and {Angerhausen}, Daniel and {Dannert}, Felix and {Garc{\'\i}a Mu{\~n}oz}, Antonio and {Quanz}, Sascha P. and {Absil}, Olivier and {Beichman}, Charles A. and {Girard}, Julien H. and {Mennesson}, Bertrand and {Meyer}, Michael R. and {Stapelfeldt}, Karl R. and {LIFE Collaboration}},
        title = "{Large Interferometer For Exoplanets (LIFE). X. Detectability of currently known exoplanets and synergies with future IR/O/UV reflected-starlight imaging missions}",
      journal = {A\&A},
         year = 2023,
        month = oct,
       volume = {678},
          eid = {A96},
        pages = {A96},
          doi = {10.1051/0004-6361/202347027},
archivePrefix = {arXiv},
       eprint = {2308.09646},
 primaryClass = {astro-ph.EP},
       adsurl = {https://ui.adsabs.harvard.edu/abs/2023A&A...678A..96C}
}

@ARTICLE{carro21,
       author = {{Carro}, Joseph M.},
        title = "{Misurazioni di Stelle Doppie Neglette. Report maggio 2021}",
      journal = {Il Bollettino delle Stelle Doppie},
         year = 2021,
        month = jun,
       volume = {33},
        pages = {12-17},
}

@book{carroll07,
        title = {An Introduction to Modern Galactic Astrophysics and Cosmology},
       author = {Carroll, Bradley W and Ostlie, Dale A and Black, A},
         year = {2007},
    publisher = {Pearson Addison Wesley}
}

@ARTICLE{carson09,
       author = {{Carson}, Joseph C. and {Hiner}, Kyle D. and {Villar}, Gregorio G., III and {Blaschak}, Michael G. and {Rudolph}, Alexander L. and {Stapelfeldt}, Karl R.},
        title = "{A Distance-Limited Imaging Survey of Substellar Companions to Solar Neighborhood Stars}",
      journal = {AJ},
         year = 2009,
        month = jan,
       volume = {137},
       number = {1},
        pages = {218-225},
          doi = {10.1088/0004-6256/137/1/218},
archivePrefix = {arXiv},
       eprint = {0810.1723},
 primaryClass = {astro-ph},
       adsurl = {https://ui.adsabs.harvard.edu/abs/2009AJ....137..218C}
}

@ARTICLE{cassan12,
       author = {{Cassan}, A. and {Kubas}, D. and {Beaulieu}, J. -P. and {Dominik}, M. and {Horne}, K. and {Greenhill}, J. and {Wambsganss}, J. and {Menzies}, J. and {Williams}, A. and {J{\o}rgensen}, U.~G. and {Udalski}, A. and {Bennett}, D.~P. and {Albrow}, M.~D. and {Batista}, V. and {Brillant}, S. and {Caldwell}, J.~A.~R. and {Cole}, A. and {Coutures}, Ch. and {Cook}, K.~H. and {Dieters}, S. and {Dominis Prester}, D. and {Donatowicz}, J. and {Fouqu{\'e}}, P. and {Hill}, K. and {Kains}, N. and {Kane}, S. and {Marquette}, J. -B. and {Martin}, R. and {Pollard}, K.~R. and {Sahu}, K.~C. and {Vinter}, C. and {Warren}, D. and {Watson}, B. and {Zub}, M. and {Sumi}, T. and {Szyma{\'n}ski}, M.~K. and {Kubiak}, M. and {Poleski}, R. and {Soszynski}, I. and {Ulaczyk}, K. and {Pietrzy{\'n}ski}, G. and {Wyrzykowski}, {\L}.},
        title = "{One or more bound planets per Milky Way star from microlensing observations}",
      journal = {Nature},
         year = 2012,
        month = jan,
       volume = {481},
       number = {7380},
        pages = {167-169},
          doi = {10.1038/nature10684},
archivePrefix = {arXiv},
       eprint = {1202.0903},
 primaryClass = {astro-ph.EP},
       adsurl = {https://ui.adsabs.harvard.edu/abs/2012Natur.481..167C}
}

@ARTICLE{chabrier97,
       author = {{Chabrier}, Gilles and {Baraffe}, Isabelle},
        title = "{Structure and evolution of low-mass stars}",
      journal = {A\&A},
         year = 1997,
        month = nov,
       volume = {327},
        pages = {1039-1053},
          doi = {10.48550/arXiv.astro-ph/9704118},
archivePrefix = {arXiv},
       eprint = {astro-ph/9704118},
 primaryClass = {astro-ph},
       adsurl = {https://ui.adsabs.harvard.edu/abs/1997A&A...327.1039C}
}

@ARTICLE{chabrier03,
       author = {{Chabrier}, Gilles},
        title = "{Galactic Stellar and Substellar Initial Mass Function}",
      journal ={PASP},
         year = 2003,
        month = jul,
       volume = {115},
       number = {809},
        pages = {763-795},
          doi = {10.1086/376392},
archivePrefix = {arXiv},
       eprint = {astro-ph/0304382},
 primaryClass = {astro-ph},
       adsurl = {https://ui.adsabs.harvard.edu/abs/2003PASP..115..763C}
}

@ARTICLE{chambers16a,
       author = {{Chambers}, K.~C. and {Magnier}, E.~A. and {Metcalfe}, N. and {Flewelling}, H.~A. and {Huber}, M.~E. and {Waters}, C.~Z. and {Denneau}, L. and {Draper}, P.~W. and {Farrow}, D. and {Finkbeiner}, D.~P. and {Holmberg}, C. and {Koppenhoefer}, J. and {Price}, P.~A. and {Rest}, A. and {Saglia}, R.~P. and {Schlafly}, E.~F. and {Smartt}, S.~J. and {Sweeney}, W. and {Wainscoat}, R.~J. and {Burgett}, W.~S. and {Chastel}, S. and {Grav}, T. and {Heasley}, J.~N. and {Hodapp}, K.~W. and {Jedicke}, R. and {Kaiser}, N. and {Kudritzki}, R. -P. and {Luppino}, G.~A. and {Lupton}, R.~H. and {Monet}, D.~G. and {Morgan}, J.~S. and {Onaka}, P.~M. and {Shiao}, B. and {Stubbs}, C.~W. and {Tonry}, J.~L. and {White}, R. and {Ba{\~n}ados}, E. and {Bell}, E.~F. and {Bender}, R. and {Bernard}, E.~J. and {Boegner}, M. and {Boffi}, F. and {Botticella}, M.~T. and {Calamida}, A. and {Casertano}, S. and {Chen}, W. -P. and {Chen}, X. and {Cole}, S. and {Deacon}, N. and {Frenk}, C. and {Fitzsimmons}, A. and {Gezari}, S. and {Gibbs}, V. and {Goessl}, C. and {Goggia}, T. and {Gourgue}, R. and {Goldman}, B. and {Grant}, P. and {Grebel}, E.~K. and {Hambly}, N.~C. and {Hasinger}, G. and {Heavens}, A.~F. and {Heckman}, T.~M. and {Henderson}, R. and {Henning}, T. and {Holman}, M. and {Hopp}, U. and {Ip}, W. -H. and {Isani}, S. and {Jackson}, M. and {Keyes}, C.~D. and {Koekemoer}, A.~M. and {Kotak}, R. and {Le}, D. and {Liska}, D. and {Long}, K.~S. and {Lucey}, J.~R. and {Liu}, M. and {Martin}, N.~F. and {Masci}, G. and {McLean}, B. and {Mindel}, E. and {Misra}, P. and {Morganson}, E. and {Murphy}, D.~N.~A. and {Obaika}, A. and {Narayan}, G. and {Nieto-Santisteban}, M.~A. and {Norberg}, P. and {Peacock}, J.~A. and {Pier}, E.~A. and {Postman}, M. and {Primak}, N. and {Rae}, C. and {Rai}, A. and {Riess}, A. and {Riffeser}, A. and {Rix}, H.~W. and {R{\"o}ser}, S. and {Russel}, R. and {Rutz}, L. and {Schilbach}, E. and {Schultz}, A.~S.~B. and {Scolnic}, D. and {Strolger}, L. and {Szalay}, A. and {Seitz}, S. and {Small}, E. and {Smith}, K.~W. and {Soderblom}, D.~R. and {Taylor}, P. and {Thomson}, R. and {Taylor}, A.~N. and {Thakar}, A.~R. and {Thiel}, J. and {Thilker}, D. and {Unger}, D. and {Urata}, Y. and {Valenti}, J. and {Wagner}, J. and {Walder}, T. and {Walter}, F. and {Watters}, S.~P. and {Werner}, S. and {Wood-Vasey}, W.~M. and {Wyse}, R.},
        title = "{The Pan-STARRS1 Surveys}",
      journal = {arXiv e-prints},
         year = 2016,
        month = dec,
          eid = {arXiv:1612.05560},
        pages = {arXiv:1612.05560},
archivePrefix = {arXiv},
       eprint = {1612.05560},
 primaryClass = {astro-ph.IM},
       adsurl = {https://ui.adsabs.harvard.edu/abs/2016arXiv161205560C}
}

@ARTICLE{chaname04,
   author = {{Chanam{\'e}}, J. and {Gould}, A.},
    title = "{Disk and Halo Wide Binaries from the Revised Luyten Catalog: Probes of Star Formation and MACHO Dark Matter}",
  journal = {ApJ},
   eprint = {astro-ph/0307434},
     year = 2004,
    month = jan,
   volume = 601,
    pages = {289-310},
      doi = {10.1086/380442},
   adsurl = {http://cdsads.u-strasbg.fr/abs/2004ApJ...601..289C}
}

@ARTICLE{chandrasekhar31a,
   author = {{Chandrasekhar}, S.},
    title = "{The density of white dwarf stars}",
  journal = {Phil. Mag.},
     year = 1931,
    month = feb,
   volume = 11,
    pages = {592-596}
}

@ARTICLE{chandrasekhar31b,
       author = {{Chandrasekhar}, S.},
        title = "{The Maximum Mass of Ideal White Dwarfs}",
      journal = {ApJ},
         year = 1931,
        month = jul,
       volume = {74},
        pages = {81},
          doi = {10.1086/143324},
       adsurl = {https://ui.adsabs.harvard.edu/abs/1931ApJ....74...81C}
}

@ARTICLE{chauvin04,
       author = {{Chauvin}, G. and {Lagrange}, A. -M. and {Dumas}, C. and {Zuckerman}, B. and {Mouillet}, D. and {Song}, I. and {Beuzit}, J. -L. and {Lowrance}, P.},
        title = "{A giant planet candidate near a young brown dwarf. Direct VLT/NACO observations using IR wavefront sensing}",
      journal = {A\&A},
         year = 2004,
        month = oct,
       volume = {425},
        pages = {L29-L32},
          doi = {10.1051/0004-6361:200400056},
archivePrefix = {arXiv},
       eprint = {astro-ph/0409323},
 primaryClass = {astro-ph},
       adsurl = {https://ui.adsabs.harvard.edu/abs/2004A&A...425L..29C}
}

@ARTICLE{chauvin10,
       author = {{Chauvin}, G. and {Lagrange}, A. -M. and {Bonavita}, M. and {Zuckerman}, B. and {Dumas}, C. and {Bessell}, M.~S. and {Beuzit}, J. -L. and {Bonnefoy}, M. and {Desidera}, S. and {Farihi}, J. and {Lowrance}, P. and {Mouillet}, D. and {Song}, I.},
        title = "{Deep imaging survey of young, nearby austral stars . VLT/NACO near-infrared Lyot-coronographic observations}",
      journal = {A\&A},
         year = 2010,
        month = jan,
       volume = {509},
          eid = {A52},
        pages = {A52},
          doi = {10.1051/0004-6361/200911716},
archivePrefix = {arXiv},
       eprint = {0906.2945},
 primaryClass = {astro-ph.EP},
       adsurl = {https://ui.adsabs.harvard.edu/abs/2010A&A...509A..52C}
}

@ARTICLE{chauvin11,
       author = {{Chauvin}, G. and {Beust}, H. and {Lagrange}, A. -M. and {Eggenberger}, A.},
        title = "{Planetary systems in close binary stars: the case of HD 196885. Combined astrometric and radial velocity study}",
      journal = {A\&A},
         year = 2011,
        month = apr,
       volume = {528},
          eid = {A8},
        pages = {A8},
          doi = {10.1051/0004-6361/201015433},
       adsurl = {https://ui.adsabs.harvard.edu/abs/2011A&A...528A...8C}
}

@ARTICLE{chen05,
       author = {{Chen}, C.~H. and {Patten}, B.~M. and {Werner}, M.~W. and {Dowell}, C.~D. and {Stapelfeldt}, K.~R. and {Song}, I. and {Stauffer}, J.~R. and {Blaylock}, M. and {Gordon}, K.~D. and {Krause}, V.},
        title = "{A Spitzer Study of Dusty Disks around Nearby, Young Stars}",
      journal ={ApJ},
         year = 2005,
        month = dec,
       volume = {634},
       number = {2},
        pages = {1372-1384},
          doi = {10.1086/497124},
       adsurl = {https://ui.adsabs.harvard.edu/abs/2005ApJ...634.1372C}
}

@ARTICLE{chen22,
       author = {{Chen}, Minghan and {Li}, Yiting and {Brandt}, Timothy D. and {Dupuy}, Trent J. and {Cardoso}, C{\'a}tia V. and {McCaughrean}, Mark J.},
        title = "{Precise Dynamical Masses of ɛ Indi Ba and Bb: Evidence of Slowed Cooling at the L/T Transition}",
      journal = {AJ},
         year = 2022,
        month = jun,
       volume = {163},
       number = {6},
          eid = {288},
        pages = {288},
          doi = {10.3847/1538-3881/ac66d2},
archivePrefix = {arXiv},
       eprint = {2205.08077},
 primaryClass = {astro-ph.SR},
       adsurl = {https://ui.adsabs.harvard.edu/abs/2022AJ....163..288C}
}

@ARTICLE{cherepashchuk21,
       author = {{Cherepashchuk}, A.~M. and {Belinski}, A.~A. and {Dodin}, A.~V. and {Postnov}, K.~A.},
        title = "{Discovery of orbital eccentricity and evidence for orbital period increase of SS433}",
      journal = {MNRAS},
         year = 2021,
        month = oct,
       volume = {507},
       number = {1},
        pages = {L19-L23},
          doi = {10.1093/mnrasl/slab083},
archivePrefix = {arXiv},
       eprint = {2107.09005},
 primaryClass = {astro-ph.SR},
       adsurl = {https://ui.adsabs.harvard.edu/abs/2021MNRAS.507L..19C}
}

@ARTICLE{chini12,
       author = {{Chini}, R. and {Hoffmeister}, V.~H. and {Nasseri}, A. and {Stahl}, O. and {Zinnecker}, H.},
        title = "{A spectroscopic survey on the multiplicity of high-mass stars}",
      journal = {MNRAS},
         year = 2012,
        month = aug,
       volume = {424},
       number = {3},
        pages = {1925-1929},
          doi = {10.1111/j.1365-2966.2012.21317.x},
archivePrefix = {arXiv},
       eprint = {1205.5238},
 primaryClass = {astro-ph.SR},
       adsurl = {https://ui.adsabs.harvard.edu/abs/2012MNRAS.424.1925C}
}

@ARTICLE{christian22,
       author = {{Christian}, Sam and {Vanderburg}, Andrew and {Becker}, Juliette and {Yahalomi}, Daniel A. and {Pearce}, Logan and {Zhou}, George and {Collins}, Karen A. and {Kraus}, Adam L. and {Stassun}, Keivan G. and {de Beurs}, Zoe and {Ricker}, George R. and {Vanderspek}, Roland K. and {Latham}, David W. and {Winn}, Joshua N. and {Seager}, S. and {Jenkins}, Jon M. and {Abe}, Lyu and {Agabi}, Karim and {Amado}, Pedro J. and {Baker}, David and {Barkaoui}, Khalid and {Benkhaldoun}, Zouhair and {Benni}, Paul and {Berberian}, John and {Berlind}, Perry and {Bieryla}, Allyson and {Esparza-Borges}, Emma and {Bowen}, Michael and {Brown}, Peyton and {Buchhave}, Lars A. and {Burke}, Christopher J. and {Buttu}, Marco and {Cadieux}, Charles and {Caldwell}, Douglas A. and {Charbonneau}, David and {Chazov}, Nikita and {Chimaladinne}, Sudhish and {Collins}, Kevin I. and {Combs}, Deven and {Conti}, Dennis M. and {Crouzet}, Nicolas and {de Leon}, Jerome P. and {Deljookorani}, Shila and {Diamond}, Brendan and {Doyon}, Ren{\'e} and {Dragomir}, Diana and {Dransfield}, Georgina and {Essack}, Zahra and {Evans}, Phil and {Fukui}, Akihiko and {Gan}, Tianjun and {Esquerdo}, Gilbert A. and {Gillon}, Micha{\"e}l and {Girardin}, Eric and {Guerra}, Pere and {Guillot}, Tristan and {K. Habich}, Eleanor Kate and {Henriksen}, Andreea and {Hoch}, Nora and {Isogai}, Keisuke I. and {Jehin}, Emmanu{\"e}l and {Jensen}, Eric L.~N. and {Johnson}, Marshall C. and {Livingston}, John H. and {Kielkopf}, John F. and {Kim}, Kingsley and {Kawauchi}, Kiyoe and {Krushinsky}, Vadim and {Kunzle}, Veronica and {Laloum}, Didier and {Leger}, Dominic and {Lewin}, Pablo and {Mallia}, Franco and {Massey}, Bob and {Mori}, Mayuko and {McLeod}, Kim K. and {M{\'e}karnia}, Djamel and {Mireles}, Ismael and {Mishevskiy}, Nikolay and {Tamura}, Motohide and {Murgas}, Felipe and {Narita}, Norio and {Naves}, Ramon and {Nelson}, Peter and {Osborn}, Hugh P. and {Palle}, Enric and {Parviainen}, Hannu and {Plavchan}, Peter and {Pozuelos}, Francisco J. and {Rabus}, Markus and {Relles}, Howard M. and {Rodr{\'\i}guez L{\'o}pez}, Cristina and {Quinn}, Samuel N. and {Schmider}, Francois-Xavier and {Schlieder}, Joshua E. and {Schwarz}, Richard P. and {Shporer}, Avi and {Sibbald}, Laurie and {Srdoc}, Gregor and {Stibbards}, Caitlin and {Stickler}, Hannah and {Suarez}, Olga and {Stockdale}, Chris and {Tan}, Thiam-Guan and {Terada}, Yuka and {Triaud}, Amaury and {Tronsgaard}, Rene and {Waalkes}, William C. and {Wang}, Gavin and {Watanabe}, Noriharu and {Wenceslas}, Marie-Sainte and {Wingham}, Geof and {Wittrock}, Justin and {Ziegler}, Carl},
        title = "{A Possible Alignment Between the Orbits of Planetary Systems and their Visual Binary Companions}",
      journal = {AJ},
         year = 2022,
        month = may,
       volume = {163},
       number = {5},
          eid = {207},
        pages = {207},
          doi = {10.3847/1538-3881/ac517f},
archivePrefix = {arXiv},
       eprint = {2202.00042},
 primaryClass = {astro-ph.EP},
       adsurl = {https://ui.adsabs.harvard.edu/abs/2022AJ....163..207C}
}

@ARTICLE{christiansen22,
       author = {{Christiansen}, Jessie L. and {Bhure}, Sakhee and {Zink}, Jon K. and {Hardegree-Ullman}, Kevin K. and {Adkins}, Britt Duffy and {Hedges}, Christina and {Morton}, Timothy D. and {Bieryla}, Allyson and {Ciardi}, David R. and {Cochran}, William D. and {Dressing}, Courtney D. and {Everett}, Mark E. and {Isaacson}, Howard and {Livingston}, John H. and {Ziegler}, Carl and {Berlind}, Perry and {Calkins}, Michael L. and {Esquerdo}, Gilbert A. and {Latham}, David W. and {Endl}, Michael and {MacQueen}, Phillip J. and {Fulton}, Benjamin J. and {Hirsch}, Lea A. and {Howard}, Andrew W. and {Weiss}, Lauren M. and {Allen}, Bridgette E. and {Berberyann}, Arthur and {Ciardi}, Krys N. and {Dunlavy}, Ava and {Glassford}, Sofia H. and {Dai}, Fei and {Hirano}, Teruyuki and {Tamura}, Motohide and {Beichman}, Charles and {Gonzales}, Erica J. and {Schlieder}, Joshua E. and {Barclay}, Thomas and {Crossfield}, Ian J.~M. and {Gilbert}, Emily A. and {Matthews}, Elisabeth C. and {Giacalone}, Steven and {Petigura}, Erik A.},
        title = "{Scaling K2. V. Statistical Validation of 60 New Exoplanets From K2 Campaigns 2-18}",
      journal = {AJ},
         year = 2022,
        month = jun,
       volume = {163},
       number = {6},
          eid = {244},
        pages = {244},
          doi = {10.3847/1538-3881/ac5c4c},
archivePrefix = {arXiv},
       eprint = {2203.02087},
 primaryClass = {astro-ph.EP},
       adsurl = {https://ui.adsabs.harvard.edu/abs/2022AJ....163..244C}
}

@ARTICLE{christy69,
       author = {{Christy}, James W. and {Walker}, R.~L., Jr.},
        title = "{MK Classification of 142 Visual Binaries}",
      journal ={PASP},
         year = 1969,
        month = oct,
       volume = {81},
       number = {482},
        pages = {643},
          doi = {10.1086/128831},
       adsurl = {https://ui.adsabs.harvard.edu/abs/1969PASP...81..643C}
}

@ARTICLE{christy22,
       author = {{Christy}, C.~T. and {Jayasinghe}, T. and {Stanek}, K.~Z. and {Kochanek}, C.~S. and {Way}, Z. and {Prieto}, J.~L. and {Shappee}, B.~J. and {Holoien}, T.~W. -S. and {Thompson}, T.~A. and {Schneider}, A.},
        title = "{Citizen ASAS-SN Data Release. I. Variable Star Classification Using Citizen Science}",
      journal = {PASP},
         year = 2022,
        month = feb,
       volume = {134},
       number = {1032},
          eid = {024201},
        pages = {024201},
          doi = {10.1088/1538-3873/ac44f0},
archivePrefix = {arXiv},
       eprint = {2111.02415},
 primaryClass = {astro-ph.SR},
       adsurl = {https://ui.adsabs.harvard.edu/abs/2022PASP..134b4201C}
}

@ARTICLE{cifuentes20,
       author = {{Cifuentes}, C. and {Caballero}, J.~A. and {Cort{\'e}s-Contreras}, M. and {Montes}, D. and {Abell{\'a}n}, F.~J. and {Dorda}, R. and {Holgado}, G. and {Zapatero Osorio}, M.~R. and {Morales}, J.~C. and {Amado}, P.~J. and {Passegger}, V.~M. and {Quirrenbach}, A. and {Reiners}, A. and {Ribas}, I. and {Sanz-Forcada}, J. and {Schweitzer}, A. and {Seifert}, W. and {Solano}, E.},
        title = "{CARMENES input catalogue of M dwarfs. V. Luminosities, colours, and spectral energy distributions}",
      journal = {A\&A},
         year = 2020,
        month = oct,
       volume = {642},
          eid = {A115},
        pages = {A115},
          doi = {10.1051/0004-6361/202038295},
archivePrefix = {arXiv},
       eprint = {2007.15077},
 primaryClass = {astro-ph.SR},
       adsurl = {https://ui.adsabs.harvard.edu/abs/2020A&A...642A.115C}
}

@ARTICLE{cifuentes21,
       author = {{Cifuentes}, Carlos and {Caballero}, Jos{\'e} A. and {Agust{\'\i}}, Sergio},
        title = "{One Is the Loneliest Number: Multiplicity in Cool Dwarfs}",
      journal = {RNAAS},
         year = 2021,
        month = jun,
       volume = {5},
       number = {5},
          eid = {129},
        pages = {129},
          doi = {10.3847/2515-5172/ac05ce},
archivePrefix = {arXiv},
       eprint = {2106.05049},
 primaryClass = {astro-ph.SR},
       adsurl = {https://ui.adsabs.harvard.edu/abs/2021RNAAS...5..129C}
}

@ARTICLE{cifuentes25,
       author = {{Cifuentes}, C. and {Caballero}, J.~A. and {Gonz{\'a}lez-Payo}, J. and {Amado}, P.~J. and {B{\'e}jar}, V.~J.~S. and {Burgasser}, A.~J. and {Cort{\'e}s-Contreras}, M. and {Lodieu}, N. and {Montes}, D. and {Quirrenbach}, A. and {Reiners}, A. and {Ribas}, I. and {Sanz-Forcada}, J. and {Seifert}, W. and {Zapatero Osorio}, M.~R.},
        title = "{CARMENES input catalogue of M dwarfs: IX. Multiplicity from close spectroscopic binaries to ultra-wide systems}",
      journal = {A\&A},
         year = 2025,
        month = jan,
       volume = {693},
          eid = {A228},
        pages = {A228},
          doi = {10.1051/0004-6361/202452527},
       adsurl = {https://ui.adsabs.harvard.edu/abs/2025A&A...693A.228C}
}

@INPROCEEDINGS{civera24,
       author = {{Civera}, T.},
        title = "{CEFCA Catalogues Portal Towards FAIR Principles}",
    booktitle = {Astromical Data Analysis Software and Systems XXXI},
         year = 2024,
       editor = {{Hugo}, B.~V. and {Van Rooyen}, R. and {Smirnov}, O.~M.},
       series = {Astronomical Society of the Pacific Conference Series},
       volume = {535},
        month = may,
        pages = {275},
       adsurl = {https://ui.adsabs.harvard.edu/abs/2024ASPC..535..275C}
}

@ARTICLE{clark22,
       author = {{Clark}, Catherine A. and {van Belle}, Gerard T. and {Ciardi}, David R. and {Lund}, Michael B. and {Howell}, Steve B. and {Everett}, Mark E. and {Beichman}, Charles A. and {Winters}, Jennifer G.},
        title = "{A Dearth of Close-in Stellar Companions to M-dwarf TESS Objects of Interest}",
      journal = {AJ},
         year = 2022,
        month = may,
       volume = {163},
       number = {5},
          eid = {232},
        pages = {232},
          doi = {10.3847/1538-3881/ac6101},
archivePrefix = {arXiv},
       eprint = {2203.12795},
 primaryClass = {astro-ph.SR},
       adsurl = {https://ui.adsabs.harvard.edu/abs/2022AJ....163..232C}
}

@ARTICLE{clark24,
       author = {{Clark}, Catherine A. and {van Belle}, Gerard T. and {Horch}, Elliott P. and {Ciardi}, David R. and {von Braun}, Kaspar and {Skiff}, Brian A. and {Winters}, Jennifer G. and {Lund}, Michael B. and {Everett}, Mark E. and {Hartman}, Zachary D. and {Llama}, Joe},
        title = "{The POKEMON Speckle Survey of Nearby M Dwarfs. III. The Stellar Multiplicity Rate of M Dwarfs within 15 pc}",
      journal = {AJ},
         year = 2024,
        month = apr,
       volume = {167},
       number = {4},
          eid = {174},
        pages = {174},
          doi = {10.3847/1538-3881/ad267d},
archivePrefix = {arXiv},
       eprint = {2401.14703},
 primaryClass = {astro-ph.SR},
       adsurl = {https://ui.adsabs.harvard.edu/abs/2024AJ....167..174C}
}

@BOOK{clerke1908,
       author = {{Clerke}, Agnes M.},
        title = "{A Popular History of Astronomy During the Nineteenth Century}",
    publisher = {A. and C. Black},
         year = 1908,
       adsurl = {https://ui.adsabs.harvard.edu/abs/1908phad.book.....C}
}

@ARTICLE{clery23,
       author = {{Clery}, Daniel},
        title = "{Future NASA scope would find life on alien worlds}",
      journal = {Science},
         year = 2023,
        month = jan,
       volume = {379},
       number = {6628},
        pages = {123-124},
          doi = {10.1126/science.adg6273},
       adsurl = {https://ui.adsabs.harvard.edu/abs/2023Sci...379..123C}
}

@ARTICLE{close90,
       author = {{Close}, Laird M. and {Richer}, Harvey B. and {Crabtree}, Dennis R.},
        title = "{A Complete Sample of Wide Binaries in the Solar Neighborhood}",
      journal = {AJ},
         year = 1990,
        month = dec,
       volume = {100},
        pages = {1968},
          doi = {10.1086/115652},
       adsurl = {https://ui.adsabs.harvard.edu/abs/1990AJ....100.1968C}
}

@INPROCEEDINGS{close03,
       author = {{Close}, Laird M. and {Siegler}, Nick and {Freed}, Melanie},
        title = "{Detection of Nine M8.0--L0.5 Binaries: The Very Low Mass Binary Population and Its Implications for Brown Dwarf Formation Theories}",
    booktitle = {Brown Dwarfs},
         year = 2003,
       editor = {{Mart{\'\i}n}, Eduardo},
       volume = {211},
        month = jun,
        pages = {249},
       adsurl = {https://ui.adsabs.harvard.edu/abs/2003IAUS..211..249C}
}

@ARTICLE{close07,
       author = {{Close}, Laird M. and {Zuckerman}, B. and {Song}, Inseok and {Barman}, Travis and {Marois}, Christian and {Rice}, Emily L. and {Siegler}, Nick and {Macintosh}, Bruce and {Becklin}, E.~E. and {Campbell}, Randy and {Lyke}, James E. and {Conrad}, Al and {Le Mignant}, David},
        title = "{The Wide Brown Dwarf Binary Oph 1622-2405 and Discovery of a Wide, Low-Mass Binary in Ophiuchus (Oph 1623-2402): A New Class of Young Evaporating Wide Binaries?}",
      journal ={ApJ},
         year = 2007,
        month = may,
       volume = {660},
       number = {2},
        pages = {1492-1506},
          doi = {10.1086/513417},
archivePrefix = {arXiv},
       eprint = {astro-ph/0608574},
 primaryClass = {astro-ph},
       adsurl = {https://ui.adsabs.harvard.edu/abs/2007ApJ...660.1492C}
}

@ARTICLE{cochran97,
       author = {{Cochran}, William D. and {Hatzes}, Artie P. and {Butler}, R. Paul and {Marcy}, Geoffrey W.},
        title = "{The Discovery of a Planetary Companion to 16 Cygni B}",
      journal = {ApJ},
         year = 1997,
        month = jul,
       volume = {483},
       number = {1},
        pages = {457-463},
          doi = {10.1086/304245},
archivePrefix = {arXiv},
       eprint = {astro-ph/9611230},
 primaryClass = {astro-ph},
       adsurl = {https://ui.adsabs.harvard.edu/abs/1997ApJ...483..457C}
}

@ARTICLE{collins18,
       author = {{Collins}, Karen A. and {Collins}, Kevin I. and {Pepper}, Joshua and {Labadie-Bartz}, Jonathan and {Stassun}, Keivan G. and {Gaudi}, B. Scott and {Bayliss}, Daniel and {Bento}, Joao and {COL{\'O}N}, Knicole D. and {Feliz}, Dax and {James}, David and {Johnson}, Marshall C. and {Kuhn}, Rudolf B. and {Lund}, Michael B. and {Penny}, Matthew T. and {Rodriguez}, Joseph E. and {Siverd}, Robert J. and {Stevens}, Daniel J. and {Yao}, Xinyu and {Zhou}, George and {Akshay}, Mundra and {Aldi}, Giulio F. and {Ashcraft}, Cliff and {Awiphan}, Supachai and {Ba{\c{s}}t{\"u}rk}, {\"O}zg{\"u}r and {Baker}, David and {Beatty}, Thomas G. and {Benni}, Paul and {Berlind}, Perry and {Berriman}, G. Bruce and {Berta-Thompson}, Zach and {Bieryla}, Allyson and {Bozza}, Valerio and {Calchi Novati}, Sebastiano and {Calkins}, Michael L. and {Cann}, Jenna M. and {Ciardi}, David R. and {Clark}, Ian R. and {Cochran}, William D. and {Cohen}, David H. and {Conti}, Dennis and {Crepp}, Justin R. and {Curtis}, Ivan A. and {D'Ago}, Giuseppe and {Diazeguigure}, Kenny A. and {Dressing}, Courtney D. and {Dubois}, Franky and {Ellingson}, Erica and {Ellis}, Tyler G. and {Esquerdo}, Gilbert A. and {Evans}, Phil and {Friedli}, Alison and {Fukui}, Akihiko and {Fulton}, Benjamin J. and {Gonzales}, Erica J. and {Good}, John C. and {Gregorio}, Joao and {Gumusayak}, Tolga and {Hancock}, Daniel A. and {Harada}, Caleb K. and {Hart}, Rhodes and {Hintz}, Eric G. and {Jang-Condell}, Hannah and {Jeffery}, Elizabeth J. and {Jensen}, Eric L.~N. and {Jofr{\'e}}, Emiliano and {Joner}, Michael D. and {Kar}, Aman and {Kasper}, David H. and {Keten}, Burak and {Kielkopf}, John F. and {Komonjinda}, Siramas and {Kotnik}, Cliff and {Latham}, David W. and {Leuquire}, Jacob and {Lewis}, Tiffany R. and {Logie}, Ludwig and {Lowther}, Simon J. and {Macqueen}, Phillip J. and {Martin}, Trevor J. and {Mawet}, Dimitri and {Mcleod}, Kim K. and {Murawski}, Gabriel and {Narita}, Norio and {Nordhausen}, Jim and {Oberst}, Thomas E. and {Odden}, Caroline and {Panka}, Peter A. and {Petrucci}, Romina and {Plavchan}, Peter and {Quinn}, Samuel N. and {Rau}, Steve and {Reed}, Phillip A. and {Relles}, Howard and {Renaud}, Joe P. and {Scarpetta}, Gaetano and {Sorber}, Rebecca L. and {Spencer}, Alex D. and {Spencer}, Michelle and {Stephens}, Denise C. and {Stockdale}, Chris and {Tan}, Thiam-Guan and {Trueblood}, Mark and {Trueblood}, Patricia and {Vanaverbeke}, Siegfried and {Villanueva}, Steven, Jr. and {Warner}, Elizabeth M. and {West}, Mary Lou and {Yal{\c{c}}{\i}nkaya}, Sel{\c{c}}uk and {Yeigh}, Rex and {Zambelli}, Roberto},
        title = "{The KELT Follow-up Network and Transit False-positive Catalog: Pre-vetted False Positives for TESS}",
      journal = {AJ},
         year = 2018,
        month = nov,
       volume = {156},
       number = {5},
          eid = {234},
        pages = {234},
          doi = {10.3847/1538-3881/aae582},
archivePrefix = {arXiv},
       eprint = {1803.01869},
 primaryClass = {astro-ph.EP},
       adsurl = {https://ui.adsabs.harvard.edu/abs/2018AJ....156..234C}
}

@ARTICLE{correaotto17,
       author = {{Correa-Otto}, J.~A. and {Gil-Hutton}, R.~A.},
        title = "{Galactic perturbations on the population of wide binary stars with exoplanets}",
      journal = {A\&A},
         year = 2017,
        month = dec,
       volume = {608},
          eid = {A116},
        pages = {A116},
          doi = {10.1051/0004-6361/201731229},
archivePrefix = {arXiv},
       eprint = {1710.00766},
 primaryClass = {astro-ph.EP},
       adsurl = {https://ui.adsabs.harvard.edu/abs/2017A&A...608A.116C}
}

@ARTICLE{correia08,
       author = {{Correia}, A.~C.~M. and {Udry}, S. and {Mayor}, M. and {Eggenberger}, A. and {Naef}, D. and {Beuzit}, J. -L. and {Perrier}, C. and {Queloz}, D. and {Sivan}, J. -P. and {Pepe}, F. and {Santos}, N.~C. and {S{\'e}gransan}, D.},
        title = "{The ELODIE survey for northern extra-solar planets. IV. HD 196885, a close binary star with a 3.7-year planet}",
      journal = {A\&A},
         year = 2008,
        month = feb,
       volume = {479},
       number = {1},
        pages = {271-275},
          doi = {10.1051/0004-6361:20078908},
       adsurl = {https://ui.adsabs.harvard.edu/abs/2008A&A...479..271C}
}

@ARTICLE{cortescontreras17b,
       author = {{Cort{\'e}s-Contreras}, M. and {B{\'e}jar}, V.~J.~S. and {Caballero}, J.~A. and {Gauza}, B. and {Montes}, D. and {Alonso-Floriano}, F.~J. and {Jeffers}, S.~V. and {Morales}, J.~C. and {Reiners}, A. and {Ribas}, I. and {Sch{\"o}fer}, P. and {Quirrenbach}, A. and {Amado}, P.~J. and {Mundt}, R. and {Seifert}, W.},
        title = "{CARMENES input catalogue of M dwarfs. II. High-resolution imaging with FastCam}",
      journal = {A\&A},
         year = 2017,
        month = jan,
       volume = {597},
          eid = {A47},
        pages = {A47},
          doi = {10.1051/0004-6361/201629056},
archivePrefix = {arXiv},
       eprint = {1608.08145},
 primaryClass = {astro-ph.SR},
       adsurl = {https://ui.adsabs.harvard.edu/abs/2017A&A...597A..47C}
}

@ARTICLE{cortescontreras24,
       author = {{Cort{\'e}s-Contreras}, M. and {Caballero}, J.~A. and {Montes}, D. and {Cardona-Guill{\'e}n}, C. and {B{\'e}jar}, V.~J.~S. and {Cifuentes}, C. and {Tabernero}, H.~M. and {Zapatero Osorio}, M.~R. and {Amado}, P.~J. and {Jeffers}, S.~V. and {Lafarga}, M. and {Lodieu}, N. and {Quirrenbach}, A. and {Reiners}, A. and {Ribas}, I. and {Sch{\"o}fer}, P. and {Schweitzer}, A. and {Seifert}, W.},
        title = "{CARMENES input catalogue of M dwarfs: VIII. Kinematics in the solar neighbourhood}",
      journal = {A\&A},
         year = 2024,
        month = dec,
       volume = {692},
          eid = {A206},
        pages = {A206},
          doi = {10.1051/0004-6361/202451585},
archivePrefix = {arXiv},
       eprint = {2411.06825},
 primaryClass = {astro-ph.SR},
       adsurl = {https://ui.adsabs.harvard.edu/abs/2024A&A...692A.206C}
}

@ARTICLE{crepp16,
       author = {{Crepp}, Justin R. and {Gonzales}, Erica J. and {Bechter}, Eric B. and {Montet}, Benjamin T. and {Johnson}, John Asher and {Piskorz}, Danielle and {Howard}, Andrew W. and {Isaacson}, Howard},
        title = "{The TRENDS High-contrast Imaging Survey. VI. Discovery of a Mass, Age, and Metallicity Benchmark Brown Dwarf}",
      journal = {ApJ},
         year = 2016,
        month = dec,
       volume = {831},
       number = {2},
          eid = {136},
        pages = {136},
          doi = {10.3847/0004-637X/831/2/136},
archivePrefix = {arXiv},
       eprint = {1604.00398},
 primaryClass = {astro-ph.SR},
       adsurl = {https://ui.adsabs.harvard.edu/abs/2016ApJ...831..136C}
}

@ARTICLE{cruzalebes19,
       author = {{Cruzal{\`e}bes}, P. and {Petrov}, R.~G. and {Robbe-Dubois}, S. and {Varga}, J. and {Burtscher}, L. and {Allouche}, F. and {Berio}, P. and {Hofmann}, K. -H. and {Hron}, J. and {Jaffe}, W. and {Lagarde}, S. and {Lopez}, B. and {Matter}, A. and {Meilland}, A. and {Meisenheimer}, K. and {Millour}, F. and {Schertl}, D.},
        title = "{A catalogue of stellar diameters and fluxes for mid-infrared interferometry}",
      journal = {MNRAS},
         year = 2019,
        month = dec,
       volume = {490},
       number = {3},
        pages = {3158-3176},
          doi = {10.1093/mnras/stz2803},
archivePrefix = {arXiv},
       eprint = {1910.00542},
 primaryClass = {astro-ph.SR},
       adsurl = {https://ui.adsabs.harvard.edu/abs/2019MNRAS.490.3158C}
}

@ARTICLE{cuntz22,
       author = {{Cuntz}, M. and {Luke}, G.~E. and {Millard}, M.~J. and {Boyle}, L. and {Patel}, S.~D.},
        title = "{An Early Catalog of Planet-hosting Multiple-star Systems of Order Three and Higher}",
      journal = {ApJS},
         year = 2022,
        month = dec,
       volume = {263},
       number = {2},
          eid = {33},
        pages = {33},
          doi = {10.3847/1538-4365/ac9302},
archivePrefix = {arXiv},
       eprint = {2209.11346},
 primaryClass = {astro-ph.SR},
       adsurl = {https://ui.adsabs.harvard.edu/abs/2022ApJS..263...33C}
}

@ARTICLE{curiel20,
       author = {{Curiel}, Salvador and {Ortiz-Le{\'o}n}, Gisela N. and {Mioduszewski}, Amy J. and {Torres}, Rosa M.},
        title = "{An Astrometric Planetary Companion Candidate to the M9 Dwarf TVLM 513-46546}",
      journal = {AJ},
         year = 2020,
        month = sep,
       volume = {160},
       number = {3},
          eid = {97},
        pages = {97},
          doi = {10.3847/1538-3881/ab9e6e},
       adsurl = {https://ui.adsabs.harvard.edu/abs/2020AJ....160...97C}
}

@ARTICLE{curiel22,
       author = {{Curiel}, Salvador and {Ortiz-Le{\'o}n}, Gisela N. and {Mioduszewski}, Amy J. and {Sanchez-Bermudez}, Joel},
        title = "{3D Orbital Architecture of a Dwarf Binary System and Its Planetary Companion}",
      journal = {AJ},
         year = 2022,
        month = sep,
       volume = {164},
       number = {3},
          eid = {93},
        pages = {93},
          doi = {10.3847/1538-3881/ac7c66},
       adsurl = {https://ui.adsabs.harvard.edu/abs/2022AJ....164...93C}
}

@ARTICLE{cushing09,
   author = {{Cushing}, M.~C. and {Looper}, D. and {Burgasser}, A.~J. and 
	{Kirkpatrick}, J.~D. and {Faherty}, J. and {Cruz}, K.~L. and 
	{Sweet}, A. and {Sanderson}, R.~E.},
    title = "{2MASS J06164006{\ndash}6407194: The First Outer Halo L Subdwarf}",
  journal = {ApJ},
archivePrefix = "arXiv",
   eprint = {0902.1059},
     year = 2009,
    month = may,
   volume = 696,
    pages = {986-993},
      doi = {10.1088/0004-637X/696/1/986},
   adsurl = {http://cdsads.u-strasbg.fr/abs/2009ApJ...696..986C}
}

@ARTICLE{cutri14,
   author = {{Cutri}, R.~M. and {et al.}},
    title = "{VizieR Online Data Catalog: AllWISE Data Release (Cutri+ 2013)}",
  journal = {VizieR Online Data Catalog},
     year = 2014,
    month = jan,
   volume = 2328,
    pages = {0},
   adsurl = {http://cdsads.u-strasbg.fr/abs/2014yCat.2328....0C}
}

@ARTICLE{daemgen09,
       author = {{Daemgen}, S. and {Hormuth}, F. and {Brandner}, W. and {Bergfors}, C. and {Janson}, M. and {Hippler}, S. and {Henning}, T.},
        title = "{Binarity of transit host stars. Implications for planetary parameters}",
      journal = {A\&A},
         year = 2009,
        month = may,
       volume = {498},
       number = {2},
        pages = {567-574},
          doi = {10.1051/0004-6361/200810988},
archivePrefix = {arXiv},
       eprint = {0902.2179},
 primaryClass = {astro-ph.SR},
       adsurl = {https://ui.adsabs.harvard.edu/abs/2009A&A...498..567D}
}

@ARTICLE{dalba21,
       author = {{Dalba}, Paul A. and {Kane}, Stephen R. and {Howell}, Steve B. and {Horch}, Elliott P. and {Li}, Zhexing and {Hirsch}, Lea A. and {Burt}, Jennifer and {Brandt}, Timothy D. and {Mo{\v{c}}nik}, Teo and {Henry}, Gregory W. and {Everett}, Mark E. and {Rosenthal}, Lee J. and {Howard}, Andrew W.},
        title = "{Speckle Imaging Characterization of Radial Velocity Exoplanet Systems}",
      journal = {AJ},
         year = 2021,
        month = mar,
       volume = {161},
       number = {3},
          eid = {123},
        pages = {123},
          doi = {10.3847/1538-3881/abd6ed},
archivePrefix = {arXiv},
       eprint = {2012.05253},
 primaryClass = {astro-ph.EP},
       adsurl = {https://ui.adsabs.harvard.edu/abs/2021AJ....161..123D}
}

@ARTICLE{dario14,
       author = {{Da Rio}, N. and {Jeffries}, R.~D. and {Manara}, C.~F. and {Robberto}, M.},
        title = "{Strong biases in estimating the time dependence of mass accretion rates in young stars}",
      journal = {MNRAS},
         year = 2014,
        month = apr,
       volume = {439},
       number = {4},
        pages = {3308-3328},
          doi = {10.1093/mnras/stu149},
archivePrefix = {arXiv},
       eprint = {1401.5062},
 primaryClass = {astro-ph.SR},
       adsurl = {https://ui.adsabs.harvard.edu/abs/2014MNRAS.439.3308D}
}

@ARTICLE{dasilva15,
       author = {{da Silva}, Ronaldo and {Milone}, Andr{\'e} de C. and {Rocha-Pinto}, Helio J.},
        title = "{Homogeneous abundance analysis of FGK dwarf, subgiant, and giant stars with and without giant planets}",
      journal ={A\&A},
         year = 2015,
        month = aug,
       volume = {580},
          eid = {A24},
        pages = {A24},
          doi = {10.1051/0004-6361/201525770},
       adsurl = {https://ui.adsabs.harvard.edu/abs/2015A&A...580A..24D}
}

@ARTICLE{david15,
       author = {{David}, Trevor J. and {Hillenbrand}, Lynne A.},
        title = "{The Ages of Early-type Stars: Str{\"o}mgren Photometric Methods Calibrated, Validated, Tested, and Applied to Hosts and Prospective Hosts of Directly Imaged Exoplanets}",
      journal ={ApJ},
         year = 2015,
        month = may,
       volume = {804},
       number = {2},
          eid = {146},
        pages = {146},
          doi = {10.1088/0004-637X/804/2/146},
archivePrefix = {arXiv},
       eprint = {1501.03154},
 primaryClass = {astro-ph.SR},
       adsurl = {https://ui.adsabs.harvard.edu/abs/2015ApJ...804..146D}
}

@ARTICLE{davison14,
       author = {{Davison}, Cassy L. and {White}, R.~J. and {Jao}, W. -C. and {Henry}, T.~J. and {Bailey}, J.~I., III and {Quinn}, S.~N. and {Cantrell}, J.~R. and {Riedel}, A.~R. and {Subasavage}, J.~P. and {Winters}, J.~G. and {Crockett}, C.~J.},
        title = "{The Closest M-dwarf Quadruple System to the Sun}",
      journal = {AJ},
         year = 2014,
        month = feb,
       volume = {147},
       number = {2},
          eid = {26},
        pages = {26},
          doi = {10.1088/0004-6256/147/2/26},
archivePrefix = {arXiv},
       eprint = {1310.6761},
 primaryClass = {astro-ph.SR},
       adsurl = {https://ui.adsabs.harvard.edu/abs/2014AJ....147...26D}
}

@ARTICLE{deacon14,
       author = {{Deacon}, Niall R. and {Liu}, Michael C. and {Magnier}, Eugene A. and {Aller}, Kimberly M. and {Best}, William M.~J. and {Dupuy}, Trent and {Bowler}, Brendan P. and {Mann}, Andrew W. and {Redstone}, Joshua A. and {Burgett}, William S. and {Chambers}, Kenneth C. and {Draper}, Peter W. and {Flewelling}, H. and {Hodapp}, Klaus W. and {Kaiser}, Nick and {Kudritzki}, Rolf-Peter and {Morgan}, Jeff S. and {Metcalfe}, Nigel and {Price}, Paul A. and {Tonry}, John L. and {Wainscoat}, Richard J.},
        title = "{Wide Cool and Ultracool Companions to Nearby Stars from Pan-STARRS 1}",
      journal = {ApJ},
         year = 2014,
        month = sep,
       volume = {792},
       number = {2},
          eid = {119},
        pages = {119},
          doi = {10.1088/0004-637X/792/2/119},
archivePrefix = {arXiv},
       eprint = {1407.2938},
 primaryClass = {astro-ph.SR},
       adsurl = {https://ui.adsabs.harvard.edu/abs/2014ApJ...792..119D}
}

@MISC{debruijne17,
       author = {{de Bruijne}, J.~H.~J. and {Abreu}, A. and {Brown}, A.~G.~A. and {Casta{\~n}eda}, J. and {Cheek}, N. and {Crowley}, C. and {De Angeli}, F. and {Fabricius}, C. and {Fleitas}, J.~M. and {Gracia}, G. and {Guerra}, R. and {Hutton}, A. and {Messineo}, R. and {Mora}, A. and {Nienartowicz}, K. and {Panem}, C. and {Siddiqui}, H.},
        title = "{Gaia DR1 documentation Chapter 1: Introduction}",
 howpublished = {Gaia DR1 documentation, European Space Agency; Gaia Data Processing and Analysis Consortium. Online at https://gaia.esac.esa.int/documentation/GDR1/index.htm, id. 1},
         year = 2017,
        month = dec,
          eid = {1},
        pages = {1},
       adsurl = {https://ui.adsabs.harvard.edu/abs/2017gdr1.reptE...1D}
}

@ARTICLE{degeus90,
       author = {{de Geus}, E.~J. and {Lub}, J. and {van de Grift}, E.},
        title = "{Walraven photometry of nearby southern OB associations.}",
      journal = {A\&AS},
         year = 1990,
        month = oct,
       volume = {85},
        pages = {915},
       adsurl = {https://ui.adsabs.harvard.edu/abs/1990A&AS...85..915D}
}

@ARTICLE{delfosse99a,
       author = {{Delfosse}, X. and {Forveille}, T. and {Beuzit}, J. -L. and {Udry}, S. and {Mayor}, M. and {Perrier}, C.},
        title = "{New neighbours. I. 13 new companions to nearby M dwarfs}",
      journal = {A\&A},
         year = 1999,
        month = apr,
       volume = {344},
        pages = {897-910},
          doi = {10.48550/arXiv.astro-ph/9812008},
archivePrefix = {arXiv},
       eprint = {astro-ph/9812008},
 primaryClass = {astro-ph},
       adsurl = {https://ui.adsabs.harvard.edu/abs/1999A&A...344..897D}
}

@ARTICLE{delfosse99b,
       author = {{Delfosse}, X. and {Forveille}, T. and {Udry}, S. and {Beuzit}, J. -L. and {Mayor}, M. and {Perrier}, C.},
        title = "{Accurate masses of very low mass stars. II. The very low mass triple system GL 866}",
      journal = {A\&A},
         year = 1999,
        month = oct,
       volume = {350},
        pages = {L39-L42},
          doi = {10.48550/arXiv.astro-ph/9909409},
archivePrefix = {arXiv},
       eprint = {astro-ph/9909409},
 primaryClass = {astro-ph},
       adsurl = {https://ui.adsabs.harvard.edu/abs/1999A&A...350L..39D}
}

@INPROCEEDINGS{delfosse04,
   author = {{Delfosse}, X. and {Beuzit}, {J.-L.} and {Marchal}, L. and {Bonfils}, X. and 
	{Perrier}, C. and {S{\'e}gransan}, D. and {Udry}, S. and {Mayor}, M. and 
	{Forveille}, T.},
    title = "{M dwarfs binaries: Results from accurate radial velocities and high angular resolution observations}",
booktitle = {Spectroscopically and Spatially Resolving the Components of the Close Binary Stars},
     year = 2004,
   series = {ASPCS},
   volume = 318,
   editor = "{R.~W.~Hilditch, H.~Hensberge, \& K.~Pavlovski}",
    month = dec,
    pages = {166-174},
   adsurl = {http://cdsads.u-strasbg.fr/abs/2004ASPC..318..166D}
}

@ARTICLE{delpino24,
       author = {{del Pino}, A. and {L{\'o}pez-Sanjuan}, C. and {Hern{\'a}n-Caballero}, A. and {Dom{\'\i}nguez-S{\'a}nchez}, H. and {von Marttens}, R. and {Fern{\'a}ndez-Ontiveros}, J.~A. and {Coelho}, P.~R.~T. and {Lumbreras-Calle}, A. and {Vega-Ferrero}, J. and {Jimenez-Esteban}, F. and {Cruz}, P. and {Marra}, V. and {Quartin}, M. and {Galarza}, C.~A. and {Angulo}, R.~E. and {Cenarro}, A.~J. and {Crist{\'o}bal-Hornillos}, D. and {Dupke}, R.~A. and {Ederoclite}, A. and {Hern{\'a}ndez-Monteagudo}, C. and {Mar{\'\i}n-Franch}, A. and {Moles}, M. and {Sodr{\'e}}, L. and {Varela}, J. and {V{\'a}zquez Rami{\'o}}, H.},
        title = "{J-PLUS: Bayesian object classification with a strum of BANNJOS}",
      journal = {A\&A},
         year = 2024,
        month = nov,
       volume = {691},
          eid = {A221},
        pages = {A221},
          doi = {10.1051/0004-6361/202450503},
archivePrefix = {arXiv},
       eprint = {2404.16567},
 primaryClass = {astro-ph.GA},
       adsurl = {https://ui.adsabs.harvard.edu/abs/2024A&A...691A.221D}
}

@ARTICLE{demarco17,
       author = {{De Marco}, Orsola and {Izzard}, Robert G.},
        title = "{Dawes Review 6: The Impact of Companions on Stellar Evolution}",
      journal = {PASA},
         year = 2017,
        month = jan,
       volume = {34},
          eid = {e001},
        pages = {e001},
          doi = {10.1017/pasa.2016.52},
archivePrefix = {arXiv},
       eprint = {1611.03542},
 primaryClass = {astro-ph.SR},
       adsurl = {https://ui.adsabs.harvard.edu/abs/2017PASA...34....1D}
}

@ARTICLE{derosa14,
       author = {{De Rosa}, R.~J. and {Patience}, J. and {Ward-Duong}, K. and {Vigan}, A. and {Marois}, C. and {Song}, I. and {Macintosh}, B. and {Graham}, J.~R. and {Doyon}, R. and {Bessell}, M.~S. and {Lai}, O. and {McCarthy}, D.~W. and {Kulesa}, C.},
        title = "{The VAST Survey - IV. A wide brown dwarf companion to the A3V star {\ensuremath{\zeta}} Delphini}",
      journal = {MNRAS},
         year = 2014,
        month = dec,
       volume = {445},
       number = {4},
        pages = {3694-3705},
          doi = {10.1093/mnras/stu2018},
archivePrefix = {arXiv},
       eprint = {1410.0005},
 primaryClass = {astro-ph.SR},
       adsurl = {https://ui.adsabs.harvard.edu/abs/2014MNRAS.445.3694D}
}

@ARTICLE{desgrange23,
       author = {{Desgrange}, C. and {Milli}, J. and {Chauvin}, G. and {Henning}, Th. and {Luashvili}, A. and {Read}, M. and {Wyatt}, M. and {Kennedy}, G. and {Burn}, R. and {Schlecker}, M. and {Kiefer}, F. and {D'Orazi}, V. and {Messina}, S. and {Rubini}, P. and {Lagrange}, A. -M. and {Babusiaux}, C. and {Matr{\`a}}, L. and {Bitsch}, B. and {Bonavita}, M. and {Delorme}, P. and {Matthews}, E. and {Palma-Bifani}, P. and {Vigan}, A.},
        title = "{Planetary system architectures with low-mass inner planets. Direct imaging exploration of mature systems beyond 1 au}",
      journal = {A\&A},
         year = 2023,
        month = dec,
       volume = {680},
          eid = {A64},
        pages = {A64},
          doi = {10.1051/0004-6361/202346863},
archivePrefix = {arXiv},
       eprint = {2310.06035},
 primaryClass = {astro-ph.EP},
       adsurl = {https://ui.adsabs.harvard.edu/abs/2023A&A...680A..64D}
}

@ARTICLE{desidera21,
       author = {{Desidera}, S. and {Chauvin}, G. and {Bonavita}, M. and {Messina}, S. and {LeCoroller}, H. and {Schmidt}, T. and {Gratton}, R. and {Lazzoni}, C. and {Meyer}, M. and {Schlieder}, J. and {Cheetham}, A. and {Hagelberg}, J. and {Bonnefoy}, M. and {Feldt}, M. and {Lagrange}, A. -M. and {Langlois}, M. and {Vigan}, A. and {Tan}, T.~G. and {Hambsch}, F. -J. and {Millward}, M. and {Alcal{\'a}}, J. and {Benatti}, S. and {Brandner}, W. and {Carson}, J. and {Covino}, E. and {Delorme}, P. and {D'Orazi}, V. and {Janson}, M. and {Rigliaco}, E. and {Beuzit}, J. -L. and {Biller}, B. and {Boccaletti}, A. and {Dominik}, C. and {Cantalloube}, F. and {Fontanive}, C. and {Galicher}, R. and {Henning}, Th. and {Lagadec}, E. and {Ligi}, R. and {Maire}, A. -L. and {Menard}, F. and {Mesa}, D. and {M{\"u}ller}, A. and {Samland}, M. and {Schmid}, H.~M. and {Sissa}, E. and {Turatto}, M. and {Udry}, S. and {Zurlo}, A. and {Asensio-Torres}, R. and {Kopytova}, T. and {Rickman}, E. and {Abe}, L. and {Antichi}, J. and {Baruffolo}, A. and {Baudoz}, P. and {Baudrand}, J. and {Blanchard}, P. and {Bazzon}, A. and {Buey}, T. and {Carbillet}, M. and {Carle}, M. and {Charton}, J. and {Cascone}, E. and {Claudi}, R. and {Costille}, A. and {Deboulb{\'e}}, A. and {De Caprio}, V. and {Dohlen}, K. and {Fantinel}, D. and {Feautrier}, P. and {Fusco}, T. and {Gigan}, P. and {Giro}, E. and {Gisler}, D. and {Gluck}, L. and {Hubin}, N. and {Hugot}, E. and {Jaquet}, M. and {Kasper}, M. and {Madec}, F. and {Magnard}, Y. and {Martinez}, P. and {Maurel}, D. and {Le Mignant}, D. and {M{\"o}ller-Nilsson}, O. and {Llored}, M. and {Moulin}, T. and {Orign{\'e}}, A. and {Pavlov}, A. and {Perret}, D. and {Petit}, C. and {Pragt}, J. and {Puget}, P. and {Rabou}, P. and {Ramos}, J. and {Rigal}, F. and {Rochat}, S. and {Roelfsema}, R. and {Rousset}, G. and {Roux}, A. and {Salasnich}, B. and {Sauvage}, J. -F. and {Sevin}, A. and {Soenke}, C. and {Stadler}, E. and {Suarez}, M. and {Weber}, L. and {Wildi}, F.},
        title = "{The SPHERE infrared survey for exoplanets (SHINE). I. Sample definition and target characterization}",
      journal = {A\&A},
         year = 2021,
        month = jul,
       volume = {651},
          eid = {A70},
        pages = {A70},
          doi = {10.1051/0004-6361/202038806},
archivePrefix = {arXiv},
       eprint = {2103.04366},
 primaryClass = {astro-ph.EP},
       adsurl = {https://ui.adsabs.harvard.edu/abs/2021A&A...651A..70D}
}

@ARTICLE{desilva06,
       author = {{De Silva}, G.~M. and {Sneden}, C. and {Paulson}, D.~B. and {Asplund}, M. and {Bland-Hawthorn}, J. and {Bessell}, M.~S. and {Freeman}, K.~C.},
        title = "{Chemical Homogeneity in the Hyades}",
      journal = {AJ},
         year = 2006,
        month = jan,
       volume = {131},
       number = {1},
        pages = {455-460},
          doi = {10.1086/497968},
archivePrefix = {arXiv},
       eprint = {astro-ph/0509241},
 primaryClass = {astro-ph},
       adsurl = {https://ui.adsabs.harvard.edu/abs/2006AJ....131..455D}
}
\newpage

%%%%% ACKNOWLEDGEMENTS %%%%%

\chapter*{Acknowledgements}
\label{ch:Acknowledgements}
\vspace{2cm}
\pagestyle{fancy}
\fancyhf{}
\lhead[\small{\textbf{\thepage}}]{\textbf{Acknowledgements}}
\rhead[\small{\textbf{Acknowledgements}}]{\small{\textbf{\thepage}}}
\addcontentsline{toc}{chapter}{Acknowledgements}

\normalsize

We acknowledge financial support from the Agencia Estatal de Investigaci\'on (AEI/10.13039/501100011033) of the Ministerio de Ciencia e Innovaci\'on and the ERDF ``A way of making Europe'' through the project PID2022-137241NB-C42.
Based on data from the SVO hot subdwarf Data Access Service at CAB (INTA-CSIC). 
This research has made use of the Spanish Virtual Observatory (\url{http://svo.cab.inta-csic.es}) supported from the Spanish MINECO/FEDER through grant AyA2014-55216. 
This publication makes use of \textit{VOSA}, developed under the Spanish Virtual Observatory project supported from the Spanish MINECO through grant AyA2017-84089.
This research has made use of the \textit{Simbad} database \citep{wenger00}, the VizieR \citep{ochsenbein00} catalogue access tool, and the \textit{Aladin} sky atlas \citep{bonnarel00} at the Centre de données astronomiques de Strasbourg (France).
This work has made use of \texttt{Topcat} \citep{taylor05}.
This work presents results from the European Space Agency (ESA) space mission \textit{Gaia}, processed by the \textit{Gaia} Data Processing and Analysis Consortium (DPAC). Funding for the DPAC is provided by national institutions, in particular the institutions participating in the \textit{Gaia} MultiLateral Agreement (MLA). The \textit{Gaia} mission website is \url{https://www.cosmos.esa.int/gaia}.
This publication makes use of data products from the Two Micron All Sky Survey, which is a joint project of the University of Massachusetts and the Infrared Processing and Analysis Center/California Institute of Technology, funded by the National Aeronautics and Space Administration and the National Science Foundation.
This publication makes use of data products from the Wide-field Infrared Survey Explorer, which is a joint project of the University of California, Los Angeles, and the Jet Propulsion Laboratory/California Institute of Technology, funded by the National Aeronautics and Space Administration. 
This research has made use of MATLAB Version: 9.9.0.2037887 (R2020b) Update 8. The UHS is a partnership between the UK STFC, The University of Hawaii, The University of Arizona, Lockheed Martin and NASA.
Funding for SDSS-III has been provided by the Alfred P. Sloan Foundation, the Participating Institutions, the National Science Foundation, and the U.S. Department of Energy Office of Science. The SDSS-III web site is \url{http://www.sdss3.org/}.
This research has made use of the Washington Double Star Catalog maintained at the U.S. Naval Observatory.
This research has made use of data obtained from or tools provided by the portal exoplanet.eu of The Extrasolar Planets Encyclopaedia.
This research made use of the NASA’s Astrophysics Data System Bibliographic Services and the Exoplanet Archive, which is operated by the California Institute of Technology, under contract with the National Aeronautics and Space Administration under the Exoplanet Exploration Program.
This document was typeset using Leslie Lamport’s \LaTeX, based on Donald E. Knuth’s \TeX{}.

\newpage

%%%%% LISTA DE TRABAJOS DEL AUTOR %%%%%

\chapter*{List of author's works}
\label{ch:list_of_authors_publications}
\vspace{2cm}
\pagestyle{fancy}
\fancyhf{}
\lhead[\small{\textbf{\thepage}}]{\textbf{List of author's works}}
\rhead[\small{\textbf{List of author's works}}]{\small{\textbf{\thepage}}}
\addcontentsline{toc}{chapter}{List of author's works}

\section*{Peer-reviewed publications}
\label{sec:peer-reviewed_publications}

\begin{enumerate}

\item \tB{``Wide companions to M and L subdwarfs with \textit{Gaia} and the Virtual Observatory''}\\
\textbf{J. González-Payo}, M. Cortés-Contreras, N. Lodieu, E. Solano, Z. H. Zhang, M.-C. Gálvez-Ortiz\\
\textit{Astronomy \& Astrophysics} 650, A190 (2021) \\
\url{https://doi.org/10.1051/0004-6361/202140493}

\item \tB{``Reaching the boundary between stellar kinematic groups and very wide binaries.\\
IV. The widest Washington Double Star systems with $\rho\geq$\,1000\,arcsec in \textit{Gaia} DR3''}\\
\textbf{J. González-Payo}, J. A. Caballero, M. Cortés-Contreras\\
\textit{Astronomy \& Astrophysics} 670, A102 (2023) \\
\url{https://doi.org/10.1051/0004-6361/202245476}

\item \tB{``Discovery of stellar binaries by Giovanni Battista Hodierna in 1654''}\\
\textbf{J. González-Payo}, J. A. Caballero \\
\textit{Monthly Notices of the Royal Astronomical Society} 533, 3379 (2024) \\
\url{https://doi.org/10.1093/mnras/stae2010}

\item \tB{``Multiplicity of stars with planets in the solar neighbourhood''}\\
\textbf{J. González-Payo}, J. A. Caballero, J. Gorgas, M. Cortés-Contreras, M.-C. Gálvez-Ortiz, C. Cifuentes \\
\textit{Astronomy \& Astrophysics} 689, A302 (2024) \\
\url{https://doi.org/10.1051/0004-6361/202450048}

\item \tB{``CARMENES input catalogue of M dwarfs.\\
IX. Multiplicity from close spectroscopic binaries to ultrawide systems''}\\
C. Cifuentes, J. A. Caballero, \textbf{J. González-Payo}, P. J. Amado, V. J. S. Béjar, A. J. Burgasser et al.
%M. Cortés-Contreras, N. Lodieu, D. Montes, A. Reiners, I. Ribas, A. Quirrenbach, J. Sanz-Forcada, W. Seifert, A. Tokovinin, M. R. Zapatero Osorio \\
\textit{Astronomy \& Astrophysics} 693, A228 (2025) \\
\url{https://doi.org/10.1051/0004-6361/202452527}

\item \tB{``Characterisation of all known multiple stellar systems within 10 pc''}\\
\textbf{J. González-Payo}, J. A. Caballero, C. Cifuentes, M. Cortés-Contreras \\
\textit{Astronomy \& Astrophysics}. Submitted (under referee review)

\end{enumerate}

\section*{Conference proceedings and scientific meetings}
\label{sec:conference_proceedings_and_scientific_mmetings}

\begin{enumerate}

\item \tB{``Wide companions to M and L subdwarfs with \textit{Gaia} DR2 and the
Virtual Observatory''}\\
\textbf{J. González-Payo}, M. Cortés-Contreras, N. Lodieu, E. Solano, Z.H. Zhang, M.-C. Gálvez-Ortiz \\
\textit{Contributions to the XIV.0 Scientific Meeting (virtual) of the Spanish Astronomical Society}. 15 July 2020 \\
\url{https://ui.adsabs.harvard.edu/abs/2020sea..confE.143G/abstract}

\item \tB{``The widest Washington Double Star systems with $\rho\geq$\,1000\,arcsec in \textit{Gaia} DR3''}\\
\textbf{J. González-Payo}, J. A. Caballero, M. Cortés-Contreras \\
\textit{EAS 2023. European Astronomical Society Annual Meeting}. 10 -- 14 July 2023. Krakow, Poland \\
\url{https://ui.adsabs.harvard.edu/#abs/2023eas..conf.1432G/abstract}

\item \tB{``Multiplicity in low-metallicity objects''}\\
M.-C. Gálvez-Ortiz, M. Cortés-Contreras, \textbf{J. González-Payo}, Z. H. Zhang, P. Cruz, H. R. A. Jones, E. Solano, D. Pinfield, M. R. Zapatero Osorio \\
\textit{EAS 2023. European Astronomical Society Annual Meeting}. 10 -- 14 July 2023. Krakow, Poland \\
\url{https://ui.adsabs.harvard.edu/#abs/2023eas..conf.1708G/abstract}

\item \tB{``Multiplicity of stars with planets in the solar neighbourhood''}\\ 
\textbf{J. González-Payo} \\
\textit{ESA's Madrid - Area exoplanet science meeting 2023 (MAESM 2023)}. 3 October 2023.\\
\url{https://nextcloud.com.uvigo.es/s/k9rEbX5s5NFRkGR}

\item \tB{``Multiplicity of stars with planets''}\\ 
\textbf{J. González-Payo} \\
\textit{16th Scientific Meeting of the Spanish Astronomical Society}. 15 -- 19 July 2024. Granada, Spain \\
\url{https://ui.adsabs.harvard.edu/abs/2025hsa..conf..182G/abstract}

\item \tB{``Search and characterisation of multiple host star systems''}\\ 
\textbf{J. González-Payo} \\
\textit{Seminars cycle for 20th Anniversary of the Spanish Virtual Observatory (SVO)}. 8 May 2025. Centro de Astrobiología (CSIC/INTA). Campus INTA. Torrejón de Ardoz, Madrid (Spain) \\
\url{https://doi.org/10.5281/zenodo.15459144}

\end{enumerate}

\section*{\textit{VizieR} data online catalogues}
\label{sec:vizier_data_online_catalogues}

\begin{enumerate}

\item \tB{``Wide companions to M and L subdwarfs''.} (Gonzalez-Payo+, 2021) \\
J/A+A/650/A190 \\
\url{https://cdsarc.cds.unistra.fr/viz-bin/cat/J/A+A/650/A190}

\item \tB{``WDS systems with rho>1000 arcsec''.} (Gonzalez-Payo+, 2023) \\
J/A+A/670/A102 \\
\url{https://cdsarc.cds.unistra.fr/viz-bin/cat/J/A+A/670/A102}

\item \tB{``Multiplicity of stars with planets in the solar neighbourhood''.} (Gonzalez-Payo+, 2024) \\
J/A+A/689/A302 \\
\url{https://cdsarc.cds.unistra.fr/viz-bin/cat/J/A+A/689/A302}

\item \tB{``CARMENES input catalogue of M dwarfs. IX''.} (Cifuentes+, 2025)\\
J/A+A/693/A228 \\
\url{https://cdsarc.cds.unistra.fr/viz-bin/cat/J/A+A/693/A228}

\end{enumerate}

\section*{Contribution to Washington Double Star (WDS) catalogue}
\label{sec:contribution_to_WDS}

A total of 27 new pairs, forming 19 multiple systems, have been added to the WDS catalogue \citep{mason01}, all with discovery codes ``JGP nn'', coming from the following articles developed in this Ph.D thesis: \citet{gonzalezpayo21}, \citet{gonzalezpayo23}, and \citet{gonzalezpayo24}. These pairs are:

\begin{itemize}
    \item Eigth wide binaries with $\rho$ between 225.7 and 953.0\,arcsec.
    \item Nineteen ultra wide binaries with $\rho\ge$ 1000\,arcsec.
\end{itemize}

\begin{figure}[h]
 \centering
 \includegraphics[width=1\linewidth, angle=0]{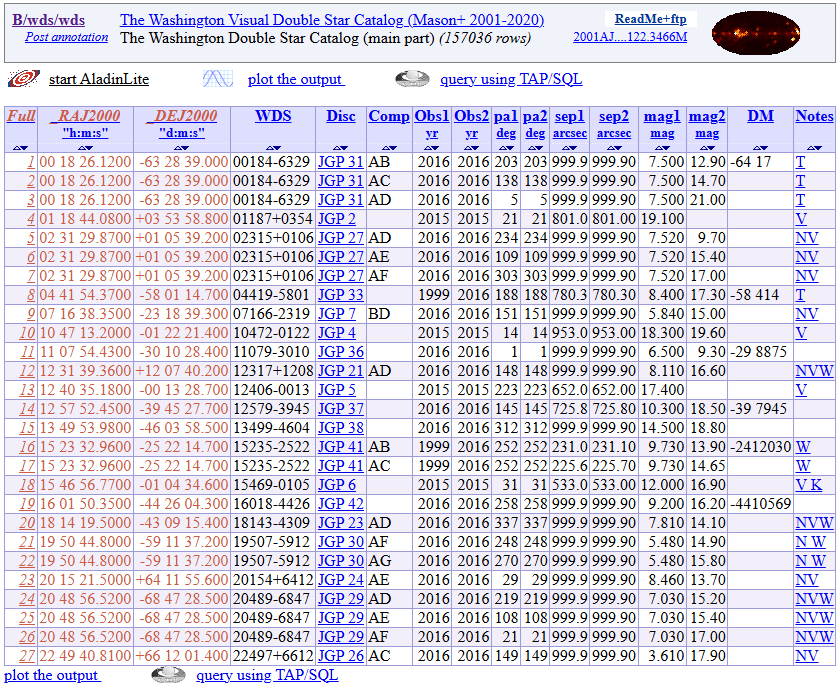}
% \caption{List of pairs added to WDS by the author (list obtained through \textit{VizieR}).}
 \label{fig:wds_author}
\end{figure}

Aditionally, $\rho$, $\theta$ and epoch of other pairs coming from the mentioned articles have been also added to the Washington Double Star Suplementary (WDSS) catalogue.

\newpage

\section*{First page of published first-author articles}
\label{sec:article_covers}

Below, the first page of each first-author article that forms the core of the doctoral thesis can be seen, arranged in chronological order of publication. The articles have been published in two high-impact journals: \textit{Astronomy \& Astrophysics} with a 2023 impact factor of 5.4; and \textit{Monthly Notices of the Royal Astronomic Society} with a 2022 impact factor of 4.8 (impact factors according to Journal Citation Reports\texttrademark{} from Clarivate, 2024 - Science Citation Index Expanded).

\vspace{1cm}
\begin{figure*}[h]
 \centering
   \includegraphics[width=0.8\hsize]{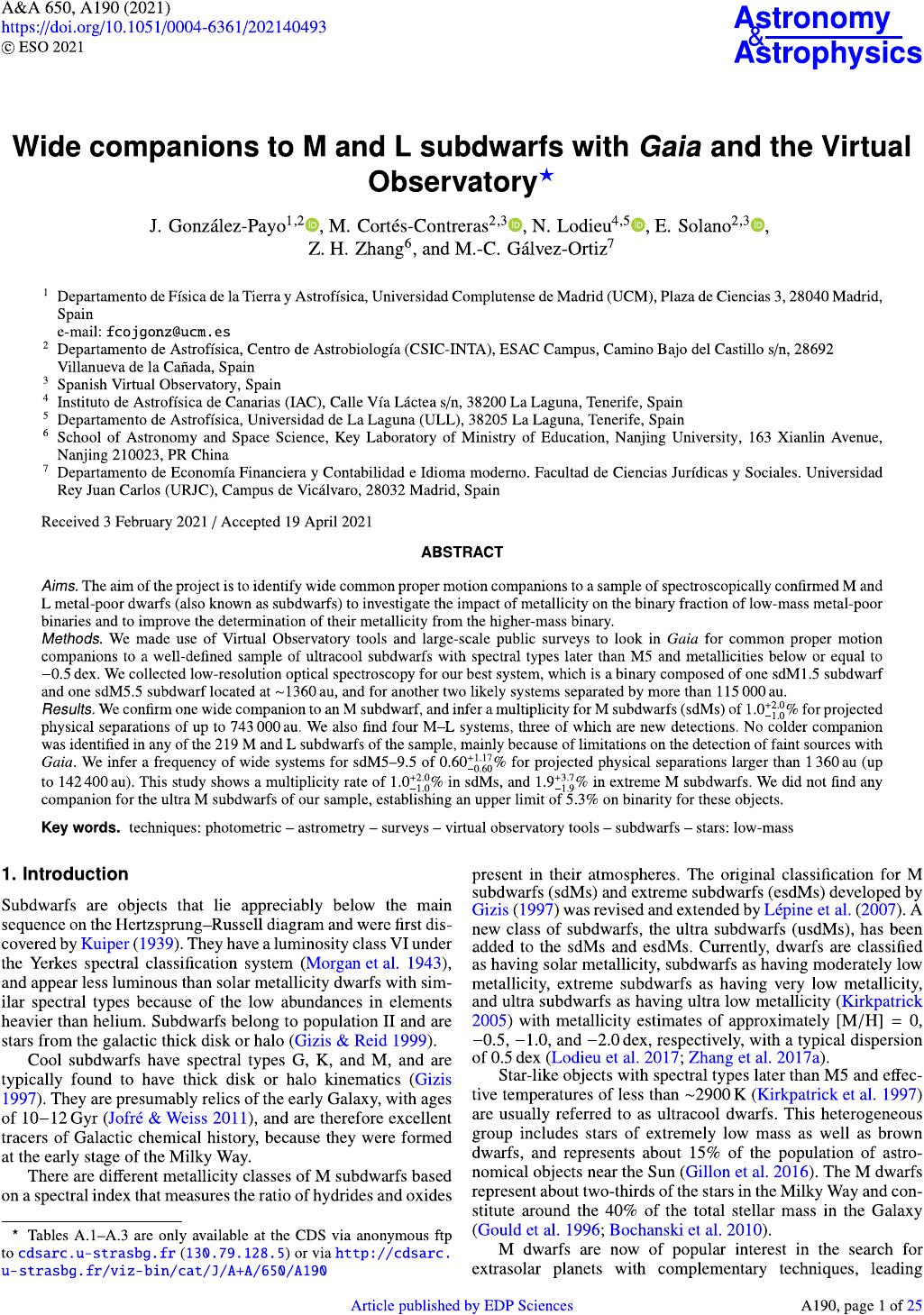}
  \label{fig:paper01}
\end{figure*}

\begin{figure*}
 \centering
   \includegraphics[width=0.8\hsize]{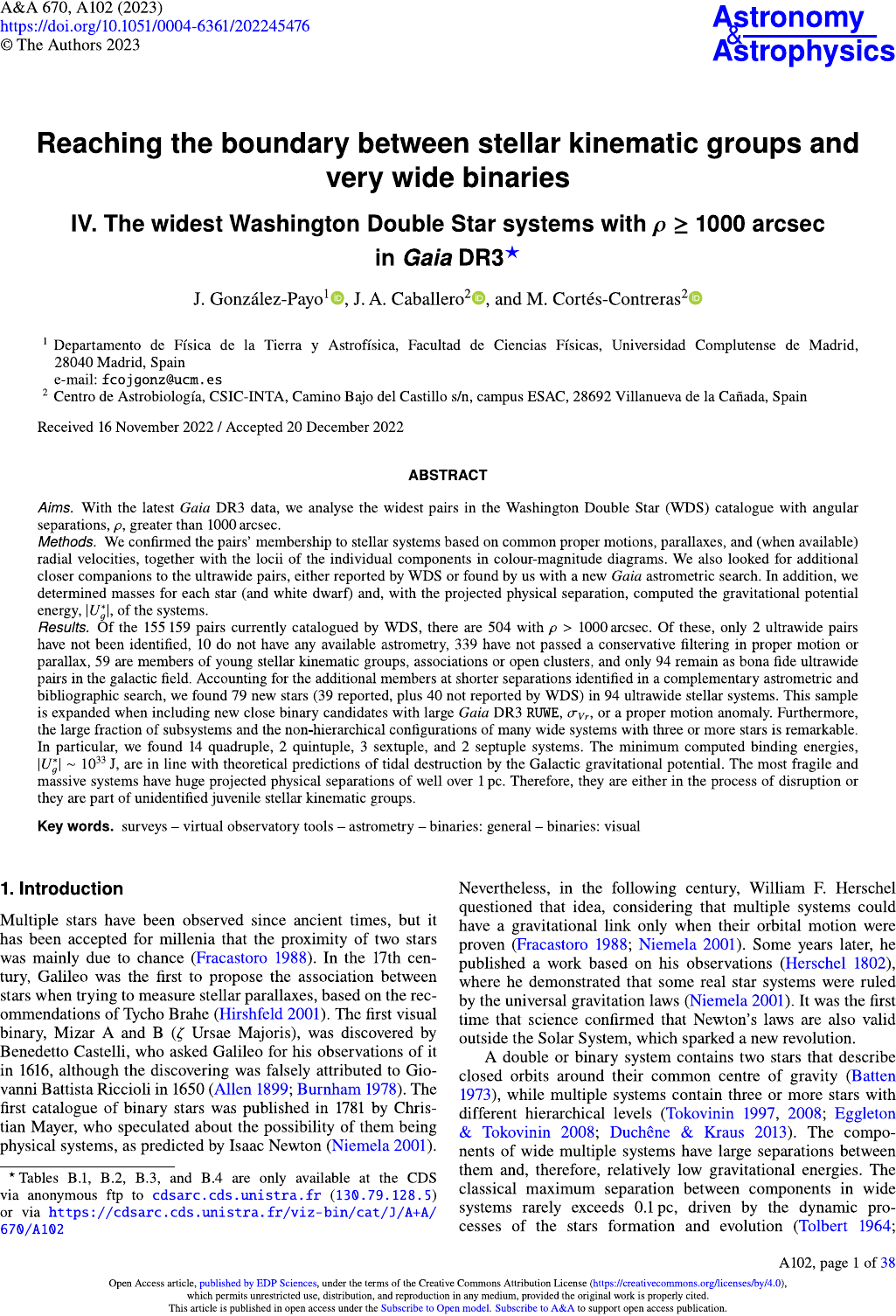}
  \label{fig:paper02}
\end{figure*}   

\begin{figure*}
 \centering 
   \includegraphics[width=0.8\hsize]{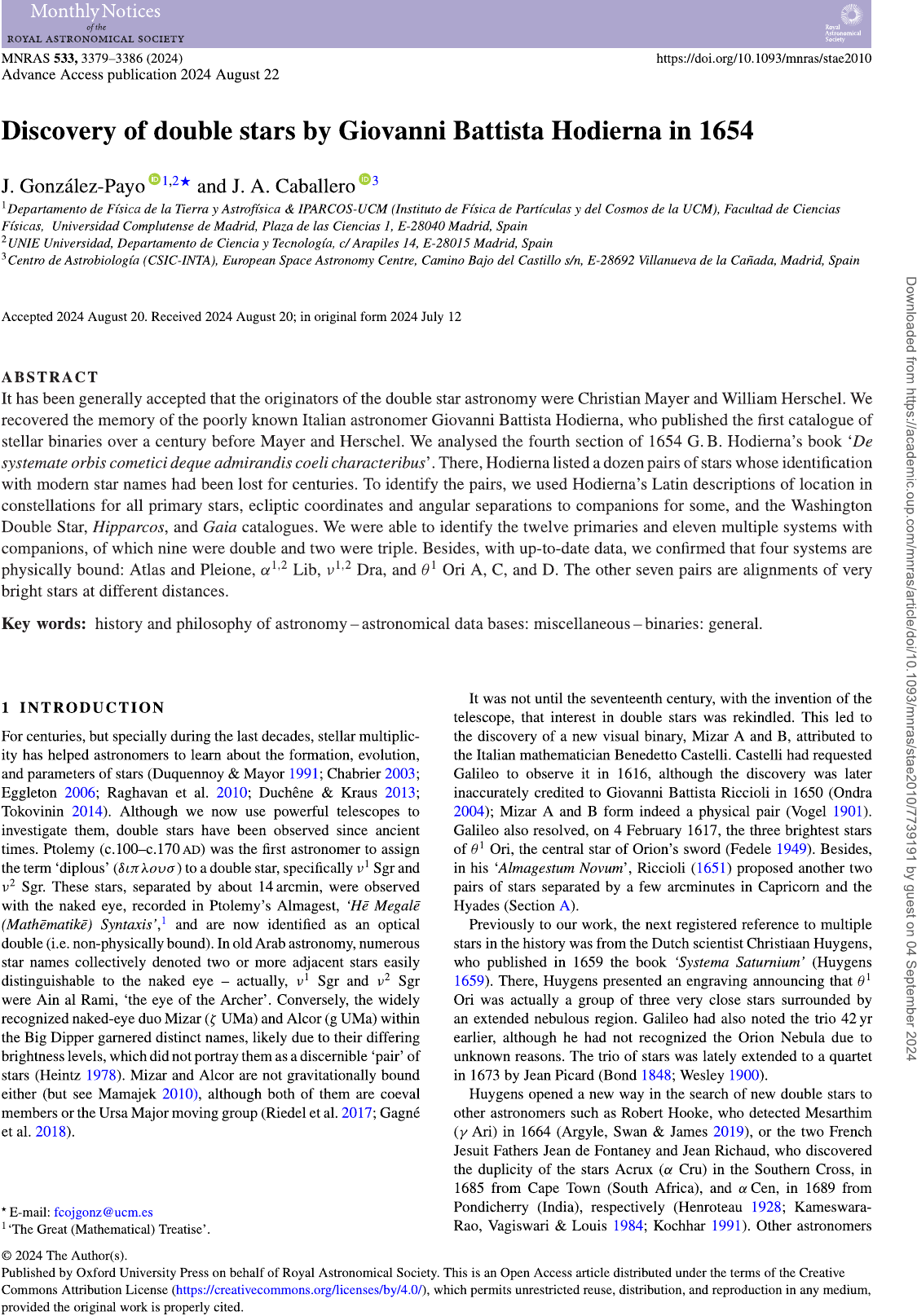}
  \label{fig:paper04}
\end{figure*}
   
\begin{figure*}
 \centering   
   \includegraphics[width=0.8\hsize]{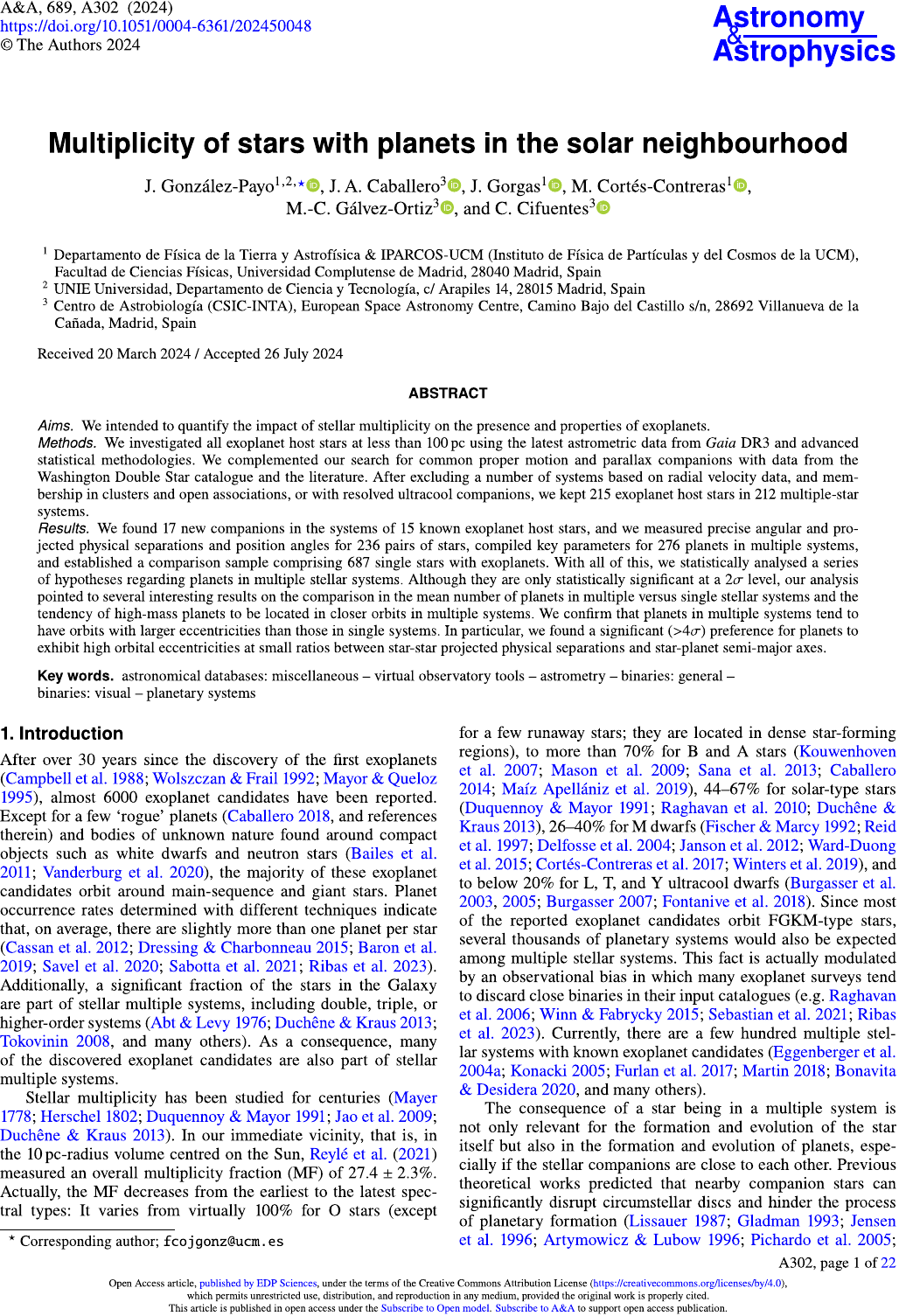}
  \label{fig:paper03}
\end{figure*}   
     
%%%%% APENDICES %%%%%

\appendix
\renewcommand{\chaptername}{}
\titleformat{\chapter}[display]
  {\normalfont\Desmesurado\centering\color{black}}{\centering\chaptertitlename\\\ \MasDesmesurado{\thechapter}}{80pt}{\bfseries\Desmesurado\color{black}}
\titlespacing*{\chapter} 
  {0pt}{10pt}{40pt}

%%%%% APENDICE A %%%%%

\renewcommand{\thetable}{A.\arabic{table}}   % Cambia el formato de numeración al tipo A.x
\begin{center}
\chapter[Appendix: Abbreviations and acronyms]{Abbreviations and acronyms}
\end{center}
\label{ch:Appendix_A}
\vspace{3cm}
\pagestyle{fancy}
\fancyhf{}
\lhead[\small{\textbf{\thepage}}]{\textbf{Appendix~A: Abbreviations and acronyms}}
\rhead[\small{\textbf{Appendix~A: Abbreviations and acronyms}}]{\small{\textbf{\thepage}}}

 \begin{table}[H]
 \caption*{}
 \label{tab:abbr1}
 \begin{tabular}{@{\hspace{10mm}}l@{\hspace{20mm}}l}
 
 1RXS & 1st ROSAT X Survey \\
 2MASS & 2 Micron All-Sky Survey  \\
 2MASSI & 2 Micron All-Sky Survey, Incremental release point source catalogue \\
 2MASSW & 2 Micron All-Sky Survey Working database \\
 $a$ & Year, semi-major axis \\
 ACAM & Auxiliary-port CAMera \\
 ADQL & Astronomical Data Query Language \\
 ADS & Aitken Double Star \\ 
 ALFOSC & Alhambra Faint Object Spectrograph and Camera \\
 AO & Adaptive Optics \\
 APMPM & Automatic Plate Measuring Proper Motion \\
 ASAS & All Sky Automated Survey \\
 au & Astronomical unit \\
 BD & Bonner Durchmusterung, Brown Dwarf \\
 BDS & Burnham Double Stars  \\
 $BP$ & Blue Photometer \\
 BPS & Beers + Preston + Shectman \\
 CAB & Centro de AstroBiología \\

\end{tabular}
\end{table}
 
 \newpage
 
 \begin{table}[H]
 \caption*{}
 \label{tab:abbr2}
 \begin{tabular}{@{\hspace{10mm}}l@{\hspace{20mm}}l} 
 
 CARMENES & Calar Alto high-Resolution search for M dwarfs with Exo-earths with \\
          & Near-infrared and optical Échelle Spectrographs \\ 
 CCD & Charge-Coupled Device \\ 
 CD & Cordoba Durchmusterung \\  
 CDF & Cumulative Distribution Function \\
 CDS & Centre de Données astronomiques de Strasbourg \\
 CEFCA & Centro de Estudios de Física del Cosmos de Aragón \\
 CFBDS & Canada-France Brown Dwarf Survey \\
 CFBDSIR& Canada-France Brown Dwarf Survey-InfraRed \\
 CHARA & Center for High Angular Resolution Astronomy \\
 CSIC &  Centro Superior de Investigaciones Científicas \\  
 CTIO & Cerro Tololo Interamerican Observatory\\
 CMD & Color-Magnitude Diagram \\
 CS & Curtis Schmidt telescope \\
 CSF & Company Star Fraction \\ 
 CWISE & CatWISE \\
 CWISEP & CWISE Preliminary catalogue \\
 $d$ & distance, days \\
 DENIS & DEep Near-Infrared Survey of the southern sky \\
 DI & Direct Imaging \\
 DPAC & Data Processing and Analysis Consortium \\ 
 DR & Data Release \\ 
 $e$ & Eccentricity \\
 EGGR & Eggen + Greenstein \\ 
 ESA & European Space Agency \\ 
 ESAC & European Space Astronomy Center \\
 ESDC & ESAC Science Data Centre \\
 ESO & European Southern Observatory \\
 esd & Extreme subdwarf \\ 
 EHT & Event Horizon Telescope \\
 EXPRES & EXtreme PREcision Spectrometer \\
 FUV & Far UltraViolet \\
 G, $G$ & Gravitational constant, Giclas, \textit{Gaia} magnitude \\
 Ga & Gigayear \\
 GALEX & GALaxy Evolution Explorer \\
 GAT & Gatewood \\
 GD & Giclas, Dwarf \\
 GJ & Gliese + Jahreiss \\
 H-R & Hertzsprung-Russell [diagram] \\
 HARPS & High Accuracy Radial velocity Planet Searcher \\
 HAT & Hungarian Automated Telescope network \\
 HCI & High-Contrast Imaging \\ 
 HD & Henry Draper \\
 HE & Hamburg-ESO [survey] \\
 HF &  Higher-order multiple systems Fraction \\
 HIRES & High Resolution Echelle Spectrometer \\
 HMC+NUTS & Hamiltonian Monte Carlo No U-Turn Sampler \\
 HRD & Hertzsprung-Russell Diagram \\
 HSOY & Hot Stuff for One Year \\

\end{tabular}
\end{table}
 
 \newpage

 \begin{table}[H]
 \caption*{}
 \label{tab:abbr3}
 \begin{tabular}{@{\hspace{10mm}}l@{\hspace{20mm}}l} 

 HST & \textit{Hubble} Space Telescope \\
 HWO & Habitable Worlds Observatory \\
 IAU & International Astronomical Union \\ 
 IDS & Index catalogue of visual Double Stars \\ 
 INTA & Instituto Nacional de Técnica Aerospacial \\
 IPAC & Infrared Processing and Analysis Center\\ 
 IRAF & Image Reduction and Analysis Facility \\ 
 IRAS & InfraRed Source \\ 
 IS & Infrared Spectroscopy \\
 IVOA & International Virtual Observatory Alliance \\
 J-PAS & Javalambre-Physics of the Accelerated Universe Astrophysical Survey \\
 J-PLUS & Javalambre Photometric Local Universe Survey \\
 JWST & \textit{James Webb} Space Telescope \\ 
 KMT & Korea Microlensing Telescope [network] \\
 K2 & Kepler Extended Mission \\
 L & Luyten \\ 
 L$_\odot$ & Solar Luminosity \\
 LAS & Large Area Survey \\ 
 LAWD & Luyten, Atlas of White Dwarfs \\ 
 LDS & Luyten Double Star \\ 
 LEHPM & Liverpool-Edinburgh High Proper Motion survey \\ 
 LHS & Luyten Half Second \\ 
 LI & Lucky Imaging \\
 LIFE & Large Interferometer For Exoplanets \\ 
 LOO-CV & Leave-One-Out Cross-Validation Criterion \\
 LP & Luyten, Palomar observatory \\  
 LSPM & Lépine and Shara Northern Stars Proper Motion \\ 
 LSR & Lépine, Shara, Rich \\
 LSST & Large Synoptic Survey Telescope \\
 $M$ & Mass \\
 M$_\oplus$ & Earth mass \\
 M$_{\text{Jup}}$ & Jupiter mass \\
 M$_\odot$ & Solar mass \\
 $M_\star$ & Star mass \\
 Ma & Megayear \\
 MAROON-X & M dwarf Advanced Radial velocity Observer Of Neighbouring \\
  & eXoplanets  \\
 mas & milliarcseconds \\
 MATLAB & MATrix LABoratory \\ 
 MCC & McCormick observatory \\
 MCMC & Markov Chain Monte Carlo \\
 MF & Multiplicity Frequency \\ 
 $\mu$FUN & Microlensing Follow-Up Network \\ 
 MIR & Mid InfraRed \\
 MOA & Microlensing Observations in Astrophysics \\ 
 NASA & National Aeronautics and Space Administration \\ 
 NICMOS & Near Infrared Camera and Multi-Object Spectrometer \\
 NIR & Near InfraRed \\
 NOT & Nordic Optical Telescope \\  

\end{tabular}
\end{table}

 \newpage
 
 \begin{table}[H]
 \caption*{}
 \label{tab:abbr4}
 \begin{tabular}{@{\hspace{10mm}}l@{\hspace{20mm}}l} 

 NUV & Near-UltraViolet \\
 OAJ & Observatorio Astronómico de Javalambre \\ 
 OGLE & Optical Gravitational Lensing Experiment \\ 
 ORB6 & Sixth Catalog of Orbits of Visual Binary Stars \\ 
 $P$ & Period \\ 
 $PA$ & Proper motion Angle, Position Angle \\  
 PARSEC & PAdova and TRieste Stellar Evolution Code \\ 
 pc & Parsec \\  
 PHL & Palomar observatory, Haro Luyten \\ 
 PLANET & Probing Lensing Anomalies NETwork \\ 
 PLATO & PLAnetary Transits and Oscillations of stars \\ 
 PM & Proper Motion \\ 
 PMD & Proper Motion Diagram \\ 
 PMO & Planetary Mass Object \\
 PS78 &  Pesch + Sanduleak \\
 R$_\text{Jup}$ & Jupiter Radius \\ 
 RAVE & Radial Velocity Experiment \\
 $RP$ & Red Photometer \\
 \texttt{RUWE} & Re-normalised Unit Weight Error \\
 RVS & Radial Velocity Spectrometer \\
 RX & ROSAT sat. (X-ray source) \\ 
 S & Sanduleak \\
 $s$ & separation \\
 SB1 & Single-lined Binary System \\ 
 SB2 & Double-lined Binary System \\
 SB3 & Triple-line Spectroscopic Triple System \\
 SCR & SuperCOSMOS-RECONS \\ 
 sd & subdwarf \\ 
 SDS & Southern Double Stars \\ 
 SDSS & Sloan Digital Sky Survey \\ 
 SED & Spectral Energy Distribution \\ 
 SI & Speckle Imaging \\
 Simbad & Set of Identifications, Measurements and Bibliography for \\
        & Astronomical Data \\  
 SIMP & Sondage Infrarouge de Mouvement Propre \\
 SIPS & Southern IR Proper motion Survey \\ 
 SKG & Stellar Kinematic Group \\  
 SO & Space Observatory \\
 SOAR & SOuthern Astrophysical Research \\
 SPHERE & Spectro-Polarimetric High-contrast Exoplanet REsearch instrument \\ 
 SPIRou & SPectropolarimètre InfraROUge \\ 
 SpT & Spectral Type \\  
 SIA & Simple Image Access \\
 SLAP & Simple Line Access Protocol \\
 SSA & Single Spectrum Access \\
 SSSPM & SuperCOSMOS Sky Survey Proper Motion \\
 ST3 & Triple-line Spectroscopic Triple System \\
 STIL & Starlink Tables Infrastructure Library \\  
 StKM & Stephenson, K and M stars \\  
\end{tabular}
\end{table}

\newpage
 
\begin{table}[H]
\caption*{}
\label{tab:abbr5}
\begin{tabular}{@{\hspace{10mm}}l@{\hspace{20mm}}l}
 SVO & Spanish Virtual Observatory \\
 TAP & Table Access Protocol \\
 \texttt{Topcat} & Tool for OPerations on Catalogues And Tables \\ 
 TESS & Transiting Exoplanet Survey Satellite \\ 
 TIC & TESS Input Catalogue \\
 TOI & TESS Object of Interest \\ 
 TT & Terrestrial Time \\
 TVLM & Tinney, Very Low Mass \\
 TWA & TW Hyades Association \\
 TYC & Tycho mission \\
 UCACx & USNO CCD Astrograph Catalog (UCAC1, UCAC2, UCAC3, UCAC4 \& UCAC5) \\ 
 UCD & Ultra-Cool Dwarf, Unified Content Descriptor \\ 
 UCM & Universidad Complutense de Madrid \\
 UGPS & UKIDSS Galactic Plane Survey \\ 
 UKIDSS & UKIRT Infrared Deep Sky Survey \\ 
 UKIRT & United Kingdom InfraRed Telescope \\ 
 ULAS & UKIDSS Large Area Survey \\ 
 UPM & U.S. Naval Observatory Proper Motion \\ 
 usd & Ultra subdwarf \\ 
 USNO & United States Naval Observatory \\
 UVES & Ultraviolet and Visual Echelle Spectrograph \\ 
 vB & van Biesbroeck \\
 VHS & VISTA Hemisphere Survey \\  
 VISTA & Visible and Infrared Survey Telescope for Astronomy \\
 VLA & Very Large Array \\
 VLBA & Very Long Baseline Array \\
 VLTI & Very Large Telescope Interferometer \\
 VO & Virtual Observatory \\
 VOQL & VO Query Language \\
 VOSA & Virtual Observatory SED Analyser \\ 
 VPH & Volume Phase Holographic \\ 
 WAIC & Watanabe-Akaike Information Criterion \\
 WASP & Wide Angle Search for Planets \\ 
 WD & White Dwarf \\ 
 WDS & Washington Double Star \\ 
 WDSS &  Washington Double Star Supplemental catalogue \\
 WFIRST & Wide Field Infrared Survey Telescope \\ 
 WHT & William Herschel Telescope \\ 
 WISE & Wide-field Infrared Survey Explorer \\ 
 WISEA & AllWISE Source Catalogue \\
 WISENF & NEOWISE Post-Cryo Release Single-exposure Source Working Database \\
 WISEP & WISE Preliminary Release Source Catalogue \\
 WISEU & unWISE Catalogue \\
 WT & Wroblewski + Torres \\ 
 XML &  eXtensible Markup Language \\
 $Z$ & Metallicity \\
 Z$_{\odot}$ & Solar metallicity \\ 
 ZTP & Zero Truncated Poisson [distribution] \\

\end{tabular}
\end{table}

%%%%% APENDICE B %%%%%

\begin{center}
\chapter[Appendix: Abbreviations for constellation names]{Abbreviations for constellation names}
\end{center}
\label{ch:Appendix_B}
\vspace{3cm}
\pagestyle{fancy}
\fancyhf{}
\lhead[\small{\textbf{\thepage}}]{\textbf{Appendix~B: Abbreviations for constellation names}}
\rhead[\small{\textbf{Appendix~B: Abbreviations for constellation names}}]{\small{\textbf{\thepage}}}
\renewcommand{\thetable}{B.\arabic{table}} % Cambia el formato de numeración al tipo B.x

 \begin{table}[H]
 \normalsize
 \caption*{}
 \label{tab:constellations1}
 \begin{tabular}{@{\hspace{35mm}}ll@{\hspace{20mm}}ll}

\textit{And} & Andromeda &\textit{Car} & Carina \\
\textit{Ant} & Antlia &\textit{Cas} & Cassiopeia \\
\textit{Aps} & Apus &\textit{Cen} & Centaurus \\
\textit{Aqr} & Aquarius &\textit{Cep} & Cepheus \\
\textit{Aql} & Aquila &\textit{Cet} & Cetus \\
\textit{Ara} & Ara &\textit{Cha} & Chamaeleon \\
\textit{Ari} & Aries &\textit{Cir} & Circinus \\
\textit{Aur} & Auriga &\textit{Col} & Columba \\
\textit{Boo} & Bo\"otes &\textit{Com} & Coma Berenices \\
\textit{Cae} & Caelum &\textit{CrA} & Corona Australis \\
\textit{Cam} & Camelopardalis &\textit{CrB} & Corona Borealis \\
\textit{Cnc} & Cancer &\textit{Crv} & Corvus \\
\textit{CVn} & Canes Venatici &\textit{Crt} & Crater \\
\textit{CMa} & Canis Major &\textit{Cru} & Crux \\
\textit{CMi} & Canis Minor & \textit{Cyg} & Cygnus \\
\textit{Cap} & Capricornus & \textit{Del} & Delphinus \\

 \end{tabular}
\end{table}

 \begin{table}[H]
 \caption*{}
 \label{tab:constellations2}
 \begin{tabular}{@{\hspace{35mm}}ll@{\hspace{20mm}}ll}

 \textit{Dor} & Dorado & \textit{Pav} & Pavo	 \\
 \textit{Dra} & Draco & \textit{Peg} & Pegasus	 \\
 \textit{Eql} & Equuleus & \textit{Per} & Perseus	 \\
 \textit{Eri} & Eridanus & \textit{Phe} & Phoenix	 \\
 \textit{For} & Fornax & \textit{Pic} & Pictor	 \\
 \textit{Gem} & Gemini & \textit{Psc} & Pisces	 \\
 \textit{Gru} & Grus & \textit{PsA} & Piscis Austrinus	 \\
 \textit{Her} & Hercules & \textit{Pup} & Puppis	 \\
 \textit{Hor} & Horologium & \textit{Pyx} & Pyxis	 \\
 \textit{Hya} & Hydra & \textit{Ret} & Reticulum	 \\
 \textit{Hyi} & Hydrus & \textit{Sge} & Sagitta	 \\
 \textit{Ind} & Indus & \textit{Sgr} & Sagittarius	 \\
 \textit{Lac} & Lacerta & \textit{Sco} & Scorpius	 \\
 \textit{Leo} & Leo & \textit{Scl} & Sculptor	 \\
 \textit{LMi} & Leo Minor & \textit{Sct} & Scutum	 \\
 \textit{Lep} & Lepus & \textit{Ser} & Serpens	 \\
 \textit{Lib} & Libra & \textit{Sex} & Sextans	 \\
 \textit{Lup} & Lupus & \textit{Tau} & Taurus	 \\
 \textit{Lyn} & Lynx & \textit{Tel} & Telescopium	 \\
 \textit{Lyr} & Lyra & \textit{Tri} & Triangulum	 \\
 \textit{Men} & Mensa & \textit{TrA} & Triangulum Australe	 \\
 \textit{Mic} & Microscopium & \textit{Tuc} & Tucana	 \\
 \textit{Mon} & Monoceros & \textit{UMa} & Ursa Major	 \\
 \textit{Mus} & Musca & \textit{UMi} & Ursa Minor	 \\
 \textit{Nor} & Norma & \textit{Vel} & Vela	 \\
 \textit{Oct} & Octans & \textit{Vir} & Virgo	 \\
 \textit{Oph} & Ophiuchus & \textit{Vol} & Volans	 \\
 \textit{Ori} & Orion & \textit{Vul} & Vulpecula	 \\

 \end{tabular}
\end{table}

\normalsize

%%%%% APENDICE C %%%%%

\begin{center}
\chapter[Appendix: Properties and plots of all subdwarfs and companions]{Properties and plots of\\all subdwarfs and companions}
\end{center}
\label{ch:Appendix_C}
\vspace{3cm}
\pagestyle{fancy}
\fancyhf{}
\lhead[\small{\textbf{\thepage}}]{\textbf{Appendix~C: Properties and plots of all subdwarfs and companions}}
\rhead[\small{\textbf{Appendix~C: Properties and plots of all subdwarfs and companions}}]{\small{\textbf{\thepage}}}
\renewcommand{\thetable}{C.\arabic{table}}   % Cambia el formato de numeración al tipo D.x

\footnotesize

\begin{longtable}{clcc@{\hspace{2mm}}c@{\hspace{2mm}}cc}
\caption[Coordinates, \textit{Gaia} DR2 identifiers, spectral types and references of the subdwarfs.]{Coordinates, \textit{Gaia} DR2 identifiers, spectral types and references of the subdwarfs (available at the CDS via \url{http://cdsarc.u-strasbg.fr/viz-bin/cat/J/A+A/650/A190}).}\\
\noalign{\smallskip}
\noalign{\hrule height 1pt}
\noalign{\smallskip}
Id & Name & $\alpha$ (J2000)$^{\text{(a)}}$ & $\delta$ (J2000)$^{\text{(a)}}$ & \textit{Gaia} DR2 & SpT & Refs.$^{\text{(b)}}$  \\
 &  & (hh:mm:ss.s) & (dd:mm:ss.s) &  &  &  \\
\noalign{\smallskip}
\hline
\noalign{\smallskip}
\endfirsthead
\caption[]{Coordinates, \textit{Gaia} DR2 identifiers, spectral types and references of the subdwarfs \textit{(continued)}.}\\
\noalign{\hrule height 1pt}
\noalign{\smallskip}
\label{tab:coordinates_subdwarfs}
Id & Name & $\alpha$ (J2000)$^{\text{(a)}}$ & $\delta$ (J2000)$^{\text{(a)}}$ & \textit{Gaia} DR2 & SpT & Refs.$^{\text{(b)}}$  \\
 &  & (hh:mm:ss.s) & (dd:mm:ss.s) &  &  &  \\
\noalign{\smallskip}
\hline
\noalign{\smallskip}
\endhead
\hline
\noalign{\smallskip}
\endfoot

1 & 2MASS J00054517+0723423 & 00:05:45.17 & +07:23:42.3 & 2745936355142216064 & sdM6.0 & 9\\
2 & LHS 1035 & 00:11:42.08 & --20:57:12.9 & 2364524086522978560 & sdM6.0 & 7\\
3 & LHS 1074 & 00:25:51.17 & --07:48:06.9 & 2430347896307698304 & sdM6.0 & 17, 17\\
4 & SDSS J00255258+0109249 & 00:25:52.58 & +01:09:24.9 & 2546981096190666752 & sdM6.0 & 23 \\
5 & SDSS J00290696+0716590 & 00:29:06.96 & +07:16:59.0 & 2748000890085870080 & sdM7.0 & 23 \\
6 & LHS 1135 & 00:43:31.34 & +20:54:31.6 & 2801587139749159168 & sdM6.5 & 7, 16, 17\\
7 & ULAS J00453997+1350327 & 00:45:39.97 & +13:50:32.7 & 2776383343942921856 & sdM7.5 & 13\\
8 & 2MASS J00524886+1712434 & 00:52:48.86 & +17:12:43.4 & 2781801290567594752 & esdM5.0 & 9\\
\end{longtable}

\vspace{-0.5cm}
\begin{justify}
\scriptsize{\textbf{\textit{Notes. }}}
\scriptsize{{$^{\text{(a)}}$ The values of the coordinates are obtained from the catalogue determined by the preffix of the names (SDSS, ULAS, 2MASS, LSR, LHS, SSSPM, and APMPM).
$^{\text{(b)}}$ References: 1. \citet{bowler10a}; 2. \citet{burgasser03b}; 3. \citet{burgasser04b}; 4. \citet{cushing09}; 5. \citet{gizis97}; 6. \citet{gizis97b}; 7. \citet{reid97c}; 8. \citet{gizis00c}; 9. \citet{kirkpatrick10}; 10. \citet{lepine03a}; 11. \citet{lepine08b}; 12. \citet{lodieu10a}; 13. \citet{lodieu12b}; 14. \citet{lodieu17a}; 15. \citet{reid04a}; 16. \citet{reid05b}; 17. \citet{riaz08a}; 18. \citet{schmidt10a}; 19. \citet{scholz04c}; 20. \citet{scholz04b}; 21. \citet{schweitzer99}; 22. \citet{sivarani09}; 23. \citet{zhang13}; 24. \citet{zhang17a}; 25. \citet{zhang18a}; 26. \citet{zhang18b}; 27. \citet{lodieu19c}; 28. \citet{zhang19g}.}}
\end{justify}

\newpage

\footnotesize
\begin{longtable}{@{\hspace{3mm}}c@{\hspace{7mm}}c@{\hspace{7mm}}c@{\hspace{7mm}}l@{\hspace{7mm}}c@{\hspace{7mm}}l@{\hspace{7mm}}c@{\hspace{2mm}}}
\label{tab:astrometric_subdwarfs}\\
\caption[Astrometric data of the subdwarfs.]{Astrometric data of the subdwarfs (available at the CDS via \url{http://cdsarc.u-strasbg.fr/viz-bin/cat/J/A+A/650/A190}).}\\
\noalign{\smallskip}
\noalign{\smallskip}
\noalign{\smallskip}
\noalign{\smallskip}
\noalign{\hrule height 1pt}
\noalign{\smallskip}
Id & $\mu_{\alpha}\cos\delta$ & $\mu_\delta$ & Ref. & $d$ & Ref. & Search radius \\
  & (mas\,a$^{-\text{1}}$) & (mas\,a$^{-\text{1}}$) &  & (pc) &  & (deg) \\
\noalign{\smallskip}
\hline
\noalign{\smallskip}
\endfirsthead
\caption[]{Astrometric data of the subdwarfs \textit{(continued)}.}\\
\noalign{\hrule height 1pt}
\noalign{\smallskip}
%\label{tab:astrometric_subdwarfs}
Id & $\mu_{\alpha}\cos\delta$ & $\mu_\delta$ & Ref. & $d$ & Ref. & Search radius \\
  & (mas\,a$^{-\text{1}}$) & (mas\,a$^{-\text{1}}$) &  & (pc) &  & (deg) \\

\noalign{\smallskip}
\hline
\noalign{\smallskip}
\endhead
\hline
\noalign{\smallskip}
\endfoot

1 & --15.154 $\pm$ 0.405 & --266.211 $\pm$ 0.223 & \textit{Gaia} DR2 & 168.0 $\pm$ 5.5 & \textit{Gaia} DR2 & 0.70 \\
2 & 613.567 $\pm$ 0.229 & --0.276 $\pm$ 0.154 & \textit{Gaia} DR2 & 47.0 $\pm$ 0.3 & \textit{Gaia} DR2 & 2.50 \\
3 & 396.789 $\pm$ 0.431 & --711.306 $\pm$ 0.225 & \textit{Gaia} DR2 & 84.4 $\pm$ 1.3 & \textit{Gaia} DR2 & 1.39 \\
4 & 158.169 $\pm$ 0.475 & --197.428 $\pm$ 0.386 & \textit{Gaia} DR2 & 140.6 $\pm$ 5.3 & \textit{Gaia} DR2 & 0.84 \\
5 & 185.494 $\pm$ 0.548 & --174.401 $\pm$ 0.337 & \textit{Gaia} DR2 & 103.7 $\pm$ 3.0 & \textit{Gaia} DR2 & 1.02 \\
6 & 628.340 $\pm$ 0.339 & --203.922 $\pm$ 0.191 & \textit{Gaia} DR2 & 55.6 $\pm$ 0.6 & \textit{Gaia} DR2 & 2.12 \\
7 & 120.750 $\pm$ 4.275 & --70.136 $\pm$ 3.519 & This work & 358.0 $\pm$ 29.6 & This work & 0.30 \\
8 & 369.858 $\pm$ 0.572 & --268.693 $\pm$ 0.395 & \textit{Gaia} DR2 & 210.9 $\pm$ 19.5 & \textit{Gaia} DR2 & 0.78 \\
\end{longtable}

\normalsize
\newpage

\footnotesize
\begin{longtable}{@{\hspace{2mm}}l@{\hspace{3mm}}c@{\hspace{3mm}}c@{\hspace{3mm}}c@{\hspace{4mm}}c@{\hspace{3mm}}c@{\hspace{4mm}}c@{\hspace{3mm}}c@{\hspace{2mm}}}
\label{tab:astrometric_subdwarfs_companions}\\
\caption[Astrometric data of the companion candidates.]{Astrometric data of the companion candidates (available at the CDS via \url{http://cdsarc.u-strasbg.fr/viz-bin/cat/J/A+A/650/A190}).}\\
\noalign{\smallskip}
\noalign{\smallskip}
\noalign{\smallskip}
\noalign{\smallskip}
\noalign{\hrule height 1pt}
\noalign{\smallskip}
Candidate & \textit{Gaia} DR2 & $\alpha$ (J2015.5) & $\delta$ (J2015.5) & $\mu_{\alpha}\cos\delta$ & $\mu_\delta$ & \textit{d} & \textit{s} \\
Id &  &  (hh:mm:ss.s) & (dd:mm:ss.s) & (mas a$^{-\text{1}}$) & (mas a$^{-\text{1}}$) & (pc) & (arcmin) \\
\noalign{\smallskip}
\hline
\noalign{\smallskip}
\endfirsthead
\caption[]{Astrometric data of the companion candidates \textit{(continued)}.}\\
\noalign{\hrule height 1pt}
\noalign{\smallskip}
%\label{tab:astrometric_subdwarfs_companions}
Candidate & \textit{Gaia} DR2 & $\alpha$ (J2015.5) & $\delta$ (J2015.5) & $\mu_{\alpha}\cos\delta$ & $\mu_\delta$ & \textit{d} & \textit{s} \\
Id &  &  (hh:mm:ss.s) & (dd:mm:ss.s) & (mas a$^{-\text{1}}$) & (mas a$^{-\text{1}}$) & (pc) & (arcmin) \\
\noalign{\smallskip}
\hline
\noalign{\smallskip}
\endhead
\hline
\noalign{\smallskip}
\endfoot
11--1 & 2562778780473565312 & 01:19:33.51 & +03:33:15.8  & 18.0 $\pm$ 0.1 &  --72.3 $\pm$ 0.1 &  172 $\pm$ 2.1  & 19.0 \\
11--2 & 2562682087874908160 & 01:18:40.78 & +03:23:43.9  & 11.1 $\pm$ 0.3 &  --85.3 $\pm$ 0.2 &  172.9 $\pm$ 6.7  & 18.2 \\
11--3 & 2562791218698851072 & 01:19:39.61 & +03:42:46.4 &  --2.3 $\pm$ 0.2 &  --23.9 $\pm$ 0.1 &  206.8 $\pm$ 7.2  & 18.7 \\
11--4 & 2562781185655249536 & 01:19:49.92 & +03:38:04.0 &  67.6 $\pm$ 0.2 &  37.9 $\pm$ 0.1 &  164.3 $\pm$ 4.6  & 21.5 \\
11--5 & 2562787129889986816 & 01:19:52.70 & +03:39:19.2 &  --61.6 $\pm$ 0.6 &  --23.4 $\pm$ 0.4 &  184.3 $\pm$ 15.9  & 22.0 \\
11--6 & 2562673880192373632 & 01:18:15.79 & +03:15:00.9 & 40.6 $\pm$ 0.1 &  --3.6 $\pm$ 0.1 &  181.7 $\pm$ 2  & 26.6 \\
11--7 & 2562790664647542144 & 01:20:01.29 & +03:42:43.9 & 68.7 $\pm$ 0.3 &  --2.8 $\pm$ 0.2 &  184.8 $\pm$ 9.1  & 24.1 \\
11--8 & 2562871070730884992 & 01:17:06.63 & +03:28:09.9 & --20.0 $\pm$ 0.1 &  --42.0 $\pm$ 0.0 &  197.3 $\pm$ 2  & 23.7 \\
\end{longtable}
\vspace{-0.7cm}
\begin{justify}
  \scriptsize{\textbf{\textit{Notes. }} $^{\text{(a)}}$ This value in arcsec.}
\end{justify}
\newpage

\normalsize
\renewcommand{\thefigure}{C.\arabic{figure}}   % Cambia el formato de numeración al tipo C.x

\begin{figure}[H]
  \caption*{\textbf{Id 11}}
    \begin{subfigure}{.5\textwidth}
    \includegraphics[width=0.91\linewidth]{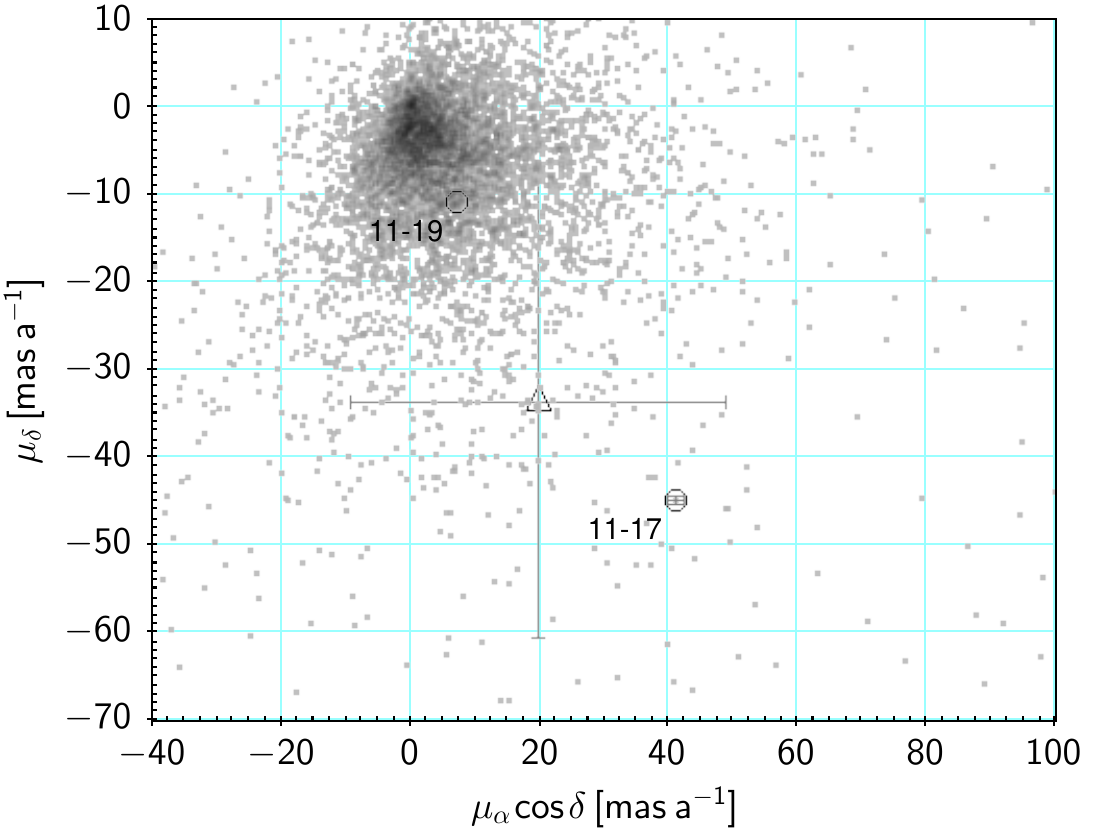}
    \end{subfigure}
    \begin{subfigure}{.5\textwidth}
      \includegraphics[width=0.91\linewidth]{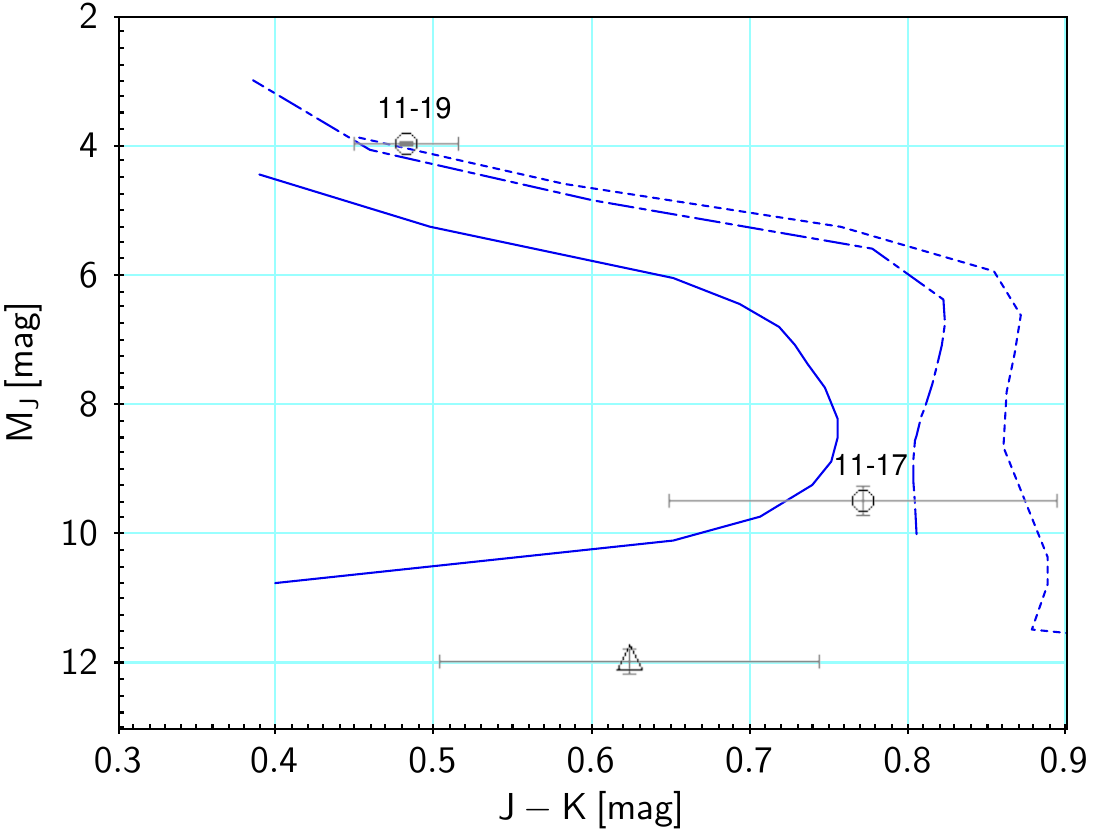}
    \end{subfigure} 
  \vspace{2mm}
  
     \begin{subfigure}{.5\textwidth}
    \includegraphics[width=0.91\linewidth]{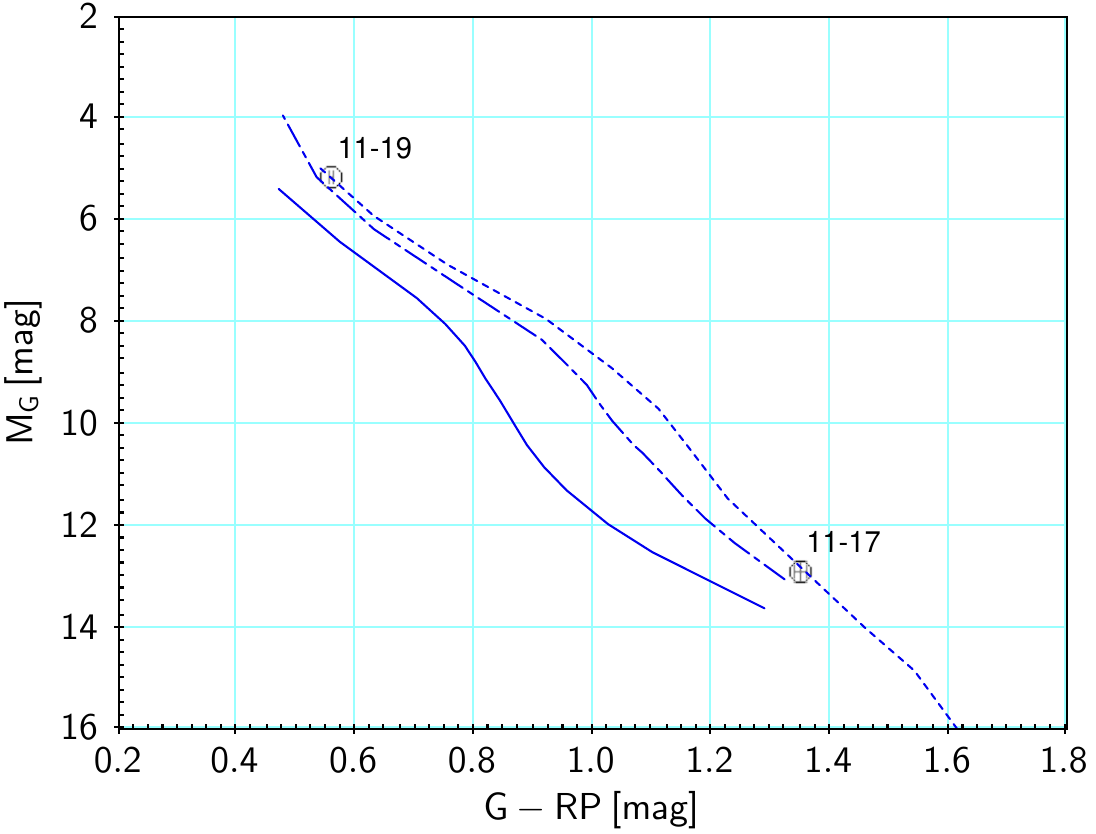}
    \end{subfigure}
    \begin{subfigure}{.5\textwidth}
      \includegraphics[width=0.91\linewidth]{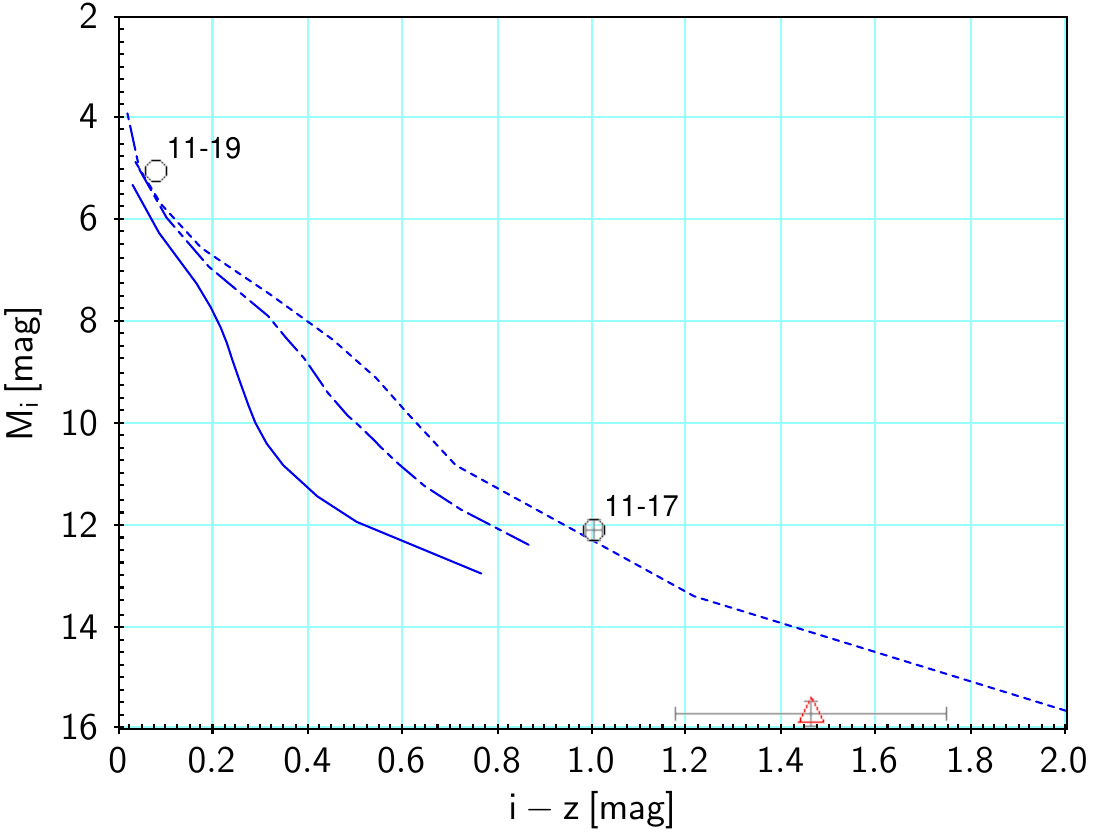}
    \end{subfigure}
    
  \vspace{2mm}  
     \begin{subfigure}{.5\textwidth}
    \includegraphics[width=0.91\linewidth]{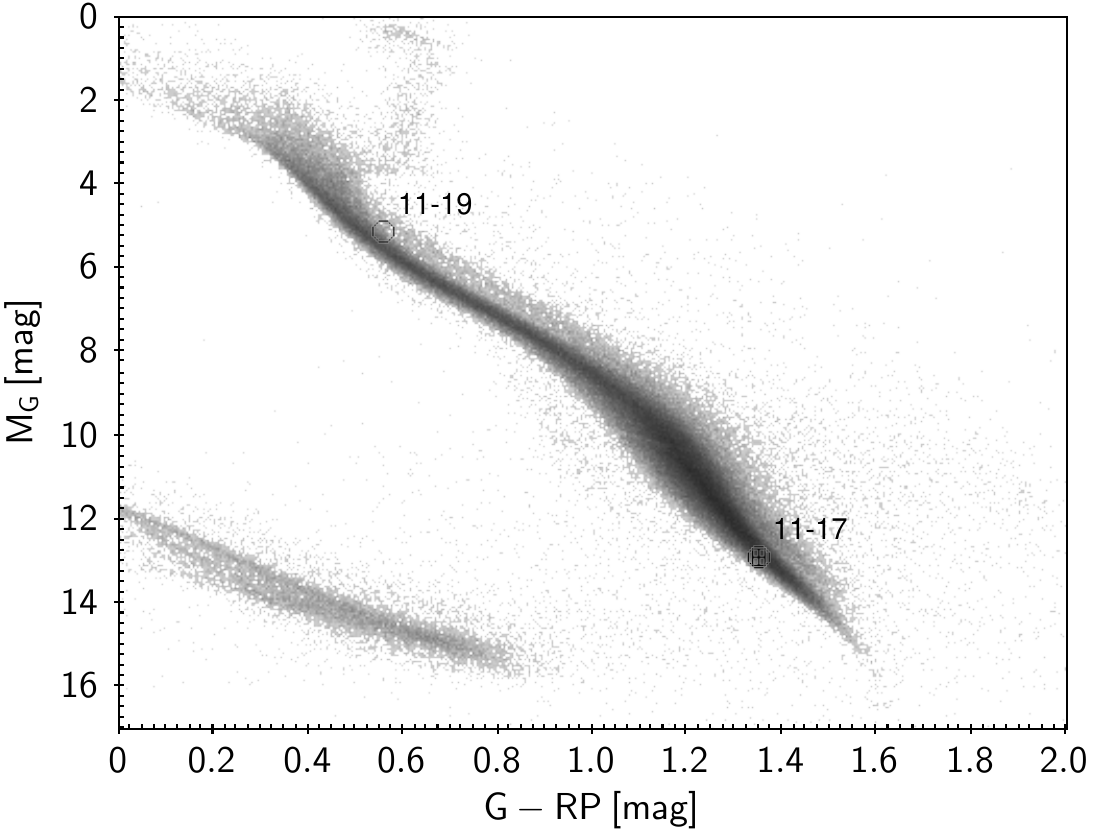}
    \end{subfigure}
    \begin{subfigure}{.5\textwidth}
      \includegraphics[width=0.91\linewidth]{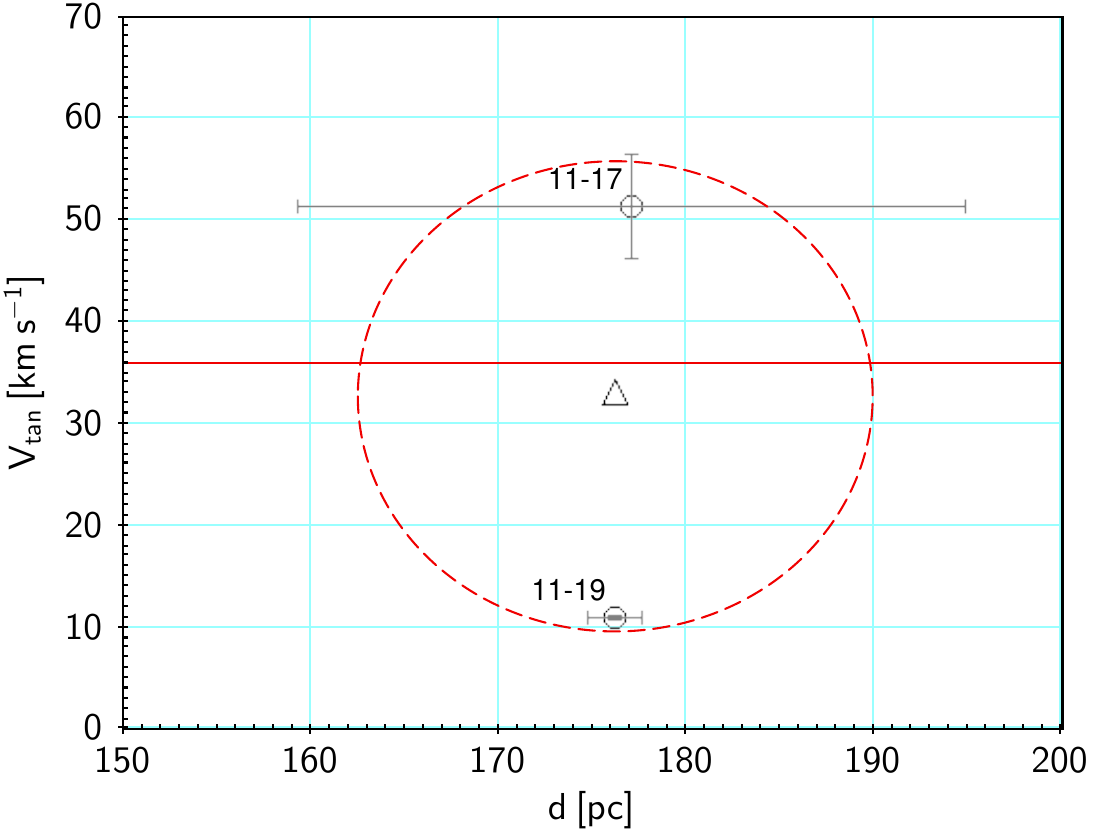}
    \end{subfigure}
  \vspace{2mm}
  \caption[PMD, CMDs, H-R diagram and tangential velocity vs distance diagram for the target Id 11 and its candidate companions.]{PMD (top left panel), CMDs (top right and mid panels), H-R diagram (bottom left panel), and tangential velocity--distance diagram (bottom right) for the target Id 11 and its candidate companions. The black triangle represents the source under study, and the numbered black circles are the companion candidates. Grey dots in the PMD represent field stars, and in the H-R diagram they are \textit{Gaia} DR2 sources with parallaxes $>$\,10\,mas used as a reference. The blue solid, dashed, and dotted lines stand for [M/H]\,=\,$-$2.0, $-$0.5, and =\,0.0 \textit{BT-Settl} isochrones in the CMDs, respectively. The red solid line in the $d/V_{tan}$ plot marks the value $V_{tan}=$\,36\,km\,s$^{-\text{1}}$ which is the mean value for field stars \citep{zhang18a}, and the red dashed ellipse around Id 11 indicates its values of $V_{tan}\pm \sigma$ and $d \pm \sigma$.}
\label{fig:plot_CMD_PM_11}
\end{figure}

\begin{figure}[H]
  \caption*{\textbf{Id 25}}
    \begin{subfigure}{.5\textwidth}
    \includegraphics[width=0.91\linewidth]{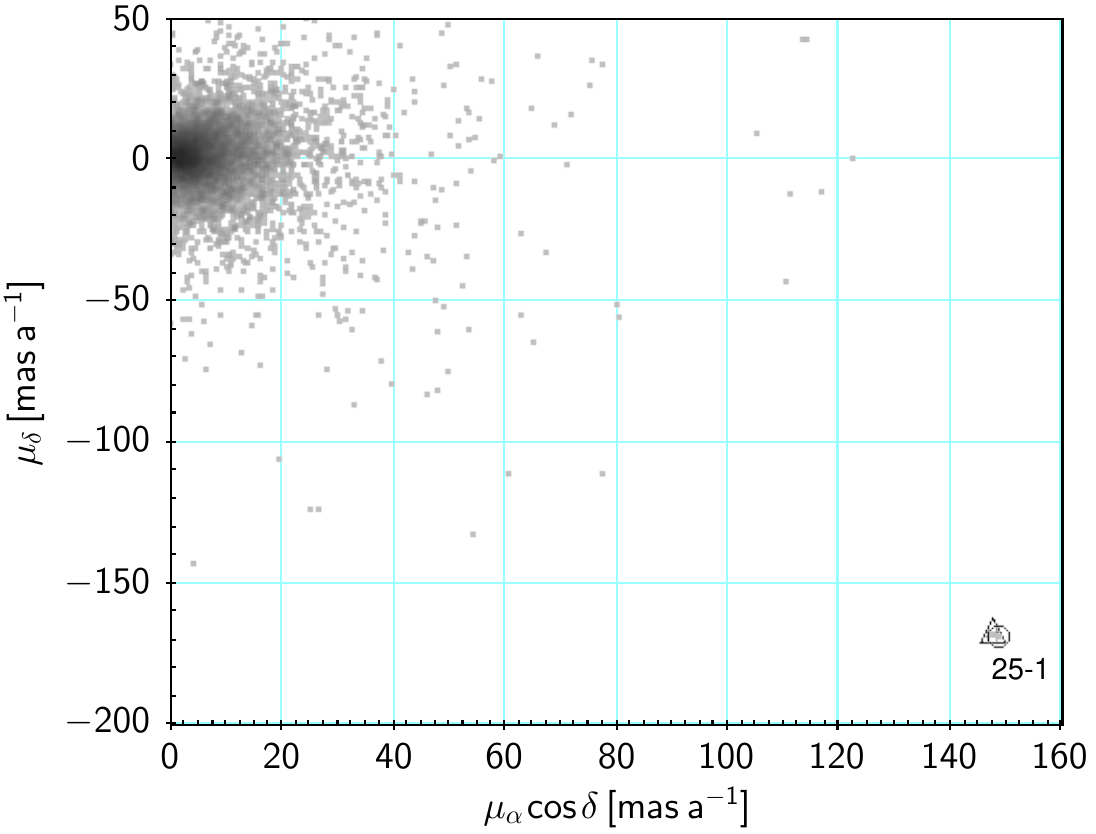}
    \end{subfigure}
    \begin{subfigure}{.5\textwidth}
      \includegraphics[width=0.91\linewidth]{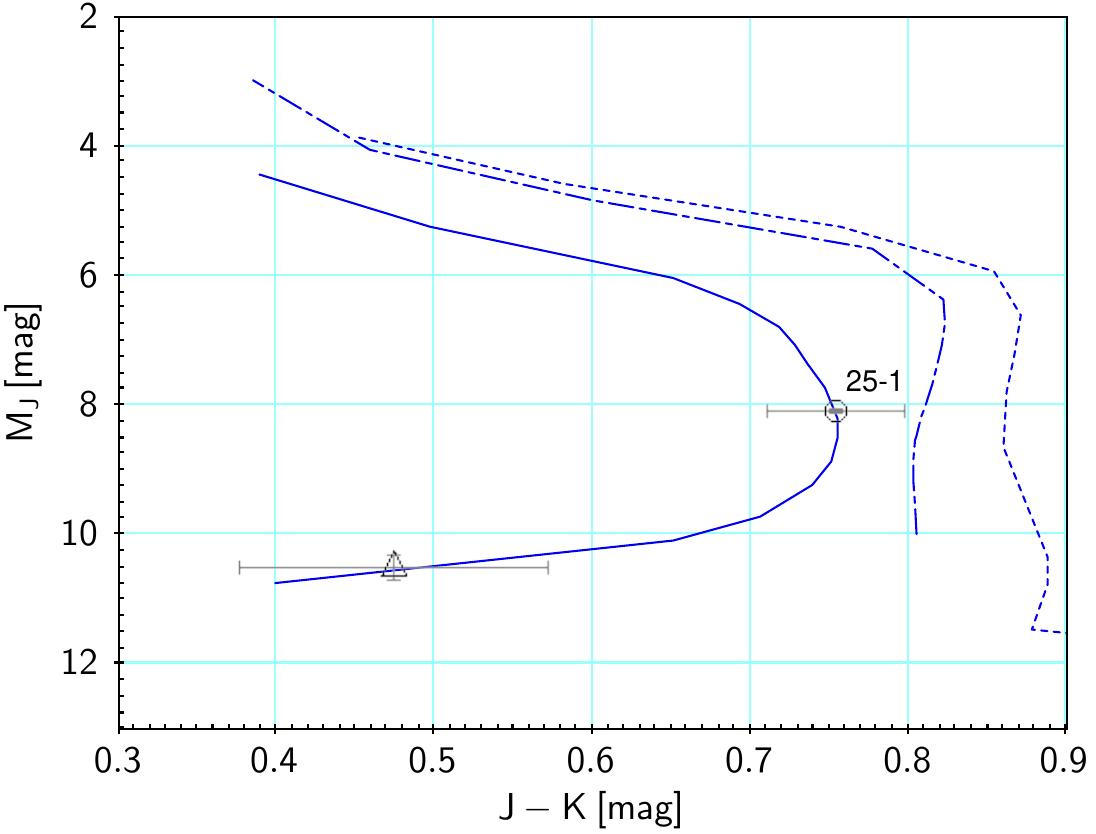}
    \end{subfigure} 
  \vspace{2mm}
  
     \begin{subfigure}{.5\textwidth}
    \includegraphics[width=0.91\linewidth]{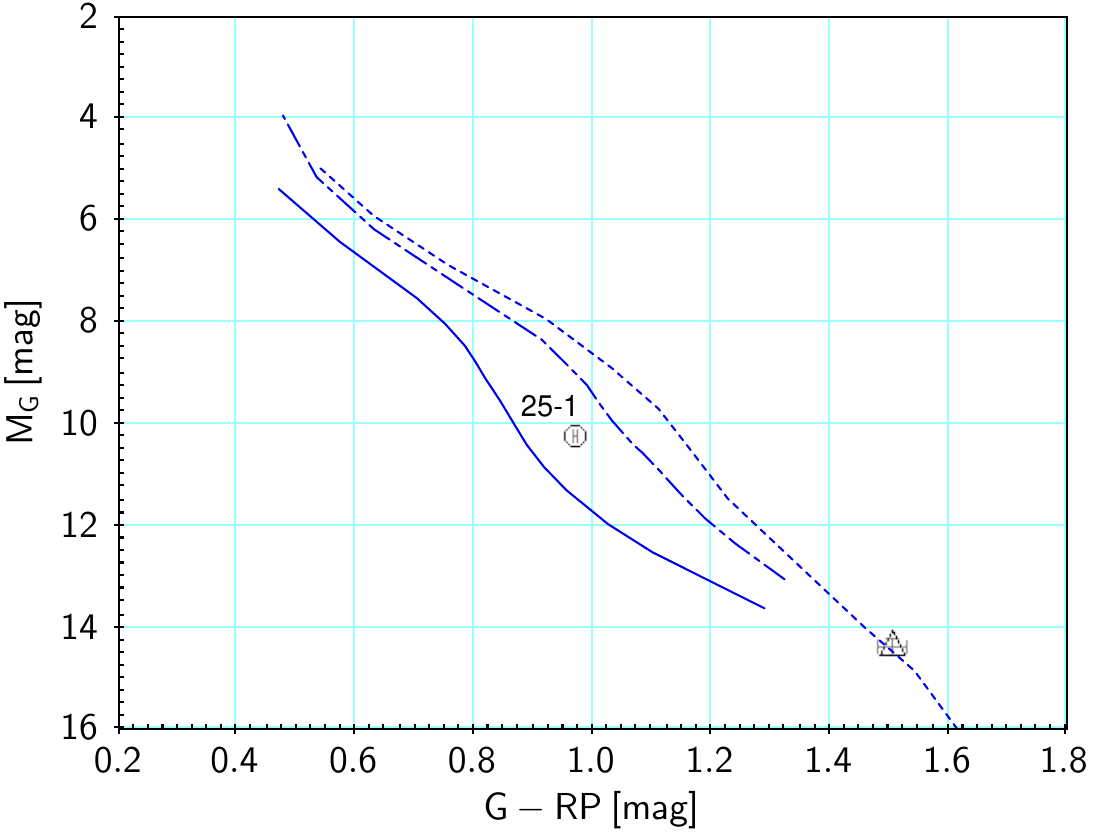}
    \end{subfigure}
    \begin{subfigure}{.5\textwidth}
      \includegraphics[width=0.91\linewidth]{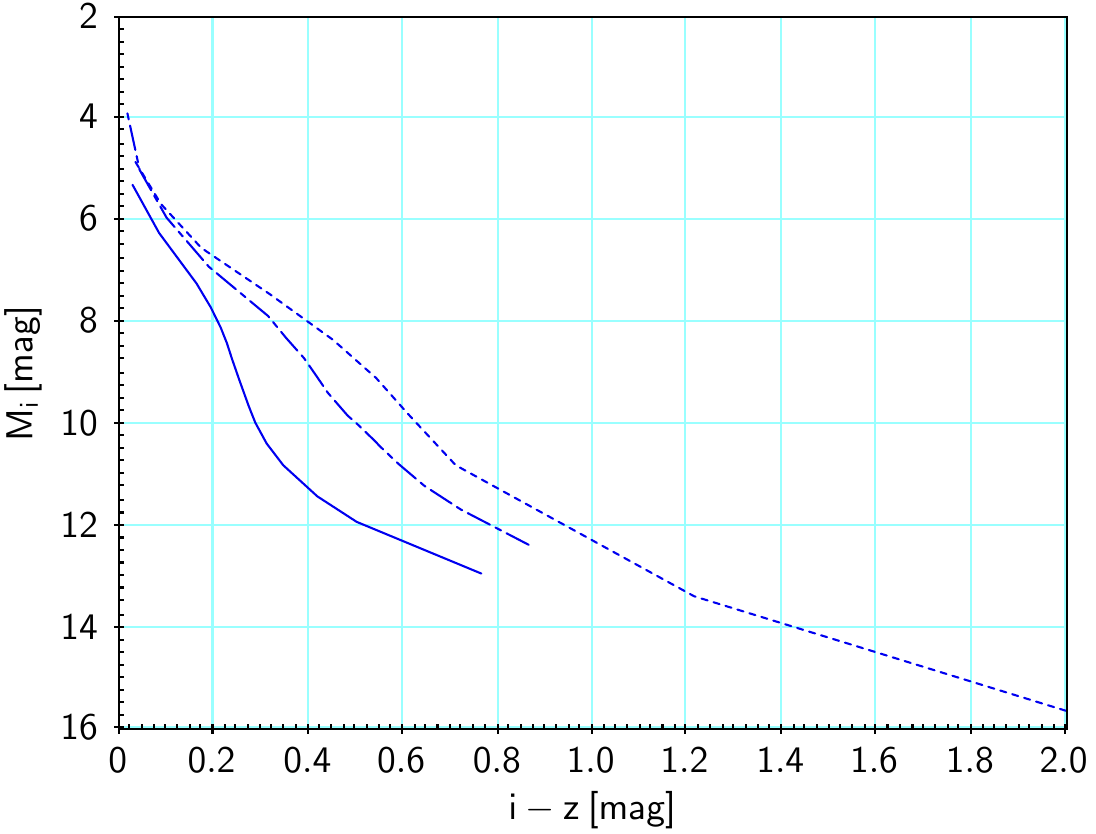}
    \end{subfigure}
    
  \vspace{2mm}  
     \begin{subfigure}{.5\textwidth}
    \includegraphics[width=0.91\linewidth]{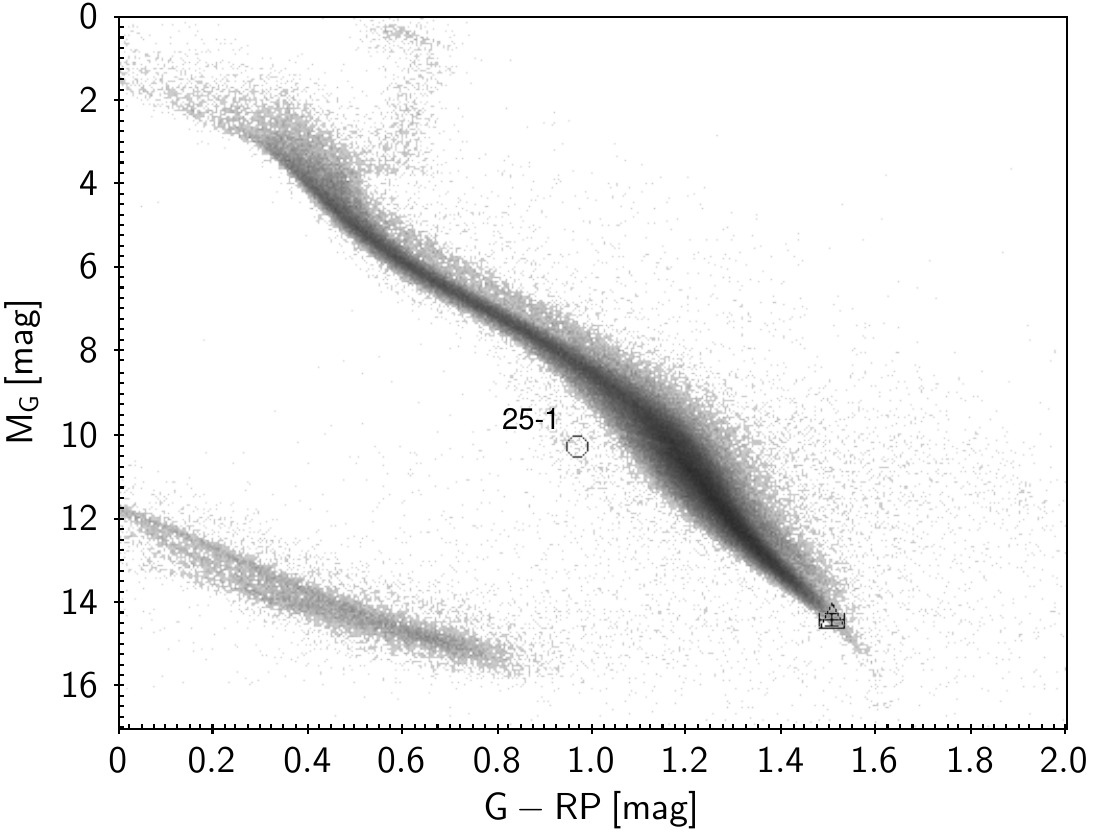}
    \end{subfigure}
    \begin{subfigure}{.5\textwidth}
      \includegraphics[width=0.91\linewidth]{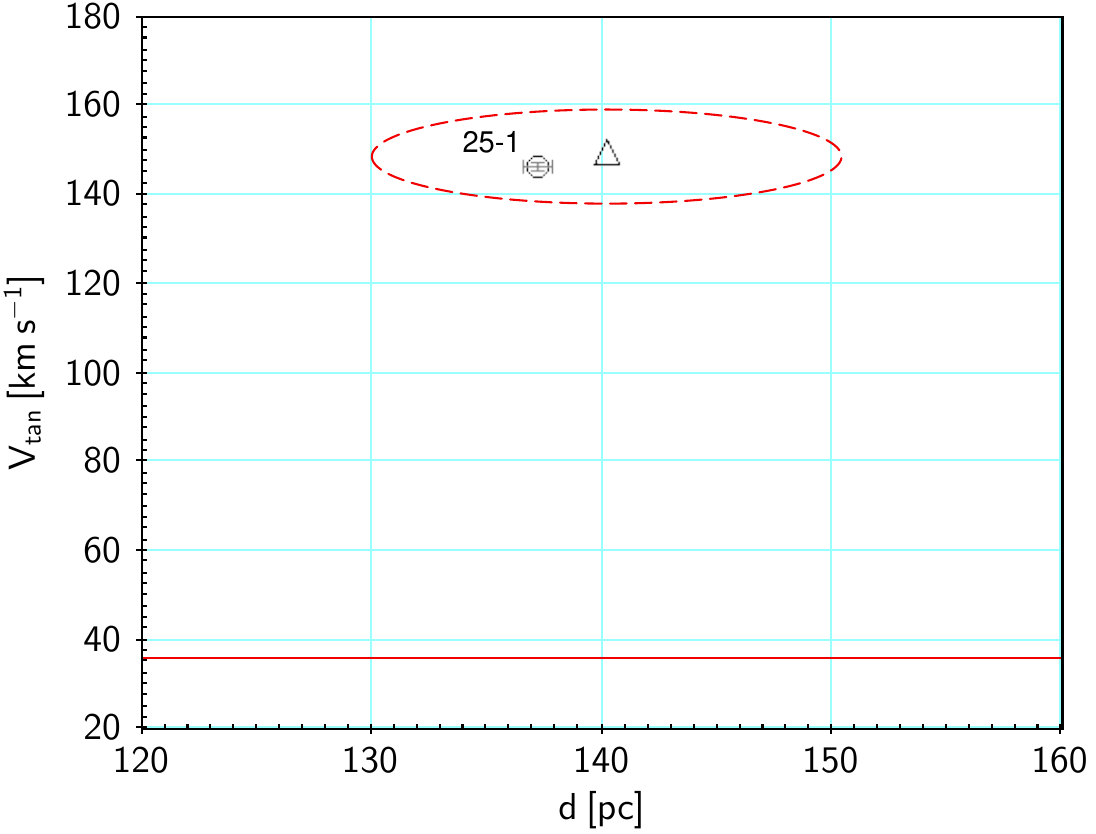}
    \end{subfigure}
  \vspace{5mm}
  \caption[PMD, CMDs, H-R diagram and tangential velocity vs distance diagram for the target Id 25 and its candidate companions.]{PMD (top left panel), CMDs (top right and mid panels), H-R diagram (bottom left panel), and tangential velocity--distance diagram (bottom right) for the target Id 25 and its candidate companion. The black triangle represents the source under study, and the numbered black circle is the companion candidate. Grey dots in the PMD represent field stars, and in the H-R diagram they are \textit{Gaia} DR2 sources with parallaxes $>$\,10\,mas used as a reference. The blue solid, dashed, and dotted lines stand for [M/H]\,=\,$-$2.0, $-$0.5, and =\,0.0 \textit{BT-Settl} isochrones in the CMDs, respectively. The red solid line in the $d/V_{tan}$ plot marks the value $V_{tan}=$\,36\,km\,s$^{-\text{1}}$ which is the mean value for field stars \citep{zhang18a}, and the red dashed ellipse around Id 25 indicates its values of $V_{tan}\pm \sigma$ and $d \pm \sigma$.}
\label{fig:plot_CMD_PM_25}
\end{figure}

\begin{figure}[H]
  \caption*{\textbf{Id 73}}
    \begin{subfigure}{.5\textwidth}
    \includegraphics[width=0.91\linewidth]{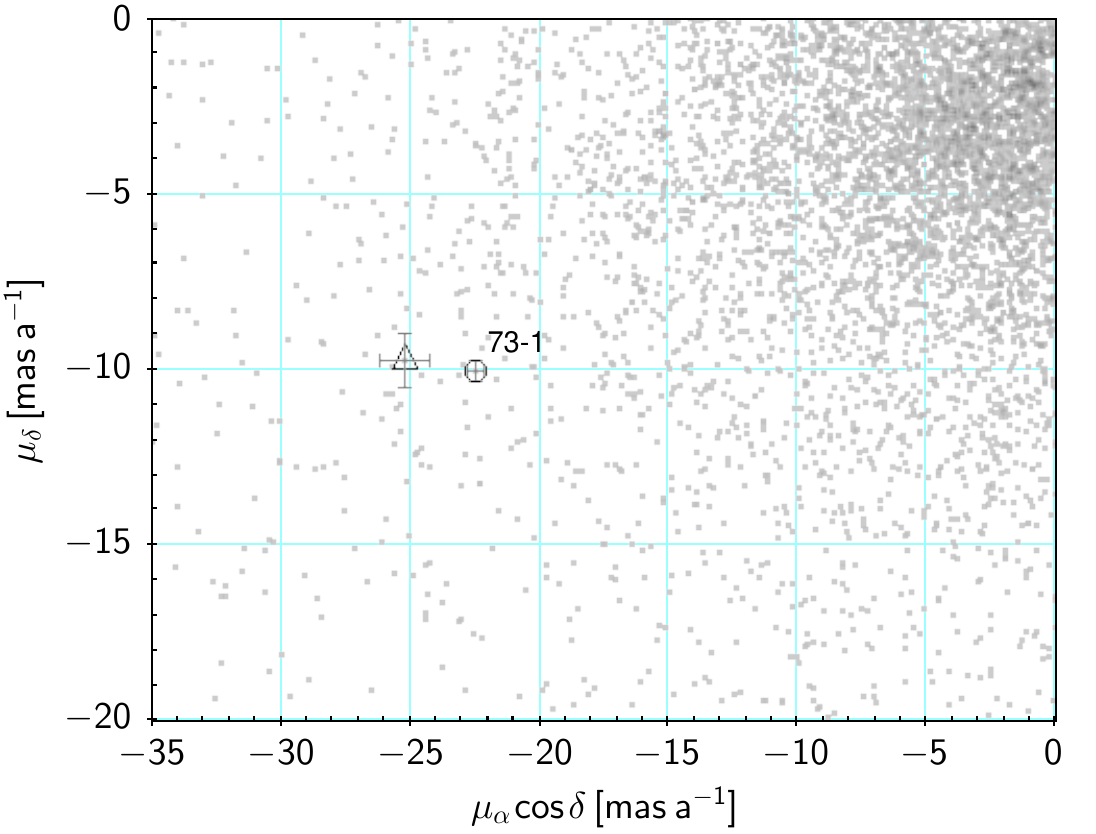}
    \end{subfigure}
    \begin{subfigure}{.5\textwidth}
      \includegraphics[width=0.91\linewidth]{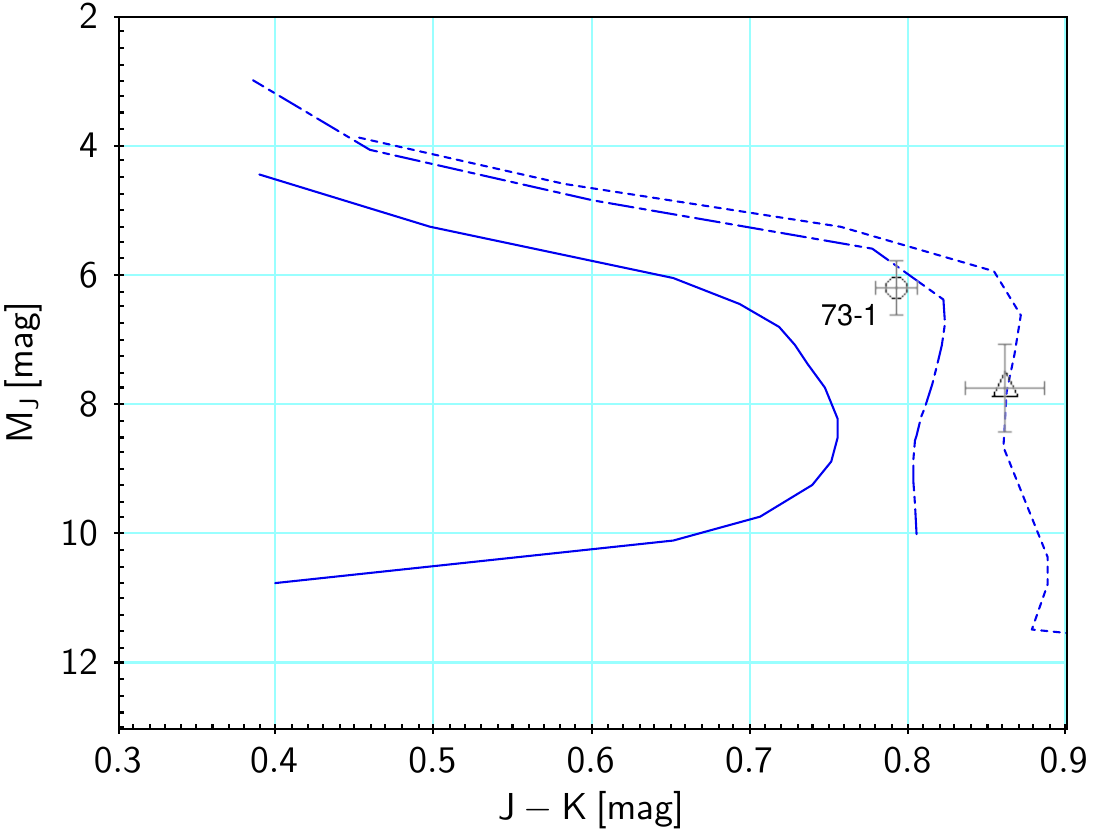}
    \end{subfigure} 
  \vspace{2mm}
  
     \begin{subfigure}{.5\textwidth}
    \includegraphics[width=0.91\linewidth]{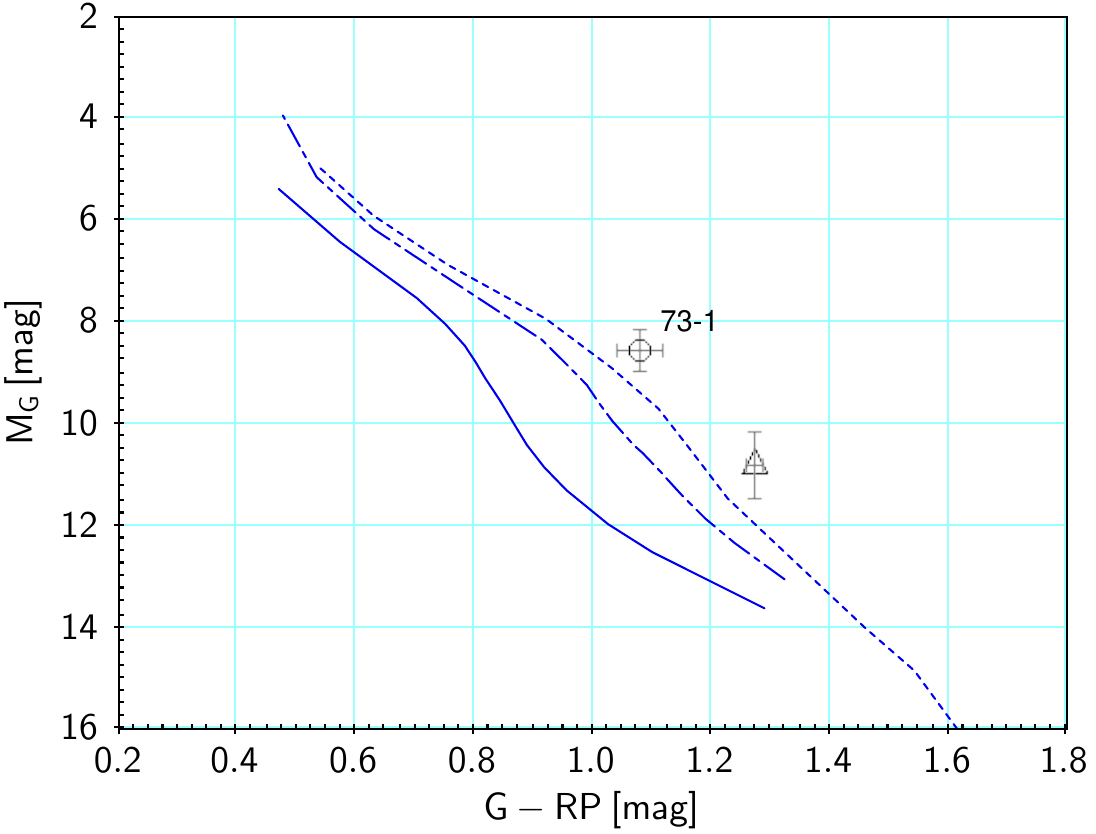}
    \end{subfigure}
    \begin{subfigure}{.5\textwidth}
      \includegraphics[width=0.91\linewidth]{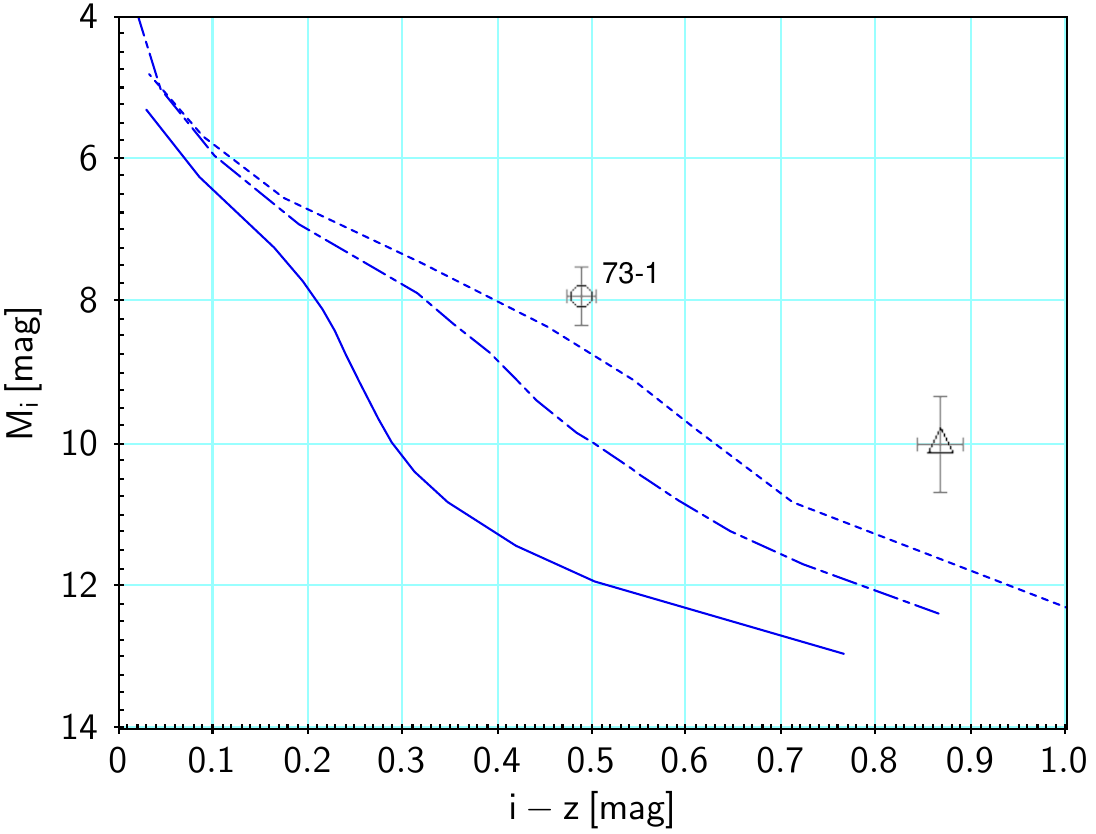}
    \end{subfigure}
    
  \vspace{2mm}  
     \begin{subfigure}{.5\textwidth}
    \includegraphics[width=0.91\linewidth]{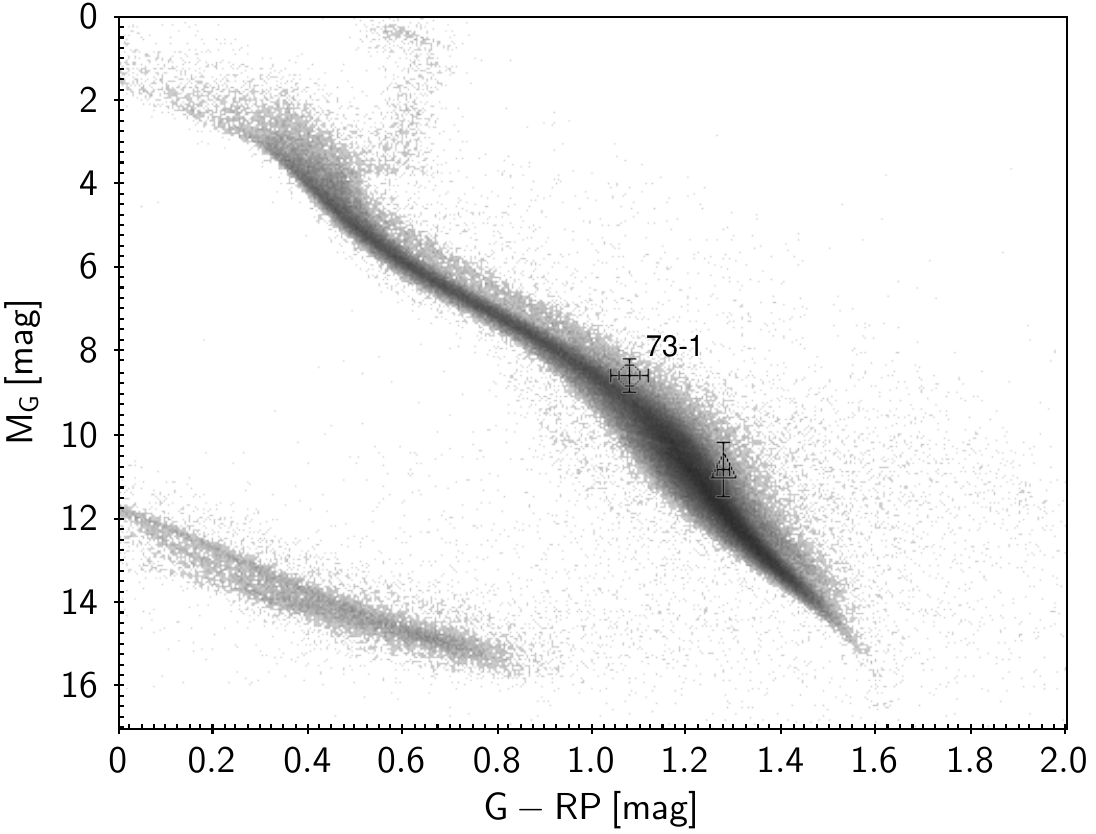}
    \end{subfigure}
    \begin{subfigure}{.5\textwidth}
      \includegraphics[width=0.91\linewidth]{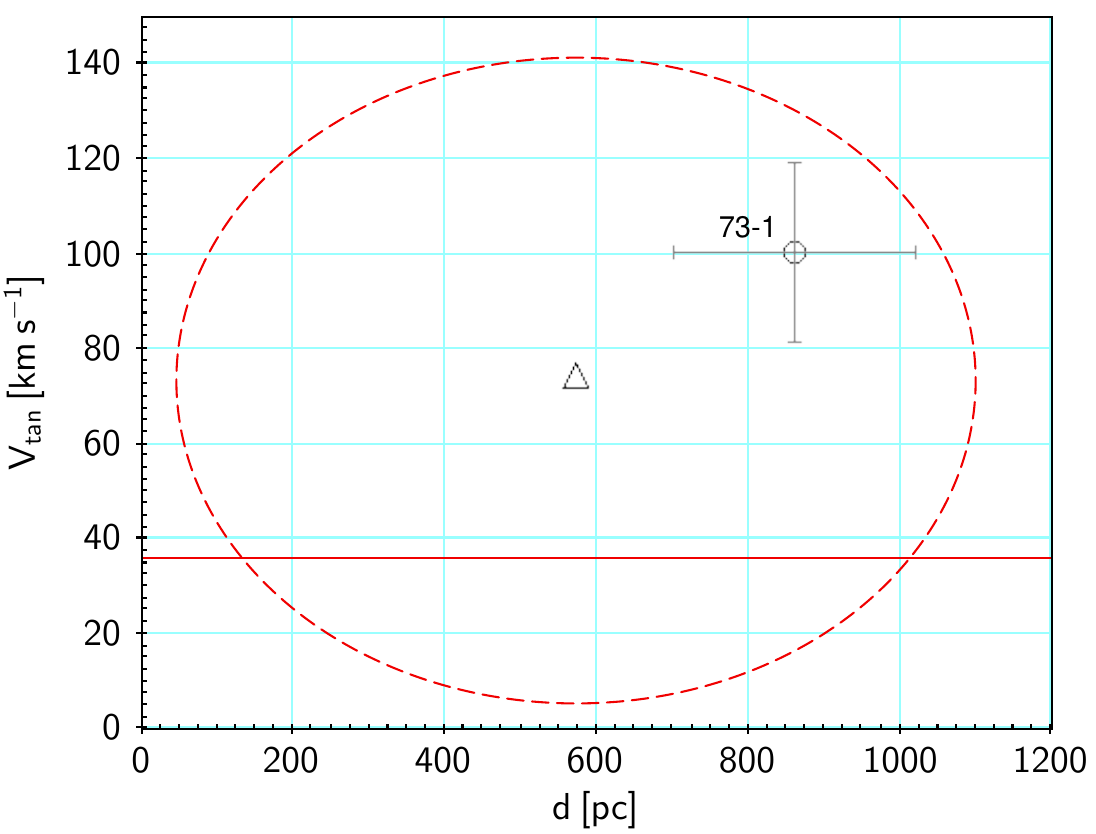}
    \end{subfigure}
  \vspace{5mm}
  \caption[PMD, CMDs, H-R diagram and tangential velocity vs distance diagram for the target Id 73 and its candidate companions.]{PMD (top left panel), CMDs (top right and mid panels), H-R diagram (bottom left panel), and tangential velocity--distance diagram (bottom right) for the target Id 73 and its candidate companion. The black triangle represents the source under study, and the numbered black circle is the companion candidate. Grey dots in the PMD represent field stars, and in the H-R diagram they are \textit{Gaia} DR2 sources with parallaxes $>$\,10\,mas used as a reference. The blue solid, dashed, and dotted lines stand for [M/H]\,=\,$-$2.0, $-$0.5, and =\,0.0 \textit{BT-Settl} isochrones in the CMDs, respectively. The red solid line in the $d/V_{tan}$ plot marks the value $V_{tan}=$\,36\,km\,s$^{-\text{1}}$ which is the mean value for field stars \citep{zhang18a}, and the red dashed ellipse around Id 73 indicates its values of $V_{tan}\pm \text{3}\sigma$ and $d \pm \text{3}\sigma$.}
\label{fig:plot_CMD_PM_73}
\end{figure}

\begin{figure}[H]
  \caption*{\textbf{Id 89}}
    \begin{subfigure}{.5\textwidth}
    \includegraphics[width=0.91\linewidth]{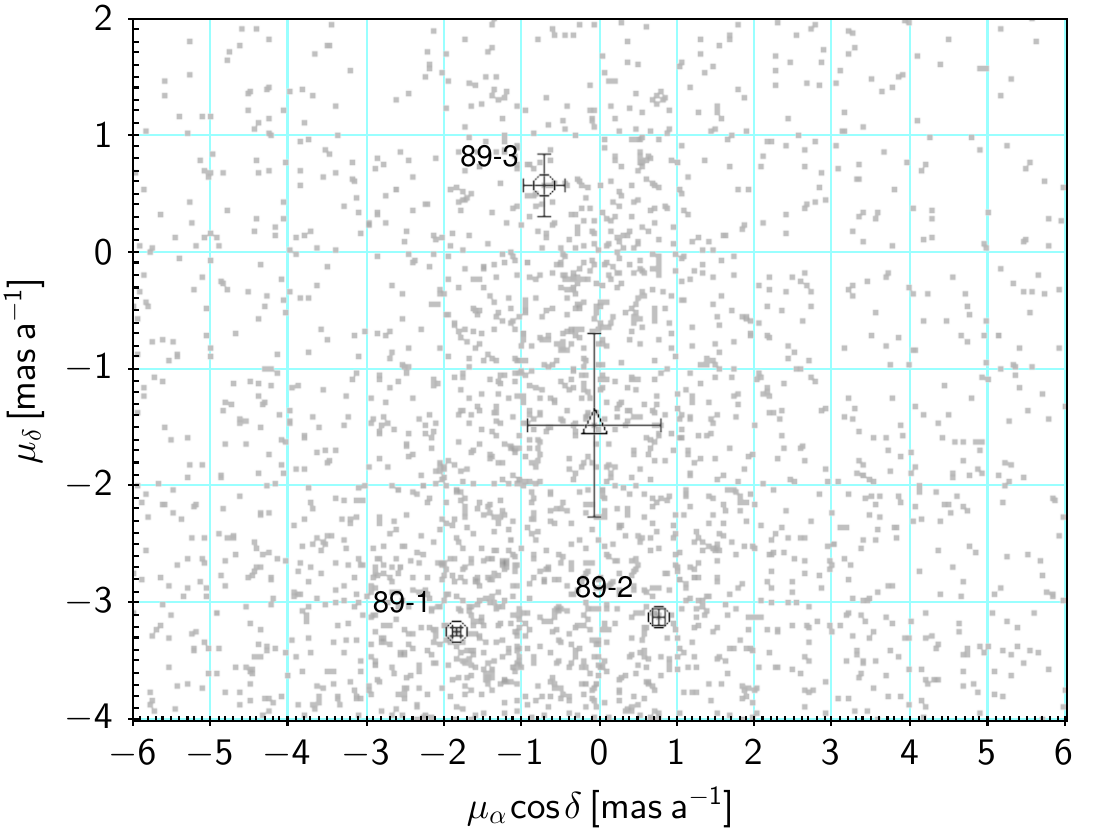}
    \end{subfigure}
    \begin{subfigure}{.5\textwidth}
      \includegraphics[width=0.91\linewidth]{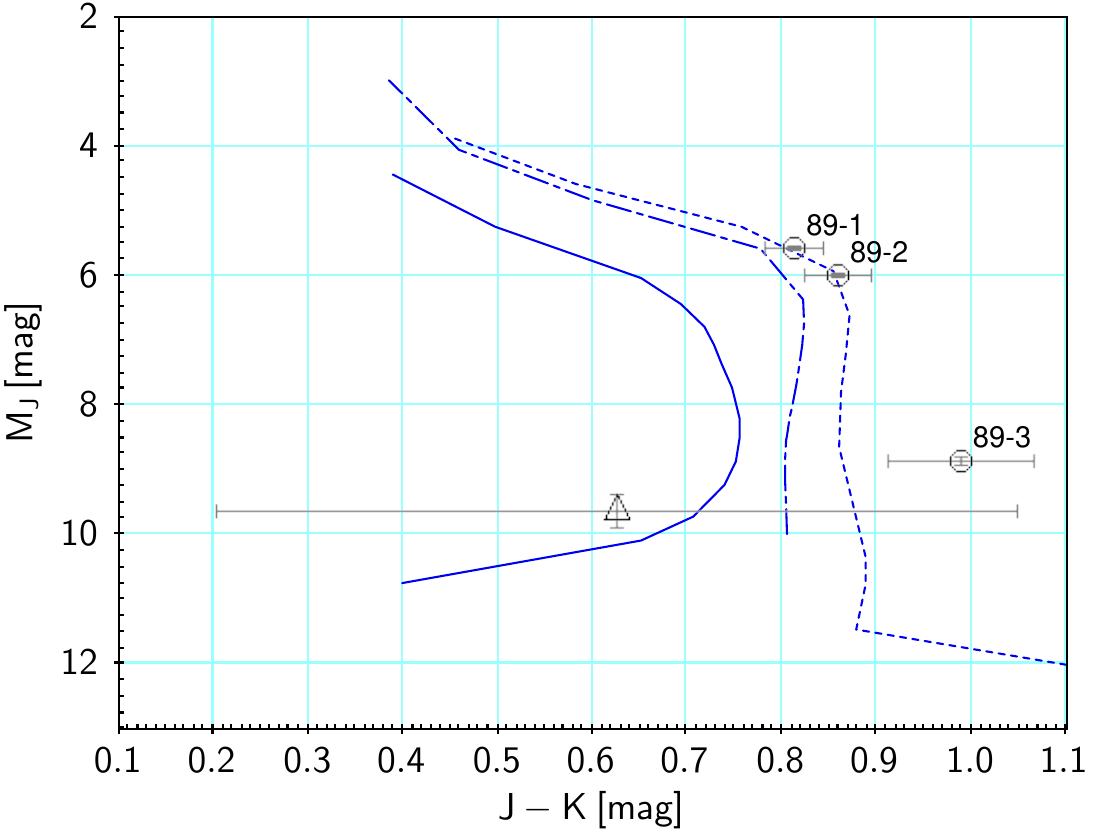}
    \end{subfigure} 
  \vspace{2mm}
  
     \begin{subfigure}{.5\textwidth}
    \includegraphics[width=0.91\linewidth]{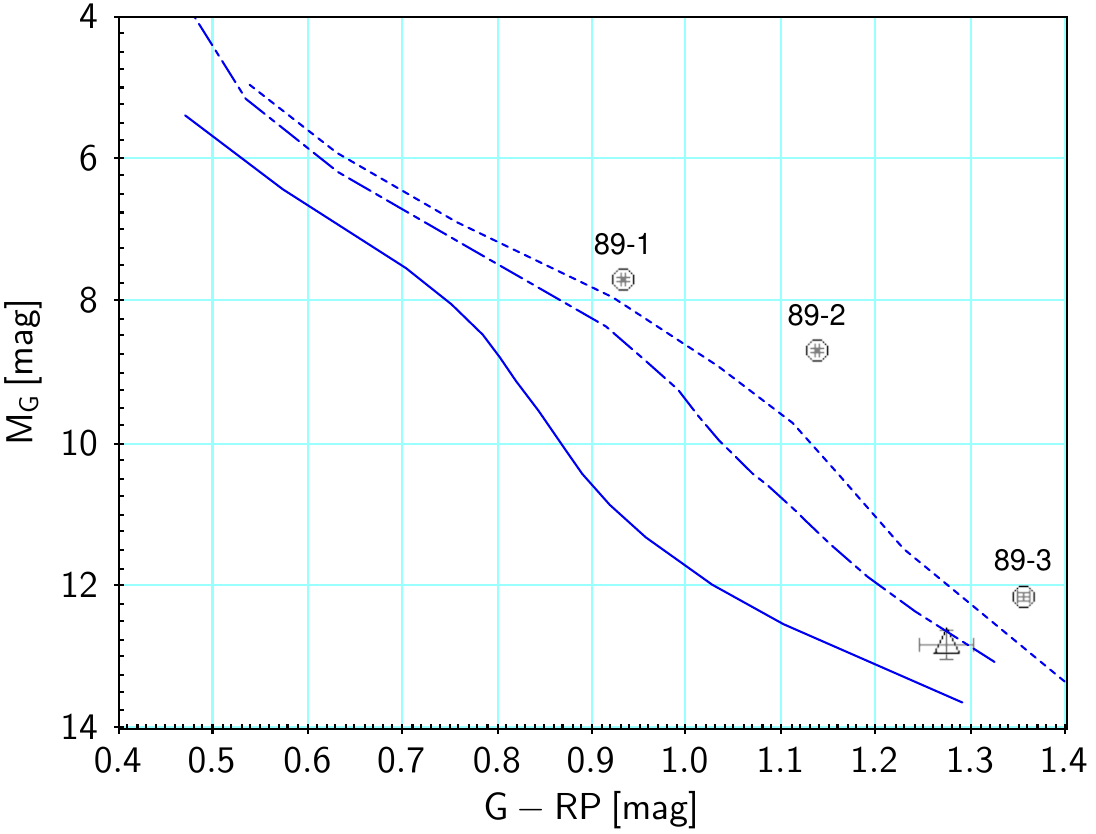}
    \end{subfigure}
    \begin{subfigure}{.5\textwidth}
      \includegraphics[width=0.91\linewidth]{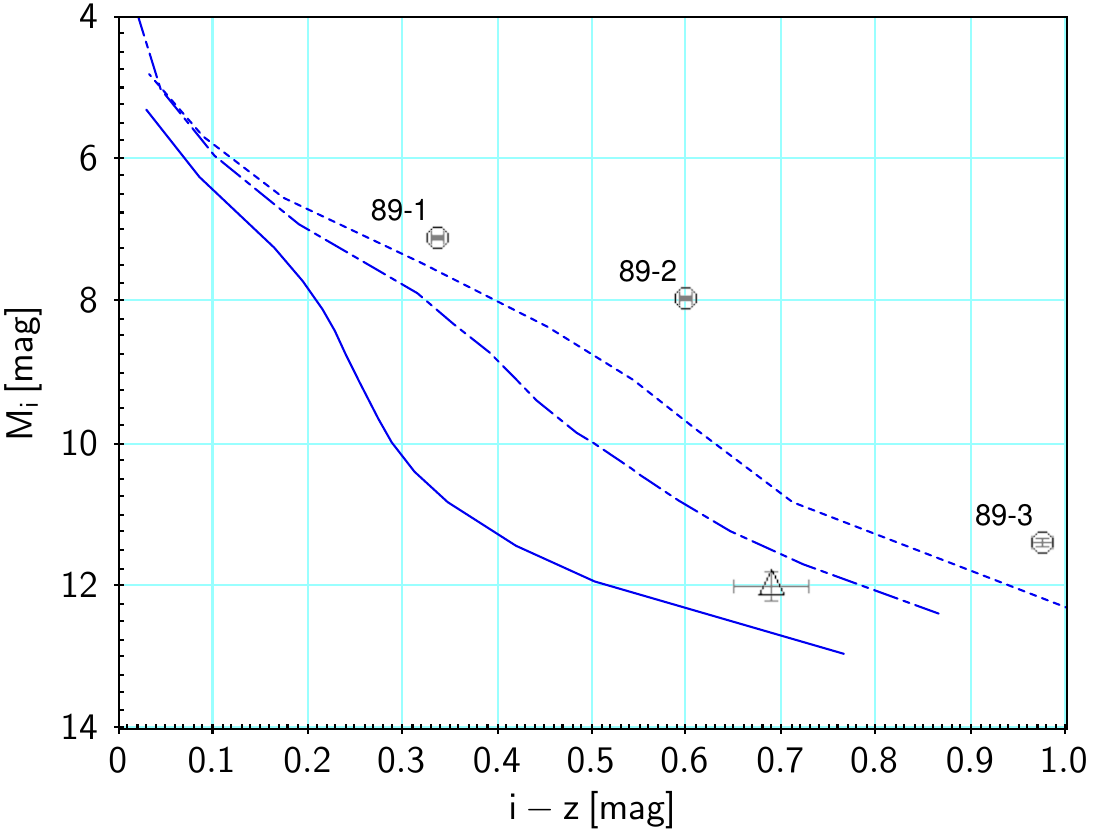}
    \end{subfigure}
    
  \vspace{2mm}  
     \begin{subfigure}{.5\textwidth}
    \includegraphics[width=0.91\linewidth]{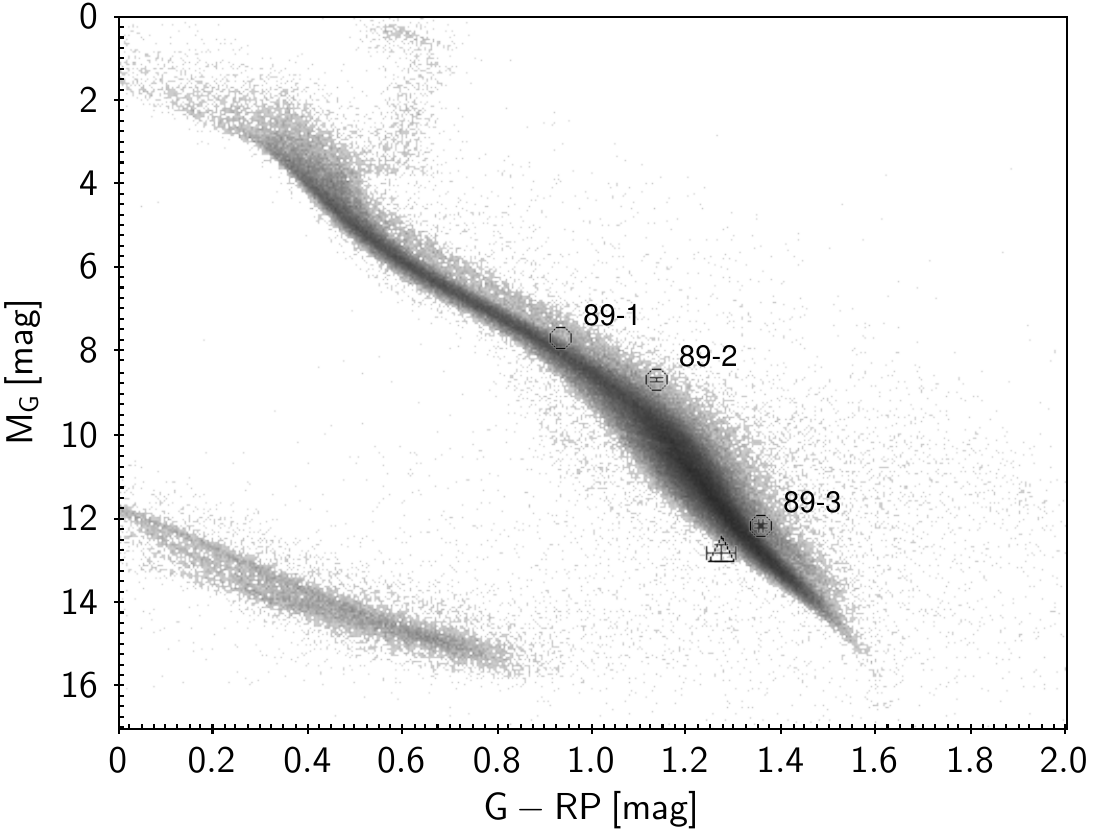}
    \end{subfigure}
    \begin{subfigure}{.5\textwidth}
      \includegraphics[width=0.91\linewidth]{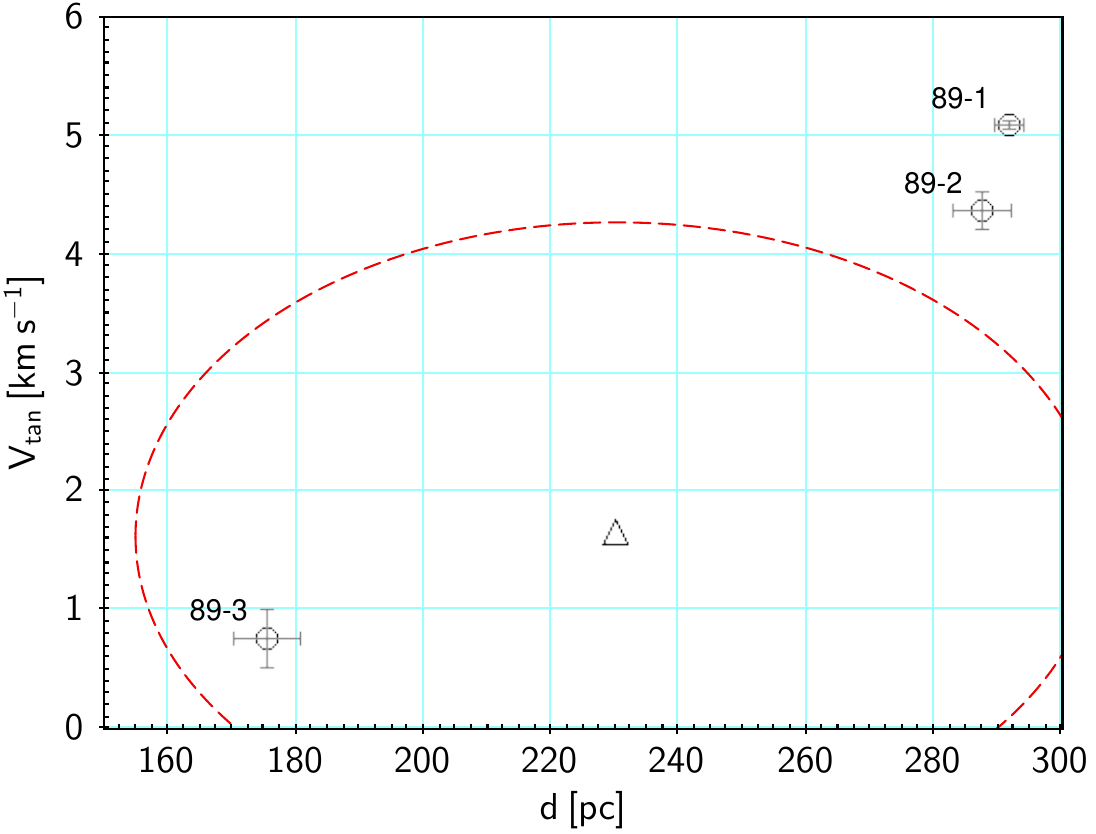}
    \end{subfigure}
  \vspace{5mm}
  \caption[PMD, CMDs, H-R diagram and tangential velocity vs distance diagram for the target Id 89 and its candidate companions.]{PMD (top left panel), CMDs (top right and mid panels), H-R diagram (bottom left panel), and tangential velocity--distance diagram (bottom right) for the target Id 89 and its candidate companion. The black triangle represents the source under study, and the numbered black circle is the companion candidate. Grey dots in the PMD represent field stars, and in the H-R diagram they are \textit{Gaia} DR2 sources with parallaxes $>$\,10\,mas used as a reference. The blue solid, dashed, and dotted lines stand for [M/H]\,=\,$-$2.0, $-$0.5, and =\,0.0 \textit{BT-Settl} isochrones in the CMDs, respectively. The red solid line in the $d/V_{tan}$ plot (not visible) marks the value $V_{tan}=$\,36\,km\,s$^{-\text{1}}$ which is the mean value for field stars \citep{zhang18a}, and the red dashed ellipse around Id 89 indicates its values of $V_{tan}\pm \text{3}\sigma$ and $d \pm \text{3}\sigma$.}
\label{fig:plot_CMD_PM_89}
\end{figure}

\begin{figure}[H]
  \caption*{\textbf{Id 107}}
    \begin{subfigure}{.5\textwidth}
    \includegraphics[width=0.91\linewidth]{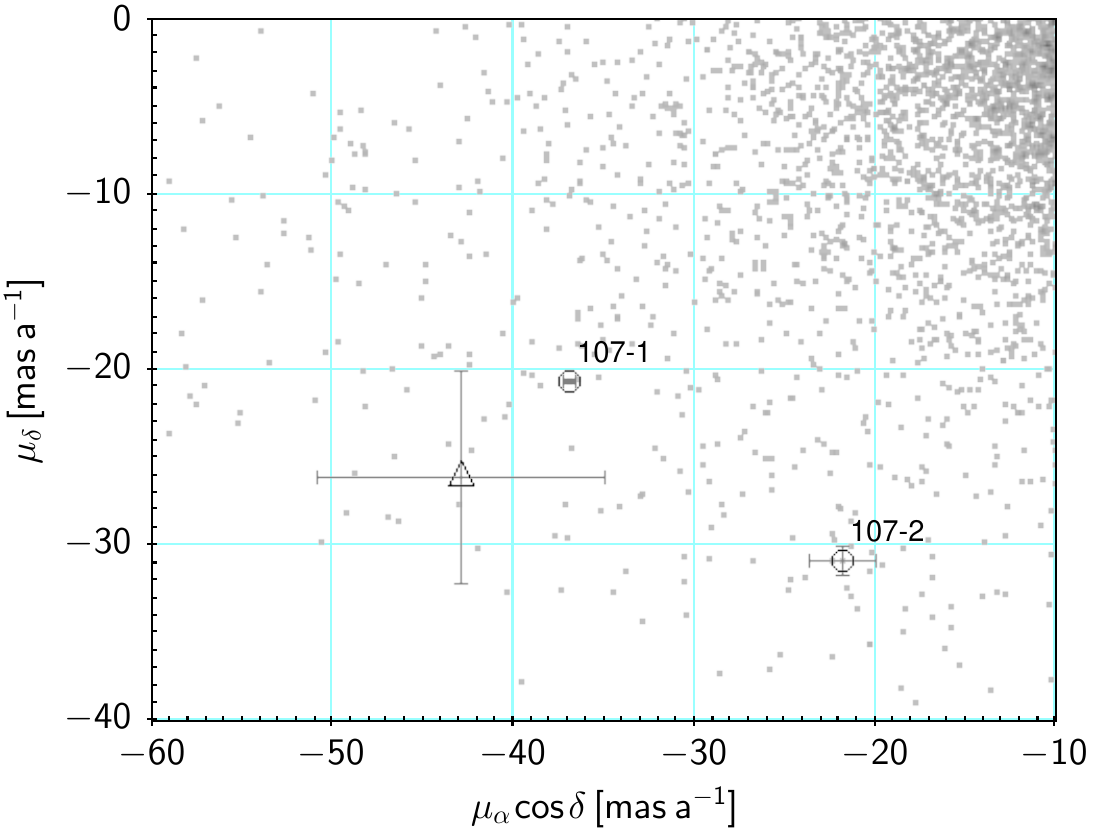}
    \end{subfigure}
    \begin{subfigure}{.5\textwidth}
      \includegraphics[width=0.91\linewidth]{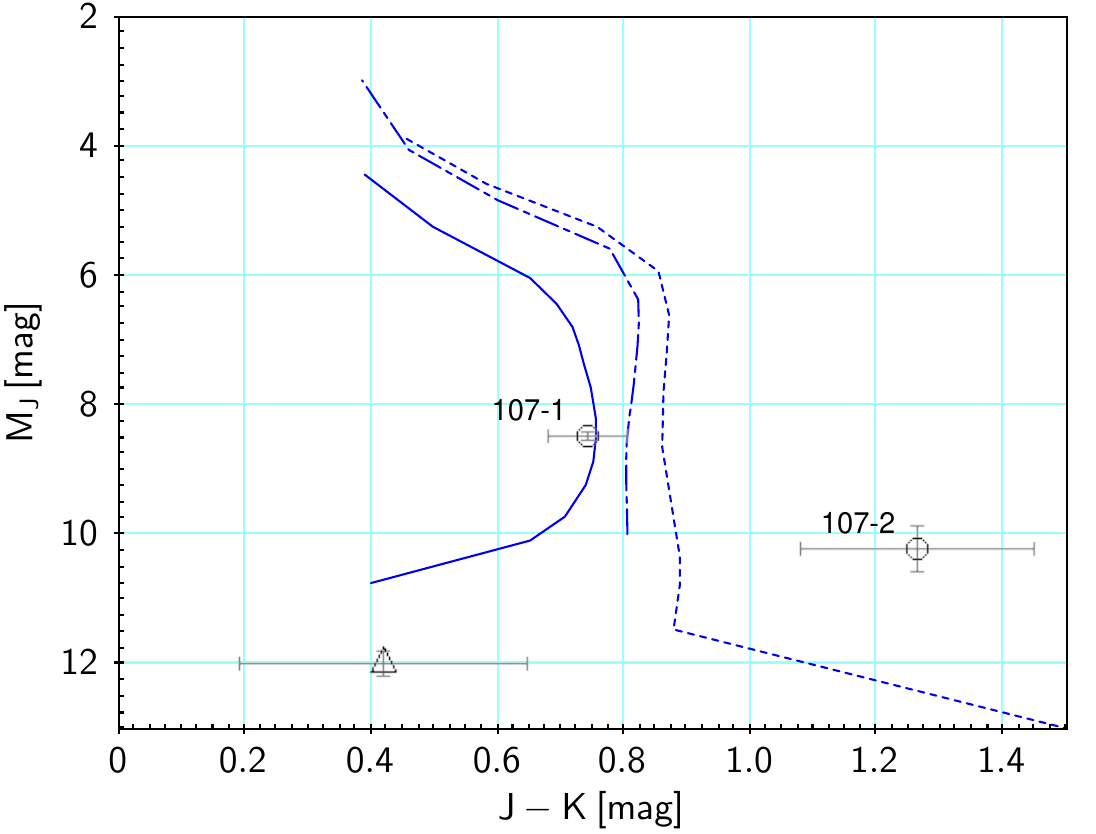}
    \end{subfigure} 
  \vspace{2mm}
  
     \begin{subfigure}{.5\textwidth}
    \includegraphics[width=0.91\linewidth]{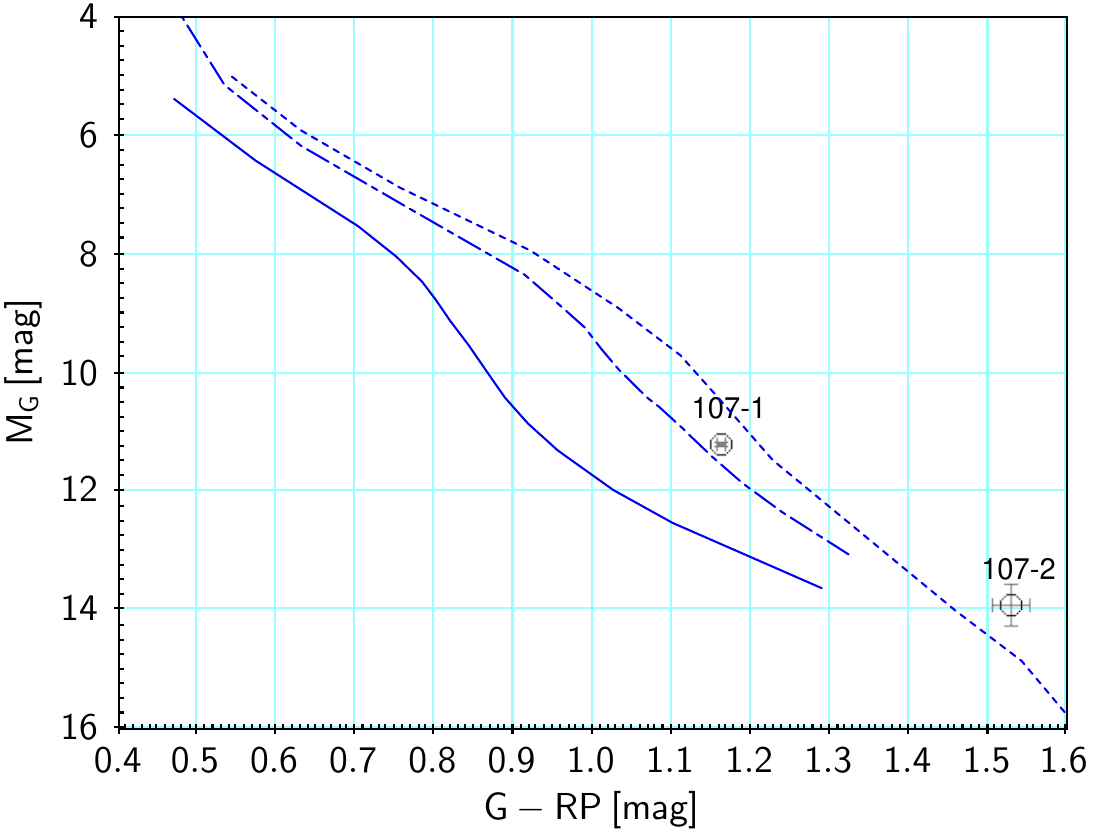}
    \end{subfigure}
    \begin{subfigure}{.5\textwidth}
      \includegraphics[width=0.91\linewidth]{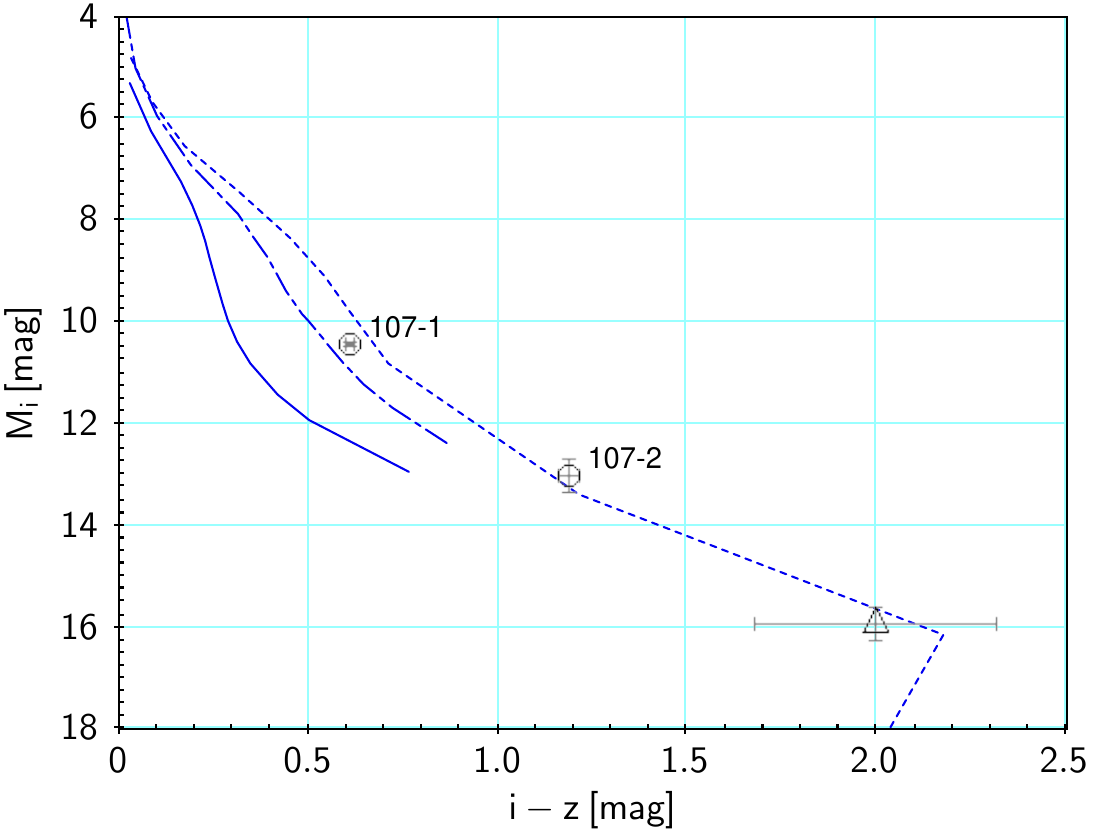}
    \end{subfigure}
    
  \vspace{2mm}  
     \begin{subfigure}{.5\textwidth}
    \includegraphics[width=0.91\linewidth]{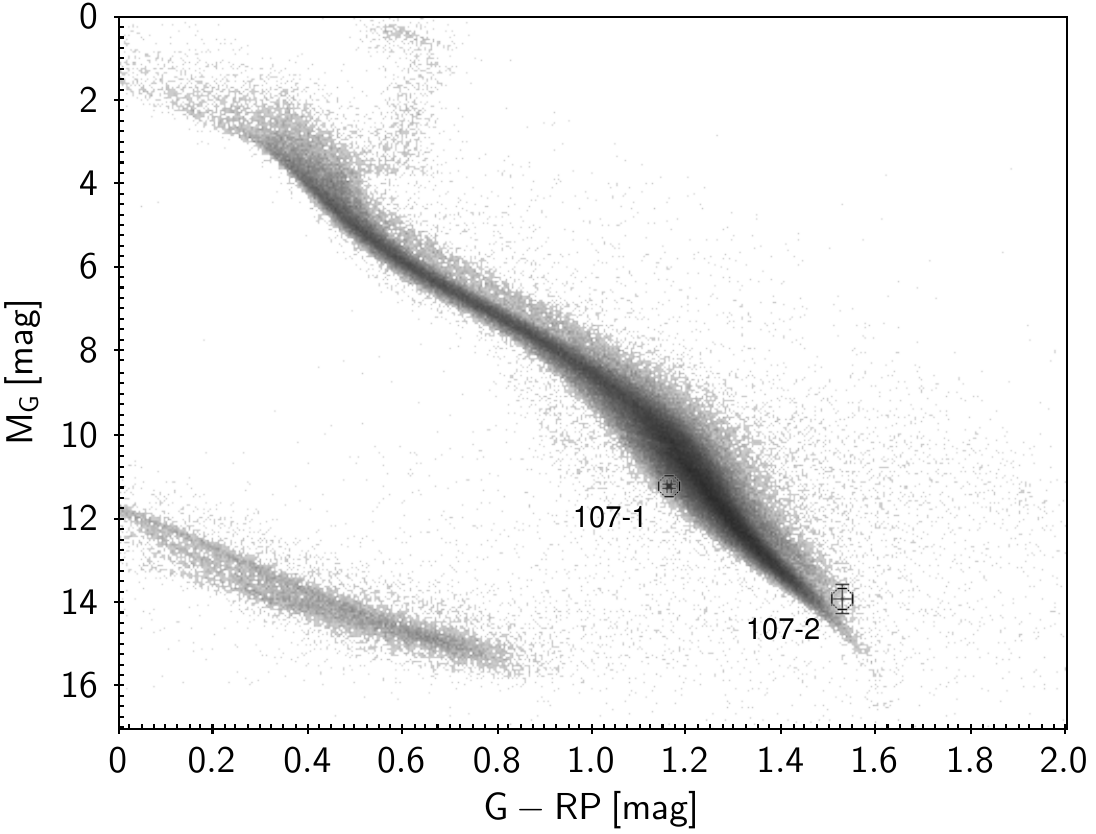}
    \end{subfigure}
    \begin{subfigure}{.5\textwidth}
      \includegraphics[width=0.91\linewidth]{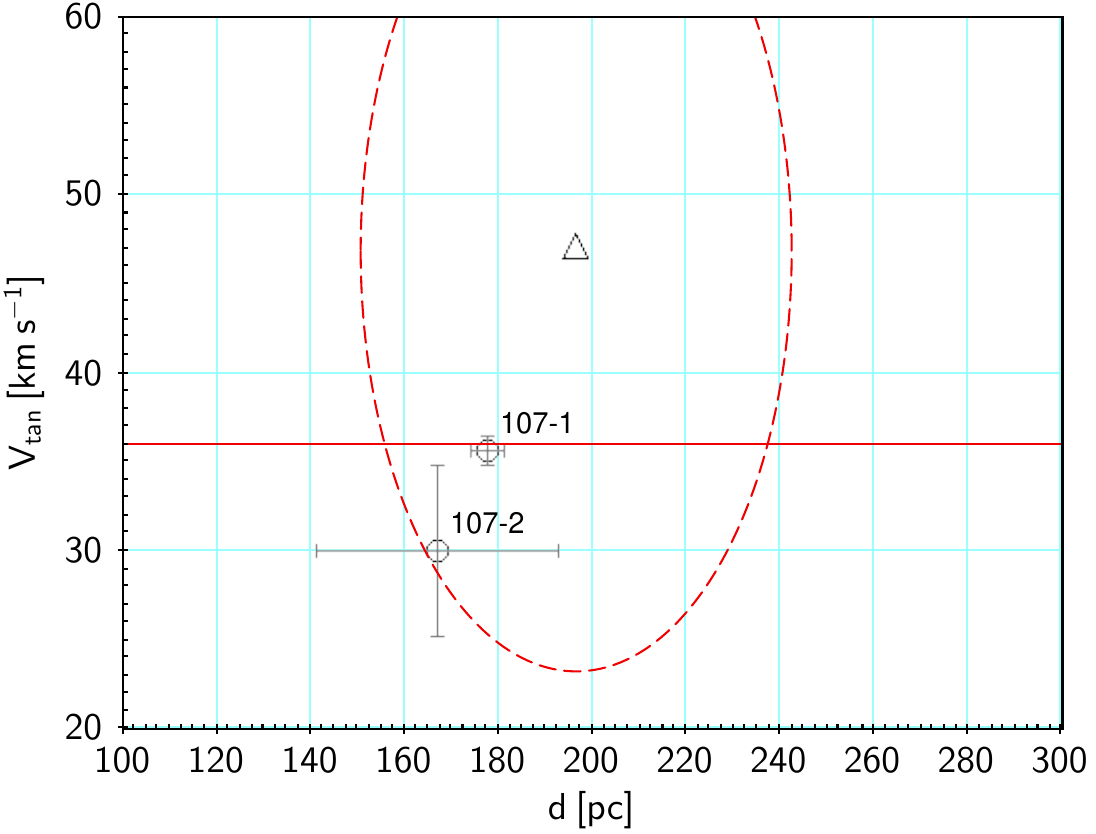}
    \end{subfigure}
  \vspace{5mm}
  \caption[PMD, CMDs, H-R diagram and tangential velocity vs distance diagram for the target Id 107 and its candidate companions.]{PMD (top left panel), CMDs (top right and mid panels), H-R diagram (bottom left panel), and tangential velocity--distance diagram (bottom right) for the target Id 107 and its candidate companion. The black triangle represents the source under study, and the numbered black circle is the companion candidate. Grey dots in the PMD represent field stars, and in the H-R diagram they are \textit{Gaia} DR2 sources with parallaxes $>$\,10\,mas used as a reference. The blue solid, dashed, and dotted lines stand for [M/H]\,=\,$-$2.0, $-$0.5, and =\,0.0 \textit{BT-Settl} isochrones in the CMDs, respectively. The red solid line in the $d/V_{tan}$ plot marks the value $V_{tan}=$\,36\,km\,s$^{-\text{1}}$ which is the mean value for field stars \citep{zhang18a}, and the red dashed ellipse around Id 107 indicates its values of $V_{tan}\pm \text{3}\sigma$ and $d \pm \text{3}\sigma$.}
\label{fig:plot_CMD_PM_107}
\end{figure}

\begin{figure}[H]
  \caption*{\textbf{Id 126}}
    \begin{subfigure}{.5\textwidth}
    \includegraphics[width=0.91\linewidth]{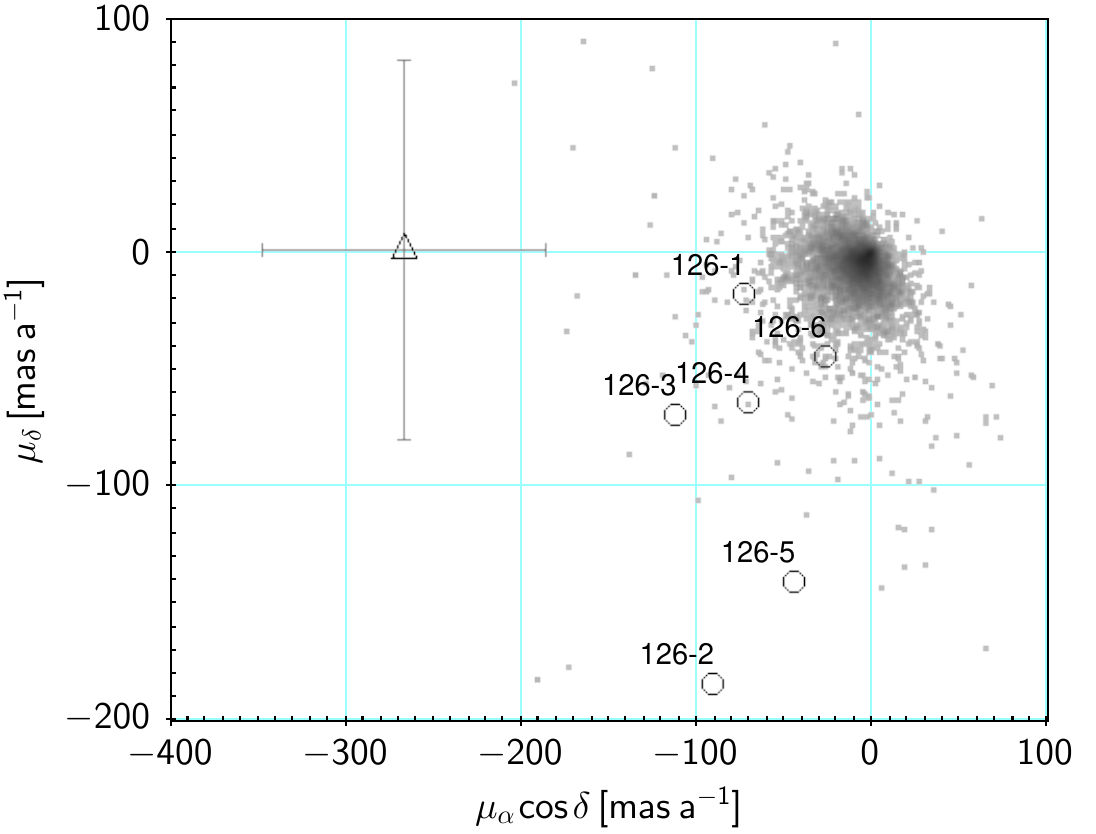}
    \end{subfigure}
    \begin{subfigure}{.5\textwidth}
      \includegraphics[width=0.91\linewidth]{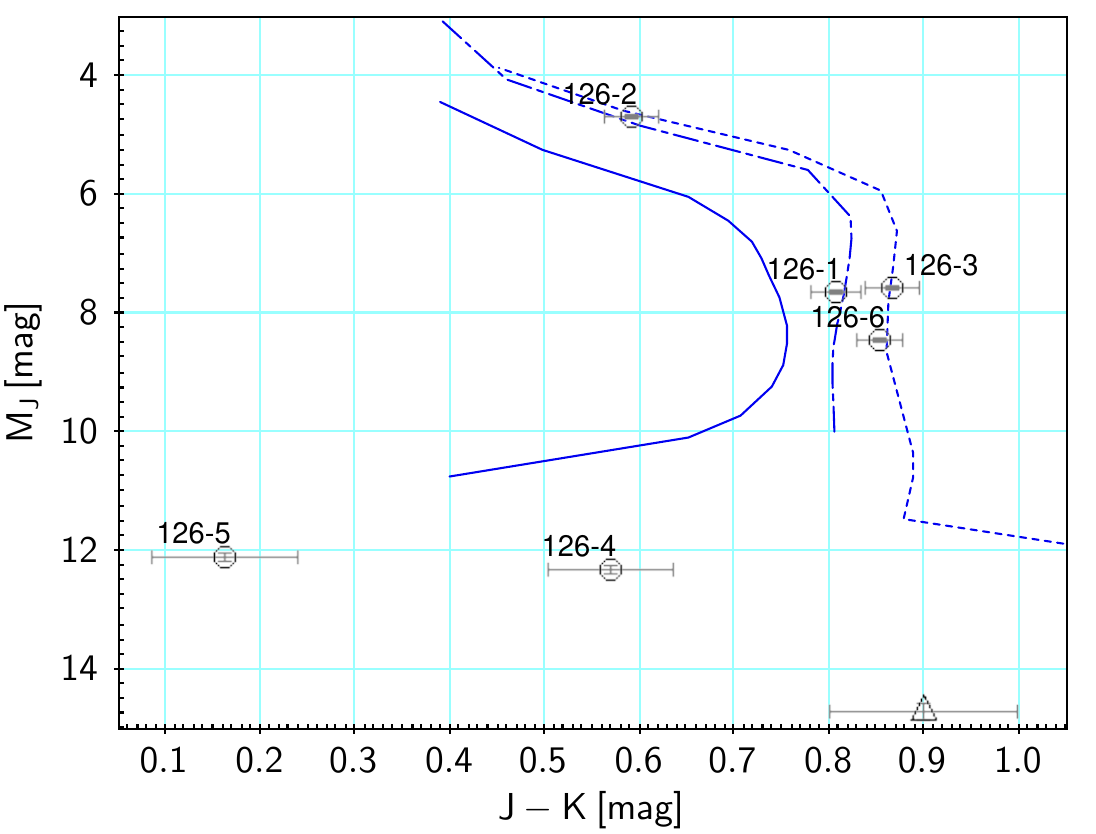}
    \end{subfigure} 
  \vspace{2mm}
  
     \begin{subfigure}{.5\textwidth}
    \includegraphics[width=0.91\linewidth]{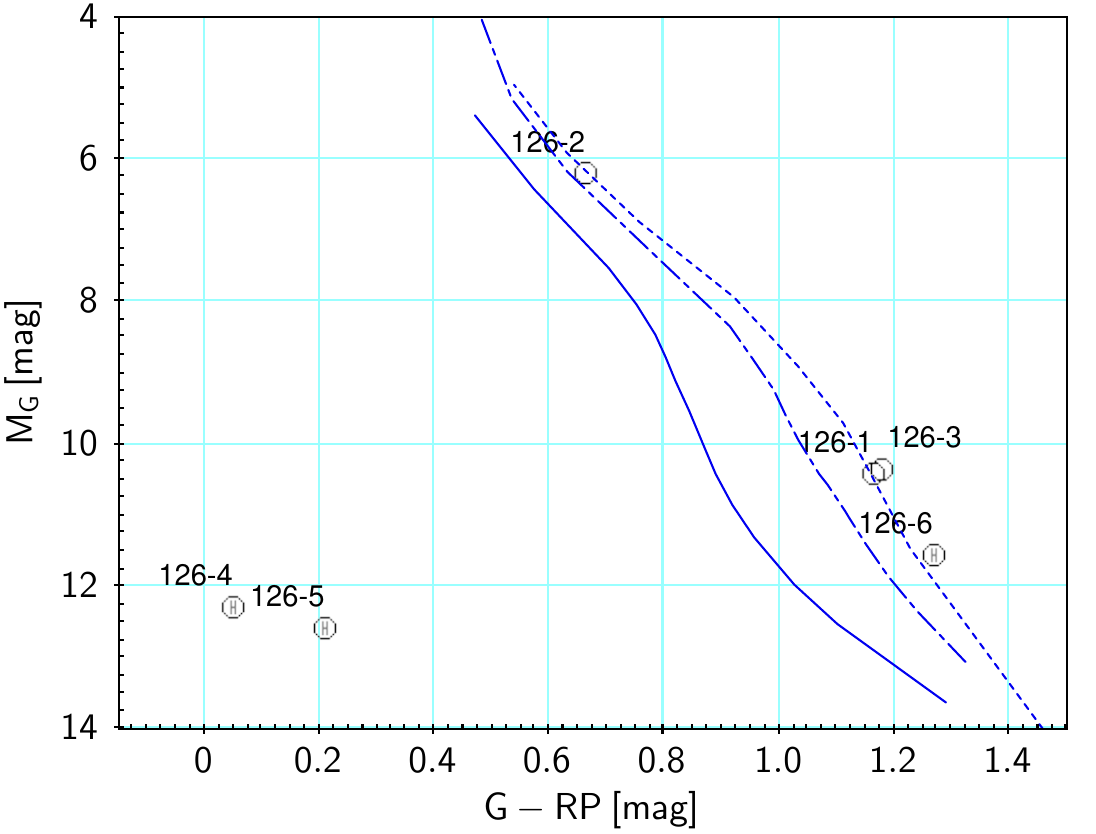}
    \end{subfigure}
    \begin{subfigure}{.5\textwidth}
      \includegraphics[width=0.91\linewidth]{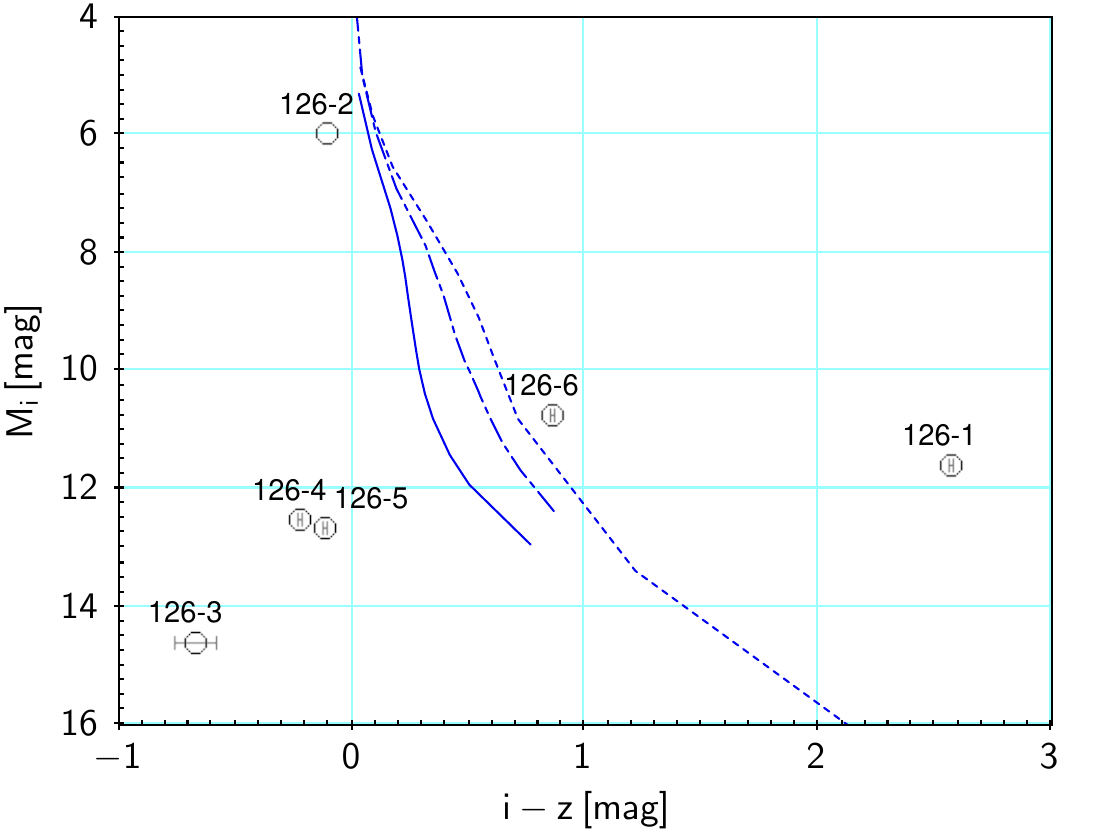}
    \end{subfigure}
    
  \vspace{2mm}  
     \begin{subfigure}{.5\textwidth}
    \includegraphics[width=0.91\linewidth]{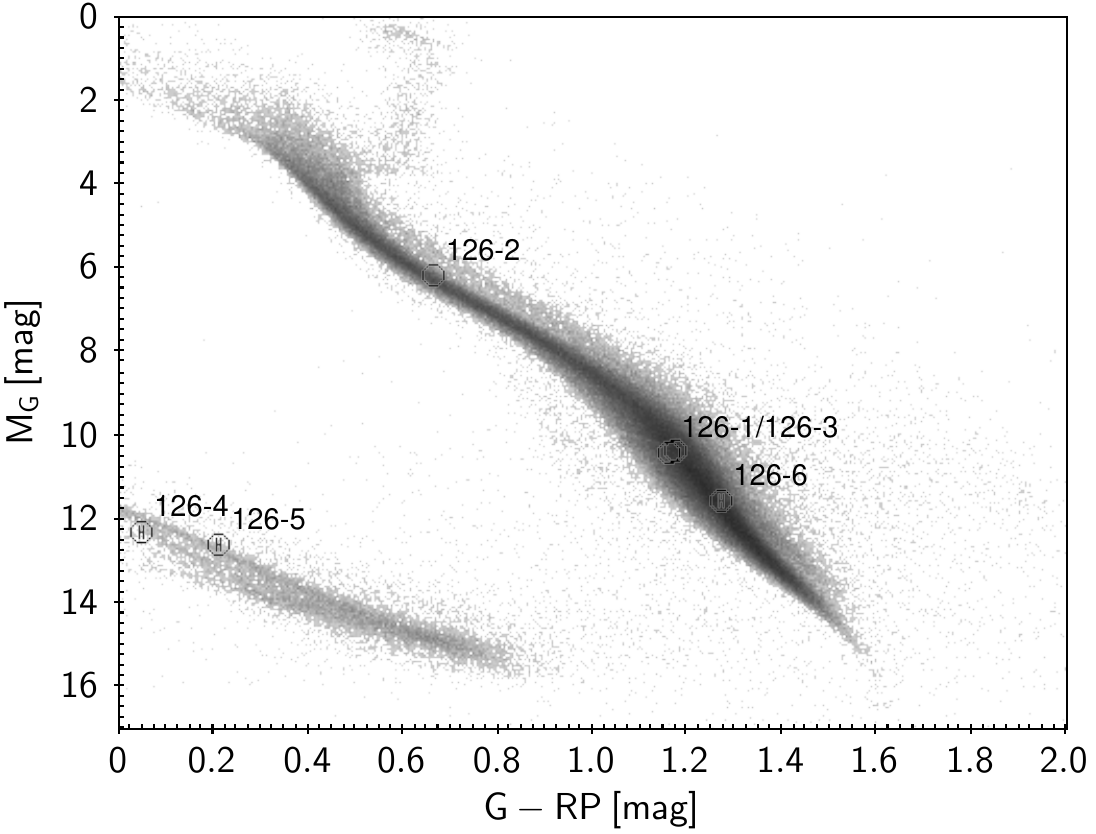}
    \end{subfigure}
    \begin{subfigure}{.5\textwidth}
      \includegraphics[width=0.91\linewidth]{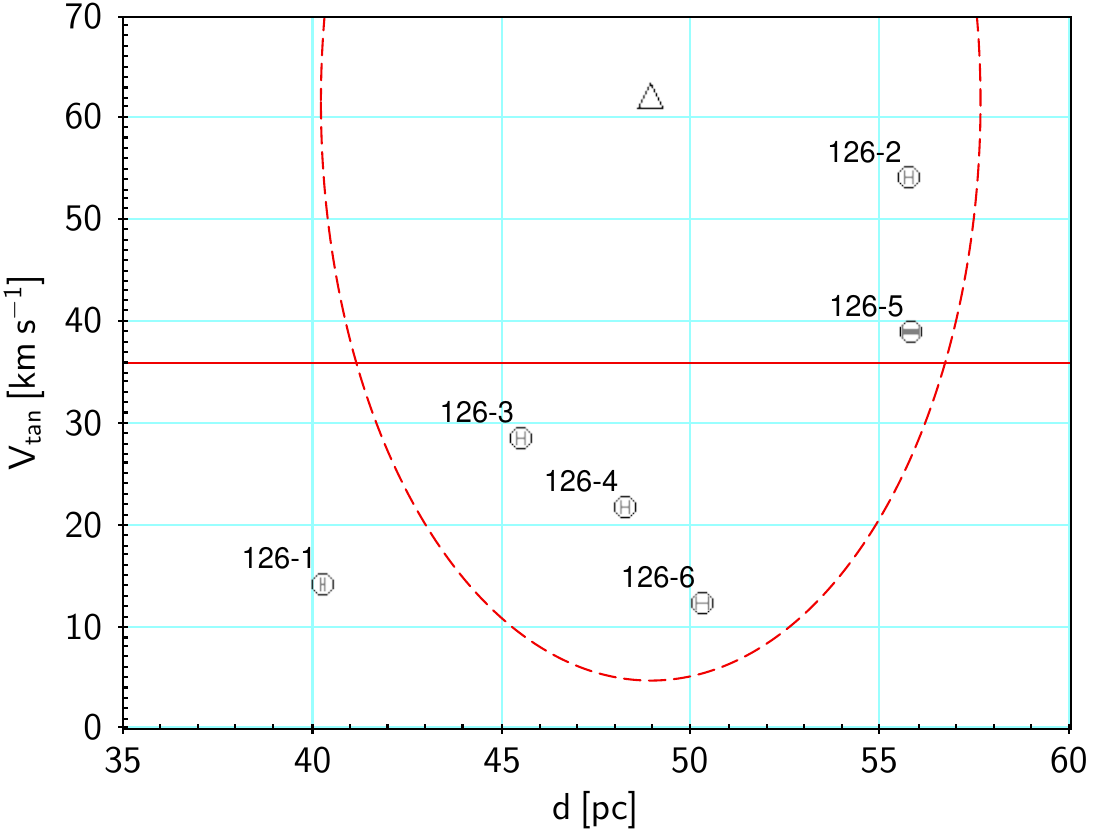}
    \end{subfigure}
  \vspace{5mm}
  \caption[PMD, CMDs, H-R diagram and tangential velocity vs distance diagram for the target Id 126 and its candidate companions.]{PMD (top left panel), CMDs (top right and mid panels), H-R diagram (bottom left panel), and tangential velocity--distance diagram (bottom right) for the target Id 126 and its candidate companion. The black triangle represents the source under study, and the numbered black circle is the companion candidate. Grey dots in the PMD represent field stars, and in the H-R diagram they are \textit{Gaia} DR2 sources with parallaxes $>$\,10\,mas used as a reference. The blue solid, dashed, and dotted lines stand for [M/H]\,=\,$-$2.0, $-$0.5, and =\,0.0 \textit{BT-Settl} isochrones in the CMDs, respectively. The red solid line in the $d/V_{tan}$ plot marks the value $V_{tan}=$\,36\,km\,s$^{-\text{1}}$ which is the mean value for field stars \citep{zhang18a}, and the red dashed ellipse around Id 126 indicates its values of $V_{tan}\pm \text{3}\sigma$ and $d \pm \text{3}\sigma$.}
\label{fig:plot_CMD_PM_126}
\end{figure}

\begin{figure}[H]
  \caption*{\textbf{Id 128}}
    \begin{subfigure}{.5\textwidth}
    \includegraphics[width=0.91\linewidth]{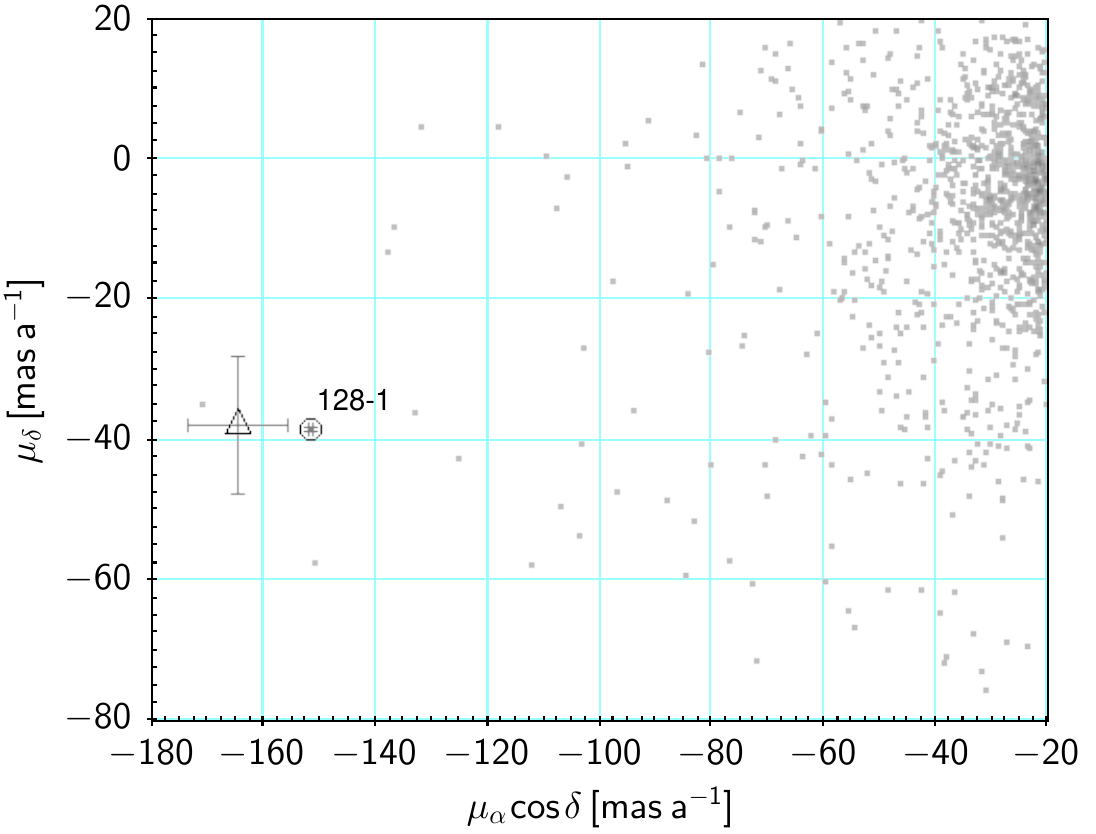}
    \end{subfigure}
    \begin{subfigure}{.5\textwidth}
      \includegraphics[width=0.91\linewidth]{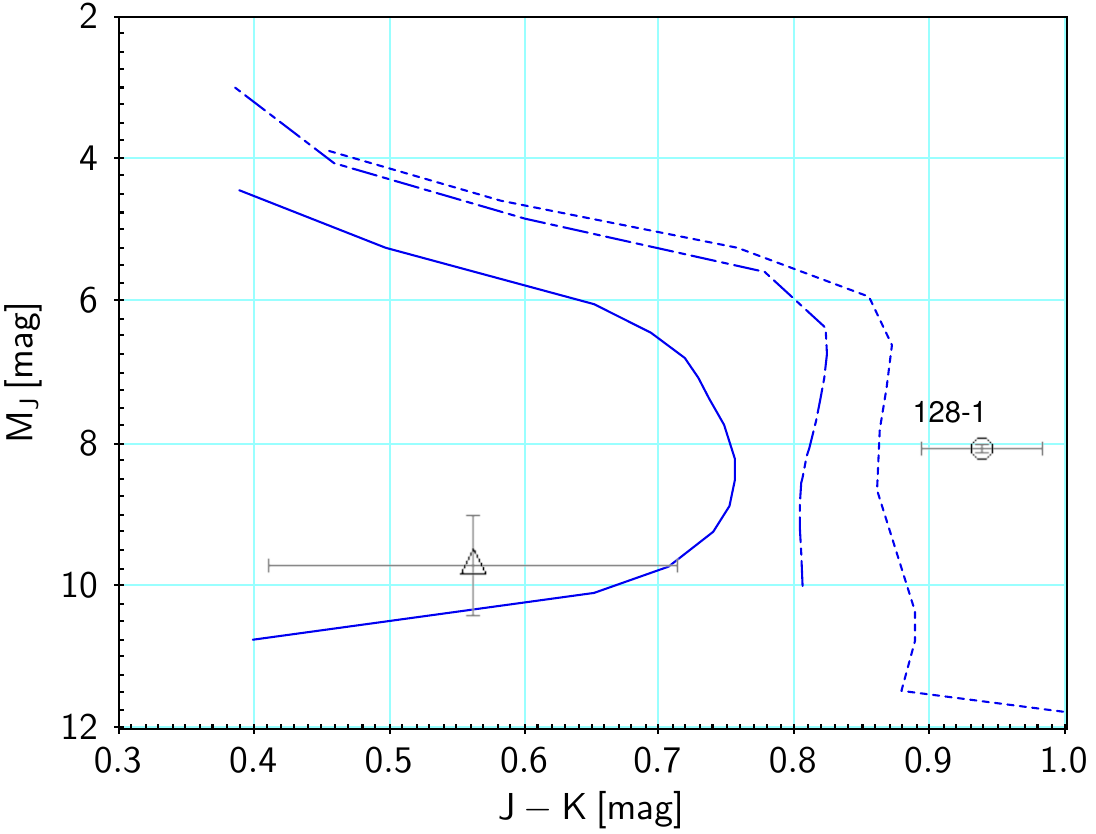}
    \end{subfigure} 
  \vspace{2mm}
  
     \begin{subfigure}{.5\textwidth}
    \includegraphics[width=0.91\linewidth]{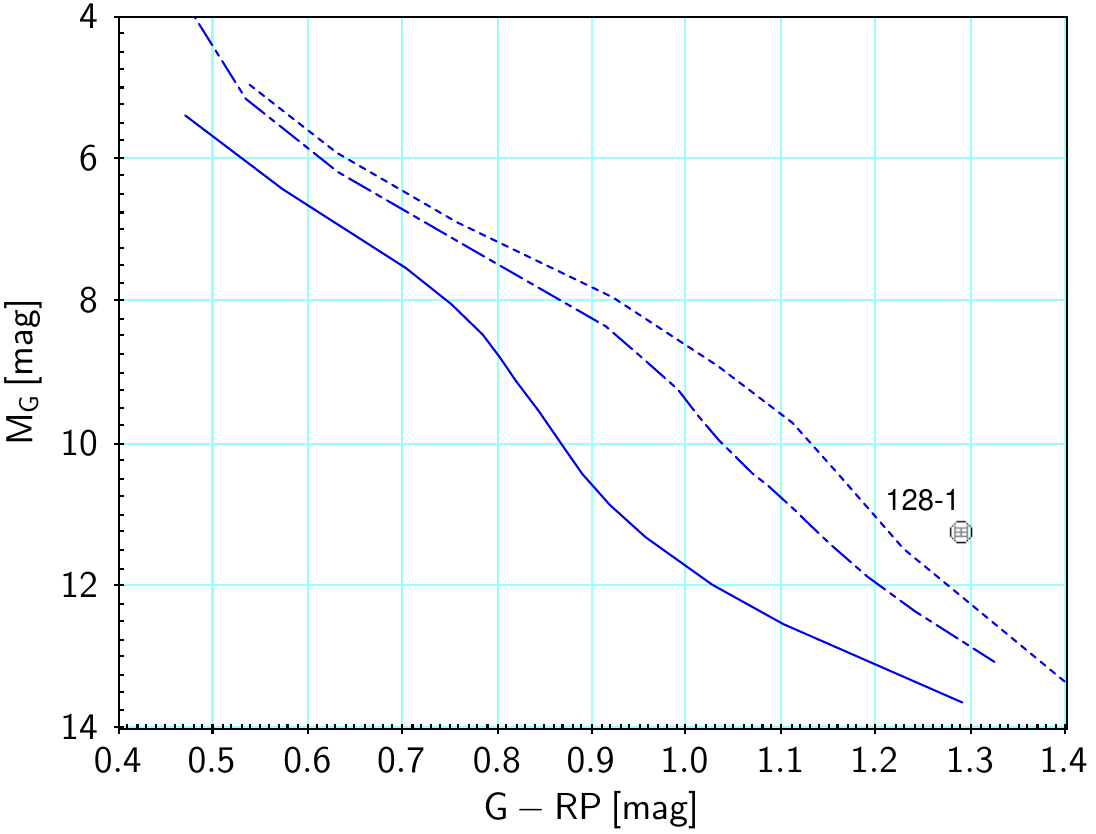}
    \end{subfigure}
    \begin{subfigure}{.5\textwidth}
      \includegraphics[width=0.91\linewidth]{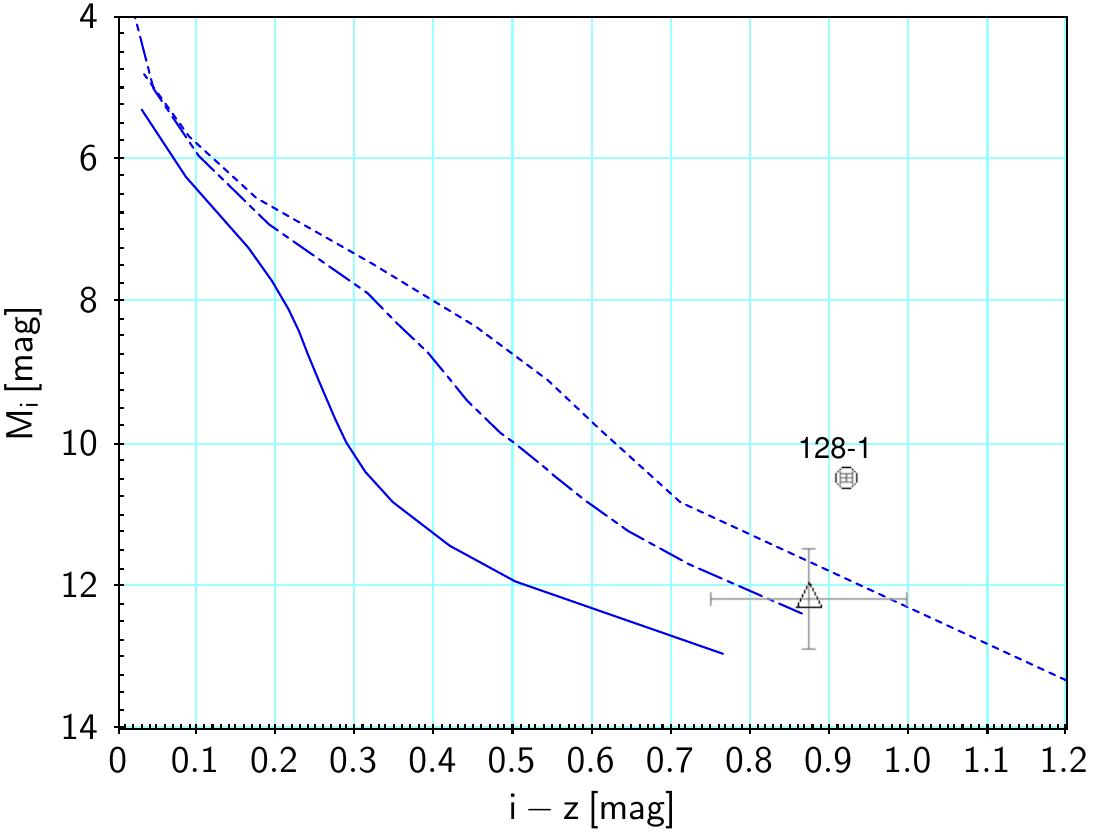}
    \end{subfigure}
    
  \vspace{2mm}  
     \begin{subfigure}{.5\textwidth}
    \includegraphics[width=0.91\linewidth]{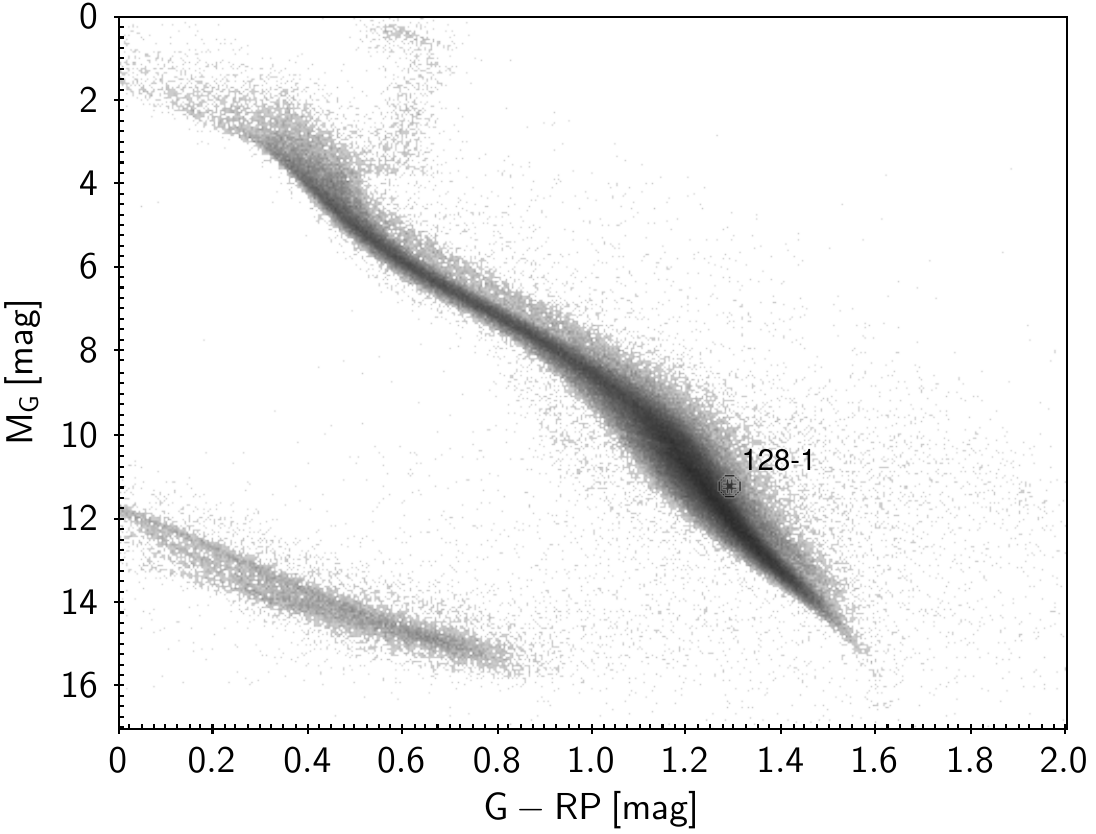}
    \end{subfigure}
    \begin{subfigure}{.5\textwidth}
      \includegraphics[width=0.91\linewidth]{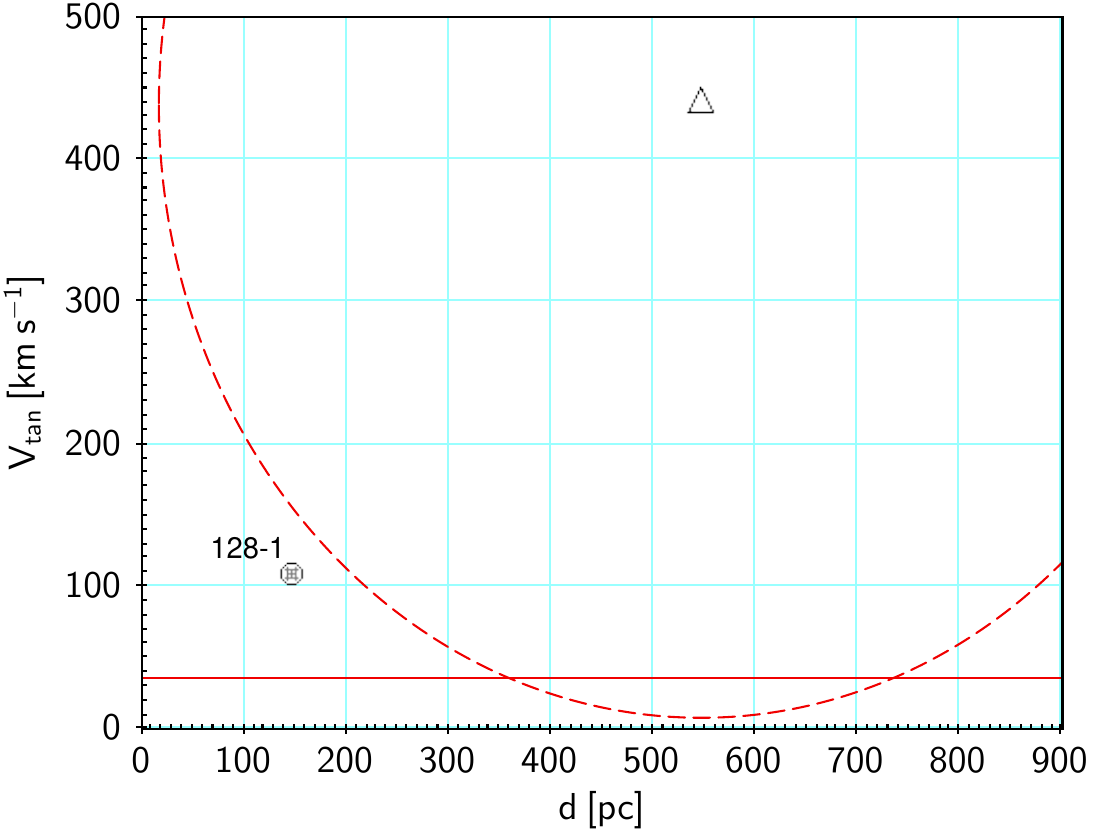}
    \end{subfigure}
  \vspace{5mm}
  \caption[PMD, CMDs, H-R diagram and tangential velocity vs distance diagram for the target Id 128 and its candidate companions.]{PMD (top left panel), CMDs (top right and mid panels), H-R diagram (bottom left panel), and tangential velocity--distance diagram (bottom right) for the target Id 128 and its candidate companion. The black triangle represents the source under study, and the numbered black circle is the companion candidate. Grey dots in the PMD represent field stars, and in the H-R diagram they are \textit{Gaia} DR2 sources with parallaxes $>$\,10\,mas used as a reference. The blue solid, dashed, and dotted lines stand for [M/H]\,=\,$-$2.0, $-$0.5, and =\,0.0 \textit{BT-Settl} isochrones in the CMDs, respectively. The red solid line in the $d/V_{tan}$ plot marks the value $V_{tan}=$\,36\,km\,s$^{-\text{1}}$ which is the mean value for field stars \citep{zhang18a}, and the red dashed ellipse around Id 128 indicates its values of $V_{tan}\pm \text{3}\sigma$ and $d \pm \text{3}\sigma$.}
\label{fig:plot_CMD_PM_128}
\end{figure}

\begin{figure}[H]
  \caption*{\textbf{Id 149}}
    \begin{subfigure}{.5\textwidth}
    \includegraphics[width=0.91\linewidth]{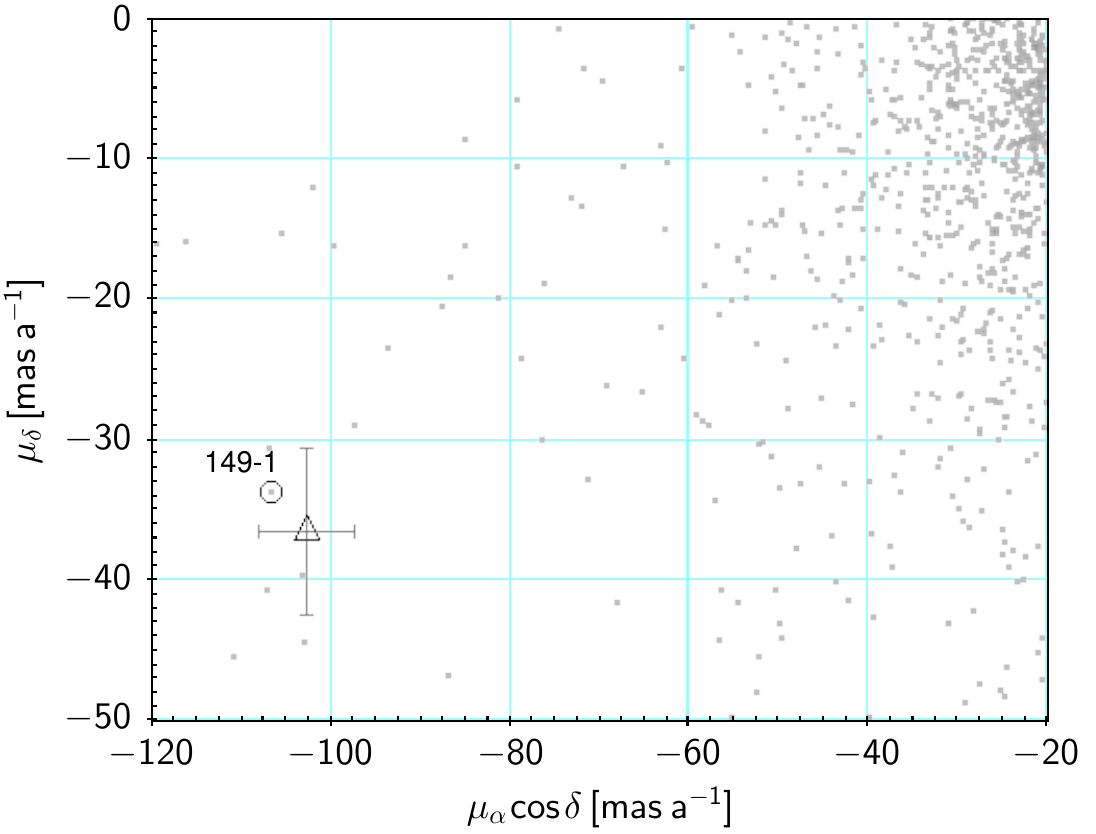}
    \end{subfigure}
    \begin{subfigure}{.5\textwidth}
      \includegraphics[width=0.91\linewidth]{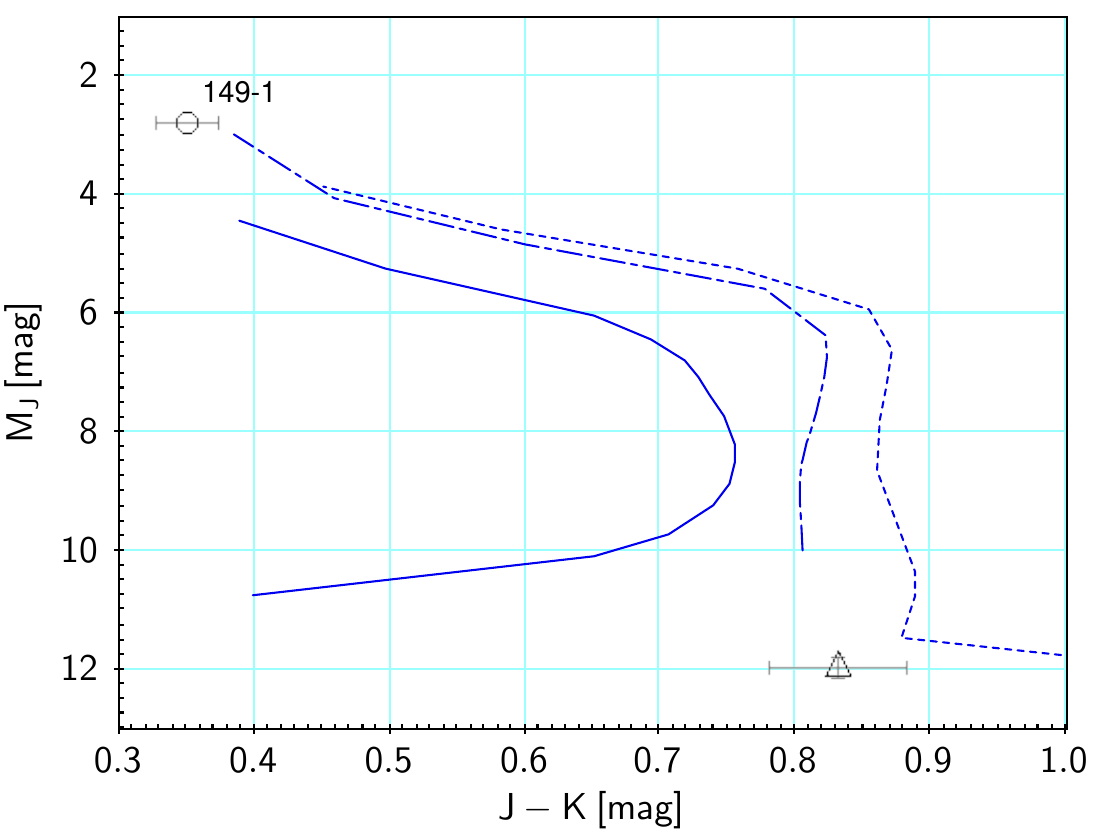}
    \end{subfigure} 
  \vspace{2mm}
  
     \begin{subfigure}{.5\textwidth}
    \includegraphics[width=0.91\linewidth]{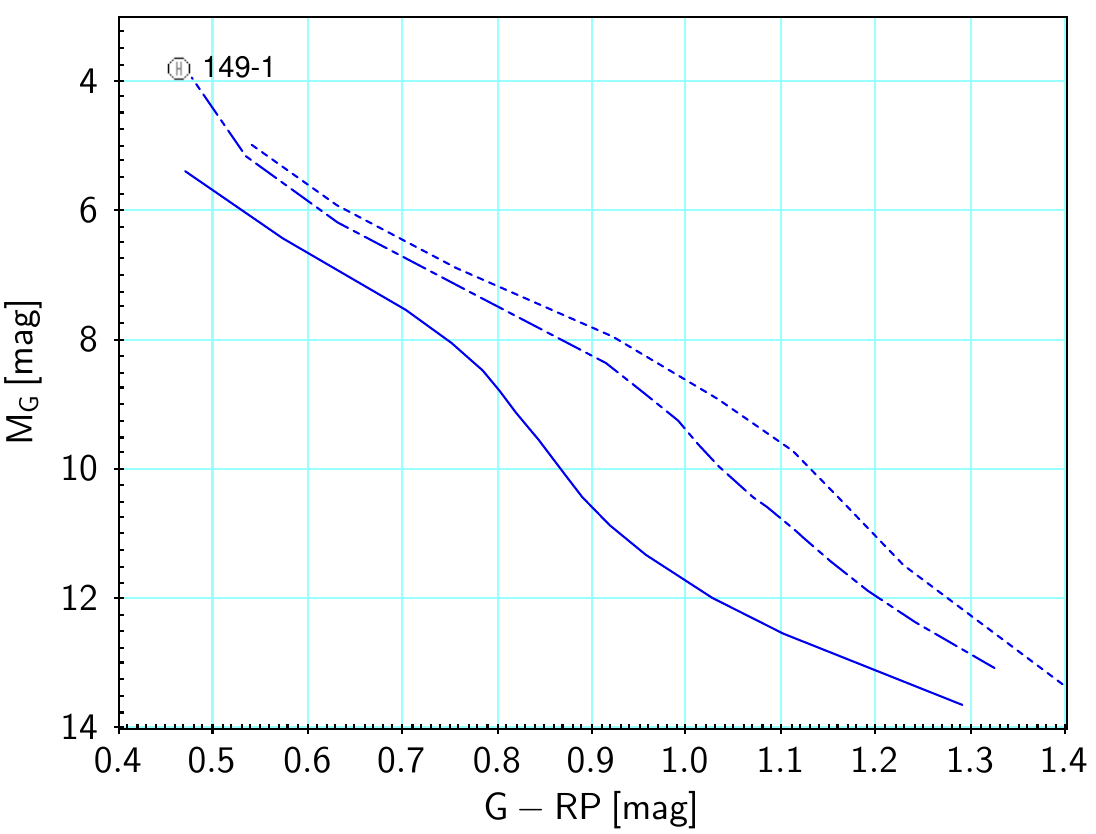}
    \end{subfigure}
    \begin{subfigure}{.5\textwidth}
      \includegraphics[width=0.91\linewidth]{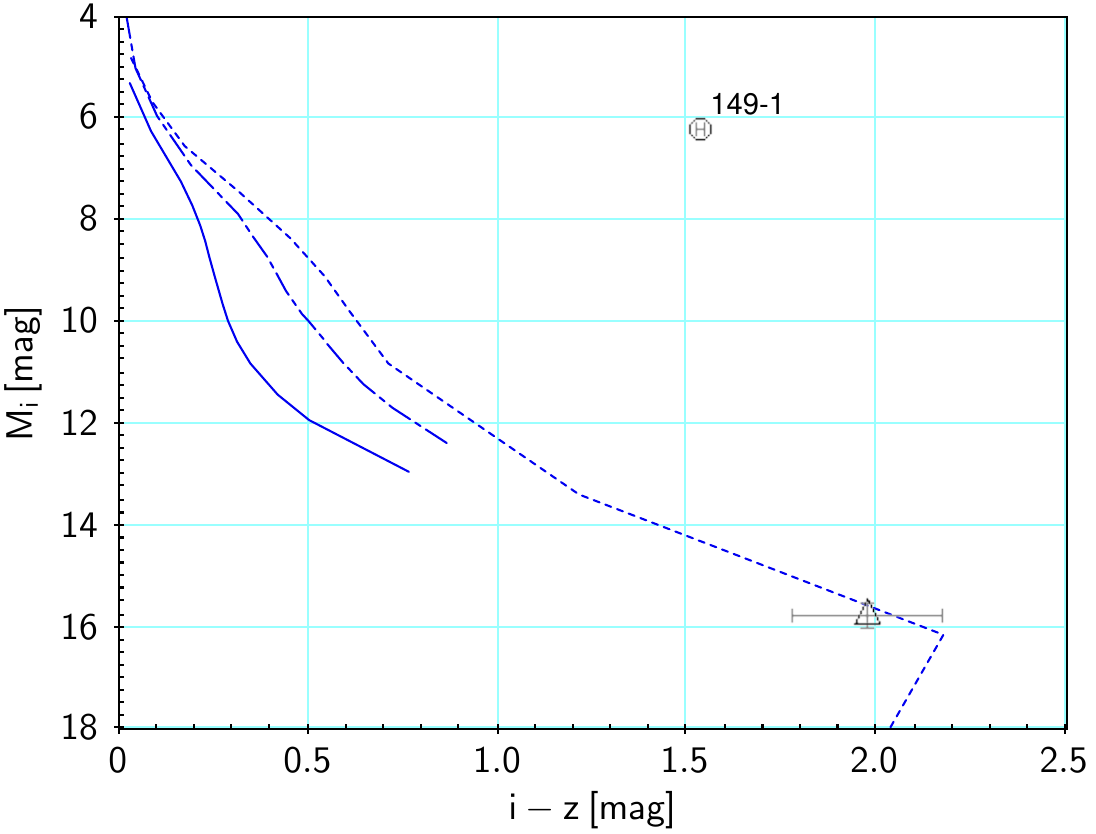}
    \end{subfigure}
    
  \vspace{2mm}  
     \begin{subfigure}{.5\textwidth}
    \includegraphics[width=0.91\linewidth]{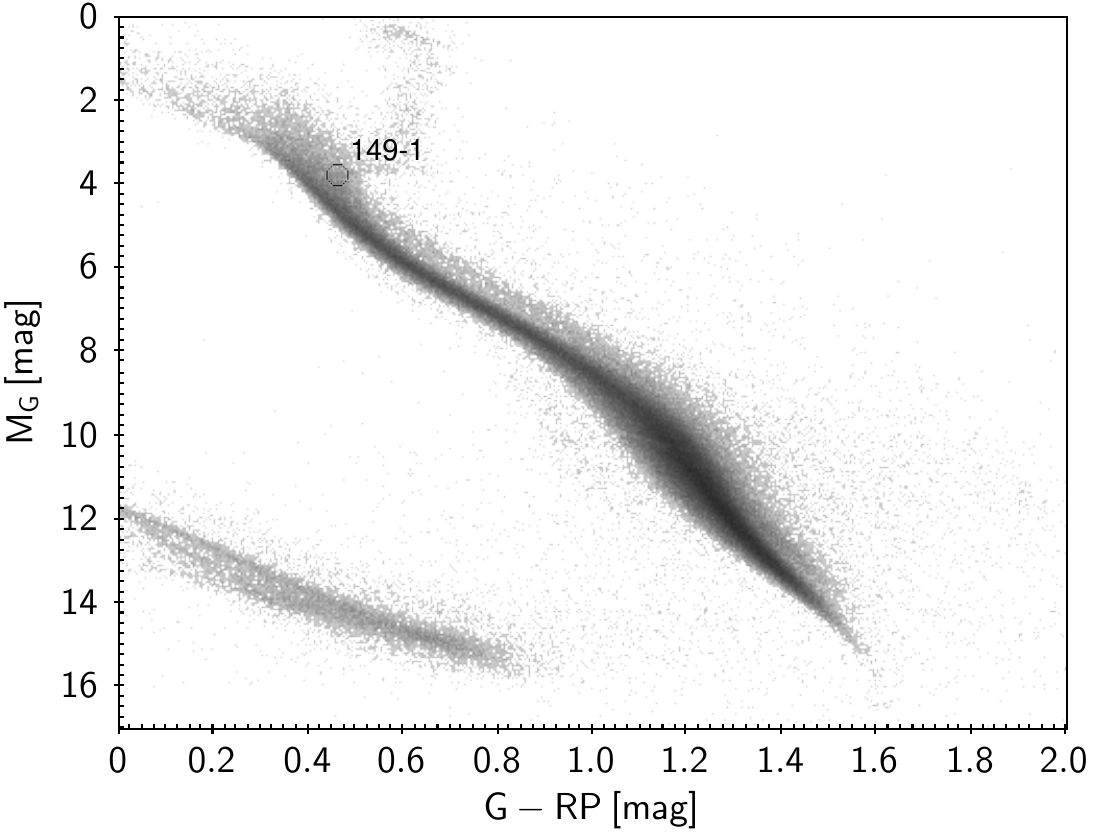}
    \end{subfigure}
    \begin{subfigure}{.5\textwidth}
      \includegraphics[width=0.91\linewidth]{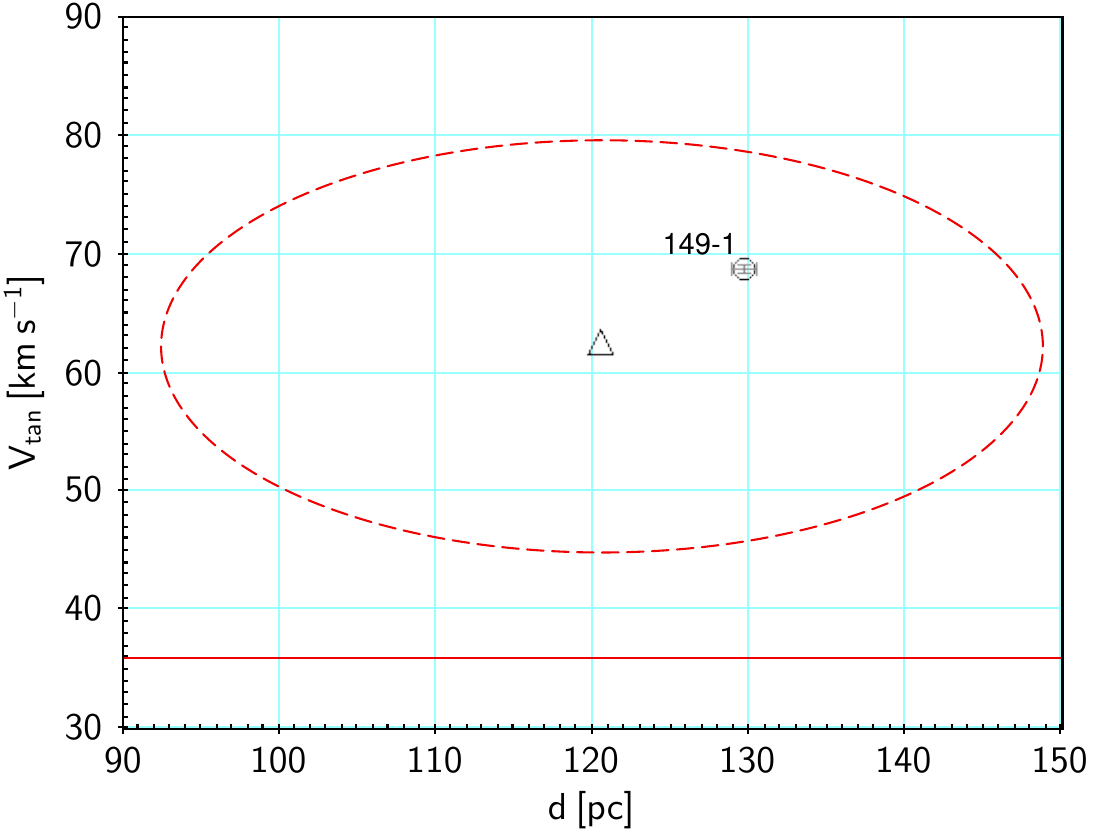}
    \end{subfigure}
  \vspace{5mm}
  \caption[PMD, CMDs, H-R diagram and tangential velocity vs distance diagram for the target Id 149 and its candidate companions.]{PMD (top left panel), CMDs (top right and mid panels), H-R diagram (bottom left panel), and tangential velocity--distance diagram (bottom right) for the target Id 149 and its candidate companion. The black triangle represents the source under study, and the numbered black circle is the companion candidate. Grey dots in the PMD represent field stars, and in the H-R diagram they are \textit{Gaia} DR2 sources with parallaxes $>$\,10\,mas used as a reference. The blue solid, dashed, and dotted lines stand for [M/H]\,=\,$-$2.0, $-$0.5, and =\,0.0 \textit{BT-Settl} isochrones in the CMDs, respectively. The red solid line in the $d/V_{tan}$ plot marks the value $V_{tan}=$\,36\,km\,s$^{-\text{1}}$ which is the mean value for field stars \citep{zhang18a}, and the red dashed ellipse around Id 149 indicates its values of $V_{tan}\pm \text{3}\sigma$ and $d \pm \text{3}\sigma$.}
\label{fig:plot_CMD_PM_149}
\end{figure}

\begin{figure}[H]
  \caption*{\textbf{Id 190}}
    \begin{subfigure}{.5\textwidth}
    \includegraphics[width=0.91\linewidth]{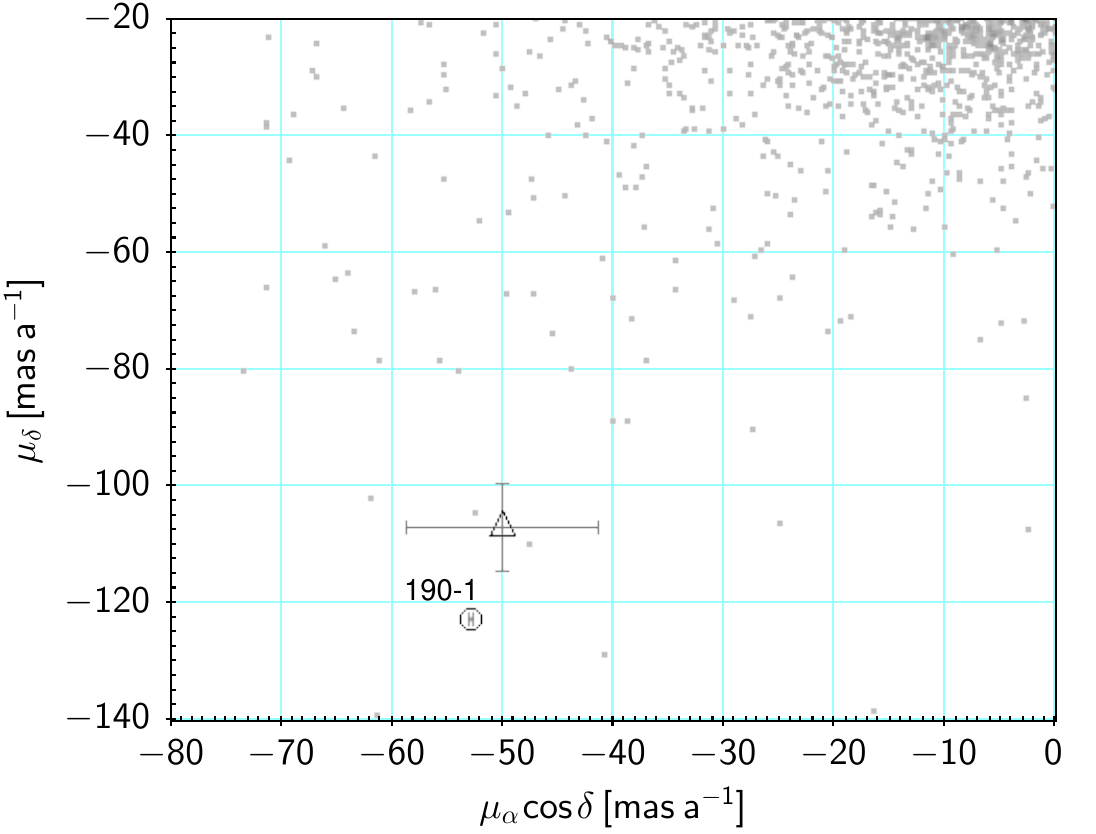}
    \end{subfigure}
    \begin{subfigure}{.5\textwidth}
      \includegraphics[width=0.91\linewidth]{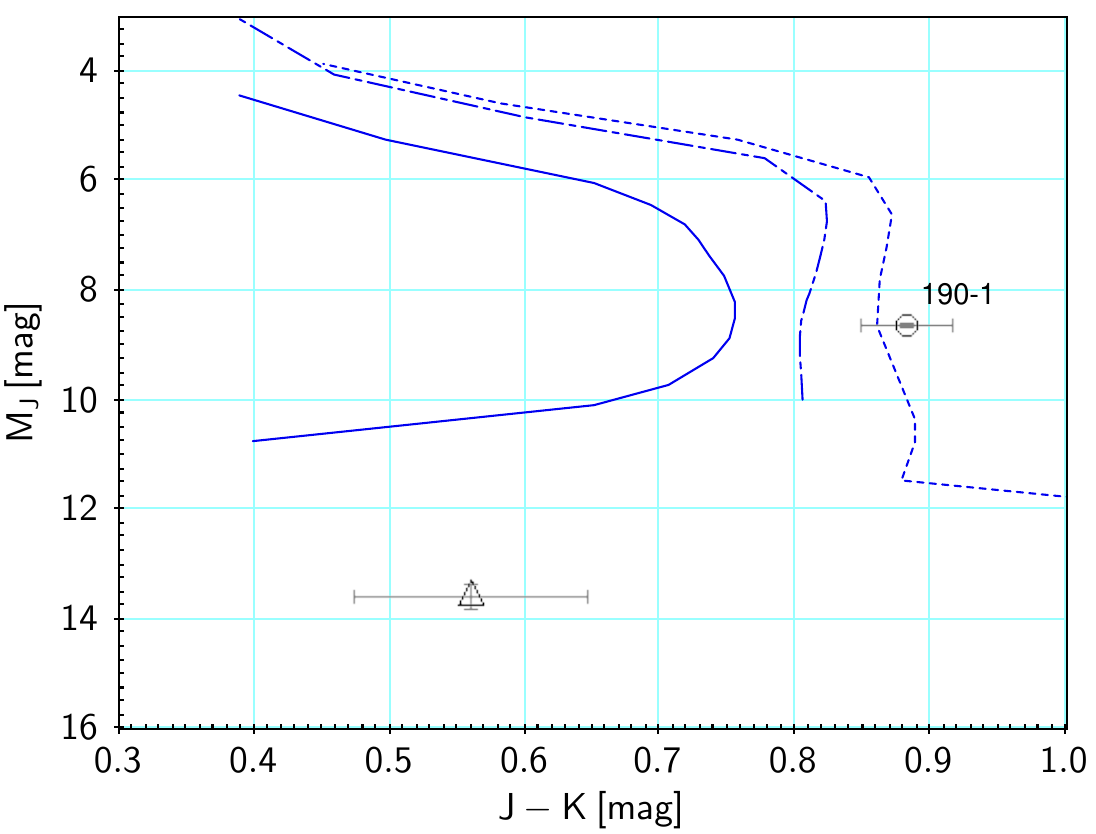}
    \end{subfigure} 
  \vspace{2mm}
  
     \begin{subfigure}{.5\textwidth}
    \includegraphics[width=0.91\linewidth]{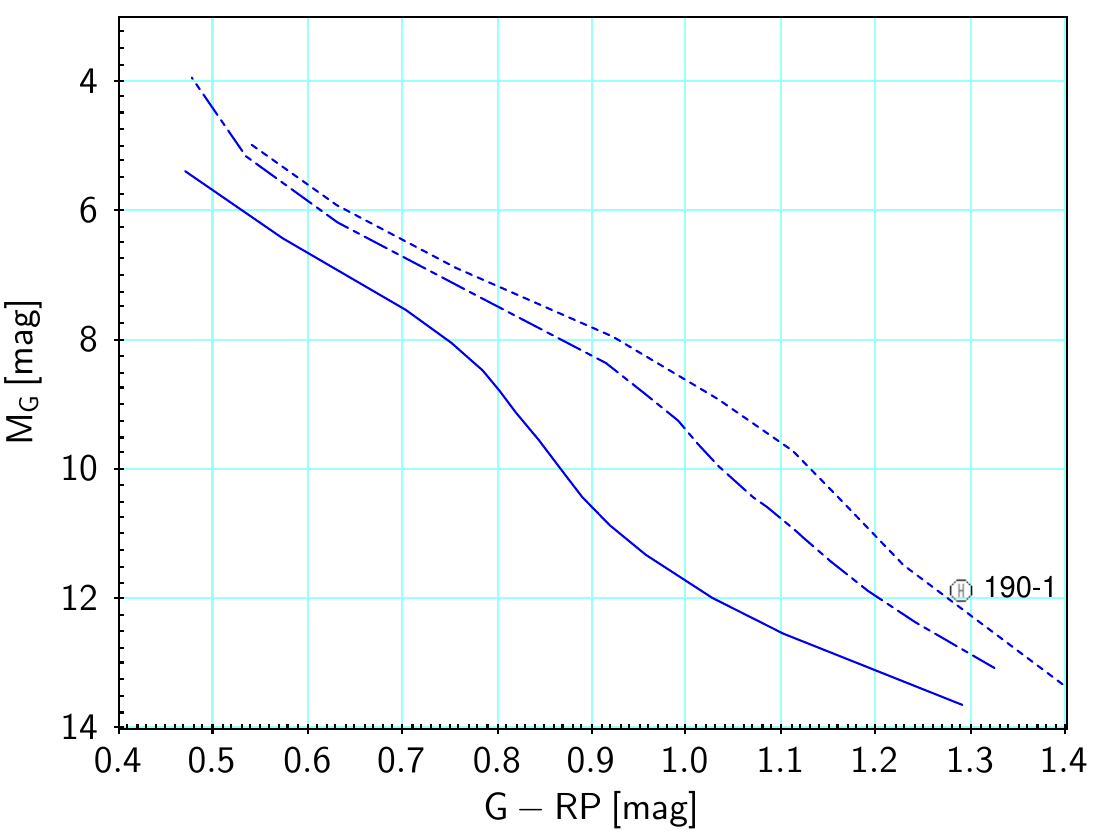}
    \end{subfigure}
    \begin{subfigure}{.5\textwidth}
      \includegraphics[width=0.91\linewidth]{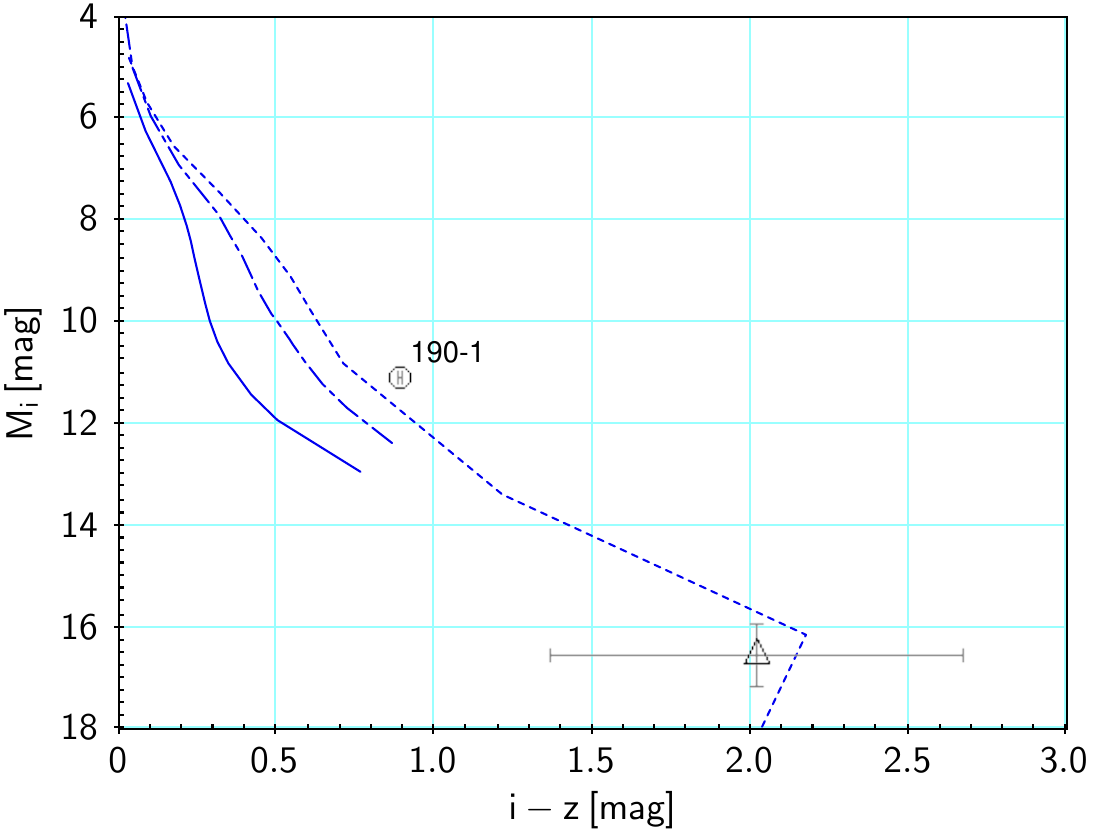}
    \end{subfigure}
    
  \vspace{2mm}  
     \begin{subfigure}{.5\textwidth}
    \includegraphics[width=0.91\linewidth]{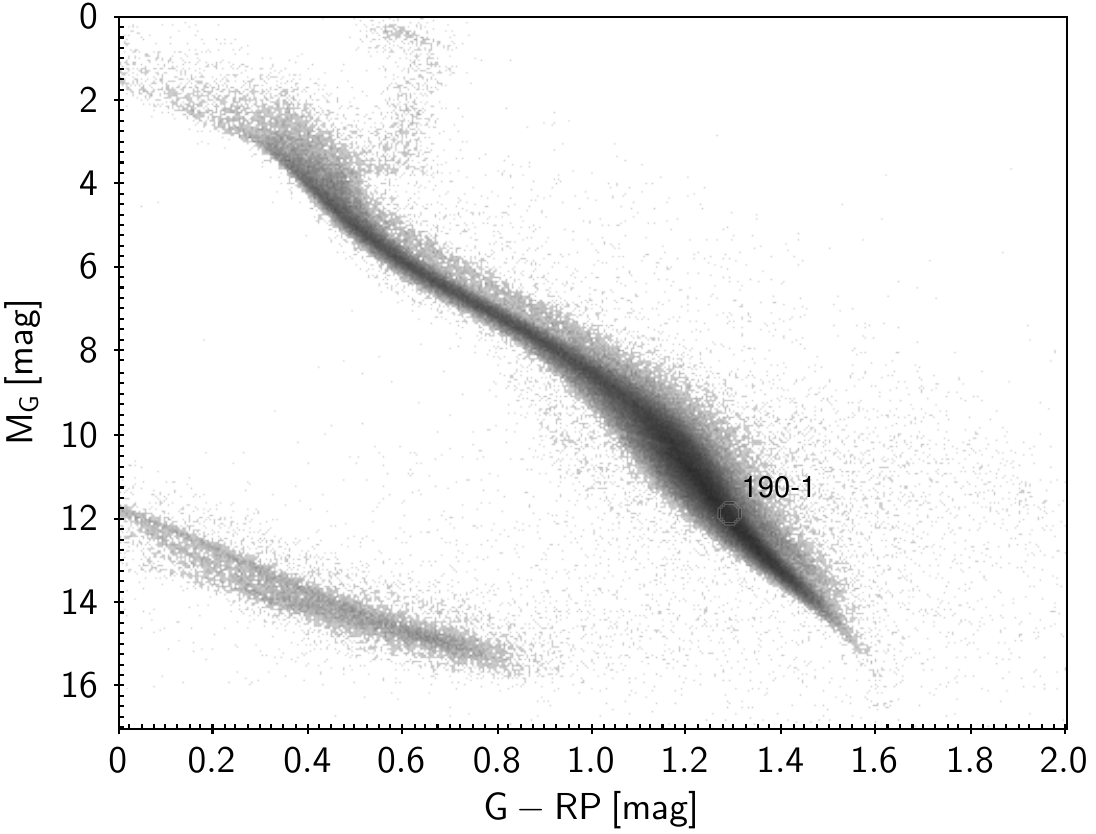}
    \end{subfigure}
    \begin{subfigure}{.5\textwidth}
      \includegraphics[width=0.91\linewidth]{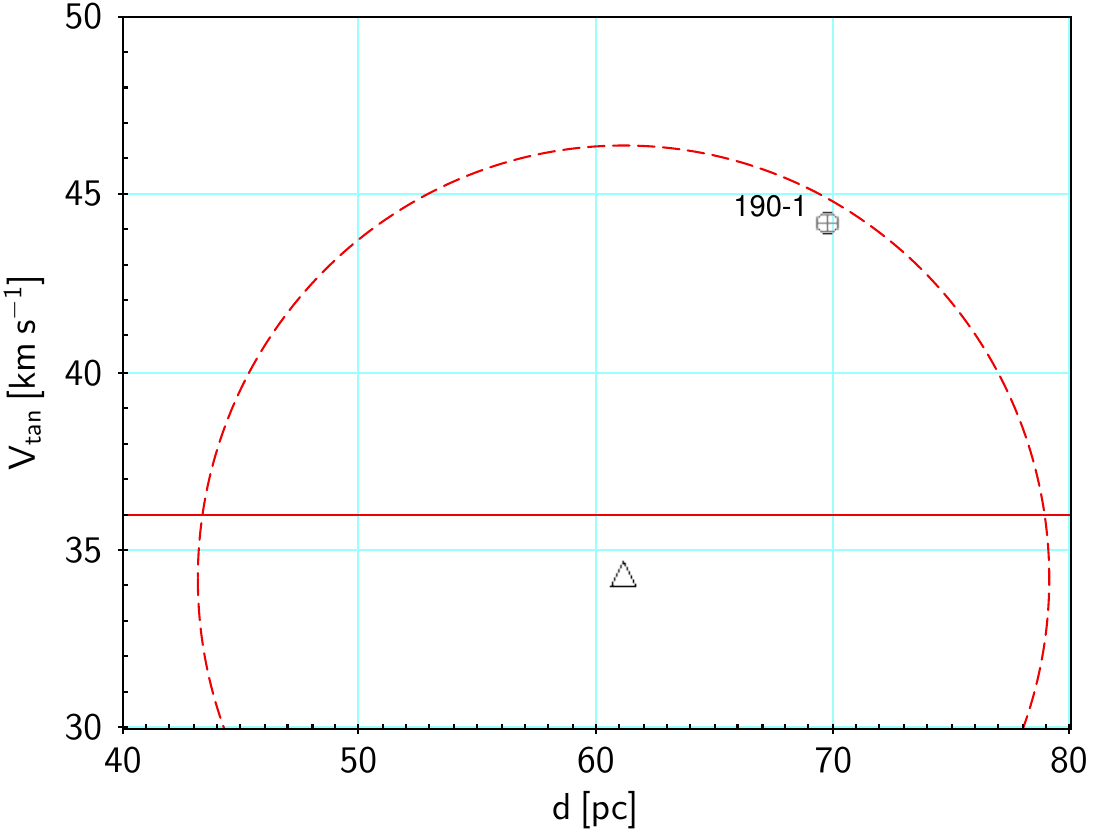}
    \end{subfigure}
  \vspace{5mm}
  \caption[PMD, CMDs, H-R diagram and tangential velocity vs distance diagram for the target Id 190 and its candidate companions.]{PMD (top left panel), CMDs (top right and mid panels), H-R diagram (bottom left panel), and tangential velocity--distance diagram (bottom right) for the target Id 190 and its candidate companion. The black triangle represents the source under study, and the numbered black circle is the companion candidate. Grey dots in the PMD represent field stars, and in the H-R diagram they are \textit{Gaia} DR2 sources with parallaxes $>$\,10\,mas used as a reference. The blue solid, dashed, and dotted lines stand for [M/H]\,=\,$-$2.0, $-$0.5, and =\,0.0 \textit{BT-Settl} isochrones in the CMDs, respectively. The red solid line in the $d/V_{tan}$ plot marks the value $V_{tan}=$\,36\,km\,s$^{-\text{1}}$ which is the mean value for field stars \citep{zhang18a}, and the red dashed ellipse around Id 190 indicates its values of $V_{tan}\pm \text{3}\sigma$ and $d \pm \text{3}\sigma$.}
\label{fig:plot_CMD_PM_190}
\end{figure}

\begin{figure}[H]
  \caption*{\textbf{Id 213}}
    \begin{subfigure}{.5\textwidth}
    \includegraphics[width=0.91\linewidth]{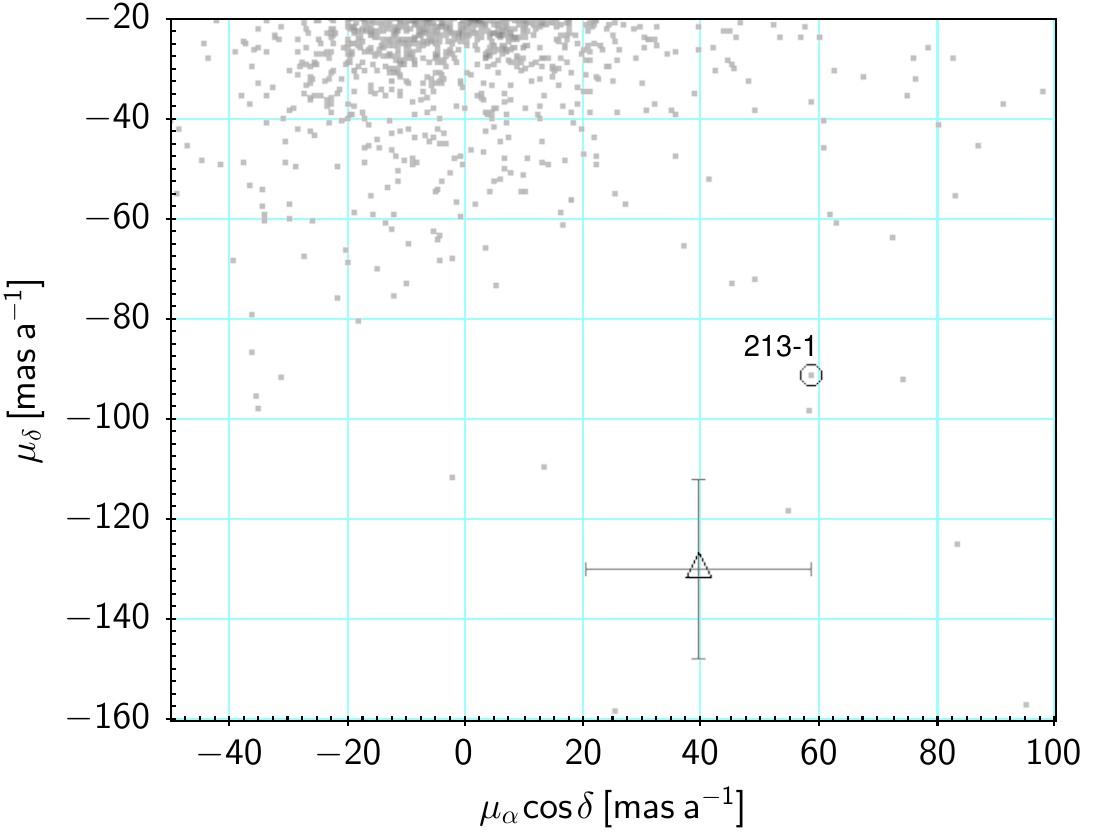}
    \end{subfigure}
    \begin{subfigure}{.5\textwidth}
      \includegraphics[width=0.91\linewidth]{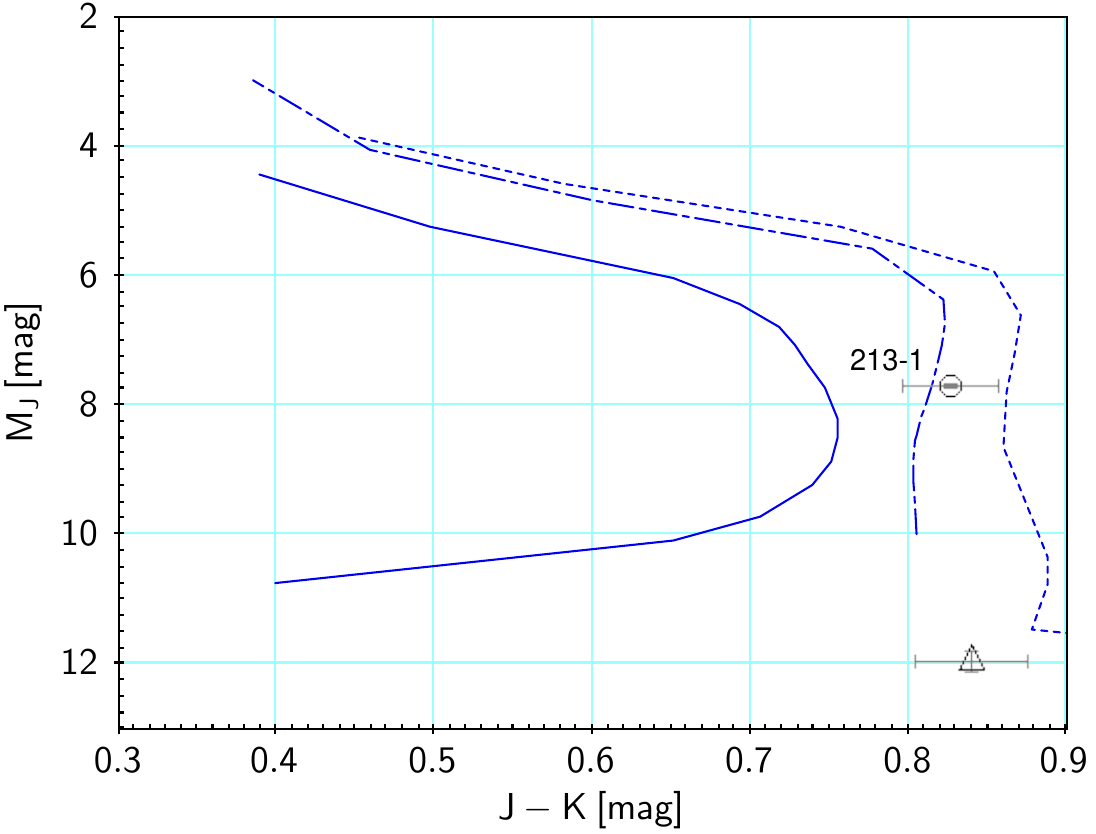}
    \end{subfigure} 
  \vspace{2mm}
  
     \begin{subfigure}{.5\textwidth}
    \includegraphics[width=0.91\linewidth]{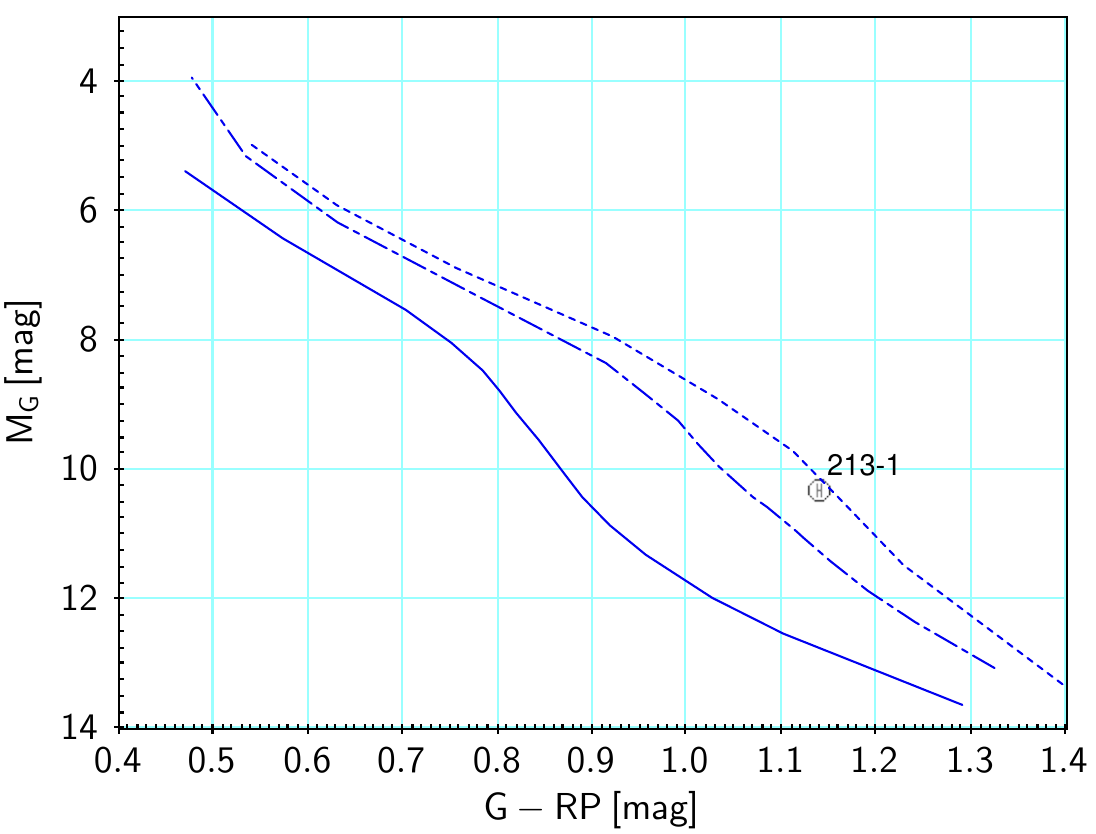}
    \end{subfigure}
    \begin{subfigure}{.5\textwidth}
      \includegraphics[width=0.91\linewidth]{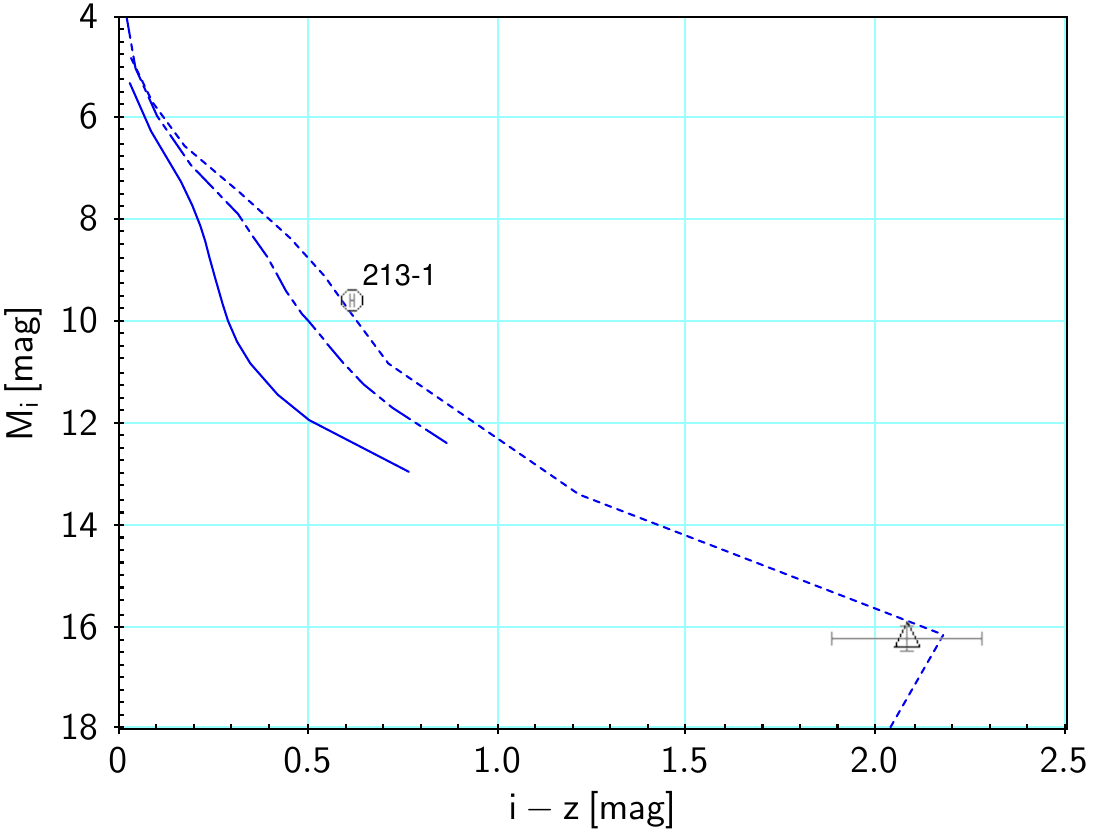}
    \end{subfigure}
    
  \vspace{2mm}  
     \begin{subfigure}{.5\textwidth}
    \includegraphics[width=0.91\linewidth]{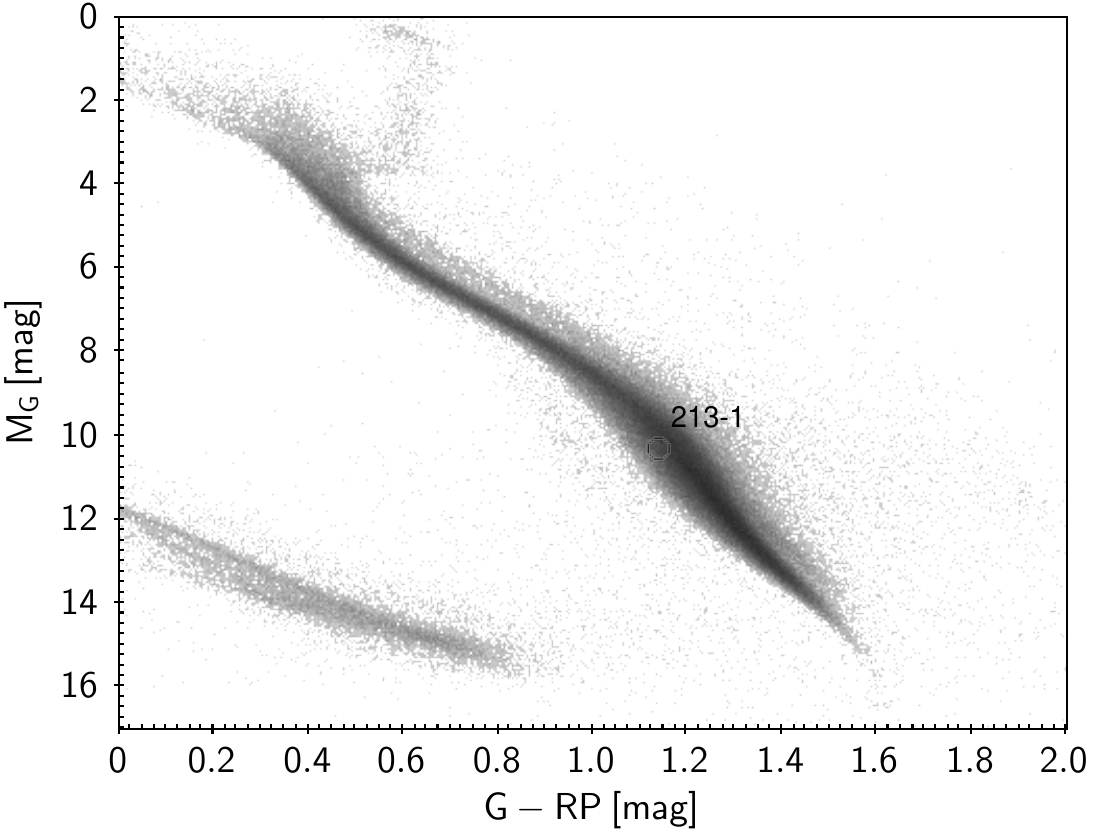}
    \end{subfigure}
    \begin{subfigure}{.5\textwidth}
      \includegraphics[width=0.91\linewidth]{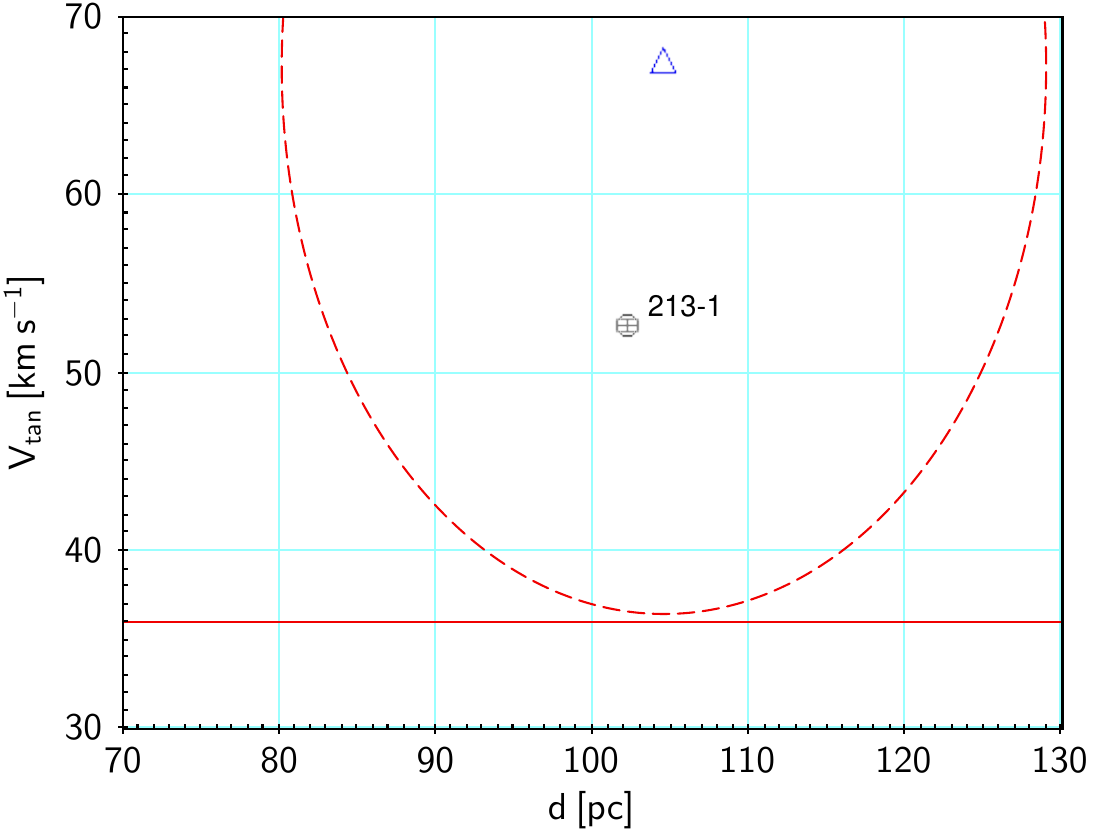}
    \end{subfigure}
  \vspace{5mm}
  \caption[PMD, CMDs, H-R diagram and tangential velocity vs distance diagram for the target Id 213 and its candidate companions.]{PMD (top left panel), CMDs (top right and mid panels), H-R diagram (bottom left panel), and tangential velocity--distance diagram (bottom right) for the target Id 213 and its candidate companion. The black triangle represents the source under study, and the numbered black circle is the companion candidate. Grey dots in the PMD represent field stars, and in the H-R diagram they are \textit{Gaia} DR2 sources with parallaxes $>$\,10\,mas used as a reference. The blue solid, dashed, and dotted lines stand for [M/H]\,=\,$-$2.0, $-$0.5, and =\,0.0 \textit{BT-Settl} isochrones in the CMDs, respectively. The red solid line in the $d/V_{tan}$ plot marks the value $V_{tan}=$\,36\,km\,s$^{-\text{1}}$ which is the mean value for field stars \citep{zhang18a}, and the red dashed ellipse around Id 213 indicates its values of $V_{tan}\pm \text{3}\sigma$ and $d \pm \text{3}\sigma$.}
\label{fig:plot_CMD_PM_213}
\end{figure}

\begin{figure}[H]
  \caption*{\textbf{Id 215}}
    \begin{subfigure}{.5\textwidth}
    \includegraphics[width=0.91\linewidth]{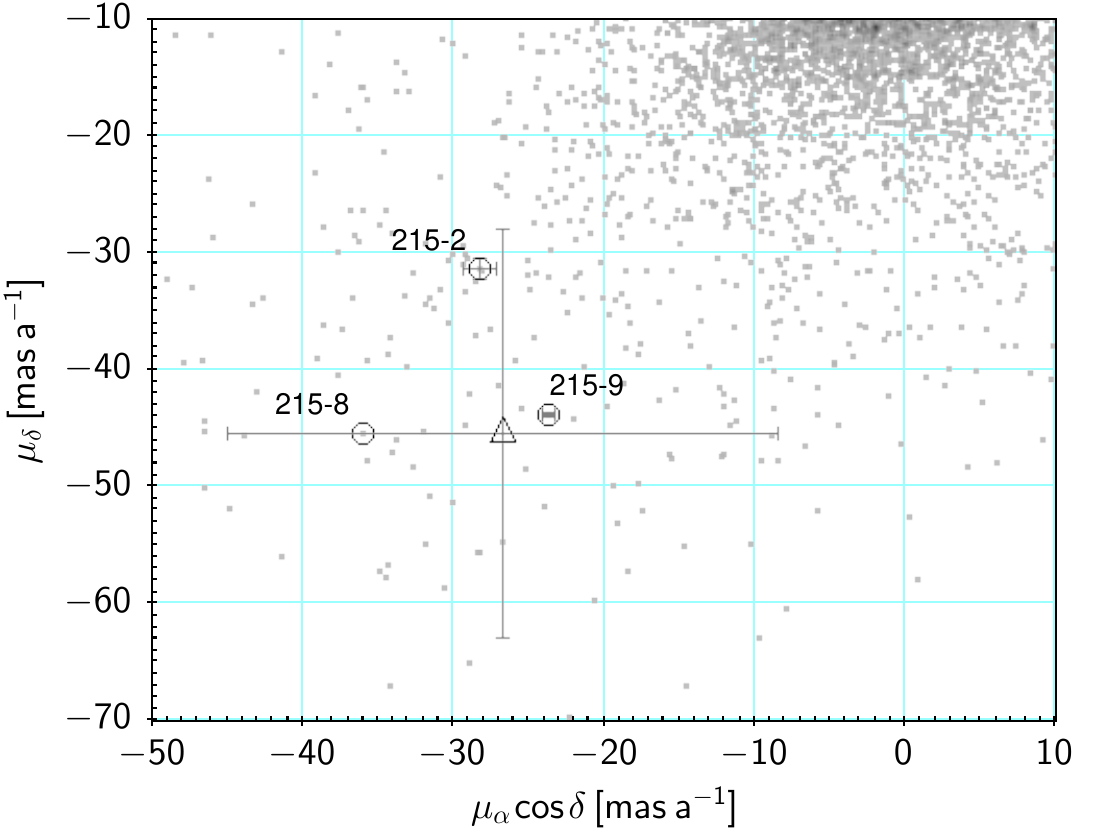}
    \end{subfigure}
    \begin{subfigure}{.5\textwidth}
      \includegraphics[width=0.91\linewidth]{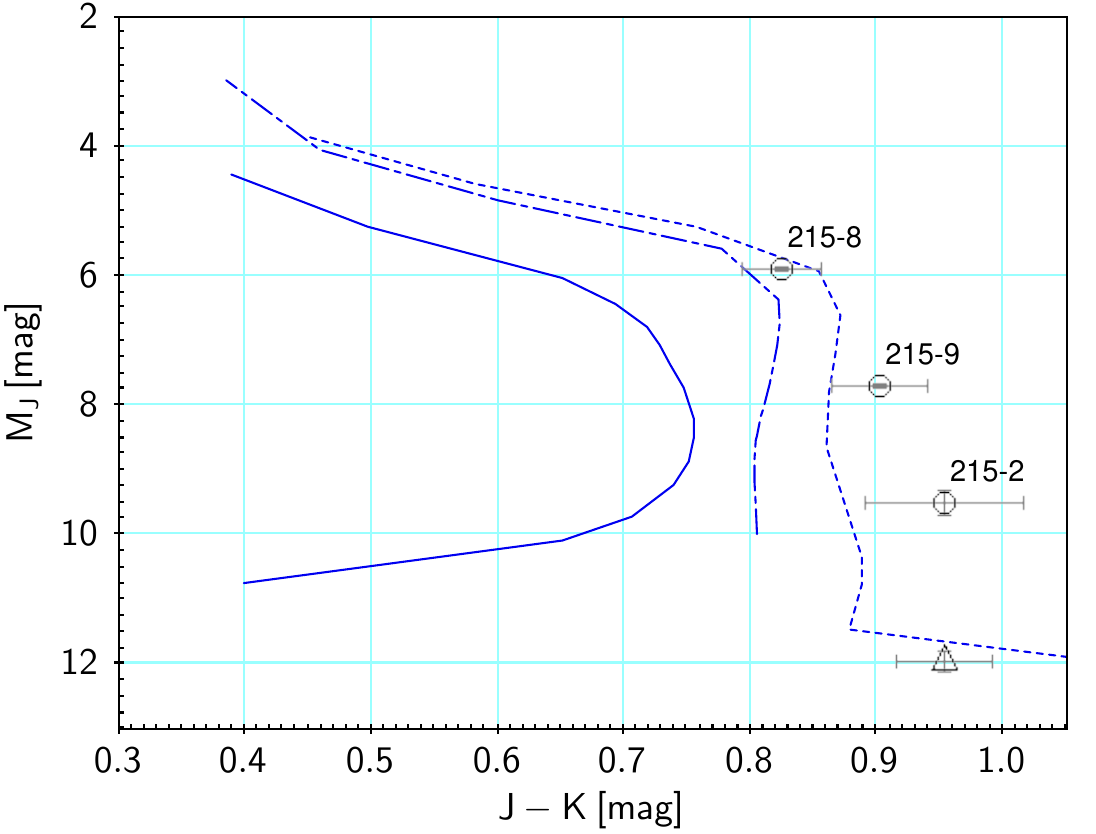}
    \end{subfigure} 
  \vspace{2mm}
  
     \begin{subfigure}{.5\textwidth}
    \includegraphics[width=0.91\linewidth]{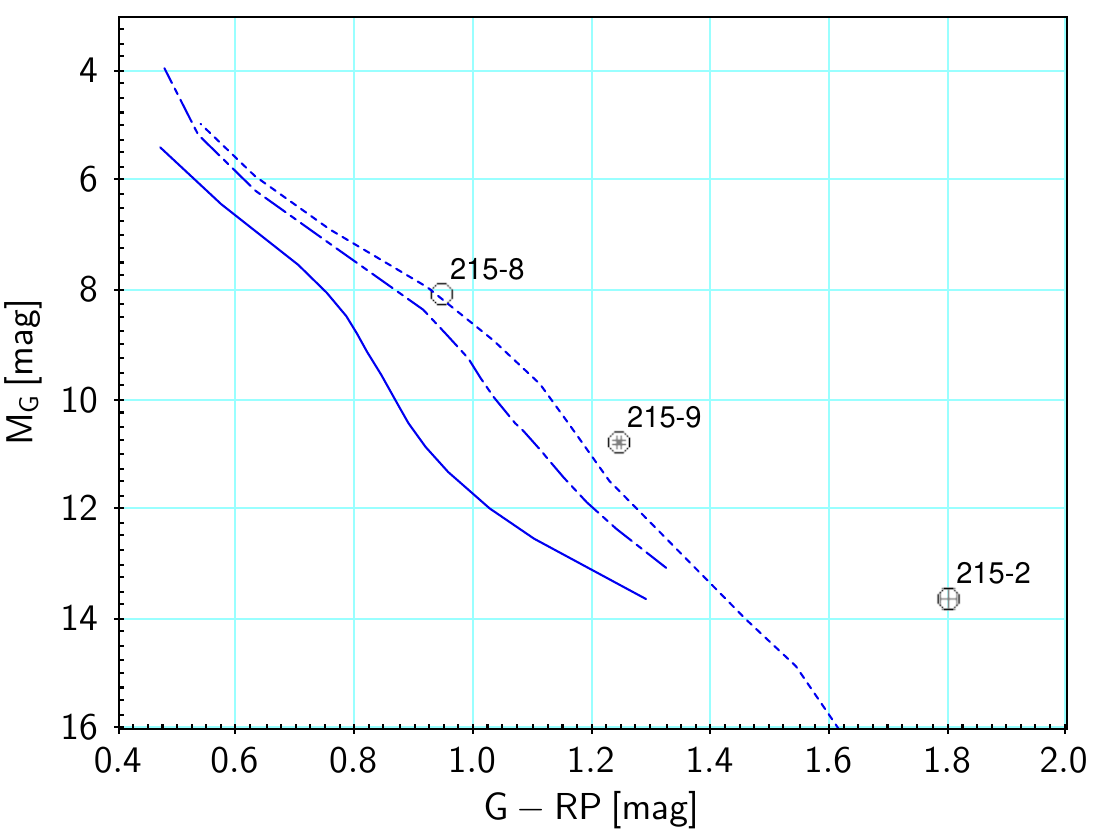}
    \end{subfigure}
    \begin{subfigure}{.5\textwidth}
      \includegraphics[width=0.91\linewidth]{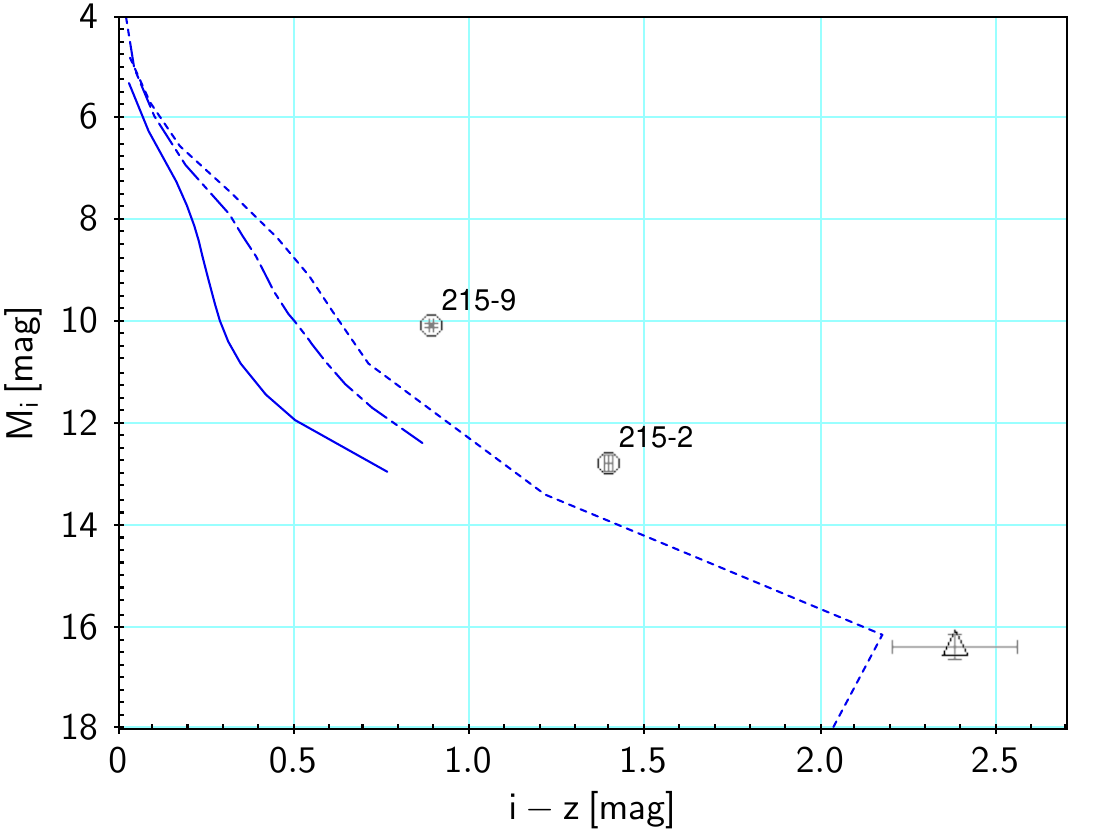}
    \end{subfigure}
    
  \vspace{2mm}  
     \begin{subfigure}{.5\textwidth}
    \includegraphics[width=0.91\linewidth]{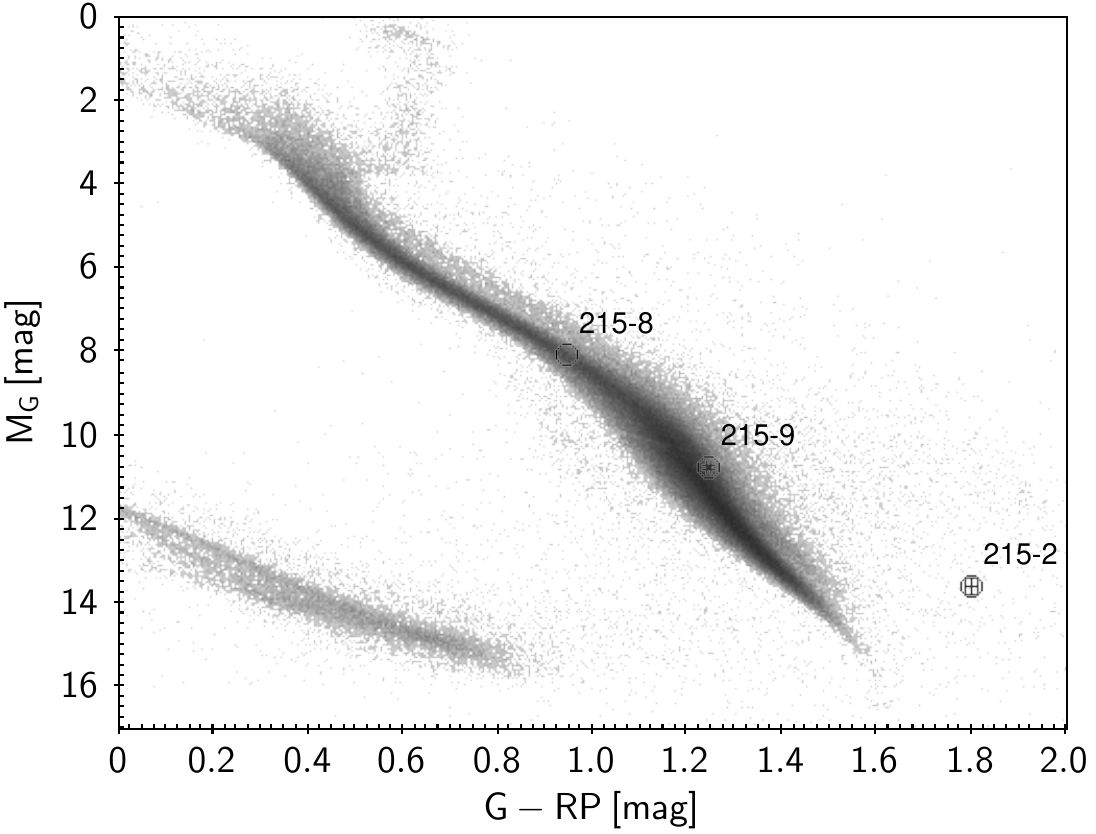}
    \end{subfigure}
    \begin{subfigure}{.5\textwidth}
      \includegraphics[width=0.91\linewidth]{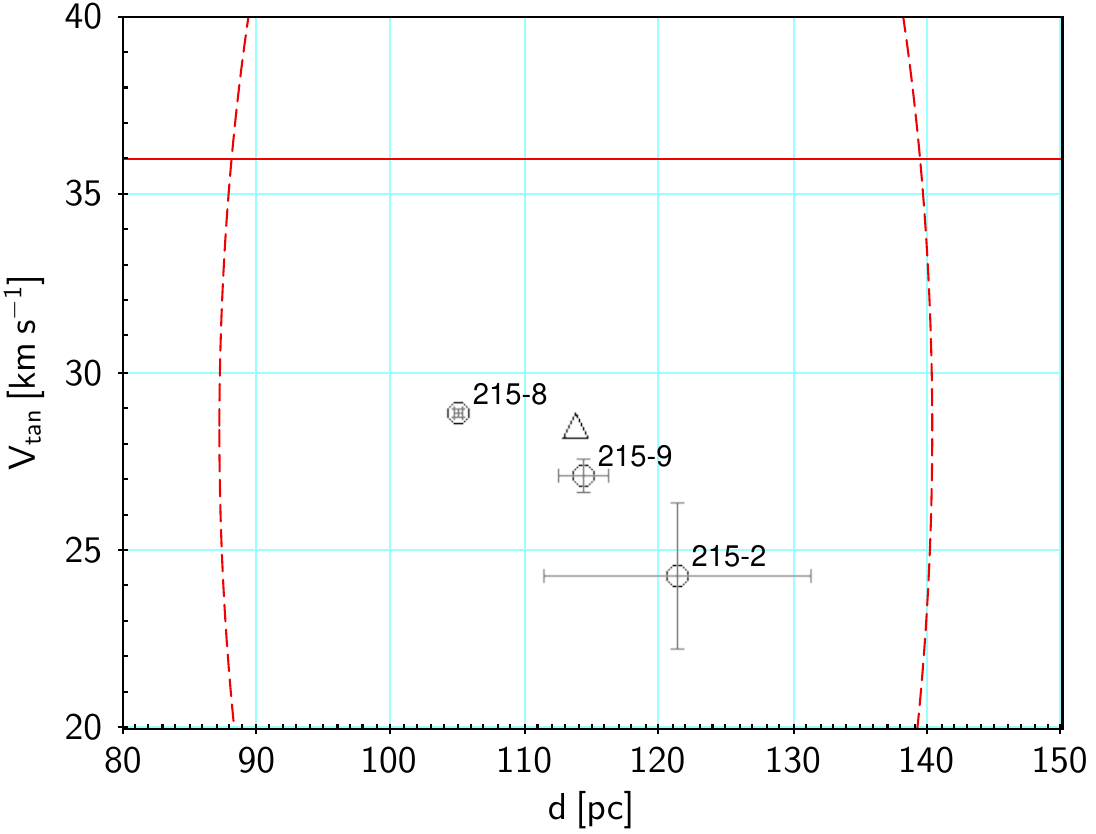}
    \end{subfigure}
  \vspace{5mm}
  \caption[PMD, CMDs, H-R diagram and tangential velocity vs distance diagram for the target Id 215 and its candidate companions.]{PMD (top left panel), CMDs (top right and mid panels), H-R diagram (bottom left panel), and tangential velocity--distance diagram (bottom right) for the target Id 215 and its candidate companion. The black triangle represents the source under study, and the numbered black circle is the companion candidate. Grey dots in the PMD represent field stars, and in the H-R diagram they are \textit{Gaia} DR2 sources with parallaxes $>$\,10\,mas used as a reference. The blue solid, dashed, and dotted lines stand for [M/H]\,=\,$-$2.0, $-$0.5, and =\,0.0 \textit{BT-Settl} isochrones in the CMDs, respectively. The red solid line in the $d/V_{tan}$ plot marks the value $V_{tan}=$\,36\,km\,s$^{-\text{1}}$ which is the mean value for field stars \citep{zhang18a}, and the red dashed ellipse around Id 215 indicates its values of $V_{tan}\pm \text{3}\sigma$ and $d \pm \text{3}\sigma$.}
\label{fig:plot_CMD_PM_215}
\end{figure}

%%%%% APENDICE D %%%%%

\begin{center}
\chapter[Appendix: Properties of the widest systems]{Properties of the widest systems}
\end{center}
\label{ch:Appendix_D}
\vspace{3cm}
\pagestyle{fancy}
\fancyhf{}
\lhead[\small{\textbf{\thepage}}]{\textbf{Appendix~D: Properties of the widest systems}}
\rhead[\small{\textbf{Appendix~D: Properties of the widest systems}}]{\small{\textbf{\thepage}}}
\renewcommand{\thetable}{D.\arabic{table}}   % Cambia el formato de numeración al tipo D.x

\footnotesize
\begin{longtable}{@{\hspace{2mm}}l@{\hspace{4mm}}l@{\hspace{2mm}}c@{\hspace{2mm}}c@{\hspace{2mm}}c@{\hspace{2mm}}c@{\hspace{2mm}}c@{\hspace{2mm}}}
\label{tab:stars_young_skg}\\
\caption[Stars in young stellar kinematic group.]{Stars in young stellar kinematic groups (available at the CDS via \url{http://cdsarc.u-strasbg.fr/viz-bin/cat/J/A+A/670/A102}).}\\
\noalign{\smallskip}
\noalign{\smallskip}
\noalign{\hrule height 01pt}
\noalign{\smallskip}
Star & Discoverer & $\alpha$\,(J2000) & $\delta$\,(J2000) & $G$ & Group$^{\text{(b)}}$ & Refs.$^{\text{(c)}}$ \\
 & code$^{\text{(a)}}$ & (hh:mm:ss.ss) & (dd:mm:ss.s) & (mag) & & \\
\noalign{\smallskip}
\hline
\noalign{\smallskip}
\endfirsthead
\caption[]{Stars in young stellar kinematic groups 
\textit{(continued)}.}\\

\noalign{\hrule height 1pt}
\noalign{\smallskip} 

Star & Discoverer & $\alpha$\,(J2000) & $\delta$\,(J2000) & $G$ & Group$^{\text{(b)}}$ & Refs.$^{\text{(c)}}$ \\
 & code$^{\text{(a)}}$ & (hh:mm:ss.ss) & (dd:mm:ss.s) & (mag) & & \\
\noalign{\smallskip}
\hline
\noalign{\smallskip}
\endhead
\noalign{\smallskip}
\hline
\endfoot 
DENIS J000657.9--643654 & ... & 00:06:57.93 & $-$64:36:54.2 & 18.2 & Tuc-Hor & 1, 2 \\
UPM J0014--6003 & ... & 00:14:47.68 & $-$60:03:47.8 & 12.7 & Tuc-Hor & 1, 3 \\
UCAC3 52--533 & ... & 00:15:27.52 & $-$64:14:54.8 & 11.9 & Tuc-Hor & 1, 3 \\
CD--60 31 & ... & 00:17:30.50 & $-$59:57:04.4 & 10.6 & Tuc-Hor & 3 \\
HD 1466 & SHY 113 G, & 00:18:26.12 & $-$63:28:39.0 & 7.3 & Tuc-Hor & 1, 4 \\
 & SHY 114 G, & & & & & \\
 & CVN 33 G & & & & & \\
2MASS J00182834--6703130 & ... & 00:18:28.33 & $-$67:03:13.0 & 20.6 & Tuc-Hor & 2 \\
\end{longtable}
\vspace{-5mm}
\begin{justify}
    \scriptsize{\textit{\textbf{Notes.}}}
    \scriptsize{$^{\text{(a)}}$ Marked with ``...'' are new young star candidates.}
    \scriptsize{$^{\text{(b)}}$ 32 Ori: 32 Orionis; AB Dor: AB Doradus; Ale13: Alessi 13; $\beta$ Pic: $\beta$ Pictoris; Car: Carina; Car-Near: Carina-Near; Car-Vela: Carina-Vela; $\gamma$ Cas: $\gamma$ Cassiopeia; Castor: Castor ($\alpha$ Geminorum); CBer: Coma Berenice; $\chi^{\text{01}}$ For: $\chi^{\text{01}}$ Fornacis; Col: Columba; $\epsilon$ Cha: $\epsilon$ Chamaelontis; G-X: [TPY2019] Group-X; Hya: Hyades; IC2391: IC2391 Supercluster; LCC: Lower Centarus Crux; SCO: Scorpio- Centaurus; Tuc-Hor: Tucana-Horologium; UCL: Upper Centaurus-Lupus; UMa: Ursa Major; VCA: Volans-Carina.}
    \scriptsize{$^{\text{(c)}}$ 1. \citet{riedel17}; 2. \citet{gagne15}; 3. \citet{kraus14}; 4. \citet{gagne18a}; 5. This work; 6. \citet{gagne18c}; 7. \citet{gagne18b}; 8. \citet{freund20}; 9. \citet{kopytova16}; 10. \citet{roser11}; 11. \citet{cantat-gaudin18}; 12. \citet{reino18}; 13. \citet{reid93}; 14. \citet{bouvier08}; 15. \citet{guenther05}; 16. \citet{bell17}; 17. \citet{goldman18}; 18. \citet{murphy13}; 19. \citet{hoogerwerf00}; 20. \citet{dezeeuw99}; 21. \citet{rizzuto11}; 22. \citet{pecaut16}; 23. \citet{stauffer21}; 24. \citet{bohn22}; 25. \citet{dopcke19}; 26. \citet{dzib18}; 27. \citet{furnkranz19}; 28. \citet{tang19}.
    $^\text{(d)}$ Not available in \textit{Gaia}.}
\end{justify}

\newpage

\footnotesize
\begin{longtable}{@{\hspace{2mm}}l@{\hspace{3mm}}l@{\hspace{2mm}}l@{\hspace{2mm}}c@{\hspace{3mm}}c@{\hspace{3mm}}c@{\hspace{3mm}}c@{\hspace{2mm}}}
\caption[Basic data of the 94 galactic field systems]{Basic data of the 94 galactic field systems (available at the CDS via \url{http://cdsarc.u-strasbg.fr/viz-bin/cat/J/A+A/670/A102}).}\\
\noalign{\smallskip}
\noalign{\hrule height 01pt}
\noalign{\smallskip}
WDS & Discoverer & Star & $\alpha$ (J2000) & $\delta$ (J2000) & $d$ & $G^{\text{(a)}}$ \\
 & code & & (hh:mm:ss.ss) & (dd:mm:ss.s) & (pc) & (mag) \\
\noalign{\smallskip}
\hline
\noalign{\smallskip}
\endfirsthead
\caption[]{Basic data of the 94 galactic field systems \textit{(continued)}.}\\
\noalign{\hrule height 01pt}
\noalign{\smallskip} 
\label{tab:galactic_field_system}
WDS & Discoverer & Star & $\alpha$ (J2000) & $\delta$ (J2000) & $d$ & $G^{\text{(a)}}$ \\
 & code & & (hh:mm:ss.ss) & (dd:mm:ss.s) & (pc) & (mag) \\
\noalign{\smallskip}
\endhead
\hline
\noalign{\smallskip}
\endfoot 
\multicolumn{7}{c}{\it Double systems}\\
 \noalign{\smallskip} 
 \hline
 \noalign{\smallskip} 
00016--0102 &  & 2MASS J00013688--0101441 & 00:01:36.89 & $-$01:01:44.2 & 60.33$\pm$0.13 & 15.2 \\
 & WIS   1 & SIPS J0000--0112 & 00:00:35.39 & $-$01:12:46.3 & 65.41$\pm$0.64 & 17.3\\
 \noalign{\smallskip}
 \hline
 \noalign{\smallskip}  
01005--1923 &  & HD 5911 & 01:00:28.01 & $-$19:23:21.8 & 81.63$\pm$0.17 & 7.9 \\
 & SHY 392 & HD 6103 & 01:02:06.90 & $-$19:40:10.8 & 81.59$\pm$0.16 & 8.1 \\
 \noalign{\smallskip}
 \hline
 \noalign{\smallskip}  
01066+1353 &  & HD 6566 & 01:06:36.52 & +13:53:04.6 & 58.08$\pm$0.10 & 7.1 \\
 & SHY 396 & HD 5433$^{\text{(b)}}$ & 00:56:13.48 & +15:39:22.2 & 63.44$\pm$0.43 & 8.5 \\
 \noalign{\smallskip}
 \hline
 \noalign{\smallskip}  
01134--3932 &  & HD 7382 & 01:13:26.52 & $-$39:32:24.3 & 59.18$\pm$0.11 & 7.2 \\
\end{longtable} 
\vspace{-7.5mm}
\begin{justify}
    \scriptsize{\textbf{\textit{Notes. }}}
    \scriptsize{$^{\text{(a)}}$ The maximum error in $G$ provided by \textit{Gaia} DR3 is less than 0.005\,mag;}
    \scriptsize{$^{\text{(b)}}$ Proper motion anomaly measured by \citet{kervella19}, \citet{brandt21} or both;}
    \scriptsize{$^{\text{(c)}}$ \texttt{RUWE} $> 10$;}
    \scriptsize{$^{\text{(d)}}$ Large $\sigma_{Vr}$ for its $G$ magnitude;}
    \scriptsize{$^{\text{(e)}}$ HD~35066[Aab] was recently resolved at SOAR with $\rho$\,=\,0.14\,arcsec (anonymous reviewer, priv.comm);}
    \scriptsize{$^{\text{(f)}}$ Value from \textit{Gaia} DR2;}
    \scriptsize{$^{\text{(g)}}$ Very shiny star, not resolved by \textit{Gaia} DR3. Astrometric data from \textit{Hipparcos};}
    \scriptsize{$^{\text{(h)}}$ \citet{pourbaix16};}
    \scriptsize{$^{\text{(i)}}$ \citet{pecaut13}.}
\end{justify}

\newpage

\footnotesize
\begin{longtable}{@{\hspace{1mm}}l@{\hspace{2mm}}l@{\hspace{2mm}}l@{\hspace{2mm}}c@{\hspace{1mm}}c@{\hspace{2mm}}c@{\hspace{0mm}}c@{\hspace{0mm}}c@{\hspace{1mm}}}
\caption[Masses, $\rho$, $\theta$, separations and gravitational binding energy of our systems.]{Masses, $\rho$, $\theta$, separations and gravitational binding energy of the systems identified in the sample of work (available at the CDS via \url{http://cdsarc.u-strasbg.fr/viz-bin/cat/J/A+A/670/A102}).}\\
\noalign{\smallskip}
\noalign{\hrule height 01pt} 
\noalign{\smallskip}
WDS & Discoverer & Star & Mass & $\rho$ & $\theta$ & \textit{s} & $|U^*_g|$ \\
 & code &  & (M$_{\odot}$) & (arcsec) & (deg) & (10$^\text{3}$\,au) & (10$^{\text{33}}$\,J) \\
\noalign{\smallskip}
\hline
\noalign{\smallskip}
\endfirsthead
\caption[]{Masses, $\rho$, $\theta$, separations, and gravitational binding energy of the systems identified in the sample of work \textit{(continued)}.}\\
\noalign{\hrule height 01pt} 
\noalign{\smallskip} 
\label{tab:masses_rho_theta_u_systems}
WDS & Discoverer & Star & Mass & $\rho$ & $\theta$ & \textit{s} & $|U^*_g|$ \\
 & code &  & (M$_{\odot}$) & (arcsec) & (deg) & (10$^\text{3}$\,au) & (10$^{\text{33}}$\,J) \\
\noalign{\smallskip}
\endhead
\hline
\noalign{\smallskip}
\endfoot 
\multicolumn{8}{c}{\textit{Double systems}}\\
\noalign{\smallskip}
 \hline
 \noalign{\smallskip}   
00016--0102 &  & 2MASS J00013688--0101441 & 0.22$\pm$0.02 &  &  &  & 0.74$\pm$0.10 \\
 & WIS   1 & SIPS J0000--0112 & 0.13$\pm$0.01 & 1135 & 234 & 68.4$\pm$0.2 &  \\
 \noalign{\smallskip} 
 \hline
 \noalign{\smallskip}  
01005--1923 &  & HD 5911 & 1.32$\pm$0.13 &  &  &  & 20.6$\pm$2.9 \\
 & SHY 392 & HD 6103 & 1.24$\pm$0.12 & 1725 & 126 & 140.8$\pm$0.3 &  \\
  \noalign{\smallskip}
  \hline
 \noalign{\smallskip}  
01066+1353 &  & HD 6566 & 1.32$\pm$0.10 &  &  &  & 3.69$\pm$0.52 \\
 & SHY 396 & HD 5433$^{\text{(a)}}$ & 1.02$\pm$0.10 & 11089 & 305 & 643.7$\pm$1.2 &  \\
  \noalign{\smallskip}
  \hline
 \noalign{\smallskip}  
01134--3932 &  & HD 7382 & 1.32$\pm$0.13 &  &  &  & 7.82$\pm$1.11 \\
\end{longtable}
\vspace{-7mm}
\begin{justify}
    \scriptsize{\textbf{\textit{Notes.}}
    $^{\text{(a)}}$ Proper motion anomaly measured by \citet{kervella19}, \citet{brandt21} or both;
    $^{\text{(b)}}$\texttt{RUWE} $> 10$;
    $^{\text{(c)}}$ Large $\sigma_{Vr}$ for its $G$ magnitude;
    $^{\text{(d)}}$ Classified as a white dwarf candidate by \citet{gentilefusillo19}, it has instead absolute magnitudes and colours of an intermediate M dwarf;
    $^{\text{(e)}}$ \citet{dasilva15};
    $^{\text{(f)}}$ \citet{tokovinin14b};
    $^{\text{(g)}}$ Dynamical mass \citep{malkov12};
    $^{\text{(h)}}$ \citet{gontcharov10};
    $^{\text{(i)}}$ \citet{mitrofanova21};
    $^{\text{(j)}}$ \citet{pourbaix16};
    $^{\text{(k)}}$ \citet{kervella19};
    $^{\text{(l)}}$ \citet{eker18};
    $^{\text{(m)}}$ \citet{feuillet16};
    $^{\text{(n)}}$ \citet{tokovinin18};
    $^{\text{(o)}}$ \citet{montes18a}.}
\end{justify}

\newpage

\footnotesize
\begin{longtable}{@{\hspace{3mm}}l@{\hspace{5mm}}c@{\hspace{5mm}}c@{\hspace{4mm}}c@{\hspace{4mm}}c@{\hspace{5mm}}c@{\hspace{5mm}}c@{\hspace{5mm}}c@{\hspace{3mm}}}
\caption[Candidate stars to be part of the $\gamma$ Cas association.]{Candidate stars to be part of the $\gamma$ Cas association, in order of the distance to the central star $\gamma$ Cas (available at the CDS via \url{http://cdsarc.u-strasbg.fr/viz-bin/cat/J/A+A/670/A102}).}\\
\noalign{\smallskip}
\noalign{\hrule height 01pt}
\noalign{\smallskip}
\label{tab:gamCas_candidates}
Star & $\alpha$\,(J2000) & $\delta$\,(J2000) & SpT & Mass & $G^{\text{(a)}}$ & $J^{\text{(b)}}$ & $\rho_{\gamma\,\text{Cas}}$  \\
 & (hh:mm:ss.ss) & (dd:mm:ss.s) & & (M$_{\odot}$) & (mag) & (mag) & (deg) \\
\noalign{\smallskip}
\hline
\noalign{\smallskip}
\endfirsthead
\caption[]{Candidate stars to be part of the $\gamma$ Cas association, in order of the distance to the central star $\gamma$ Cas \textit{(continued).}}\\
\noalign{\hrule height 01pt}
\noalign{\smallskip} 

Star & $\alpha$\,(J2000) & $\delta$\,(J2000) & SpT & Mass & $G^{\text{(a)}}$ & $J^{\text{(b)}}$ & $\rho_{\gamma\,\text{Cas}}$  \\
 & (hh:mm:ss.ss) & (dd:mm:ss.s) & & (M$_{\odot}$) & (mag) & (mag) & (deg) \\
\noalign{\smallskip}
\hline
\noalign{\smallskip}
\endhead
\noalign{\smallskip}
\hline
\noalign{\smallskip}
\endfoot 
$\gamma$ Cas & 00:56:42.53 & +60:43:00.3 & B0.5IV & $\sim$\,13$^{\text{(c)}}$ & 2.3 & 2.0 & 0.000  \\
Gaia DR3 426558563962119808 & 00:56:26.00 & +60:41:55.5 &  & 0.80 $\pm$ 0.08 & 12.3 & 10.8 & 0.038  \\
Gaia DR3 426559693524626816 & 00:55:55.21 & +60:45:44.5 &  & 0.16 $\pm$ 0.02 & 18.7 & 15.0 & 0.107  \\
UCAC4 752--011208 & 00:56:44.80 & +60:22:46.3 & M4Ve & 0.46 $\pm$ 0.05 & 15.4 & 12.4 & 0.337  \\
HD 5408[Aa] & \multirow{3}{*}{\bigg\} 00:56:46.97} & & B7V & 3.40 $\pm$ 0.10$^{\text{(d)}}$ & \multirow{3}{*}{\bigg\} 6.0} & &  \\
HD 5408[Ab] &  & +60:21:46.2 & B9V & 4.10 $\pm$ 0.10$^{\text{(d)}}$ & & 5.6 & 0.354  \\
HD 5408[B] & & & A1V & 3.40 $\pm$ 0.10$^{\text{(d)}}$ & & &  \\
Gaia DR3 426494169508443520 & 00:59:29.59 & +60:28:43.0 &  & 0.14 $\pm$ 0.01 & 19.4 & 15.4 & 0.417  \\
Gaia DR3 426656106950424960 & 00:58:49.75 & +61:07:34.5 &  & 0.19 $\pm$ 0.02 & 18.3 & 14.7 & 0.484  \\
Gaia DR3 426907040426100352 & 00:52:12.81 & +60:28:25.5 &  & 0.45 $\pm$ 0.05 & 15.3 & 12.9 & 0.603  \\
Gaia DR3 426117827290256896 & 00:53:27.10 & +60:01:58.1 &  & 0.75 $\pm$ 0.08 & 12.7 & 11.1 & 0.794  \\
Gaia DR3 426696758828470144 & 00:59:12.34 & +61:38:10.3 &  & 0.28 $\pm$ 0.03 & 17.3 & 14.2 & 0.967  \\
Gaia DR3 426391055933615872 & 01:05:01.74 & +60:30:11.1 &  & 0.11 $\pm$ 0.01 & 20.4 & 16.3 & 1.04  \\
Gaia DR3 427444254930463744 & 00:52:19.64 & +61:37:04.8 &  & 0.24 $\pm$ 0.02 & 17.4 & 13.9 & 1.04  \\
Gaia DR3 426937483155345280 & 00:48:27.00 & +60:20:51.1 &  & 0.43 $\pm$ 0.04 & 15.7 & 12.9 & 1.08  \\
Gaia DR3 427465454889829248 & 00:55:51.79 & +61:52:51.7 &  & 0.35 $\pm$ 0.03 & 16.5 & 13.4 & 1.17  \\
Gaia DR3 426833373147610240 & 00:48:42.59 & +60:03:34.6 &  & 0.39 $\pm$ 0.04 & 15.9 & 12.7 & 1.19  \\
Gaia DR3 426416177204072448 & 01:06:29.72 & +60:53:50.6 &  & 0.40 $\pm$ 0.04 & 16.0 & 12.9 & 1.21  \\
Gaia DR3 426831375980796800 & 00:48:27.33 & +59:58:28.6 &  & 0.11 $\pm$ 0.01 & 20.1 & 16.2 & 1.26  \\
Gaia DR3 522715116413884416 & 01:04:25.80 & +61:38:25.1 &  & 0.38 $\pm$ 0.04 & 16.3 & 13.3 & 1.31  \\
Gaia DR3 426835125494086144 & 00:47:19.08 & +60:00:51.0 &  & 0.14 $\pm$ 0.01 & 19.2 & 15.8 & 1.36  \\
Gaia DR3 426326116027684480 & 01:07:11.44 & +60:15:44.2 &  & 0.25 $\pm$ 0.03 & 17.6 & 14.4 & 1.37  \\
Gaia DR3 425868517332283776 & 01:01:36.89 & +59:29:25.4 &  & 0.41 $\pm$ 0.04 & 15.9 & 12.9 & 1.37  \\
TYC 4021--505--1 & 01:02:27.29 & +62:01:55.3 & G5V & 0.91 $\pm$ 0.09 & 11.4 & 10.0 & 1.48  \\
Gaia DR3 522555034397334912 & 01:06:20.90 & +61:43:55.1 &  & 0.78 $\pm$ 0.08 & 12.4 & 10.9 & 1.54  \\
Gaia DR3 522765865748123648 & 01:02:51.30 & +62:07:44.7 &  & 0.34 $\pm$ 0.03 & 16.6 & 13.4 & 1.59  \\
Gaia DR3 427035030443328000 & 00:44:21.55 & +60:11:20.1 &  & 0.18 $\pm$ 0.02 & 18.7 & 15.6 & 1.61  \\
Gaia DR3 522537644061868544 & 01:08:10.45 & +61:35:06.6 &  & 0.21 $\pm$ 0.02 & 18.2 & 14.9 & 1.63  \\
Gaia DR3 427202843412335104 & 00:43:54.51 & +61:23:23.2 &  & 0.32 $\pm$ 0.03 & 16.9 & 13.8 & 1.69  \\
HD 236617 & 01:05:43.02 & +59:23:26.8 & G5 & 1.28 $\pm$ 0.13 & 9.9 & 8.6 & 1.74  \\
Gaia DR3 427169445745848832 & 00:41:53.36 & +60:57:58.8 &  & 0.23 $\pm$ 0.02 & 17.7 & 14.2 & 1.82  \\
Gaia DR3 523592384950148992 & 00:57:02.17 & +62:39:28.4 &  & 0.22 $\pm$ 0.02 & 17.7 & 14.7 & 1.94  \\
Gaia DR3 426197133869019520 & 01:09:50.64 & +59:30:48.7 &  & 0.27 $\pm$ 0.03 & 17.4 & 14.7 & 2.03  \\
Gaia DR3 523605720834117632 & 00:56:04.78 & +62:46:10.7 &  & 0.34 $\pm$ 0.03 & 16.6 & 13.7 & 2.05  \\
Gaia DR3 523606820345403520 & 00:55:29.74 & +62:49:28.9 &  & 0.48 $\pm$ 0.05 & 15.6 & 13.1 & 2.11  \\
TYC 4021--815--1 & 00:59:20.89 & +62:48:42.6 &  & 0.99 $\pm$ 0.10 & 11.0 & 9.9 & 2.12  \\
Gaia DR3 427108354134806784 & 00:39:17.10 & +60:36:55.3 &  & 0.56 $\pm$ 0.06 & 14.5 & 11.4 & 2.14  \\
V761 Cas & 01:13:09.85 & +61:42:22.3 & B9V & 3.22 $\pm$ 0.32 & 6.5 & 6.2 & 2.21  \\
Gaia DR3 522574580780909440 & 01:13:19.43 & +61:40:00.1 &  & 0.19 $\pm$ 0.02 & 18.4 & 14.8 & 2.22  \\
Gaia DR3 425198223255334400 & 00:48:24.65 & +58:45:19.6 &  & 0.16 $\pm$ 0.02 & 19.1 & 15.8 & 2.22  \\
Gaia DR3 427777647475959680 & 00:41:34.65 & +62:31:42.5 &  & 0.28 $\pm$ 0.03 & 17.3 & 14.2 & 2.55  \\
Gaia DR3 430106477517090560 & 00:35:44.63 & +60:49:57.6 &  & 0.79 $\pm$ 0.08 & 12.5 & 10.9 & 2.56  \\
Gaia DR3 510588362155056384 & 01:16:30.81 & +61:41:01.5 &  & 0.19 $\pm$ 0.02 & 18.4 & 14.9 & 2.57  \\
Gaia DR3 425164005257167232 & 00:43:17.19 & +58:42:44.3 &  & 0.27 $\pm$ 0.03 & 17.5 & 14.4 & 2.62  \\
Gaia DR3 430133793510727040 & 00:35:08.90 & +60:59:03.0 &  & 0.17 $\pm$ 0.02 & 18.8 & 15.5 & 2.64  \\
Gaia DR3 510353788221523584 & 01:18:57.90 & +60:10:51.9 &  & 0.31 $\pm$ 0.03 & 16.8 & 13.7 & 2.80  \\
Gaia DR3 425348547108352768 & 00:41:07.58 & +58:40:29.8 &  & 0.24 $\pm$ 0.02 & 17.6 & 14.6 & 2.83  \\
Gaia DR3 428588949621015296 & 00:33:48.86 & +60:21:42.7 &  & 0.15 $\pm$ 0.01 & 19.4 & 15.8 & 2.84  \\
Gaia DR3 523346541025484160 & 01:03:22.37 & +63:28:35.3 &  & 0.12 $\pm$ 0.01 & 19.9 & 16.2 & 2.87  \\
Gaia DR3 430258798542242176 & 00:33:41.47 & +61:36:28.7 &  & 0.26 $\pm$ 0.03 & 17.6 & 14.5 & 2.91  \\
Gaia DR3 428608599088304512 & 00:32:59.63 & +60:30:54.1 &  & 0.13 $\pm$ 0.01 & 19.6 & 15.0 & 2.92  \\
Cl* NGC 129 SS 525 & 00:32:36.63 & +60:41:19.2 &  & 0.96 $\pm$ 0.10 & 11.2 & 10.1 & 2.95  \\
BD+62 170 & 00:55:40.97 & +63:40:51.8 & F8 & 1.24 $\pm$ 0.12 & 9.7 & 8.8 & 2.97  \\
Gaia DR3 430199734155452160 & 00:32:04.70 & +61:14:49.8 &  & 0.29 $\pm$ 0.03 & 16.9 & 14.1 & 3.03  \\
Gaia DR3 523899664093745408 & 00:43:41.06 & +63:20:19.7 &  & 0.17 $\pm$ 0.02 & 18.8 & 15.6 & 3.03  \\
Gaia DR3 414097283284109952 & 01:16:18.96 & +58:55:15.9 &  & 0.44 $\pm$ 0.04 & 15.7 & 12.7 & 3.05  \\
Gaia DR3 523114788890667776 & 01:11:43.52 & +63:12:14.3 &  & 0.39 $\pm$ 0.04 & 16.1 & 13.2 & 3.05  \\
Gaia DR3 428606507444503552 & 00:31:41.43 & +60:32:29.1 &  & 0.32 $\pm$ 0.03 & 16.8 & 13.7 & 3.07  \\
Gaia DR3 428651789276927872 & 00:31:28.31 & +60:30:36.4 &  & 0.18 $\pm$ 0.02 & 18.8 & 15.1 & 3.10  \\
Gaia DR3 430206674822463744 & 00:31:16.38 & +61:27:47.3 &  & 0.26 $\pm$ 0.03 & 17.3 & 13.9 & 3.16  \\
Gaia DR3 424138538865527168 & 00:57:44.28 & +57:31:32.6 &  & 0.76 $\pm$ 0.08 & 12.8 & 11.1 & 3.19  \\
Gaia DR3 523891933149775616 & 00:46:04.81 & +63:39:46.4 &  & 0.23 $\pm$ 0.02 & 17.8 & 14.8 & 3.20  \\
Gaia DR3 430351707277551616 & 00:33:23.98 & +62:18:00.1 &  & 0.44 $\pm$ 0.04 & 15.7 & 12.9 & 3.20  \\
Gaia DR3 413591954612999296 & 01:06:23.15 & +57:44:02.9 &  & 0.28 $\pm$ 0.03 & 17.2 & 14.0 & 3.23  \\
NGC 129 48 & 00:30:33.45 & +60:17:27.3 & F6V & 1.27 $\pm$ 0.13 & 9.5 & 8.8 & 3.25  \\
Gaia DR3 424242404052127872 & 00:54:06.70 & +57:27:25.2 &  & 0.22 $\pm$ 0.02 & 17.7 & 14.7 & 3.28  \\
Gaia DR3 523918424511630208 & 00:40:28.49 & +63:26:04.7 &  & 0.10 $\pm$ 0.01 & 20.4 & 16.6 & 3.32  \\
Gaia DR3 424738421233795072 & 00:42:39.06 & +57:52:50.3 &  & 0.20 $\pm$ 0.02 & 17.9 & 14.7 & 3.36  \\
BD+61 258 & 01:23:33.25 & +61:46:05.2 & F2 & 1.44 $\pm$ 0.14 & 9.6 & 8.9 & 3.39  \\
Gaia DR3 523413271935393280 & 01:09:59.65 & +63:45:58.1 &  & 0.39 $\pm$ 0.04 & 16.1 & 13.1 & 3.42  \\
Gaia DR3 524005221511877120 & 00:49:59.25 & +64:05:45.1 &  & 0.32 $\pm$ 0.03 & 16.9 & 13.8 & 3.47  \\
Gaia DR3 523228549680118656 & 01:14:29.82 & +63:34:44.6 &  & 0.22 $\pm$ 0.02 & 17.6 & 14.4 & 3.54  \\
Gaia DR3 428677391586484480 & 00:27:30.80 & +60:57:36.1 &  & 0.42 $\pm$ 0.04 & 16.0 & 12.7 & 3.56  \\
Gaia DR3 424897712980359296 & 00:41:06.06 & +57:45:14.0 &  & 0.20 $\pm$ 0.02 & 18.0 & 14.6 & 3.57  \\
Gaia DR3 424073525939801984 & 01:01:54.61 & +57:11:58.5 &  & 0.19 $\pm$ 0.02 & 18.4 & 15.1 & 3.58  \\
Gaia DR3 424031851879846912 & 00:57:35.08 & +57:07:43.0 &  & 0.52 $\pm$ 0.05 & 15.1 & 12.9 & 3.59  \\
Gaia DR3 430861399634552448 & 00:37:36.63 & +63:35:12.6 &  & 0.59 $\pm$ 0.06 & 14.3 & 12.0 & 3.63  \\
Gaia DR3 428484045032883456 & 00:28:38.41 & +59:39:27.0 &  & 0.15 $\pm$ 0.02 & 19.2 & 15.9 & 3.64  \\
Gaia DR3 430518764324231040 & 00:28:32.52 & +62:06:39.1 &  & 0.52 $\pm$ 0.05 & 14.8 & 12.0 & 3.64  \\
HD 4810 & 00:51:11.44 & +64:19:29.7 & A2 & 1.87 $\pm$ 0.19 & 8.4 & 8.1 & 3.66  \\
TYC 4024--250--1 & 00:51:04.61 & +64:20:17.9 &  & 1.07 $\pm$ 0.11 & 10.7 & 9.7 & 3.68  \\
Gaia DR3 413481488050839424 & 01:16:18.59 & +57:58:51.2 &  & 0.80 $\pm$ 0.08 & 12.7 & 10.5 & 3.70  \\
Gaia DR3 510825478709127424 & 01:26:26.82 & +61:49:43.0 &  & 0.33 $\pm$ 0.03 & 17.0 & 14.1 & 3.74  \\
Gaia DR3 423826513780557952 & 00:53:28.82 & +56:54:47.2 &  & 0.11 $\pm$ 0.01 & 20.5 & 16.9 & 3.83  \\
Gaia DR3 424630050620861312 & 00:45:16.82 & +57:06:56.8 &  & 0.23 $\pm$ 0.02 & 17.4 & 14.4 & 3.89  \\
Gaia DR3 524282706455103744 & 01:02:14.82 & +64:37:47.9 &  & 0.23 $\pm$ 0.02 & 17.6 & 14.3 & 3.96  \\
Gaia DR3 423823494422951040 & 00:53:26.81 & +56:44:57.9 &  & 0.57 $\pm$ 0.06 & 14.3 & 12.2 & 3.99  \\
Gaia DR3 524957222478303616 & 01:10:05.51 & +64:25:50.8 &  & 0.19 $\pm$ 0.02 & 18.1 & 14.7 & 4.02  \\
TYC 4015--1647--1 & 00:23:32.15 & +60:38:28.4 &  & 0.97 $\pm$ 0.10 & 11.2 & 10.0 & 4.06  \\
Gaia DR3 427905706211506560 & 00:30:32.34 & +58:20:28.2 &  & 0.14 $\pm$ 0.01 & 19.5 & 16.2 & 4.08  \\
Gaia DR3 512646373042841984 & 01:24:00.84 & +63:21:27.8 &  & 0.79 $\pm$ 0.08 & 12.5 & 11.1 & 4.15  \\
Gaia DR3 524328581007238656 & 00:57:51.06 & +64:52:55.5 &  & 0.40 $\pm$ 0.04 & 16.2 & 13.4 & 4.17  \\
Gaia DR3 423947460059430912 & 01:03:02.76 & +56:37:16.3 &  & 0.10 $\pm$ 0.01 & 20.5 & 16.2 & 4.18  \\
Gaia DR3 524704236026503552 & 01:17:29.40 & +64:09:57.5 &  & 0.26 $\pm$ 0.03 & 17.5 & 14.4 & 4.20  \\
Gaia DR3 526998195239910656 & 00:34:15.41 & +64:02:07.1 &  & 0.39 $\pm$ 0.04 & 16.3 & 13.5 & 4.21  \\
Gaia DR3 526998195239908736 & 00:34:13.63 & +64:02:13.1 &  & 0.79 $\pm$ 0.08 & 12.6 & 11.2 & 4.22  \\
HD 6822 & 01:10:16.31 & +64:38:45.2 & A0 & 1.89 $\pm$ 0.19 & 8.2 & 7.8 & 4.22  \\
Gaia DR3 431081851713650944 & 00:30:03.36 & +63:38:14.5 &  & 0.51 $\pm$ 0.05 & 15.2 & 12.8 & 4.26  \\
TYC 4031--2224--1 & 01:31:39.03 & +60:30:44.6 & F5 & 0.99 $\pm$ 0.10 & 11.0 & 9.9 & 4.29  \\
Gaia DR3 512575969937525376 & 01:27:42.20 & +62:58:26.6 &  & 0.25 $\pm$ 0.03 & 17.3 & 14.2 & 4.29  \\
Gaia DR3 510790844092027008 & 01:31:20.08 & +61:52:29.1 &  & 0.20 $\pm$ 0.02 & 18.4 & 15.7 & 4.31  \\
Gaia DR3 424842569903670656 & 00:34:37.72 & +57:24:57.8 &  & 0.32 $\pm$ 0.03 & 16.8 & 13.9 & 4.35  \\
Gaia DR3 428283418521147392 & 00:23:09.47 & +59:12:54.8 &  & 0.29 $\pm$ 0.03 & 17.1 & 14.0 & 4.45  \\
Gaia DR3 428172299125461120 & 00:24:08.03 & +58:53:49.3 &  & 0.25 $\pm$ 0.03 & 17.4 & 14.1 & 4.48  \\
BD+59 37 & 00:20:47.67 & +60:03:39.4 &  & 1.26 $\pm$ 0.13 & 9.8 & 6.8 & 4.48  \\
Gaia DR3 423769102956471680 & 00:53:28.68 & +56:12:18.6 &  & 0.44 $\pm$ 0.04 & 15.6 & 12.7 & 4.53  \\
Gaia DR3 430603907757603584 & 00:22:05.36 & +62:52:09.2 &  & 0.46 $\pm$ 0.05 & 15.4 & 12.7 & 4.62  \\
Gaia DR3 428342762091655168 & 00:19:49.86 & +59:36:50.2 &  & 0.53 $\pm$ 0.05 & 14.7 & 11.9 & 4.71  \\
Gaia DR3 527493250347583872 & 00:42:12.63 & +65:10:36.1 &  & 0.32 $\pm$ 0.03 & 16.8 & 13.6 & 4.75  \\
Gaia DR3 423535250579327488 & 00:59:56.20 & +55:51:15.4 &  & 0.27 $\pm$ 0.03 & 17.5 & 14.8 & 4.88  \\
Gaia DR3 411948871922993024 & 01:10:06.75 & +56:09:03.8 &  & 0.47 $\pm$ 0.05 & 15.5 & 13.1 & 4.89  \\
Gaia DR3 512501989120374144 & 01:32:08.47 & +63:18:56.2 &  & 0.19 $\pm$ 0.02 & 18.4 & 14.8 & 4.90  \\
Gaia DR3 512789034676716032 & 01:27:08.42 & +64:13:28.1 &  & 0.22 $\pm$ 0.02 & 17.6 & 14.2 & 4.96  \\
Gaia DR3 509912226919850880 & 01:37:26.63 & +60:48:22.2 &  & 0.19 $\pm$ 0.02 & 18.4 & 14.9 & 4.97  \\
Gaia DR3 524616854909023488 & 00:45:39.31 & +65:32:40.3 &  & 0.22 $\pm$ 0.02 & 17.8 & 14.5 & 4.99  \\
Gaia DR3 527148725254317312 & 00:35:20.74 & +65:04:29.5 &  & 0.62 $\pm$ 0.06 & 14.3 & 12.1 & 4.99  \\
Gaia DR3 421818053931122176 & 00:29:34.63 & +57:07:03.5 &  & 0.74 $\pm$ 0.07 & 12.8 & 11.2 & 5.02  \\
Gaia DR3 524934716850804992 & 01:16:26.04 & +65:13:51.0 &  & 0.82 $\pm$ 0.08 & 12.3 & 10.9 & 5.04  \\
Gaia DR3 418461897767667456 & 00:45:39.92 & +55:52:59.2 &  & 0.22 $\pm$ 0.02 & 17.9 & 14.7 & 5.05  \\
Gaia DR3 421811697379570688 & 00:29:48.73 & +57:01:45.2 &  & 0.13 $\pm$ 0.01 & 19.2 & 12.8 & 5.06  \\
Gaia DR3 524938084107230720 & 01:16:01.32 & +65:18:17.2 &  & 0.10 $\pm$ 0.01 & 20.6 & 16.7 & 5.08  \\
Gaia DR3 429704571660226176 & 00:15:13.72 & +61:46:23.3 &  & 0.18 $\pm$ 0.02 & 18.3 & 15.0 & 5.09  \\
Gaia DR3 512451931283034112 & 01:35:47.24 & +63:06:48.3 &  & 0.53 $\pm$ 0.05 & 15.0 & 12.0 & 5.18  \\
HD 4948 & 00:52:36.09 & +65:53:30.3 & A2 & 1.89 $\pm$ 0.19 & 8.3 & 8.1 & 5.20  \\
Gaia DR3 527534653834163072 & 00:41:10.00 & +65:46:21.3 &  & 0.21 $\pm$ 0.02 & 18.0 & 14.8 & 5.35  \\
Gaia DR3 411492437156666368 & 01:05:30.07 & +55:27:14.2 &  & 0.60 $\pm$ 0.06 & 14.3 & 12.3 & 5.39  \\
Gaia DR3 422967112302952192 & 00:15:35.09 & +58:59:21.4 &  & 0.55 $\pm$ 0.06 & 14.9 & 12.7 & 5.44  \\
Gaia DR3 421980919092603392 & 00:22:46.22 & +57:25:23.3 &  & 0.49 $\pm$ 0.05 & 15.2 & 12.6 & 5.46  \\
Gaia DR3 431392360672275584 & 00:15:18.31 & +63:14:19.4 &  & 0.22 $\pm$ 0.02 & 17.9 & 14.9 & 5.47  \\
TYC 4028--969--1 & 00:48:42.53 & +66:07:10.6 &  & 0.86 $\pm$ 0.09 & 11.9 & 10.6 & 5.48  \\
Gaia DR3 421990608538759808 & 00:21:17.48 & +57:27:37.2 &  & 0.30 $\pm$ 0.03 & 17.2 & 14.3 & 5.59  \\
Gaia DR3 512489211600057600 & 01:37:48.00 & +63:36:33.3 &  & 0.53 $\pm$ 0.05 & 15.0 & 12.6 & 5.59  \\
Gaia DR3 422982677264653440 & 00:13:53.86 & +59:04:25.8 &  & 0.36 $\pm$ 0.04 & 16.5 & 13.6 & 5.61  \\
Gaia DR3 508974691392427776 & 01:37:12.24 & +58:23:19.4 &  & 0.39 $\pm$ 0.04 & 16.3 & 13.2 & 5.63  \\
TYC 3673--1289--1 & 01:18:51.10 & +55:49:08.2 &  & 1.28 $\pm$ 0.13 & 9.7 & 8.8 & 5.69  \\
Gaia DR3 429623693140336512 & 00:09:54.06 & +61:10:34.9 &  & 0.21 $\pm$ 0.02 & 18.2 & 12.2 & 5.69  \\
Gaia DR3 421959169377531264 & 00:21:46.47 & +57:08:46.7 &  & 0.15 $\pm$ 0.02 & 18.9 & 15.7 & 5.74  \\
Gaia DR3 421945356762784640 & 00:22:33.71 & +56:59:31.6 &  & 0.34 $\pm$ 0.03 & 16.9 & 13.7 & 5.77  \\
Gaia DR3 526042891426149888 & 01:01:11.94 & +66:32:38.6 &  & 0.11 $\pm$ 0.01 & 20.4 & 16.2 & 5.85  \\
Gaia DR3 421933021616821504 & 00:23:06.48 & +56:47:52.9 &  & 0.64 $\pm$ 0.06 & 14.0 & 11.9 & 5.85  \\
Gaia DR3 526137930469180288 & 00:51:34.57 & +66:35:09.8 &  & 0.71 $\pm$ 0.07 & 13.2 & 11.4 & 5.90  \\
Gaia DR3 509511974622906496 & 01:43:21.47 & +59:34:09.2 &  & 0.21 $\pm$ 0.02 & 18.2 & 14.5 & 5.91  \\
Gaia DR3 509092884599480704 & 01:41:43.57 & +58:49:22.1 &  & 0.14 $\pm$ 0.01 & 19.6 & 16.4 & 5.97  \\
Gaia DR3 431890542519116928 & 00:15:25.61 & +64:20:23.2 &  & 0.53 $\pm$ 0.05 & 14.9 & 12.3 & 5.97  \\
Gaia DR3 422021291786053120 & 00:18:16.50 & +57:20:46.3 &  & 0.47 $\pm$ 0.05 & 15.4 & 12.6 & 5.97  \\
\noalign{\smallskip} 
\noalign{\hrule height 01pt}
\vspace{-5.5mm}
\end{longtable}
\vspace{-8mm}
\begin{justify}
    \scriptsize{\textbf{\textit{Notes. }}
    $^{\text{(a)}}$ The maximum error in $G$ provided by \textit{Gaia} DR3 is less than 0.005\,mag;
    $^{\text{(b)}}$ The maximum error in $J$ provided by 2MASS is less than 0.2\,mag;
    $^{\text{(c)}}$ \citet{nemravova12};
    $^{\text{(d)}}$ \citet{tokovinin21a}.}
\end{justify}
\normalsize

%%%%% APENDICE E %%%%%

\begin{center}
\chapter[Appendix: Properties of multiple systems with exoplanets]{Properties of multiple systems\\with exoplanets}
\end{center}
\label{ch:Appendix_E}
\vspace{3cm}
\pagestyle{fancy}
\fancyhf{}
\lhead[\small{\textbf{\thepage}}]{\textbf{Appendix~E: Properties of multiple systems with exoplanets}}
\rhead[\small{\textbf{Appendix~E: Properties of multiple systems with exoplanets}}]{\small{\textbf{\thepage}}}
\renewcommand{\thetable}{E.\arabic{table}}   % Cambia el formato de numeración al tipo E.x

\footnotesize

\begin{longtable}{@{\hspace{10mm}}l@{\hspace{20mm}}c@{\hspace{20mm}}c@{\hspace{20mm}}c@{\hspace{10mm}}}
\caption[Non-duplicated exoplanet host stars at $d <$\,100\,pc.]{Non-duplicated exoplanet host stars at $d <$\,100\,pc.}\\
\noalign{\smallskip}
\noalign{\hrule height 01pt}
\noalign{\smallskip}
Star & $\alpha$ (J2000) & $\delta$ (J2000) & $d$  \\ 
  & (hh:mm:ss.ss) & (dd:mm:ss.s) & (pc) \\
\noalign{\smallskip}
\hline
\noalign{\smallskip}
\endfirsthead
\caption[]{Non-duplicated exoplanet host stars at $d <$ 100\,pc. \textit{(continued).}}\\
\noalign{\hrule height 01pt}
\noalign{\smallskip} 
\label{tab:sample_host_stars}
Star & $\alpha$ (J2000) & $\delta$ (J2000) & $d$  \\ 
  & (hh:mm:ss.ss) & (dd:mm:ss.s) & (pc) \\
\noalign{\smallskip}
\hline
\noalign{\smallskip}
\endhead
\noalign{\smallskip}
\hline
\endfoot 

CD--55 9423 & 00:00:44.91 & --54:49:49.9 & 61.9 \\
HD 142 & 00:06:19.18 & --49:04:30.7 & 26.2 \\
G 158--27 & 00:06:43.20 & --07:32:17.0 & 4.8 \\
HD 457 & 00:08:59.68 & --39:44:13.8 & 55.3 \\
CD--27 16  & 00:09:04.36 & --27:07:19.7 & 23.8 \\
HD 564 & 00:09:52.82 & --50:16:04.2 & 51.2 \\
CD--39 31 & 00:11:41.65 & --39:13:47.8 & 34.9 \\
HD 984 & 00:14:10.25 & --07:11:56.8 & 45.7 \\
\end{longtable}

\newpage
\footnotesize

\begin{longtable}{l@{\hspace{1mm}}c@{\hspace{2mm}}c@{\hspace{2mm}}c@{\hspace{2mm}}c@{\hspace{2mm}}l}
 \caption[Multiple systems with one or more ultra-cool dwarfs.]{Multiple systems with one or more ultra-cool dwarfs.}\\
 \noalign{\smallskip}
 \noalign{\hrule height 01pt}
 \noalign{\smallskip}
 Star & $\alpha$\,(J2000) & $\delta$\,(J2000) & \textit{G} & Spectral &  Reference of the UCD discovery \\
 & (hh:mm:ss.ss) & (dd:mm:ss.s) & (mag) & type & \\
 \noalign{\smallskip}
 \hline
 \noalign{\smallskip} 
 \endfirsthead
 \caption[]{Multiple systems with one or more ultra-cool dwarfs \textit{(continued).}}
 \label{tab:UCD_systems}\\
 \noalign{\hrule height 01pt}
 \noalign{\smallskip}
 Star & $\alpha$\,(J2000) & $\delta$\,(J2000) & \textit{G} & Spectral &  Reference of the UCD discovery \\
 & (hh:mm:ss.ss) & (dd:mm:ss.s) & (mag) & type & \\
\noalign{\smallskip}
\endhead
\hline
\noalign{\smallskip}
\endfoot
 HD 984[A] & 00:14:10.25 & --07:11:56.8 & 7.2 & F7 & \\
 HD 984[B] & 00:14:10.25 & --07:11:56.8 & ... & M6.5 & \citealt{johnsongroh17} \\
\noalign{\smallskip}
\hline
\noalign{\smallskip}
 DENIS J004135.3-562112[A] & 00:41:35.39 & --56:21:12.9  & 16.4 & M6.5 & \\
 DENIS J004135.3-562112[B] & 00:41:35.39 & --56:21:12.9  & ... & M9.0 & \citealt{reiners10} \\
\noalign{\smallskip}
\hline
\noalign{\smallskip}
 HD 4747[A] & 00:49:26.77 & --23:12:44.9 & 7.0 & G8/K0 & \\
 HD 4747[B] & 00:49:26.77 & --23:12:44.9 & ... & T1 & \citealt{crepp16} \\
\noalign{\smallskip}
\hline
\noalign{\smallskip}
 SCR J0103--5515[A] & 01:03:35.83 & --55:15:57.0 & 13.9 & M5.0 &  \\
 SCR J0103--5515[B] & 01:03:35.83 & --55:15:57.0 & ... & M5.0 &  \\
\end{longtable}
\vspace{-8mm} 
  \begin{justify}
     \scriptsize{\textbf{\textit{Notes.}} $^{\text{(a)}}$ This brown star was initially classified as a circumbinary planet due to its orbit around the other two stars in the system.}
  \end{justify}

\newpage

\begin{table}[h]
 \centering
 \caption[Astrometry of systems with new common $\mu$ and $\pi$ companions.]{Astrometry of systems with new common $\mu$ and $\pi$ companions.}
 \footnotesize
  \medskip
  \medskip
 \scalebox{1}[1]{
 \begin{tabular}{lccccc}
 \noalign{\hrule height 01pt}
 \noalign{\smallskip}
 Star & $\pi$ & $\mu_{\alpha}\cos\delta$ & $\mu_\delta$ & $\mu_{\rm ratio}$ & $\Delta$PA \\
  & (mas) & (mas\,a$^{-\text{1}}$) & (mas\,a$^{-\text{1}}$) &  & (deg) \\
 \noalign{\smallskip}
 \hline
 \noalign{\smallskip} 
* HD 1466 & 23.36 $\pm$ 0.02  &  90.05 $\pm$ 0.02  &  --59.21 $\pm$ 0.02  &    &    \\
UCAC3 52--533 & 20.97 $\pm$ 0.01  &  79.94 $\pm$ 0.02  &  --54.33 $\pm$ 0.02  &  0.10  &  0.88  \\
UCAC3 53--724 & 23.18 $\pm$ 0.08  &  91.47 $\pm$ 0.08  &  --54.92 $\pm$ 0.08  &  0.04  &  2.34  \\
2MASS J00191296--6226005 & 20.60 $\pm$ 1.67  &  88.10 $\pm$ 1.63  &  --54.50 $\pm$ 1.43  &  0.05  &  1.58  \\
 \noalign{\smallskip}
 \hline
 \noalign{\smallskip} 
* HD 27196 & 17.40 $\pm$ 0.01  &  --25.77 $\pm$ 0.01  &  --116.43 $\pm$ 0.02  &    &    \\
2MASS J04164536--2646205 & 17.40 $\pm$ 0.03  &  --24.52 $\pm$ 0.03  &  --118.49 $\pm$ 0.03  &  0.02  &  0.79  \\
\end{tabular}
 } % \scalebox
 \label{tab:new_detected_systems_astrometry}  
\end{table}

\newpage

\scriptsize

\begin{longtable}{@{\hspace{0mm}}l@{\hspace{0.5mm}}l@{\hspace{0.5mm}}l@{\hspace{0mm}}c@{\hspace{0mm}}c@{\hspace{0.5mm}}c@{\hspace{0mm}}c@{\hspace{0.5mm}}c@{\hspace{0.5mm}}c@{\hspace{1mm}}l@{\hspace{0mm}}}
\caption[Basic data of the identified multiple systems with planets.]{Basic data of the identified multiple systems with planets.}\\
\noalign{\smallskip}
\noalign{\hrule height 01pt}
\noalign{\smallskip}
WDS & Disc. code / & Star$^{\text{(b)}}$ & $\alpha$ (J2000) & $\delta$ (J2000) & $d$ & $\rho$ & $\theta^{\text{(c)}}$ & \textit{s} & Refs.$^{\text{(a)}}$ \\
 & Disc. ref.$^{\text{(a)}}$ & & (hh:mm:ss.ss) & (dd:mm:ss.s) & (pc) & (arcsec) & (deg) & (au) & \\
\noalign{\smallskip}
\hline
\noalign{\smallskip}
\endfirsthead
\caption[]{Basic data of the identified multiple systems with planets \textit{(continued).}}\\
\noalign{\hrule height 01pt}
\noalign{\smallskip} 
\label{tab:basic_data_identified_hosts} 
WDS & Disc. code / & Star$^{\text{(b)}}$ & $\alpha$ (J2000) & $\delta$ (J2000) & $d$ & $\rho$ & $\theta^{\text{(c)}}$ & \textit{s} & Refs.$^{\text{(a)}}$ \\
 & Disc. ref.$^{\text{(a)}}$ & & (hh:mm:ss.ss) & (dd:mm:ss.s) & (pc) & (arcsec) & (deg) & (au) & \\
\noalign{\smallskip}
\endhead
\hline
\noalign{\smallskip}
\endfoot 
00007--5450 &  & * CD--55 9423[A] & 00:00:44.91 & --54:49:49.9 & 61.87$\pm$0.04 &  &  &  &  \\
 & TOI 129 & CD--55 9423B & 00:00:44.65 & --54:49:46.8 & 61.34$\pm$0.26 & 3.813 & 322 & 235.9 & This work \\
 & KPP2838 & Gaia DR3 4923848227231122688 & 00:00:47.59 & --54:50:59.8 & 63.74$\pm$0.71 & 73.62 & 162 & 4554 & This work \\
\noalign{\smallskip}
\hline 
\noalign{\smallskip}
00063--4905 &  & * HD 142 & 00:06:19.18 & --49:04:30.7 & 26.18$\pm$0.02 &  &  & &   \\
 & HDO 180 & HD 1428 & 00:06:19.15 & --49:04:34.9 & 26.16$\pm$0.02 & 3.883 & 186 & 101.67 & This work \\
\noalign{\smallskip}
\hline 
\noalign{\smallskip}
00163--7951 &  & * HD 1237[A] & 00:16:12.68 & --79:51:04.2 & 17.55$\pm$0.01 &  &  &  & \\
 & GJ 3021 & HD 1237B & 00:16:12.57 & --79:51:00.4 & 17.59$\pm$0.03 & 3.985 & 352 & 69.96 & This work \\
\noalign{\smallskip}
\hline 
\noalign{\smallskip}
00184+4401 &  & * GX And & 00:18:22.89 & +44:01:22.6 & 3.56$\pm$0.01 &  &  &  &   \\
\end{longtable}
\vspace{-8mm} 
 \begin{justify}
    \textbf{\textit{Notes. }} $^{\text{(a)}}$ WDS discoverer codes are written in uppercase, and literature references in lowercase. References: 
    Aka23: \citet{akana23};
    App23: \citet{apps23};
    Bai50: \citet{baize50}; 
    Bar20: \citet{barnes20}; 
    Ber23: \citet{bertini23}; 
    Bru98: \citet{bruch98}; 
    Cam15: \citet{campante15};
    Car20: \citet{carleo20}; 
    Cha11: \citet{chauvin11}; 
    Chr22: \citet{christian22}; 
    Cla22: \citet{clark22}; 
    Dea14: \citet{deacon14};
    Des23: \citet{desidera23}; 
    Egg07: \citet{eggenberger07}; 
    Elb21: \citet{elbadry21}; 
    Fah12: \citet{faherty12}; 
    Far14; \citet{farrington14}; 
    Fei19: \citet{feinstein19}; 
    Fen22: \citet{feng22};
    Fon21: \citet{fontanive21};  
    Gai21: \citet{gaiacollaboration21b};
    Hei74: \citet{heintz74}; 
    Jus19: \citet{justesen19}; 
    Ker19: \citet{kervella19}; 
    Krz84: \citet{krzeminski84};
    Ma16 : \citet{ma16}; 
    Mal12: \citet{malkov12}
    Mar20: \citet{marocco20}; 
    Mic21: \citet{michel21}; 
    Mug19: \citet{mugrauer19}; 
    Mug20: \citet{mugrauer20}; 
    Neu07: \citet{neuhauser07}; 
    Oh17: \citet{oh17}; 
    Pie23: \citet{pierens23}; 
    Pou04: \citet{pourbaix04};
    Prs11: \citet{prsa11}; 
    Ram09: \citet{ramm09};
    Rob15: \citet{roberts15}; 
    Rod16: \citet{rodigas16}; 
    Roe12: \citet{roell12}; 
    Ser22: \citet{serrano22};
    Su21:  \citet{su21}; 
    Udr02: \citet{udry02};
    Van19: \citet{vanderburg19}; 
    Wit16: \citet{wittrock16}; 
    Zho22: \citet{zhou22}; 
    Zie20: \citet{ziegler20};
    Zuc03: \citet{zucker03}.
    $^{\text{(b)}}$ The exoplanet(s) host star is marked with an asterisk.
    $^{\text{(c)}}$ The value is not tabulated when $\rho<1$\,arcsec.
    $^{\text{(d)}}$ These stars are within Tucana-Horologium association. Only the four that have a separation less than one parsec are shown.    
    $^{\text{(e)}}$ The separation between HD 2638 A and B is 0.512 arcsec (25$\pm$2\,au).
    $^{\text{(f)}}$ System with circumbinary planets. We do not tabulate $\rho$ or $\theta$ for this type of systems.
    $^{\text{(g)}}$ Binary (this work).
    $^{\text{(h)}}$ The value, obtained by \citet{finch18}, is not correct because of the distortion exerted by its companion.
  \end{justify}
\normalsize

%%%%% APENDICE F %%%%%

\begin{center}
\chapter[Appendix: Properties of multiple star systems within 10\,pc]{Properties of multiple star systems\\ within 10\,pc}
\end{center}
\label{ch:Appendix_F}
\vspace{3cm}
\pagestyle{fancy}
\fancyhf{}
\lhead[\small{\textbf{\thepage}}]{\textbf{Appendix~F: Properties of multiple star systems within 10\,pc}}
\rhead[\small{\textbf{Appendix~F: Properties of multiple star systems within 10\,pc}}]{\small{\textbf{\thepage}}}
\renewcommand{\thetable}{F.\arabic{table}}   % Cambia el formato de numeración al tipo F.x

\footnotesize

\begin{longtable}{l@{\hspace{6mm}}l@{\hspace{7mm}}c@{\hspace{7mm}}c@{\hspace{7mm}}c@{\hspace{7mm}}c@{\hspace{5mm}}c}
\caption{Stars and brown dwarfs at $d <$ 10\,pc.}
\label{tab:sample_10pc}\\
\noalign{\smallskip}
\noalign{\hrule height 01pt}
\noalign{\smallskip}
Star or brown dwarf & GJ$^{(\text{a})}$ & $\alpha$ (J2000) & $\delta$ (J2000) & $\mu_{\alpha}\cos\delta$ & $\mu_\delta$ & $d$ \\
 &  & (hh:mm:ss.ss) & (dd:mm:ss.s) & (mas\,a$^{-\text{1}}$) & (mas\,a$^{-\text{1}}$) & (pc)  \\
\noalign{\smallskip}
\hline
\noalign{\smallskip}
\endfirsthead
\caption[]{Stars and brown dwarfs at $d <$ 10\,pc \textit{(continued).}}\\
\noalign{\hrule height 01pt}
\noalign{\smallskip} 
Star or brown dwarf & GJ$^{(\text{a})}$ & $\alpha$ (J2000) & $\delta$ (J2000) & $\mu_{\alpha}\cos\delta$ & $\mu_\delta$ & $d$ \\
 &  & (hh:mm:ss.ss) & (dd:mm:ss.s) & (mas\,a$^{-\text{1}}$) & (mas\,a$^{-\text{1}}$) & (pc)  \\
\noalign{\smallskip}
\hline
\noalign{\smallskip}
\endhead
\hline
\noalign{\smallskip}
\endfoot 

LAWD 96 & 915 & 00:02:10.72 & --43:09:55.4 & 613.79 & --686.99 & 8.33 \\
WISE J000517.48+373720.5 & ... & 00:05:17.48 & +37:37:20.5 & 997.30 & --271.60 & 7.88 \\
HD 225213 & 1 & 00:05:24.43 & --37:21:26.5 & 5633.44 & --2334.72 & 4.35 \\
G 158--27 & 1002 & 00:06:43.20 & --07:32:17.0 & --811.57 & --1893.25 & 4.85 \\
GAT 1367 & ... & 00:11:31.82 & +59:08:39.9 & --905.72 & --1166.83 & 9.31 \\
G 158--50 [AB] & 1005 AB & 00:15:28.77 & --16:08:11.7 & 731.83 & --607.73 & 6.00 \\
HD 1326 [A] & 15 A & 00:18:22.88 & +44:01:22.6 & 2891.52 & 411.83 & 3.56 \\
HD 1326B & 15 B & 00:18:25.83 & +44:01:38.1 & 2862.80 & 336.43 & 3.56 \\
\end{longtable}
\vspace{-4mm} 

\begin{justify}
\scriptsize{\textbf{\textit{Notes. }}}
\scriptsize{$^{\text{(a)}}$ For homogeneity and readability, we write ``GJ'' \citep{gliese79,gliese91} even for stars in the catalogue of nearby stars of \citet{gliese69}, which originally had a ``Gl'' identifier. We refrain from tabulating stars with ``GJ'' identifier $\ge$ 9000, as they were not part of any Gliese or Gliese-Jahreiss catalogues.}
\end{justify}

\newpage

%\pagebreak[4]
%   \global\pdfpageattr\expandafter{\the\pdfpageattr/Rotate 90}

\makeatletter
\newcommand\Tammedio{\@setfontsize\Tammedio{8.3}{8.3}}
\makeatother
\Tammedio

\begin{landscape}
\begin{longtable}{@{\hspace{0mm}}l@{\hspace{1mm}}l@{\hspace{0mm}}l@{\hspace{1mm}}c@{\hspace{0mm}}c@{\hspace{1mm}}c@{\hspace{1mm}}c@{\hspace{1mm}}l@{\hspace{1mm}}l@{\hspace{1mm}}c@{\hspace{1mm}}c@{\hspace{0mm}}c@{\hspace{-2mm}}c@{\hspace{1mm}}l@{\hspace{0mm}}c@{\hspace{-2mm}}c@{\hspace{-1mm}}c@{\hspace{0mm}}}
\caption[All resolved multiple systems at $d <$ 10\,pc.]{\small{All resolved multiple systems at $d <$ 10\,pc.}}\\
\noalign{\smallskip}
\noalign{\hrule height 01pt}
\noalign{\smallskip}
Star or brown dwarf & WDS & Disc. & $d$ & $\rho$ & $\theta$ & $s$ & Ref. & Spectral & $G$ & $M$ & Ref. $M^{(\text{a})}$ & $P$ & Ref.$P^{(\text{a})}$ & $|U^*_g|$ & Planets & System \\
 &  & code & (pc) & (arcsec) & (deg) & (au) & $\rho,\theta,s^{\text{(a)}}$ & type & (mag) & (M$_\odot$) &  & (a) &  & (10$^{33}$\,J) &  & diagram$^{\text{(b)}}$ \\
\noalign{\smallskip}
\hline
\noalign{\smallskip}
\endfirsthead
\caption[]{\small{All resolved multiple systems at $d <$ 10\,pc \textit{(continued).}}}\\
\noalign{\hrule height 01pt}
\noalign{\smallskip}
Star or brown dwarf & WDS & Disc. & $d$ & $\rho$ & $\theta$ & $s$ & Ref. & Spectral & $G$ & $M$ & Ref. $M^{(\text{a})}$ & $P$ & Ref.$P^{(\text{a})}$ & $|U^*_g|$ & Planets & System \\
 &  & code & (pc) & (arcsec) & (deg) & (au) & $\rho,\theta,s^{\text{(a)}}$ & type & (mag) & (M$_\odot$) &  & (a) &  & (10$^{33}$\,J) &  & diagram$^{\text{(b)}}$ \\
\noalign{\smallskip}
\noalign{\smallskip}
\endhead
\noalign{\smallskip}
\endfoot
   G 158--50 [A] & 00155--1608 &  & 5.90 &  &  &  &  & M4.0 V & 10.27 & 0.179$\pm$0.002 & Kir24 &  &  &  &  & \multirow{2}{*}{\includegraphics[width=10.3mm]{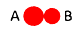}}\\
   G 158--50 [B] &  & HEI 299 & ... & 0.3765 & ... & 1.79 & Man19 & M4.5: V & 10.7: & 0.112$\pm$0.001 & Kir24 & (4557.26$^{+\text{0.75}}_{-\text{0.74}}$)$\cdot$10$^{-\text{3}}$ & Ben16 & 20\,000 &  & \\
 \noalign{\smallskip}
 \hline
 \noalign{\smallskip}
   HD 1326 [A] (GJ 15 A) & 00184+4401 &  & 3.56 &  &  &  &  & M1.0 V & 7.22 & 0.399$\pm$0.012 & Kir24 &  &  &  & b, c & \multirow{2}{*}{\includegraphics[width=6.5mm]{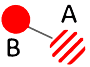}}\\
   HD 1326B &  & GRB 34 & 3.56 & 34.35 & 65 & 122.37 & This work & M3.5 V & 9.69 & 0.165$\pm$0.005 & Kir24 & 1226$\pm$236 & Rom21 & 950 &  & \\
 \noalign{\smallskip}
 \hline
 \noalign{\smallskip}
  LP 881--64 [A] & 00247--2653 &  & 7.73 &  &  &  &  & M5.5 V & 13.16 & 0.112$\pm$0.001 & Kir24 &  &  &  &  & \multirow{3}{*}{\includegraphics[width=7mm]{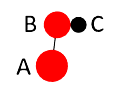}}\\
  LP 881--64 [B] &  & LEI 1 & ... & 1.07 & 353 & 8.27 & Cla24 & M9.5 V & 16.5: & 0.084$\pm$0.001 & Kir24 & 88.63 & Kna20 &  &  & \\
  LP 881--64 [C] &  & LEI 1 & ... & 0.267$^{\text{(c)}}$ & ... & 2.06$^{\text{(c)}}$ & Cla24 & L0 &  16.8: & 0.080$\pm$0.001 & Kir24 & (17\,296$\pm$11)$\cdot$10$^{-\text{3}}$ & Man19 & 17\,000 &  & \\ 
\noalign{\smallskip}
\hline
\noalign{\smallskip}
  BD+66 34A[a] & 00321+6715 &  & 9.89 &  &  &  &  & M2.0 V & 9.57 & 0.405$\pm$0.008 & Kir24 &  &  &  &  & \multirow{3}{*}{\includegraphics[width=9.5mm]{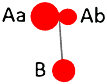}}\\
\end{longtable}
\vspace{-12mm}
\begin{justify}
\scriptsize{\textbf{\textit{Notes. }}}
\scriptsize{$^{\text{(a)}}$ References:
      Aga15: \citet{agati15};
      Ake21: \citet{akeson21};      
      Bed24: \citet{bedin24};
      Ben16: \citet{benedict16};      
      Bil06: \citet{biller06};
      Bon15: \citet{bond15};
      Bon17: \citet{bond17};
      Bon20: \citet{bond20};      
      Bra21: \citet{brandt21};
      Bur00: \citet{burgasser00};      
      Bur15: \citet{burgasser15};
      Cab09: \citet{caballero09};
      Cal24: \citet{calamari24};
      Che22: \citet{chen22};
      Cif25: \citet{cifuentes25};      
      Cla24: \citep{clark24};
      Dav14: \citet{davison14};      
      Del99a: \citet{delfosse99a};
      Del99b: \citet{delfosse99b};
      Dup17: \citet{dupuy17};      
      Dup19: \citet{dupuy19};
      Far10: \citet{farrington10};      
      Fen21: \citet{feng21};
      Fen22: \citet{feng22};      
      For99: \citet{forveille99};
      Fuh08: \citet{fuhrmann08};     
      Gel11: \citet{gelino11};     
      Gon20: \citet{gonzalezalvarez20};     
      Gon23: \citet{gonzalezpayo23};
      Gra24: \citet{gravitycollaboration24};
      Han02: \citet{han02};
      Hei96: \citet{heintz96}; 
      Hen18: \citet{henry18};     
      Hor19: \citet{horch19};
      Hor20: \citet{horch20};
      Irw96: \citet{irwin96};
      Izm19: \citet{izmailov19};
      Izm21: \citet{izmailov21};     
      Kir24: \citet{kirkpatrick24};
      Kna20: \citet{knapp20};     
      Kon02: \citet{konig02};
      Leg19: \citet{leggett19};
      Liu12: \citet{liu12};
      Mal24: \citet{mallorquin24};
      Mam13: \citet{mamajek13};
      Man19: \citet{mann19};
      Mas95: \citet{mason95};
      Mas01: \citet{mason01};
      Mas18: \citet{mason18};
      Mas21a: \citet{mason21a};
      Mas21b: \citet{mason21b};
      Mas21c: \citet{mason21c};      
      Maz01: \citet{mazeh01};      
      Mon06: \citet{montagnier06};
      Obr24: \citet{obrien24};      
      Pop13: \citet{pope13};
      Pou00: \citet{pourbaix00};
      Pri14: \citet{prieur14};
      Rom21: \citet{romanenko21};      
      Sal21: \citet{salama21};
      Seg00: \citet{segransan00};      
      Sim06: \citet{simon06};
      Spe19: \citet{sperauskas19};
      Tok15: \citet{tokovinin15};     
      Tok20: \citet{tokovinin20};
      Tok21: \citet{tokovinin21};  
      Tok23: \citet{tokovinin23};
      Tur01: \citep{turner01};
      Vig12: \citet{vigan12}; 
      Vri22: \citet{vrijmoet22};
      Win19: \citet{winters19};
      Win21: \citet{winters21};
      Woi00: \citet{woitas00};      
      Wri13: \citet{wright13};
      Xua24: \citet{xuan24};
      Zha21: \citet{zhang21}.}
\scriptsize{$^{\text{(b)}}$ Colours indicate spectral types (dark blue: A, light blue: B, yellow: G, orange: K, red: black: M, L, T, and Y, white: WD). Concentric circles indicate unresolved binaries (Table~\ref{tab:unresolved_systems_10pc}). Stripped circles indicate exoplanet host stars. Symbol sizes are approximate proportional to stellar masses, but are not to scale to the physical separations. When separation is about 1\,arcsec or less, the black connection line is not drawn.}  
\scriptsize{$^{\text{(c)}}$ $\rho$, $\theta$, and $s$ measured from second or third component of the system.} 
\scriptsize{$^{\text{(d)}}$ WDS~08589+0829 --
G 41-14A[ab]: Sum of estimated masses of A[a] (0.200$\pm$0.020\,M$_\odot$, \citealt{kirkpatrick24}) and A[b] (0.164$\pm$0.016\,M$_\odot$,  \citealt{kirkpatrick24}).}
\scriptsize{$^{\text{(e)}}$ WDS~11182+3132 --
$\xi$ UMa B[ab]: Sum of estimated massed of B[a] (0.880$\pm$0.088\,M$_\odot$, \citealt{kirkpatrick24}) and B[b] (0.140$\pm$0.090\,M$_\odot$, \citealt{fuhrmann08}).}
\scriptsize{$^{\text{(f)}}$ WDS~16555--0820 --
HD~152751~[Bab]: Sum of estimated mass of [Ba] (0.336$\pm$0.016\,M$_\odot$, \citealt{mazeh01}) and [Bb] (0.304$\pm$0.014\,M$_\odot$, \citealt{mazeh01}).}
\scriptsize{$^{\text{(g)}}$ WDS~22388--2037 --
FK~Aqr~[AB]: Sum of estimated masses of [A] (0.457$\pm$0.046\,M$_\odot$, \citealt{kirkpatrick24}) and [B] (0.457$\pm$0.046\,M$_\odot$,  \citealt{kirkpatrick24}).
FL~Aqr~[AB]: Sum of estimated mass of [A] (0.29$\pm$0.06\,M$_\odot$, \citealt{kirkpatrick24}) and minimum mass of [B] (0.061$\pm$0.007\,M$_\odot$, \citealt{davison14}).}
\end{justify}     
\end{landscape}

%   \global\pdfpageattr\expandafter{\the\pdfpageattr/Rotate 0}
%   \pagebreak[4]

%%%%% LISTA DE FIGURAS %%%%%

\normalsize
\hypersetup{linkcolor=black} % Pone color negro en las listas de figuras y tablas.

\newpage
\thispagestyle{plain}
\vspace{9cm}
\lhead[\small{\textbf{\thepage}}]{\textbf{List of figures}}
\rhead[\small{\textbf{List of figures}}]{\small{\textbf{\thepage}}}
\phantomsection
\addcontentsline{toc}{chapter}{List of Figures}
\listoffigures

%%%%% LISTA DE TABLAS %%%%%

\newpage
\thispagestyle{plain}
\vspace{9cm}
\lhead[\small{\textbf{\thepage}}]{\textbf{List of tables}}
\rhead[\small{\textbf{List of tables}}]{\small{\textbf{\thepage}}}
\phantomsection
\addcontentsline{toc}{chapter}{List of Tables}
\listoftables

%%%%%%% PAGINA FINAL %%%%%%%%

\newpage
\thispagestyle{empty}

\begin{figure}[h]
 \centering \includegraphics[width=0.9\linewidth, angle=0]{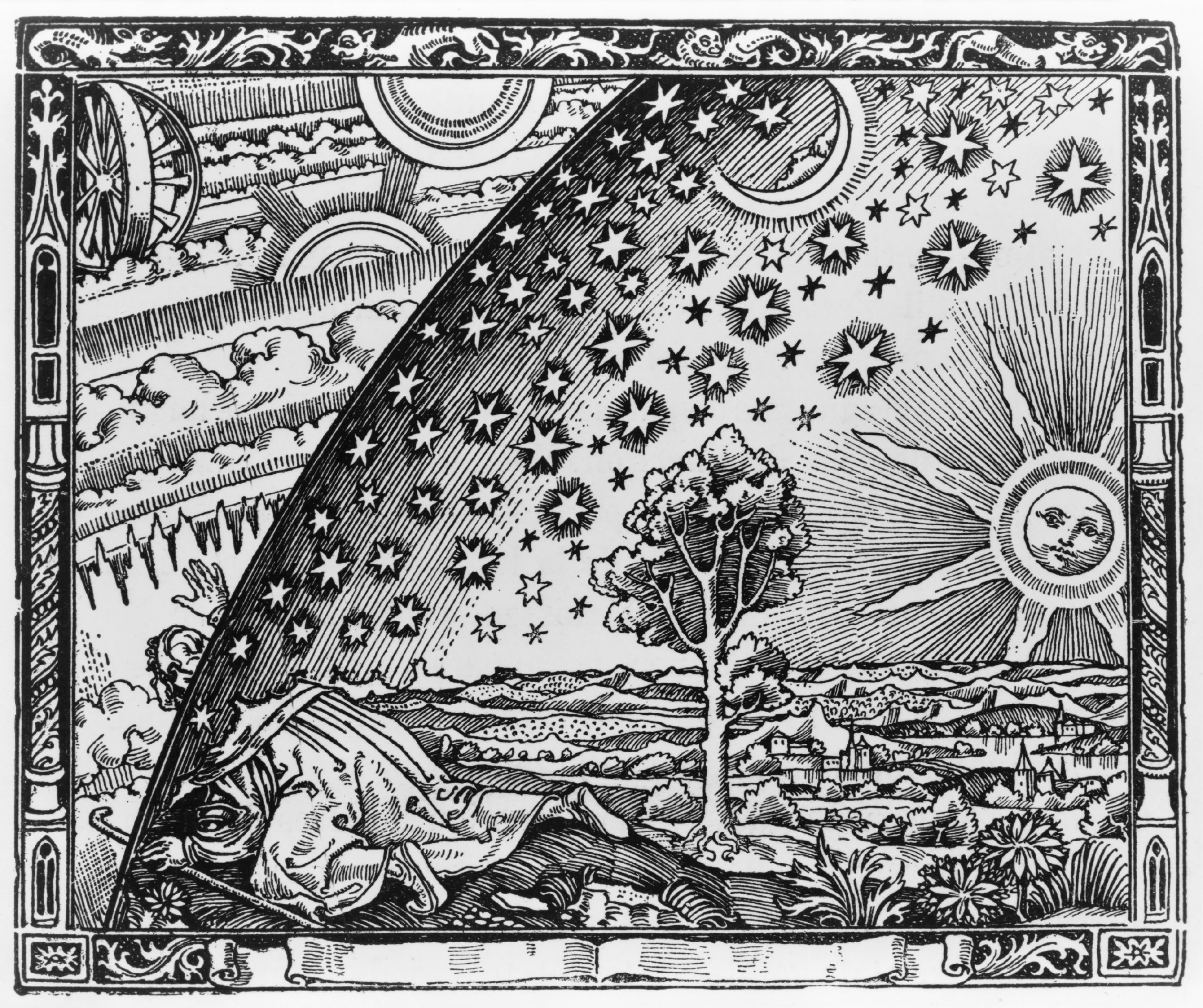}
 \label{fig:final_image}
\end{figure}
%\vspace{-7mm}
\footnotesize
    The Flammarion engraving. Unknown artist. Its first documented appearance is in the book ``L'atmosphère: météorologie populaire'', published in 1888 by the French astronomer and writer Camille Flammarion.

\vspace{20mm}
\begin{center}
 \begin{Large} % \textgoth, \textswab, \textfrak
   ``Multiplicity of stellar systems in the solar neighbourhood,\\
   wide binaries, and planet-hosting stars''\\
   Ph.D. thesis\\
 \end{Large}
\end{center}

\begin{center}
 \begin{Large} % \textgoth, \textswab, \textfrak
   \textbf{Francisco Javier González Payo}\\
   \textsc{Universidad Complutense de Madrid}\\
   July 2025\\
 \end{Large}
\end{center}

\begin{figure}[h]
 \centering \includegraphics[width=0.1\linewidth, angle=0]{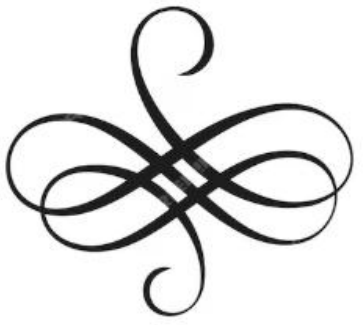}
 \label{fig:filigrana}
\end{figure}

\clearpage\hbox{}\thispagestyle{empty}\newpage %Borrar esta línea si salen número par de páginas sin contar esta.

%%%%%%%%%%%%%%%%%%%%%%%%%%%%%%%%%%%%%%%%%%%%%%%%%%%%%%%%%%%%%%%%
%%%  - Revisar como se mencionan las tablas de los apéndices en 
%%%    flujogramas.
%%%  - Dejar número par de páginas totales
%%%  - Actualizar fuentes WDS en sección 1.3.2
%%%  - Actualizar fuentes exoplanetas en sección 1.4.5.1
%%%  - Quitar comillas simples (especialmente Capítulo 6)
%%%  - Comprobar interrogaciones por si hay algo no referenciado
%%%%%%%%%%%%%%%%%%%%%%%%%%%%%%%%%%%%%%%%%%%%%%%%%%%%%%%%%%%%%%%%

\end{document}